\documentclass[a4paper,11pt]{article}
\usepackage{jheppub} 
\newcommand{\tabincell}[2]{\begin{tabular}{@{}#1@{}}#2\end{tabular}}
\usepackage{rotating}                     
\usepackage{etoolbox}
    \makeatletter
    \patchcmd{\maketitle}{\@fpheader}{}{}{}
    \makeatother
\usepackage{amsmath,amssymb,amsthm,amscd,amsfonts}
\usepackage{enumerate}
\usepackage{ytableau}
\usepackage{mathrsfs}
\usepackage{graphics,graphicx,epsfig,bbm,verbatim}
\usepackage{epstopdf,subfigure}
\usepackage{mathtools}
\usepackage{caption}
\usepackage{booktabs}
\usepackage{tabulary}
\usepackage{longtable}
\usepackage{xcolor}
\usepackage{indentfirst}
\usepackage{setspace}

\usepackage[all,cmtip,arrow]{xy}
\usepackage{pb-diagram,pb-xy}%
\input xy
\xyoption{all}

\def\localelliptic{elliptic non-compact }

\allowdisplaybreaks[1]

\numberwithin{equation}{section}

\input epsf.sty
\ytableausetup{mathmode, boxsize=0.4em}

\graphicspath{{figs/}}

\usepackage{supertabular}

\def\cO{\mathcal{O}}

\def\IC{\mathbb{C}}
\def\IE{\mathbb{E}}
\def\IF{\mathbb{F}}

\def\IN{\mathbb{N}}
\def\IP{\mathbb{P}}

\def\IR{\mathbb{R}}

\def\IT{\mathbb{T}}
\def\IZ{\mathbb{Z}}

\def\fn{{n}}

\newcommand{\mfr}{\mathfrak}

\def\q{\mathfrak{q}}

\def\mfr{\mathfrak}

\def\fg{\mathfrak{g}}
\def\fh{\mathfrak{h}}

\def\mb{\mathbb}
\def\mc{\mathcal}
\def\md{\mathbf}
\def\mf{\mathfrak}

\theoremstyle{definition}

\newcommand{\bitem}{\begin{itemize}}
\newcommand{\eitem}{\end{itemize}}
\newcommand{\be}{\begin{equation}}
\newcommand{\ee}{\end{equation}}
\newcommand{\ba}{\begin{aligned}}
\newcommand{\ea}{\end{aligned}}
\newcommand{\ben}{\begin{eqnarray}\displaystyle}
\newcommand{\een}{\end{eqnarray}}

\def\cO{\mathcal{O}}

\def\cS{\mathcal{S}}

\def\eq{{\epsilon_1}}
\def\et{{\epsilon_2}}

\newcommand{\br}{{r}}

\newcommand{\ep}{\epsilon}

\newcommand{\Qtau}{q}
\newcommand{\dualCox}{h_G^{\vee}}

\renewcommand{\d}{\partial}

\def\Det{{\rm Det}}
\def\half {{1\over 2}}

\def\vev#1{\langle #1 \rangle}

\def\({\left(}
\def\){\right)}

\newcommand{\mass}{{m}}

\def\vev#1{\left\langle #1 \right\rangle}

\def\({\left(}
\def\){\right)}
\def\[{\left[}
\def\]{\right]}

\newcommand{\wh}[1]{\widehat{#1}}
\newcommand{\wb}[1]{\overline{#1}}

\newcommand{\re}{{\rm e}}
\newcommand{\ri}{{\mathsf{i}}}
\newcommand{\rd}{{\rm d}}
\newcommand{\nn}{\nonumber \\}
\newcommand{\hg}{h^\vee_{\fg}}
\newcommand{\hG}{{h^\vee_{G}}}
\newcommand{\w}{\omega}
\newcommand{\wv}{\omega^\vee}
\newcommand{\wG}{\omega_G}
\newcommand{\wF}{\omega_F}
\newcommand{\lG}{\lambda_G}
\newcommand{\lF}{\lambda_F}
\newcommand{\lGF}{\lambda_{G,F}}
\newcommand{\lU}{\lambda_{\mf{u}(1)}}
\newcommand{\lo}{\lambda_0}
\newcommand{\mG}{m_G}
\newcommand{\mF}{m_F}
\newcommand{\mGF}{m_{G,F}}
\newcommand{\mU}{m_{\mf{u}(1)}}
\newcommand{\und}[1]{{#1}}
\newcommand{\av}{\alpha^\vee}
\newcommand{\Pv}{P^\vee}
\newcommand{\Qv}{Q^\vee}

\DeclareMathOperator{\Ker}{Ker}

\DeclareMathOperator{\Tr}{Tr}
\DeclareMathOperator{\tr}{tr}

\DeclareMathOperator{\ind}{ind}
\DeclareMathOperator{\Ind}{Ind}
\DeclareMathOperator{\imag}{Im}

\newcommand{\bs}{\begin{split}}
\newcommand{\es}{\end{split}}

\usepackage{amsmath}
\newcommand{\bdm}{\begin{math}}
\newcommand{\edm}{\end{math}}

\newcommand{\xdownarrow}[1]{%
  {\left\downarrow\vbox to #1{}\right.\kern-\nulldelimiterspace}
}

\usepackage{tikz}
\usetikzlibrary{decorations.pathmorphing,arrows,decorations.markings,calc}

\newdimen\tableauside\tableauside=1.0ex
\newdimen\tableaurule\tableaurule=0.4pt
\newdimen\tableaustep
\def\phantomhrule#1{\hbox{\vbox to0pt{\hrule height\tableaurule width#1\vss}}}
\def\phantomvrule#1{\vbox{\hbox to0pt{\vrule width\tableaurule height#1\hss}}}
\def\sqr{\vbox{%
  \phantomhrule\tableaustep
  \hbox{\phantomvrule\tableaustep\kern\tableaustep\phantomvrule\tableaustep}%
  \hbox{\vbox{\phantomhrule\tableauside}\kern-\tableaurule}}}
\def\squares#1{\hbox{\count0=#1\noindent\loop\sqr
  \advance\count0 by-1 \ifnum\count0>0\repeat}}
\def\tableau#1{\vcenter{\offinterlineskip
  \tableaustep=\tableauside\advance\tableaustep by-\tableaurule
  \kern\normallineskip\hbox
    {\kern\normallineskip\vbox
      {\gettableau#1 0 }%
     \kern\normallineskip\kern\tableaurule}%
  \kern\normallineskip\kern\tableaurule}}
\def\gettableau#1{\ifnum#1=0\let\next=\null\else
\squares{#1}\let\next=\gettableau\fi\next}

\rightline{BONN-TH-2020-05}

\title{\boldmath Elliptic Blowup Equations for 6d SCFTs. IV: Matters}

\author[a]{Jie Gu,}
\author[b]{Babak Haghighat,}
\author[c,d]{Albrecht Klemm,}
\author[e,f]{Kaiwen Sun}
\author[c,g]{and Xin Wang}
\affiliation[a]{D\'epartement de Physique Th\'eorique et Section
  de Math\'ematiques, Universit\'e de Gen\`eve, Gen\`eve, CH-1211 Switzerland}
\affiliation[b]{Yau Mathematical Sciences Center, Tsinghua University, Beijing, 100084, China}
\affiliation[c]{Bethe Center for Theoretical Physics, Universit\"at Bonn, D-53115, Bonn, Germany}
\affiliation[d]{Hausdorff
  Center for Mathematics, Endenicher Allee 62, D-53115, Bonn, Germany}
\affiliation[e]{Scuola Internazionale Superiore di Studi Avanzati (SISSA), via Bonomea 265, 34136, Trieste, Italy}
\affiliation[f]{INFN, Sezione di Trieste}
\affiliation[g]{Max Planck Institute for Mathematics, Vivatsgasse 7, D-53111 Bonn, Germany}

\emailAdd{jie.gu@unige.ch}
\emailAdd{babak@math.tsinghua.edu.cn}
\emailAdd{aklemm@th.physik.uni-bonn.de}
\emailAdd{ksun@sissa.it}
\emailAdd{wxin@mpim-bonn.mpg.de}

\abstract{Given the recent geometrical classification of 6d $(1,0)$
  SCFTs, a major question is how to compute for this large class their
  elliptic genera.  The latter encode the refined BPS spectrum of the
  SCFTs, which determines geometric invariants of the associated
  \localelliptic Calabi-Yau threefolds.  In this paper we establish
  for \emph{all} 6d $(1,0)$ SCFTs in the atomic classification blowup
  equations that fix these elliptic genera to large extent. The latter
  fall into two types: the unity-- and the vanishing blowup
  equations. For almost all rank one theories, we find unity blowup
  equations which determine the elliptic genera completely. We develop
  several techniques to compute elliptic genera and BPS invariants
  from the blowup equations, including a recursion formula with
  respect to the number of strings, a Weyl orbit expansion, a refined
  BPS expansion and an $\epsilon_1,\epsilon_2$ expansion. For
  higher-rank theories, we propose a gluing rule to obtain all their
  blowup equations based on those of rank one theories. For example,
  we explicitly give the elliptic blowup equations for the three
  higher-rank non-Higgsable clusters, ADE chain of $-2$ curves and
  conformal matter theories. We also give the toric construction for
  many \localelliptic Calabi-Yau threefolds which engineer 6d $(1,0)$
  SCFTs with various matter representations.}

\begin{document}
\maketitle
\section{Introduction}
\label{sec:introduction}

In the absence of a Lagrangian description, six dimensional theories
with $(2,0)$ and $(1,0)$ superconformal symmetries are generally
difficult to study.  The data of BPS spectra, especially the one
associated to non-critical strings wrapping the torus, thus becomes
crucial for understanding the properties of this important class of
theories and their compactification to lower dimensions.  Among the
techniques to obtain the BPS spectrum are the calculation of the
elliptic genera in a dual 2d quiver gauge theory
\cite{Haghighat:2014vxa}, the refined holomorphic anomaly equation and
the closely related modular bootstrap
approach~\cite{Haghighat:2014pva,Gu:2017ccq,DelZotto:2017mee,DelZotto:2018tcj},
the application of the topological vertex~\cite{Haghighat:2013gba} and
the elliptic blowup equations \cite{Gu:2018gmy,Gu:2019dan,Gu:2019pqj}.
The purpose of this paper is to generalise the last approach to
generic \localelliptic Calabi-Yau spaces, which geometrically engineer
six dimensional superconformal field theories with geometrically
realisable gauge symmetries and matters in generic flavour
representations.  Geometrically these theories can be viewed as a
generalisation of the theories associated to non-higgsable clusters
discussed in~\cite{Gu:2018gmy,Gu:2019dan} to include matter by
considering additional singularities of the elliptic fibration over
the non-compact directions in the base.  In this paper, we write down
elliptic blowup equations for all the rank one 6d superconformal
theories with one tensor multiplet.  These theories are geometrically
realised as non-compact elliptic fibrations with an isolated curve of
self intersection $-n$ in the base.  We demonstrate that for almost
all rank one theories, with the exception of 12 theories (see the
beginning of Section~\ref{sc:sol} for a list of these theories), the
BPS spectrum or equivalently the elliptic genera can be computed from
the elliptic blowup equations.  Our results go beyond all previous
results in the literature, see for instance
\cite{Haghighat:2013tka,Haghighat:2014vxa,Kim:2015fxa,Yun:2016yzw,Kim:2018gak,Kim:2018gjo,DelZotto:2018tcj,Kim:2019dqn},
in that firstly, in many cases our methods can be pushed to compute
BPS invariants with arbitrary base degrees, or equivalently elliptic
genera with arbitrary numbers of strings, and secondly, we can produce
results with both gauge and flavor fugacities turned on.  In contrast,
the existing techniques either produce complete solutions only for a
limited set of theories (localisation on quiver construction), or are
limited to one-string elliptic genera (analysis of worldsheet theory),
or cannot turn on all fugacities (modular bootstrap).  We review all
known computational methods and previous results on elliptic genera,
if they exist, in detail in Section~\ref{sec:known-comp-meth}.  In the
case of higher rank theories, we formulate the generic \emph{gluing
  rules}, by which we can build elliptic blowup equations for them
using as ingredients elliptic blowup equations of rank one theories.
However for almost all higher rank theories, elliptic blowup equations
are not strong enough to solve the corresponding elliptic genera
uniquely. Nevertheless these novel equations do give interesting
structural constraints on the elliptic genera in particular on their
modular properties.  We leave it as an open problem how to supplement
in particular the theories with only vanishing elliptic blowup equations 
by further data so that one can determine their  elliptic genera and the BPS degeneracies completely.

The original blowup equations were derived in order to confirm the
gauge theory instanton corrections to the so called Seiberg-Witten
prepotential of $\mathcal{N}=2$ four dimensional supersymmetric gauge
theory.  In particular Nakajima and Yoshioka confirmed the Nekrasov
partition function, which depends on the $\eq,\et$ background
parameters, encodes the prepotential in the $\eq,\et \rightarrow 0$
limit~\cite{Nakajima:2003pg}.  The same authors generalised them to
K-theoretic invariants of five dimensional gauge theories and obtained
the K-theoretic blowup equations~\cite{Nakajima:2005fg}, which count
K-theoretic Donaldson invariants as discussed in a paper with
G\"ottsche~\cite{Gottsche:2006bm}. The K-theoretic invariants are
closely related to the refined topological string or refined BPS
invariants on local Calabi-Yau spaces, which engineer these 5d gauge
theories. In~\cite{Huang:2017mis} it was shown that the five
dimensional blowup equations can be used to calculate these refined
BPS invariants $N_{j_l,j_r}^\beta$ on arbitrary toric Calabi-Yau
spaces.  Mathematically the $N_{j_l,j_r}^\beta$ are defined as refined
stable pair invariants~\cite{Choi:2012jz}. Unrefined stable pair
invariants were introduced in~\cite{MR2545686} and were explicitly
calculated using localisation on toric Calabi-Yau
manifolds~\cite{MR2497313}. It is in this context that refined
invariants could also be calculated using localisation
in~\cite{Choi:2012jz}. However their definition as invariants is
suggested by physical arguments, only when there is a global
$\mf{u}(1)_R$ symmetry in five dimensions, which geometrically is
induced by a $\mf{u}(1)$ isometry of the Calabi-Yau space.
Mathematically one must have an orientation on the moduli space of
gauge theory --- or here specifically stable pair
invariants~\cite{Nekrasov:2014nea}.  Both conditions are naturally
realised on non-compact Calabi-Yau spaces.

In \cite{Huang:2017mis} a generalisation of the K-theoretic blowup
equations of Nakajima and Yoshioka was proposed that calculates the
refined BPS invariants also for those non-compact toric Calabi-Yau
spaces that do not by themselves define gauge theories like local
$\mathbb{P}^2$.  Later \cite{Gu:2017ccq,Huang:2017mis}, based on
\cite{Sun2016,Grassi:2016nnt}, generalised them to elliptic blowup
equations for \localelliptic Calabi-Yau threefolds associated to 6d
SCFTs.  In particular, consistency checks of blowup equations with the
E-string BPS spectrum, which corresponds to refined BPS invariants on
the \localelliptic half K3, which is not toric, were made
in~\cite{Gu:2017ccq,Huang:2017mis}.  In~\cite{Gu:2018gmy} the rank one
non-higgsable clusters with $A$ or $D$ type gauge groups were solved
using the elliptic blowup equations, while this was extended
in~\cite{Gu:2019dan} to include all rank one non-higgsable clusters
except for the $E_7$ theory with a half hypermultiplet.  In
\cite{Gu:2019pqj} the blowup equations for E-- and M-- string theories
and chains of these were discussed.  The current paper completes these
results by generalising elliptic blowup equations to all 6d
superconformal theories covered in the atomic classification
\cite{Heckman:2015bfa,Heckman:2013pva}.

The discussion of elliptic blowup equations falls into three
steps. First the formulation of the elliptic blowup equations for each
model.  For rank one theories, this is neatly summarised in
Section~\ref{sec:ellipt-blow-equat}.  They also serve as basic
building blocks.  Together with the ``gluing rules'' proposed in
Section~\ref{sec:glue}, they can be used to construct elliptic blowup
equations of any higher rank 6d SCFT, whose forms are given in
Section~\ref{sc:arebe}.  It is important to classify elliptic blowup
equations into the \emph{unity} type and the \emph{vanishing} type,
according to whether integral or fractional gauge flux is summed over,
or equivalently whether the right hand side of the equation vanishes
identically.  Almost all rank one theories have unity blowup
equations, except for 12 theories, which have only vanishing blowup
equations. On the contrary, the majority of higher rank theories have
no unity equation.  This fact is important for the solvability of
these theories.

Second, the proof that the elliptic genera which satisfy elliptic
blowup equations are forms with modular transformation in the elliptic
parameter, the K\"ahler parameter of the fibre class $\tau$, as well
as with Jacobi-form transformations with a prescribed \emph{index of
  the elliptic parameters} given by the two $\epsilon_\pm$
deformations as well as by the masses of Cartan gauge bosons $m_{G_i}$
and flavour masses $m_{F_i}$.  This is discussed in Section
\ref{sc:mod} for the rank one cases and in Section
\ref{Ssec:ModularityII} for the higher rank cases as well as commented
in more detail in the examples.  This \emph{index} can be inferred
either from the anomaly polynomial of the 6d chiral SCFT, as reviewed
in Section~\ref{sec:gauge-anomalies}, or from the geometric
auto-equivalences acting on the central charges of the derived
category of quasi coherent sheaves of the Calabi-Yau
space~\cite{HKKprep,Katz:2016SM,Schimannek:2019ijf,Cota:2019cjx}.
Both data are related ultimately to classical intersections on the
Calabi-Yau space.  It has been also related to the Casimir energy of
the 2d $(0, 2)$ supersymmetric quiver theory that depends
quadratically on the flavour
fugacities~\cite{Bobev:2015kza,DelZotto:2016pvm,DelZotto:2017mee}.  On
the physics side we explain the interesting structural changes of the
blowup equations along the Higgsing trees in Section~\ref{sc:Hig} and
give a direct heuristic physical interpretation and partial derivation
of them in Section~\ref{sec:phys-interpr}.
We also discuss the relation between the elliptic blowup equations of
rank one 6d SCFTs and the K-theoretic five dimensional blowup
equations in Section~\ref{sec:5dblowup}.  We point out that the 5d
limit $q=\re^{2 \pi \ri \tau}\rightarrow 0$ is quite non-trivial. In
particular, as explained in Section \ref{sec:5dblowup}, not all
K-theoretic blowup equations in 5d can be obtained from the 6d
elliptic blowup equations.

The last question is the actual solvability of BPS invariants or
elliptic genera from the elliptic blowup equations.  This depends
critically on whether the theory has unity blowup equations or only
vanishing blowup equations.  We call the latter case class $\bf C$.
In the former case, we distinguish further by whether the theory has
enough unity blowup equations so that a recursion formula in the
spirit of \cite{Keller:2012da} can be written down, or not. We call
these two cases classes $\bf A$ and $\bf B$ respectively.  The case of
class $\bf A$ is clear cut, and we can write down elliptic genera with
arbitrary number of strings using the recursion formula presented in
Section~\ref{sc:rec}.  In class $\bf B$, using various methods
presented in Sections~\ref{vexpansion}, \ref{sc:refBPS}, \ref{sc:e12},
in practice elliptic genera or BPS invariants can still be solved
order by order from blowup equations.
In these two cases, compared with all the other techniques to solve
the refined BPS invariants mentioned above, the method of the elliptic
blowup equations is generically the most efficient one to obtain
results for arbitrary spins and arbitrary curve degrees.  In class
$\bf C$, the theory cannot be solved completely.  Nevertheless we can
still get partial results and interesting structural insight from the
vanishing blowup equations.
%
%
Particularly interesting are the leading degree $\theta$-functions
identities that follow from the vanishing blowup equations.  The
prototypical example is given for the
$n=1,G=\mf{su}(3), F=\mf{su}(12)$ theory in (\ref{thetaindentI}).
Many more non-trivial ones are scattered in Sections \ref{sc:ex},
\ref{sec:3nhc} and \ref{sec:arbitraryrank}, and summarised in Appendix
\ref{app:list}.  Also it is quite notable that we can 
derive (\ref{Ginst2}) general \emph{exact} formulas
of the $v$ expansion of 5d one-instanton Nekrasov partition functions with matters. This is a generalization for the pure gauge case -- the Hilbert
series (\ref{Ginst1}) of the reduced moduli spaces of one $G$ instanton
\cite{Benvenuti:2010pq,Keller:2011ek}.

To save readers' precious time, we provide a quick guide for the paper
here. Readers merely interested in the elliptic genera of rank-one 6d
$(1,0)$ SCFTs can directly go to Appendix \ref{app:list}, check the
list and find the links to the results for individual
theories. Readers interested in the structure of elliptic blowup
equations can first go to the main equation (\ref{eq:ebeq}) and Tables
\ref{tb:ubeq-1}, \ref{tb:ubeq-2}, \ref{tb:vbeq-1}, \ref{tb:vbeq-2}, \ref{tb:vbeq-3},
which contain the information for all elliptic blowup equations --
both unity and vanishing -- for all rank-one theories. Then, if one is
also interested in higher-rank theories, one can go to Section
\ref{sec:3nhc} for the three familiar higher-rank non-Higgsable
clusters and Section \ref{sec:arbitraryrank} for the gluing rules and
general form of elliptic blowup equations for all $(1,0)$ SCFTs in
atomic classification, especially the explicit examples in Section
\ref{sec:exhigh} for ADE chains of $-2$ curves, conformal matter
theories and blown-up of some $-n$ curves. Readers interested in how
to solve blowup equations can go to Section \ref{sc:sol} where four
different methods are provided, each with its own merits and range of
application. Then in Section \ref{sc:ex}, details are given on blowup
equations and elliptic genera of some of the most interesting rank one
theories. Readers interested in toric constructions for elliptic
noncompact Calabi-Yau threefolds, exact $v$ expansion formulas for 5d
one-instanton Nekrasov partition functions, vanishing theta identities are suggested
to check Appendix \ref{app:list} for a list of results.

\section{6d $(1,0)$ SCFTs and Calabi-Yau geometries}
\label{sec:6d-1-0}

This section provides the necessary background for the paper.  We
first give a quick review of 6d $(1,0)$ SCFTs, including in particular
a discussion of anomaly cancellation, which will be useful for the
formulation of elliptic blowup equations, and the atomic
classification.  We then proceed to describe the objects of interest,
the elliptic genera of 6d SCFT, and summarise the current status of
the computational results.  Then we describe the semi-classical pieces
of the free energy, which will be crucial for deriving the elliptic
blowup equations.  We finally discuss the toric construction of the
non-compact Calabi-Yau geometries, whose classical data can be used to
reaffirm the semi-classical free energy. Together with mirror symmetry
one can compute the quantum corrected prepotential, which provides
further useful initial data in order to solve all BPS invariants by
the elliptic blowup equations.

\subsection{Review of 6d $(1,0)$ SCFTs}
\label{sec:review-6d-1}

A (1,0) 6d SCFT has superconformal algebra $\mf{osp}(6,2|1)$,
which has a bosonic sub-algebra
\begin{equation}
  \mf{osp}(6,2|1) \supset \mf{so}(6,2)\times \mf{sp}(1),
\end{equation}
where $\mf{so}(6,2)$ is the conformal algebra, and
$\mf{sp}(1)\cong \mf{su}(2)$ the R-symmetry.  Massless states are
labeled by representations of the sub-algebra
$\mf{so}(4)\cong\mf{su}(2)\times\mf{su}(2) \subset \mf{so}(6,2)$.
They can be grouped into the following three types of (1,0)
supermultiplets:
\begin{itemize}
\item Tensor multiplets: Each has an anti-self-dual tensor field $B_i$
  of spin $(1,0)$, a scalar field $\phi_i$ of spin $(0,0)$, and two
  fermions of spin $(\frac{1}{2},0)$.
\item Vector multiplets: Each has a vector field of spin
  $(\frac{1}{2},\frac{1}{2})$, and two fermions of spin
  $(0,\frac{1}{2})$.
\item Hypermultiplets: Each has four scalars of spin $(0,0)$, and two
  fermions of spin $(\frac{1}{2},0)$.
\end{itemize}
In addition, there are massless BPS strings sourced by the tensor
fields.  The charges of these strings $n=(n_i)$ can be computed by
integrating the flux $H_i$ of tensor fields
\begin{equation}
  n_i = \int_{M_4^{\perp}} \rd H_i \in \IZ_{\geq 0}
\end{equation}
over  the four dimensional hypersurface  $M_4^\perp$ transverse to
the worldsheet of the BPS string.  The lattice of string charges
$\Lambda$ is equipped with a symmetric pairing
\begin{equation}
  \vev{n,n'} = \sum_{i,j} A_{ij}n_i n'_j
\end{equation}
analogous to the Dirac pairing in 4d electromagnetism.  Dirac
quantisation condition requires that $A_{ij}$ is integral.

6d SCFT can be constructed by F-theory compactification on an elliptic
Calabi-Yau threefold whose base is an orbifold singularity of the type
\begin{equation}
  \mc{B}_{\text{sing}} = \IC^2/\Gamma_{\mf{u}(2)}\ ,
\end{equation}
where $\Gamma_{\mf{u}(2)}$ is some discrete subgroup of $\mf{u}(2)$
\cite{Heckman:2013pva}.  In order to have a better understanding of
the 6d SCFT, it is beneficial to move to the tensor branch of  the moduli
space. This corresponds in geometry to resolving the base singularity by
successive blow-ups.  In a generic point of the full tensor branch,
the base surface is smooth and the compact curves inside are rational
curves which intersect with each other in such a way that the
intersection matrix
\begin{equation}
\label{eq:intersectionbase}
  A_{ij} = \Sigma_i\cap \Sigma_j
\end{equation}
is negative definite. This implies that all of them can be blown down by the Theorem
of Grauert~\cite{MR137127}.  In addition, every elliptic fiber over a base curve should be
of Kodaira-Tate type.  All these geometric conditions allow a systematic
classification of 6d SCFTs \cite{Heckman:2015bfa,Heckman:2013pva} as
we will later review in Section~\ref{sec:classification}\footnote{A
  handful of 6d SCFTs with the so-called ``frozen singularities'' do
  not have valid geometric construction \cite{Witten:1997bs,
    Tachikawa:2015wka, Bhardwaj:2015oru, Apruzzi:2017iqe,
    Bhardwaj:2018jgp, Bhardwaj:2019hhd}.  We will treat these special
  6d SCFTs with our methods in a later work.}.

After resolution of the base singularity, massless fields and tensionless
strings have clear geometric origin.  Tensor multiplets come from
dimensional reduction of type IIB fields on compact base curves.  The
number of these base curves gives the number of tensor mulipltes,
which is also called the rank of a 6d SCFT, while the volumes of these
curves are identified with the imaginary parts of the tensor moduli.
Vector multiples come from string modes on seven branes wrapping the
discriminant loci $\Delta$ of elliptic fibration.  The irreducible
components in $\Delta$ can be both compact or non-compact.
Correspondingly the associated vector fields are either dynamic or
fixed as background fields, and they induce non-Abelian
gauge and flavor symmetries respectively.  We split
$\Delta = \Delta_c\cup \Delta_n$, where $\Delta_{c,n}$ are the unions
of compact and non-compact components respectively.  There could also
be Abelian flavor symmetries, which are not localised but are rather
associated to additional sections of the elliptic fibration
\cite{Lee:2018ihr,Apruzzi:2020eqi}.  Furthermore, charged
hypermultiplets are localised at intersection loci of two base curves,
at least one of which is a compact curve in $\Delta$.  They come from
the zero modes of strings stretched between the seven branes wrapping
the two base curves.  Finally, D3 branes of type IIB can wrap compact
base curves and give rise to BPS strings.  It is clear that the
pairing of strings should be identified with intersection matrix of
compact base curves (\ref{eq:intersectionbase}).  Furthermore
the tension of strings is proportional to the volumes of base curves, i.e.~the tensor moduli.
The BPS strings thus become tensionless precisely at the origin of the
tensor branch where all compact base curves are blown down.

\subsubsection{Gauge anomalies}
\label{sec:gauge-anomalies}

From the field theory point of view, a 6d SCFT in its tensor branch is
simply a weakly coupled 6d gauge theory, whose vector multiplets are
coupled to tensor multiplets by the following kinetic terms in the
Lagrangian
\begin{equation}
  \mc{L} \supset \phi_i \tr F_i^{\mu\nu} F_{i,\mu\nu},
\end{equation}
where the tensor moduli $\phi_i$ serve as inverse gauge couplings.  It
is clear that this 6d field theory must be anomaly free.  Symmetry
anomalies of a 6d field theory are encoded in a closed and gauge
invariant eight form $I_{\text{tot}}$.  We distinguish two different
scenarios.  With the absence of  a current from the BPS strings,
$I_{\text{tot}}$ can be written as
\begin{equation}
  I_{\text{tot}} = I_{\text{1-loop}} + I_{\text{GS}}
\end{equation}
where $I_{\text{1-loop}}$ are 1-loop contributions from massless
fields, while $I_{\text{GS}}$ are contributions from Green-Schwarz
counter-terms \cite{Sadov:1996zm,Green:1984bx,Sagnotti:1992qw}.  The
latter can be written as
\begin{equation}
  I_{\text{GS}} = \frac{1}{2}\sum_{i,j} (A^{-1})_{ij} X_i X_j.
  \label{eq:IGS}
\end{equation}
The four-forms $X_i$ read
\cite{Sadov:1996zm,Green:1984bx,Sagnotti:1992qw}
\begin{equation}
  X_i = \frac{1}{4} a_i p_1(M_6) + b_i c_2(I) + \sum_{k'}b_{i,k'} c_2(\mf{g}_{k'}) +
  \frac{1}{2} \sum_{m,n}b_{i,mn} c_1(\mf{u}(1)_m) c_1(\mf{u}(1)_n).
\end{equation}
Here $p_1(M_6)$ is the first Pontryagin class of the tangent bundle of
the six dimensional spacetime, $c_2(I)$ and $ c_2(\mf{g}_{k'})$ are
the second Chern classes of the bundle of the $\mf{su}(2)$ R-symmetry,
and the bundles of non-Abelian gauge or flavor symmetries
respectively.  We also include in the last term the contributions from
the first Chern classes of the flavor $\mf{u}(1)$ bundles.  The
anomaly coefficients actually have a beautiful geometric meanings
\cite{Grassi:2011hq,Grassi:2000we,Sadov:1996zm} and determine the
index of the elliptic genus~\cite{Gu:2017ccq,Cota:2019cjx}.  The
four-form $X_i$ is related to the compact base curve $\Sigma_i$, $b_i$
is conjectured to be minus the dual Coxeter number $h^\vee_{\fg_i}$ of
the symmetry algebra supported on $\Sigma_i$
\cite{Ohmori:2014kda}\footnote{If $\Sigma_i$ is not part of the
  discriminant and there is no non-trivial symmetry algebra,
  $h^\vee_{\fg_i}$ is replaced by 1.}, and we can interpret the
coefficients $a_i,b_{i,k'}$ as
\begin{equation}
  a_i = \Sigma_i\cap(-K),\quad b_{i,k'} = \Sigma_i\cap \Sigma_{k'}=:A_{ik'},\quad
  k'\in \Delta = \Delta_{c}\cup \Delta_{n}.
\end{equation}
Here $-K$ is the anti-canonical divisor of the base.  Since the base
surface is non-compact, we would like to interpret $-K$ as the
Poincar\'e dual of the cohomology class, which is locally meaningful.
In practice, the neighborhood in the base of the curve $\Sigma_i$ with
self-intersection number $-\fn:=A_{ii}$ can be locally replaced by the
Hirzebruch surface $\IF_\fn$, so that we find readily
$a_i = 2-\fn = 2+A_{ii}$.  $\Sigma_{k'}$ is an irreducible component
of the discriminant $\Delta = \Delta_{c}\cup \Delta_{n}$ which
supports the symmetry algebra $\mf{g}_{k'}$.  In the same formula we
have extended the definition of $A_{ik'}$ to accommodate possible
intersection with non-compact base curves.  Finally $b_{i,mn}$ can
also be interpreted as intersection numbers of the vertical divisor
pulled back from $\Sigma_i$ with the additional sections that
induce the Abelian symmetries~\cite{Cota:2019cjx}.

Every term concerning gauge symmetry in $I_{\text{tot}}$ must be
canceled.  This includes not only pure gauge anomaly, but also mixed
gauge-flavor and mixed gauge-gravity anomalies in order to preserve
superconformal invariance
\cite{Cordova:2018cvg,Cordova:2019wns}.  Let us
define the fiducial trace
$\tr = \frac{1}{2\ind_{\square}}\tr_{\square}$, where $\square$ is the
defining representation, so that
\begin{equation}
  \tr F^2 = \frac{1}{2\hg} \tr_{\mf{adj}} F^2,
\end{equation}
and the following Lie algebraic constants\footnote{The second
  definition implies
  \begin{equation}\label{eq:cnt-ex}
    \tr_R F_1F_2F_3F_4 = B_R \tr F_1F_2F_3F_4
    +\frac{1}{3} C_R \left(\tr F_1 F_2\tr F_3 F_4 + \tr F_1 F_3\tr F_2
      F_4 +\tr F_1 F_4\tr F_2 F_4 \right).
  \end{equation}}
\begin{equation}\label{eq:Lcnt}
  \tr_R F^2 = 2\ind_R \tr F^2,\quad
  \tr_R F^4 = x_R \tr F^4 + y_R (\tr F^2)^2,\quad
  \tr_R F^3 = z_R \tr F^3.
\end{equation}
These constants for common representations of simple Lie algebras in
our convention can be found
in~\cite{Grassi:2011hq,Grassi:2000we,DelZotto:2018tcj}.  We then have
the following anomaly cancellation conditions \cite{Sadov:1996zm,Green:1984bx,Sagnotti:1992qw}:
\begin{itemize}
\item Mixed gauge-gravity anomaly cancellation:
 \begin{equation}
   \ind_{\mf{adj}_i} - \sum_{R_i} n_{R_i} \ind_{R_i} = -3(A_{ii}+2).
   \label{eq:anm-A}
 \end{equation}
\item Pure gauge anomaly cancellation:
  \begin{align}
    x_{\mf{adj}_i}- \sum_{R_i} n_{R_i} x_{R_i}
    &= 0,
      \label{eq:anm-B}\\
    y_{\mf{adj}_i} - \sum_{R_i} n_{R_i} y_{R_i}
    &= -3 A_{ii},
      \label{eq:anm-C}
  \end{align}
\item Mixed gauge-gauge anomaly cancellation\footnote{Implicit in this
    condition is that two irreducible components of $\Delta_c$ can at
    most intersect once transversely, which is required for a 6d SCFT,
    but not generally true for a 6d field theory.}:
  \begin{equation}
    \sum_{R_i,R_j} n_{R_i,R_j} \ind_{R_i}\ind_{R_j} = \frac{1}{4}.
    \label{eq:anm-1}
  \end{equation}
\item Mixed gauge-flavor anomaly cancellation:
  \begin{align}
    \sum_{R_i,R_{\ell'}} n_{R_i,R_{\ell'}} \ind_{R_i}\ind_{R_{\ell'}}
    &=\frac{1}{4} b_{i,\ell'}\, ,
      \label{eq:anm-kF}\\
    \sum_{R_i,q_m,q_n} n_{R_i,q_m,q_n} q_m q_n \ind_{R_i}
    &= \frac{1}{2} b_{i,mn}\, ,
      \label{eq:anm-k1}\\
    \sum_{R_i,q_m} n_{R_i,q_m} q_m z_{R_i}
    &= 0.
      \label{eq:anm-E}
  \end{align}
\end{itemize}
Here $i,j$ label gauge symmetries, and $\ell'$ a non-Abelian flavor
symmetry.  $n_{R_i}$, $n_{R_i,R_j}$, $n_{R_i,R_{\ell'}}$,
$n_{R_i,q_m}$, $n_{R_i,q_m,q_n}$ are the numbers of charged
hypermultiplets respectively transforming in symmetry representations
with $\mf{u}(1)$ charges $q_m,q_n$.

On the other hand, if the current of BPS strings is present, it
induces additional contribution to $I_{\text{tot}}$, which must be
canceled by the anomaly on the world-sheet theory of BPS strings
through the anomaly inflow mechanism
\cite{Shimizu:2016lbw,Kim:2016foj} (see \cite{DelZotto:2018tcj} for a
nice summary).  This determines the 't Hooft anomaly four-form $I_4$
on the worldsheet theory wrapping the base curve
$S = \sum_i d_i \Sigma_i$ as\footnote{We use the normalisation that
  $c_2(\mf{g}) = \frac{1}{4}\tr F^2$ and
  $c_1(\mf{u}(1)) = \tr F_{\mf{u}(1)}$.}
\begin{align}
&\phantom{--}  I_4 =
  -\frac{1}{2}\sum_{i,j}A_{ij} d_i d_j (c_2(L) - c_2(R))\nonumber\\[-1mm]
  &+\sum_id_i\Big(h^\vee_{\mf{g}_i}c_2(I) -
    \frac{2+A_{ii}}{4}(p_1(T_2)-2c_2(L)-2c_2(R))-\frac{1}{4}b_{i,k'}\tr
    F_{k'}^2- \frac{1}{2}b_{i,mn}\tr F_{\mf{u}(1)_m}F_{\mf{u}(1)_n}\Big).
    \label{eq:I4}
\end{align}
Here $c_2(L), c_2(R)$ refer to the second Chern classes of the bundles
associated to the global $\mf{su}(2)_L,\mf{su}(2)_R$ symmetry of $\IR^4$
perpendicular to the string worldsheet $M_2$ in 6d, and we have used
the identity
\begin{equation}
  p_1(M_6) = p_1(M_2) - 2c_2(L) - 2c_2(R).
\end{equation}
$F_{k'}$ are the field strength of non-Abelian symmetries and we sum
over both gauge and flavor symmetries, while
$F_{\mf{u}(1)_m}, F_{\mf{u}(1)_n}$ are the field strength of Abelian
flavor symmetries.  In the case of flavor symmetries, the
coefficients $b_{i,k'}$ and $b_{i,mn}$ are interpreted as the
levels\footnote{The level of $\mf{u}(1)$ is interpreted as the radius
  of the compact boson that realises the current algebra squared
  \cite{DelZotto:2018tcj}.} of the associated current algebras
\cite{DelZotto:2018tcj}, and are sometimes denoted as $k_F$.

\subsubsection{Classification}
\label{sec:classification}

The geometric constraints on the elliptic Calabi-Yau threefold
associated to 6d SCFTs in the tensor branch discussed in the beginning
of this section, as well as the anomaly cancellation conditions
discussed previously, allow for a possible classification of 6d SCFTs.
This program was systematically carried out in
\cite{Heckman:2015bfa,Heckman:2013pva}\footnote{As mentioned before,
  this program misses a handful of 6d SCFTs with ``frozen
  singularities'', which will not be discussed in this paper.}  (see
also \cite{Bhardwaj:2015xxa,Bhardwaj:2019hhd}), and we quickly review
the salient points here.

The classification scheme is divided into two steps.  In the first
step, all possible configurations of compact base curves are
classified.  There are three types of basic configurations called
``atoms''
\begin{itemize}
\item A single $-1$ curve, i.e. a single rational curve (a
  $\mathbb{P}^1$) with self-intersection $-1$.
\item Configuration of $-2$ curves intersecting according to
  appropriate $ADE$ Dynkin diagrams\footnote{This class includes a
    single $-2$ curve.}.
\item Non-Higgsible clusters \cite{Morrison:2012np}, which include: a
  single $-\fn$ curve, i.e. a rational curve with self-intersection
  $-n$ with $n=3,\ldots, 8, 12$~\footnote{For a single $\IP^1$ with
    self-intersection $-9,-10,-11$, the elliptic fiber is not of
    Kodaira-Tate type, and additional blow-ups are required.}, and
  three higher rank cases.
\end{itemize}
The $-1$ curve in the first category comes equipped with an $E_8$
flavor symmetry.  The chain of $-2$ curves of type $A$ in the second
category always has an overall $\mf{u}(1)$ flavor
symmetry\footnote{This overall flavor symmetry is enhanced to
  $\mf{su}(2)$ when the fibers over the $-2$ curves are not very
  singular.}.  The non-Higgsible clusters in the last category
distinguish themselves in that elliptic fibers over them have minimal
non-trivial singularity (hence the name non-Higgsible), and they are
tabulated in Tab.~\ref{tb:NHCs}.  A larger configuration of base
curves $\{\Sigma_i\}$ is then built by gluing the last two categories
of ``atomic'' configurations using $-1$ curves subject to certain
constraints,\footnote{The others are i) three curves cannot intersect
  in a point, ii) two curves cannot intersect tangentially, iii)
  intersection graphs contain no loops, iv) $-1$ curves can intersect at
  most two other curves, v) two $-1$ curves $\Sigma ,\Sigma'$ have
  $\Sigma \cdot \Sigma'=0$.}  the most important of which is the
gluing gluing condition that the minimal algebras $\mf{g}_L,\mf{g}_R$
carried by two curves glued by a $-1$ curve must satisfy
$\mf{g}_L\times\mf{g}_R \subset E_8$.

\begin{table}
  \centering%
  \resizebox{\linewidth}{!}{
  \begin{tabular}{c*{10}{>{$}c<{$}}}
    \toprule
   -n  curves: n
    & 3 & 4 & 5
    & 6 & 7 & 8 & 12
    & 3,2
    & 2,3,2
    & 3,2,2\\
    algebra
    & \mf{su}(3) & \mf{so}(8) & F_4
    & E_6 & E_7 & E_7 & E_8
    & G_2\oplus\mf{su}(2)
    & \mf{su}(2)\oplus\mf{so}(7)\oplus\mf{su}(2)
    & G_2\oplus\mf{su}(2)\oplus\emptyset\\
    hypers
    & - & - & -
    & - & \frac{1}{2}\md{56} & - & -
    & \frac{1}{2}(\md7+\md1,\md2)
    & \frac{1}{2}(\md2,\md8,\md1)\oplus\frac{1}{2}(\md1,\md8,\md2)
    & \frac{1}{2}(\md7+\md1,\md2,\md1)\\
    \bottomrule
  \end{tabular}
}
  \caption{All possible non-Higgsible clusters with minus the
    self-intersection numbers $\fn$ of curves, the symmetry algebras
    of the minimal singularities of elliptic fibers, and possible
    charged hypers.}
  \label{tb:NHCs}
\end{table}

All such configurations can be classified, and they generally fit into
the pattern of a generalised quiver \cite{Heckman:2015bfa}.  A
``node'' in such a quiver is a $- \fn$ curve with $\fn=4,6,7,8,9$
which supports a minimal symmetry algebra of
$D$- or $E$-type.  A ``link'' is an appropriate  configuration  of
curves which do not involve any nodes.  All possible links are listed
in \cite{Heckman:2015bfa}.  The simplest links are called the minimal
conformal matters, some examples of which are listed below
\begin{align}
  & [\mf{so}(8)]\; 1 \;[\mf{so}(8)] \\
  & [E_6]\; 1,3,1 \;[E_6] \\
  & [E_7]\; 1,2,3,2,1 \;[E_7] \\
  & [E_8]\; 1,2,2,3,1,5,1,3,2,2,1 \;[E_8]
\end{align}
Here the symmetry algebras wrapped in square brackets are flavor
symmetries, and they are also the symmetry algebras carried by the
nodes that can be connected to the links, while the chains of integers $\fn$
in the middle represent intersecting $-n$-curves.  These configurations are so named because they come from
resolving the singularity at the intersection of two seven branes that
carry $D$- or $E$-type symmetry algebras, similar to conventional
matters which can be found at the intersection of $A$-type seven
branes.  They can also be realised in M-theory as a M5 brane probing
$D$- or $E$-type singularity $\IC^2/\Gamma_{DE}$.  A complete list of
minimal conformal matters can be found in \cite{DelZotto:2014hpa}.
Some of the more complicated link configurations can be obtained by
joining two minimal conformal matters and gauging the common flavor
symmetry, or by performing Higgs branch RG flow
\cite{DelZotto:2014hpa,Heckman:2016ssk}.

The second step of classification is to assign suitable singular
fibers so that the total space of fibration is a Calabi-Yau threefold.
In particular, one has to make sure that every elliptic fiber is of
the Kodaira-Tate type, which is in general equivalent to the condition
of gauge anomaly cancellation discussed in
Section~\ref{sec:gauge-anomalies}.  This step can also be done in two
parts.  The first part involves the classification of singular fibers
over a single base curve or equivalently the associated symmetry
algebra,\footnote{Some symmetry algebras could be realised by
  different singular fibers in the tensor branch
  \cite{Bertolini:2015bwa}.  Nevertheless it was argued
  \cite{Heckman:2016ssk} that at the origin of tensor branch they
  correspond to the same SCFT.} i.e.~the classification of rank one 6d
SCFTs.  The minimal symmetry algebras have been given in
Tab.~\ref{tb:NHCs}, and they can be enhanced by making worse the
singularity of elliptic fibers.  At the same time the numbers of
charged hypermultiplets increase.  Their numbers as well as the
representations of symmetry algebras  under which they transform
are completely determined by the anomaly cancellation conditions
\eqref{eq:anm-A},\eqref{eq:anm-B},\eqref{eq:anm-C}.

If there are multiple hypermultiplets in the same gauge representation
$R$, they support a non-trivial flavor symmetry $F$.  The type of the flavor
symmetry is determined by the number $m$ of hypermultiplets and the
nature of $R$ (see for instance \cite{Gaiotto:2009we}).  If $R$ is
complex, the hypermultiplets transform in the representation $\md{m}$
of the flavor symmetry $\mf{su}(m)$.  To be more precise, each
hypermultiplet can be regarded as consisting of two half-hypers, and
the $2m$ half-hypers transform in the gauge-flavor bi-representation
$(R,\wb{\md{m}})\oplus (\wb{R},\md{m})$.  If $R$ is real, the flavor
symmetry is enhanced to the quaternionic representation of
$\mf{sp}(m)$;\footnote{We use the convention that the Lie group
  $\mf{sp}(m)$ has rank $m$.} if $R$ is pseudo-real, the flavor
symmetry is enhanced to the vector representation of $\mf{so}(2m)$.
In the last case, half-hypers can exist by themselves and thus the
number $m$ of hypermultipets can be half integers.  The flavor
symmetry determined in this field theoretic way is expected to hold at
the superconformal fixed point as well, except for some rare cases.
The eight half-hypers of the $\fn=2, G=\mf{su}(2)$ theory were found
to be in the spinor representation of $\mf{so}(7)$ instead of the
vector representation of $\mf{so}(8)$ \cite{Ohmori:2015pia}.
Furthermore, for a handful of $\fn=1,2$ theories,
i.e.~$\fn=2,G=\mf{so}(11)$, $\fn=1,G=\mf{so}(11)$, and
$\fn=1,G=\mf{so}(12)_b$, the flavor symmetries deduced by the field
theoretic method do not seem to yield a consistent current algebra on
the worldsheet of BPS string \cite{DelZotto:2018tcj}.

We also comment that a rank one 6d SCFT may also have Abelian flavor
symmetry \cite{Heckman:2016ssk}, which can be uncovered by either
subjecting the candidate Abelian symmetry that accompanies complex
representations to the test of the ABJ anomaly cancellation condition
\eqref{eq:anm-E} \cite{Apruzzi:2020eqi}, or by studying the current
algebras on the worldsheet theory of BPS strings
\cite{DelZotto:2018tcj}.  These Abelian flavor symmetries are
interpreted as the weak coupling limit of Abelian gauge symmetries in
supergravity when gravity is turned off \cite{Lee:2018ihr}.  Finally,
once the flavor symmetry is known, the associated anomaly coefficients
$b_{i,k'},b_{i,mn}$ can be computed by
\eqref{eq:anm-kF},\eqref{eq:anm-k1}.
With these caveats taken into account, all possible gauge symmetries
and flavor symmetries of rank one 6d SCFTs are given in
\cite{Heckman:2015bfa,DelZotto:2018tcj}, and we reproduce it in
Tabs.~\ref{tb:rk1-1},\ref{tb:rk1-2}.

\begin{table}[h]
\centering
\begin{tabular}{|c|l|l|c|}\hline
  $\fn$
  &$G$
  &$F$
  & $(R_G,R_F)$
  \\ \hline
  {12}&$E_8$&$-$
  & $-$
  \\\hline
  8&$E_7$ & $-$
  & $-$
  \\\hline
  7&$E_7$&$-$
  & $(\md{56},\md 1)$
  \\\hline
  6&$E_6$&$-$&$-$
  \\
  6&$E_7$&$\mf{so}(2)_{12}$
  & $(\md{56},\md 2)$
  \\\hline
  5&$F_4$&$-$ & $-$
  \\
  5&$E_6$&$\mf{u}(1)_6$
  & $\md{27}_{-1}\oplus c.c.$
  \\
  5&$E_7$&$\mf{so}(3)_{12}$
  & $(\md{56},\md 3)$
  \\\hline
  4&$\mf{so}(8)$&$-$&$-$
  \\
  4&$\mf{so}(N\geq 9)$&$\mf{sp}(N-8)_1$
  & $(\md{N},\md{2(N-8)})$
  \\
  4&$F_4$&$\mf{sp}(1)_3$
  & $(\md{26},\md 2)$
  \\
  4 &$E_6$&$\mf{su}(2)_{6}\times \mf{u}(1)_{12}$
  & $({\bf 27},\wb{\md 2})_{-1}\oplus c.c.$
  \\
  4&$E_7$&$\mf{so}(4)_{12}$
  & $(\md{56},\md 2\oplus\md 2)$
  \\\hline
  3&$\mf{su}(3)$&$-$& $-$
  \\
  3&$\mf{so}(7)$&$\mf{sp}(2)_1$ & $(\md 8,\md 4)$
  \\
  3&$\mf{so}(8)$&$\mf{sp}(1)_1^a\times \mf{sp}(1)_1^b\times \mf{sp}(1)_1^c$
  & $(\md 8_v\oplus\md 8_c\oplus\md 8_s,\md 2)$
  \\
  3&$\mf{so}(9)$&$\mf{sp}(2)_1^a\times \mf{sp}(1)_2^b$
  & $(\md 9,\md 4^a)\oplus(\md{16},\md 2^b)$
  \\
  3&$\mf{so}(10)$&$\mf{sp}(3)_1^a\times (\mf{su}(1)_4\times \mf{u}(1)_4)^b$
  & $(\md{10},\md 6^a)\oplus[(\md{16}_s)^b_1\oplus c.c.]$
  \\
  3&$\mf{so}(11)$&$\mf{sp}(4)_1^a\times \text{Ising}^b$
  & $(\md{11},\md 8^a)\oplus(\md{32},\md 1_s^b)$
  \\
  3&$\mf{so}(12)$&$\mf{sp}(5)_1$
  & $(\md{12},\md{10})\oplus(\md{32}_s,\md{1})$
  \\
  3&$G_2$&$\mf{sp}(1)_1$ & $(\md{7},\md 2)$
  \\
  3&$F_4$&$\mf{sp}(2)_3$
  & $(\md{26},\md 4)$
  \\
  3 &$E_6$&$\mf{su}(3)_{6}\times \mf{u}(1)_{18}$
  & $({\bf 27},\wb{\md 3})_{-1}\oplus c.c.$
  \\
  3&$E_7$&$\mf{so}(5)_{12}$
  & $(\md{56},\md 5)$
  \\\hline
\end{tabular}
\caption{Gauge, flavor symmetries and charged matter contents of rank
  one 6d SCFTs with $\fn\geq 3$ \cite{DelZotto:2018tcj}.  The
  subscript in a flavor symmetry algebra indicates the level of the
  associated current algebra.  When a flavor symmetry has multiple
  simple components, superscripts are used to distinguish them and
  their representations.  Matters are presented as the gauge and
  flavor representations by which the half-hypermultiplets transform.
  If there is an Abelian flavor symmetry, the Abelian charge is given as
  subscript.}
\label{tb:rk1-1}
\end{table}

\begin{table}[h]
  \centering
  \begin{tabular}{|c|l|l|c|}\hline
    $\fn$&$G$&$F$
    & $(R_G,R_F)$
    \\ \hline
    2&$\mf{su}(1)$&$\mf{su}(2)_1$&$-$
    \\
    2&$\mf{su}(2)$&$\mf{so}(7)_1\times\text{Ising}$
    & $(\md 2,\md 8_s\times \md 1_s)$
    \\
    2&$\mf{su}(N\geq 3)$&$\mf{su}(2N)_1$
    & $(\md N,\wb{2\md N})\oplus c.c.$
    \\
    2&$\mf{so}(7)$&$\mf{sp}(1)_1^a\times \mf{sp}(4)_1^b$
    & $(\md 7,\md 2^a)\oplus(\md 8,\md 8^b)$
    \\
    2&$\mf{so}(8)$&$\mf{sp}(2)_1^a\times \mf{sp}(2)_1^b\times \mf{sp}(2)_1^c$
    & $(\md 8_v,\md 4^a)\oplus (\md 8_s,\md 4^b)\oplus (\md
      8_c,\md 4^c)$
    \\
    2&$\mf{so}(9)$&$\mf{sp}(3)_1^a\times \mf{sp}(2)_2^b$
    & $(\md 9,\md 6^a)\oplus(\md{16},\md 4^b)$
    \\
    2&$\mf{so}(10)$&$\mf{sp}(4)_1^a\times (\mf{su}(2)_4\times \mf{u}(1)_{8})^b$
    & $(\md{10},\md 8^a)\oplus[(\md{16}_s,\md 2^b)_1\oplus c.c.]$
    \\
    2&$\mf{so}(11)$&$\mf{sp}(5)_1^a\times ?^b$
    & $(\md{11},\md{10}^a)\oplus(\md{32},\md 2^b)$
    \\
    2&$\mf{so}(12)_a$&$\mf{sp}(6)_1^a\times \mf{so}(2)_8 $
    & $(\md{12},\md{12}^a)\oplus(\md{32}_s,\md 2^b)$
    \\
    2&$\mf{so}(12)_b$&$\mf{sp}(6)_1^a\times \text{Ising}^b\times \text{Ising}^c$
    & $(\md{12},\md{12}^a)\oplus(\md{32}_s,\md
      1^b_s)\oplus(\md{32}_c,\md 1_s^c)$
    \\
    2&$\mf{so}(13)$&$\mf{sp}(7)_1$
    & $(\md{13},\md{14})\oplus(\md{64},\md 1)$
    \\
    2&$G_2$&$\mf{sp}(4)_1$
    & $(\md 7,\md 8)$
    \\
    2&$F_4$&$\mf{sp}(3)_3$
    & $(\md{26},\md 6)$
    \\
    2 &$E_6$&$\mf{su}(4)_{6}\times \mf{u}(1)_{24}$
    & $(\md{27},\wb{\md 4})_{-1}\oplus c.c.$
    \\
    2&$E_7$&$\mf{so}(6)_{12}$
    & $(\md{56},\md 6)$
    \\\hline
    1&$\mf{sp}(0)$&$(E_8)_1$
    & $-$
    \\
    1&$\mf{sp}(N\geq 1)$&$\mf{so}(4N+16)_1$
    & $(\md{2N},\md{4N+16})$
    \\
    1&$\mf{su}(3)$&$\mf{su}(12)_1$
    & $(\md 3,\wb{\md{12}})_1\oplus c.c.$
    \\
    1&$\mf{su}(4)$&$\mf{su}(12)_1^a\times \mf{su}(2)_1^b$
    & $[(\md 4,\wb{\md{12}}_1^a)\oplus c.c.]\oplus (\md 6,\md
      2^b)$
    \\
    1 &$\mf{su}(N\geq 5)$&$\mf{su}(N\!+\!8)_1\!\times\! \mf{u}(1)_{2N(N-1)(N+8)}$
    &$[(\md{N},\wb{\md{N+8}})_{-N+4}\oplus(\md{\Lambda}^2,\md{1})_{N+8}]\oplus
      c.c.$
    \\
    1&$\mf{su}(6)_*$&$\mf{su}(15)_1$
    & $[(\md{6},\overline{\md{15}})\oplus c.c.]\oplus
      (\md{20},\md{1})$
    \\
    1&$\mf{so}(7)$&$\mf{sp}(2)_1^a\times \mf{sp}(6)_1^b$
    & $(\md{7},\md{4}^a)\oplus(\md{8},\md{12}^b)$
    \\
    1&$\mf{so}(8)$&$\mf{sp}(3)_1^a\times \mf{sp}(3)_1^b\times \mf{sp}(3)_1^c$
    & $(\md{8}_v,\md{6}^a)\oplus(\md{8}_s,\md{6}^b)\oplus(\md{8}_c,\md{6}^c)$
    \\
    1&$\mf{so}(9)$&$\mf{sp}(4)_1^a\times \mf{sp}(3)_2^b$
    & $(\md{9},\md{8}^a)\oplus(\md{16},\md{6}^b)$
    \\
    1&$\mf{so}(10)$&$\mf{sp}(5)_1^a\times (\mf{su}(3)_4\times \mf{u}(1)_{12})^b$
    & $(\md{10},\md{10}^a)\oplus[(\md{16}_s,\md{3}^b)_1\oplus c.c.]$
    \\
    1 &$\mf{so}(11)$&$\mf{sp}(6)_1^a\times ?^b$
    & $(\md{11},\md{12}^a)\oplus(\md{32},\md{3}^b)$
    \\
    1&$\mf{so}(12)_a$&$\mf{sp}(7)_1^a\times \mf{so}(3)_8^b$
    & $(\md{12},\md{14}^a)\oplus(\md{32}_s,\md{3}^b)$
    \\
    1&$\mf{so}(12)_b$&$\mf{sp}(7)_1^a\times ?^b\times ?^c$
    & $(\md{12},\md{14}^a)\oplus(\md{32}_s,\md{2}^b)\oplus(\md{32}_c,\md{1}^c)$
    \\
    1&$G_2$&$\mf{sp}(7)_1$
    & $(\md{7},\md{14})$
    \\
    1&$F_4$&$\mf{sp}(4)_3$
    & $(\md{26},\md{8})$
    \\
    1 &$E_6$&$\mf{su}(5)_{6}\times \mf{u}(1)_{30}$
    & $(\md{27},\overline{\md{5}})_{-1}\oplus c.c.$
    \\
    1&$E_7$&$\mf{so}(7)_{12}$
    & $(\md{56},\md{7})$
    \\\hline
  \end{tabular}
  \caption{Gauge, flavor symmetries and charged matter contents of
    rank one 6d SCFTs with $\fn=1,2$ \cite{DelZotto:2018tcj}.
    $\Lambda^2$ is the rank-two anti-symmetric representation.  $?$
    means the flavor symmetry predicted by field theoretic
    considerations cannot be realised consistently on the worldsheet
    of BPS strings \cite{DelZotto:2018tcj}.}
\label{tb:rk1-2}
\end{table}

The second part of fiber classification is to consider mixed
representation of two gauge algebras, which are further constrained by
the anomaly cancellation condition \eqref{eq:anm-1}.  There are only
five possibilities \cite{Heckman:2018jxk}
\begin{itemize}
\item $\mf{g}_a = \mf{su}(n_a),\mf{g}_b = \mf{su}(n_b), R =
  (\md{n}_a,\md{n}_b)$
\item $\mf{g}_a = \mf{su}(n_a),\mf{g}_b = \mf{sp}(n_b), R =
  (\md{n}_a,\md{2n}_b)$
\item $\mf{g}_a = \mf{sp}(n_a),\mf{g}_b = \mf{so}(n_b), R =
  \frac{1}{2}(\md{2n}_a,\md{n}_b)$
\item $\mf{g}_a = \mf{sp}(n_a),\mf{g}_b = \mf{so}(n_b), n_b=7,8, R =
  \frac{1}{2}(\md{2n}_a,\md{8}_{s,c})$
\item $\mf{g}_a = \mf{sp}(n_a),\mf{g}_b = G_2, R =
  \frac{1}{2}(\md{2n}_a,\md{7})$
\end{itemize}

At the end of this subsection, we comment that 6d SCFTs in this
classification could be related to each other by RG flows.  There are
two types of RG flows, the tensor branch and the Higgs branch flows.
The former simply corresponds to blowing up or blowing down base
curves, while the latter are related to complex structure deformation
and they do not change the rank of 6d SCFTs.  RG flows of 6d SCFTs
have been intensively studied in
\cite{DelZotto:2014hpa,Heckman:2015ola,Heckman:2016ssk,Heckman:2018pqx}.
In this paper, we will mainly be interested in RG flows of rank one 6d
SCFTs.  All rank one 6d SCFTs with the same $\fn$ are connected to
each other by Higgs branch RG flows, which are summarised in Section
2.4 of \cite{DelZotto:2018tcj}.

\subsection{Elliptic genera}
\label{sec:elliptic-genera}

We are interested in the partition function of 6d SCFT on the  tensor
branch on the 6d $\Omega$ background.\footnote{A 6d SCFT directly
  compactified on a $S^1$ is also known as a 5d KK or marginal theory
  \cite{Jefferson:2018irk,Apruzzi:2019opn}.  One can also consider a
  twisted circle compactification of a 6d SCFT by modding out a
  discrete symmetry on the string charge lattice $\Lambda$ or/and the
  affinised fibral symmetry algebra \cite{Bhardwaj:2019fzv}.  This is
  also one way of realising the ``frozen singularity''
  \cite{Bhardwaj:2015oru,Apruzzi:2017iqe,Bhardwaj:2019fzv}.  We will
  consider these constructions in a later work.}  The latter is a
curved spacetime background, which is topologically $T^2\times \IR^4$ with
the metric \cite{Losev:2003py}
\begin{equation}
  \rd s^2 =
  \rd z \rd \bar{z} + (\rd x^\mu+\Omega^\mu \rd z+\bar{\Omega}^\mu \rd
  \bar{z})^2 ,
  \quad \mu=1,2,3,4
\end{equation}
where $z,\bar{z}$ are coordinates on $T^2$ and $x^\mu$ coordinates on
$\IR^4$. The $\Omega^\mu$ satisfy
\begin{equation}
  \rd \Omega = \ep_1\, \rd x^1\wedge \rd x^2 - \ep_2 \,\rd x^3 \wedge
  \rd x^4,
\end{equation}
and $\ep_{L,R} = (\ep_1\mp \ep_2)/2$ are the background field
strengths for the spacetime symmetry
$\mf{su}(2)_L \times \mf{su}(2)_R $ acting on $\IR^4$.  The
compactification on $T^2$ allows access to the BPS states on BPS
strings, encoded in the Ramond-Ramond elliptic genera, which are the
generalised Witten index on the worldsheet theory of BPS strings.  The
BPS strings wrapped on $T^2$ would appear as instantons on $\IR^4$,
and the curvature on $\IR^4$ serves as the IR regulator analogous to
the 4d $\Omega$ background \cite{Nekrasov:2002qd}.  The partition
function of 6d SCFT is then a finite quantity and it splits as
follows:
\begin{equation}
  Z(\phi,\tau,\mGF,\ep_{1,2}) = Z^{\text{cls}}(\phi,\tau,\mGF,\ep_{1,2})
  Z^{\text{1-loop}}(\tau,\mGF,\ep_{1,2})
  \Big(1+\sum_d
    \re^{\ri 2\pi \phi\cdot d}
    \IE_d(\tau,\mGF,\ep_{1,2})\Big).
\end{equation}
Here $Z^{\text{cls}}, Z^{\text{1-loop}}$ are semi-classical
contributions, and one-loop contributions from tensor, vector and
hypermultiplets respectively.  $\IE_d$ is the RR elliptic genus of the
BPS strings with string charge $d = (d_i) \in \Lambda$ associated to
the base curve $S = \sum_i d_i \Sigma_i$.  $\phi=(\phi_i),\tau$ are
respectively the tensor moduli and the complex structure of $T^2$. We
have turned on the vevs $\mGF$ of Wilson loops of gauge and flavor
vector fields along 1-cycles in $T^2$, also called the gauge and
flavor fugacities.  They take value in the complexified Cartan
subalgebra of the corresponding symmetry algebra, where a Weyl
invariant bilinear form $(\bullet,\bullet)$ is defined.  See
Appendix~\ref{sc:Lie} for our Lie algebraic convention.
We will also use the notation of the \emph{reduced} $d$-string elliptic genus:
\begin{equation}
  \IE_d^{\text{red}}(\tau,\mGF,\epsilon_1,\epsilon_2) =
    \IE_d(\tau,\mGF,\epsilon_1,\epsilon_2)/\IE_{c.m.}(\tau,\eq,\et),
\end{equation}
where the the contribution from the center of mass free hypermultiplet
\be
\IE_{c.m.}(\tau,\eq,\et)=\frac{\eta(\tau)^2}{\theta_1(\tau,\eq)\theta_1(\tau,\et)}
\ee
is factored out \cite{DelZotto:2016pvm}. This brings certain simplification for elliptic genera especially for the one-string case.

The 6d $\Omega$ background has the additional advantage of allowing
for connection with topological string theory.  F-theory
compactified on an elliptic Calabi-Yau threefold $X$ and $T^2$ is dual
to M-theory compactified on the same threefold $X$ and the
M-theory circle $S^1$, where the volume of elliptic fiber in $X$ is
inversely proportional to the volume of $T^2$.  Turning on Wilson
loops of gauge and flavor vector fields amounts to resolving singular
elliptic fibers so that the threefold $X$ is smooth.  M-theory BPS states are
computed in this setup by topological string theory which
encodes in particular the numbers of BPS states of M2 branes wrapping
2-cycles in $X$.   Note that the elliptic genera contain BPS states which
wrap the base curves non-trivially.  One can therefore use  topological string
theory techniques  to get  information about the $\IE_d$ and in
particular initial data for the recursive blow up equations.

One important property of the elliptic genera is how they transform
under the action of the modular group $SL(2,\IZ)$.  Thanks to the
non-trivial 't Hooft anomalies, the elliptic genera are not invariant,
but transform as meromorphic Jacobi forms of weight zero but
non-trivial index, where both the gauge/flavor fugacities and the
parameters of the $\Omega$ background transform as elliptic
parameters.
\begin{equation}
  \IE_d(\frac{a\tau+b}{c\tau+d},\frac{\mGF}{c\tau+d},\frac{\ep_{1,2}}{c\tau+d})
  = \re^{\frac{c}{c\tau+d}\Ind {\IE_d}(\mGF,\ep_{1,2})}
  \IE_d(\tau,\mGF,\ep_{1,2}).
\end{equation}
Here $\Ind {\IE_d}(\mGF,\ep_{1,2})$, called the modular index
polynomial, is a quadratic polynomial.  The index polynomial can be
given by an equivariant integral of the 't Hooft anomaly four-form
\cite{Bobev:2015kza}, and it boils down to the following replacement
rules \cite{DelZotto:2017mee,DelZotto:2016pvm}
\begin{equation}
  p_1(M_2) \to 0,\quad c_2(L)\to -\epsilon_L^2, \quad c_2(R),c_2(I)\to
  -\epsilon_R^2,\quad \tr F_{k'}^2 \to -2 (m_{k'},m_{k'}),
  \quad \tr F_{\mf{u}(1)} \to \ri\, m_{\mf{u}(1)}.
\end{equation}
Applying these rules on \eqref{eq:I4} yields the following modular
index polynomial
\begin{align}
  \Ind {\IE_d}(\mGF,\ep_{1,2}) =
  &-\frac{1}{4}(\eq+\et)^2\sum_i(2+A_{ii} +h^\vee_{\fg_i})d_i +
    \frac{1}{2}\eq\et\Big(\sum_i(2+A_{ii})d_i-\sum_{i,j}A_{ij}d_id_j\Big)\nonumber \\[-1mm]
  &+\frac{1}{2}
    \sum_{i,k'}b_{i,k'}d_i(m_{k'},m_{k'})
    +\frac{1}{2}\sum_{i,\ell,n}b_{i,\ell n} m_{\ell}m_n.
    \label{eq:IndEd}
\end{align}

\subsubsection{Known computational methods}
\label{sec:known-comp-meth}

In this section, we summarize all known results on the elliptic genera
of 6d $(1,0)$ SCFTs, in particular all rank one theories. For the
minimal SCFTs which are the pure gauge rank one theories, a thorough
summary on the results from all kinds of approaches has been presented
in the introduction of \cite{Gu:2019pqj}. Here we focus more on
theories with matters. Three methods with relatively wide range of
application are 2d quiver gauge theories, modular ansatz and refined
topological vertex. In the following, we briefly introduce each
method, list the theories it can solve and comment on its advantages
and disadvantages.

In the spirit of the ADHM construction for 4d/5d instantons, certain 6d
$(1,0)$ SCFTs are known to correspond to 2d quiver gauge
theories. Once the 2d quiver construction is found, one can use
localization -- Jeffrey-Kirwan residue to exactly compute the elliptic
genera to arbitrary number of strings. However, like in the ADHM
construction, such correspondence normally just exists for classical
gauge groups, but difficult to generalize to exceptional gauge
groups. All rank one $(1,0)$ theories with known 2d quiver
construction are listed below:
\begin{itemize}
\item $n=1$, $G=\mf{sp}(N)$ \cite{Kim:2015fxa,Yun:2016yzw}
\item $n=1$, $G=\mf{su}(N)$ \cite{Kim:2015fxa}
\item $n=2$, $G=\mf{su}(N)$ \cite{Haghighat:2013tka}
\item $n=3$, $G=\mf{su}(3),G_2$ and $\mf{so}(7)$ \cite{Kim:2018gjo}
\item $n=4$, $G=\mf{so}(8+N)$ \cite{Haghighat:2014vxa,DelZotto:2018tcj}
\end{itemize}
For all these theories, we use the known elliptic genera from quiver
formulas to check against our elliptic blowup equations and find
perfect agreement.

The modular ansatz method exploits the Jacobi-form transformations of the
elliptic genera as well as their pole structures and can be very constraining sometimes.
For the reduced one string elliptic genus with all gauge and flavor fugacities turned off, the modular ansatz
has a particularly simple form and was extensively studied
in \cite{DelZotto:2018tcj}. For example, using the constraints from the spectral flow relation between RR and NSR elliptic genus, such ansatze were fixed in \cite{DelZotto:2018tcj} for all rank-one theories except for
\begin{itemize}
\item $n=1,2,3,4$, $G=E_7$
\item $n=1,2$, $G=E_6,\ \mf{so}(11)$ and $\mf{so}(12)_b$
\end{itemize}
These results provide an excellent testing ground for our blowup
equations. Indeed, for all the theories we have studied where
the modular ansatz is fixed in \cite{DelZotto:2018tcj}, we find
agreement for the one-string elliptic genera.
Besides, we are able to use blowup equations to further determine the
modular ansatz for $n=2,4$ $E_7$ theories, $n=1,2$ $E_6$ theories and
$n=2$ $\mf{so}(11)$ theory and make cross checks. The modular ansatz
method also extends to the situation with gauge and flavor fugacities
turned on, where Weyl-invariant Jacobi forms are involved and the
computation becomes much more complicated. Still, the ansatz for the
one-string elliptic
genus 
with gauge fugacities turned on for $n=3\ \mf{su}(3)$ and $n=4\
\mf{so}(8)$ theories was determined in \cite{DelZotto:2017mee}, and
for $n=1$ $\mf{sp}(1)$, $n=2\ \mf{su}(2)$, $n=3\ \mf{su}(3)$ and $n=3\
G_2$ theories was determined in \cite{Kim:2018gak}. Note the modular ansatz method works even for compact elliptically fibred Calabi-Yau
manifolds~\cite{Huang:2015sta}.

The refined topological vertex and the brane-webs can also compute the
elliptic genera of some 6d theories with matters. For example, the
brane web construction was known for $n=1,
G=\mf{sp}(N)$ theories \cite{Hayashi:2015zka}, $n=1,
G=\mf{su}(N)$ theories \cite{Hayashi:2015zka},
$n=1,G=\mf{su}(6)_{*}$ theory \cite{Hayashi:2019yxj}, a family of
$n=2,3$
$\mf{so}(N)$ theories \cite{Kim:2019dqn}, the D-type conformal matter
theories \cite{Hayashi:2015fsa}. See also
\cite{Hayashi:2015vhy,Hayashi:2016abm}. The brane construction for
theories with non
$\mf{su}$ type gauge symmetry or complicated matter representations
typically involves orientifold 7-plane and O5-planes.

It is also worthwhile to point out some relevant 5d results.  For
example, the 5d Nekrasov partition functions of $n=2, G=\mf{su}(N)$
theories were well known long time ago, see
\cite{Nekrasov:2004vw,Shadchin:2004yx,Benvenuti:2010pq}.  The 5d
blowup equations with matters were initially studied in
\cite{Nakajima:2009qjc}. Recently, the 5d unity blowup equations for
all possible gauge and matter content were studied in
\cite{Kim:2019uqw}. For a lot of 5d theories, their Nekrasov partition
functions can be solved from these blowup equations recursively with
respect to the instanton numbers. Such blowup equations can be
regarded as the 5d limit of our elliptic blowup equations in this
paper. Besides, the brane web construction for 5d $G_2$ theories with
a fundamental matter was also obtained recently in
\cite{Hayashi:2018bkd}. These results provide consistency checks for
the elliptic genera we solved from elliptic blowup equations when
taking $q\to 0$ limit.

For higher rank 6d SCFTs, the known results on elliptic
genera are only for some special theories. For example, the 2d quiver constructions are known for the three
higher-rank non-Higgsable clusters \cite{Kim:2018gjo}, ADE chain of $(-2)$
curves with gauge symmetry \cite{Gadde:2015tra}, and notably $(E_6,E_6)$ conformal matter theory \cite{Kim:2016foj}. The
modular ansatz has been studied for higher rank E-string and M-string
theories in \cite{Gu:2017ccq}. Beside, the elliptic genera of of A-type chain of $(-2)$
curves can be computed by refined topological vertex \cite{Haghighat:2013tka} or from the viewpoint of 2d sigma model \cite{MR1746311,Haghighat:2013tka}. The recently proposed elliptic topological vertex can also compute the partition function of these theories \cite{Foda:2018sce,Kimura:2019gon}.

\subsection{Semiclassical free energy}
\label{sec:geom-engin-matt}

We consider a 6d SCFT as the F-theory compactification on an
elliptic-fibered Calabi-Yau threefold $X$, where the base $B$ is a
non-compact two dimensional surface. Because of the duality between
M-theory and F-theory, the refined BPS spectrum is captured by refined
topological string theory on such a non-compact
elliptic-fibered Calabi-Yau threefold $X$.  In order to
obtain the elliptic blowup equations for a rank one 6d SCFT, we first
write down the blowup equation for refined topological string on $X$,
and then transform it to our preferred form -- in terms of elliptic
genera, more details can be found in Appendix~\ref{sc:dbe}. As described in \cite{Gu:2018gmy}, we only need the
semiclassical pieces of the genus zero and the genus one free
energies, and the one-loop contributions from BPS particles.  The
latter can be readily read off from the vector and hypermultiplet
spectrum \cite{Gu:2018gmy}.  There are also recent results on how to
compute the BPS particle spectrum from elliptic Calabi-Yau geometry
\cite{Kashani-Poor:2019jyo,Paul-KonstantinOehlmann:2019jgr}.  In this
section, we consider the computation of semiclassical free energies.
We focus on the case of rank one 6d SCFTs, and relegate some results
on higher rank theories to the  Appendix~\ref{sc:hrp}.


We start from the results of \cite{Bhardwaj:2019fzv}.  The elliptic non-compact
Calabi-Yau threefold associated to a 6d SCFT on $S^1$ is locally the
neighborhood of a union of compact surfaces
\begin{equation}
  \cS = \cup_{i,a} \cS_{a,i}\, .
\end{equation}
Here $i$ is the index of base curves, and the compact surfaces
$\cS_{a,i}$ with fixed $i$ project to the same base curve.  They
intersect with each other according to the affine Dynkin diagram of a
Lie algebra, and we denote the divisor associated to the affine node
by $\cS_{0,i}$.  The K\"ahler class is then parameterized by
\begin{equation}
  J = -\sum_{i,a} \phi_{a,i} \cS_{a,i},
\end{equation}
where the K\"ahler parameters $\phi_{a,i}$ measure the volumes of the
divisors $\cS_{a,i}$.  The semiclassical prepotential of the 6d SCFT
on $S^1$ is given by the triple intersection $J\cdot J\cdot J/6$
together with Chern-Simons terms
coming from the tree-level circle reduction of the Green-Schwarz
counter-terms.
One finds that the prepotential is
\begin{equation}
  {F}_{(0,0)}^{\rm cls} = -\frac{1}{12}\bigg(
    \sum_{\alpha\in\Delta}|\alpha\cdot \phi|^3
    -\sum_{f=1}^{N_f} \sum_{\omega\in{R_f}} |\omega\cdot \phi+m_f|^3
  \bigg)
  -\frac{1}{2}\sum_{i,j}\Omega_{ij}\phi_{0,i}
  \sum_{a,b}K_j^{ab}\phi_{a,j}\phi_{b,j}.
\end{equation}
Here $\Omega_{ij}$ is the negative base intersection matrix;
$K^{ab}_j$ is the Killing form for the Lie algebra associated to the
intersecting divisors over the base curve $j$, so that the last term
can be written as $\phi_{,j}\cdot\phi_{,j}$.  In the case of a rank
one theory, we simply have
\begin{equation}
  \Omega_{ii} = \fn, \quad \Omega_{ij} = -k_{F_j},\quad \text{for fixed }i.
\end{equation}
In order to incorporate flavor symmetry, we also introduce K\"ahler
parameters formally associated to non-compact vertical divisors,
denoted collectively by $\phi'$.  Then the prepotential can be written
as
\begin{equation}\label{F00tree}
  {F}_{(0,0)}^{\rm cls} = -\frac{1}{12}\bigg(
    \sum_{\alpha\in\Delta}|\alpha\cdot \phi|^3
    -\frac{1}{2}\sum_{\omega_{G,F}\in R_{G,F}}
    |\omega_G\cdot \phi+\omega_F\cdot \phi'|^3
  \bigg)
  -\frac{1}{2}\phi_{0}\bigg(\fn \phi\cdot\phi - \sum_j
    k_{F_j}\phi'_j\cdot\phi'_j\bigg).
\end{equation}

This formula is still incomplete, as we are still missing terms from
the intersections of the base divisor with associated K\"ahler
parameter $\phi_B$.  The non-vanishing triple intersection numbers
involving the base divisor $\cS_B$ are
\begin{equation}
  \cS_B\cdot \cS_B\cdot \cS_0 = (B\cdot B)_{\cS_0} = \fn-2,\quad
  \cS_B\cdot \cS_0\cdot \cS_0 = (B\cdot B)_{\cS_B} = -\fn
\end{equation}
and the corresponding additional terms to prepotential are
\begin{equation}
  \frac{\fn\phi_B\phi_0^2-(\fn-2)\phi_B^2\phi_0}{2}.
\end{equation}

Finally, we need to change bases and use the K\"ahler parameters
measuring curve volumes instead.  Using the curve-divisor intersection
numbers, we find the following identification
\begin{equation}
  t_B = -(\fn-2)\phi_B +\fn \phi_0,\quad \tau = -\phi_B, \quad m_G =
  \phi, \quad m_F = \phi'.
\end{equation}
For later convenience, we also define\footnote{$t_B$ is more natural in the description of the geometries, we will always use $t_B$ in the discussion of the geometries.}
\begin{equation}
  t_{\text{ell}} = t_B - \frac{\fn-2}{2}\tau.
\end{equation}
Therefore, the final form of the semiclassical prepotential for a rank
one theory with $-n$ base curve and gauge symmetry $G$ as well as
flavor symmetry $F = \otimes_j F_j$ is
\begin{equation}
\label{conjF0}
\begin{split}
  {F}_{(0,0)}^{\rm cls}
  =&-\frac{1}{6}\sum_{\alpha\in\Delta^+}(\alpha\cdot\und{m}_G)^3 +
  \frac{1}{12}\sum_{\w_{G,F}\in R_{G,F}^{m+}}
  (\w_G\cdot\und{m}_G+\w_F\cdot\und{m}_F)^3\\[-1mm]
  &+\frac{t_{\text{ell}}-(\fn-2)\frac{\tau}{2}}{2\fn}
  (-\fn\und{m}_G\cdot\und{m}_G+k_F\und{m}_F\cdot\und{m}_F)
  -\frac{1}{2\fn}t_{\text{ell}}^2 \tau
  +\mathcal{O}(\tau^3).
\end{split}
\end{equation}
where we have defined
\begin{equation}
  R^{{m+}}_{G,F}  = \{\omega_G \in R_G,\omega_F\in R_F\,;\,
  \omega_G\cdot m_G+\omega_F \cdot m_F \geq 0\}.
\end{equation}
We ignore all terms only in $m_G,m_F,\tau$, as they depend on the
embedding of the associated Calabi-Yau in a compact geometry and thus
are not inherent properties of the 6d SCFT.  In addition, we can also
fix the semiclassical pieces of genus one free energies from 5d results
\cite{Nekrasov:2002qd,Shadchin:2005mx,Kim:2019uqw}, as well as
modularity of elliptic blowup equations as discussed in
Section~\ref{sc:mod}.  We find
\begin{align}
  {F}_{(0,1)}^{\rm cls}=
  &-\frac{1}{12}\sum_{\alpha \in \Delta^+}
    \alpha\cdot\und{m}_G + \frac{1}{24}\sum_{\w_{G,F} \in
    R_{G,F}^{m+}}(\w_G\cdot\und{m}_G+\w_F\cdot\und{m}_F)+
    \frac{\fn-2}{2\fn}t_{\text{ell}},\\
  {F}_{(1,0)}^{\rm cls}=
  &\frac{1}{12}\sum_{\alpha \in
    \Delta^+}\alpha\cdot\und{m}_G+\frac{1}{48}\sum_{{\w_{G,F} \in
    R_{G,F}^{m+}}}(\w_G\cdot\und{m}_G+\w_F\cdot\und{m}_F)+
    \frac{\fn-2-{h^\vee_G}}{4\fn}t_{\text{ell}}.
\end{align}
Note here all the summations over roots and weights only sum over half
sets of them, and the one loop contributions of BPS particles have to
have the same half sets of them. The choices of the half weights do not
have effects on our final result, since they are the same under
analytic continuation. In the language of geometries, different
choices of half weights reflect different choices of Calabi-Yau
phases, and they are connected by flop transitions.

\subsection{Calabi-Yau construction}
\label{sec:calabi-yau-constr}
In this section, we construct the elliptic non-compact Calabi-Yau three-folds
directly. Our basic strategy is to first construct a smooth toric base
which has the correct intersection numbers, and then add elliptic
fibers. We embed the whole geometry into a four dimensional
non-compact toric variety, then the non-compact Calabi-Yau three-fold
is a hypersurface inside the toric variety, described by a 4d
polytope. The mirror construction for compact Calabi-Yau hypersurfaces
requires that the polytope is reflexive, and thus necessarily has an unique
inner lattice point \cite{MR1269718}. However, for our direct
construction of non-compact Calabi-Yau threefolds, the polytope does
not have any inner lattice point, and it can not be reflexive. Here,
we relax the reflexive condition by requiring that the dual polytope
is still a lattice polytope, and have the same origin point, but it is
not necessarily bounded.  In fact, the dual polytope always has
infinitely many points.

We claim here the mirror construction for compact Calabi-Yau
hypersurfaces in \cite{Hosono:1993qy} still works in this setting. There should
be always a compact geometry\footnote{The compact geometry can be
  recovered by completing the rays in the toric base and then
  resolving all toric singularities.} to start with, which is a
hypersurface or a complete intersection Calabi-Yau.  After taking the
volume of some curves to infinity, we get our non-compact
Calabi-Yau. A proper minimal combination of the Mori cone generators
of the compact Calabi-Yau gives us the Mori cone of the non-compact
geometry. From the perspective of lattice polytopes, this limit is
equivalent to removing some lattice points, after which the simplices separated by these lattice points merge to
give a triangulation of the new polytope. Alternatively if we start
from a non-compact toric variety directly, we can simply triangulate
it, and get all the Mori cone generators of the ambient space. We then
use the method in \cite{Berglund:1995gd} to find the Mori cone of the
hypersurface. Note that such a naive construction sometimes does not give the
correct phase that a 6d SCFT wants. There could be some finite curves
flop out of the physical curves, we then have to modify the invariants according to the
rules for the flop transitions. This phenomenon happens for the geometries of NHC $3,2$, NHC $3,2,2$ and NHC $2,3,2$.
We will point out these degrees in the example sections.
\subsubsection*{Construction of the bases}
The intersections of divisors in base can be realised easily in toric
geometry.  For a {\it{smooth}} toric surface, if we put the ray
generators $\{u_i\}$ in clockwise order, the intersection numbers
of the divisors $\{D_i\}$ are \cite{MR2810322}
\begin{equation}
  D_i\cdot D_j=
  \left\{
    \begin{aligned}
      0,\quad & |i-j|>1,\\
      1,\quad&|i-j|=1,\\
      -\fn_i,\quad&i=j,\\
    \end{aligned}
  \right.
\end{equation}
where $\fn_i$ is defined in
\begin{equation}
u_{i-1}+u_{i+1}=\fn_i u_i,
\end{equation}
and is minus the self-intersection of the divisor $D_i$.

\begin{figure}
  \centering
  \begin{picture}(200,110)(- 93,- 55)
    \thicklines
    \put(-175, 25){\line(1,0){50}}
    \put(-150,32){\makebox(0,0){$-m$}}
    \put(-150,-33){\makebox(0,0){\small (a) ${\mf{su}}(3), {\mf{so}}(8), {F}_4$}}
    \put(-150,-47){\makebox(0,0){\small ${E}_6,{E}_7,{E}_8$}}
    \put(-70,55){\line(1,-1){40}}
    \put(-30,35){\line(-1,-1){40}}
    \put(-50,45){\makebox(0,0){$-3$}}
    \put(-50,5){\makebox(0,0){$-2$}}
    \put(-50,-40){\makebox(0,0){\small (b)  ${G}_2 \oplus {\mf{su}}(2)$}}
    \put(30,70){\line(1,-1){40}}
    \put(30,20){\line(1,-1){40}}
    \put(70,45){\line(-1,-1){40}}
    \put(45,65){\makebox(0,0){$-3$}}
    \put(44,31){\makebox(0,0){$-2$}}
    \put(60, 0){\makebox(0,0){$-2$}}
    \put(50,-40){\makebox(0,0){\small (c)  ${G}_2 \oplus {\mf{su}}(2)$}}
    \put(130,70){\line(1,-1){40}}
    \put(130,20){\line(1,-1){40}}
    \put(170,45){\line(-1,-1){40}}
    \put(145,65){\makebox(0,0){$-2$}}
    \put(144,31){\makebox(0,0){$-3$}}
    \put(160, 0){\makebox(0,0){$-2$}}
    \put(165,-40){\makebox(0,0){\small (d) ${\mf{su}}(2) \oplus {\mf{so}}(7)
        \oplus {\mf{su}}(2)$}}
  \end{picture}
  \caption{Intersection numbers of bases of non-Higgsible clusters.}
  \label{fg:NHCbase}
\end{figure}
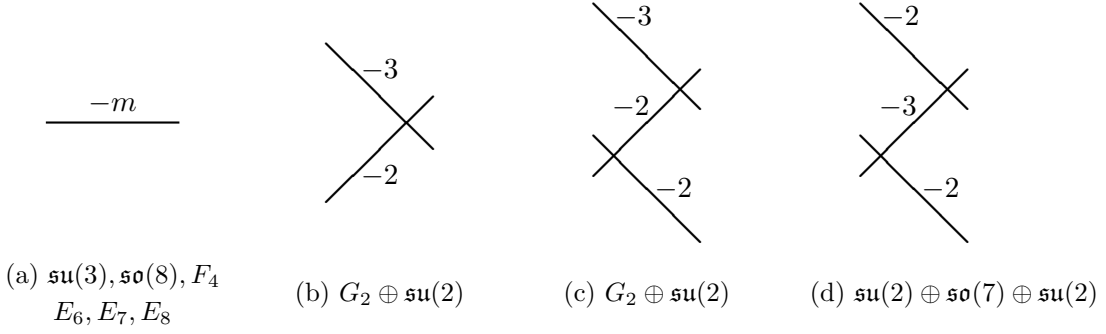

\begin{table}
  \centering
  \begin{tabular}{crr|rrr|}
    $D$ &\multicolumn{2}{c}{$\nu_i^*$}     &$l^{(1)}$& \\
    $D_u$    &         1&   0&  1 &\\
    $S$    &        0&  $-1$&  $ -m$ &   \\
    $D_v$    &         $-1$&  $-m$&   1 &   \\
    \noalign{\vskip 2mm} \multispan{3} (a)\\
  \end{tabular}
  \hskip 15pt
  \begin{tabular}{crr|rrrr|}
    $D$ &\multicolumn{2}{c}{$\nu_i^*$}     &$l^{(1)}$&$l^{(2)}$& \\
    $D_u$    &         1&   0&  1 &   0&\\
    $S_1$    &        0&  $-1$&  $-3$ &   1&\\
    $S_2$    &         $-1$& $ -3$&   1 &   $-2$&\\
    $D_v$    &         $-2$&  $-5$&   0 &   1&\\
     \noalign{\vskip 2mm} \multispan{3} (b)\\
  \end{tabular}
  \vskip 15pt
  \begin{tabular}{crr|rrrrr|}
    $D$ &\multicolumn{2}{c}{$\nu_i^*$}     &$l^{(1)}$&$l^{(2)}$&$l^{(3)}$& \\
    $D_u$    &         1&   0&  1 &   0&0&\\
    $S_1$    &        0&  $-1$&  $ -3$ &   1&0&\\
    $S_2$    &         $-1$&  $-3$&   1 &   $-2$&1&\\
    $S_3$   &         $-2$&  $-5$&   0 &   1& $-2$&\\
    $D_v$    &         $-3$&  $-7$&   0 &   0&1&\\
    \noalign{\vskip 2mm} \multispan{4} (c)\\
  \end{tabular}
  \hskip 15pt
  \begin{tabular}{crr|rrrrr|}
    $D$ &\multicolumn{2}{c}{$\nu_i^*$}     &$l^{(1)}$&$l^{(2)}$&$l^{(3)}$& \\
    $D_u$    &         1&   0&  1 &   0&0&\\
    $S_1$    &        0&  $-1$&  $ -2$ &   1&0&\\
    $S_2$    &         $-1$& $ -2$&   1 &  $ -3$&1&\\
    $S_3$   &         $-3$&  $-5$&   0 &   1&$-2$&\\
    $D_v$    &        $ -5$&  $-8$&   0 &   0&1&\\
    \noalign{\vskip 2mm} \multispan{4} (d)\\
  \end{tabular}
  \caption[x]{Toric realisation of non-Higgsable clusters.}\label{tb:NHCbase}
\end{table}

With these rules, we can write down the toric construction of the
$A$-type bases for non-compact Calabi-Yau threefolds.  We list in
particular the bases of non-Higgsible clusters in
Figure~\ref{fg:NHCbase} and their toric construction in
Table~\ref{tb:NHCbase}.  The other $A$-type bases can be constructed
in a similar way.  For bases of $D,E$-type chain of $(-2)$ curves,
there is no toric construction, but they can still be embedded in
toric varieties as hypersurfaces.

\subsubsection*{Adding elliptic fibers}
The fiber of an elliptic Calabi-Yau threefold is an elliptic curve,
which can be embedded into the weighted projective space
$\mathbb{P}^{2,3,1}(x,y,z)$, where the generic form of the elliptic
curve is of the Tate form
\begin{equation}
  y^2+x^3+a_1(s_i)x y z+a_2(s_i)x^2 z^2+a_3(s_i) y z^3+a_4(s_i) x
  z^4+a_6(s_i) z^6=0.
\end{equation}
We can promote the coordinates $x,y,z$ and the coefficients $a_j(s_i)$
to sections of line bundles over the base, and the equation then
defines the entire elliptic Calabi-Yau threefold.
The variables $s_i$ are the local coordinates of the base, and the
equation $s_i=0$ defines a vertical divisor pulled back from the
corresponding base curve.
In general, this vertical divisor can be singular, signaling the
singularity of elliptic fibers supported on the base curve.  The type
of singularity can be determined by the Tate's algorithm
\cite{Horodecki:1997nx,Bizet:2014uua}.  Depending on whether the
supporting base curve is compact or not, the singularity type is
identified with either gauge symmetry or flavor symmetry.  The two
cases can be converted to each other by tuning the volume of base
curves, corresponding to gauging flavor symmetry and turning off
coupling of gauge symmetry.
We can also make the Calabi-Yau smooth by performing the crepant
resolution on singular vertical divisors, e.g.~\cite{Esole:2019ynq},
corresponding to turning on gauge or flavor fugacities.

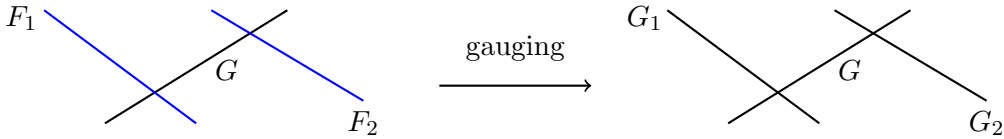
\begin{figure}
  \begin{center}
    \begin{tikzpicture}
      \def\xvalue{8.2} \draw[-,thick] (-2.4,-1.5)edge(0,0);%
      \draw[-,blue,thick] (-1,0)edge(1,-1.2);%
      \draw[-,blue,thick] (-3.2,0)edge(-1.2,-1.5);%
      \draw (-0.8,-0.8) node {$G$};%
      \draw (1.0,-1.5) node {$F_2$};%
      \draw (-3.5,-0.1) node {$F_1$};%
      \draw (3.0,-0.5) node {gauging};%
      \draw[->,thick] (2,-1)--(4,-1);%
      \draw[-,thick] (\xvalue-2.4,-1.5)edge(\xvalue+0,0);%
      \draw[-,thick] (\xvalue-1,0)edge(\xvalue+1,-1.2);%
      \draw[-,thick] (\xvalue-3.2,0)edge(\xvalue-1.2,-1.5);%
      \draw (\xvalue-0.8,-0.8) node {$G$};%
      \draw (\xvalue+1.0,-1.5) node {$G_2$};%
      \draw (\xvalue-3.5,-0.1) node {$G_1$};%
    \end{tikzpicture}
  \end{center}
  \caption[x]{\footnotesize An illustration of gauging flavor
    symmetries, with two non-compact curves intersecting with one
    compact curve.}
  \label{fig:gauging}
\end{figure}

Note that the smooth toric base can not be the correct base for
theories with $k_{F_i}>1$. From (\ref{F00tree}), we know that $k_{F_i}$
has a geometric meaning as the intersection number of compact and
non-compact base curves. If $k_{F_i}>1$, there is no compact smooth
base support the intersection number.

In the following we give the example of the toric construction of NHC
$3,2$ and its rank one limit.
More examples can be found in Appendix~\ref{app:cy}.

\subsubsection*{NHC 3,2}

\begin{equation} \label{NHC32polytope}
 \begin{array}{crrrr|rrrrrr|rrrrrrrrrr|}
    &\multicolumn{4}{c}{\nu_i^*}     &l^{(1)} & l^{(2)} & l^{(3)} & l^{(4)} & l^{(5)} & l^{(6)} & l^{(0)}_{\mathfrak{g}_2}&l^{(1)}_{\mathfrak{g}_2} & l^{(2)}_{\mathfrak{g}_2} & l^{(0)}_{\mathfrak{su}(2)} & l^{(1)}_{\mathfrak{su}(2)} & l_{B_1} & l_{B_2}  \\
    D_{0}& 0 & 0 & 0 & 0\phantom{,}  & -1 & 0 & 0 & 0 & 0 & -1\phantom{,} &   0 &   -3&0&-2&  -4 &   0&0& \\
D_{x}& -1 & 0 & 0 & 0\phantom{,}& 0 & 0 & 0 & 0 & 0 & 1\phantom{,} &	 0 &   1&0& 0&  2 &   0&0&\\
D_{y}& 0 & -1 & 0 & 0\phantom{,} & 0 & 0 & 0 & 0 & 1 & 0\phantom{,} &	0 &   0&1&1&  2 &   0&0&\\
D_{z}& 2 & 3 & 0 & 0\phantom{,} & 0 & 1 & 0 & 0 & 0 & 0\phantom{,} & 	1 &   0&0&1&  0 &   1&0&\\
D_{u}& 2 & 3 & 1 & 0\phantom{,} & 0 & 0 & 0 & 1 & 0 & 0\phantom{,} & 	  0 &   0&0&0&  0 &   1&0&\\
S_{1}& 2 & 3 & 0 & -1\phantom{,} & 0 & -2 & 1 & -1 & 0 & 0\phantom{,} & 	-2 &   1&0&0&  0 &   -3&1& \\
S_{1}^{'}& 2 & 3 & 0 & -2\phantom{,} & -1 & 1 & 0 & -1 & 1 & 0\phantom{,} &  	1 &   -2&1&0&  0 &   0&0&\\
S_{1}^{''}& 1 & 1 & 0 & -1\phantom{,} & 1 & 0 & 0 & 0 & -2 & 1\phantom{,} & 	0 &   3&-2&0&  0 &   0&0&\\
S_{2}& 2 & 3 & -1 & -3\phantom{,} & 1 & 0 & -2 & 1 & 0 & 0\phantom{,} & 	 0 &   0&0&-2&  2 &   1&-2&\\
S_{2}^{'}& 1 & 2 & -1 & -3\phantom{,} & 1 & 0 & 0 & 0 & 0 & -2\phantom{,} & 	0& 0&0&2&  -2 &   0&0&\\
D_{v}& 2 & 3 & -2 & -5\phantom{,} & -1 & 0 & 1 & 0 & 0 & 1\phantom{,} & 	0 &   0&0&0&  0 &   0&1&\\
\end{array} \
\end{equation}
The compact geometry of the NHC $3,2$ was studied in
\cite{Esole:2018mqb}. For our non-compact case, we have the polytope
(\ref{NHC32polytope}), and we choose one phase with Mori cone vectors
$l^{(i)},i=1,\cdots,6$. The triple intersection numbers behave like
the ones of an orbifold, with the intersection ring
\begin{equation}
  \begin{aligned}\label{32triple}
    \mathcal{R}
    &=-\frac{1}{5}(8 J_1^3+8 J_3 J_1^2+8 J_4 J_1^2+6 J_5
      J_1^2+4 J_6 J_1^2+8 J_3^2 J_1+8 J_4^2 J_1+12 J_5^2
      J_1+2 J_6^2 J_1+8 J_3 J_4 J_1 \\[-1mm]
    +\,
    &6 J_3 J_5 J_1+6 J_4 J_5 J_1+4 J_3 J_6 J_1+4 J_4 J_6
      J_1+3 J_5 J_6 J_1+2 J_2^3+12 J_3^3+24 J_5^3+6 J_6^3+3 J_2 J_3^2\\
    +\,
    &2 J_2 J_4^2+4 J_3 J_4^2+12 J_3 J_5^2+12 J_4 J_5^2+2 J_3
      J_6^2+2 J_4 J_6^2+4 J_5 J_6^2+J_2^2 J_3+2 J_2^2
      J_4+6 J_3^2 J_4+J_2 J_3 J_4 \\
    +\,
    &6 J_3^2 J_5+6 J_4^2 J_5+6 J_3 J_4 J_5+4 J_3^2 J_6+4 J_4^2
      J_6+6 J_5^2 J_6+4 J_3 J_4 J_6+3 J_3 J_5 J_6+3 J_4 J_5 J_6),
  \end{aligned}
\end{equation}
where $J_i$ are the K\"ahler cone generators.  However, the
intersection ring does not match the prepotential as we predicted in
(\ref{conjF0}). By computing the genus zero Gopakumar-Vafa invariants,
we found that there are two irreducible rational curve classes of
degrees
\begin{equation}
  \und{\beta}_1\cdot
  \und{t}=t_{{\mathfrak{g}_2},1}+t_{{\mathfrak{g}_2},2}
  -\frac{1}{2}t_{{\mathfrak{su}(2)},1}-t_{B_2},\quad
  \und{\beta}_2\cdot
  \und{t}=t_{{\mathfrak{g}_2},1}+2t_{{\mathfrak{g}_2},2}
  -\frac{1}{2}t_{{\mathfrak{su}(2)},1}-t_{B_2},
\end{equation}
with multiplicity one; in particular, the degree of one base curve is
negative.
Such a phenomenon means our toric construction is not exactly in the
phase corresponding to the rank two SCFT, we have to flop these two
curve classes to reach the correct geometric phase. As explained in
\cite{Witten:1996qb,Gu:2019pqj}, the flop of a rational curve
$\und{\beta}\cdot \und{t}\rightarrow -\und{\beta}\cdot \und{t}$ with
multiplicity one
changes the semiclassical prepotential by subtracting a term
$\frac{1}{6}(\und{\beta}\cdot \und{t})^3$. After the flop, we indeed
get the correct semiclassical prepotential.

We can reduce the current geometry to that of the rank one $\fn=3$
theory with gauge symmetry $G_2$ and flavor symmetry $\mf{su}(2)$ by
removing the lattice point $(2,3,-2,-5)$, and the corresponding
homogeneous coordinate is sent to zero.  In this process, some
K\"ahler parameters are sent to positive infinity, and some sent to
negative infinity; their appropriate linear combinations remain finite
and become the new K\"ahler parameters of the reduced geometry.  In
our current example, the volumes of the curve classes
$l^{(2)},l^{(1)}+l^{(3)},l^{(4)},l^{(5)},l^{(1)}+l^{(6)}$ remain
finite, and they are the Mori cone generators of the reduced geometry.

We list the result in (\ref{n3G2polytope2}).
\begin{equation} \label{n3G2polytope2}
  \begin{array}{crrrr|rrrrrrr|}
    &\multicolumn{4}{c}{\nu_i^*}
    &l^{(1)} & l^{(2)} & l^{(3)}
    &  l^{(4)} & l^{(5)} &   \\
    D_{0}& 0 & 0 & 0 & 0\phantom{,} & 0 & -1 & 0 & 0 & -2 & \\
    D_{x}& -1 & 0 & 0 & 0\phantom{,} & 0 & 0 & 0 & 0 & 1 & \\
    D_{y}& 0 & -1 & 0 & 0\phantom{,} & 0 & 0 & 0 & 1 & 0 & \\
    D_{z}& 2 & 3 & 0 & 0\phantom{,} & 1 & 0 & 0 & 0 & 0 & \\
    D_{u}& 2 & 3 & 1 & 0\phantom{,} & 0 & 0 & 1 & 0 & 0 & \\
    S_{1}& 2 & 3 & 0 & -1\phantom{,} & -2 & 1 & -1 & 0 & 0 & \\
    S_{1}^{'}& 2 & 3 & 0 & -2\phantom{,} & 1 & -1 & -1 & 1 & -1 & \\
    S_{1}^{''}& 1 & 1 & 0 & -1\phantom{,} & 0 & 1 & 0 & -2 & 2 & \\
    S_{2}& 2 & 3 & -1 & -3\phantom{,} & 0 & -1 & 1 & 0 & 1 & \\
    S_{2}^{'}& 1 & 2 & -1 & -3\phantom{,} & 0 & 1 & 0 & 0 & -1 & \\
\end{array} \
\end{equation}
The intersection ring is
\begin{equation}\label{n3G2triple}
  \ba
    \mathcal{R}=&\,-\frac{1}{3}(J_1^3+J_3 J_1^2+J_3^2 J_1+2 J_2^3+16
    J_4^3+2 J_2 J_3^2+7 J_2 J_4^2+7 J_3 J_4^2+2 J_2^2
    J_3+3 J_2^2 J_4\\
    &+3 J_3^2 J_4+3 J_2 J_3 J_4+2 J_2^2 J_5+2 J_3^2 J_5+2 J_4^2 J_5+2
    J_2 J_3 J_5+J_2 J_4 J_5+J_3 J_4 J_5).
 \ea
\end{equation}

\section{Elliptic blowup equations}
\label{sec:ellipt-blow-equat}

We present the elliptic blowup equations for all rank one 6d SCFTs on
the 6d Omega background and discuss various properties of these
equations in this section.  Since the derivation of the elliptic
blowup equations from the blowup equations for the topological string
theory is extremely similar to those given in
\cite{Gu:2018gmy,Gu:2019dan,Gu:2019pqj}, we refer to those papers for
details of the derivation and only provide some important points in
Appendix~\ref{sc:dbe}.  The additional information required in this
process includes the semiclassical free energies, which we have
discussed in Section~\ref{sec:geom-engin-matt}, and the one-loop
partition functions coming from vector and hypermultiplets, whose
formulas can be found in \cite{Hayashi:2016abm,Gu:2018gmy}.

We first write down the explicit form of the elliptic blowup equations
and list various constraints on and properties of the equations. We
then discuss in details two of these properties, the modularity and
the possibility to transform equations along the Higgsing trees.  We
then give some physical arguments for the elliptic blowup equations,
generalising the arguments presented in \cite{Gu:2019pqj} for theories
with no gauge symmetry.  Finally, we illustrate the relation between
the K-theoretic blowup equations and our elliptic blowup equations.

\subsection{Elliptic blowup equations for all rank one theories}
\label{sec:ellipt-blow-equat-1}

Consider a rank one 6d SCFT with tensor branch coefficient $\fn$,
gauge symmetry $G$, flavor symmetry $F$, and half-hypermultiplets
transforming in the representations $(R_{G},R_{F})$.  The flavor
symmetry induces a current algebra of level $k_F$ on the worldsheet of
BPS strings.  Then the elliptic genera $\IE_d(\tau,\mGF,\ep_{1,2})$
satisfy the following elliptic blowup equations
\begin{align}
  &\sum_{\lG\in\phi_{\lo}(Q^\vee(G))}^{\frac{1}{2}||\lG||^2+d'+d''=d+\delta}
    (-1)^{|\phi_{\lo}^{-1}(\lG)|}\nn
  &\phantom{===}\times\theta_{i}^{[a]}(\fn\tau,-\fn\lG\cdot\mG +
    k_F\lF\cdot\mF+(y-\small{\frac{n}{2}}||\lG||^2)(\eq+\et)
    -\fn d'\eq-\fn d''\et)\nonumber \\[-1mm]
  &\phantom{===}\times
    A_{{V}}(\tau,\mG,\lG)
    A_{{H}}(\tau,\mG,\mF,\lG,\lF)\nn
  &\phantom{===}\times
    \IE_{d'}(\tau,\mG+\eq\lG,\mF+\eq\lF,\eq,\et-\eq)
    \IE_{d''}(\tau,\mG+\et\lG,\mF+\et\lF,\eq-\et,\et)\nn
  &\phantom{==}=\Lambda(\delta)\,
    \theta_{i}^{[a]}(\fn\tau,k_F\lF\cdot\mF+y(\eq+\et))
    \IE_d(\tau,\mG,\mF,\eq,\et),\qquad d=0,1,2,\ldots
        \label{eq:ebeq}
\end{align}
where\footnote{We set $\hg = 1$ if gauge symmetry is trivial.}
\begin{equation}\label{eq:y}
  y = \frac{\fn-2+\hg}{4}+\frac{k_F}{2}(\lF\cdot\lF),
\end{equation}
and
\begin{equation}
  \Lambda(\delta) =
  \begin{cases}
    1, & \delta =0,\\
    0, & \delta >0.
  \end{cases}
\end{equation}
In the generalised theta function $\theta_{i}^{[a]}$, the subscript
$i$ is 3 if $\fn$ is even and 4 if $\fn$ is odd, and the
characteristic $a$ of the theta function can be one of the following
$\fn$ numbers
\begin{equation}\label{eq:a}
  a = \frac{1}{2} - \frac{k}{\fn},\quad k =0,1,\ldots,\fn-1.
\end{equation}
Besides, if there is an Abelian factor in the flavor symmetry, the
argument $k_F\lF\cdot \mF$ should be extended to
\begin{equation}
  k_F\lF\cdot \mF \to k_F\lF\cdot\mF+k_{\mf{u}(1)}\lU \mU.
\end{equation}
The summation index $\lG$ is a coweight vector\footnote{This is
  sometimes called a magnetic weight vector in the literature,
  e.g.~\cite{Kapustin:2005py}.  See Appendix~\ref{sc:Lie} for our Lie
  algebraic convention.} of $G$; to be more precise, it takes value in
the shifted coroot lattice defined by the embedding through a coweight
vector $\lo$
\begin{align}
  \phi_{\lo} :
  &Q^\vee \hookrightarrow P^\vee \nn
  &\alpha^\vee \to\alpha^\vee+\lo,\quad \lo\in P^\vee.
\end{align}
The index $\lG$ in fact consists of components of the so-called
$r$-field\footnote{Up to a factor of $1/2$.} in the blowup equations
of topological string, and different $\lG$ correspond to $r$-fields
which are equivalent to each other.  On the other hand, there can be
different embeddings.  The number of different embeddings is the index
of $Q^\vee$ as an Abelian subgroup of $P^\vee$, which is also the
determinant of the Cartan matrix of $G$.  There is a special embedding
where the shift $\lo$ is a coroot vector.  $\delta$ is the smallest
norm in the shifted coroot lattice; it is zero in the special
embedding and positive otherwise.  The inverse $\phi_{\lo}^{-1}$ pulls
back the coweight $\lG$ to the coroot lattice, and $|\bullet|$ in the
sign factor sums up the coefficients in its decomposition in terms of
simple coroots.  We say the blowup equation is of the \emph{unity}
type if the embedding is the special embedding so that $\Lambda$ is
unity.  Otherwise, the r.h.s.~of the blowup equation vanishes
identically and we say the blowup equations are of the
\emph{vanishing} type. Clearly if $P^\vee \cong Q^\vee$, which happens
for $G_2,F_4$ and $E_8$, there can be no vanishing blowup equation.

The components $A_{{V}}$ and $A_{{H}}$ are contributions from
vector and hypermultiplets respectively.  They have the form
\begin{align}
  A_{{V}}(\tau,\mG,\lG)
  &=
    \prod_{\beta\in\Delta_+}
    \breve{\theta}_V(\beta\cdot\mG,\beta\cdot\lG)
    \ , \label{eq:AV}\\
  A_{{H}}(\tau,\mGF,\lGF)
  &=
    \prod_{\w_{G,F}\in R_{G,F}^{+}}
    \breve{\theta}_H(\wG\cdot\mG+\wF\cdot\mF, \wG\cdot\lG+\wF\cdot\lF) \ .
    \label{eq:AH}
\end{align}
Here $R_{G,F}^+$ is half of the total weight space.  For unity
blowup equations
\begin{equation}\label{eq:uR+}
  R_{G,F}^+ = \{\wG\in R_{G},\wF\in R_{F}\,|\, \wF\cdot \lF = +{1}/{2}\},
\end{equation}
and for vanishing blowup equations
\begin{equation}\label{eq:uR+}
  R_{G,F}^+ = \{\wG\in R_{G},\wF\in R_{F}\,|\, \wG\cdot\lG+\wF\cdot \lF >0\}.
\end{equation}
Furthermore, the $\breve{\theta}$ functions are defined as
\begin{align}
  &\breve{\theta}_V(z,R) :=
    \prod_{\substack{m,n\geq 0\\m+n\leq |R|-1}}
  \frac{\eta}{\theta_1(z+s m\eq+s n\et)}
  \prod_{\substack{m,n\geq 0\\m+n\leq |R|-2}}
  \frac{\eta}{\theta_1(z+s(m+1)\eq+s(n+1)\et)} ,
   \quad R\in \IZ ,\label{eq:thetaV}\\
  &\breve{\theta}_H(z,R) :=
    \prod_{\substack{m,n\geq 0\\m+n\leq |R|-3/2}}
  \frac{(-1)^R\theta_1(z+s(m+1/2)\eq+s(n+1/2)\et)}{\eta} \ ,
  \quad R\in
  \frac{1}{2}+\IZ ,\label{eq:thetaH*}
\end{align}
with $s$ the sign of $R$. For unity blowup equations, the factor $(-1)^{R}$ in \eqref{eq:thetaH*} is a constant under the choice \eqref{eq:uR+}, so that we drop it in later computations.

There is still one free parameter $\lF$. It is a
coweight vector of the flavor symmetry, and in fact also consists of
components of the so-called $r$-field\footnote{Up to a factor of
  $1/2$.}  \cite{Gu:2017ccq,Huang:2017mis,Gu:2018gmy}.  The value of
$\lF$ can be determined by the following constraints:
\begin{itemize}
\item Checker board pattern: The second arguments on
  the r.h.s.~of \eqref{eq:AV},\eqref{eq:AH} are one half of the $r$-field
  component associated to
  the refined BPS states of vector and hypermultipets, and thus they must
  satisfy the conditions \cite{Gu:2018gmy}
  \begin{gather}
    \beta\cdot \lG \in \IZ ,\quad \beta\in \Delta,
    \label{eq:RV}\\
    \wG\cdot\lG + \wF \cdot\lF \in \frac{1}{2}+\IZ,\quad \wG\in
    R_{G},\wF\in R_{F}.
    \label{eq:RH}
  \end{gather}
  The first condition only confirms that $\lG$ is a coweight vector of
  $G$.  The second condition constrains that $\lF$ takes value in a
  subset of the coweight lattice of $F$ depending on the domain of
  $\phi_{\lo}$.  In the case of unity equations, $\lG$ is a coroot
  vector of $G$ and $\wG\cdot\lG$ is an integer, \eqref{eq:RH} reduces
  to
  \begin{equation}
    \wF\cdot\lF \in \frac{1}{2} + \IZ,\quad \wF \in R_{F}.
    \label{eq:uRH}
  \end{equation}
  The above conditions are also called the \emph{$B$ field condition}.
\item Modularity: We observe that the elliptic blowup equations
  \eqref{eq:ebeq} are identities of Jacobi forms.  An important
  consistency condition for \eqref{eq:ebeq} is that every term on the
  l.h.s.~should have the same modular weight and modular index, and
  when the r.h.s.  does not vanish, they coincide with the modular
  weight and modular index of the r.h.s.~as well.  The condition on
  modular weight is trivially satisfied as every term has weight one
  half.  The condition on modular index, on the other hand, is highly
  nontrivial and very constraining.  As we will see in
  Section~\ref{sc:mod}, this condition puts strong constraints on
  $\lF$, especially in the case of unity blowup equations.
\item Higgsing: Rank one 6d SCFTs with the same tensor branch
  parameter $\fn$ are related to each other via the Higgs branch RG
  flows.
  As we will discuss in Section~\ref{sc:Hig} unity blowup equations of
  rank one 6d SCFTs can be transformed to each other along the
  Higgsing trees, and in addition the Higgsing process puts some mild
  constraints on $\lF$ for unity blowup equations as well.
\item Leading degree identities: The degree $d$ is the degree of the
  shifted base curve $t_{\text{ell}}$.  In the leading degree with
  $d=d'=d''=0$,
  the elliptic genera do not contribute and the blowup equations
  become identities of Jacobi theta functions.  For unity blowup
  equations the leading order identities are trivial, while for
  vanishing blowup equations the leading order identities are very
  non-trivial and they can be used to constrain the parameter $\lF$.
\end{itemize}

We list below the values of the parameter $\lF$ satisfying all the
four constraints for each rank one model and the corresponding $y$
parameter.  The coweight vectors $\lF$ are presented by their Dynkin
labels. Note that such a coweight vector can be mapped to a weight
vector by the isomorphism $\varphi$ defined in \eqref{eq:iso}.  We are
sometimes sloppy in the main text and refer to $\lF$ as weights, by
which we actually mean the images of $\varphi$. Besides in the main
text we often directly write the factor
$(-1)^{|\phi_{\lo}^{-1}(\lG)|}$ in (\ref{eq:ebeq}) as $(-1)^{|\lG|}$.
We will later test the corresponding elliptic blowup equations in
Sections~\ref{sc:sol} and \ref{sc:ex} by checking them explicitly with
known results of elliptic genera, and by solving unknown elliptic
genera as well as refined BPS invariants from them.

There is another convenient form of elliptic blowup equations, in
which we replace $(y-\small{\frac{n}{2}}||\lG||^2)$ by $(\bar{y}-nd)$
in (\ref{eq:ebeq}). The advantage is that $d$ is always integer for
both unity and vanishing cases, and $\bar{y}$ are typically simpler
numbers than $y$. On the other hand, the merit of the current form
(\ref{eq:ebeq}) is that the modularity proof of both unity and
vanishing blowup equations can be combined together. We will also use
the notion of $d$ and $\bar{y}$ in the example sections, where
$\bar{y}$ and $y$ are related by $y=\bar{y}+n\delta$ in the vanishing
cases and naturally $y=\bar{y}$ in the unity cases.


\subsubsection{Unity blowup equations}
\label{sec:unity-blow-equat}

We tabulate in Tables~\ref{tb:ubeq-1}, \ref{tb:ubeq-2} the coweight
vectors $\lF$ and the associated parameter $y$ for unity blowup
equations which satisfy the four constraints discussed above.  We note
that if a coweight vector $\lF$ is valid, all the vectors in the same
Weyl orbit should be valid as well, and we only list in
Tables~\ref{tb:ubeq-1}, \ref{tb:ubeq-2} the dominant coweight
vectors. We commment that for ease of computation, we have used the
isomorphism of algebras
\begin{equation}
  \mf{so}(2) \cong \mf{sp}(1),\quad \mf{so}(4)\cong \mf{sp}(1)\times
  \mf{sp}(1).
\end{equation}
Whenever possible, we prefer the notation $\mf{sp}(1)$ instead of
$\mf{su}(2)$ as it is more similar to other $C$-algebras instead of
$A$-algebras.

Note that the following theories have unpaired half-hypers and they do
not have unity blowup equations.  Technically this is because their
flavor weight spaces have zero weight, with which the checkerboard
pattern constraint \eqref{eq:uRH} cannot be satisfied.
\begin{itemize}
\item $\fn=1$: $G = \mf{su}(6)_*, \mf{so}(11), \mf{so}(12)_{a,b}, E_7$;
\item $\fn=2$: $G = \mf{so}(12)_b, \mf{so}(13)$;
\item $\fn=3$: $G = \mf{so}(11), \mf{so}(12), E_7$;
\item $\fn=5,7$: $G = E_7$.
\end{itemize}

\begin{table}[h]
\centering
\begin{tabular}{|c|l|l|c|c|c|}\hline
  $\fn$
  &$G$
  &$F$
  &$\#$
  &$y$
  &$\lF$
  \\ \hline
  {12}&$E_8$&$-$
  & 1 & 10 & $\emptyset$ \\\hline
  8&$E_7$ & $-$
  & 1 & 6 & $\emptyset$ \\\hline
  7&$E_7$&$-$
  & 0 & $-$ & $-$ \\\hline
  6&$E_6$&$-$
  & 1 & 4 & $\emptyset$ \\
  6&$E_7$&$\mf{so}(2)_{12} = \mf{sp}(1)_6$
  & 2 & 7 & (1) \\\hline
  5&$F_4$&$-$
  & 1 & 3 & $\emptyset$ \\
  5&$E_6$&$\mf{u}(1)_6$
  & 2 & 9/2  & $\pm 1/2$\\
  5&$E_7$&$\mf{so}(3)_{12}$
  & 0 & $-$ & $-$\\\hline
  4&$\mf{so}(8)$&$-$
  & 1  & 2 & $\emptyset$ \\
  4&$\mf{so}(N\geq 9)$&$\mf{sp}(N-8)_1$
  & $2^{N-8}$ & $(N-4)/2$ & $(0\ldots01)$ \\
  4&$F_4$&$\mf{sp}(1)_3$
  & 2 & 7/2  & $(1)$ \\
  4 &$E_6$&$\mf{su}(2)_{6}\times \mf{u}(1)_{12}$
  & 4 & 5 & $(1)_0$ or $(0)_{\pm \frac{1}{2}}$  \\
  4&$E_7$&$\mf{so}(4)_{12} =\mf{sp}(1)_6\times \mf{sp}(1)_6$
  & 4 & 8 & (1),(1)  \\\hline
  3&$\mf{su}(3)$&$-$
  & 1 & 1 & $\emptyset$ \\
  3&$\mf{so}(7)$&$\mf{sp}(2)_1$
  & 4 & 2 & (01) \\
  3&$\mf{so}(8)$&$\mf{sp}(1)_1^a\times \mf{sp}(1)_1^b\times \mf{sp}(1)_1^c$
  & 8 & 5/2 & (1),(1),(1) \\
  3&$\mf{so}(9)$&$\mf{sp}(2)_1^a\times \mf{sp}(1)_2^b$
  & 8 & 3 & (01),(1) \\
  3&$\mf{so}(10)$&$\mf{sp}(3)_1^a\times (\mf{su}(1)_4\times \mf{u}(1)_4)^b$
  & 16 & 7/2 & (001),$\pm 1/2$ \\
  3&$\mf{so}(11)$&$\mf{sp}(4)_1^a\times \text{Ising}^b$
  & 0 & $-$ & $-$ \\
  3&$\mf{so}(12)$&$\mf{sp}(5)_1$
  & 0 & $-$ & $-$ \\
  3&$G_2$&$\mf{sp}(1)_1$
  & 2 & 3/2  & (1) \\
  3&$F_4$&$\mf{sp}(2)_3$
  & 4 & 4 & (01) \\
  3 &$E_6$&$\mf{su}(3)_{6}\times \mf{u}(1)_{18}$
  & 8  & 11/2
  & $\pm(01)_{\frac{1}{6}}$ or $(00)_{\pm \frac{1}{2}}$ \\
  3&$E_7$&$\mf{so}(5)_{12}$
  & 0  & $-$  & $-$ \\\hline
\end{tabular}
\caption{The parameters $y, \lF$ of unity blowup equations for rank
  one models with $\fn\geq 3$.  \# is the number of unity equations
  with fixed characteristic $a$.}
\label{tb:ubeq-1}
\end{table}

\begin{table}[h]
  \centering%
  \resizebox{\linewidth}{!}{
  \begin{tabular}{|c|l|l|c|c|c|c|}\hline
    $\fn$&$G$&$F$
    & $\#$
    & $y$
    & $\lF$
    \\ \hline
    2&$\mf{su}(1)$&$\mf{su}(2)_1$
    & 2 & 1/2 & (1) \\
    2&$\mf{su}(2)$&$\mf{so}(8)_1\to \mf{so}(7)_1\times\text{Ising}$
    & 6 & 1 & (100) \\
    2&$\mf{su}(N\geq 3)$&$\mf{su}(2N)_1$
    & ${2N\choose N}$ & N/2 & $(0\ldots010\ldots0)$ \\
    2&$\mf{so}(7)$&$\mf{sp}(1)_1^a\times \mf{sp}(4)_1^b$
    & 32 & 5/2 & (1),(0001) \\
    2&$\mf{so}(8)$&$\mf{sp}(2)_1^a\times \mf{sp}(2)_1^b\times \mf{sp}(2)_1^c$
    & 64 & 3 & (01),(01),(01) \\
    2&$\mf{so}(9)$&$\mf{sp}(3)_1^a\times \mf{sp}(2)_2^b$
    & 32 & 7/2 & (001),(01) \\
    2&$\mf{so}(10)$&$\mf{sp}(4)_1^a\times (\mf{su}(2)_4\times \mf{u}(1)_{8})^b$
    & 64 & 4 & (0001) and $(1)_0$ or $(0)_{\pm\frac{1}{2}}$ \\
    2&$\mf{so}(11)$&$\mf{sp}(5)_1^a\times (?\to\mf{so}(2)_8)^b$
    & 64 & 9/2 & (00001),(1) \\
    2&$\mf{so}(12)_a$&$\mf{sp}(6)_1^a\times \mf{so}(2)_8^b$
    & 128 & 5 & (000001),(1) \\
    2&$\mf{so}(12)_b$&$\mf{sp}(6)_1^a\times \text{Ising}^b\times \text{Ising}^c$
    & 0 & $-$ & $-$ \\
    2&$\mf{so}(13)$&$\mf{sp}(7)_1$
    & 0 & $-$ &  $-$ \\
    2&$G_2$&$\mf{sp}(4)_1$
    & 16 & 2 & (0001) \\
    2&$F_4$&$\mf{sp}(3)_3$
    & 8 & 9/2 & (001) \\
    2 &$E_6$&$\mf{su}(4)_{6}\times \mf{u}(1)_{24}$
    & 16 & 6 & $(010)_{0}$ or
               $\pm(001)_{\frac{1}{4}}$ or $(000)_{\pm \frac{1}{2}}$\\
    2&$E_7$&$\mf{so}(6)_{12}$
    & 8 & 9 & (001) or (010) \\\hline
    1&$\mf{sp}(0)$&$(E_8)_1$
    & 240 & 1 & $(10\ldots0)$  \\
    1&$\mf{sp}(N\geq 1)$&$\mf{so}(4N+16)_1$
    & $2^{2N+7}$ & $(N+2)/2$
    & $(0\ldots01)$ \\
    1&$\mf{su}(3)$&$\mf{su}(12)_1$
    & 924 & 2 & $(0\ldots010\ldots0)$ \\
    1&$\mf{su}(4)$&$\mf{su}(12)_1^a\times \mf{su}(2)_1^b$
    & 1848 & 5/2 & $(0\ldots010\ldots0),(1)$ \\
    1 &$\mf{su}(N\geq 5)$&$\mf{su}(N\!+\!8)_1\!\times\! \mf{u}(1)_{2N(N-1)(N+8)}$
    & $2{N+8\choose 6}$ & $(N+1)/2$
    & $(0000010\ldots)_{-\frac{1}{2(N+8)}}$ or minus\\
    1&$\mf{su}(6)_*$&$\mf{su}(15)_1$
    & 0 & $-$ & $-$  \\
    1&$\mf{so}(7)$&$\mf{sp}(2)_1^a\times \mf{sp}(6)_1^b$
    & 256 & 3 & (01),(000001)  \\
    1&$\mf{so}(8)$&$\mf{sp}(3)_1^a\times \mf{sp}(3)_1^b\times \mf{sp}(3)_1^c$
    & 512 & 7/2 & (001),(001),(001) \\
    1&$\mf{so}(9)$&$\mf{sp}(4)_1^a\times \mf{sp}(3)_2^b$
    & 128 & 4 & (0001),(001) \\
    1&$\mf{so}(10)$&$\mf{sp}(5)_1^a\times (\mf{su}(3)_4\times \mf{u}(1)_{12})^b$
    & 256 & 9/2 & (00001), and $\pm(01)_{\frac{1}{6}}$ or $(00)_{\pm\frac{1}{2}}$  \\
    1 &$\mf{so}(11)$&$\mf{sp}(6)_1^a\times ?^b$
    & 0 & $-$ & $-$  \\
    1&$\mf{so}(12)_a$&$\mf{sp}(7)_1^a\times \mf{so}(3)_8^b$
    & 0 & $-$ & $-$ \\
    1&$\mf{so}(12)_b$&$\mf{sp}(7)_1^a\times ?^b\times ?^c$
    & 0 & $-$ & $-$  \\
    1&$G_2$&$\mf{sp}(7)_1$
    & 128 & 5/2 & $(0\ldots01)$ \\
    1&$F_4$&$\mf{sp}(4)_3$
    & 16 & 5 & $(0\ldots01)$  \\
    1 &$E_6$&$\mf{su}(5)_{6}\times \mf{u}(1)_{30}$
    & 32 & 13/2 & $\pm(0001)_{\frac{3}{10}}$ or
                  $\pm(0010)_{\frac{1}{10}}$ or $(0000)_{\pm\frac{1}{2}}$ \\
    1&$E_7$&$\mf{so}(7)_{12}$
    & 0 & $-$ & $-$  \\\hline
\end{tabular}}
\caption{The parameters $y, \lF$ of unity blowup equations for rank
  one models with $\fn=1,2$.  $\#$ is the number of unity equations
  with fixed characteristic $a$.}
\label{tb:ubeq-2}
\end{table}

\subsubsection{Vanishing blowup equations}
\label{sec:vanish-blow-equat}

We tabulate in Tables~\ref{tb:vbeq-1}, \ref{tb:vbeq-2},
\ref{tb:vbeq-3} the values of $\lF$ and the associated parameter $y$
for vanishing equations that satisfy the constraints discussed in the
beginning of this section.  In particular, we have tested the leading
degree identities for all the vanishing blowup equations up to order
20 in $\Qtau = \exp(2\pi\ri\tau)$. We find unlike the unity
$\lambda_F$ fields which form Weyl orbits, the admissible vanishing
$\lambda_F$ fields typically form representations rather than just
Weyl orbits. To be precise, the admissible vanishing $\lambda_F$
fields are all coweight vectors inside the representation whose
highest coweight is given in Dynkin label in Tables~\ref{tb:vbeq-1},
\ref{tb:vbeq-2}, \ref{tb:vbeq-3}. Note one representation in general
contains many Weyl orbits.
Besides, different Weyl orbits inside one representation in general
have different associated $y$ which are easily computable with
equation (\ref{eq:y}).  Thus for the situation where several values of
$y$ are involved, we leave $\cdots$ in Tables~\ref{tb:vbeq-1},
\ref{tb:vbeq-2}, \ref{tb:vbeq-3}.

\begin{table}[h]
\centering
\resizebox{\linewidth}{!}{
\begin{tabular}{|c|l|l|c|c|c|}\hline
  $n$&$G$&$F$
  &$\lo$ & $y$ & $\lF$
  \\ \hline
  {12}&$E_8$&$-$
  & $-$ & $-$ & $-$ \\\hline
  8&$E_7$ & $-$
  &(0000010)&6&$\emptyset$ \\\hline
  7&$E_7$&$-$
  &(0000010)&23/4&$\emptyset$ \\\hline
  6&$E_6$&$-$
  &(100000)&4&$\emptyset$ \\
  &&&(000010)&4&$\emptyset$\\
  6&$E_7$&$\mf{so}(2)_{12} = \mf{sp}(1)_6$
  &(0000010)&11/2&(0) \\\hline
  5&$F_4$&$-$
  &$-$&$-$&$-$ \\
  5&$E_6$&$\mf{u}(1)_6$
  &(100000)&23/6 &$-1/6$  \\
     &&&(000010)&23/6 & $1/6$  \\
  5&$E_7$&$\mf{so}(3)_{12}$
  &(0000010)&21/4&(0) \\\hline
  4&$\mf{so}(8)$&$-$
  &all three&2&$\emptyset$ \\
  4&$\mf{so}(2N), N\geq 5$&$\mf{sp}(2N-8)_1$
  &($10\ldots0$)&$N-2$&($0\ldots01$) \\
     &&&($\ldots010$)&$\cdots$&$[N-2,0\ldots00]$\\
     &&&($\ldots001$)&$\cdots$&$[N-2,0\ldots00]$\\
  4&$\mf{so}(2N-1), N\geq 5$&$\mf{sp}(2N-9)_1$
  &$(10\ldots0)$&$(2N-5)/2$&$(0\ldots01)$\\
  4&$F_4$&$\mf{sp}(1)_3$
  &$-$&$-$&$-$ \\
  4 &$E_6$&$\mf{su}(2)_{6}\times \mf{u}(1)_{12}$
  & (100000)&11/3&$(0),-1/6$ \\
     &&&(000010)&11/3&$(0),1/6$ \\
  4&$E_7$&$\mf{so}(4)_{12} =\mf{sp}(1)_6\times \mf{sp}(1)_6$
  &(0000010)&5&(0),(0) \\\hline
  3&$\mf{su}(3)$&$-$
  &(10) or (01)&1&$\emptyset$ \\
  3&$\mf{so}(7)$&$\mf{sp}(2)_1$
  &(100)& $\cdots$ &[10] \\
  3&$\mf{so}(8)$&$\mf{sp}(1)_1^a\times \mf{sp}(1)_1^b\times \mf{sp}(1)_1^c$
  &(1000)&  $\cdots$  &(1),(0),[2] or (1),[2],(0) \\
     &&&(0010)&  $\cdots$   &[2],(1),(0) or (0),(1),[2] \\
     &&&(0001)&  $\cdots$  &(0),[2],(1) or [2],(0),(1) \\
  3&$\mf{so}(9)$&$\mf{sp}(2)_1^a\times \mf{sp}(1)_2^b$
  &(1000)&5/2&(01),(0) \\
  3&$\mf{so}(10)$&$\mf{sp}(3)_1^a\times (\mf{su}(1)_4\times \mf{u}(1)_4)^b$
  & (10000)&3&(001),(0),0 \\
     &&&(00010)&$\cdots$&$[j00],(0),-1/4+\ell$:
                                                    see text\\
     &&&(00001)&$\cdots $&$[j00],(0),1/4-\ell$:
                                                    see text\\
  3&$\mf{so}(11)$&$\mf{sp}(4)_1^a\times \text{Ising}^b$
  &(10000)&7/2&(0001) \\
  3&$\mf{so}(12)$&$\mf{sp}(5)_1$
  &(100000)&4&(00001) \\
     &&&(000001)&$\cdots$&$[30000]$\\
     &&&(000010)&$-$&$-$\\
  3&$G_2$&$\mf{sp}(1)_1$
  &$-$&$-$&$-$ \\
  3&$F_4$&$\mf{sp}(2)_3$
  &$-$&$-$&$-$ \\
  3 &$E_6$&$\mf{su}(3)_{6}\times \mf{u}(1)_{18}$
  &(100000)&7/2&$(00),-1/6$ \\
     &&&(000010)&7/2&$(00),1/6$\\
  3&$E_7$&$\mf{so}(5)_{12}$
  &(0000010)&19/4&(00)\\\hline
\end{tabular}}
\caption{The parameters $y, \lF$ of vanishing blowup equations for
  rank one models with $\fn\geq 3$. In the column of $\lambda_F$, the representations are labeled by their highest coweights. When a representation is composed by many Weyl orbits, we use $[*]$ instead of $(*)$ to stress the difference.}
\label{tb:vbeq-1}
\end{table}

\begin{table}[h]
\centering
\resizebox{\linewidth}{!}{
  \begin{tabular}{|c|l|l|c|c|c|}\hline
    $n$&$G$&$F$
    &$\lo$ & $y$ & $\lF$
    \\ \hline
    2&$\mf{su}(1)$&$\mf{su}(2)_1$
    &$-$&$-$&$-$ \\
    2&$\mf{su}(2)$&$\mf{so}(8)_1\to \mf{so}(7)_1\times\text{Ising}$
    &(1)&1/2&(000) \\\cline{4-6}
    2&$\mf{su}(N\geq 3)$&$\mf{su}(2N)_1$
    &\multicolumn{3}{c|}{see text} \\\cline{4-6}
    2&$\mf{so}(7)$&$\mf{sp}(1)_1^a\times \mf{sp}(4)_1^b$
    &(100)& $\cdots$  &(1),[1000] \\
    2&$\mf{so}(8)$&$\mf{sp}(2)_1^a\times \mf{sp}(2)_1^b\times \mf{sp}(2)_1^c$
    &(1000)& $\cdots$ &(01),(00),[10] or (01),[10],(00) \\
       &&&(0010)&$\cdots$&[10],(01),(00) or (00),(01),[10]\\
       &&&(0001)&$\cdots$&(00),[10],(01) or [10],(00),(01)\\
    2&$\mf{so}(9)$&$\mf{sp}(3)_1^a\times \mf{sp}(2)_2^b$
    &(1000)&5/2&(001),(00) \\
    2&$\mf{so}(10)$&$\mf{sp}(4)_1^a\times (\mf{su}(2)_4\times \mf{u}(1)_{8})^b$
    &(10000)&3&(0001),(0),0 \\
       &&&(00010)&$\cdots$&$[j000],(0),-1/4+\ell$:
                                                   see text\\
       &&&(00001)&$\cdots$&$[j000],(0),1/4-\ell$:
                                                   see text\\
    2&$\mf{so}(11)$&$\mf{sp}(5)_1^a\times (?\to\mf{so}(2)_8)^b$
    &(10000)&7/2&(00001),(0) \\
    2&$\mf{so}(12)_a$&$\mf{sp}(6)_1^a\times \mf{so}(2)_8^b$
    &(100000)&4&(000001),(0) \\
       &&&(000001)&$\cdots$&$[300000],(0)$\\
       &&&(000010)&$\cdots$&$[200000],(1)$\\
    2&$\mf{so}(12)_b$&$\mf{sp}(6)_1^a\times \text{Ising}^b\times \text{Ising}^c$
    &(100000)&4&(000001) \\
       &&&(000001)&$-$&$-$\\
       &&&(000010)&$-$&$-$\\
    2&$\mf{so}(13)$&$\mf{sp}(7)_1$
    &(100000)&9/2&(0000001) \\
    2&$G_2$&$\mf{sp}(4)_1$
    &$-$&$-$&$-$ \\
    2&$F_4$&$\mf{sp}(3)_3$
    &$-$&$-$&$-$ \\
    2 &$E_6$&$\mf{su}(4)_{6}\times \mf{u}(1)_{24}$
    &(100000)&10/3& $(000),-1/6$\\
       &&&(000010)&10/3& $(000),1/6$\\
    2&$E_7$&$\mf{so}(6)_{12}$
    &(0000010)& 9/2 & (000) \\\hline
    \end{tabular}}
\caption{The parameters $y, \lF$ of vanishing blowup equations for
  rank one models with $\fn=2$. In the column of $\lambda_F$, the representations are labeled by their highest coweights. When a representation is composed by many Weyl orbits, we use $[*]$ instead of $(*)$ to stress the difference.
  See the main text
  for more discussion.}
\label{tb:vbeq-2}
\end{table}
\begin{table}[h]
\centering
\resizebox{\linewidth}{!}{
  \begin{tabular}{|c|l|l|c|c|c|}\hline
    $n$&$G$&$F$
    &$\lo$ & $y$ & $\lF$
    \\ \hline
    1&$\mf{sp}(0)$&$(E_8)_1$
    &$\emptyset$&0&($0\ldots0$) \\
    1&$\mf{sp}(N\geq 1)$&$\mf{so}(4N+16)_1$
    &$(0\ldots01)$&$\cdots$&$[N,0\ldots0]$\\
    1&$\mf{su}(3)$&$\mf{su}(12)_1$
    &(10) or (01)& $\cdots$ &[$\ldots02$] or [$20\ldots$] \\
    1&$\mf{su}(4)$&$\mf{su}(12)_1^a\times \mf{su}(2)_1^b$
    &(100) or (001)& $\cdots$ &[$\ldots03$] or [$30\ldots0$],(0) \\
       &&&(010)& $\cdots$ & [$10\ldots01$],(1)\\
    \cline{4-6}
    1 &$\mf{su}(N\geq 5)$&$\mf{su}(N\!+\!8)_1\!\times\! \mf{u}(1)_{2N(N-1)(N+8)}$
    &\multicolumn{3}{c|}{see text}\\
    \cline{4-6}
    1&$\mf{su}(6)_*$&$\mf{su}(15)_1$
    &(10000) or (00001)& $\cdots$ &[$\ldots05$] or [$50\ldots$] \\
           & &  &(00100)&$\cdots$ & $[20\ldots02]$\\
    1&$\mf{so}(7)$&$\mf{sp}(2)_1^a\times \mf{sp}(6)_1^b$
    &(100)& $\cdots$&(01),[100000] \\
    1&$\mf{so}(8)$&$\mf{sp}(3)_1^a\times \mf{sp}(3)_1^b\times \mf{sp}(3)_1^c$
    &(1000)&$\cdots$ &(001),(000),[100] or (001),[100],(000)\\
           &&&(0010)& $\cdots$ & [100],(001),(000) or (000),(001),[100]\\
       &&&(0001)& $\cdots$ & (000),[100],(001) or [100],(000),(001)\\
    1&$\mf{so}(9)$&$\mf{sp}(4)_1^a\times \mf{sp}(3)_2^b$
    &(1000)&5/2&(0001),(000) \\
    1&$\mf{so}(10)$&$\mf{sp}(5)_1^a\times (\mf{su}(3)_4\times \mf{u}(1)_{12})^b$
    &(10000)&3&(00001),(0),0 \\
       &&&(00010)&$\cdots$&$[20000],(0),-1/4$\\
       &&&(00001)&$\cdots$&$[20000],(0),1/4$\\
    1 &$\mf{so}(11)$&$\mf{sp}(6)_1^a\times (?\to{\mf{sp}(1)_6})^b$
    &(10000)&7/2&(000001),(0) \\
    1&$\mf{so}(12)_a$&$\mf{sp}(7)_1^a\times \mf{so}(3)_8^b$
    &(100000)&4&(0000001),(0) \\
       &&&(000001)&$\cdots$&$[3000000],(0)$ \\
       &&&(000010)&$-$&$-$ \\
    1&$\mf{so}(12)_b$&$\mf{sp}(7)_1^a\times (?\to{\mf{sp}(1)_4})^b\times ?^c$
    &(100000)&4&(0000001),(0) \\
       &&&(000001)&$-$&$-$ \\
       &&&(000010)&$\cdots$&$[2000000],(1)$ \\
    1&$G_2$&$\mf{sp}(7)_1$
    &$-$&$-$&$-$ \\
    1&$F_4$&$\mf{sp}(4)_3$
    &$-$&$-$&$-$ \\
    1 &$E_6$&$\mf{su}(5)_{6}\times \mf{u}(1)_{30}$
    &(100000)&19/6&$(0000),-1/6$ \\
       &&&(000010)&19/6&$(0000),1/6$\\
    1&$E_7$&$\mf{so}(7)_{12}$
    &(0000010)&17/4 & (000) \\\hline
\end{tabular}}
\caption{The parameters $y, \lF$ of vanishing blowup equations for
  rank one models with $\fn=1$. In the column of $\lambda_F$, the representations are labeled by their highest coweights. When a representation is composed by many Weyl orbits, we use $[*]$ instead of $(*)$ to stress the difference. We make assumption for flavor symmetries of the $G=\mf{so}(11)$ and $\mf{so}(12)_b$   models which allow for vanishing blowup equations;
  see the main text
  for more discussion.}
\label{tb:vbeq-3}
\end{table}

In the following we discuss some special cases in
Tables~\ref{tb:vbeq-1}, \ref{tb:vbeq-2}, \ref{tb:vbeq-3} in more
detail.
\begin{itemize}
\item $G=\mf{so}(7)$: For $\fn=3$, the representation $[10]$ of
  $\lambda_{\mf{sp}(2)}$ has two Weyl orbits generated by coweights
  $(00)$ and $(10)$, whose associated $y$ are 3/2 and 5/2
  respectively.  For $\fn=2$, the representation $[1000]$ of
  $\lambda_{\mf{sp}(4)}$ has two Weyl orbits generated by coweights
  $(0000)$ and $(1000)$, whose associated $y$ are 3/2 and 5/2
  respectively.  For $\fn=1$, the representation $[100000]$ of
  $\lambda_{\mf{sp}(6)}$ has two Weyl orbits generated by coweights
  $(000000)$ and $(100000)$, whose associated $y$ are 3/2 and 5/2
  respectively.
\item $G=\mf{so}(8)$: For $\fn=3$, the representation $[2]$ of
  $\lambda_{\mf{sp}(1)}$ has two Weyl orbits $(0)$ and $(2)$, whose
  associated $y$ are 2 and 3 respectively.  For $\fn=2$, the
  representation $[10]$ of $\lambda_{\mf{sp}(2)}$ has two Weyl orbits
  generated by coweights $(00)$ and $(10)$, whose associated $y$ are 2
  and 3 respectively.  For $\fn=1$, the representation $[100]$ of
  $\lambda_{\mf{sp}(3)}$ has two Weyl orbits generated by coweights
  $(000)$ and $(100)$, whose associated $y$ are 2 and 3 respectively.
\item $G=\mf{so}(10)$: For $\fn=3$, the possibilities are
  \begin{equation}
    \begin{aligned}
      &\ell=0,\;\;j=2,\\
      &\ell=-1,\;\;j=0,\\
      &\ell=1,\;\;j=1.
    \end{aligned}
  \end{equation}
  For $\fn=2$, the possibilities are
  \begin{equation}
    \begin{aligned}
      &\ell=0,\;\;j=2,\\
      &\ell=1,\;\;j=0.
    \end{aligned}
  \end{equation}
  For $\fn=1$, the possibilities are
  \begin{equation}
    \begin{aligned}
      \ell=0,\;\;j=2.
    \end{aligned}
  \end{equation}
\item $\fn=2, G=\mf{su}(N), N\geq 3$: when
  $\lo = {\wv_j}^{\mf{su}(N)}$, $j=1,2,\dots,N-1$,
  \begin{equation}\label{eq:rho-1}
    \lF^{\mf{su}(2N)} \in  [j-1,0,\dots,0,(N-1-j)].
  \end{equation}
  For fixed $N$ and $j$, this is a very large representation which
  contains many Weyl orbits, each of which has its own associated
  $y$. We do not list all of them since they are easily computable
  from equation (\ref{eq:y}). Instead, we just point out one
  a particularly simple Weyl orbit inside such representation. For
  example, if $N$ is odd,
  \begin{equation}\label{eq:rho-1}
    \lo = {\wv_j}^{\mf{su}(N)}, \quad \lF \in{\cO_{N+2j}}^{\mf{su}(2N)},
    \quad y = \frac{N^2-2j^2}{2N},\quad j=1,\ldots,(N-1)/2,
  \end{equation}
  and
  \begin{equation}\label{eq:rho-2}
    \lo = {\wv_{N-j}}^{\mf{su}(N)},\quad \lF \in {\cO_{N-2j}}^{\mf{su}(2N)},
    \quad y =\frac{N^2-2j^2}{2N},\quad j=1,\ldots,(N-1)/2,
  \end{equation}
  where $\cO_{i}$ is the Weyl orbit generated by the $i$-th
  fundamental coweight.
  If $N$ is even, $j$ runs from 1 up to $N/2$ in the equations
  (\ref{eq:rho-1}) and (\ref{eq:rho-2}). We will explicitly show the
  leading degree vanishing identities for these Weyl orbits in Section
  (\ref{sec:exn2su}).
\item $\fn=1, G=\mf{su}(N), N\geq 5$: For
  $k\leq \lfloor N/2\rfloor$, we have
  \begin{equation}
    \lo = {\wv_k}^{\mf{su}(N)},\quad \lF^{U(1)} =
    \frac{4k-N}{2N(N+8)},\quad \lF^{\mf{su}(N+8)} \in[k-1,0,\dots,0,N+1-2k] ,
  \end{equation}
  The associated parameters $y$ are
  computed by \eqref{eq:y} as,
  \begin{equation}
    y = \frac{N-1}{4}+\frac{1}{2}
    (\lF^{\mf{su}(N+8)},\lF^{\mf{su}(N+8)})
    + N(N-1)(N+8) (\lF^{U(1)})^2.
  \end{equation}
  The cases of
  $k > \lfloor N/2 \rfloor$ can be obtained by complex conjugation.
\item $\fn=1, G=\mf{so}(11),\mf{so}(12)_b$: the flavor
  symmetries consistent at the level of worldsheet theory are not
  known for these two models \cite{DelZotto:2018tcj}, especially the
  component governing the three half-hypers in spinor representation
  of $\mf{so}(11)$ in the first model, and the component governing the
  two half-hypers in spinor and and one half-hyper in conjugate spinor
  representations of $\mf{so}(12)$ in the second model.  We find that
  if we assume the three half-hypers in the first model transform as
  $\md{3}$ of flavor symmetry $\mf{sp}(1)$, and the two half-hypers in
  the second model transform as $\md{2}$ of flavor symmetry
  $\mf{sp}(1)$, we can find $\lF$ of vanishing equations which satisfy
  all the constraints discussed in the beginning of this section.  In
  particular, we have checked the leading base degree identities also
  up to order 20 in $q$.
\end{itemize}

Let us give a simple example of the leading base degree
identities. Consider $n=1,G=\mf{su}(3),F=\mf{su}(12)$ theory with
matter representation $(\mathbf{3},\mathbf{\overline{12}})$. Let us
look at the situation with $\lG=(10)_{\mf{su}(3)}=\mathbf{3}$. The
admissible $\lF$ form representation $[\ldots002]_{\mf{su}(12)}$ which
has two Weyl orbits $(\ldots010)$ and $(\ldots002)$.
%
The first Weyl orbit itself is a representation
$\mathbf{\overline{66}}$. In this case, the leading base degree of the
vanishing blowup equations gives the following identity:
$\forall \lambda\in \mathbf{\overline{66}}$,
\begin{equation}\label{thetaindentI}
  \sum_{w\in
    \mathbf{3}}(-1)^{|w|}\theta_1(-m_w+m_{\lambda}+2\ep_+)
  \prod_{\beta\in\Delta(\mf{su}(3))}^{w\cdot\beta=1}\frac{1}{\theta_1(m_{\beta})}
  \prod_{\mu\in\mathbf{\overline{12}}}^{\mu\cdot\lambda=5/6}
  {\theta_1(m_{w}+m_{\mu}+\ep_+)}=0.
\end{equation}
We have checked this identity to $\mathcal{O}(q^{20})$. To write it
more explicitly, we have
\begin{equation}\label{thetaindentIeasy}
  \begin{aligned}
    \phantom{+}
    &\frac{\theta_1(-a_1+x_i+x_j)\theta_1(a_1+x_i)\theta_1(a_1+x_j)}{\theta_1(a_1-a_2)\theta_1(a_1-a_3)}
    +\frac{\theta_1(-a_2+x_i+x_j)\theta_1(a_2+x_i)\theta_1(a_2+x_j)}{\theta_1(a_2-a_1)\theta_1(a_2-a_3)}\\+
    &\frac{\theta_1(-a_3+x_i+x_j)\theta_1(a_3+x_i)\theta_1(a_3+x_j)}{\theta_1(a_3-a_1)\theta_1(a_3-a_2)}=0,\quad\quad
    \textrm{for } a_1+a_2+a_3=0,\ 1\le i< j\le12.
\end{aligned}
\end{equation}
Here $a_k,k=1,2,3$ are the $\mf{su}(3)$ fugacities and
$x_i=m_i+\epsilon_+,i=1,2,\dots,12$ where $m_i$ are the symmetric
$\mf{su}(12)$ fugacities. The modularity here means that each among
the three terms in the above equation has the same index
$-(a_1^2+a_2^2+a_3^2)/2$.  The leading base degree identities from the
other set of vanishing blowup equations are just similar.  We have
tons of vanishing theta identities like this involving the root and
weight lattices of Lie algebras from the leading base degree of
vanishing blowup equations. We present some of them in Section
\ref{sc:ex} and make a summary in Appendix \ref{app:list}.
%

\subsection{Modularity}
\label{sc:mod}

Here we discuss the modularity constraint.  For both unity and
vanishing elliptic blowup equations, every term on the l.h.s.~of
\eqref{eq:ebeq} should have the same modular index independent of the
summation index $\lG,d',d''$.  The modular index polynomial for the
generalised theta function $\theta_{i}^{[a]}(\fn\tau,z)$ is
\begin{equation}
  \Ind \theta_{i}(n\tau,z) = \frac{1}{2\fn}z^2.
\end{equation}
The modular index polynomial for $d$-string elliptic genus for a rank
one model can be deduced from \eqref{eq:IndEd}
\cite{DelZotto:2016pvm,DelZotto:2017mee}
\be
\ba
  \Ind \IE_d(\eq,\et,\mG,\mF) =
  &-\Big(\frac{\eq+\et}{2}\Big)^2(2-\fn+\hg)d +
    \frac{\eq\et}{2}(\fn d^2 +(2-\fn)d) \nn
  & +\frac{d}{2} (- \fn\,\mG\cdot\mG +
    k_F\,\mF\cdot\mF ) \ .
    \label{eq:Ed-index}
\ea
\ee
Here if the flavor
  symmetry has a $\mf{u}(1)$ factor, we should replace
  \begin{equation}
    k_F \mF\cdot\mF \to k_F
    \mF\cdot\mF+k_{\mf{u}(1)}\mU^2.
  \end{equation}
Finally the index polynomials of $A_{ V}$, $A_{ H}$ can be calculated from
their definitions \eqref{eq:AV}, \eqref{eq:AH}, \eqref{eq:thetaV},
\eqref{eq:thetaH*}; in particular, the following results are useful
\begin{align}
  \Ind \breve{\theta}_V(z,R) =
  &-\frac{R^2z^2}{2}- \frac{(R-1)R(R+1)}{3}z(\eq+\et) -
    \frac{(R-1)R^2(R+1)}{12}(\ep_1^2+\eq\et+\ep_2^2),\\[-2mm]
  \Ind \breve{\theta}_H(z,R) =
  &\,\frac{(R+1/2)(R-1/2)}{4}z^2 + \frac{R(R-1/2)(R+1/2)}{6}
    z(\eq+\et)\nn
  +\,&\frac{(R-1/2)(R+1/2)(R^2-3/4)}{24}(\ep_1^2+\ep_2^2)
    +\frac{(R-1/2)(R+1/2)(R^2+3/4)}{24} \eq\et.
\end{align}
If we compute the modular index polynomial of an arbitary term on the
l.h.s.~subtracted by that of the r.h.s., we find that the dependence
on $\lG,d',d''$ cancel completely thanks to the choice of $y$
\eqref{eq:y} and the constraints on the number of hypermultiplets
imposed by the anomaly cancellation conditions \eqref{eq:anm-A},
\eqref{eq:anm-B}, \eqref{eq:anm-C}, \eqref{eq:anm-kF},
\eqref{eq:anm-E}, \eqref{eq:anm-k1}.  What remains is a quadratic
polynomial of the following form
\begin{equation}
  \Ind(\delta,d,\mG,\ep_{1,2},\lF) = \delta^2 P_2(d,\mG,\ep_{1,2},\lF)+ \delta P_1(d,\mG,\ep_{1,2},\lF)+ P_0(d,\mG,\ep_{1,2},\lF).
\end{equation}
where
\begin{align}
  P_0(d,\mG,\ep_{1,2},\lF) =
  &\frac{1}{8}n_{R_G,R_F}(2\ind_{R_G}(\mG\cdot\mG)\,\text{id}_1 + \dim R_G\,\text{id}_2)
    +\frac{1}{12}n_{R_G,R_F} \dim R_G \,\text{id}_3 (\eq+\et)\nn
  &+\frac{1}{48}n_{R_G,R_F}\dim R_G (\text{id}_4-\text{id}_1) (\eq+\et)^2\nn
  &-\Big(\frac{1}{2}n_{R_G,R_F} \ind_{R_G}d\,\text{id}_1+\frac{1}{96}n_{R_G,R_F}\dim
    R_G(2\text{id}_4-5\text{id}_1)\Big)\eq\et
\end{align}
and
\begin{align}
  &\text{id}_1 = \sum_{\w\in R_F} \Big((\w\cdot\lF)^2 -
    \frac{1}{4}\Big),
  \label{eq:ucdn1}\\
  &\text{id}_2 = \sum_{\w\in R_F} \Big((\w\cdot\lF)^2(\w\cdot\mF)^2 -
    \frac{1}{4} (\w\cdot\mF)^2\Big),
  \label{eq:ucdn2}\\
  &\text{id}_3 = \sum_{\w\in R_F} \Big((\w\cdot\lF)^3(\w\cdot\mF) -
    \frac{1}{4} (\w\cdot\lF)(\w\cdot\mF)\Big),
  \label{eq:ucdn3}\\
  &\text{id}_4 = \sum_{\w\in R_F} \Big((\w\cdot\lF)^4 -
    \frac{1}{4} (\w\cdot\lF)^2\Big).
    \label{eq:ucdn4}
\end{align}
In the case of unity blowup equations with $\delta=0$, the index
polynomial $\Ind(0,d,\mG,\ep_{1,2},\lF)$ should vanish identically for
arbitrary values of $d,\ep_{1,2},\mG$.  This implies the additional
conditions
\begin{equation}
  \text{id}_1 = \text{id}_2=\text{id}_3=\text{id}_4=0.
\end{equation}

The checkerboard pattern condition \eqref{eq:uRH} together with the
first condition \eqref{eq:ucdn1} above lead to the identity
\begin{equation}\label{eq:lFhalf}
  \w\cdot \lF = \pm\frac{1}{2},\quad \w\in R_F,
\end{equation}
with which the other three conditions above
\eqref{eq:ucdn2},\eqref{eq:ucdn3},\eqref{eq:ucdn4} are automatically
satisfied.  The identity \eqref{eq:lFhalf} fixes the norm of $\lF$
\begin{equation}
  \lF\cdot\lF = \frac{1}{2\ind_{R_F}} \sum_{\w\in R_F}(\w\cdot \lF)^2 =
  \frac{\hg-3\fn+6}{2k_F},
\end{equation}
where we have used \eqref{eq:anm-A},\eqref{eq:anm-kF}.  The expression
\eqref{eq:y} can then be simplified to
\begin{equation}\label{eq:uy}
  y = \frac{\hg -\fn+2}{2}.
\end{equation}

The identity \eqref{eq:lFhalf} completely characterises the coweight
vector $\lF$ for unity blowup equations.  There are only four types of
flavor symmetry.  The half-hypers transform in either $\md{2n}$ of
$\mf{sp}(n)$, or $\md{2n}$ of $\mf{so}(2n)$\footnote{Flavor symmetries
  of type $\mf{so}(2n+1)$ only appears in rank one models with no
  unity blowup equations and thus we do not have to consider them
  here.}, or $\md{n}\oplus\wb{\md{n}}$ of $\mf{u}(n)$ or $\mf{su}(n)$.
In the case of $F = \mf{sp}(n)$, the representation $\md{2n}$ has
weights\footnote{The weights $\w^{(i)}$ are the standard basis of
  $\IR^n$.}
\begin{equation}\label{eq:wi}
  \pm\w^{(i)},\quad i=1,\ldots,n
\end{equation}
where $w^{(i)}$ are all independent.
The coweight $\lF$ is characterised by
\begin{equation}\label{eq:lFi}
  \lF^{(i)}:= \w^{(i)}\cdot\lF = \pm \frac{1}{2}.
\end{equation}
This implies that when presented as a vector in $\IR^n$ with $e^*_i$
as the standard basis, the coweight $\lF$ is
\begin{equation}
  \lF = \sum_i \lF^{(i)}e^*_i = \pm \frac{1}{2}e^*_1 \pm \ldots
  \pm\frac{1}{2} e^*_n
\end{equation}
and there are $2^{n}$ of them.
The Weyl group of $\mf{sp}(n)$ permutes the components $\lF^{(i)}$ and
flips signs of $\lF^{(i)}$, and therefore all the coweights $\lF$ are
in a single Weyl orbit whose dominant element is
\begin{equation}
  \lF = \frac{1}{2}e^*_1+\ldots +\frac{1}{2}e^*_n = (0\ldots01),
\end{equation}
where in the last equality we give the Dynkin labels of $\lF$.  In the
case of $F=\mf{so}(2n), \mf{u}(n)$, it is completely the same with the
characterisation \eqref{eq:lFi} of $\lF$ and there are $2^n$ of them.
The difference is that for $F=\mf{so}(2n)$ the Weyl group permutes
$\lF^{(i)}$ and flips \emph{pairs} of $\lF^{(i)}$ and thus all $\lF$
are in two Weyl orbits whose dominant elements are
\begin{equation}
  \lF = \frac{1}{2}e^*_1+\ldots+\frac{1}{2}e^*_{n-1}+\frac{1}{2}e^*_n
  =  (0\ldots01),
\end{equation}
and
\begin{equation}
  \lF = \frac{1}{2}e^*_1 + \ldots +\frac{1}{2}e^*_{n-1} -
  \frac{1}{2}e^*_n = (0\ldots10).
\end{equation}
while the flavor group $F=\mf{u}(n)$ is usually presented as the
product $F=\mf{su}(n)\times \mf{u}(1)$, and $\lF$ is presented as a
coweight of $\mf{su}(n)$ plus a $\mf{u}(1)$ charge.  Finally in the
case of $F=\mf{su}(n)$, the representation $\md{n}\oplus\wb{\md{n}}$
has weights \eqref{eq:wi} subject to the constraint
$\w^{(1)} + \ldots + \w^{(n)} = 0$.  The coweight $\lF$ is then
characterised by \eqref{eq:lFi} with the constraint
\begin{equation}\label{eq:lF=0}
  \lF^{(1)} + \ldots + \lF^{(n)} = 0.
\end{equation}
The number of such $\lF$ is ${n\choose n/2}$.
They are all in a single Weyl orbit whose dominant element is
\begin{equation}
  \lF = (0\ldots010\ldots0).
\end{equation}
Note that when the flavor symmetry is $\mf{su}(n)$, $n$ is always an
even integer and therefore the number ${n\choose n/2}$ makes sense.
All the $\lF$ for unity blowup equations in Tables~\ref{tb:ubeq-1},
\ref{tb:ubeq-2} can be determined in this way, except for the models
$\fn=2, G=\mf{su}(2)$ and $\fn=1, G=\mf{sp}(N), N\geq 1$, where the
Higgsing procedure discussed in Section~\ref{sc:Hig} imposes further
constraints.

\subsection{Blowup equations along Higgsing tree}
\label{sc:Hig}

Let us consider a rank one 6d SCFT $\mc{T}$ with gauge symmetry $G$
Higgsed to a daughter theory $\mc{T}'$ with gauge symmetry $G'$.  We
expect that the weight lattice of $G$ is projected by a surjective map
$f$ to the weight lattice of $G'$.  In particular, the vector
multiplets transform in the adjoint representation of $G$ whose weight
space decomposes under the projection by
\begin{equation}
  f: \Delta \to \Delta' \oplus R'.
\end{equation}
In addition there is a set of hypermultiplets transforming in a
representation $R$ whose weight space decomposes under the projection
by
\begin{equation}
  f: R \to R' \oplus \md{1}.
\end{equation}
To trigger the Higgsing, we should give non-zero vev to the scalar in
the hypermultiplet $(H_0)$ transforming in the singlet after the
projection.  The hypermultipets transforming in $R'$ then become
massive.  They get eaten by the vector multiplets transforming in $R'$
and decouple together.  Concretely this operation is realised as
follows.  The vev of a hypermultiplet with gauge and flavor charges
$\wG,\wF$ is
\begin{equation}
  \wG\cdot\mG+\wF\cdot\mF.
\end{equation}
First of all, the gauge charge carried by the hypermultiplet $(H_0)$
should be in $\Ker(f)\cap R$, and thus we set
\begin{equation}\label{eq:mGHig}
  \wG\cdot\mG = 0,\quad \wG\in \Ker(f)\cap R.
\end{equation}
Secondly, let the hypermultiplet $(H_0)$ carry certain flavor charge
$\wF\in R_F$.\footnote{More precisely, the hyper $H_0$ consists of two
  half-hypers carrying flavor charges $\pm \wF$.}  We set values for
the following components of $\mF$
\begin{equation}\label{eq:mFHig}
  \wF\cdot \mF = \mp \frac{\eq+\eq}{2},\quad \text{for}\;
  \wF\cdot \lF = \pm \frac{1}{2}.
\end{equation}

It has been shown in the literature (for instance \cite{Kim:2018gjo})
that elliptic genera reduce properly along the Higgsing tree through
the limits of parameters \eqref{eq:mGHig}, \eqref{eq:mFHig}.  We find
that all the other components of unity elliptic blowup equations in
\eqref{eq:ebeq} also reduce properly and therefore the unity blowup
equations transform consistently along the Higgsing trees.
\begin{itemize}
\item The contributions to $A_{{V}}$ from vector multiplets
  transforming in $R'$ cancel with the contributions to $A_{{H}}$
  from hypermultiplets transforming in the same representation.  Pick
  a pair $\pm \wG\in R$ which is mapped to $R'$ and let $+\wG$ be the
  weight in $\Delta_+$.  If $\wG\cdot\lG>0$, the vector charged
  with $\wG$ contributes
  \begin{equation}\label{eq:vectorwG-1}
    \prod_{\substack{k,l\geq 0\\ k+l \leq \wG\cdot\lG-1}}
    \frac{\eta}{\theta_1(\wG\cdot\mG+k\eq+l\et)}
    \prod_{\substack{k,l\geq 0\\ k+l \leq \wG\cdot\lG-2}}
    \frac{\eta}{\theta_1(\wG\cdot\mG+(k+1)\eq+(l+1)\et)},
  \end{equation}
  and the hypers charged with $\pm\wG$ and $\wF$ contribute
  \begin{equation}\label{eq:hyperwGwF-1}
  \ba
       \phantom{\times}&\prod_{\substack{k,l\geq 0\\k+l\leq \wG\cdot\lG+\wF\cdot\lF-\frac{3}{2}}}
     \frac{\theta_1(\wG\cdot\mG+\wF\cdot\mF+(k+\frac{1}{2})\eq+(l+\frac{1}{2})\et)}{\eta}\\
      \times&
      \prod_{\substack{k,l\geq 0\\k+l\leq \wG\cdot\lG-\wF\cdot\lF-\frac{3}{2}}}
      \frac{\theta_1(-\wG\cdot\mG+\wF\cdot\mF-(k+\frac{1}{2})\eq-(l+\frac{1}{2})\et)}{\eta}.
      \end{aligned}
  \end{equation}
  If $\wG\cdot\lG<0$, the vector charged with $\wG$
  contributes
  \begin{equation}\label{eq:vectorwG-2}
    \prod_{\substack{k,l\geq 0\\ k+l \leq -\wG\cdot\lG-1}}
    \frac{\eta}{\theta_1(\wG\cdot\mG-k\eq-l\et)}
    \prod_{\substack{k,l\geq 0\\ k+l \leq -\wG\cdot\lG-2}}
    \frac{\eta}{\theta_1(\wG\cdot\mG-(k+1)\eq-(l+1)\et)},
  \end{equation}
  and the hypers charged with $\pm\wG$ and $\wF$ contribute
  \begin{equation}\label{eq:hyperwGwF-2}
    \ba
      \phantom{\times}&\prod_{\substack{k,l\geq 0\\k+l\leq -\wG\cdot\lG+\wF\cdot\lF-\frac{3}{2}}}
      \frac{\theta_1(-\wG\cdot\mG+\wF\cdot\mF+(k+\frac{1}{2})\eq+(l+\frac{1}{2})\et)}{\eta}\\
     \times& \prod_{\substack{k,l\geq 0\\k+l\leq -\wG\cdot\lG-\wF\cdot\lF-\frac{3}{2}}}
      \frac{\theta_1(\wG\cdot\mG+\wF\cdot\mF-(k+\frac{1}{2})\eq-(l+\frac{1}{2})\et)}{\eta}
      \end{aligned}
  \end{equation}
  In both cases, the contributions of vectors and hypers cancel up to
  a sign if we set \eqref{eq:mFHig}.
\item The coroot lattice $Q^\vee$ over which the summation index $\lG$
  runs reduces properly to the coroot sub-lattice $Q^{\vee,'}$ of the
  daughter theory.  As discussed in Appendix~\ref{sc:br}, the
  projection $f$ induces an injection $f^*$ from $Q^{\vee,'}$ to
  $Q^\vee$, which preserves the norms of coroot vectors.  If $\lG$ is
  in the image of $f^*$, we know from \eqref{eq:imagf*} that
  $\wG\cdot\lG = 0$ for $\wG\in \Ker(f)\cap R$ projected to the
  singlet by $f$.  The contribution to $A_{\text{H}}$ by the hyper in
  the singlet collapses to 1, and the term corresponding to $\lG$
  survives in the limit \eqref{eq:mGHig}, \eqref{eq:mFHig}.  If,
  however, $\lG\in Q^\vee$ is not in the image of $f^*$,
  $\wG\cdot\lG \neq 0$ for $\wG\in \Ker(f)\cap R$.  The hyper in the
  singlet contributes to $A_{\text{H}}$ by at least
  \begin{equation}
    \frac{\theta_1(\mG\cdot
      \wG+\wF\cdot\mF \pm\frac{1}{2}(\eq+\et))}{\eta}
    ,\quad \wG \in \Ker(f)\cap R
  \end{equation}
  which becomes zero in the limit \eqref{eq:mGHig}, \eqref{eq:mFHig},
  thus annihilating the corresponding term.
\item The only terms in the argument of the generalised theta function
  affected by Higgsing procedure are
  \begin{align}
    &k_F \lF\cdot \mF +  y (\eq+\et)\nn =
    &\,\ind_{R_G}\sum_{\wF\in R_F}\Big(
      (\wF\cdot\lF)(\wF\cdot\mF)+
      \frac{\eq+\et}{2}((\wF\cdot\lF)(\wF\cdot\lF)+\frac{1}{4})
      )\Big)\nn =
    &\,\ind_{R_G}\sum_{\wF\in
      R_F}\Big((\wF\cdot\lF)(\wF\cdot\mF)+\frac{\eq+\et}{4}\Big)
  \end{align}
  which reduces properly to the argument of the daughter under the
  limit \eqref{eq:mFHig}.
\item There could be extra signs coming from the map of
  $(-1)^{|\lG|}$, from projection of positive roots\footnote{Even if a
    positive root of $G$ is mapped to a root of $G'$, it is not
    necessary still \emph{positive}.  We may have to flip its sign.},
  and from cancellation of one-loop contributions.  We checked with
  concrete examples that the extra sign is always positive.
\end{itemize}

Let us comment that the case of vanishing blowup equations is very
different: it is in general not possible to transform vanishing blowup
equations along Higgsing trees.  There are two arguments for this.
First, recall that in unity blowup equations the summation index $\lG$
takes value in the unshifted coroot lattice and it is crucial that in
the Higgsing process from a mother theory $\mc{T}$ with coroot lattice
$Q^\vee$ to a daughter theory $\mc{T}'$ with coroot sub-lattice
$Q^{\vee,'}$ there exists an injection $f^*$ from the coroot lattice
$Q^{\vee,'}$ to $Q^\vee$ induced by the projection $f$.  In vanishing
blowup equations, on the other hand, the summation index $\lG$ takes
value in the coweight lattice, or more precisely in the shifted coroot
lattice.  Unfortunately, a similar map from $P^{\vee,'}$ of a daughter
theory to $P^{\vee}$ of a mother theory does not exist as discussed in
Appendix~\ref{sc:br}.  Another way to see this is that the leading
order identities of vanishing blowup equations do not transform to
each other properly along the Higgsing trees in the limit
\eqref{eq:mGHig}, \eqref{eq:mFHig}.

In the remainder of the section, we discuss in detail some typical
examples of the Higgsing procedure for unity blowup equations as well
as some mild constraints on $\lF$ this procedure entails.

\subsubsection{Examples of Higgsing unity blowup equations}
\label{sec:exampl-higgs-unity}

\begin{itemize}
\item $\fn=6, G = E_7 \to E_6$: Representations of gauge symmetry
  decompose by
  \begin{align}
    \md{133}
    &\to \md{78} \oplus \textcolor{red}{\md{27}} \oplus
      \textcolor{red}{\wb{\md{27}}} \oplus \md{1} \quad(V)\nn
      \md{56}
    &\to \textcolor{red}{\md{27}} \oplus \textcolor{red}{\wb{\md{27}}}
      \oplus 2\cdot\md{1}\quad (H)
      \label{eq:E7-E6}
  \end{align}
  Here $(V)$ means vectors and $(H)$ means hypers.  We mark in red the
  multiplets whose 1-loop contributions cancel with each other after
  Higgsing.  In these models components $\lF^{(i)}$ of $\lF$ are free.
  Since only one hyper is used in the Higgsing, one component of $\lF$
  is fixed and the number of different $\lF$ is
  reduced by half.\\
  Most other models follow the same procedure of Higgsing.
\item $\fn=4, G = E_7\supset E_6$: The gauge symmetry branching rules
  are the same as \eqref{eq:E7-E6}, and naively the number of $\lF$ is
  reduced by half from 4 to 2.  On the other hand, the remaining hyper
  in $\md{56}$ also branches as in the second line of \eqref{eq:E7-E6}
  and it splits to two hypers with opposite charges, which have
  opposite flavor masses and $\lF$ components.  In the daughter
  theory, the two flavor masses can actually be made independent and
  the flavor symmetry is enhanced.  We can use the enhanced Weyl group
  to generate full Weyl orbits from the reduced $\lF$ and increase the
  number of $\lF$ from 2 to 4.\\[+2mm]
  Many other models are Higgsed by a similar procedure including
  \begin{itemize}
  \item $\fn = 3, G = \mf{so}(9) \to \mf{so}(8), F_4\to \mf{so}(8)$,
  \item
    $\fn = 2, G = E_7 \to E_6, \mf{so}(11)\to \mf{so}(10),
    \mf{so}(9)\to \mf{so}(8),
    F_4\to \mf{so}(8), G_2\to \mf{su}(3)$,
  \item $\fn = 1, G = \mf{so}(9) \to \mf{so}(8), F_4\to \mf{so}(8),
    G_2\to \mf{su}(3)$.
  \end{itemize}
\item $\fn=2, G = \mf{su}(N)\to \mf{su}(N-1)$ for $N\geq 3$: The gauge
  symmetry branching rules for the cancelling vectors and hypers are
  \begin{align}
    \Delta_N
    &\to \Delta_{N-1} \oplus \textcolor{red}{\md{(N-1)}}
      \oplus \textcolor{red}{\wb{\md{(N-1)}}}\oplus \md{1} \quad(V)\nn
      \md{N}
    &\to \textcolor{red}{\md{(N-1)}}\oplus \md{1} \quad(H) \nn
      \wb{\md{N}}
    &\to \textcolor{red}{\wb{\md{(N-1)}}}\oplus \md{1} \quad(H)
      \label{eq:suN}
  \end{align}
  while the remaining $2N-2$ hypers also branch by the last two lines
  of \eqref{eq:suN}.  In these models, the flavor symmetry is of type
  $\mf{su}(2N)$ and the components $\lF^{(i)}$ are always subject to
  the constraint \eqref{eq:lF=0}.  In the Higgsing procedure, two
  hypers are decoupled and thus two components of $\lF$ should be
  fixed.  In order to maintain the condition \eqref{eq:lF=0}, we
  should fix two components of $\lF$ as in \eqref{eq:mFHig} with
  opposite signs.  The number of $\lF$ is then reduced from
  ${2N \choose N}$ to
  ${2N-2\choose N-1}$.\\
  The Higgsing procedure for $\fn=1, G=\mf{su}(N)\to \mf{su}(N-1)$
  with $N\geq 3$ is similar.
\item $\fn=2, G = \mf{su}(2)\to \emptyset$: We follow the same
  branching rules in the gauge sector as \eqref{eq:suN}.  The $\lF$
  obtained by Higgsing should still follow \eqref{eq:lF=0}, and the
  number of $\lF$ of the daughter theory, which is the M-string
  theory, is
  therefore ${2\choose 1}=2$.\\
  Note that in the mother theory with $G = \mf{su}(2)$, since the
  fundamental and the anti-fundamental representations of $\mf{su}(2)$
  are isomorphic, one usually expects the flavor symmetry is enhanced
  from $\mf{su}(4)$ to $\mf{so}(8)$, and consequently the lifting of
  the constraint \eqref{eq:lF=0}.  This however would mean too many
  $\lF$ that will be reduced to the M-string theory.  Since there are
  only two $\lF$ for the latter \cite{Gu:2019pqj}, we conclude the
  condition \eqref{eq:lF=0} cannot be lifted for the mother theory.
\item $\fn=1, G=\mf{sp}(N)\to \mf{sp}(N-1)$ with $(N\geq 1)$: Recall
  that the analysis of modularity constraint in Section~\ref{sc:mod}
  together with checkerboard pattern constraint indicates that $\lF$
  for unity equations should be
  \begin{equation}\label{eq:lFSO}
    \lF =
    \begin{cases}
      \frac{1}{2}e^*_1+\ldots+\frac{1}{2}e^*_{n-1}+\frac{1}{2}e^*_n =
      (0\ldots01),\\
      \frac{1}{2}e^*_1+\ldots+\frac{1}{2}e^*_{n-1}-\frac{1}{2}e^*_n =
      (0\ldots10),
      \end{cases}
  \end{equation}
  up to Weyl transformations.  For unity blowup equations to be
  Higgsed properly, one of the two possibilities has to be eliminated.
  Upon Higgsing, the gauge symmetry branching rules of canceling
  vectors and hypers are
  \begin{align}
    \Delta_N
    &\to \Delta_{N-1} \oplus \textcolor{red}{2(\md{2N-2})} \oplus
      3\cdot\md{1} \quad
      (V)\nn
      \md{2N}
    &\to \textcolor{red}{(\md{2N-2})} \oplus 2\cdot \md{1} \quad (H)
  \end{align}
  Here we need two hypers to cancel with massive vectors, and we set
  the corresponding components of $\mF$ by \eqref{eq:mFHig} depending
  on the value of $\lF$ components.  On the other hand, the reduced
  one-string elliptic genus of the $\fn=1, G=\mf{sp}(N)$ theory reads
  \cite{Yun:2016yzw}
  \begin{equation}\label{n1SpNreducedZ1}
    \IE_1^{\rm red}(v,q,\und{m}_G,\und{m}_F) = \frac{1}{2}\sum_{j=1,2,3,4}
    \bigg(\prod_{i=1}^{8+2N}\frac{\theta_j(m_{F}^i)}{\eta}\bigg)
    \bigg(\prod_{i=1}^N\frac{\eta^2}{\theta_j(\ep_++ m^i_G)\theta_j(\ep_+-m^i_G)}\bigg).
  \end{equation}
  Clearly the elliptic genus can only be properly Higgsed if a pair of
  $m_F^i,m_F^j$ take the limit \eqref{eq:mFHig} with the same sign,
  and correspondingly we should fix two components of $\lF$ with the
  same value.
  This indicates the following chain of Higgsing for $\lF$
  \begin{align}
    \lF:
    \begin{cases}
      (\underbrace{0\ldots001}_{2N+8}) \to
      (\underbrace{0\ldots001}_{2N+6}) \to\ldots\to
      (\underbrace{0\ldots01}_8) =
      \frac{1}{2}e^*_1 + \ldots +\frac{1}{2}e^*_8 \\
      (\underbrace{0\ldots010}_{2N+8}) \to
      (\underbrace{0\ldots010}_{2N+6}) \to\ldots\to
      (\underbrace{0\ldots10}_8) = \frac{1}{2}e^*_1 + \ldots
      -\frac{1}{2}e^*_8
    \end{cases}
    \label{eq:lFSO}
  \end{align}
  At the end of the Higgsing chain, we find the E-string theory with
  $G = \mf{sp}(0)$.  The flavor symmetry is in fact enhanced from
  $\mf{so}(16)$ to $E_8$.  The (co)weight lattice of $E_8$,
  however, is a sub-lattice with index two of the (co)weight lattice
  of $\mf{so}(16)$.  Between the two coweights at the end of the chain
  in \eqref{eq:lFSO} only $\lF = (0\ldots01)$ can be lifted to a
  coweight vector of $E_8$, which we can use together with the Weyl group of
  $E_8$ to generate the full Weyl orbit $\cO_{2,240}$.  This
  means going back the Higgsing chain the $\lF$ can \emph{never} take
  the value of $(0\ldots10)$ for $\fn=1, G=\mf{sp}(N)$ models.
\end{itemize}


\subsection{Physical interpretation}
\label{sec:phys-interpr}

In our previous paper \cite{Gu:2019pqj}, we proposed a physical
explanation for the elliptic blowup equations of rank one 6d SCFTs
with no gauge symmetry, namely, the E-string and the M-string
theories. We generalise this argument to cover 6d SCFTs with gauge
symmetry as well.

Let us first briefly summarise the salient points of the argument in
our previous paper. For 6d SCFTs with no gauge symmetry, the elliptic
blowup equations read\footnote{$\lF$ here equals $\frac{1}{2}r_m$ in
  \cite{Gu:2019pqj}.}
\begin{align}
  \sum_{d'+d'' = d}
  &\theta^{[a]}_i(\fn \tau,\lF\cdot
    \mF+y(\eq+\et)-\fn(d'\eq+d''\et))\nn[-3mm]
  &\ \ \cdot\IE_{d'}(\tau,\mF+\lF \eq,\eq,\et-\eq)
    \IE_{d''}(\tau,\mF+\lF \et,\eq-\et,\et)\nn =
  &\,\,\theta^{[a]}(\fn\tau,\lF+y(\eq+\et)) \IE_d(\tau,\mF,\eq,\et)
    \label{eq:blowup-EM}
\end{align}
where
\begin{equation}
  y = \frac{\fn-1}{4}+\frac{1}{2}(\lF,\lF).
  \label{eq:y-EM}
\end{equation}
To explain this equation, the idea is to compute the Nekrasov
partition function of the 6d SCFT on $T^2\times\wh{\IC}^2$, first with
finite size of the exceptional divisor $\IP^1$ of the blowup space
$\wh{\IC}^2$ (left hand side), and then compute the same partition
function with the exceptional $\IP^1$ blown down (right hand
side). Since the partition function does not depend on the size of the
$\IP^1$, we can identify the partition functions on both sides.

We first consider the left hand side of the equation.  There are
essentially two types of configurations with finite energy.  The first
type corresponds to the worldsheet of non-critical string wrapping the
torus.  The strings appear as solitons in $\wh{\IC}^2$ and they are
localised at the north pole and the south pole of the $\IP^1$, whose
neighborhoods locally resemble the Omega background with twist
parameters $(\eq,\et-\eq)$ and $(\eq-\et,\et)$ respectively.  Suppose
there are $d'$ and $d''$ solitons localised at the north pole and the
south pole respectively, the one-loop contribution of this saddle
point configuration gives rise to the elliptic genera
\begin{equation}
  \IE_{d'}(\tau,\mF+\lF \eq,\eq,\et-\eq)
  \IE_{d''}(\tau,\mF+\lF \et,\eq-\et,\et)
\end{equation}
The shifted flavor mass is to account for the embedding of the
$SU(2)_R$-symmetry in the flavor symmetry.

The second type of finite energy configuration corresponds to the flux
of the self-dual 3-form $C$, which is the field strength of the tensor
field, through $\IP^1\times S^1$. Either by an argument of computing the
partition function of 4d theory resulting from torus reduction of 6d
theory, or by a holographic argument, we found
\cite{Gu:2019pqj} that this type of configuration gives rise to a
generalised theta function of the form
\begin{equation}
  \Theta_{\Omega}^{[a]}(\tau,z)=\sum_{n_i\in\IZ}\exp\left(
    \frac{1}{2}\Omega_{ij}(n_i+a_i)(n_j+a_j)\tau
    +\Omega_{ij}(n_i+a_i)z_j\right),
\end{equation}
where $\Omega_{ij} = -A_{ij}$ is the opposite of the base intersection
matrix.
These theta functions with different characteristics $a$ are sections
of a line bundle over the torus
\begin{equation}
  \IT = \IC^r/(\Omega \IZ^r \oplus \tau\Omega \IZ^r),
\end{equation}
where $r$ is the number of base curves, and the total number of
independent sections is equal to the determinant of $\Omega$.  The
elliptic parameter of the theta function is given by
\begin{equation}
 z_i = \int_{S_A^1\times\IP^1} C_i + \tau\int_{S_B^1\times \IP^1} C_i
\end{equation}
where $S_{A,B}^1$ are the two 1-cycles on the torus, and the 4-form
fluxes $C_i$ are identified with anti-derivative of the four-forms
$X_i$ in the Green-Schwarz counter-term \eqref{eq:IGS}.  In a rank one
6d SCFT, the anomaly four-form reads
\begin{equation}
  X = -\frac{\fn}{4}\Tr F_{\mf{g}}^2+\frac{k_F}{4}\Tr F_{\mf{f}}^2 +
  h^\vee_{\mf{g}} c_2(R) - \frac{2-\fn}{4}p_1(M_6).
  \label{eq:X}
\end{equation}
Here $F_{\mf{g}}$ and $h^\vee_{\mf{g}}$ are the field strength and the
dual Coxeter number of gauge symmetry. In a theory with trivial gauge
symmetry, we suppress $F_{\mf{g}}$ and set $h^\vee_{\mf{g}}$ to 1.
$F_{\mf{f}}$ is the flavor symmetry field strength, and $k_F$ is the
level of the associated current algebra on the string worldsheet,
which is 1 in the E-, M-string theories.  We expect therefore the
elliptic parameter to be
\begin{equation}
  z = \int_{S^1\times\IP^1} \frac{k_F}{4} \omega_{\mf{f}} + \omega_R -
  \frac{2-\fn}{4}
  p_1^{-1}(M_6),
  \label{eq:z}
\end{equation}
where
\begin{equation}
  \omega_{\mf{f}} =
  \Tr\left(\frac{2}{3}A_{\mf{f}}^3+A_{\mf{f}}\wedge\rd
    A_{\mf{f}}\right),
  \quad
  \omega_R = \Tr\left(\frac{2}{3}A_R^3 + A_R\wedge\rd A_R\right)
\end{equation}
and $p_1^{-1}$ is the anti-derivative of the Pontryagin class.
We compute
\begin{equation}
  \frac{k_F}{4}\int_{S^1\times\IP^1}\omega_{\mf{f}} = k_F \lF\cdot \mF
  \label{eq:omegaF}
\end{equation}
and
\begin{equation}
  \int_{S^1\times \IP^1} \omega_R = \epsilon_+/2
\end{equation}
as well as
\begin{equation}
  \int_{S^1\times\IP^1} p_1^{-1}(M_6) = \eq+\et = 2\epsilon_+.
\end{equation}
Due to the embedding of the R-symmetry, we shift
\begin{equation}
  \mF \to \mF + \lF \epsilon_+.
  \label{eq:mFshift}
\end{equation}
Putting all these pieces together, we thus recover the main part of
elliptic parameter of theta function in \eqref{eq:blowup-EM},
\eqref{eq:y-EM}. Finally, the coupling to the number of strings in the
elliptic parameter is due to the modification of the three form flux
$C$ in the presence of string source, in which case \cite{Gu:2019pqj}
\begin{equation}
  C = X + d' \chi_{4}(N') + d''\chi_{4}(N'')
\end{equation}
where $\chi_4(N'),\chi_4(N'')$ are the Euler classes of the normal
bundles of the strings localised at the north pole and the south pole
of $\IP^1$. Integrating the anti-derivative
$e_3^{(0)} = \chi_4^{-1}(N)$ of the Euler class through the
enveloping 3-manifold of the string worldsheet, which is
$\IP^1\times S^1$, gives us the coupling terms $\fn(\eq d'+\et d'')$.

In order to obtain the right hand side of the blowup equation, we
simply choose the exceptional $\IP^1$ to be very small. Looking from
far away, the $d=d'+d''$ solitons of strings are clustered at the
origin of $\IC^2$. The enveloping 3-manifold of these strings becomes
$S^3$ which does not intersect the exceptional $\mathbb{P}^1$ and
hence the integral of $e_3^{(0)}$ vanishes.

Now we would like to generalise this argument. For simplicity we focus
on unity equations of rank one theories.  The presence of gauge
symmetry entails three modifications to the argument above.  First,
now we need to use the proper value of $h^\vee_{\mf{g}}$, and turn on
the gauge flux $F_{\mf{g}}$ in \eqref{eq:X}.  Following a similar
computation as \eqref{eq:omegaF}, we find
\begin{equation}
  -\frac{\fn}{4}\int_{S^1\times\IP^1}\omega_{\mf{g}}
  = -\fn \left(\frac{1}{4}\int_{\IP^1}F_{\mf{g}}\right)\cdot \mG
  = -\fn \lG\cdot\mG.
\end{equation}
Since both gauge and flavor symmetries of the 6d SCFT are global
symmetries on the string worldsheet, we need to embed $SU(2)_R$ in the
gauge symmetry as well, which leads to a similar shift as
\eqref{eq:mFshift}
\begin{equation}
  \mG \to \mG + \lG\epsilon_+.
\end{equation}
These extra ingredients allow us to recover the elliptic parameter of
the theta function in \eqref{eq:ebeq},\eqref{eq:y}.  Second, the gauge
flux is coupled to the tensor field by the following term in the
Lagrangian
\begin{equation}
  \mc{L}\supset \int\phi\, \frac{1}{4}\Tr F_{\mf{g}}\wedge \star F_{\mf{g}}
\end{equation}
This means with the gauge flux is present, the tensor modulus is
weighted not only by non-critical strings but in addition by
$||\lG||^2/2$, which explains the balancing condition
\begin{equation}
  ||\lG||^2/2 + d' + d'' = d
\end{equation}
in \eqref{eq:ebeq}.


Finally, we would like to explain the appearance of the factors
$A_V(\tau,m_G,\lambda_G)$, $A_H(\tau, m_{G,F}, \lambda_{G,F})$ whose
expressions are given by \eqref{eq:AV}, \eqref{eq:AH},
\eqref{eq:thetaV}, \eqref{eq:thetaH*}.  Let us first look at $A_V$
which is given as a product of theta functions in the denominator.  We
interpret each such theta function as arising from the oscillator
modes of a complex bosonic coordinate or two real such coordinates in
the target space of a 2d sigma model with the torus of complex
structure $\tau$ being the worldsheet \cite{MR1746311}.  Next, we want
to count the bosonic dimension of this target space. To this end,
observe that the modular index of each ingredient is
\begin{equation}
  \textrm{Ind}\, \breve{\theta}_V(z,R) = - R^2 \frac{z^2}{2} + \ldots
\end{equation}
Thus, the total modular index of $A_V$ becomes
\begin{equation}
  \textrm{Ind} A_V = \sum_{\beta \in \Delta_+} - (\beta \cdot \lambda_G)^2 \frac{z^2}{2} + \ldots
\end{equation}
Now we know that the modular index of a single theta function in the
denominator is
\begin{equation}
  \textrm{Ind}~\frac{1}{\theta_1(z)} = -\frac{z^2}{2}.
\end{equation}
Thus we conclude that the total number of theta-functions in $A_V$ is
\begin{equation}
  \sum_{\beta \in \Delta_+} R^2 = \sum_{\beta \in \Delta_+}
  (\beta \cdot \lambda_G) (\beta \cdot \lambda_G) = \half
  \sum_{\beta \in \Delta} (\beta \cdot \lambda_G) (\beta \cdot
  \lambda_G) = h^{\vee}_G ||\lambda_G||^2.
\end{equation}
Using $d_0 = \half ||\lambda_G||^2$, we see that the total bosonic
dimension of our target space $\mathcal{M}$ is
\begin{equation}
  \textrm{dim}_{\mathbb{R}}\mathcal{M} = 2 ||\lambda_G||^2
  h^{\vee}_G = 4 h^{\vee}_G d_0.
\end{equation}
But this is nothing else than the dimension of the moduli space of
$d_0$ $G$-instantons, denoted by $\mathcal{M}_{G,d_0}$. The arguments
in the theta-functions in (\ref{eq:thetaV}) are obtained from
equivariant localization on $\mathcal{M}_{G,d_0}$, see
\cite{Nakajima:2003pg}. A similar argument can be worked out for $A_H$
describing the fermionic coordinates on the moduli space. Thus we see
that the factors $A_V$ and $A_H$ arise from path integrals over
collective coordinates of strings moving in the moduli space of $d_0$
$G$-instantons! Therefore, the instanton contributions on the
left-hand side of (\ref{eq:ebeq}) split into three pieces: strings
localized at the north pole of the exceptional $\mathbb{P}^1$, strings
localized at the south pole, and non-localized strings on the
$\mathbb{P}^1$. This mirrors and generalizes the three contributions
for the instanton moduli space on the blowup geometry found in
\cite{Nakajima:2003pg}.

\subsection{K-theoretic blowup equations}
\label{sec:5dblowup}

When taking the K-theoretic limit $\Qtau\to 0$, the elliptic genera of
a 6d theory $T_{\rm 6d}$ in general reduce to the K-theoretic
instanton partition function of the 5d theory $T_{\rm 5d}$ with the
same gauge, flavor group and the same matter contents on a circle of
radius one,\footnote{The radius can be easily recovered to arbitrary
  $\beta$ by dimensional analysis.} and the elliptic blowup equations
in general reduce to the K-theoretic blowup equations. For example, it
is easy to find that the unity elliptic blowup equations
(\ref{eq:ebeq}) with $n\ge 3$ in the $\Qtau\to 0$ limit naturally
reduce to the 5d blowup equations with matters proposed in
\cite{Kim:2019uqw}. However, there are several subtle points.
\begin{itemize}
\item The elliptic blowup equation with characteristic $a=-1/2$ could
  split to \emph{two} K-theoretic blowup equations in the $\Qtau\to 0$
  limit. This was already observed for the minimal $\mf{su}(3)$ and
  $\mf{so}(8)$ SCFTs in \cite{Gu:2017ccq}. For other characteristics
  $a$, each elliptic blowup equation will reduce to one K-theoretic
  blowup equation.
\item For $n=2$ theories, the elliptic genera in the $\Qtau\to 0$
  limit give the 5d Nekrasov partition function with an extra term
  which are neutral with respect to $G$. To obtain the precise 5d
  blowup equations from 6d, one needs to factor out the extra term
  which possibly contributes to the $\Lambda$ factor.
\item All $n=1$ theories in the $\Qtau\to 0$ limit just reduce to the
  theory of a free hypermultiplet, whose associated Calabi-Yau space
  is simply the resolved conifold \cite{DelZotto:2018tcj}. The reduced
  one-string elliptic genera all have leading $\Qtau$ order as just
  $q^{-1}$, and the 5d gauge theory information are encoded in the
  $q^{0}$ coefficient. It is easy to see this works along well with
  elliptic blowup equations. In fact, the unity elliptic blowup
  equations for all $n=1$ theories at the $\Qtau$ leading order just
  give the blowup equation of the resolved conifold:
  \cite{Huang:2017mis}
  \begin{equation}
    S(\et)S(m+(y_{\rm u}-1)\eq+y_{\rm u}\et)-S(\eq)S(m+y_{\rm
      u}\eq+(y_{\rm u}-1)\et)=S(\et-\eq)S(m+y_{\rm u}(\eq+\et)),
  \end{equation}
  where we denote $S(x)=e^{\frac{x}{2}}-e^{-\frac{x}{2}}$ and
  $m=\lambda_F\cdot m_F$. It is easy to check this identity. After
  factoring out the $q^{-1}$ term and a gauge natural term similar
  with the $n=2$ situation, one can obtain the 5d blowup equations
  from the order $q^{0}$ of 6d ones.
\item More importantly, we find in general, \emph{not all} K-theoretic
  blowup equations are reduced from elliptic blowup equations. In
  particular, the admissible range of the shifts for the 5d instanton
  counting parameter $\mathfrak q$ can be larger than the admissible
  range of the shifts for the 6d string number counting parameter
  $Q_{\rm ell}$. This makes some $n=2$ theories such as
  $G=\mf{su}(N),F=\mf{su}(2N)$ theory not recursively solvable in 6d,
  but recursively solvable in 5d.
\end{itemize}

Let us discuss the pure gauge minimal 6d $(1,0)$ SCFTs as examples. In
the $\Qtau\to 0$ limit, the elliptic genera directly reduce to the 5d
Nekrasov-partition functions. We find all possible 5d blowup equations
for the pure gauge theory with $G=A_2,D_4,F_4,E_{6,7,8}$. The 5d $\br$
fields and $\Lambda$ factors and their 6d origins are listed in Table
\ref{tb:5d6dtable}.
\begin{table}[h]
\centering
\begin{tabular}{|c|c|c|c|c|}
  \hline
  & 5d $\br$ & 5d $\Lambda$ & from 6d $\br$ & 6d $\Lambda$ \\
  \hline
  $j=1,2,\dots,n-1$ & $({0},-n+2j)$  & 1 & $(0,{0},-n+2j)$ & $\Lambda^{[1/2-j/n]}$\\
  \hline
  & $({0},\pm n)$  & 1 & $(0,{0},n)$ & $\Lambda^{[-1/2]}$\\
  \hline
  $j=1,2,\dots,n-3$ & $({0},\pm(n+2j))$  & 1 &  &\\
  \hline
  & $({0},\pm (3n-4))$  & $1-(-1)^ne^{\pm\frac{3(n-2)}{2}(\eq+\et)}\q$ &  &\\
  \hline
  $j=1,2,\dots,n-1$ & $(2{w},-n+2j)$  & 0& $(0,2{w},-n+2j)$ &0\\
  \hline
  & $(2{w},\pm n)$  & 0 & $(0,2{w},n)$ & 0\\
  \hline
  $j=1,2,\dots,n-4$ & $(2{w},\pm(n+2j))$  & 0&  & \\
  \hline
  & $(2{w},\pm (3n-6))$  & $e^{\pm\frac{3(n-2)^2}{2n}(\eq+\et)}\q^{\frac{n-2}{n}}$ &  & \\
  \hline
\end{tabular}
\caption{The 5d and 6d blowup equations for pure gauge theories with
  $G=A_2,D_4,F_4,E_{6,7,8}$ and corresponding $n=3,4,5,6,8,12$. The 5d
  $\br$ fields are denoted as $(r_{m_G},r_{\log\q})$, with weight
  ${w}\in (P^\vee\backslash Q^{\vee})_G$ and 6d $\br$ fields are
  denoted as $(r_{\tau},r_{m_G},r_{\log Q_{\rm ell}})$.}
\label{tb:5d6dtable}
\end{table}
Note the first three rows were given by Keller-Song's K-theoretic
blowup equations \cite{Keller:2012da}.

In fact,
we further find for all simple Lie groups $G$, there exist
$\dualCox+3+2(r_{\rm c}-1)$ non-equivalent 5d unity $\br$ fields and
$(\dualCox-1)(r_{\rm c}-1)$ non-equivalent 5d vanishing $\br$ fields,
where $r_{\rm c}$ is the rank of the center of $G$ with
$r_{\rm c}=|P^\vee/Q^{\vee}|$. We summarize the corresponding $\Lambda$
factors in Table \ref{tb:all5d}, in which
$\delta=({w},{w})$.
\begin{table}[h]
  \centering
  \begin{tabular}{|c|c|c|}
    \hline
    & 5d $\br$ & 5d $\Lambda$  \\
    \hline
    $j=0,1,\dots,\dualCox$ & $({0},-\dualCox+2j)$  & 1 \\
    \hline
    & $({0},\pm (\dualCox+2))$  & $1-(-1)^{\dualCox} \exp\(\pm\frac{1}{2}\dualCox(\eq+\et)\)\q$ \\
    \hline
    $j=1,2,\dots,\dualCox-1$ & $(2{w},-\dualCox+2j)$  & 0 \\
    \hline
    & $(2{w},\pm \dualCox)$  & $\exp\(\pm\frac{3}{2}\delta_G\dualCox(\eq+\et)\)\q^{\delta_G}$ \\
    \hline
  \end{tabular}\\
  \caption{5d blowup equations for all simple Lie group $G$.}
\label{tb:all5d}
\end{table}
Note the first row was conjectured by Nakajima-Yoshioka
\cite{Nakajima:2003pg} and explicitly checked by Keller-Song
\cite{Keller:2012da}. The second row is beyond Nakajima-Yoshioka's
range for flux $d$ for $\mf{su}(N)$ geometries. The existence of unity
blowup equations of such type was already noticed for local
$\IP^1\times\IP^1$ Calabi-Yau geometries in \cite{Huang:2017mis}. The
third and fourth rows agree with Nakajima-Yoshioka's K-theoretic
blowup equations for $\mf{su}(N)$ gauge group and Chern class
$c=1,2,\dots,N-1$ \cite{Nakajima:2005fg}.

\section{Solving blowup equations}
\label{sc:sol}

In this section, we discuss how to solve the blowup equations.  By
solving we mean extracting refined BPS invariants of the local
Calabi-Yau threefold, or equivalently computing elliptic genera in the
case of 6d theories and instanton partition functions in the case of
5d theories.
Although it was believed that for general local Calabi-Yau threefolds
the blowup equations always uniquely determine the refined topological
string partition function \cite{Huang:2017mis}, the question remains
what is the minimal set of required input data.
It was at first conjectured \cite{Huang:2017mis} that merely the
classical intersection numbers of the Calabi-Yau geometries as input
data already allow for a complete solution of the blowup equations,
but it turns out not to be the case in some examples such as
the massless half K3 associated to the massless E-string theory
\cite{Gu:2019pqj}.
This gives a flavor of the complexity of the problem of solving the
blowup equations in general.

To discuss solving the blowup equations for 6d $(1,0)$ SCFTs, it is
convenient to divide all these theories into three classes according
to the difficulty of solving their associated blowup equations:
\begin{itemize}
\item [$\md A$] These theories have unity blowup equations and
  possibly vanishing blowup equations as well; there are enough unity
  equations so that recursion formulas \`a la
  \cite{Nakajima:2005fg,Keller:2012da} can be written down and the
  blowup equations can be solved immediately.
  \begin{itemize}
  \item In the case of rank one 6d SCFTs, these are the theories with
    $\fn\geq 3$ and without unpaired half-hypermultiplets\footnote{The
      theory of $\fn=3, G=\mf{su}(3)$ is a bit special.  The recursion
      formulas do not work for the \emph{one}-string elliptic genus,
      as the latter enjoys an enhanced symmetry so that the number of
      independent unity equations is reduced; the recursion formulas,
      nevertheless, still work for elliptic genera of more than one
      string \cite{Gu:2019dan}.
    }. There are infinitely many theories in this class.
  \end{itemize}
\item [$\md B$] These theories have unity blowup equations and
  possibly also vanishing blowup equations; the number of unity blowup
  equations is not sufficient to allow for recursion formulas.
  Nevertheless in practise it is still possible to solve blowup
  equations order by order using other methods.
  \begin{itemize}
  \item In the case of rank one 6d SCFTs, these are the theories with
    $\fn =1,2$ and without unpaired half-hypermultiplets.  There are
    also infinitely many theories in this class.
  \end{itemize}
\item [$\md C$] These theories have only vanishing blowup equations
  but no unity blowup equations. There is currently no
  algorithm to solve these equations completely.
  \begin{itemize}
  \item In the case of rank one theories, these are the theories with
    unpaired half-hypermultiplets.  There are in total 12 theories in
    this class which are $n=1,3,5,7$ $G=E_7$ theories, $n=1,3$
    $G=\mf{so}(11)$ theories, $n=3$ $G=\mf{so}(12)$ theory, $n=2$
    $G=\mf{so}(12)_b,\mf{so}(13)$ and $n=1$
    $G=\mf{su}(6)_{\star},\mf{so}(12)_{a,b}$ theories.
  \end{itemize}
\end{itemize}
In this section and the next section of examples, we will focus on
rank one theories. In later sections, we will see that all
\emph{higher-rank} theories belong to classes $\mathbf{ B}$ or
$ \mathbf{ C}$.

We discuss four methods to solve blowup equations, summarised in
Table~\ref{tab:solvability}.  The first two methods, the recursion
formulas and the Weyl orbit expansion, are designed to compute
elliptic genera.  Since they are based on elliptic blowup equations,
they require implicitly the semiclassical intersection numbers of
Calabi-Yau and the one-loop partition function as input data.  The
recursion formulas, as a generalisation of
\cite{Keller:2012da}, has the least scope of applicability among the first two methods;
but when it applies, it is the most powerful, as it calculates explicitly
elliptic genera of arbitrary numbers of strings.  The Weyl-orbit
expansion, initiated in \cite{Gu:2019pqj} and fully developed and
exploited in this paper, has a wider range of applicability.  The last
two methods, the refined BPS expansion and the $\eq,\et$ expansion,
are designed to compute refined BPS invariants or refined free
energies.  They are in fact applied to general refined topological
string theory \cite{Huang:2017mis}, and therefore require a slightly
different set of input data.
We comment that although theories in class $\md{C}$ cannot be solved
completely, there are some examples, for instance the $\fn=7, G=E_7$
model as we see in Section~\ref{E7theories}, where one can use the BPS
expansion method to solve the majority of refined BPS invariants below
any degree bound.
We also need to point out that although the method of $\eq,\et$
expansion seems to apply to all three classes, the necessary initial
data are sometimes rather difficult to come by. Here it is only used
to discuss the solvability of the blowup equations associated to
different classes of theories.

We explain individual methods in turn in the following subsections.
We give an inventory of all our results in Appendix~\ref{app:list},
and present some of these results explicitly in Section~\ref{sc:ex}
and Appendix~\ref{app:D},~\ref{app:bps}; more results can be found on
the website \cite{kl}.

\begin{table}
  \centering
  \begin{tabular}{cccc}
    \hline
    methods & solvable classes
    & input data
    & output results\\
    \hline
    recursion formulas & $\md{A}$ & semiclassical, one-loop
    & elliptic genera\\
    Weyl orbit expansion & $\md{A},\md{B}$ & semiclassical, one-loop
    & elliptic genera\\
    refined BPS expansion & $\md{A},\md{B}$, partially $\md{C}$
    & semiclassical,
      prepotential
    & BPS invariants\\
    $\eq,\et$ expansion & $\md{A},\md{B},\md{C}$
    & depends, see Section~\ref{sc:e12}
    & free energies\\
    \hline
  \end{tabular}
  \caption{Summary of methods to solve blowup equations.}
  \label{tab:solvability}
\end{table}

\subsection{Recursion formula}
\label{sc:rec}
From (\ref{eq:ebeq}), the unity blowup equation can be written as
\begin{align}
  &\phantom{---} \theta_{i}^{[a]}(\fn\tau,k_F\lF\cdot
    \mF+\fn y(\eq+\et))
    \IE_d(\mG,\mF,\eq,\et) \nn
  &\phantom{--} -\theta_{i}^{[a]}(\fn\tau,k_F\lF\cdot \mF+\fn(y(\eq+\et)-d
    \eq))\IE_{d}(\mG,\mF+\eq\lF,\eq,\et-\eq) \nn
  &\phantom{--} -\theta_{i}^{[a]}(\fn\tau,k_F\lF\cdot \mF+\fn(y(\eq+\et)-d
    \et))\IE_d(\mG,\mF+\et\lF,\eq-\et,\et) \nn =
  &{\sum_{\lG,d_1,d_2}}' (-1)^{|\alpha^\vee|}
    \theta_{i}^{[a]}(\fn\tau,-\fn\,\lG\cdot\mG+k_F\,\lF\cdot
    \mF+\fn((y-d_0)(\eq+\et)-d_1\eq-d_2\et)) \nn[-4mm]
  &\phantom{------}\times A_V(\tau,\mG,\lG) A_H(\tau,\mGF,\lGF)\nn
  &\phantom{--}\times
    \IE_{d_1}(\tau,\mGF+\eq\lGF,\eq,\et-\eq)
    \IE_{d_2}(\tau,\mGF+\et\lGF,\eq-\et,\et),
    \label{eq:ubeq}
\end{align}
where ${\sum'_{\lG,d_1,d_2}}$ means the summation over all
$\lG\in Q^\vee, d_0 = \frac{1}{2}||\lG||_G^2$ and $0\le d_{1,2}<d$
with $d_0+d_1+d_2=d$.
All the instances of $d$-string elliptic genus are collected on the
l.h.s., and there are only less than $d$-string elliptic genera on the
r.h.s..  With the characteristic $a$ taking value as in \eqref{eq:a}, the
number of such equations with fixed $d$ and $\lF$ is $\fn$.  For
models with $\fn\geq 3$, we can choose three arbitrary characteristics
$a_1,a_2,a_3$ and solve the $d$-string elliptic genus from
\eqref{eq:ubeq} as
\begin{align}
  &\IE_d(\tau,\mG,\mF,\eq,\et) = \nn
    {\sum_{\lG,d_1,d_2}}'
  & (-1)^{|\lG|} [{D/D}]^{\text{red}}_{\fn,(a_1,a_2,a_3)}
    (\mGF,\epsilon_{1,2},\lGF)\nn[-2ex]
    \times
  &\frac{\theta_1((d-d_2)\eq-d_1\et-\lG\cdot\mG)
    \theta_1((d-d_1)\et-d_2\eq-\lG\cdot\mG)}
    {\theta_1(d\eq)\theta_1(d\et)} \nn
    \times
  &A_{{V}}(\tau,\mG,\lG)
    A_{{H}}(\tau,\mGF,\lGF)\nn
    \times
  &\IE_{d_1}(\tau,\mGF+\eq \lGF,\eq,\et-\eq)
    \IE_{d_2}(\tau,\mGF+\et\lGF,\eq-\et,\et)
    \label{eq:Edrec}
\end{align}
where we define
\begin{align}
  &[{D/D}]^{\text{red}}_{\fn,(a_1,a_2,a_3)}
    (\mGF,\epsilon_{1,2},\lGF)\nn
    =
  &\frac{{D}^{\text{red}}_{\fn,(a_1,a_2,a_3)}
    (\lG\cdot\mG-d_0(\eq+\et)-d_1\eq-d_2\et,-d\eq,-d\et
    ;k_F \lF\cdot \mF + \fn y (\eq+\et))}
    {{D}^{\text{red}}_{\fn,(a_1,a_2,a_3)}
    (0,-d\eq,-d\et
    ;k_F \lF\cdot\mF + \fn y (\eq+\et))} \label{eq:DD}
\end{align}
with a reduced version of determinant
\begin{align}\label{redD}
  {D}^{\text{red}}_{\fn,(a_1,a_2,a_3)}(z_1,z_2,z_3; z_{\fn})=
  \frac{\det\(\theta_{i}^{[a_j]}(\fn\tau,z_{\fn}+\fn
  z_k)_{j,k=1,2,3}\)}
  {\theta_1(\tau,z_1-z_2)\theta_1(\tau,z_2-z_3)\theta_1(\tau,z_3-z_1)}.
\end{align}
In particular, the one-string elliptic genus can be simply obtained as
\begin{align}\label{E1universal}
  \IE_1(\tau,\mGF,\epsilon_{1,2}) =
  &\sum_{||\lG||^2=2}
    (-1)^{|\lG|}
    A_V(\tau,\mG,\lG)A_H(\tau,\mGF,\lGF)
    \frac{\theta_1(\eq-\lG\cdot\mG)\theta_1(\et-\lG\cdot\mG)}
    {\theta_1(\eq)\theta_1(\et)}\nn
  &\times\frac{{D}^{\text{red}}_{\fn,(a_1,a_2,a_3)}
    (\lG\cdot\mG-(\eq+\et),-\eq,-\et
    ;k_F \lF\cdot \mF + \fn y (\eq+\et))}
    {{D}^{\text{red}}_{\fn,(a_1,a_2,a_3)}
    (0,-\eq,-\et
    ;k_F \lF\cdot\mF + \fn y (\eq+\et))}    .
\end{align}
Note that in the definition of ${D}^{\text{red}}_{\fn,(a_1,a_2,a_3)}$
the determinant has zeros at $z_1 - z_2 = z_2 - z_3 = z_3 - z_1 = 0$,
which cancel with the zeros of the theta functions in the
denominator.\footnote{One can also define the determinant
  ${D}_{\fn,(a_1,a_2,a_3)}$ without the denominator in (\ref{redD}),
  in which case the recursion formulas (\ref{eq:Edrec}) and
  (\ref{E1universal}) will be cleaner with the line of Jacobi
  $\theta_1$ functions disappearing. In the meantime, one should be
  careful about the pole cancellation.}  Furthermore, similar to the
results of \cite{Gu:2019dan}, the final result of $\IE_d$ does not
depend on the choice of $a_{1,2,3}$.

\subsection{Weyl orbit expansion}
\label{vexpansion}

The problem to solve class $ \mathbf{ B}$ theories was already
encountered in \cite{Gu:2019pqj} for E-strings and M-strings. Without
sufficient number of unity blowup equations with different
characteristics, one can not have explicit recursion formulas for
$\IE_d$. A new method based on Weyl orbit expansion of elliptic genera
was initiated in \cite{Gu:2019pqj} using which the one-string elliptic
genera of E-string and M-string were successfully solved. Here we
further develop this method for all theories in class $ \mathbf{ A}$
and $ \mathbf{ B}$. Although we do not have a proof, explicit
computation for many examples shows that the unity blowup equations
are sufficient to uniquely determine the elliptic genera.\footnote{We
  exclude E-string theory in the following discussion. As there is no
  gauge symmetry, the unity blowup equations can only determine the
  elliptic genus up to a free function of $q$. See more details in our
  last paper \cite{Gu:2019pqj}.} This method is particularly efficient
for \emph{one}-string elliptic genus and \emph{small} flavor group
such as $\mf{su}(2)$ or $\mf{u}(1)$. The reason is that the reduced
one-string elliptic genus only depends on $v=\re^{(\eq+\et)/2}$ but
not on $x=\re^{(\eq-\et)/2}$, while reduced higher string elliptic
genera do depend on $x$ thus the flavor group is effectively
$F\times \mf{su}(2)_x$.

Let us focus on the reduced one-string elliptic genus
$\IE_{1}^{\rm red}(v,q,m_G,m_F)$. We can always write it in the
following Weyl orbit expansion of $F$, or $v$ expansion in other
words:\footnote{For a Weyl orbit $\mathcal{O}_{p,k}$, we adopt the
  common notation that $p$ is the length of its elements and $k$ is
  the number of its elements.}
\begin{equation}
  \label{eq:vansatz}
  \IE_{1}^{\rm red}(v,q,m_G,m_F)=\sum
c_{n,m,p,k}(m_G) q^n v^m\mathcal{O}_{p,k}^F.
\end{equation}
Here $c_{n,m,p,k}(m_G)$ are some $G$ Weyl-invariant rational functions
of $\exp({m_G})$. Note the order $n$ of $q$ has a known \emph{lower}
bound, and for each $n$, the order $m$ of $v$ also has a \emph{lower}
bound, while for each $(n,m)$ pair, the lengths of Weyl orbit elements
of flavor group $p$ have an \emph{upper} bound, i.e. there are only
finitely many different flavor Weyl orbits. One can further decompose
$c_{n,m,p,k}(m_G)$ into the Weyl orbit expansion w.r.t.\ the gauge
group. In this case, for fixed $(n,m)$ and flavor Weyl orbit
$\mathcal{O}_{p,k}^F$, there could be in general infinitely many gauge
Weyl orbits.\footnote{These properties can also be easily seen from
  the unity elliptic blowup equations. For example, the flavor
  parameters only appear in the nominators of blowup equations, which
  determine that there exist only finitely many different flavor Weyl
  orbits for elliptic genus at each fixed order of $q$ and $v$.}  Our
strategy here is to use the unity blowup equations to solve the
coefficient functions $c_{n,m,p,k}$.\footnote{It is proposed in
  \cite{DelZotto:2018tcj} that one should be able to
  $\IE_{1}^{\rm red}(v,q,m_G,m_F)$ in terms of representations of $G$
  and $F$ rather than just Weyl orbits. Nevertheless, from the
  viewpoint of solving blowup equations, the most natural setting is
  Weyl orbit expansion.}  Remember that the full one-string elliptic
genus is
\begin{equation}
  \IE_{1}(q,m_G,m_F,\eq,\et)=\frac{\eta^2}{\theta_1(\eq)\theta_1(\et)}
  \IE_{1}^{\rm red}(\re^{(\eq+\et)/2},q,m_G,m_F).
\end{equation}
For class $\bf A$ and $\bf B$ theories, the unity blowup equation for
one-string elliptic genus reads
\begin{align}
  &\phantom{-}\ \theta_{i}^{[a]}(\fn\tau,k_F\lF\cdot
    \mF+\fn y(\eq+\et))
    \IE_1(\mG,\mF,\eq,\et) \nn
  & -\theta_{i}^{[a]}(\fn\tau,k_F\lF\cdot \mF+\fn(y(\eq+\et)-
    \eq))\IE_{1}(\mG,\mF+\eq\lF,\eq,\et-\eq) \nn
  & -\theta_{i}^{[a]}(\fn\tau,k_F\lF\cdot \mF+\fn(y(\eq+\et)-
    \et))\IE_1(\mG,\mF+\et\lF,\eq-\et,\et) \nn =
  &\sum_{\lG\in Q_G^{\vee}} (-1)^{|\lG|}
    \theta_{i}^{[a]}(\fn\tau,-\fn\lG\cdot\mG+k_F\lF\cdot
    \mF+\fn((y-1)(\eq+\et))) \nn[-4mm]
  &\phantom{----}\times A_V(\tau,\mG,\lG) A_H(\tau,\mGF,\lGF).
    \label{eq:ubeq1}
\end{align}
By substituting the $v$ expansion ansatz (\ref{eq:vansatz}) into the
above equation, it is expected that all the coefficient functions
$c_{n,m,p,k}(m_G)$ can be determined. Note the shift such as
$\mF+\eq\lF$ in the elliptic genus will break the flavor Weyl orbits
into pieces, and the blowup equations work in a miraculous way such
that all such pieces in each term of the l.h.s.~are reorganized again
into a Weyl invariant on the r.h.s..

The complexity of solving Weyl orbit expansion in blowup equations
increases  with the complexity of the Weyl orbits of the  flavor group
$F$. Therefore, the solution process can be complicated (but still feasible)
when the flavor group is large. Practically, one can
normally just turn on a subgroup $\mf{su}(2)$ or $\mf{u}(1)$ of the
flavor group to make use of the blowup equations and still obtain the
elliptic genera with lots of useful information. In particular, if one
does not need the gauge fugacities, the computation can be even easier
where one can firstly turn off the gauge fugacities in unity blowup
equations (\ref{eq:ubeq1}) and then solve $c_{*,*,*,*}$ as numbers. In
such a situation, we can even withdraw the $v$ expansion and arrive in
the following useful ansatz for the reduced one-string elliptic genus of a
6d SCFT from a $-n$ curve:
\begin{equation}
  \IE_{1}^{\rm
    red}(v,q,m_G=0,m_F)=\delta_{n,1}\,
  q^{-\frac{1}{3}}+q^{\frac{1}{6}-\frac{n-2}{2}}\sum_{m,p,k}\frac{P_{m,p,k}(v^2)}{(1-v^2)^{2\dualCox-2}}q^m\mathcal{O}_{p,k}^F.
\end{equation}
Here the $P_{m,p,k}(v^2)$ are, up to an overall factor like $v^N$, $N\in\IZ$,
\emph{palindromical polynomial} functions of $v^2$ with integral
coefficients. We use this ansatz to solve many class $\bf B$ theories.

The leading $q$ order of
$\IE_{1}^{\rm red}(v,q,m_G,m_F)$,\footnote{The subleading order for
  $n=1$ theories.} i.e.~the 5d reduced one-instanton partition
function is expected to yield \emph{exact} formulae as $v$ expansions.
For example, it is well-known that the reduced one $G$-instanton
Nekrasov partition function in 5d, i.e.~the Hilbert series of the reduced
moduli space of one $G$-instanton has the following exact formula
\cite{Benvenuti:2010pq,Keller:2011ek}
\begin{equation}\label{Ginst1}
\sum_{n=0}^{\infty}v^{\dualCox-1+2n}\chi^G_{n\theta},\quad\quad\quad
\text{where $\theta$ is the Dynkin label for $\mathfrak{adj}_G$.}
\end{equation}
For theories with matter and flavor group $F$, the one-instanton
Nekrasov partition function still have an exact but more complicated $v$
expansion formula as follows
\begin{equation}\label{Ginst2}
\sum_{i\in I}\pm v^{k_i}\chi^G_{a_i}\chi^F_{b_i}+\sum_{j\in
  J}\Big(\pm\sum_{n=0}^{\infty}v^{h_j+2n}\chi^G_{c_j+n\theta}\chi^F_{d_j}\Big).
\end{equation}
Here both $I$ and $J$ are finite sets, $k_i$ and $h_j$ are integers,
$a_i$ and $c_j$ are Dynkin labels of $G$, while $b_i$ and $d_j$ are
Dynkin labels of $F$. Besides, $\pm$ means the coefficient can only be
either $+1$ or $-1$. For lots of 6d $(1,0)$ SCFTs, this type of exact
formulae were conjectured or found in \cite{DelZotto:2018tcj} and
\cite{Kim:2019uqw}. Benefiting from the results of one-string elliptic
genera solved from the blowup equations, we are able to confirm them
and obtain such exact formulas for more theories, see Section
\ref{sc:ex} and Appendix \ref{app:D}. We hope these exact $v$
expansion formulas could have an interpretation from 3d monopole
formulas \cite{Benvenuti:2010pq} in the future.

\subsection{Refined BPS expansion}
\label{sc:refBPS}

The one-loop partition function and the elliptic genera of a 6d SCFT
together can be identified with the worldsheet instanton partition
function $Z^{\text{inst}}$ of the topological string theory on the
associated non-compact Calabi-Yau threefold, i.e.
\begin{equation}
  Z^{\text{inst}} = Z^{\text{1-loop}}\Big(1+\sum_{d} \re^{\ri
    2\pi\phi\cdot d}\IE_d\Big).
\end{equation}
On the other hand, $Z^{\text{inst}}$ can be expressed in terms of
refined BPS invariants \cite{Iqbal:2007ii}
\begin{equation}
  Z^{\text{inst}} = \exp{\Bigg[
  \sum_{j_l,j_r=0}^\infty\sum_{\beta}\sum_{w=1}^\infty
  \frac{N^{\beta}_{j_l,j_r}}{w} f_{(j_l,j_r)}(q_1^w,q_2^w)Q^{w\beta}
  \Bigg]}.
\end{equation}
Here we define
\begin{equation}\label{fjljr}
  f_{(j_l,j_r)}(q_1,q_2) = \frac{\chi_{j_l}(q_l)\chi_{j_r}(q_r)}
  {(q_1^{1/2}-q_1^{-1/2})(q_2^{1/2}-q_2^{-1/2})},
\end{equation}
where $\chi_j(q)$ is the $\mf{su}(2)$ character given by
\begin{equation}
  \chi_j(q) = \frac{q^{2j+1}-q^{-2j-1}}{q-q^{-1}}.
\end{equation}
Define
\begin{equation}\label{Bldef}
  Bl_{(j_l,j_r,R)}(q_1,q_2) = f_{(j_l,j_r)}(q_1,q_2/q_1)q_1^R +
  f_{(j_l,j_r)}(q_1/q_2,q_2)q_2^R - f_{(j_l,j_r)}(q_1,q_2),
\end{equation}
where $R$ is the shift of the K\"ahler parameter $t$ in the blowup
equation, the blowup equation of topological string can be
reformulated as \cite{Gu:2019pqj}
\begin{equation}\label{eq:detail}
  \sum_{\und{n}\in \IZ^{b_4^c}}(-1)^{|\und{n}|}
  \re^{f_0(\und{n})(\eq+\et)+\sum_{i=1}^{b_2^c} f_i(\und{n})t_i}
  \exp\Bigg[-\sum_{j_l,j_r,\beta}\sum_{w=1}^\infty
  N^\beta_{j_l,j_r}
  \frac{Q^{w\beta}}{w}Bl_{(j_l,j_r,R)}(q_1^w,q_2^w)\Bigg]
  =\Lambda(\eq,\et,\und{m}),
\end{equation}
where $b_2^c, b_4^c$ are numbers of linearly independent compact
curves and surfaces in the Calabi-Yau geometry, and $f_0(n), f_i(n)$
are respectively some cubic and quadratic polynomials, whose
expressions can be found in \cite{Gu:2019pqj}.  For any fixed curve
degree $\beta$, comparison of coefficients on both sides of the
equation gives
\begin{equation}\label{eq:Qbeta-coefs}
  \sum_{j_l,j_r} N^\beta_{j_l,j_r}
  Bl_{(j_l,j_r,R(\beta,\und{n}_0))}(q_1,q_2)
  = I^\beta(q_1,q_2),
\end{equation}
where $I^{\beta}(q_1,q_2)$ consists only of invariants of lower curve
degrees.  The BPS invariants $N^\beta_{j_l,j_r}$ can be regarded as
the coefficients of $Bl_{(j_l,j_r,R(\beta,\und{n}_0))}(q_1,q_2)$, and
they can be fixed if the decomposition of $I^{\beta}(q_1,q_2)$ in
terms of $Bl_{(j_l,j_r,R(\beta,\und{n}_0))}(q_1,q_2)$ is unique.  It
was proved in \cite{Gu:2019pqj,Huang:2017mis} to be indeed the case
except for spin $(0,0)$ and spin $(0,1/2)$ invariants, and the
degeneracy can be lifted if genus zero BPS invariants are available.
We conclude therefore that if genus zero BPS invariants are provided
as initial input, we can always use one unity blowup equation to solve
all the refined BPS invariants order by order.  In practice, if we
have a toric construction of the Calabi-Yau geometry, we can use
mirror symmetry techniques \cite{Hosono:1993qy} to compute the genus
zero invariants to furnish the necessary input data.
In this paper, we constructed the geometries for $\fn=3, \mf{so}(7)$
and $\fn=1,2,3, G_2$ theories to get the genus zero invariants. We
list partial results of BPS invariants for these geometries in
Appendix \ref{app:bps}.

\subsection{$\eq,\et$ expansion}
\label{sc:e12}

One particular useful approach to study blowup equations is the
$\eq,\et$ expansion. This method applies to general local Calabi-Yau
threefolds and has been analyzed in detail in
\cite{Huang:2017mis}. Expanding both blowup equations and refined free
energy
\begin{equation}
  F({t},\epsilon_1,\epsilon_2)=
  \sum_{n,g=0}^{\infty}(\epsilon_1+\epsilon_2)^{2n}
  (\epsilon_1\epsilon_2)^{g-1}{F}_{(n,g)}({t})
\end{equation}
we can obtain lots of algebraic/differential equations for all
$F_{n,g}(t)$ which are graded by the ``total genus'' $n+g$. For
example, for total genus one free energies, blowup equation of one $r$
field in general gives
\begin{equation}\label{eq:u00}
\sum_{ n \in \mb Z^{g}}\Theta(n,r) :=\sum_{ n \in \mb Z^{g}}   (-)^{| n|} \exp \left(-\frac{1}{2} R^2 F_{(0,0)}''+F_{(0,1)}-F_{(1,0)}\right)=\Lambda_{0}(r),
\end{equation}
and
\begin{equation}\label{second}
\sum_{ n \in \mb Z^{g}} \left(-\frac{1}{6} R^3 F_{(0,0)}^{(3)}+R\left(F_{(0,1)}'+F_{(1,0)}'\right)\right)\Theta(n,r)=\Lambda_{1}(r).
\end{equation}
Here $g$ is the mirror curve genus, $R:=C\cdot n+r/2$ where $C$ is the
intersection matrix between the $b$ curve classes and the $g$
irreducible compact divisor classes. $R^k f^{(k)}$ is defined as
$\sum_{i_1,\dots,i_k=1}^b R_{i_1}\dots
R_{i_k}\partial_{t_{i_1}}\dots\partial_{t_{i_k}} f(t)$. Moreover,
$\Lambda_{0},\Lambda_1$ are the leading $\eq,\et$ expansion
coefficients of the full $\Lambda$ factor with
$\Lambda=\Lambda_0+(\eq+\et)\Lambda_1+\dots$. Consider a local
Calabi-Yau with one K\"ahler parameter and mirror curve of genus one,
it is easy to see that if there exists one unity blowup equation, the
component equation (\ref{eq:u00}) can be solved for  $F_{(0,1)}-F_{(1,0)}$,
while component equation (\ref{second}) can be solved for $F_{(0,1)}+F_{(1,0)}$,
thus we obtain $F_{(0,1)}$ and $F_{(1,0)}$ at the same time\footnote{We assume the linear coefficients $b_{\rm GV}$ and
  $b_{\rm NS}$ are already known, which fix the integration constants
  here.}.  In general, by counting the number of independent equations
for $F_{n,g}$, we can find the following conclusions.  Let $w_u$ and
$w_v$ be the number of unity and vanishing blowup equations
respectively.
\begin{itemize}
\item For a generic local Calabi-Yau with $b$ K\"ahler parameters, if
  $w_{\rm u}\ge 1$ and $w_{\rm u}+w_{\rm v}\ge b$, then given
  $F_{(0,0)}$, one can solve all $F_{(n,g)}$ with $n,g\ge0$ from
  blowup equations.
\item For a generic local Calabi-Yau with $b$ K\"ahler parameters, if
  $w_{\rm v}\ge b$, then given NS free energy or the self-dual free
  energy, i.e. all $F_{(n,0)}$ or all $F_{(0,g)}$, one can solve all
  $F_{(n,g)}$ with $n,g\ge0$ from the  blowup equations.
\end{itemize}
For example, all rank one theories in class $\mathbf{ B}$ can be
solved according to the first conclusion, and all rank one theories in
class $\mathbf{ C}$ except for the four $G=E_7$ theories can be solved
according to the second conclusion, if the necessary input data are
provided.  Since the full NS free energy or the full self-dual free
energy are themselves difficult to compute, and besides for 6d SCFTs,
we are more interested in the elliptic genera and BPS invariants
rather than $F_{n,g}$ themselves, we only use this method to discuss
the solvability of 6d theories in different classes, but do not
further explore this method.

\section{Examples}
\label{sc:ex}
In this section, we choose some of the most interesting rank one
theories to explicitly show the $\lF$ parameter
and the elliptic blowup equations. The chosen theories with gauge
symmetry of classical type all have known 2d quiver theory
correspondence, therefore the elliptic genera are exactly computable
via Jeffrey-Kirwan residue of localization. We checked against them
our results from blowup equations and found perfect agreement, mostly
for one-string elliptic genera and some up to two-string. For theories
with exceptional gauge symmetries, we explicitly show our
computational results on the elliptic genera for most of
them.\footnote{To shorten the paper, we usually only show the elliptic
  genera with gauge and flavor fugacities turned off or partially
  turned on. Readers interested in more detailed results for certain
  theories are welcome to send requests to us or visit the website \cite{kl}.}  Sometimes to specify a theory with base curve $-n$
and gauge group $G$, we denote the reduced $k$-string elliptic genus
as
\begin{equation}
  \IE_{h_{n,G}^{(k)}}(q,v,x,m_G,m_F)=
  \frac{\theta_1(\tau,\eq)\theta_1(\tau,\et)}{\eta(\tau)^2}
  \IE_{k}(\tau,m_G,m_F,\eq,\et).
\end{equation}
Recall $v=\re^{(\eq+\et)/2}=\re^{\epsilon_+}$, $x=\re^{(\eq-\et)/2}$ and
reduced one-string elliptic genus does not depend on $x$.

We also show some interesting theta identities coming from the leading degree of vanishing blowup equations. Although we have checked the leading degree identities for all the vanishing blowup equations in Tables~\ref{tb:vbeq-1},\ref{tb:vbeq-2},\ref{tb:vbeq-3}, here we only explicitly written down a small part of them, in particular with $\lambda_F$ in small representations.

\subsection{$n=1$ $\mf{sp}(N)$ theories}

The $n=1$ $G=\mf{sp}(N)$ theories have $8+2N$ fundamental
hypermultiplets and flavor symmetry $\mf{so}(16+4N)$. For $N=0$, it
specializes to E-string theory, with flavor symmetry $\mf{so}(16)$
enhanced to $E_8$. The 2d quiver description for these theories was
proposed in \cite{Kim:2015fxa,Yun:2016yzw}, therefore their elliptic
genera can be exactly computed from localization. For example, the
reduced one-string elliptic genus has been shown in equation
(\ref{n1SpNreducedZ1}).
The index of $d$-string elliptic genus of $\mf{sp}(N)$ theory is known
to be
\begin{equation}
  \mathrm{Ind}_{\mathbb{E}_d}=
  -\frac{N+2}{4}(\eq+\et)^2d+\eq\et\frac{d^2+d}{2}
  -\frac{d}{2}(m,m)_{\mf{sp}(N)}+\frac{d}{2}(m,m)_{\mf{so}(16+4N)}.
\end{equation}

Let us first discuss the vanishing blowup equations. Since for $C$
type Lie algebra $(P^\vee\slash Q^{\vee})_{C_n}\cong\IZ_2$, there
could exist one vanishing equation when the paramter $\lF$ and the
characteristic $a$ are fixed with $\lambda_G$ taking value in
$(P^\vee\backslash Q^{\vee})_{C_n}$. Denote the smallest Weyl orbit in
$(P^\vee\backslash Q^{\vee})_{C_n}$ as $\mathcal{O}_{\rm min}$, which
is just $\mathcal{O}_{[00\cdots01]}^{\mf{sp}(N)}$. Note
$|\mathcal{O}_{\rm min}|=2^N$. Then for $N\ge 2$, the leading base
degree of the vanishing blowup equations with $\lambda_F=0$ can be
universally written as
\begin{equation}\label{spvanishj0}
\begin{aligned}
  \sum_{w\in\mathcal{O}_{\rm min}}(-1)^{|w|}\theta_1(\tau,m_w)\times
  \prod_{\beta\in\Delta(C_N)}^{w\cdot\beta=
    1}\frac{1}{\theta_1(\tau,m_{\beta})}=0,\quad\quad N\ge2.
\end{aligned}
\end{equation}
We have checked this identity up to $\mathcal{O}(q^{20})$ for
$N=2,3,4,5$. Note there are $(N+1)N/2$ Jacobi $\theta_1$ functions
in the denominator.

The $G=\mf{sp}(1),F=\mf{so}(20)$ case is peculiar due to the Lie
algebra isomorphism $C_1\cong A_1$. In fact, it is easy to check
(\ref{spvanishj0}) \emph{does not} hold for $N=1$. The correct
$\lambda_F$ in this case belongs to vector representation $\bf 20_v$
of $\mf{so}(20)$.  The leading base degree of the vanishing blowup
equations turn out to be the following trivial identity
\begin{equation}\label{sp1vanish}
\begin{aligned}
\theta_1(m_w+\lambda_F\cdot m_F+\epsilon_+)\theta_1(-m_w+\lambda_F\cdot m_F+\epsilon_+)-(w\to -w)=0.
\end{aligned}
\end{equation}
Here $w$ is the fundamental weight of $\mf{sp}(1)$, the first
$\theta_1$ comes from the contribution of perturbative part, the
second $\theta_1$ comes from the contribution of hypermultiplet and we
have factored out the contribution from vector multiplet.


Now let us turn to unity blowup equations. All the unity $\lambda_F$
fields are just the weights of the spinor representation
$S = [0,0,\dots,0,1]$ of $\mf{so}(16+4N)$. There are $2^{7+2N}$ of
them. The matters are in representation
$([1,0,\dots,0,0],[1,0,\dots,0,0])$ of
$\mf{sp}(N)\times \mf{so}(16+4N)$, i.e.~$({\bf 2N, 16+4N})$. The
following fact about the weight system of $S$ is
crucial for blowup equations to hold: $\forall w\in [0,0,\dots,0,1]$,
there are precisely $8+2N$ weights $w'\in\mathbf{16+4N}$ such that
$w\cdot w'=1/2$, and the rest $8+2N$ weights $w'\in\mathbf{16+4N}$ are
such that $w\cdot w'=-1/2$. Besides, $w\cdot w=2+N/2$. Note the
conjugate spinor representation $C=[0,0,\dots,1,0]$ of
$\mf{so}(16+4N)$ if serving as $\lambda_F$ is not correct, although
they satisfy the modularity of unity blowup equations!

The unity elliptic blowup equations for $G=\mf{sp}(N)$ theory with
$\lambda_F=\lambda\in S_{\mf{so}(16+4N)}$ can be written as
\begin{equation}
\begin{aligned}
  \sum_{\displaystyle \substack{d_0=\frac{1}{2}||\alpha^{\vee}||_{\mf{sp}(N)}}}^{d_0+d_1+d_2=d}&(-1)^{|\alpha^{\vee}|}\theta_{1}\Big(\tau,-\alpha^{\vee}\cdot m_{\mf{sp}(N)}+{\lambda\cdot m_{\mf{so}(16+4N)}}+\Big(\frac{N+2}{2}-d_0\Big)\(\eq+\et\)-d_1\eq-d_2\et\Big)\\[-3mm]
  &\times A_{V}^{\mf{sp}(N)}(\alpha^{\vee},\tau,m_{\mf{sp}(N)})A^{\frac{1}{2}(\mathbf{2N,4N+16}) }_{H}(\alpha^{\vee},\tau,m_{\mf{sp}(N)},m_{\mf{so}(16+4N)},\lambda)\\ &\times \IE_{d_1}\big(\tau,m_{\mf{sp}(N)}+\eq\alpha^{\vee},m_{\mf{so}(16+4N)}+\eq \lambda,\eq,\et-\eq\big)\\
  &\times
  \IE_{d_2}\big(\tau,m_{\mf{sp}(N)}+\et\alpha^{\vee},m_{\mf{so}(16+4N)}+\et
  \lambda,\eq-\et,\et\big)\\
  =\,\theta_1&\Big(\tau,{\lambda\cdot
    m_{\mf{so}(16+4N)}}+\frac{N+2}{2}(\eq+\et)\Big)\,
  \IE_{d}\big(\tau,m_{\mf{sp}(N)},m_{\mf{so}(16+4N)},\eq,\et\big).\label{unityn1spN}
\end{aligned}
\end{equation}
For $N=0$ case, there is no summation over coroots, and the above
equation goes back to the unity blowup equations of E-strings given in
\cite{Gu:2019pqj}. For $N=1$ case, using the 2d quiver formula for
one-string elliptic genus (\ref{n1SpNreducedZ1}), we have verified the unity blowup equations
for all possible $\lambda_F$ up to $\mathcal{O}(\Qtau^{10})$.

Let us have a closer look at the $N=1$ case. From the 2d quiver formula (\ref{n1SpNreducedZ1}),
it is easy to find the following expansion
\begin{align}
  \IE_{h_{1,\mf{sp}(1)}^{(1)}}=&\,q^{-1/3}+\Big(\frac{\chi_{(2)}^{\mf{sp}(1)}}{v^2}-\frac{\chi_{(1)}^{\mf{sp}(1)}\cdot \mathbf{20}_{\mathrm{v}}}{v}+\chi_{(2)}^{\mf{sp}(1)}+1+\mathbf{190}-\chi_{(1)}^{\mf{sp}(1)}\cdot\mathbf{20}_{\mathrm{v}}v
  +\chi_{(2)}^{\mf{sp}(1)}v^2\nn
  &\ +\sum_{n=0}^{\infty}\Big[\chi_{(2n)}^{\mf{sp}(1)}\cdot\mathbf{512}_{\mathrm{s}}\, v^{1+2n}-\chi_{(1+2n)}^{\mf{sp}(1)}\cdot\mathbf{512}_{\mathrm{c}}\,v^{2+2n}\Big]\Big)q^{2/3}
  +\mathcal{O}(q^{5/3}).\label{E1n1sp1first}
\end{align}
Here the bold numbers are the representations for flavor symmetry $\mf{so}(20)$
or its character based on the context. We can also check the unity
blowup equation using this expansion or use the Weyl orbit expansion
method to solve elliptic genus in this
form from unity blowup equation (\ref{unityn1spN}).
We summarize some useful information on the intersection distribution
relevant to Weyl orbit expansion in Table \ref{tb:n1sp1}. Combining
equation (\ref{E1n1sp1first}) and Table \ref{tb:n1sp1}, one can
already understand why weights in the spinor representation
$\bf 512_s$ can serve as $\lambda_F$ fields while weights in the
conjugate spinor representation $\bf 512_c$ cannot. This is because in
(\ref{E1n1sp1first}) the coefficients of each $v^n$ should have
$\lambda_F$ shifts all even or all odd to preserve the $B$ field
condition.
\begin{table}[h]
  \centering
  \begin{tabular}{c|c|c|c|c|c|c|c|c|c|c|c|c}\hline
   $\lambda$ & $w$ &$-5/2$ & $-2$ & $-3/2$ & $-1$ & $-1/2$ & 0 & $1/2$ &  1 & $3/2$ & $2$ & $5/2$ \\ \hline
   $\bf 512_c$ & $\bf 512_s$ & & 10 & & 120 & & 252 &  & 120 &  & 10 & \\ \hline
   $\bf 512_c$ & $\bf 20_v$ & & &  & &  10 & & 10 & &  & & \\ \hline
   $\bf 512_c$ & $\bf 512_c$ & $1$ &  & 45 &  & 210 & & 210 & & 45 & & 1  \\ \hline
   $\bf 512_s$ & $\bf 512_s$ & $1$ &  & 45 &  & 210 & & 210 & & 45 & & 1 \\ \hline
   $\bf 512_s$ & $\bf 20_v$ & &  & &  & 10 & & 10 & &  & & \\ \hline
   $\bf 512_s$ & $\bf 512_c$ &  & 10 & & 120 & & 252 &  & 120 &  & 10 &  \\ \hline
     \end{tabular}
     \caption{For any $\lambda$ in a fixed reprentation, the numbers
       of weights $w$ in another representation with $\lambda\cdot w$
       equal to a given number.} \label{tb:n1sp1}
\end{table}

Similarly, from (\ref{n1SpNreducedZ1}), the reduced one-string elliptic genus of $G=\mf{sp}(2),F=\mf{so}(24)$ theory has the following expansion
\begin{align}
  \IE_{h_{1,\mf{sp}(2)}^{(1)}}=
  &\,q^{-1/3}+\Big(\frac{\chi_{(20)}^{\mf{sp}(2)}}{v^2}
    -\frac{\chi_{(10)}^{\mf{sp}(2)}\cdot \mathbf{24}_{\mathrm{v}}}{v}
    +\chi_{(20)}^{\mf{sp}(2)}+1+\mathbf{276}
    -\chi_{(10)}^{\mf{sp}(2)}\cdot \mathbf{24}_{\mathrm{v}}v
    +\chi_{(20)}^{\mf{sp}(2)}v^2\nn
  &\ +\sum_{n=0}^{\infty}\Big[\chi_{(2n,0)}^{\mf{sp}(2)}\cdot
    \mathbf{2048}_{\mathrm{s}}\, v^{2+2n}
    -\chi_{(1+2n,0)}^{\mf{sp}(2)}\cdot\mathbf{2048}_{\mathrm{c}}
    \,v^{3+2n}\Big]\Big)q^{2/3}
    +\mathcal{O}(q^{5/3}),\label{E1n1sp2first}
\end{align}
which we also reconfirm by solving the unity blowup equations in Weyl
orbit expansion with the fugacity of one subalgebra
$\mf{so}(3)$ of the flavor symmetry turned on.

\subsection{$n=1$ $\mf{su}(N)$ theories}

All $n=1$ $\mf{su}(N)$ theories with $N\ge 2$ have known universal 2d
quiver gauge constructions in \cite{Kim:2015fxa}, therefore the
elliptic genera are exactly computable via Jeffrey-Kirwan residue. For example,
the reduced one-string elliptic genus can be universally written as
\begin{equation}\label{n1suNquiver}
  \begin{aligned}
\,&-\sum_{i=1}^N\frac{\prod_{j=1}^{N+8}\theta_1(m_i-\ep_+-\mu_j)}{\eta^8\theta_1(2m_i-3\ep_++\mu_{N+9})}\prod_{1\le j\le N}^{j\neq i}\frac{\theta_1(m_i+m_j-\ep_++\mu_{N+9})}{\theta_1(m_i-m_j)\theta_1(2\ep_+-(m_i-m_j))}\\
&-\frac{1}{2\eta^8}\bigg(\frac{\prod_{j=1}^{N+8}\theta_1(\frac{\ep_+-\mu_{N+9}}{2}-\mu_j)}{\prod_{i=1}^{N}\theta_1(\frac{3\ep_+-\mu_{N+9}}{2}-m_i)}+(-1)^N\sum_{k=2}^4
\frac{\prod_{j=1}^{N+8}\theta_k(\frac{\ep_+-\mu_{N+9}}{2}-\mu_j)}{\prod_{i=1}^{N}\theta_k(\frac{3\ep_+-\mu_{N+9}}{2}-m_i)}\bigg).
\end{aligned}
\end{equation}
Here $m_i,i=1,2,\dots,N$ are the symmetric $\mf{su}(N)$ fugacities and
$\mu_j,j=1,2,\dots,N+9$ are the symmetric $\mf{su}(N+9)$
fugacities. Note this formula is from UV 2d gauge theory, where the IR
global symmetry i.e. the true flavor symmetry does not manifest
itself. One can convert $\mu_j$ into the fugacities of the
true flavor symmetry according to the matter representations. Note all
these theories are on one single branch of the E-string Higgsing tree,
which is also easy to see from the above elliptic genus
formula.\footnote{See some recent discussion on this Higgsing tree
  using Hasse diagrams in \cite{Bourget:2019aer}.} Besides, for the
$N=2$ case, the flavor symmetry is enhanced to $\mf{so}(20)$ which is
just the $n=1, G=\mf{sp}(1)$ theory we have discussed in the last
subsection.

One additional case is the $G=\mf{su}(6)_*, F=\mf{su}(15)$ theory,
where there is a half hypermultiplet in the 3-antisymmetric
representation $\Lambda^3_{\mf{su}(6)}=\bf 15$. This theory does not
have known 2d quiver gauge construction, but has a brane web
construction, thus the topological string partition function can be
computed by refined topological vertex \cite{Hayashi:2019yxj}. Due to
the presence of half hypermultiplet, this theory does not have unity
blowup equation. This is the single case with $\mf{su}$ gauge symmetry
where we could not solve elliptic genera from blowup equations.

Let us first discuss the vanishing blowup equations. We have shown
some leading degree vanishing identities for
$G=\mf{su}(3),F=\mf{su}(15)$ theory in (\ref{thetaindentI}) and
(\ref{thetaindentIeasy}) with $\lambda_G=\mathbf{3}$.  In fact, the
vanishing theta identity (\ref{thetaindentIeasy}) can be generalized
to all $N\ge2$:
\begin{equation}\label{thetaindentn1suN}
  \begin{aligned}
\sum_{i=1}^N\frac{\theta_1(-a_i+\sum_{k=1}^{N-1}x_k)\prod_{k=1}^{N-1}\theta_1(a_i+x_k)}{\prod_{1\le j\le N}^{j\neq i}\theta_1(a_i-a_j)}
=0,\quad\quad
    \textrm{for } \sum_{i=1}^N a_i=0.
\end{aligned}
\end{equation}
These identities come from the leading base degree of vanishing blowup
equations for $G=\mf{su}(N)$ theories with
$\lambda_G=\omega_1\in\mathbf{N}$, i.e. the first fundamental weight
that induces the fundamental representation.  Similarly, for
$\lambda_G=\omega_2\in\Lambda^2$ i.e. the second fundamental weight
that induces the anti-symmetric representation, we find the leading
degree of vanishing blowup equations result in the following
identities for arbitrary $N\ge 4$,
\begin{equation}\label{thetaindentn1suNw2}
  \begin{aligned}
\sum_{1\le i<j\le N}\frac{\theta_1(-a_i-a_j+y+\sum_{k=1}^{N-4}x_k)\theta_1(a_i+a_j+y)\prod_{k=1}^{N-4}\theta_1(a_i+x_k)\theta_1(a_j+x_k)}{\prod_{1\le k\le N}^{k\neq i,j}\theta_1(a_i-a_k)\theta_1(a_j-a_k)}
=0.
\end{aligned}
\end{equation}
More generally, for $\lambda_G=\omega_k$, we find the leading degree
of vanishing blowup equations result in the following identities for
arbitrary $N\ge 3k-2$,
\begin{equation}\label{thetaindentn1suNwk}
  \begin{aligned}
\sum_{1\le i_1<i_2< \dots< i_k\le N}&\frac{\theta_1(-\sum_{s=1}^ka_{i_s}+(k-1)y+\sum_{h=1}^{N-3k+2}x_h)\prod_{1\le s<s'\le k}\theta_1(a_{i_s}+a_{i_{s'}}+y)
}{\prod_{1\le l\le N}^{l\neq i_1,\dots,i_k}\prod_{s=1}^k\theta_1(a_{i_s}-a_l)}\\
&\times\prod_{h=1}^{N-3k+2}\prod_{s=1}^k\theta_1(a_{i_s}+x_h)
=0.
\end{aligned}
\end{equation}
Here still $\sum_{i=1}^N a_i=0$ and $y$ and $x_k$ are arbitrary
numbers. We have checked this identity for many different $(N,k)$ up
to very high order of $q$. Note the second line in
(\ref{thetaindentn1suNwk}) comes from the contribution of
hypermultiplets in $(\md{N},\wb{\md{N+8}})_{-N+4}$, while the product
$\prod_{1\le s<s'\le k}\theta_1(a_{i_s}+a_{i_{s'}}+y)$ in the first
line comes from the contribution of hypermultiplets in
$(\md{\Lambda}^2,\md{1})_{N+8}$.  Besides, we also find the following
identity for arbitrary $N\ge 1$:
\begin{equation}\label{thetaindentn1suNwmiddle}
\sum_{1\le i_1<i_2< \dots< i_N\le
  2N}\frac{\theta_1(-\sum_{s=1}^Na_{i_s}+\ep_+)\theta_1(\sum_{s=1}^Na_{i_s}+\ep_+)}{\prod_{1\le
    l\le 2N}^{l\neq
    i_1,\dots,i_N}\prod_{s=1}^N\theta_1(a_{i_s}-a_l)}=0,\quad\quad
\textrm{for } \sum_{i=1}^{2N} a_i=0.
\end{equation}
This identity is related to the situation where matters are in the
middle representation of gauge group such as the $\mf{su}(6)_*$ theory
with 3-antisymmetric representation. For example, taking $N=2$ it
gives the leading base degree of vanishing blowup equation of
$n=1,G=\mf{su}(4),F=\mf{su}(12)\times \mf{su}(2)$ theory with
$(\lG,\lF) = (\mathbf{6},\mathbf{1})$, and taking $N=3$ gives the one
of $n=1,G=\mf{su}(6)_*,F=\mf{su}(15)$ theory with
$(\lG,\lF) = (\mathbf{20},\mathbf{1})$.

Now we turn to the unity blowup equations for all $\mf{su}(N)$
theories with $N\ge 3$. Since flavor enhancement does not matter here,
let us use the symmetric $\mf{su}(N+9)$ fugacities
$\mu_j,j=1,2,\dots,N+9$ in (\ref{n1suNquiver}) to make the form of
blowup equations universal. Consider the flavor decomposition
$\mf{su}(N+8)\oplus \mf{u}(1)\subset\mf{su}(N+9)$ according to
\begin{equation}
(\nu_1+\nu_0,\nu_2+\nu_0,\dots,\nu_{N+8}+\nu_0,-(N+8)\nu_0)
\end{equation}
such that $\nu_{j},j=1,2,\dots N+8$ are the symmetric $\mf{su}(N+8)$
fugacities and $\nu_0$ is the $\mf{u}(1)$ fugacity. Then the unity $r$
fields have two possibilities
\begin{equation}
  \lambda=(\lambda^{\mf{su}(N+8)},\lambda^{\mf{u}(1)})=\Big(\omega_6,-\frac{1}{2(N+8)}\Big)\text{ or }\Big(\omega_{N+2},\frac{1}{2(N+8)}\Big).
\end{equation}
The unity blowup equations can be universally written as
\begin{equation}
\begin{aligned}
  \sum_{\displaystyle \substack{d_0=\frac{1}{2}||\alpha^{\vee}||_{\mf{su}(N)}}}^{d_0+d_1+d_2=d}&(-1)^{|\alpha^{\vee}|}\theta_{1}\Big(\tau,-\alpha^{\vee}\cdot m_{\mf{su}(N)}+{\lambda\cdot \mu_{\mf{su}(N+9)}}+\Big(\frac{N+1}{2}-d_0\Big)\(\eq+\et\)-d_1\eq-d_2\et\Big)\\[-3mm]
  &\times A_{V}^{\mf{su}(N)}(\alpha^{\vee},\tau,m_{\mf{su}(N)})A^{\mathfrak{R}}_{H}(\alpha^{\vee},\tau,m_{\mf{su}(N)},\mu_{\mf{su}(N+9)},\lambda)\\ &\times \IE_{d_1}\big(\tau,m_{\mf{su}(N)}+\eq\alpha^{\vee},\mu_{\mf{su}(N+9)}+\eq \lambda,\eq,\et-\eq\big)\\
  &\times
  \IE_{d_2}\big(\tau,m_{\mf{su}(N)}+\et\alpha^{\vee},\mu_{\mf{su}(N+9)}+\et
  \lambda,\eq-\et,\et\big)\\
  =\,\theta_1&\Big(\tau,{\lambda\cdot
    \mu_{\mf{su}(N+9)}}+\frac{N+1}{2}(\eq+\et)\Big)\,
  \IE_{d}\big(\tau,m_{\mf{su}(N)},\mu_{\mf{su}(N+9)},\eq,\et\big).\label{unityn1suN}
\end{aligned}
\end{equation}
For $N=4$, it is easy to find the two copies of $\lambda$ combined
together form the middle representation
$\chi_{(0000001000000)}=\mathbf{924}$ of flavor $\mf{su}(12)$. Indeed,
for arbitrary one of the 924 $\lambda$ fields, we have used the quiver
formula (\ref{n1suNquiver}) to check the above unity blowup equations
up to $\mathcal{O}(q^{10})$.  Conversely, we also used blowup equation
(\ref{unityn1suN}) to solve elliptic genus independently. In the
following we show two examples. As the quiver formulas are powerful
enough for computational purposes in these cases, we only turn on a
small subgroup of the flavor to solve blowup equations and only to the
subleading $q$ order which contains the information of 5d
one-instanton partition functions.

\subsubsection*{$\mathbf{n=1,\, G=\mf{su}(3),\,F=\mf{su}(12)}$}
Using the Weyl orbit expansion, we turn on a subgroup $\mf{su}(2)$ of
the flavor group to compute the elliptic genus. We obtain the reduced
one-string elliptic genus as
\begin{equation}\label{n1su3E1}
  \mathbb{E}_{h_{1,\mf{su}(3)}^{(1)}}(\Qtau,v,m_{\mf{su}(3)}=0,m_{\mf{su}(12)}=0)=\Qtau^{-1/3}+\Qtau^{2/3}v^{-2}\sum_{n=0}^{\infty}\Qtau^n\frac{P_n(v)}{(1 + v)^{4}},
\end{equation}
where
\begin{equation}\nonumber
\begin{aligned}
P_0(v)=8 (1 - 5 v - 11 v^2 + 81 v^3 + 364 v^4 + 81 v^5 - 11 v^6 - 5 v^7 +
   v^8).
\end{aligned}
\end{equation}
This agrees with the modular ansatz in \cite{DelZotto:2018tcj}.
Using the result with flavor fugacities turned on, we obtain the following exact $v$ expansion formula for the subleading $q$ order coefficient, which contains the 5d one-instanton Nekrasov partition function:
\begin{align} \nonumber
\,&\chi_{(1,1)}^{\mf{su}(3)}v^{-2}-(\chi_{(1,0)}^{\mf{su}(3)}\chi^{\mf{su}(12)}_{(00000000001)}+c.c.)v^{-1}+\chi_{(1,1)}^{\mf{su}(3)}
+1+\chi^{\mf{su}(12)}_{(10000000001)}\nn
&\,+(\chi^{\mf{su}(12)}_{(00100000000)}v-\chi_{(1,0)}^{\mf{su}(3)}\chi^{\mf{su}(12)}_{(01000000000)}v^2
+\chi_{(2,0)}^{\mf{su}(3)}\chi^{\mf{su}(12)}_{(10000000000)}v^3-\chi_{(3,0)}^{\mf{su}(3)}v^4+c.c\,)\nn
+&\sum_{n=0}^{\infty}\,\Big[\chi_{(n,n)}^{\mf{su}(3)}\chi^{\mf{su}(12)}_{(00000100000)}v^{2+2n}+\Big(
-\chi_{(n+1,n)}^{\mf{su}(3)}\chi^{\mf{su}(12)}_{(00001000000)}v^{3+2n}+\chi_{(n+2,n)}^{\mf{su}(3)}\chi^{\mf{su}(12)}_{(00010000000)}v^{4+2n}\nn
&\qquad-\chi_{(n+3,n)}^{\mf{su}(3)}\chi^{\mf{su}(12)}_{(00100000000)}v^{5+2n}+\chi_{(n+4,n)}^{\mf{su}(3)}\chi^{\mf{su}(12)}_{(01000000000)}v^{6+2n}\nn
&\qquad-\chi_{(n+5,n)}^{\mf{su}(3)}\chi^{\mf{su}(12)}_{(10000000000)}v^{7+2n}+\chi_{(n+6,n)}^{\mf{su}(3)}v^{8+2n}+c.c.\Big)
\Big].\label{n1SU3HS}
\end{align}
Here $c.c.$ means complex conjugate. We have checked this agrees with the localization formula
(\ref{n1suNquiver}) from 2d quiver gauge theory.

\subsubsection*{$\mathbf{n=1,\, G=\mf{su}(4),\,F=\mf{su}(12)_a\times\mf{su}(2)_b}$}
Using the Weyl orbit expansion, we turn on a subgroup
$\mf{su}(2)\times \mf{su}(2)$ of the full flavor to compute the
elliptic genus. We obtain the reduced one-string elliptic genus as
\begin{equation}\label{n1su4E1}
  \mathbb{E}_{h_{1,\mf{su}(4)}^{(1)}}(\Qtau,v,m_{\mf{su}(4)}=0,m_{F}=0)=\Qtau^{-1/3}+\Qtau^{2/3}v^{-2}\sum_{n=0}^{\infty}\Qtau^n\frac{P_n(v)}{(1 + v)^{6}},
\end{equation}
where
\begin{equation}\nonumber
  \begin{aligned}
    P_0(v)=15 - 18 v - 261 v^2 - 72 v^3 + 2934 v^4 + 10676 v^5 + 2934
    v^6 - 72 v^7 - 261 v^8 - 18 v^9 + 15 v^{10}.
\end{aligned}
\end{equation}
This agrees with the modular ansatz in \cite{DelZotto:2018tcj}.  Using
the result with flavor fugacities turned on, we obtain the following
exact $v$ expansion formula for the subleading $q$ order coefficient,
which contains the 5d one-instanton Nekrasov partition function:
\begin{align} \nonumber
\,&\chi_{(1,0,1)}^{\mf{su}(4)}v^{-2}-(\chi_{(1,0,0)}^{\mf{su}(4)}\chi^{F}_{(00000000001)_a}+c.c.)v^{-1}-\chi_{(0,1,0)}^{\mf{su}(4)}\chi^{F}_{(1)_b}v^{-1}
+\chi_{(1,0,1)}^{\mf{su}(4)}+\chi^{F}_{(10000000001)_a\oplus(2)_b}\nn
&+1 +\chi^F_{(1)_b}(\chi^{F}_{(01000000000)_a}v-\chi_{(1,0,0)}^{\mf{su}(4)}\chi^{F}_{(10000000000)_a}v^2+\chi_{(2,0,0)}^{\mf{su}(4)}v^3+c.c.)
+                  \chi_{(0,1,0)}^{\mf{su}(4)}\chi^F_{(1)_b}v\nn
& -\chi_{(1,0,1)}^{\mf{su}(4)}v^2 +( \chi^{F}_{(00010000000)_a}v^2-\chi_{(1,0,0)}^{\mf{su}(4)}\chi^{F}_{(00100000000)_a}v^3
+\chi_{(2,0,0)}^{\mf{su}(4)}\chi^{F}_{(01000000000)_a}v^4\nn
&-\chi_{(3,0,0)}^{\mf{su}(4)}\chi^{F}_{(10000000000)_a}v^5 +\chi_{(4,0,0)}^{\mf{su}(4)}v^6+c.c.)\nn
+\,&\sum_{n=0}^{\infty}\Big[\chi^F_{(1)_b}\Big(
\chi_{(n,0,n)}^{\mf{su}(4)}\chi^{F}_{(00000100000)_a}v^{3+2n}+(-\chi_{(n+1,0,n)}^{\mf{su}(4)}\chi^{F}_{(00001000000)_a}v^{4+2n}\nonumber\\[-1mm]
+\,&\chi_{(n+2,0,n)}^{\mf{su}(4)}\chi^{F}_{(00010000000)_a}v^{5+2n}-\chi_{(n+3,0,n)}^{\mf{su}(4)}\chi^{F}_{(00100000000)_a}v^{6+2n}
+\chi_{(n+4,0,n)}^{\mf{su}(4)}\chi^{F}_{(01000000000)_a}v^{7+2n}\nn
-\,&\chi_{(n+5,0,n)}^{\mf{su}(4)}\chi^{F}_{(10000000000)_a}v^{8+2n}
+\chi_{(n+6,0,n)}^{\mf{su}(4)}v^{9+2n}+c.c.)\Big)+\Big(-\chi_{(n,1,n)}^{\mf{su}(4)}\chi^{F}_{(00000100000)_a}v^{4+2n}\nn
+\,&(\chi_{(n+1,1,n)}^{\mf{su}(4)}\chi^{F}_{(00001000000)_a}v^{5+2n}-\chi_{(n+2,1,n)}^{\mf{su}(4)}\chi^{F}_{(00010000000)_a}v^{6+2n}
+\chi_{(n+3,1,n)}^{\mf{su}(4)}\chi^{F}_{(00100000000)_a}v^{7+2n}\nn
-\,&\chi_{(n+4,1,n)}^{\mf{su}(4)}\chi^{F}_{(01000000000)_a}v^{8+2n}+\chi_{(n+5,1,n)}^{\mf{su}(4)}\chi^{F}_{(10000000000)_a}v^{9+2n}
-\chi_{(n+6,1,n)}^{\mf{su}(4)}v^{10+2n}+c.c.)
\Big)
\Big].\label{n1SU4HS}
\end{align}
We have checked this agrees with the localization formula (\ref{n1suNquiver}) from 2d quiver gauge theory.

We also used the Weyl orbit expansion to compute the reduced one-string elliptic genus of $G=\mf{su}(5),\,F=\mf{su}(13)_a\times\mf{u}(1)_b$ theory and found consistency with the modular ansatz in \cite{DelZotto:2018tcj}.

\subsection{$n=2$ $\mf{su}(N)$ theories}
\label{sec:exn2su}
The $n=2$ $\mf{su}(N)$ theories with flavor $\mf{su}(2N)$ and matter
in bi-representation
$(R^G,R^F)=({\bf N},{\bf \overline{2N}})$ are the most
familiar SCFTs. The theory at $N=2$ is special, as the flavor symmetry
is enhanced to $\mf{so}(7)$. Nevertheless, as the flavor enhancement
does not affect blowup equations, one can still use $\mf{su}(4)$
effectively. Besides, the $N=1$ case is just the M-string. All these
theories are on one single Higgsing tree, and the elliptic genus of
$\mf{su}(N)$ theory can be obtained by Higgsing from the elliptic
genus of $\mf{su}(N+1)$ theory.
The 2d quiver construction is a slight modification of the $A_1$
string chain with $\mf{su}(N)$ gauge group proposed in
\cite{Gadde:2015tra}. By Jeffrey-Kirwan residue, the $n$-string elliptic genus can
be computed as\footnote{Here we adopt the same notation as in \cite{Gadde:2015tra} to make the formula simple. The variable of theta functions are  multiplicative. Deformation parameters $t,d=\re^{\epsilon_{1,2}}$. The coordinates of the boxes in a Young
  diagram start from 0 rather than 1.  }
\begin{equation}
\begin{aligned}\label{n2sunquiver}
  \,& \IE^{N}_{n} =\sum_{\sum_{\ell=1}^{N}\vert Y_{\ell}
    \vert=n}\Bigg( \prod_{\ell,m=1}^{N} \prod_{\substack{(x_1,y_1) \in
      Y_\ell\\(x_2,y_2) \in
      Y_m}}\frac{\theta_1(\frac{s_{\ell}}{s_{m}}t^{{x_1}-{x_2}}d^{{y_1}-{y_2}})
    \theta_1(\frac{s_{\ell}}{s_{m}}t^{{x_1}-{x_2}+1}d^{{y_1}-{y_2}+1})}{\theta_1(\frac{s_{\ell}}{s_{m}}t^{{x_1}-{x_2}+1}d^{{y_1}-{y_2}})
    \theta_1(\frac{s_{\ell}}{s_{m}}t^{{x_1}-{x_2}}d^{{y_1}-{y_2}+1})}\Bigg)\\[-1mm]
  \times \,& \Bigg( \prod_{\ell,m=1}^{N} \prod_{(x,y) \in
    Y_\ell}\theta_1(\frac{s_{\ell}}{s_{m}}t^{x}d^{y})\theta_1(\frac{s_{\ell}}{s_{m}}t^{x+1}d^{y+1})
  \Bigg)^{-1} \Bigg( \prod_{\ell=1}^{N}\prod_{m=1}^{2N} \prod_{(x,y)
    \in
    Y_\ell}\theta_1(\frac{s_{\ell}}{f_{m}}t^{x+\frac{1}{2}}d^{y+\frac{1}{2}})\Bigg).
\end{aligned}
\end{equation}
Here $s_{\ell},\ell=1,\dots,N$ are gauge parameters for $\mf{su}(N)$,
and $f_{m},m=1,\dots,2N$ are the flavor parameters for
$\mf{su}(2N)$. Note the products include various factors of
$ \theta_1(1) $, which however completely cancel against each
other. The index of $d$-string elliptic genera of $\mf{su}(N)$ theory
is known to be
\begin{equation}
  \mathrm{Ind}_{\mathbb{E}_d}=-\frac{Nd}{4}(\eq+\et)^2+d^2\eq\et-d
  (s,s)_{su(N)}+\frac{d}{2}(f,f)_{su(2N)}.
\end{equation}

For $A$ type Lie algebra
$(P^\vee\slash Q^{\vee})_{A_n}\cong\IZ_{n+1}$. The zero element,
i.e.~the coroot lattice $Q^{\vee}$ is labeled by trivial
representation and results in unity blowup equations, while the $n$
other elements each labeled by one of the $n$ fundamental weights
$\omega_i$, $i=1,2,\dots,n$ result in vanishing blowup equations. The
checkerboard pattern condition of blowup equations is guaranteed by
the following Lie algebra facts.  For $\mf{su}(N)$ algebra,
i.e.~$A_{N-1}$, we denote by $\cO_{\omega_i},i=1,2,\dots,N-1$ the Weyl
orbit containing the fundamental weight $\omega_i$.  Note
$|\mathcal{O}_{\omega_i}|={N \choose i}$.  Then
$\forall w'\in\mathcal{O}_{\omega_i}$, $w'$ intersects with $i$
weights and $(N-i)$ weights of $\mathbf{N}=\mathcal{O}_{\omega_1}$
with intersection numbers $(N-i)/N$ and $-i/N$ respectively.
Similarly, $w'$ intersects with $i$ weights and $(N-i)$ weights of
$\mathbf{\overline{ N}}=\mathcal{O}_{\omega_{N-1}}$ with intersection
numbers $-(N-i)/N$ and $i/N$ respectively.

Let us first discuss some vanishing blowup equations.
For odd $N$ and
$i=1,2,\dots,({N-1})/{2}$, the leading base degree of the vanishing
blowup equations with $\lambda_F$ as
$\lambda\in\mathcal{O}_{\omega_{N+2i}}(\mf{su}({2N}))$ can be universally
written as
\begin{equation}\label{suvanishodd1}
\begin{aligned}
  \phantom{-}&\sum_{w'\in\mathcal{O}_{\omega_i}(\mf{su}({N}))}(-1)^{|w'|}
  \theta_3^{[a]}(2\tau,-2m_{w'}+m_{\lambda}+(N-2i)\ep_+)
  \times\prod_{\beta\in\Delta({\mf{su}({N})})}^{w'\cdot\beta=
    1}\frac{1}{\theta_1(m_{\beta})}\\[-2mm]
  &\qquad\quad\times\prod_{\mu\in\bf{N}}^{\mu\cdot w'=1-\frac{i}{N}}\
  \prod_{\nu\in\bf{\overline{2N}}}^{\nu\cdot
    \lambda=\frac{1}{2}+\frac{i}{N}}{\theta_1(m_{\mu}+m_{\nu}+\epsilon_+)}=0,
  \quad\quad\quad
  a=-1/2,0.
\end{aligned}
\end{equation}
For $i=(N+1)/2,\dots,N-2,N-1$, the leading base degree of the
vanishing blowup equations with $\lambda_F$ as
$\lambda\in\mathcal{O}_{\omega_{2N-1-2i}}(\mf{su}({2N}))$ can be
universally written as
\begin{equation}\label{suvanishodd2}
\begin{aligned}
  \phantom{-}&\sum_{w'\in\mathcal{O}_{\omega_i}(\mf{su}({N}))}(-1)^{|w'|}
  \theta_3^{[a]}(2\tau,-2m_{w'}+m_{\lambda}+(2i-N)\ep_+)\times
  \prod_{\beta\in\Delta(\mf{su}({N}))}^{w'\cdot\beta= 1}\frac{1}{\theta_1(m_{\beta})}\\[-2mm]
  &\qquad\quad\times\prod_{\mu\in\bf{N}}^{\mu\cdot w'=-\frac{i}{N}}\
  \prod_{\nu\in\bf{\overline{2N}}}^{\nu\cdot
    \lambda=-\frac{3}{2}+\frac{i}{N}}
  {\theta_1(m_{\mu}+m_{\nu}-\epsilon_+)}=0,\quad\quad\quad
  a=-1/2,0.
\end{aligned}
\end{equation}
Note in the denominator there are $i(N-i)$ Jacobi $\theta_1$, while in
nominator there are $i(N-2i)$ Jacobi $\theta_1$ if $i\le N/2$ or
$(N-i)(2i-N)$ Jacobi $\theta_1$ if $i\ge N/2$. For even $N$, the
leading base degree of the vanishing blowup equations look almost the
same with the above formulas, except the two cases are divided by
$i=N/2$. In fact, we find for all integers $N\ge 2$ and
$1\le i\le N/2$, the leading base degree of vanishing blowup equations
result in the following mathematical identity:
\begin{equation}\label{suvanish2}
\begin{aligned}
  \sum_{\substack{\sigma\subset I_N\\|\sigma|=i}}
  \frac{\theta_3^{[a]}(2\tau,-2\sum_{j=1}^i  m_{\sigma_j}
    +\sum_{k=1}^{N-2i}y_k)\prod_{j=1}^{i}\prod_{k=1}^{N-2i}
    {\theta_1(m_{\sigma_j}+y_k)}}{\prod_{j=1}^{i}\prod_{s\in
      I_N\backslash \sigma} \theta_1(m_{\sigma_j}-m_s)}=0, \quad\quad
  \textrm{for}\, \sum_{i=1}^Nm_i=0.
\end{aligned}
\end{equation}
Here $\sigma=(\sigma_1,\ldots,\sigma_i)$ runs over all unordered
subsets of size $i$ of $I_N = (1,2,\ldots,N)$.  Note $y_k$ are
arbitrary numbers. We have verified this identity for lots of $N$ and
$i$ pair up to $\mathcal{O}(q^{20})$. For example, for $i=1$, the
above identity gives
\begin{equation}\label{suvanishomega1}
\begin{aligned}
  \phantom{-}&\sum_{i=1}^N\frac{\theta_3^{[a]}(2\tau,-2m_{i}
    +\sum_{k=1}^{N-2}y_k)\prod_{k=1}^{N-2}{\theta_1(m_{i}+y_k)}}{\prod_{j\neq
      i}{\theta_1(m_i-m_j)}}=0,\quad\quad \textrm{for}\,
  \sum_{i=1}^Nm_i=0.
\end{aligned}
\end{equation}

All the unity $\lambda_F$ fields are just the weights of
representation $[0,\dots,0,1,0,\dots,0]$ of $\mf{su}(2N)$, which is
the largest representation generated by fundamental weights. There are
${2N \choose N}=\frac{(2N)!}{N!N!}$ of them, i.e. the sums of arbitrary $N$ fundamental weights among the
total $2N$ fundamental weights. Note
$\forall w\in {\bf 2N}$ and $w'\in \chi_{[0,\dots,0,1,0,\dots,0]}^{\mf{su}(2N)}$,
\begin{equation}
w\cdot w'=\left\{
\begin{array}{l}
  1/2\ \ \ \ \text{if $w$ is among the $N$ weights that sum up to $w'$,}\\
  -1/2\ \ \text{otherwise.}
\end{array}
\right.
\end{equation}
Besides, for $\mf{su}(N)$, any vector $\alpha^{\vee}$ in the coroot
lattice and any fundamental weight $w$, there always is
$\alpha^{\vee}\cdot w\in \IZ$. These two properties are necessary for
$A_{H}$ to have correct $R$ shift.

The unity elliptic blowup equations for $G=\mf{su}(N),F=\mf{su}(2N)$ theory with $\lambda_F\in \chi_{[0,\dots,0,1,0,\dots,0]}^{\mf{su}(2N)}$ can be written as
\begin{align}
\sum_{\displaystyle \substack{d_0=\frac{1}{2}||\alpha^{\vee}||_{\mf{su}(N)}}}^{d_0+d_1+d_2=d}(-1)^{|\alpha^{\vee}|}\theta_{3}^{[a]}&\Big(2\tau,2\Big(-\alpha^{\vee}\cdot m_{G}+{\lambda_F\cdot m_{F}}+\big(\frac{N}{4}-d_0\big)\(\eq+\et\)-d_1\eq-d_2\et\Big)\Big)\nonumber\\[-4mm]
 &\times A_{V}(\alpha^{\vee},\tau,m_{G})A^{\mathfrak{R} }_{H}(\alpha^{\vee},\tau,m_{G},m_{F},\lambda_F)\nn \times \IE_{d_1}\big(\tau,m_{G}+\eq\alpha^{\vee}&,m_{F}+\eq \lambda_F,\eq,\et-\eq\big)\cdot \IE_{d_2}\big(\tau,m_{G}+\et\alpha^{\vee},m_{F}+\et
 \lambda_F,\eq-\et,\et\big)\nn
=\,\theta_{3}^{[a]}&\Big(2\tau,2{\lambda_F\cdot m_{F}}+\frac{N}{2}(\eq+\et)\Big)\, \IE_{d}\big(\tau,m_{G},m_{F},\eq,\et\big).\label{n2su2blowup}
\end{align}
Using the quiver formula (\ref{n2sunquiver}) for one-string elliptic
genus, we have checked the above unity blowup equations hold for
$G=\mf{su}(3)$ theory for all fifteen $\lambda_F$ and $a=-1/2,0$ up to
$\mathcal{O}(q^{10})$. The $G=\mf{su}(2)$ case is more subtle, we
leave the check of blowup equations later.  Conversely, we also used
the Weyl orbit expansion method to solve one-string elliptic genus
from above unity blowup equations at $a=0$ for
$G=\mf{su}(2),\mf{su}(3)$ and obtained consistent results with the
quiver formulas.

\subsubsection*{$\mathbf{n=2,\, G=\mf{su}(2),\,F=\mf{so}(7)}$}
The $G=\mf{su}(2)$ case is special because the flavor symmetry
$\mf{su}(4)$ is enhanced to $\mf{so}(7)$. In \cite{DelZotto:2018tcj},
an inspiring exact formula for the reduced one-string elliptic genus
was proposed in which it is found the flavor fugacities are even
naturally arranged in $\mf{so}(8)$ characters:
\begin{align}\label{eq:esu2ch}
\mathbb{E}_{h_{2,\mf{su}(2)}^{(1)}}(q,v,\mass_{\mf{su}(2)},m_{\mf{so}(8)}) &= \widehat\chi^{\mf{so}(8)}_{\bf{0}}(\mass_{\mf{so}(8)},q)\xi^{2,\mf{su}(2)}_{\bf{0}}(\mass_{\mf{su}(2)},v,q)\nonumber\\[-1mm]
&+\widehat\chi^{\mf{so}(8)}_{\bf{c}}(\mass_{\mf{so}(8)},q)\xi^{2,\mf{su}(2)}_{\bf{c}}(\mass_{\mf{su}(2)},v,q)\nonumber\\
&+\widehat\chi^{\mf{so}(8)}_{\bf{v}}(\mass_{\mf{so}(8)},q)\xi^{2,\mf{su}(2)}_{\bf{v}}(\mass_{\mf{su}(2)},v,q),
\end{align}
where the affine characters of $\mf{so}(8)$ representations are defined as
\begin{align}
\widehat\chi^{\mf{so}(8)}_{\bf{1}}(m_{\mf{so}(8)}) &= \frac{1}{2}\sum_{j=3}^4\prod_{i=1}^{4}\frac{\theta_j(m_i)}{\eta},\quad\quad\widehat\chi^{\mf{so}(8)}_{\bf{v}}(m_{\mf{so}(8)}) = \frac{1}{2}\sum_{j=3}^4(-1)^{j+1}\prod_{i=1}^{4}\frac{\theta_j(m_i)}{\eta},\nonumber\\
\widehat\chi^{\mf{so}(8)}_{\bf{s}}(m_{\mf{so}(8)}) &= \frac{1}{2}\sum_{j=1}^2\prod_{i=1}^{4}\frac{\theta_j(m_i)}{\eta},\quad\quad\widehat\chi^{\mf{so}(8)}_{\bf{c}}(m_{\mf{so}(8)}) = \frac{1}{2}\sum_{j=1}^2(-1)^j\prod_{i=1}^{4}\frac{\theta_j(m_i)}{\eta},
\end{align}
and
\begin{align}\nonumber
\xi^{2,\mf{su}(2)}_{\bf{0}}&=\frac{1}{q^{1/6}\prod_{j=1}^\infty(1\!-\!q^j)\widetilde\Delta_{\mf{su}(2)}(\mass_{\mf{su}(2)},q)} \,\sum_{k\geq 0}\frac{q^{k+1/2}(v^{2k+1}\!+\!v^{-2k-1})}{1-q^{2k+1}}\chi^{\mf{su}(2)}_{(2k)}(\mass_{\mf{su}(2)}),\\ \nonumber
\xi^{2,\mf{su}(2)}_{\bf{c}}&=-\frac{1}{q^{1/6}\prod_{j=1}^\infty(1\!-\!q^j)\widetilde\Delta_{\mf{su}(2)}(\mass_{\mf{su}(2)},q)} \,\sum_{k\geq 0}\frac{v^{2k+1}+q^{2k+1}v^{-2k-1}}{1-q^{2k+1}}\chi^{\mf{su}(2)}_{(2k)}(\mass_{\mf{su}(2)}),\\ \nonumber
\xi^{2,\mf{su}(2)}_{\bf{v}}&=\frac{1}{q^{1/6}\prod_{j=1}^\infty(1\!-\!q^j)\widetilde\Delta_{\mf{su}(2)}(\mass_{\mf{su}(2)},q)} \,\sum_{k\geq 0}\frac{v^{2k+2}-q^{k+1}v^{-2k-2}}{1+q^{k+1}}\chi^{\mf{su}(2)}_{(2k+1)}(\mass_{\mf{su}(2)}),
\end{align}
and a modified version of Weyl-Kac determinant
\begin{equation}
\widetilde\Delta_{G}(\mass_G,q) = \prod_{j=1}^
\infty (1-q^j)^{\text{rank}(G)}\prod_{\alpha\in \Delta^G_+}(1-q^j m_\alpha)(1-q^j m_\alpha^{-1}).
\end{equation}
Using the above formula for one string elliptic genus, we have checked the unity elliptic blowup equations (\ref{n2su2blowup}) hold only for $F=\mf{so}(7)$ but not $\mf{so}(8)$. For arbitrary $m_{\mf{so}(7)}$, we checked the $6\times 2$ unity blowup equations up to $\mathcal{O}(\Qtau^{10})$.

\subsection{$n=3$ $\mf{so}(7)$ theory}

The $n=3$, $G=\mf{so}(7)$ theory has flavor symmetry $F=\mf{sp}(2)$ and
matter representation $\bf 8$. This theory is particularly interesting
because it has a known 2d quiver description and can be Higgsed to the
$n=3,G = G_2$ theory, making which the first exactly computable
exceptional 6d SCFT \cite{Kim:2018gjo}. The elliptic genera of this
theory were computed via Jeffrey-Kirwan residue of localization in
\cite{Kim:2018gjo}. For example, the reduced one-string elliptic genus
can be expressed as
\begin{equation}
  \mathbb{E}_{h_{3,\mf{so}(7)}^{(1)}}(\tau,\epsilon_{1,2},m_i,\mu_k)=
  \sum_{i=1}^3\frac{\theta(4\epsilon_+-2m_i)\prod_{k=1}^2
  \theta(\mu_k\pm(m_i-\epsilon_+))}
  {\prod_{j\neq i}\theta(m_{ij})\theta(2\epsilon_+-m_{ij})
  \theta(2\epsilon_+-m_i-m_j)},
\end{equation}
where $\theta(z)=\theta_1(\tau,z)/\eta(\tau)$, $m_{ij}\equiv m_i-m_j$,
and $m_i,i=1,2,3$ are the $\mf{so}(7)$ fugacities such that
${\bf 7}_{\bf v }^{\mf{so}(7)}=1+\sum_{i=1}^3(m_i+m_i^{-1})$ and
$\mu_{k},k=1,2$ are associated to each $\mf{sp}(1)$ in flavor
decomposition $\mf{sp}(2)\to \mf{sp}(1)\times \mf{sp}(1)$.

Let us first discuss the vanishing blowup equations. Since
$(P^\vee\slash Q^{\vee})_{\mf{so}(7)}\cong\IZ_2$, there should exist
vanishing blowup equations with $\lambda_G$ taking value in
$(P^\vee\backslash Q^{\vee})_{\mf{so}(7)}$. For flavor fugacities, we
find $\lambda_F$ has five possible values, weights of representation
$\bf 1$ or $\bf 4$ of $\mf{sp}(2)$. The checkerboard pattern
condition of $A_{V}$ is guaranteed by the Lie algebra fact
$\forall \alpha\in\Delta(\mf{so}(7)),w\in (P^\vee\backslash Q^{\vee})_{\mf{so}(7)}$,
the intersection $\alpha\cdot w\in \IZ$. On the other hand, the
checkerboard pattern condition of $A_{H}$ is guaranteed by the Lie
algebra fact
$\forall \omega'\in\mathbf{8},w\in (P^\vee\backslash Q^{\vee})_{\mf{so}(7)}$,
the intersection $\omega'\cdot w\in \IZ+1/2$.

Note the smallest Weyl orbit in
$(P^\vee\backslash Q^{\vee})_{\mf{so}(7)}$ is $\mathcal{O}_{{1/2},6}$,
which is contained in the weight space of the vector representation
${\bf 7}_{\bf v }^{\mf{so}(7)}=1+\mathcal{O}_{{1}/{2},6}$. We find the
leading base degree of the vanishing blowup equations with
$\lambda_F=0$ can be written as
\begin{equation}\label{n3so7v1}
\begin{aligned}
\sum_{w\in\mathcal{O}_{{1}/{2},6}}(-1)^{|w|}\theta^{[a]}_{4}(3\tau,3m_w)\times\prod_{\beta\in\Delta(\mf{so}(7))}^{w\cdot\beta= 1}\frac{1}{\theta_1(\tau,m_{\beta})}=0,
\end{aligned}
\end{equation}
where $a=-1/2$ and $\pm 1/6$. We have checked this identity up to $\mathcal{O}(q^{20})$. Here the hypermultiplets do not contribute to the leading base degree equation, since $\forall w\in \mathcal{O}_{1/2,6},w'\in \mathbf{8}$, $w\cdot w'=\pm 1/2$. On the other hand,
the leading base degree of the vanishing blowup equations with $\lambda_F\in \bf 4$ is
\begin{equation}\label{n3so7v2}
\begin{aligned}
\sum_{w\in\mathcal{O}_{1/2,6}}(-1)^{|w|}\theta^{[a]}_{4}(3\tau,3m_w+2x)\prod_{\beta\in\Delta(\mf{so}(7))}^{w\cdot\beta= 1}\frac{1}{\theta_1(m_{\beta})}\prod_{\omega'\in\mathbf{8}}^{w\cdot\omega'=-1/2}{\theta_1(m_{\omega'}+x)}=0,
\end{aligned}
\end{equation}
where $x=\pm m_{\mf{sp}(1)}+\epsilon_+$ is an arbitrary number. We also checked this identity up to $\mathcal{O}(q^{20})$. For higher base degrees, we checked all five vanishing blowup equations from the viewpoint of Calabi-Yau.

For unity blowup equations, $\lF$ has four choices which are just the
four short roots of $\mf{sp}(2)$, or explicitly $(\pm1,\pm 1)$ if we
view the effective flavor group as $\mf{sp}(1)_a\times
\mf{sp}(1)_b$. Therefore, all the 12 unity blowup equations with
$\lambda_F=\lambda\in\mathcal{O}_{[01]}^{\mf{sp}(2)}$ can be written as
\begin{align}
  \sum_{\displaystyle \substack{d_0=\frac{1}{2}||\alpha^{\vee}||_{\mf{so}(7)}}}^{d_0+d_1+d_2=d}(-1)^{|\alpha^{\vee}|}\theta_{4}^{[a]}&\Big(3\tau,3\Big(-\alpha^{\vee}\cdot m_{\mf{so}(7)}+{\lambda\cdot m_{\mf{sp}(2)}}+\big(\frac{2}{3}-d_0\big)\(\eq+\et\)-d_1\eq-d_2\et\Big)\Big)\nonumber\\[-4mm]
  &\times A_{V}^{\mf{so}(7)}(\alpha^{\vee},\tau,m_{\mf{so}(7)})A^{(\mathbf{8},\frac{1}{2}\mathbf{4})
  }_{H}(\alpha^{\vee},\tau,m_{\mf{so}(7)},m_{\mf{sp}(2)},\lambda)\nn \times
  \IE_{d_1}\big(\tau,m_{\mf{so}(7)}+\eq\alpha^{\vee}&,m_{\mf{sp}(2)}+\eq
  \lambda,\eq,\et-\eq\big)\cdot
  \IE_{d_2}\big(\tau,m_{\mf{so}(7)}+\et\alpha^{\vee},m_{\mf{sp}(2)}+\et
  \lambda,\eq-\et,\et\big)\nn
  =\,\theta_{4}^{[a]}&\Big(3\tau,3\lambda\cdot m_{\mf{sp}(2)}
  +{2}(\eq+\et)\Big)\,
  \IE_{d}\big(\tau,m_{\mf{so}(7)},m_{\mf{sp}(2)},\eq,\et\big).
\end{align}
Here $a=-1/2,\pm 1/6$. All four possible $\lambda$ just give
$\lambda\cdot m_{\mf{sp}(2)}=\pm m_{\mf{sp}(1)_a}\pm m_{\mf{sp}(1)_b}$. Fix arbitrary
one $\lambda$, there are three unity blowup equations with different
characteristics from which one can solve elliptic genera
recursively. For example, using the recursion formula, we computed the
one-string elliptic genus to $\mathcal{O}(\Qtau^3)$. Our result agrees
precisely with the quiver formula in \cite{Kim:2018gjo} and the
modular ansatz in \cite{DelZotto:2018tcj}, therefore we just present
the first few $\Qtau$ orders with all gauge and flavor fugacities
turned off. For example, denote the reduced one-string elliptic genus
as
\begin{equation}\label{n3SO7E1}
  \mathbb{E}_{h_{3,\mf{so}(7)}^{(1)}}(\Qtau,v,m_{\mf{so}(7)}=0,m_{\mf{sp}(2)}=0)=\Qtau^{-1/3}v^{4}\sum_{n=0}^{\infty}\Qtau^n\frac{P_n(v)}{(1
    - v)^{4} (1 + v)^{8}}.
\end{equation}
We obtain
\begin{align}
  P_0(v)=
  &-(5 - 12 v + 22 v^2 - 12 v^3 + 5 v^4),\nonumber\\
  P_1(v)=
  &\,v^{-6}(1 + 4 v + 2 v^2 - 12 v^3 - 18 v^4 + 4 v^5 + 158 v^6 - 316 v^7 + 418 v^8 - \dots+ v^{16}).\nonumber
\end{align}
Note that the polynomials in the parentheses are palindromic.  The full
expression of $P_1(v)$ can be recovered from this property.  We also
computed the two-string elliptic genus using the recursion formula and
find agreement with the quiver formula in \cite{Kim:2018gjo}. For
example,
\begin{equation}\label{n3SO7E2}
\mathbb{E}_{h_{3,\mf{so}(7)}^{(2)}}(\Qtau,v,x=1,m_{\mf{so}(7)}=0,m_{\mf{sp}(2)}=0)=-\Qtau^{-5/6}v^9\sum_{n=0}^{\infty}\Qtau^n\frac{P_n^{(2)}(v)}{(1
  - v)^{10} (1 + v)^{10} (1 + v + v^2)^9},
\end{equation}
where
\begin{equation}\nonumber
\begin{aligned}
P_0^{(2)}&(v)=(14 + 18 v - 3 v^2 + 69 v^3 + 298 v^4 + 295 v^5 + 175
v^6 + 684 v^7 + 1426 v^8 + 1132 v^9\\ &\phantom{---.} + 660 v^{10} +
\dots +
14 v^{20}),\\
P_1^{(2)}&(v)=v^{-6}(5 + 23 v + 68 v^2 + 135 v^3 + 216 v^4 + 273 v^5 + 649 v^6 + 838 v^7 -
 117 v^8 - 407 v^9\\& + 3496 v^{10} + 6341 v^{11} + 6252 v^{12} + 12839 v^{13} +
 24595 v^{14} + 23918 v^{15} + 19272 v^{16}+ \dots + 5 v^{32}).
\end{aligned}
\end{equation}
Again the full expressions of the polynomials in the parentheses can
be recovered by their palindromic properties.

\subsection{$n=4$ $\mf{so}(N+8)$ theories}

The $n=4$, $G=\mf{so}(N+8)$ theories have flavor group $F=\mf{sp}(N)$
and matter representation
$(R^G,R^F)=(\mathbf{N+8},\mathbf{2N})$. For even $N=2p$, such
theories can be realized by type IIB superstring theory with
orientfold. The Kodaira elliptic singularity of type $I^{\star}_p$
here is due to the presence of $4 + p$ D7-branes wrapping the base
$\IP^1$ together with an orientifold 7-plane. This picture results in
a quiver gauge theory description which makes the elliptic genera
exactly computable via Jeffrey-Kirwan residues
\cite{Haghighat:2014vxa}. For example, the reduced one-string elliptic
genus can be computed as
\begin{equation}
  \IE_{h^{(1)}_{4,\mf{so}(8+2p)}} = \frac{1}{2}\sum_{i=1}^{4+p}\bigg[\frac{\theta(2\epsilon_++ 2{m_i})\theta(4\epsilon_++ 2{m_i})\prod_{j=1}^{2p}\theta(\epsilon_++  {m_i}\pm {\mu_j})}{\prod_{j\neq i}\theta({m_i}\pm {m_j})\theta(2\epsilon_++ {m_i}\pm {m_j})}  +({m_i}\to -{m_i})\bigg]. \\ \label{eq:n4soZ1}
\end{equation}
Here $\theta(z)=\theta_1(\tau,z)/\eta(\tau)$, $m_i$ and $\mu_j$ are
fugacities of gauge $\mf{so}(8+2p)$ and flavor $\mf{sp}(2p)$. For odd
$N$ cases, the 2d quiver description also
exists similarly and was discussed in Appendix D of
\cite{DelZotto:2018tcj}. For example, the reduced one-string elliptic
genus for $G=\mf{so}(9+2p),\ F=\mf{sp}(1+2p)$ theory is
\begin{equation}
  \IE_{h^{(1)}_{4,\mf{so}(9+2p)}} = \frac{1}{2}\sum_{i=1}^{4+p}\bigg[\frac{\theta(2\epsilon_++ 2{m_i})\theta(4\epsilon_++ 2{m_i})\prod_{j=1}^{2p+1}\theta(\epsilon_++  {m_i}\pm {\mu_j})}{\theta({m_i})\theta(2\epsilon_++ {m_i})\prod_{j\neq i}\theta({m_i}\pm {m_j})\theta(2\epsilon_++ {m_i}\pm {m_j})}  +({m_i}\to -{m_i})\bigg]. \\ \label{eq:n4sooddZ1}
\end{equation}
Still $m_i$ and $\mu_j$ are gauge and flavor fugacities respectively.

Let us first discuss the vanishing blowup equations. As is well-known
in Lie algebra, $(P^\vee\slash Q^{\vee})_{B_n}\cong\IZ_2$ and
$(P^\vee\slash Q^{\vee})_{D_n}\cong\IZ_4$. Consider the vanishing blowup
equations with $\lambda_G$ taking value in
$\mathcal{O}^{\mf{so}(8+N)}_{[10\cdots00]}$, i.e. the Weyl orbit
associated to the vector representation. For flavor fugacities, we
find $\lambda_F$ can always take value in Weyl orbit
$\mathcal{O}^{\mf{sp}(N)}_{[00\cdots01]}$. 
Let us denote the smallest Weyl orbit in
$(P^\vee\backslash Q^{\vee})_{\mf{so}(8+N)}$ as $\mathcal{O}_{\rm min}$. It
has relation with the vector representation of $\mf{so}(8+N)$ as
\be
\bf (8+N)_v=\begin{cases}
\mathcal{O}_{\rm min},\quad & \text{for even }N,\\
{\bf 1}+\mathcal{O}_{\rm min},\quad &\text{for odd }N,\\
\end{cases}
\end{equation}
Then the leading base degree of the vanishing blowup equations of $G=\mf{so}(8+N)$ theory with $\lambda_F\in\mathcal{O}^{\mf{sp}(N)}_{[00\cdots01]}$ can be universally written as
\begin{equation}\label{n4sov}
\begin{aligned}
\sum_{w\in\mathcal{O}_{\rm min}}(-1)^{|w|}\theta^{[a]}_{3}(4\tau,4m_w+Nx)\theta_1(-m_w+x)^N\times\prod_{\beta\in\Delta(\mf{so}(8+N))}^{w\cdot\beta= 1}\frac{1}{\theta_1(\tau,m_{\beta})}=0,\quad\quad N\ge0.
\end{aligned}
\end{equation}
Here $a=-1/2,-1/4,0,1/4$ and $x=\lambda_F\cdot m_F+\epsilon_+$. We
have checked this identity up to $\mathcal{O}(q^{20})$ for several
$N$. Note there are $N+6$ Jacobi $\theta_1$ functions
in the denominator.

For even $N$ cases, there exist more vanishing blowup equations with
$\lambda_G$ taking value in
$\mathcal{O}^{\mf{so}(8+N)}_{[00\cdots01]}$ and
$\mathcal{O}^{\mf{so}(8+N)}_{[00\cdots10]}$, which coincide with the
spinor and conjugate spinor representations. For example, the leading base degree
of the vanishing blowup equations with $\lambda_F=0$ can be
universally written as
\begin{equation}\label{n4sov2}
\begin{aligned}
  \sum_{w\in
    S}(-1)^{|w|}\theta^{[a]}_{3}(4\tau,4m_w)\times\prod_{\beta\in\Delta(\mf{so}(8+N))}^{w\cdot\beta=
    1}\frac{1}{\theta_1(\tau,m_{\beta})}=0,\quad\quad N\ge0,\, N\equiv
  0\, (\mathrm{mod}\, 2).
\end{aligned}
\end{equation}
Here $S$ is the spinor representation of $\mf{so}(8+N)$ which can
also be replaced by its conjugate representation. We
have checked this identity up to $\mathcal{O}(q^{20})$ for several
even $N$. Note there are $(N+6)(N+8)/8$ Jacobi $\theta_1$
functions in the denominator.

The unity $\lambda_F$ fields of $\mf{so}(N+8)$ theories all take value
in the Weyl orbit $\mathcal{O}^{\mf{sp}(N)}_{[00\cdots01]}$. There are
$2^N$ of them.
The unity elliptic blowup equations for $G=\mf{su}(8+N),F=\mf{sp}(N)$
theory with $\lambda$ short for $\lambda_F$ can be written as
\begin{equation}
\begin{aligned}
\sum_{\displaystyle \substack{d_0=\frac{1}{2}||\alpha^{\vee}||_{\mf{so}(8+N)}}}^{d_0+d_1+d_2=d}&(-1)^{|\alpha^{\vee}|}\theta_{3}^{[a]}\Big(4\tau,4(-\alpha^{\vee}\cdot m_{\mf{so}(8+N)}+{\lambda\cdot m_{\mf{sp}(N)}}+(\frac{N+4}{8}-d_0)(\eq+\et)-d_1\eq-d_2\et)\Big)\\[-4mm]
 &\times A_{V}^{\mf{so}(8+N)}(\alpha^{\vee},\tau,m_{\mf{so}(8+N)})A^{\frac{1}{2}(\mathbf{8+N,2N}) }_{H}(\alpha^{\vee},\tau,m_{\mf{so}(8+N)},m_{\mf{sp}(N)},\lambda)\\
 &\times \IE_{d_1}\big(\tau,m_{\mf{so}(8+N)}+\eq\alpha^{\vee},m_{\mf{sp}(N)}+\eq \lambda,\eq,\et-\eq\big)\\
 &\times \IE_{d_2}\big(\tau,m_{\mf{so}(8+N)}+\et\alpha^{\vee},m_{\mf{sp}(N)}+\et
 \lambda,\eq-\et,\et\big)\\
=\,\theta_{3}^{[a]}&\Big(4\tau,{4\lambda\cdot m_{\mf{sp}(N)}}+\frac{N+4}{2}(\eq+\et)\Big)\, \IE_{d}\big(\tau,m_{\mf{so}(8+N)},m_{\mf{sp}(N)},\eq,\et\big).
\end{aligned}
\end{equation}
Here $a=-1/2,-1/4,0,1/4$. Fix arbitrary one $\lambda$ and choose arbitrary three characteristics $a$, one can use the three unity blowup equations to solve  elliptic genera recursively.

In the following, we present some of our computational results on
one-string and two-string elliptic genera from recursion formula. To
save space, we turn off both gauge and flavor fugacities. For
$G=\mf{so}(9),\,F=\mf{sp}(1)$ theory, let us denote the reduced
one-string elliptic genus as
\begin{equation}\label{n4SO9E1}
\mathbb{E}_{h_{4,\mf{so}(9)}^{(1)}}(\Qtau,v,m_{\mf{so}(9)}=0,m_{\mf{sp}(1)}=0)=\Qtau^{-5/6}v^{6}\sum_{n=0}^{\infty}\Qtau^n\frac{P_n(v)}{(1 - v)^{10} (1 + v)^{12}}.
\end{equation}
We obtain
\begin{align}
  P_0(v)=
  &2 - 5 v + 36 v^2 - 46 v^3 + 130 v^4 - 90 v^5 + 130 v^6 - 46 v^7 +
    36 v^8 - 5 v^9 + 2 v^{10}, \nonumber\label{n4SO9E1P0}\\
  P_1(v)=
  &4 (19 - 52 v + 270 v^2 - 368 v^3 + 815 v^4 - 648 v^5 + 815 v^6 -
    368 v^7 + 270 v^8 - 52 v^9 + 19 v^{10}).\nonumber
\end{align}
This agrees precisely with the quiver formula (\ref{eq:n4sooddZ1}) and the modular ansatz result in \cite{DelZotto:2018tcj}. Using recursion formula, we also computed the reduced two-string elliptic genus with all gauge and flavor fugacities turned off. Denote
\begin{equation}\label{n4SO9E2}\nonumber
\mathbb{E}_{h_{4,\mf{so}(9)}^{(2)}}(\Qtau,v,x=1,m_{\mf{so}(9)}=0,m_{\mf{sp}(1)}=0)=-\Qtau^{-11/6}v^{13}\sum_{n=0}^{\infty}\Qtau^n\frac{P^{(2)}_n(v)}{(1 - v)^{22} (1 + v)^{16} (1 + v + v^2)^{13}},
\end{equation}
we obtain
\begin{align}
P_0^{(2)}(v)=&\,3 + 5 v + 41 v^{2 }+ 184 v^{3 }+ 623 v^{4 }+ 1987 v^{5 }+ 6119 v^{6 }+
 16024 v^{7 }+ 38003 v^{8 }+ 84127 v^{9 }\nn
 &+ 170974 v^{10 }+ 315783 v^{11 }+
 541464 v^{12 }+ 864989 v^{13 }+ 1277738 v^{14 }+ 1747831 v^{15 }\nn&+
 2235019 v^{16 }+ 2666784 v^{17 }+ 2956416 v^{18 }+ 3054876 v^{19 }+ \dots+ 3 v^{38},\nn
P_1^{(2)}(v)=&\,2(62 + 193 v + 1031 v^2 + 4553 v^3 + 16024 v^4 + 49985 v^5 +
 146893 v^6 + 383794 v^7 \nn
 &+ 904569 v^8 + 1962488 v^9 + 3926557 v^{10} +
 7208099 v^{11} + 12237790 v^{12} + 19308839 v^{13}\nn
 & + 28304443 v^{14} +
 38563232 v^{15} + 49018799 v^{16} + 58173759 v^{17} + 64417144 v^{18}\nn
 & +
 66611780 v^{19} + \dots + 62 v^{38}).
 \end{align}

 Again, the full polynomials can be recovered from their palindromic
 properties.

For $G=\mf{so}(10),\,F=\mf{sp}(2)$ theory, let us denote the reduced one-string elliptic genus as
\begin{equation}\label{n4SO10E1}
\mathbb{E}_{h_{4,\mf{so}(10)}^{(1)}}(\Qtau,v,m_{\mf{so}(10)}=0,m_{\mf{sp}(2)}=0)=\Qtau^{-5/6}v^{7}\sum_{n=0}^{\infty}\Qtau^n\frac{P_n(v)}{(1 - v)^{10} (1 + v)^{14}}.
\end{equation}
We obtain
\begin{equation}\nonumber
\begin{aligned}
P_0(v)&\,=-(5 - 20 v + 99 v^2 - 184 v^3 + 370 v^4 - 360 v^5 + 370 v^6 -
  184 v^7 + 99 v^8 - 20 v^9 + 5 v^{10}),\\
P_1(v)&\,=v^{-2}(1 + 4 v - 249 v^2 + 1024 v^3 - 3873 v^4 + 7172 v^5 - 12223 v^6 +
  12688 v^7 - \dots + v^{14}).
 \end{aligned}
\end{equation}
This agrees precisely with the quiver formula in (\ref{eq:n4soZ1}) and the modular ansatz in \cite{DelZotto:2018tcj}. Using recursion formula, we also computed the reduced two-string elliptic genus. Denote
\begin{equation}\label{n4SO10E2}\nonumber
  \mathbb{E}_{h_{4,\mf{so}(10)}^{(2)}}(\Qtau,v,x=1,m_{\mf{so}(10)}=0,m_{\mf{sp}(2)}=0)=
  -\Qtau^{-11/6}v^{15}\sum_{n=0}^{\infty}\Qtau^n\frac{P^{(2)}_n(v)}{(1
  - v)^{22} (1 + v)^{20} (1 + v + v^2)^{15}},
\end{equation}
we obtain
\begin{align}
  P_0^{(2)}(v)&=14 + 42 v + 174 v^{2 }+ 840 v^{3 }+ 3180 v^{4 }+ 9606
  v^{5 }+ 28723 v^{6 }+ 80545 v^{7 }+ 200547 v^{8 }\nn
  &+ 453260 v^{9 }+
  967049 v^{10 }+ 1923811 v^{11 }+ 3524339 v^{12 }+ 6005020 v^{13 }+
  9637502 v^{14 }\nn
  &+ 14497632 v^{15 }+ 20342110 v^{16 }+ 26767114
  v^{17 }+ 33232318 v^{18 }+ 38795360 v^{19 }\nn
  &+ 42443836 v^{20 }+
  43677620 v^{21 }+ \dots+ 14 v^{42},\nn
P_1^{(2)}(v)&=-v^{-2}(5 + 35 v - 566 v^2 - 2413 v^3 - 9796 v^4 - 43257 v^5 - 166563 v^6 -
 516948 v^7\nn
 & - 1493092 v^8 - 4045182 v^9 - 9976992 v^{10} -
 22346950 v^{11} - 46615056 v^{12} - 90796062 v^{13}\nn
 & - 164272366 v^{14} -
 276641406 v^{15} - 437103585 v^{16} - 648567657 v^{17} - 902450252 v^{18}\nn
 & -
 1179498629 v^{19} - 1452843842 v^{20} - 1686000677 v^{21} -
 1841747735 v^{22} - 1895883244 v^{23}\nn
 & +\dots + 5 v^{46}).
 \end{align}

\subsection{$G_2$ theories}
\label{sec:g2}
$G=G_2$ theories on base curve $(-n)$, $n=1,2,3$ have flavor group
$F=\mf{sp}(10-3n)$ and $n_f=(10-3n)$ hypermultiplets in fundamental
representation ${\bf 7}$ of gauge symmetry. There only
exist unity blowup equations but no vanishing due to the Lie algebra
fact $Q^{\vee}\cong P^\vee$ for $G_2$. The unity $\lF$ fields are just
all the elements of the Weyl orbit $[0,0,\dots,0,1]$ of
$\mf{sp}(10-3n)$ or in other word take value $\pm 1$ for each
$\mf{sp}(1)$ with decomposition
$\mf{sp}(10-3n)\rightarrow \mf{sp}(1)^{10-3n}$. There are in total
$n\times 2^{10-3n}$ unity blowup equations when different choices of
the characteristic are also taken into account.

\subsubsection*{$\mathbf{n=3,\, G=G_2,\,F=\mf{sp}(1)}$}
This theory can be Higgsed from the
$n=3,\, G=\mf{so}(7),\,F=\mf{sp}(2)$ theory and to the
$n=3,\, G=\mf{su}(3)$ minimal SCFT. The 2d quiver description was
found in \cite{Kim:2018gjo}, therefore the elliptic genus can be
computed exactly via localization. For example, the reduced one-string
elliptic genus of such theory is given in \cite{Kim:2018gjo} as
\begin{equation}
  \IE_{h^{(1)}_{3,G_2}}=\sum_{i=1}^3\frac{\theta(2m_i-4\epsilon_+)\theta({m_{\mf{sp}(1)}\pm (m_i-\epsilon_+)})}
  {\theta({m_i-2\epsilon_+})\prod_{j\neq i}\theta({m_{ij}})
    \theta({2\epsilon_+-m_{ij}}) \theta({2\epsilon_++m_j}),
  }
\end{equation}
where $\theta(z)=\theta_1(\tau,z)/\eta(\tau)$ and $m_{1,2,3}$ are the
embedding of $G_2$ into $\mf{su}(3)$ with
$m_1+m_2+m_3=0$ and $m_{ij}=m_i-m_j$.

Using the recursion formula from blowup equations, we computed the
one-string elliptic genus to $\mathcal{O}(\Qtau^3)$. Our result agrees
precisely with the quiver formula in \cite{Kim:2018gjo} and the
modular ansatz in \cite{Kim:2018gak} and \cite{DelZotto:2018tcj},
therefore we just present the first few $\Qtau$ orders with all gauge
and flavor fugacities turned off. For example, denote the reduced
one-string elliptic genus as
\begin{equation}\label{n3G2E1}
\mathbb{E}_{h_{3,G_2}^{(1)}}(\Qtau,v,m_{G_2}=0,m_{\mf{sp}(1)}=0)=\Qtau^{-1/3}v^{3}\sum_{n=0}^{\infty}\Qtau^n\frac{P_n(v)}{(1 - v)^{4} (1 + v)^{6}}.
\end{equation}
We obtain
\begin{align}
  P_0(v)=
  &2 - 3 v + 8 v^2 - 3 v^3 + 2 v^4, \label{n3G2E1P0}\\
  P_1(v)=
  &\,v^{-5}(1 + 2 v - 3 v^2 - 8 v^3 + 2 v^4 + 44 v^5 - 60 v^6 +
    92 v^7 +\dots + v^{14}), \label{n3G2E1P1}\\
  P_2(v)=
  &v^{-7}(14 + 14 v - 52 v^2 - 34 v^3 + 85 v^4 - 8 v^5 - 105 v^6 + 396
    v^7\nn
  & \phantom{==}-542 v^8 + 728 v^9 - \dots + 14 v^{18}). \label{n3G2E1P2}
\end{align}

We also computed the two-string elliptic genus using the recursion formula and find perfect agreement with the quiver formula in \cite{Kim:2018gjo}. For example,
\begin{equation}\label{n3G2E2}
\mathbb{E}_{h_{3,G_2}^{(2)}}(\Qtau,v,x=1,m_{G_2}=0,m_{\mf{sp}(1)}=0)=-\Qtau^{-5/6}\sum_{n=0}^{\infty}\Qtau^n\frac{P_n^{(2)}(v)}{(1 - v)^{10} (1 + v)^6 (1 + v + v^2)^7},
\end{equation}
where
\begin{equation}
\begin{aligned}
P_0^{(2)}(v)=&\,v^7(3 - 3 v + 8 v^2 + 21 v^3 + 17 v^4 + 16 v^5 + 89 v^6 + 71 v^7 +
 42 v^8 +\dots + 3 v^{16}),\\
P_1^{(2)}(v)=&\,v^2(2 + 3 v + 11 v^2 + 9 v^3 + 20 v^4 + 46 v^5 - 24 v^6 + 19 v^7 +
 313 v^8 + 442 v^9\\
 & + 569 v^{10} + 1364 v^{11} + 1473 v^{12} + 1226 v^{13} + \dots+ 2 v^{26}).
 \end{aligned}
\end{equation}
\subsubsection*{$\mathbf{n=2,\, G=G_2,\,F=\mf{sp}(4)}$}
We study this theory from the viewpoint of both Weyl orbit expansion
and \localelliptic Calabi-Yau. Let us first just turn on the fugacity
of a subalgebra $\mf{sp}(1)\cong \mf{su}(2)$ of the flavor symmetry
$\mf{sp}(4)$. We gave a toric construction for the \localelliptic
Calabi-Yau threefold corresponding to this configuration, and computed
the triple intersection numbers and genus zero Gromov-Witten
invariants. Using the unity blowup equations in refined BPS expansion,
we computed the refined BPS invariants to very high degrees. We
describe our toric construction in
Appendix~\ref{n2G2polytope} and list some low degree BPS invariants in
the geometric bases in Appendix~\ref{app:bps}.

On the other hand, using the Weyl orbit expansion method elaborated in
section \ref{vexpansion} and the unity blowup equation with
characteristic $a=0$, we solved the reduced one-string elliptic genus
with flavor subalgebra $\mf{su}(2)$ fugacity $q_m$ at leading $q$
order, and found it to be
\begin{equation}\label{n2g2E1}
\mathbb{E}_{h_{2,G_2}^{(1)}}(\Qtau,v,m_{G_2}=0,q_m)=\Qtau^{1/6}v^{-1}\sum_{n=0}^{\infty}\Qtau^n\frac{P_n(v,q_m)}{(1 - v^2)^{6}},
\end{equation}
where
\begin{equation}\label{n2g2E1P0}
\begin{aligned}
P_0(v,q_m)=\,&-(q_m^{-4}+q_m^4)(v^4 + 8 v^6 + 8 v^8 + v^{10})+(q_m^{-3}+q_m^3)(28 v^5 + 88 v^7 + 28 v^9)\\
&-(q_m^{-2}+q_m^2)(10 v^4 + 242 v^6 + 242 v^8 + 10 v^{10})\\
&-(q_m^{-1}+q_m)(4 v^3 - 164 v^5 - 688 v^7 - 164 v^9 + 4 v^{11})\\
&+(1 - 6 v^2 + 2 v^4 - 627 v^6 - 627 v^8 + 2 v^{10} - 6 v^{12} + v^{14}),
\end{aligned}
\end{equation}
and
\begin{align}
v^2P_1(v,q_m)=\,&(q_m^{-5}+q_m^5)(28 v^7 + 88 v^9 + 28 v^{11})-(q_m^{-4}+q_m^4)(31 v^6 + 653 v^8 + 653 v^{10} + 31 v^{12})\nn
&+(q_m^{-3}+q_m^3)(4 v^3 - 44 v^5 + 1048 v^7 + 3888 v^9 + 1048 v^{11} - 44 v^{13} + 4 v^{15})\nn
&+(q_m^{-2}+q_m^2)(10 v^2 - 60 v^4 - 20 v^6 - 7562 v^8 - 7562 v^{10} - 20 v^{12} - 60 v^{14} +
 10 v^{16})\nn
&-(q_m^{-1}+q_m)(28 v - 188 v^3 + 692 v^5 - 5132 v^7 - 17008 v^9 - 5132 v^{11} +
 692 v^{13}\nn
 & - 188 v^{15} + 28 v^{17})
+(14 - 53 v^2 + 31 v^4 - 85 v^6 - 15531 v^8 - 15531 v^{10} - 85 v^{12}\nn
& +
 31 v^{14} - 53 v^{16} + 14 v^{18}).
\end{align}
When the flavor fugacity is turned off, i.e. $q_m=1$, the above result
agrees with the modular ansatz in \cite{DelZotto:2018tcj}. Besides, at
leading $\Qtau$ order, the reduced one-string elliptic genus given by (\ref{n2g2E1}) and (\ref{n2g2E1P0}) has the following expansion
\be
\begin{aligned}
v^{-1}&-4(q_m+q_m^{-1})v^2-(q_m^{-4}+10q_m^{-2}+13+10q_m^2+q_m^4)v^3\\
&+(28q_m^{-3}+140q_m^{-1}+140q_m+28q_m^3)v^4+\mathcal{O}(v^5).
\end{aligned}
\end{equation}
It is easy to check this agrees with the exact expression for reduced 5d one-instanton partition function proposed in \cite{DelZotto:2018tcj}
\begin{align}
\,&v^{-1}-\chi^{\mf{sp}(4)}_{(1000)}v^2+\chi^{G_2}_{(10)}v^3+\sum_{n=0}^\infty \Big[-\chi^{G_2}_{(0n)}\chi^{\mf{sp}(4)}_{(0001)}v^{3+2n}+ \chi^{G_2}_{(1n)}\chi^{\mf{sp}(4)}_{(0010)}v^{4+2n}\\[-1mm]
&\qquad\quad-\chi^{G_2}_{2n}\chi^{\mf{sp}(4)}_{(0100)}v^{5+2n}+ \chi^{G_2}_{(3n)}\chi^{\mf{sp}(4)}_{(1000)}v^{6+2n}- \chi^{G_2}_{(4n)}v^{7+2n}\Big]\nonumber\\[+0mm]
\,& =v^{-1}-\mathbf{8}^{\mf{sp}(4)}v^2+(\mathbf{7}^{G_2}-\mathbf{42}^{\mf{sp}(4)})v^3+\mathbf{7}^{G_2}\cdot\mathbf{48}^{\mf{sp}(4)}v^4+\mathcal{O}(v^5),
  \end{align}
with flavor symmetry $\mf{sp}(4)$ restricted to
$\mf{su}(2)_{q_m}$. One can also turn on full flavor fugacity and
gauge fugacity and push the computation to higher $\Qtau$ orders and
higher number of strings. For example, we obtained the subleading $q$
order of the reduced one-string elliptic genus as
\begin{equation}
\begin{aligned}
  \mathbf{14}v^{-3}-\mathbf{7}\cdot\chi^{\mf{sp}(4)}_{(1000)}v^{-2}+(\mathbf{14}+1+\chi^{\mf{sp}(4)}_{(2000)})v^{-1}+\chi^{\mf{sp}(4)}_{(0100)}+\mathbf{7}v+\mathcal{O}(v^2).
\end{aligned}
\end{equation}
Here and below bold letters in the $v$ expansion represent
characters of representations of gauge symmetry.

\subsubsection*{$\mathbf{n=1,\, G=G_2,\,F=\mf{sp}(7)}$}
We study this theory from the viewpoint of both Weyl orbit expansion
and \localelliptic Calabi-Yau. Let us just turn on a subgroup
$\mf{sp}(1)$ of the flavor $\mf{sp}(7)$. We constructed toric
embedding of the \localelliptic Calabi-Yau corresponding to this
configuration, and computed the triple intersection numbers and genus
zero Gromov-Witten invariants. Using the unity blowup equations in
refined BPS expansion, we computed the refined BPS invariants up to
high degrees. We describe our toric construction in
Appendix~\ref{n1G2polytope} and list some low degree BPS invariants in
geometric bases in Appendix~\ref{app:bps}.

Using the Weyl orbit expansion method and the unity blowup equation
with characteristic $a=1/2$, we solved the reduced one-string elliptic
genus with flavor subgroup $\mf{sp}(1)$ at leading $q$ order as
\begin{equation}\label{n1g2E1}
\mathbb{E}_{h_{1,G_2}^{(1)}}(\Qtau,v,m_{G_2}=0,q_m)=\Qtau^{-1/3}+\Qtau^{2/3}v^{-2}\sum_{n=0}^{\infty}\Qtau^n\frac{P_n(v,q_m)}{(1 - v^2)^{6}},
\end{equation}
where $q_m$ is the $\mf{sp}(1)$ flavor fugacity and
\begin{align}
P_0(v,q_m)=\,&(q_m^{-7}+q_m^7)(v^5 + 8 v^7 + 8 v^9 + v^{11})-
(q_m^{-6}+q_m^6)(49 v^6 + 154 v^8 + 49 v^{10})\nn
&+(q_m^{-5}+q_m^5)(28 v^5 + 791 v^7 + 791 v^9 + 28 v^{11})\nn
&+(q_m^{-4}+q_m^4)(35 v^4 - 994 v^6 - 4634 v^8 - 994 v^{10} + 35 v^{12})\nn
&+(q_m^{-3}+q_m^3)(35 v^3 - 259 v^5 + 9233 v^7 + 9233 v^9 - 259 v^{11} + 35 v^{13})\nn
&+(q_m^{-2}+q_m^2)(28 v^2 + 56 v^4 - 3787 v^6 - 28630 v^8 - 3787 v^{10} + 56 v^{12} + 28 v^{14})\nn
&-(q_m^{-1}+q_m)(49 v - 434 v^3 + 2163 v^5 - 28805 v^7 - 28805 v^9 + \dots + 49 v^{15})\nn
&+(14 - 20 v^2 + 218 v^4 - 5800 v^6 - 50600 v^8 - 5800 v^{10} + \dots+ 14 v^{16}).\label{n1g2E1P0}
\end{align}
When the flavor fugacity is turned off, i.e. $q_m=1$, the above result agrees with the modular ansatz in \cite{DelZotto:2018tcj}. Besides, if turning on both gauge and flavor fugacities, we find the following $v$ expansion for the subleading $\Qtau$ order of reduced one-string elliptic genus:
\be
\begin{aligned}
\,&\mathbf{14}v^{-2}-\mathbf{7}\cdot\chi_{(1000000)}^{\mf{sp}(7)} v^{-1}+\mathbf{14}+\chi_{(2000000)}^{\mf{sp}(7)}+1+\chi_{(0010000)}^{\mf{sp}(7)}v+\chi_{(0001000)}^{\mf{sp}(7)}v^2\\
&+ (\chi_{(0000001)}^{\mf{sp}(7)}-\mathbf{7}\cdot\chi_{(0010000)}^{\mf{sp}(7)} -\mathbf{14}\cdot\chi_{(1000000)}^{\mf{sp}(7)} )v^3+\mathcal{O}(v^4).
\end{aligned}
\end{equation}
In fact, we find the following exact formula of the $v$ expansion:
\begin{align}\nonumber
\ &\chi_{(0001000)}^{\mf{sp}(7)}v^2+\chi_{(0010000)}^{\mf{sp}(7)}(v-\chi_{(10)}^{G_2}v^3)+\chi_{(20)}^{G_2}\chi_{(0100000)}^{\mf{sp}(7)}v^4 -\chi_{(1000000)}^{\mf{sp}(7)}
(\chi_{(10)}^{G_2}v^{-1}\\ \nonumber
&+\chi_{(01)}^{G_2}v^3+\chi_{(30)}^{G_2}v^5)+ \chi_{(01)}^{G_2}v^{-2}+\chi_{(01)}^{G_2}+\chi_{(2000000)}^{\mf{sp}(7)}+1+\chi_{(11)}^{G_2}v^4+\chi_{(40)}^{G_2}v^6+ \\[-1mm] \nonumber
\sum_{n=0}^{\infty}\Big[&\chi_{(0n)}^{G_2}\chi_{(0000001)}^{\mf{sp}(7)}v^{3+2n}-\chi_{(1n)}^{G_2}\chi_{(0000010)}^{\mf{sp}(7)}v^{4+2n}
+\chi_{(2n)}^{G_2}\chi_{(0000100)}^{\mf{sp}(7)}v^{5+2n}-\chi_{(3n)}^{G_2}\chi_{(0001000)}^{\mf{sp}(7)}v^{4+2n}\\[-1mm]
+&\chi_{(4n)}^{G_2}\chi_{(0010000)}^{\mf{sp}(7)}v^{7+2n}-\chi_{(5n)}^{G_2}\chi_{(0100000)}^{\mf{sp}(7)}v^{8+2n}
+\chi_{(6n)}^{G_2}\chi_{(1000000)}^{\mf{sp}(7)}v^{9+2n}-\chi_{(7n)}^{G_2}v^{10+2n}\Big].\label{n1g2hs}
\end{align}

\subsection{$F_4$ theories}
$G=F_4$ theories on base curve $(-n)$, $n=1,2,3,4,5$ have flavor group
$F=\mf{sp}(5-n)$ and $n_f=(5-n)$ hypermultiplets in the fundamental
representation ${\bf 26}$ of gauge symmetry. There only
exist unity blowup equations but no vanishing equations due to the Lie
algebra fact $Q^{\vee}\cong P^\vee$ for $F_4$. The corresponding
Calabi-Yau geometries with flavor fugacities turned off were
constructed in \cite{Haghighat:2014vxa,Kashani-Poor:2019jyo}. The
unity $\lambda_F$ fields of these theories are just all the elements
of the Weyl orbit $[0,0,\dots,0,1]$ of $\mf{sp}(5-n)$. For $n=3,4,5$
cases, we can use the recursion formula to exactly compute the
elliptic genera to arbitrary numbers of strings. For $n=1,2$ cases, we
used the Weyl orbit expansion to compute them. The $n=5$ case belongs
to minimal 6d SCFTs and was discussed in detail in our previous paper
of this series \cite{Gu:2019dan}. In the following, we discuss the
$n=1,2,3,4$ cases individually.

\subsubsection*{$\mathbf{n=4,\, G=F_4,\,F=\mf{sp}(1)}$}
There exist 8 unity blowup equations in total with $\lambda_F=\pm
1$. Using the recursion formula, we computed the one-string elliptic
genus to $\mathcal{O}(\Qtau^3)$. Our result agrees precisely with the
modular ansatz in \cite{DelZotto:2018tcj}, therefore we just present
the first few $\Qtau$ orders. Denote the reduced one-string elliptic
genus as
\begin{equation}\label{n4F4E1}
\mathbb{E}_{h_{4,F_4}^{(1)}}(\Qtau,v,m_{F_4}=0,m_{\mf{sp}(1)}=0)=\Qtau^{-5/6}v^{7}\sum_{n=0}^{\infty}\Qtau^n\frac{P_n(v)}{(1 - v)^{10} (1 + v)^{16}}.
\end{equation}
We obtain
\begin{align}
  P_0(v)=\,
  &1 + 10 v - 49 v^2 + 266 v^3 - 549 v^4 + 1068 v^5 - 1110 v^6
    +\dots + v^{12}, \label{n4F4E1P0}\\
  P_1(v)=\,
  &2  (28 + 277 v - 1552 v^2 + 6305 v^3 - 13020 v^4 + 21834 v^5 -
    23904 v^6 + \dots + 28 v^{12}).\label{n4F4E1P1}
\end{align}
One can also keep all flavor and gauge fugacities in the recursion
formula to compute the full elliptic genus. Indeed, as the leading $q$
order of elliptic genus, we confirm the conjectural formula of the
reduced 5d one-instanton partition function in (H.36) of
\cite{DelZotto:2018tcj}:
\begin{equation}\nonumber
v^{7}+\sum_{n=0}^\infty\Big[- \chi^{F_4}_{(n000)}\chi^{\mf{sp}(1)}_{(3)}v^{8+2n}+ \chi^{F_4}_{(n001)}\chi^{\mf{sp}(1)}_{(2)}v^{9+2n}- \chi^{F_4}_{(n010)}\chi^{\mf{sp}(1)}_{(1)}v^{10+2n}+ \chi^{F_4}_{(n100)}v^{11+2n}\Big].
 \end{equation}
For the subleading $q$ order of the reduced one-string elliptic genus, we obtain the following $v$ expansion
\begin{equation}\nonumber
\begin{aligned}
\,&(\mathbf{52}+1+\chi^{\mf{sp}(1)}_{(2)})v^7+((\mathbf{52}+2)\chi^{\mf{sp}(1)}_{(3)}+\chi^{\mf{sp}(1)}_{(1)})v^8\\
-\,(\mathbf{26}&\cdot\chi^{\mf{sp}(1)}_{(4)}+(\chi_{(1001)}^{F_4}+\mathbf{273}+3\cdot\mathbf{26})\chi^{\mf{sp}(1)}_{(2)}+\mathbf{324}+\mathbf{26})v^9+\mathcal{O}(v^{10})
\end{aligned}
\end{equation}

Using the recursion formula, we also computed the two-string elliptic
genus to the subleading order of $\Qtau$. For example, denote the
reduced two-string elliptic genus as
\begin{equation}\label{n4F4E2}\nonumber
\mathbb{E}_{h_{4,F_4}^{(2)}}(\Qtau,v,x=1,m_{F_4}=0,m_{\mf{sp}(1)}=0)=-\Qtau^{-11/6}v^{15}\sum_{n=0}^{\infty}\Qtau^n\frac{P^{(2)}_n(v)}{(1 - v)^{22} (1 + v)^{16} (1 + v + v^2)^{17}},
\end{equation}
we obtain
\begin{align}
P_0(v)=\,&1 + 15 v + 34 v^{2 }+ 97 v^{3 }+ 715 v^{4 }+ 2022 v^{5 }+ 4997 v^{6 }+
 15039 v^{7 }+ 41395 v^{8 }+ 87572 v^{9 }\nn
 &+ 180994 v^{10 }+ 376306 v^{11 }+
 700157 v^{12 }+ 1152469 v^{13 }+ 1848360 v^{14 }+ 2846743 v^{15 }\nn
 &+
 3983439 v^{16 }+ 5139498 v^{17 }+ 6428973 v^{18 }+ 7611291 v^{19 }+
 8253543 v^{20 }\nn
 &+ 8388168 v^{21 }+ \dots+ v^{42},\nn
P_1(v)=\,&2  ( 30 + 480 v + 1478 v^{2 }+ 4015 v^{3 }+ 20963 v^{4 }+ 63895 v^{5 }+
 157718 v^{6 }+ 414969 v^{7 }\nn
 &+ 1079969 v^{8 }+ 2315076 v^{9 }+ 4619079 v^{10 }+
 9059109 v^{11 }+ 16530696 v^{12 }+ 27157331 v^{13 }\nn
 &+ 42451387 v^{14 }+
 63499177 v^{15 }+ 88251928 v^{16 }+ 113833998 v^{17 }+ 140332628 v^{18 }\nn
 &+
 163891834 v^{19 }+ 178266540 v^{20 }+ 182276136 v^{21}+ \dots+ v^{42}).
 \end{align}
\subsubsection*{$\mathbf{n=3,\, G=F_4,\,F=\mf{sp}(2)}$}
Using the recursion formula, we computed the one-string elliptic genus to $\mathcal{O}(\Qtau^3)$. Our result agrees precisely with the modular ansatz in \cite{DelZotto:2018tcj}, therefore we just present the first few $\Qtau$ orders with all gauge and flavor fugacities turned off. Denote the reduced one-string elliptic genus as
\begin{equation}\label{n3F4E1}
\mathbb{E}_{h_{3,F_4}^{(1)}}(\Qtau,v,m_{F_4}=0,m_{\mf{sp}(2)}=0)=\Qtau^{-1/3}v^{6}\sum_{n=0}^{\infty}\Qtau^n\frac{P_n(v)}{(1 - v)^{4} (1 + v)^{16}},
\end{equation}
we obtain
\begin{equation}\label{n3F4E1P0}
\begin{aligned}
  P_0(v)= &\,5 +\, 80 v + 268 v^2 - 1232 v^3 + 2142 v^4 - 1232 v^5 +
  268 v^6 +
  80 v^7 + 5 v^8,\\
  P_1(v)= &\,v^{-8}( 1 + 12 v + 62 v^2 + 172 v^3 + 237 v^4 - 20 v^5 -
  722 v^6 - 1472 v^7 - 1357 v^8\\
  &\phantom{==} + 4812 v^9 + 21908 v^{10} - 72624 v^{11} + 101054
  v^{12} +\dots + v^{24}).
 \end{aligned}
\end{equation}
Keeping all flavor and gauge fugacities in the recursion formula to
compute the full elliptic genus. Indeed, as the leading $q$ order of
elliptic genus, we confirm the conjectural formula of reduced 5d
one-instanton partition function in (H.36) of
\cite{DelZotto:2018tcj}. For example, the first few terms are
\begin{equation}\nonumber
\begin{aligned}
\,&\chi^{\mf{sp}(2)}_{(01)}v^{6}+\chi^{\mf{sp}(2)}_{(30)}v^7+(\chi^{\mf{sp}(2)}_{(03)}-\mathbf{52}-\mathbf{26}\cdot\chi^{\mf{sp}(2)}_{(20)})v^8
+(\mathbf{273}\cdot\chi^{\mf{sp}(2)}_{(10)}
-\mathbf{26}\cdot\chi^{\mf{sp}(2)}_{(12)})v^9\\
&+(\mathbf{52}\cdot\chi^{\mf{sp}(2)}_{(03)}+\mathbf{273}\cdot\chi^{\mf{sp}(2)}_{(21)}+\mathbf{324}\cdot\chi^{\mf{sp}(2)}_{(02)}-\mathbf{1274})v^{10}
+\mathcal{O}(v^{11}).
\end{aligned}
 \end{equation}
For the subleading $q$ order the reduced one-string elliptic genus, we obtain the following $v$ expansion
\begin{equation}\nonumber
\begin{aligned}
v^{-2}&-\chi^{\mf{sp}(2)}_{(10)}v^3-\chi^{\mf{sp}(2)}_{(20)}v^4+\Big(\chi^{\mf{sp}(2)}_{(21)}+\chi^{\mf{sp}(2)}_{(20)}+(\mathbf{52}+\mathbf{26}+2)\chi^{\mf{sp}(2)}_{(01)}\Big)v^6\\
&-(\chi^{\mf{sp}(2)}_{(31)}+\chi^{\mf{sp}(2)}_{(12)}+\chi^{\mf{sp}(2)}_{(11)}+\chi^{\mf{sp}(2)}_{(10)}+(\mathbf{52}+2)\chi^{\mf{sp}(2)}_{(30)} )v^7+\mathcal{O}(v^8)
\end{aligned}
\end{equation}

We also computed the two-string elliptic genus to the subleading order of $\Qtau$. For example, denote the reduced two-string elliptic genus as
\begin{equation}\label{n3F4E2}\nonumber
\mathbb{E}_{h_{3,F_4}^{(2)}}(\Qtau,v,x=1,m_{F_4}=0,m_{\mf{sp}(2)}=0)=-\Qtau^{-5/6}v^{13}\sum_{n=0}^{\infty}\Qtau^n\frac{P^{(2)}_n(v)}{(1 - v)^{10} (1 + v)^{16} (1 + v + v^2)^{17}},
\end{equation}
we have
\begin{align}
P_0^{(2)}(v)=\,&15 + 449 v + 5327 v^{2 }+ 30906 v^{3 }+ 101183 v^{4 }+ 187889 v^{5 }+
 183238 v^{6 }+ 180121 v^{7 }\nn
 &+ 820970 v^{8 }+ 2527029 v^{9 }+ 3954101 v^{10 }+
 3268018 v^{11 }+ 2502062 v^{12 }+ 6631296 v^{13 }\nn
 &+ 14672455 v^{14 }+
 17834663 v^{15 }+ 12802905 v^{16 }+ 8758778 v^{17 }+\dots+ 15 v^{34},\nn
P_1^{(2)}(v)=\,&v^{-8}  (5 + 145 v + 1763 v^{2 }+ 11722 v^{3 }+ 53549 v^{4 }+ 182991 v^{5 }+
 493575 v^{6 }+ 1078556 v^{7 }\nn
 &+ 1935972 v^{8 }+ 2865208 v^{9 }+
 3665294 v^{10 }+ 5010010 v^{11 }+ 8956794 v^{12 }+ 15093412 v^{13 }\nn
 &+
 14295923 v^{14 }- 2110395 v^{15 }- 13976451 v^{16 }+ 18409580 v^{17 }+
 78794748 v^{18 }\nn
 &+ 85716318 v^{19 }+ 44817687 v^{20 }+ 102304199 v^{21 }+
 290636920 v^{22 }+ 388309453 v^{23 }\nn
 &+ 271239229 v^{24 }+ 167708226 v^{25}+ \dots+ 5 v^{50}).
 \end{align}

\subsubsection*{$\mathbf{n=2,\, G=F_4,\,F=\mf{sp}(3)}$}
We used the Weyl orbit expansion to solve the one-string elliptic
genus. Let us first just turn on the fugacity of a subalgebra $\mf{sp}(1)$
of the flavor symmetry $\mf{sp}(3)$. Using the unity blowup equation with characteristic $a=0$, we solved the reduced one-string elliptic genus with flavor subgroup $\mf{sp}(1)$ fugacity $q_m$ as
\begin{equation}\label{n2F4E1}
\mathbb{E}_{h_{2,F_4}^{(1)}}(\Qtau,v,m_{F_4}=0,q_m)=\Qtau^{1/6}v^{-1}\sum_{n=0}^{\infty}\Qtau^n\frac{P_n(v,q_m)}{(1 - v^2)^{12}},
\end{equation}
where
\begin{align}
\,&P_0(v,q_m)=-v^9f_9 (1 + 36 v^2 + 341 v^4 + 1208 v^6 + 1820 v^8 + \dots + v^{16})\nn
&+39 v^{10}f_8 (1 + v^2) (2 + 47 v^2 + 274 v^4 + 506 v^6 +\dots+ 2 v^{12})\nn
&-3 v^9f_7 (2 + 942 v^2 + 14439 v^4 + 62278 v^6 + 99270 v^8 +
   \dots + 2 v^{16})\nn
&+   2 v^8f_6 (-5 + 314 v^2 + 29213 v^4 + 264959 v^6 + 723887 v^8 +
   \dots - 5 v^{18})\nn
   &-3 v^7f_5 (2 - 155 v^2 + 7792 v^4 + 238244 v^6 + 1250686 v^8 +
   2098702 v^{10} +\dots +
   2 v^{20})\nn
&+3 v^6f_4(-1 + 29 v^2 - 1795 v^4 + 135107 v^6 + 1736758 v^8 +
   5258478 v^{10} + \dots - v^{22})\nn
&-v^5f_3 (1 + 20 v^2 - 2077 v^4 + 34266 v^6 + 3666667 v^8 +
   22762210 v^{10} + 39749314 v^{12} + \dots + v^{24})\nn
&-6 v^6 f_2(3 - 138 v^2 + 5599 v^4 - 176466 v^6 - 3072141 v^8 -
   9995641 v^{10} - \dots + 3 v^{22})\nn
&-3 v^5 f_1(2 - 10 v^2 - 761 v^4 - 8900 v^6 + 2607160 v^8 +
   17861126 v^{10} + 31896078 v^{12} + \dots + 2 v^{24})\nn
&+(1 - 16 v^2 + 120 v^4 - 588 v^6 + 3293 v^8 - 59309 v^{10} +
 1403134 v^{12} + 27648874 v^{14} + 92360011 v^{16}\nn
  &\phantom{---} +\dots + v^{34}).\label{n2F4E1P0}
\end{align}
Here $f_n=q_m^{-n}+q_m^n$.
When the flavor fugacity is turned off, i.e. $q_m=1$, the above result
agrees with the modular ansatz in \cite{DelZotto:2018tcj}. Besides, at
leading $\Qtau$ order, the reduced one-string elliptic genus given by (\ref{n2F4E1}) and (\ref{n2F4E1P0}) has the following expansion
\begin{equation}
\begin{aligned}
  v^{-1}&-v^4(q_m^3+6q_m+6q_m^{-1}+q_m^{-3})-v^5(3q_m^{-4}+18q_m^{-2}+28+18q_m^2+3q_m^4)\\
  &-6v^6(q_m^{-5}+6q_m^{-3}+11q_m^{-1}+11q_m+6q_m^3+q_m^5)+\mathcal{O}(v^7).
\end{aligned}
\end{equation}
It is easy to check with flavor symmetry $\mf{sp}(3)$ restricted to $\mf{sp}(1)_{q_m}$ this agrees with the exact formula of reduced 5d one-instanton partition function conjectured in (H.26) of \cite{DelZotto:2018tcj}. For example, the first few terms are
\begin{align}
v^{-1}&-v^4\chi^{\mf{sp}(3)}_{(001)}-v^5\chi^{\mf{sp}(3)}_{(101)}-v^6\chi^{\mf{sp}(3)}_{(201)}+v^7(\mathbf{52}\cdot\chi^{\mf{sp}(3)}_{(010)}+\mathbf{26}\cdot\chi^{\mf{sp}(3)}_{(101)}
-\chi^{\mf{sp}(3)}_{(030)})\nonumber\\
 &+v^8(\mathbf{52}\cdot\chi^{\mf{sp}(3)}_{(300)}-\mathbf{273}\cdot\chi^{\mf{sp}(3)}_{(001)}+\mathbf{26}\cdot\chi^{\mf{sp}(3)}_{(120)})+\mathcal{O}(v^9).
  \end{align}
One can also turn on full flavor fugacity and gauge fugacity and push the computation to higher $\Qtau$ orders and higher number of strings. For example, for the subleading $q$ order of reduced one-string elliptic genus, we obtain
\begin{equation}
\begin{aligned}
  \,&\mathbf{52}v^{-3}-\mathbf{26}\cdot\chi^{\mf{sp}(3)}_{(100)}v^{-2}+(\mathbf{52}+\chi^{\mf{sp}(3)}_{(200)}+1)v^{-1}+\chi^{\mf{sp}(3)}_{(300)}\\
  &+\chi^{\mf{sp}(3)}_{(020)}v+\chi^{\mf{sp}(3)}_{(011)}v^2+(\chi^{\mf{sp}(3)}_{(002)}+\mathbf{26}\cdot\chi^{\mf{sp}(3)}_{(010)})v^3+\mathcal{O}(v^4)
\end{aligned}
\end{equation}

\subsubsection*{$\mathbf{n=1,\, G=F_4,\,F=\mf{sp}(4)}$}
Let us first turn on the fugacity of a subalgebra $\mf{sp}(1)$ of the
full flavor symmetry $\mf{sp}(4)$. Using the Weyl orbit expansion
method and the unity blowup equation with characteristic $a=1/2$, we
solved the reduced one-string elliptic genus with flavor subalgebra
$\mf{sp}(1)$ at leading $q$ order as
\begin{equation}\label{n1F4E1}
  \mathbb{E}_{h_{1,F_4}^{(1)}}(\Qtau,v,m_{F_4}=0,q_m)=\Qtau^{-1/3}+\Qtau^{2/3}v^{-2}\sum_{n=0}^{\infty}\Qtau^n\frac{P_n(v,q_m)}{(1 - v^2)^{16}},
\end{equation}
where $q_m$ is the $\mf{sp}(1)$ flavor fugacity and
\begin{align}
P_0(v,q_m)\,&=v^{10} (q_m^{-12}+q_m^{12})(1 + 36 v^2 + 341 v^4 + 1208 v^6 + 1820 v^8 + \dots + v^{16})\nn
&-
52 v^{11}(q_m^{-11}+q_m^{11}) (2 + 49 v^2 + 321 v^4 + 780 v^6 + \dots + 2 v^{14})+\cdots\nn
&-4 v(q_m^{-1}+q_m) (26 - 426 v^2 + 3215 v^4 - 14760 v^6 + 58005 v^8 - 494529 v^{10} +
   2024378 v^{12}\nn
   & + 306868947 v^{14} + 1249149000 v^{16} +
   \dots + 26 v^{34})+(52 - 763 v^2 + 5256 v^4 - 21590 v^6\nn
   & + 39900 v^8 + 421246 v^{10} -
 13984964 v^{12} + 300172490 v^{14} + 3270987324 v^{16}\nn
 & + 6383908850 v^{18} +
\dots + 52 v^{36}).
\end{align}
When the $\mf{sp}(1)$ fugacity is turned off,
\begin{equation}\nonumber
\begin{aligned}
P_0(v,1)=&\,(1-v)^{16}(52 + 624 v + 3001 v^2 + 5704 v^3 - 8932 v^4 - 81464 v^5 -
  210244 v^6 - 145256 v^7\\& + 896624 v^8 + 3964136 v^9 + 7404438 v^{10} +\dots + 52 v^{20}).
\end{aligned}
\end{equation}
We checked this agrees with the modular ansatz in
\cite{DelZotto:2018tcj}. We also turned on all $\mf{sp}(4)$ flavor
fugacities to perform the Weyl orbit expansion, from which we found an
exact formula for the Weyl orbit expansion of the subleading $\Qtau$
order of the reduced one-string elliptic genus, which will be given in
Appendix (\ref{n1F4exactE1}). For example, the first few terms are
\begin{equation}
\begin{aligned}
  \,&\mathbf{52}\,v^{-2}-\mathbf{26}\cdot\chi_{(1000)}^{\mf{sp}(4)}
  v^{-1}+\mathbf{52}+\chi_{(2000)}^{\mf{sp}(4)}+1
  +\chi_{(3000)}^{\mf{sp}(4)}v+\chi_{(0200)}^{\mf{sp}(4)}v^2\\
  &+\chi_{(0110)}^{\mf{sp}(4)}v^3+
  (\chi_{(0020)}^{\mf{sp}(4)}+\chi_{(2001)}^{\mf{sp}(4)}
  -\mathbf{26}\cdot\chi_{(0001)}^{\mf{sp}(4)} )v^4+\mathcal{O}(v^5).
\end{aligned}
\end{equation}
This contains the information of the 5d Nekrasov partition function of
the $G=F_4,F=\mf{sp}(4)$ theory.

\subsection{$E_6$ theories}

$G=E_6$ theories on base curve $(-n)$ have flavor symmetry
$F=\mf{su}(6-n)_6\times \mf{u}(1)_{6(6-n)}$ and $n_f=(6-n)$
hypermultiplets in the bi-representation
$({\bf 27},({\bf 6-n})_1)$. Note ${\bf 6-n}$ is the fundamental
representation of flavor symmetry $\mf{su}(6-n)$, and
$n=1,2,3,4,5,6$. There are $2n$ vanishing blowup equations with
$\lambda_{\mf{su}(6-n)}=0$ and $\lambda_{\mf{u}(1)}=\pm 1/6$.

The reason there are two copies of vanishing equations is that the
Dynkin diagram of $E_6$ is axisymmetric, in particular there exist two
fundamental representations of $E_6$: $\mathbf{27}$ and
$\overline{\mathbf{27}}$. For any two weights
$w_1,w_2\in \mathbf{27}$, $w_1\cdot
w_2=4/3,1/3,-2/3$.
The same for
$\overline{\mathbf{27}}$. While for $w_1\in \mathbf{27}$ and
$w_2\in \overline{\mathbf{27}}$, one has $w_1\cdot
w_2=-4/3,-1/3,2/3$. Since $(P^\vee/Q^{\vee})_{E_6}=\IZ_3$, accordingly
let us denote $P^\vee=Q^{\vee}\oplus\Lambda\oplus\overline{\Lambda}$,
such that $\mathbf{27}\subset\Lambda$ and
$\overline{\mathbf{27}}\subset\overline{\Lambda}$. For any
$w_1\in\mathbf{27}$, $w_2\in\overline{\mathbf{27}}$,
$\lambda_1\in\Lambda $ and $\lambda_2\in\overline{\Lambda}$, always
\begin{equation}
\begin{aligned}
  w_1\cdot\lambda_1\in\IZ+1/3,&\quad\quad w_1\cdot\lambda_2\in\IZ-1/3,\\
  w_2\cdot\lambda_1\in\IZ-1/3,&\quad\quad w_2\cdot\lambda_2\in\IZ+1/3.\\
\end{aligned}
\end{equation}

It is easy to find the leading base degree of one copy of the
vanishing blowup equations
\begin{equation}\label{E6vanish0}
  \sum_{w\in{\bf 27}}(-1)^{|w|}\theta_i^{[a]}(n\tau,nm^{E_6}_w+(6-n)\epsilon_+')
  \prod_{w'\in\mathbf{6-n}}\theta_1(m^{E_6}_w+m_{w'}^{su(6-n)}-\epsilon_+')
  \prod^{w\cdot \alpha=1}_{\alpha\in \Delta(E_6)}\frac{1}{\theta_1(m^{E_6}_\alpha)}=0,
\end{equation}
where we denote $\epsilon_+'=m_{u(1)}+\epsilon_+$. We have verified
this identity up to $\Qtau^{10}$ for all $n=1,2,\dots,6$. Note this
identity contains $m_{E_6}$, $m_{su(6-n)}$, $m_{u(1)}$ and
$\epsilon_+$ as free parameters, thus are highly nontrivial. By
setting $m_{F}$ as zero, one obtains
\begin{equation}\label{E6vanish00}
  \sum_{w\in{\bf 27}}(-1)^{|w|}\theta_i^{[a]}(n\tau,nm_w+(6-n)\epsilon_+)\big(\theta_1(m_w-\epsilon_+)\big)^{6-n}
  \prod^{w\cdot \alpha=1}_{\alpha\in \Delta(E_6)}\frac{1}{\theta_1(m_\alpha)}=0,
\end{equation}
which may be easier in case interested readers want to give a direct
proof.

For the $n=5$ case where the flavor is just $\mf{u}(1)$ itself, we
find there exist two more vanishing $r$ fields with
$\lambda_{\mf{u}(1)}=\pm 5/6$. For example, for
$(\lambda_{E_6},\lambda_{\mf{u}(1)})=(\mathbf{27},5/6)$, the leading
degree vanishing identities can be written as
\begin{equation}\label{n5E6vanish0more}
  \sum_{w\in{\bf 27}}(-1)^{|w|}\theta_4^{[a]}(5\tau,5m^{E_6}_w+5\epsilon_+')
  \prod_{w'\in{\bf 27}}^{w'\cdot w=-2/3}\theta_1(m^{E_6}_{w'}-\epsilon_+')
  \prod^{w\cdot \alpha=1}_{\alpha\in \Delta{(E_6)}}\frac{1}{\theta_1(m^{E_6}_\alpha)}=0.
\end{equation}
Here again $\epsilon_+'=m_{u(1)}+\epsilon_+$. Note the hypermultiplet
contribution i.e. the first product contains ten $\theta_1$
functions. Although we do not find vanishing $r$ fields with
$\lambda_{\mf{u}(1)}=\pm 5/6$ suitable for $n=4,3,2,1$ theories, we
indeed find one kind of generalization of (\ref{n5E6vanish0more})
which is
\begin{equation}\label{allnE6vanish0more}
  \sum_{w\in{\bf 27}}(-1)^{|w|}\theta_i^{[a]}(n\tau,nm^{E_6}_w)
  \prod_{w'\in{\bf 27}}^{w'\cdot w=-2/3}\theta_1(m^{E_6}_{w'})^{6-n}
  \prod^{w\cdot \alpha=1}_{\alpha\in \Delta(E_6)}\frac{1}{\theta_1(m^{E_6}_\alpha)}=0.
\end{equation}
Here $n=1,2,3,4,5,6$, while $i$ and $a$ are defined accordingly by the
general rule of blowup equations.

For unity blowup equations, there are $2^{6-n}$ choices for
$\lambda_F$ fields. In fact, they form the Weyl orbit
$\mathcal{O}_{[00\dots01]}^{\mf{sp}(6-n)}$ if we embed $\mf{su}(6-n)\times \mf{u}(1)$ into
$\mf{sp}(6-n)$. Note there always exist $\lambda_F$ fields
$(\lambda_{su(6-n)},\lambda_{u(1)})=(0,\pm 1/2)$. For $n=3,4,5,6$, one
can choose arbitrary one $\lambda_F$ and three unity blowup equations
with different characteristics $a$ to solve elliptic genera
recursively. The $n=6$ case belongs to minimal 6d SCFTs and was
discussed in detail in the previous paper of this series
\cite{Gu:2019dan}. In the following, we discuss the $n=1,2,3,4,5$
cases.

\subsubsection*{$\mathbf{n=5,\, G=E_6,\,F=\mf{u}(1)}$}
There exist 5 unity blowup equations with $r_F=0$. Using the recursion
formula, we computed the one-string elliptic genus to
$\mathcal{O}(\Qtau)$. Our result agrees precisely with the modular
ansatz in \cite{DelZotto:2018tcj}, therefore we just present the first
few $\Qtau$ orders. For example, denote the reduced one-string
elliptic genus with all gauge and flavor fugacities turned off as
\begin{equation}\label{n5E6E1}
\mathbb{E}_{h_{5,E_6}^{(1)}}(\Qtau,v,m_{E_6}=0,m_{u(1)}=0)=\Qtau^{-4/3}v^{10}\sum_{n=0}^{\infty}\Qtau^n\frac{P_n(v)}{(1 - v)^{16} (1 + v)^{22}},
\end{equation}
we obtain
\begin{align}
  P_0(v)=\,
  &1 + 8 v - 43 v^2 + 456 v^3 - 1436 v^4 + 5116 v^5 - 9848 v^6 + 19504 v^7 - 24164 v^8\nn
  & + 30016 v^9 +\dots+v^{18}, \nn
  P_1(v)=\,
  &2 (40 + 320 v - 2072 v^2 + 16128 v^3 - 51094 v^4 + 155036 v^5 -
    297317 v^6 + 530598 v^7\nn
  & - 670889 v^8 + 785764 v^9 - \dots + 40 v^{18}),\nn
  P_2(v)=\,
  &-2v^{-1}(27 - 1498 v - 13658 v^2 + 95382 v^3 - 590835 v^4 + 1824915 v^5 -  4912446 v^6 \nn
  &+ 9187979 v^7 - 15230210 v^8 + 19237562 v^9 -
    21771556 v^{10} +\dots+27v^{20}).\label{n5E6E1P2}
\end{align}
By keeping the gauge and flavor fugacities in the recursion formula
and taking the leading $\Qtau$ order, we confirm the conjectural
formula of reduced 5d one-instanton partition function in (H.38) of
\cite{DelZotto:2018tcj}:
\begin{align}\label{ZLn5E6}\nonumber
\,& v^{10}+\sum_{n=0}^\infty\Big[ \chi^{E_6}_{(00000n)}\chi^{u(1)}_{(3)\oplus (-3)}v^{11+2n}- (\chi^{E_6}_{(00001n)}\chi^{u(1)}_{(-2)}+\chi^{E_6}_{(10000n)}\chi^{u(1)}_{(2)})v^{12+2n}\nonumber\\[-1mm]
 &\qquad + (\chi^{E_6}_{(00010n)}\chi^{u(1)}_{(-1)}+\chi^{E_6}_{(01000n)}\chi^{u(1)}_{(1)})v^{13+2n}- \chi^{E_6}_{(00100n)}v^{14+2n}\Big].
 \end{align}
For the subleading $\Qtau$ order, we obtain
\begin{align}\nonumber
\,&(\mathbf{78}+2) v^{10}+(\mathbf{78}+2)\chi^{u(1)}_{(3)\oplus (-3)}v^{11}-\Big(\chi^{E_6}_{(100000)}\chi^{u(1)}_{(-4)}+\chi^{E_6}_{(000010)}\chi^{u(1)}_{(4)}\\
+\,&\chi^{E_6}_{(000011)\oplus(010000)\oplus3(000010)}\chi^{u(1)}_{(-2)}+\chi^{E_6}_{(100001)\oplus(000100)\oplus3(100000)}\chi^{u(1)}_{(2)}
+\chi^{E_6}_{(100010)}\Big)v^{12}+\mathcal{O}(v^{13}).\nonumber
 \end{align}

Using recursion formula, we also computed the two-string elliptic
genus to the subleading order of $\Qtau$ which will be given in
Appendix \ref{sec:n5E6E2}.

\subsubsection*{$\mathbf{n=4,\, G=E_6,\,F=\mf{su}(2)\times \mf{u}(1)}$}
Using the recursion formula, we computed the one-string elliptic genus
to $\mathcal{O}(\Qtau^2)$. Our result agrees precisely with the
modular ansatz in \cite{DelZotto:2018tcj}, therefore we just present
the first few $\Qtau$ orders. Denote the reduced one-string elliptic
genus as
\begin{equation}\label{n4E6E1}
\mathbb{E}_{h_{4,E_6}^{(1)}}(\Qtau,v,m_{E_6}=0,m_F=0)=\Qtau^{-5/6}v^{9}\sum_{n=0}^{\infty}\Qtau^n\frac{P_n(v)}{(1 - v)^{10} (1 + v)^{22}},
\end{equation}
We obtain
\begin{equation}\label{n4E6E1P0}\nonumber
\begin{aligned}
P_0(v)=\,&-(3 + 44 v + 33 v^2 - 1052 v^3 + 6513 v^4 - 17404 v^5 + 31905 v^6 -
  37432 v^7 + \dots + 3 v^{14}),
 \end{aligned}
\end{equation}
\begin{equation}\label{n4E6E1P1}\nonumber
\begin{aligned}
P_1(v)=\,&v^{-2} (3 + 36 v - 135 v^2 - 4000 v^3 - 3894 v^4 + 106168 v^5 - 500700 v^6 +
  1239080 v^7\phantom{----}\ \\
  & - 2078322 v^8 + 2430488 v^9 - \dots+ 3 v^{18}).
 \end{aligned}
\end{equation}

One can also keep all flavor and gauge fugacities in blowup equations
to compute the full elliptic genus. In \cite{DelZotto:2018tcj}, the
Weyl orbit expansion of reduced 5d one-instanton partition function
was conjectured up to $v^{11}$. Using the recursion formula from
blowup equations, we find the following exact formula where
$F=\mf{su}(2)_a\times \mf{u}(1)_b$:
\begin{align}\nonumber
\,&-v^9\chi^F_{(2)_a}-v^{10}\chi^F_{(3)_a\otimes((3)_b\oplus(-3)_b)}+v^{11}(\chi^{E_6}_{(100000)}
\chi^F_{(2)_a\otimes(2)_b}+c.c.)+v^{11}\chi^{E_6}_{(000001)}\\ \nonumber
&-v^{12}(\chi^{E_6}_{(010000)}\chi^F_{(1)_a\otimes(1)_b}+c.c)
+v^{13}\chi^{E_6}_{(001000)}+\nonumber\\[-1mm]
\sum_{n=0}^{\infty}\Big[&-v^{11+2n}\chi^{E_6}_{(00000n)}(\chi^F_{(6)_b\oplus(-6)_b}+\chi^F_{(6)_a})
+v^{12+2n}(\chi^{E_6}_{(10000n)}\chi^F_{(1)_a\otimes(5)_b}+\chi^{E_6}_{(00001n)}\chi^F_{(5)_a\otimes(1)_b}+c.c.)\nonumber\\[-1mm]
-&\,v^{13+2n}\Big((\chi^{E_6}_{(01000n)}\chi^F_{(2)_a\otimes(4)_b}+\chi^{E_6}_{(00010n)}\chi^F_{(4)_a\otimes(2)_b}
+\chi^{E_6}_{(20000n)}\chi^F_{(4)_b}+c.c.)
+\chi^G_{(10001n)}\chi^F_{(4)_a}\Big)\nonumber\\
+&\,v^{14+2n}\Big(\chi^{E_6}_{(00100n)}\chi^F_{(3)_a\otimes((3)_b\oplus(-3)_b)}+(\chi^{E_6}_{(11000n)}\chi^F_{(1)_a\otimes(3)_b}
+\chi^{E_6}_{(10010n)}\chi^F_{(3)_a\otimes(1)_b}+c.c.)\Big)\nonumber\\
-&\,v^{15+2n}\Big((\chi^{E_6}_{(10100n)}\chi^F_{(2)_a\otimes(2)_b}+\chi^{E_6}_{(02000n)}\chi^F_{(-2)_b}+c.c.)
+\chi^{E_6}_{(01010n)}\chi^F_{(2)_a}\Big)\nonumber\\
+&\,v^{16+2n}(\chi^{E_6}_{(01100n)}\chi^F_{(1)_a\otimes(1)_b}+c.c.)
-v^{17+2n}\chi^{E_6}_{(00200n)}\Big].\label{n4E6HS}
\end{align}
This formula can be reconfirmed by the Weyl dimension formula of representation of $E_6$ and $\mf{su}(2)$, where one can obtain the rational function of $v$ as in (\ref{n4E6E1}). For the subleading $\Qtau$ order of reduced one-string elliptic genus, we obtain
\begin{align}\nonumber
\,& \chi^F_{(2)_a}v^{7}-(\chi^F_{(4)_a}+(\mathbf{78}+3)\chi^F_{(2)_a}+\chi^{E_6}_{(000010)}\chi^F_{(-2)_b}+\chi^{E_6}_{(100000)}\chi^F_{(2)_b}+1)v^9\\
&-(\chi^F_{(5)_a}+(\mathbf{78}+3)\chi^F_{(3)_a}+\chi^F_{(1)_a})\chi^F_{(-3)_b\oplus(3)_b}v^{10}+\mathcal{O}(v^{11}).\nonumber
 \end{align}

Using the recursion formula, we also computed the two-string elliptic
genus to the subleading order of $\Qtau$ which will be given in
Appendix \ref{sec:n4E6E2}.

\subsubsection*{$\mathbf{n=3,\, G=E_6,\,F=\mf{su}(3)\times \mf{u}(1)}$}
Using the recursion formula, we computed the one-string elliptic genus to $\mathcal{O}(\Qtau^3)$. Our result agrees precisely with the modular ansatz in \cite{DelZotto:2018tcj}, therefore we just present the first few $\Qtau$ orders. Denote the reduced one-string elliptic genus as
\begin{equation}\label{n3E6E1}
\mathbb{E}_{h_{3,E_6}^{(1)}}(\Qtau,v,m_{E_6}=0,m_F=0)=\Qtau^{-1/3}v^{7}\sum_{n=0}^{\infty}\Qtau^n\frac{P_n(v)}{(1 - v)^{4} (1 + v)^{12}},
\end{equation}
We obtain
\begin{align}
  P_0(v)=\,
  &2 (1 + 28 v + 356 v^2 + 2045 v^3 + 1583 v^4 - 19638 v^5 +
    36572 v^6 - \dots +
    v^{12}), \nn
  P_1(v)=\,
  &v^{-9} (1 + 18 v + 149 v^{2 }+ 744 v^{3 }+ 2454 v^{4 }+ 5412 v^{5 }+ 7230 v^{6 }+ 2216 v^{7 }- 14256 v^{8 }\nn
  &- 39160 v^{9 }- 61154 v^{10 }- 18988 v^{11 }+  372829 v^{12 }+
    642294 v^{13 }- 3309245 v^{14 }\nn
  &+ 4904064 v^{15 }+\dots+ v^{30}). \label{n3E6E1P1}
\end{align}

We can also turn on all gauge and flavor fugacities. Using recursion
formula from blowup equations, we find the exact formula for the
leading $\Qtau$ order of reduced one-string elliptic genus with
$F=\mf{su}(3)_a\times \mf{u}(1)_b$, which will be presented in Appendix
(\ref{n3E6exactE1}). The first few terms are
\begin{align}
  \,&v^7\chi^F_{(3)_b\oplus(-3)_b}+v^{8}(\chi^F_{(30)_a}+\chi^F_{(03)_a})+v^9\chi^F_{(22)_a\otimes((3)_b\oplus(-3)_b)}\nonumber\\
    &-v^{10}(\chi^G_{(100000)}\chi^F_{(12)_a\otimes(2)_b}+\chi^G_{(000010)}\chi^F_{(21)_a\otimes(-2)_b}+\chi^G_{(000001)}\chi^F_{(11)_a}\nonumber\\
    &-\chi^F_{(30)_a\otimes(- 6)_b}-\chi^F_{(03)_a\otimes(6)_b}-\chi^F_{(33)_a})+\mathcal{O}(v^{12}),
\end{align}
which were already conjectured in \cite{DelZotto:2018tcj}. For the
subleading $\Qtau$ order of reduced one-string elliptic genus, we
obtain
\begin{align}\nonumber
  v^{-2}-\chi^F_{(11)_a}v^4-\chi^F_{(11)_a\otimes((3)_b\oplus(-3)_b)} v^5-\chi^F_{(22)_a}v^6+(\mathbf{78}+\chi^F_{(11)_a}+2)\chi^F_{(3)_b\oplus(-3)_b}v^7+\mathcal{O}(v^8).\nonumber
\end{align}

Using recursion formula, we also computed the two-string elliptic
genus to the leading order of $\Qtau$ which will be given in Appendix
\ref{sec:n4E6E2}.

\subsubsection*{$\mathbf{n=2,\, G=E_6,\,F=\mf{su}(4)\times \mf{u}(1)}$}
We use Weyl orbit expansion to solve elliptic genus for this
theory. Let us first turn off the $\mf{su}(4)$ fugacities and only
keep $\mf{u}(1)$ and make use of the unity blowup equations with
nonzero $\lF$ only on $\mf{u}(1)$. Then the reduced one-string
elliptic genus can be computed as
\begin{equation}\label{n2E6E1}
\mathbb{E}_{h_{2,E_6}^{(1)}}(\Qtau,v,m_{E_6}=0,m_F=0)=\Qtau^{1/6}v^{-1}\sum_{n=0}^{\infty}\Qtau^n\frac{P_n(v)}{ (1 + v)^{22}},
\end{equation}
where
\begin{equation}\nonumber
\begin{aligned}
P_0(v)=&\,(1 - v)^2 (1 + 24 v + 278 v^2 + 2072 v^3 + 11181 v^4 +
   46624 v^5 + 156660 v^6 + 436728 v^7\\
   & + 1030043 v^8 + 2066568 v^9 +
   3435967 v^{10} + 4315392 v^{11} + 3435967 v^{12}+\dots+v^{22}),\\
P_1(v)=&\,v^{-2}(78 + 1500 v + 13361 v^2 + 72354 v^3 + 260839 v^4 + 631520 v^5 +
 910434 v^6 + 142972 v^7\\
 & - 2884243 v^8 - 7465814 v^9 - 7830327 v^{10} +
 5820340 v^{11} + 30116822 v^{12} + 14704216 v^{13}\\
 & - 68988104 v^{14} +
 14704216 v^{15} +\dots+78v^{28}).
\end{aligned}
\end{equation}
We have cross-checked our result against the modular ansatz in
\cite{DelZotto:2018tcj}.\footnote{In \cite{DelZotto:2018tcj}, the
  modular ansatz for this theory is determined up to three unfixed
  parameters. Using our result from blowup equations, we are able to
  determine their three unfixed parameters as \be
  a_1=\frac{6581939}{638959998741245853696},a_2=
  -\frac{12286901}{5111679989929966829568},a_3=\frac{16984805}{5750639988671212683264}.
\end{equation}} Let us denote
\begin{equation}
  \mathbb{E}_{h_{2,E_6}^{(1)}}(\Qtau,v,m_{E_6}=0,m_F=0)=
  \Qtau^{1/6}v^{-1}\sum_{i,j}c_{i,j} v^j(\Qtau/v^2)^i.
\end{equation}
Then we have the following table \ref{tb:n2E6one} for the coefficients
$c_{ij}$. Note the red numbers in the first column are just the
dimensions of representations ${k\theta}$ of $E_6$ where $\theta$ is
the adjoint representation. The blue numbers in the second column are
eight times of the dimensions of representations $\square+k\theta$ of
$E_6$, where the eight is the double of the dimension of matter
representation $\bf 4$ of flavor $\mf{su}(4)$. The orange number 95 in
the third column is given by
$\mathrm{dim}(E_6)+\dim(\mf{su}(4)\times \mf{u}(1))+1=78+16+1=95$. These
are the constraints predicted in \cite{DelZotto:2018tcj} by analyzing
the spectral flow to Neveu-Schwarz-Ramond elliptic genus.
\begin{table}[h]
 \begin{center}
\begin{small}
\begin{tabular}{c| cccccccccccccccc }$i,j$&  0 & 1 & 2 & 3 & 4 & 5 & 6 & 7 & 8 & 9\\
\hline
0& \color{red}1& 0& 0& 0& 0& 0& -20& -72& -319& 240 \\
1 & \color{red}78& \color{blue}-216& \color{orange}95& 40& 84& 120& 195& 1248& -2155& -11488\\
2& \color{red}2430& \color{blue}-13824& 28392& -20520& -1555& -3760& 3102& 12264& 17277&
  166800 \\
3& \color{red}43758& \color{blue}-370656& 1334745& -2526856&
  2380950& -587824& -213080& -601120& -339398& 510992\\
\hline
\end{tabular}
\end{small}
\caption{Series coefficients $ c_{i,j} $ for the one-string elliptic genus of $\fn =2$ $E_6$ model.}\label{tb:n2E6one}
\end{center}
\end{table}

We also computed the elliptic genus with all flavor
$\mf{su}(4)_a\times \mf{u}(1)_b$ fugacities turned on and
gauge fugacities turned off. For example, the $\Qtau$
leading order of reduced one-string elliptic genus has $v$ expansion
as
\begin{equation}\nonumber
\begin{aligned}
  \,&\frac{1}{v}-\chi^{F}_{(020)_a}v^5-(\chi^F_{(102)_a\otimes(3)_b}+c.c.)v^6-\Big((\chi^F_{(200)_a\oplus(6)_b}+27\chi^F_{(4)_b}+c.c.)
  +\chi^{F}_{(400)_a\oplus(004)_a\oplus(121)_a}\Big)v^7\\
  +&\,(27\chi^F_{(100)_a\otimes(5)_b}+78\chi^F_{(001)_a\otimes(3)_b}-\chi^F_{((130)_a\oplus(203)_a)\otimes(3)_b}+c.c.)v^8\\
  -&\,\Big((\chi^F_{(022)_a\otimes(6)_b}+351\chi_{(4)_b}-27\chi^F_{((030)_a\oplus(103)_a)\otimes(2)_b}+c.c)+\chi^F_{(222)_a}-78\chi^F_{(210)_a\oplus(012)_a}\Big)v^{9}+\mathcal{O}(v^{10}),
\end{aligned}
\end{equation}
or in the descending order of the absolute value of $\mf{u}(1)$ charge as
\begin{equation}\label{n2E6HS}
\begin{aligned}
  \sum_{n=0}^\infty\Big[-\chi^F_{(-12)_b\oplus(12)_b}\chi^{E_6}_{(00000n)}v^{11+2n}+(\chi^F_{(001)a\otimes(11)_b}\chi^{E_6}_{(10000n)}+c.c.)v^{12+2n}+\dots\Big].
\end{aligned}
\end{equation}

\subsubsection*{$\mathbf{n=1,\, G=E_6,\,F=\mf{su}(5)\times \mf{u}(1)}$}
We use Weyl orbit expansion to solve elliptic genus for this
theory. Let us first turn off the $\mf{su}(5)$ fugacities and only keep
$\mf{u}(1)$ and make use of the unity blowup equations with nonzero
$\lF$ only on $\mf{u}(1)$. Then the reduced one-string
elliptic genus with all gauge and flavor fugacities turned off can be
computed as
\begin{equation}\label{n1E6E1}
\mathbb{E}_{h_{1,E_6}^{(1)}}(\Qtau,v,m_{E_6}=0,m_F=0)=\Qtau^{-1/3}+\Qtau^{2/3}v^{-2}\sum_{n=0}^{\infty}\Qtau^n\frac{P_n(v)}{ (1 + v)^{22}},
\end{equation}
where
\begin{align}
  P_0(v)=
  &\,78 + 1446 v + 12182 v^2 + 60108 v^3 + 180534 v^4 + 260152 v^5 -
    365242 v^6 - 3157324 v^7\nn
  & - 9013936 v^8 - 13246110 v^9 +
    3729696 v^{10} + 83186464 v^{11} + 255829040 v^{12}\nn
  & + 405233216 v^{13} +
    \dots + 78 v^{26},\nn
  P_1(v)=
  &\,v^{-2}(2430 + 36180 v + 222432 v^2 + 630204 v^3 + 69266 v^4 - 5565632 v^5 -  17594496 v^6\nn
  -&\, 11700192 v^7 + 74362142 v^8 + 245593684 v^9 +
     202313896 v^{10} - 730064340 v^{11} - 2618359266 v^{12}\nn
  -&\, 2448587624 v^{13} + 5677163436 v^{14} + 16560265456 v^{15} +
     \dots + 2430 v^{30}).
\end{align}
We have cross-checked our result against the modular ansatz in
\cite{DelZotto:2018tcj}.\footnote{In \cite{DelZotto:2018tcj}, the
  modular ansatz for this theory is determined up to three unfixed
  parameters. Using our result from blowup equations, we are able to
  determine their three unfixed parameters as \be
  a_1=-\frac{14389465}{359414999291950792704},a_2=-\frac{227027173}{11501279977342425366528}
  ,a_3=\frac{146734631}{34503839932027276099584}.
\end{equation}} Let us further denote
\begin{equation}
  \mathbb{E}_{h_{1,E_6}^{(1)}}(\Qtau,v,m_{E_6}=0,m_F=0)=\Qtau^{-1/3}\sum_{i,j}c_{i,j} v^j(\Qtau/v^2)^i.
\end{equation}
Then we have the following table \ref{tb:n1E6one} for the coefficients
$c_{ij}$. Note the red numbers in the first column are just the
dimensions of representations ${k\theta}$ of $E_6$ where $\theta$ is
the adjoint representation. The blue numbers in the second column are
10 times of the dimensions of representations $\square+k\theta$ of
$E_6$, where the 10 is the double of the dimension of matter
representation $\bf 5$ of flavor $\mf{su}(5)$. The orange number 104
in the third column is given by
$\mathrm{dim}(E_6)+\dim(\mf{su}(5)\times
\mf{u}(1))+1=78+25+1=104$. These are the constraints predicted in
\cite{DelZotto:2018tcj} by analyzing the spectral flow to
Neveu-Schwarz-Ramond elliptic genus.
\begin{table}[h]
 \begin{center}
\begin{small}
\begin{tabular}{c| cccccccccccccccc }$i,j$&  0 & 1 & 2 & 3 & 4 & 5 & 6 & 7 & 8 \\
\hline
0& \color{red}1& 0& 0& 0& 0& 0& 0& 0& 0\\
1 & \color{red}78& \color{blue}-270& \color{orange}104& 70 & 200 & 420 & 1124 & 5220 & 3468  \\
2& \color{red}2430& \color{blue}-17280 & 41262 & -28080 & -8746 & -18640 & -10490 &
 7680 & 35296  \\
3 & \color{red}43758& \color{blue}-463320 & 1999296 & -4254770 & 3930732 & -200322 & -14660 & -1987042 & -3198410\\
\hline
\end{tabular}
\end{small}
\caption{Series coefficients $ c_{i,j} $ for the one-string elliptic genus of $\fn =1$ $E_6$ model.}\label{tb:n1E6one}
\end{center}
\end{table}

Let us also show some results with all flavor
$\mf{su}(5)_a\times \mf{u}(1)_b$ fugacities turned on. For example, the
$\Qtau$ subleading order of reduced one-string elliptic genus with
$m_{E_6}=0$ is
\begin{equation}\nonumber
  \begin{aligned}
    {78}&\,v^{-2}-(27\chi^F_{(1000)_a\oplus(1)_b}+c.c.)v^{-1}+\chi^F_{(1001)_a}+80+(\chi^F_{(3000)_a\oplus(3)_b}+c.c.)v\\
    &\ \,+\chi^F_{(2002)_a}v^2+(\chi^F_{(0201)_a\oplus(3)_b}+c.c.)v^3+\mathcal{O}(v^4),
\end{aligned}
\end{equation}
or in the descending order of the absolute value of $\mf{u}(1)$ charge as
\begin{equation}\label{n1E6exactE1}
  \begin{aligned}
    \sum_{n=0}^\infty\Big[\chi^F_{(-15)_b\oplus(15)_b}\chi^{E_6}_{(00000n)}v^{11+2n}-(\chi^F_{(0001)a\otimes(14)_b}\chi^{E_6}_{(10000n)}+c.c.)v^{12+2n}+\dots\Big].
  \end{aligned}
\end{equation}

\subsection{$E_7$ theories}\label{E7theories}
$G=E_7$ theories on base curve $(-n)$ have flavor symmetry
$F=\mf{so}(8-n)$ and $n_f=(8-n)/2$ hypermultiplets in
bi-representation $\frac{1}{2}({\bf 56},{\bf
  8-n})$. Note ${\bf 8-n}$ is the fundamental representation of flavor
group and $n=1,2,3,\dots,7,8$. There are $n$ vanishing blowup
equations with $\lF=0$. Using the fact that the minimal Weyl orbit of
$(P^\vee\backslash Q^{\vee})_{E_7}$ consists just of
weights of $\bf 56$, it is easy to find the leading base degree of the
vanishing blowup equations can be written as
\begin{equation}\label{E7vanish0}
  \sum_{w\in{\bf 56}}(-1)^{|w|}\theta_i^{[a]}
  (n\tau,nm^{E_7}_w+(8-n)\epsilon_+)
  \prod_{w'\in\mathbf{8-n}}\theta_1(m^{E_7}_w+m_{w'}^{\mf{so}(8-n)}-\epsilon_+)
  \prod_{\displaystyle \substack{\alpha\in \Delta(E_7)}}^{w\cdot
      \alpha= 1}
  \frac{1}{\theta_1(m^{E_7}_\alpha)}=0,
\end{equation}
which we have checked to be correct up to $\Qtau^{20}$ for all
$n$. Note these identities contain $m_{E_7}$, $m_{\mf{so}(8-n)}$ and
$\epsilon_+$ as free parameters, thus are highly nontrivial. By
setting $m_{\mf{so}(8-n)}$ as zero, one obtains
\begin{equation}\label{E7vanish00}
  \sum_{w\in{\bf 56}}(-1)^{|w|}\theta_i^{[a]}
  (n\tau,nm_w+(8-n)\epsilon_+)\big(\theta_1(m_w-\epsilon_+)\big)^{8-n}
  \prod_{\displaystyle \substack{\alpha\in \Delta(E_7)}}^{w\cdot
      \alpha= 1}
  \frac{1}{\theta_1(m_\alpha)}=0,
\end{equation}
which may be easier in case interested readers want to give a direct
proof.

The unity blowup equations for $G=E_7$ theories only exist for even
$n$, because for odd $n$ the theory involves half-hyper. In the
following, we discuss the cases $n=7,6,4,2$ individually.

\subsubsection*{$\mathbf{n=7,\, G=E_7}$}
This theory is a minimal SCFT with a half-hypermultiplet in $\bf
56$. The associated Calabi-Yau geometry was constructed in
\cite{Haghighat:2014vxa} by slightly modifying the $\hat{E}_7$
resolution on $\mathcal{O}(-8)\to \IP^1$ as $\mathcal{O}(-7)$. We
provide a non-compact toric construction in (\ref{n7E7polytope}),
using which we computed the triple intersection numbers and the genus
zero Gromov-Witten invariants.

There are in total seven non-equivalent vanishing blowup equations for
this theory, which can be written as
\begin{equation}\label{n7vanish}
\begin{aligned}
  \sum_{\lambda\in (P^\vee\backslash Q^\vee)_{E_7}}^{d_{1,2}\ge 0}&(-1)^{|\lambda|}\theta_4^{[a]}\big(7\tau,-7m_{\lambda}-(7d_0-1/2)(\eq+\et)-7d_1\eq-7d_2\et\big)A^{E_7}_{V}(m,\lambda)A_{H}^{\frac{1}{2}\bf 56}(m,\lambda)\\[-2mm]
  &\phantom{--}\times\IE_{d_1}(\tau,m+\eq\lambda,\eq,\et-\eq)\IE_{d_2}(\tau,m+\et\lambda,\eq-\et,\et)=0,
\end{aligned}
\end{equation}
where $\lambda\cdot\lambda/2=d_0+3/4$, $\,d_0\in\IZ$ and $a=1/2-i/7$,
$i=1,2,\dots,7$. In base degree $d$ expansion, the numbers of
$\lambda$ one needs to sum over and the Weyl orbits they
lie in are summarized in the following table:
\begin{center}
  \begin{tabular}{c|c|c|c|c|c}
    \hline
    $d$  & 0 & 1 & 2 & 3 & $\cdots$ \\
    \hline
    Weyl orbit & [0000010] & [0000001] & [1000010] & [0000011]  & $\cdots$ \\
    \hline
    $\#\{\lambda\cdot\lambda/2=d+3/4\}$ & 56 & 576 & 1512  & 4032  & $\cdots$ \\
    \hline
  \end{tabular}\\
\end{center}
For example, for the leading base degree, i.e. $d_0=d_1=d_2=0$,
$\lambda$ are just all the weights of fundamental representation
$\bf 56$. Thus we have the following nontrivial identity:
\begin{equation}\label{n7vanish0}
  \sum_{w\in{\bf 56}}(-1)^{|w|}\theta_4^{[a]}(7\tau,-7m_w+\epsilon_+)\theta_1(m_w+\epsilon_+)\prod_{\alpha\in \Delta(E_7) }^{w\cdot \alpha=1}\frac{1}{\theta_1(m_\alpha)}=0.
\end{equation}
For higher base degrees, we study the vanishing blowup equations from the
viewpoint of local Calabi-Yau geometry. We find the seven vanishing
blowup equations with the input of prepotential $F_{(0,0)}$ can
determine most of the refined BPS invariants, although not all of
them. We list some refined BPS invariants solved from blowup equations
in Table \ref{tb:n7, E_7-BPS} in Appendix \ref{app:bps}.

\subsubsection*{$\mathbf{n=6,\, G=E_7,\,F=\mf{so}(2)}$}
There are 12 unity blowup equations with $\lF=(\pm 1)$. Using the
recursion formula, we computed the one-string elliptic genus with
flavor fugacities turned off to $\mathcal{O}(\Qtau^1)$. Our result
agrees precisely with the modular ansatz in \cite{DelZotto:2018tcj},
therefore we just present the first few $\Qtau$ orders. Denote the
reduced one-string elliptic genus as
\begin{equation}\label{n6E7E1}
  \mathbb{E}_{h_{6,E_7}^{(1)}}(\Qtau,v,m_{E_7}=0,m_{\mf{so}(2)}=0)=\Qtau^{-11/6}v^{15}\sum_{n=0}^{\infty}\Qtau^n\frac{P_n(v)}{(1 - v)^{22} (1 + v)^{34}},
\end{equation}
We obtain
\begin{equation}\nonumber
\begin{aligned}
  P_0&(v)=-(2 + 24 v - 43 v^{2 }+ 52 v^{3 }+ 8027 v^{4 }- 53360 v^{5
  }+ 279039 v^{6 }-
  950972 v^{7 }+ 2698740 v^{8 }\\
  &- 5898532 v^{9 }+ 10988680 v^{10 }- 16600348 v^{11 }+ 21616127
  v^{12 }- 23243264 v^{13 }+\dots+v^{26}).
 \end{aligned}
\end{equation}
and
\begin{equation}\nonumber
\begin{aligned}
  P_1(v)=\,&v^{-2}(1 + 12 v - 226 v^{2 }- 3284 v^{3 }+ 8157 v^{4 }+
  28752 v^{5 }- 1098207 v^{6 }+
  6964508 v^{7 }\\
  &- 32103023 v^{8 }+ 103825488 v^{9 }- 273840598 v^{10 }+
  575865704 v^{11 }- 1024745731 v^{12 }\\
  &+ 1517074676 v^{13 }-
  1931373701 v^{14 }+ 2077804192 v^{15 }+\dots+v^{30}),\\
  P_2(v)=\,&v^{-4}(-1 - 12 v + 91 v^2 + 1776 v^3 - 10620 v^4 - 236256
  v^5 + 594632 v^6 + 4166640 v^7 \\&- 76480778 v^8 + 449325704 v^9 -
  1870890749 v^{10} + 5714898268 v^{11} - 14169525888 v^{12}\\& +
  28626262964 v^{13} - 49011331352 v^{14} + 70988810780 v^{15} -
  88777609823 v^{16}\\& +
  95280766576 v^{17} - 88777609823 v^{18}+\dots-v^{34}).\\
 \end{aligned}
\end{equation}
With gauge and flavor fugacities turned on, we confirm the conjectural
exact formula for the reduced 5d one-instanton partition function in
(H.40) of \cite{DelZotto:2018tcj}. For example, the leading $q$ order
of (\ref{n6E7E1}) is
\begin{equation}
\begin{aligned}
  -\,&\chi_{(2)\oplus(-2)}^Fv^{15}-(\chi_{(6)\oplus(-6)}^F-\mathbf{133})v^{17}-(\mathbf{912}\cdot\chi_{(1)\oplus(-1)}^F-\mathbf{56}\cdot\chi_{(5)\oplus(-5)}^F)v^{18}\\
  &\,+(\mathbf{8645}-\mathbf{133}\cdot\chi_{(6)\oplus(-6)}^F-\mathbf{1539}\cdot\chi_{(4)\oplus(-4)}^F)v^{19}+\mathcal{O}(v^{20}),
\end{aligned}
\end{equation}
and the subleading $q$ order is
\begin{equation}
\begin{aligned}
  \,&v^{13}-(\mathbf{133}+2)\chi_{(4)\oplus(-4)}^Fv^{15}+(-(\mathbf{133}+2)\chi_{(6)\oplus(-6)}^F+\mathbf{1539}\cdot\chi_{(2)\oplus(-2)}^F\\
  &\,+
  \mathbf{8645}+\mathbf{7371}+\mathbf{1539}+3\cdot\mathbf{133}+1)v^{17}+\mathcal{O}(v^{18}).
\end{aligned}
\end{equation}

\subsubsection*{$\mathbf{n=4,\, G=E_7,\,F=\mf{so}(4)}$}
There are 16 unity blowup equations with $\lambda_F=(\pm 1,\pm1)$ if
we regard $\mf{so}(4)\cong \mf{su}(2)\times \mf{su}(2)$. Using the
recursion formula, we computed the one-string elliptic genus with
flavor fugacities turned off to $\mathcal{O}(\Qtau^4)$. Denote the
reduced one-string elliptic genus as
\begin{equation}\label{n4E7E1}
  \mathbb{E}_{h_{4,E_7}^{(1)}}(\Qtau,v,m_{E_7}=0,m_{\mf{so}(4)}=0)=\Qtau^{-5/6}v^{11}\sum_{n=0}^{\infty}\Qtau^n\frac{P_n(v)}{(1 - v)^{10} (1 + v)^{34}},
\end{equation}
We obtain
\begin{equation}\label{n4E7E1P0}\nonumber
\begin{aligned}
  P_0&(v)=-(1 + 24 v + 305 v^2 + 2720 v^3 + 14385 v^4 + 10328 v^5 -
  213107 v^6 + 227936 v^7\\ & + 3681535 v^8 - 15349240 v^9 + 32121373
  v^{10} - 40005232 v^{11} + 32121373 v^{12}+\dots+v^{22}).
 \end{aligned}
\end{equation}
\begin{equation}\label{n4E7E1P1}\nonumber
\begin{aligned}
  P_1(v)=\,&v^{-2}(9 + 216 v + 2296 v^2 + 13704 v^3 + 35681 v^4 -
  191536 v^5 -
  2195202 v^6 - 3469024 v^7\\
  & + 34360924 v^8 + 12656096 v^9 -
  543596903 v^{10} + 1892316824 v^{11} - 3595032965 v^{12}\\
  & + 4390454000 v^{13} +\dots+9v^{26}).
 \end{aligned}
\end{equation}
The leading $q$ order exactly agrees with the reduced 5d one-instanton
partition function in (A.20) of \cite{Kim:2019uqw}.
We record our higher order results for in Appendix~\ref{app:D} in
equations (\ref{n4E7E1P2},\ref{n4E7E1P3},\ref{n4E7E1P4}). Let us
denote
\begin{equation}
  \mathbb{E}_{h_{4,E_7}^{(1)}}(\Qtau,v,m_{E_7}=0,m_{\mf{so}(4)}=0)
  =\Qtau^{-5/6}v^{11}\sum_{i,j}c_{i,j} v^j(\Qtau/v^2)^i.
\end{equation}
Then we have the following Table \ref{tb:n4E7one} for the coefficients
$c_{ij}$. Note the red numbers in the first column are just the
dimensions of representations ${k\theta}$ of $E_7$ where $\theta$ is
the adjoint representation. The blue numbers in the second column are
four times the dimensions of representations $\square+k\theta$ of
$E_7$, where the four is the dimension of matter representation
$\bf 4$ of flavor $\mf{so}(4)$. The orange number 140 in the third
column is given by
$\mathrm{dim}(E_7)+\dim(\mf{so}(4))+1=133+6+1=140$. These are the
constraints given in \cite{DelZotto:2018tcj} by analyzing the spectral
flow to Neveu-Schwarz-Ramond elliptic genus which our result satisfies
perfectly. By combining our result and the constraints from NSR
elliptic genus at even higher $q$ order, we are able to determine the
modular ansatz of $\mathbb{E}_{h_{4,E_7}^{(1)}}(\Qtau,v)$, which will
be given in the \texttt{Mathematica} file on the website \cite{kl}.
\begin{table}[h]
  \begin{center}
    \resizebox{\linewidth}{!}{
      \begin{tabular}{c| *{13}{>{$}c<{$}} }
        $i,j$& -10 & -9 &-8&-7&-6&-5&-4&-3 & -2 & -1 & 0 & 1 & 2\\
        \hline
        0& 0& 0& 0& 0& 0& 0& 0& 0& 0& 0& -1& 0& -39 \\
        1 & 0 & 0 & 0 & 0 & 0 & 0 & 0 & 0& 0& 0& 9& 0& -98 \\
        2& \color{red}1 & 0 & 0 & 0 & 0 & 0 & 0 & 0 & -9 & 0 & 0 & 0 & 1330 \\
        3& \color{red}133 & \color{blue}-224 & \color{orange}140 & 0 & 25 & 0 & 14 & 0 & -42 & 224 & -1463 & 0 & 0\\
        4& \color{red}7371 & \color{blue}-25920 & 41249 & -31360
                              & 10010 & -2688 & 3500 & 0 & 2050 &  2688 & -7419 & 31360 & -127480\\
        \hline
      \end{tabular}}
    \caption{Series coefficients $ c_{i,j} $ for the one-string
      elliptic genus of the $\fn =4$ $E_7$ model.}\label{tb:n4E7one}
\end{center}
\end{table}

If turning on all gauge $E_7$ and flavor
$\mf{so}(4)\cong \mf{su}(2)\times \mf{su}(2)$ fugacities, we find the
leading $\Qtau$ order of reduced one-string elliptic genus, i.e. the
reduced 5d Nekrasov partition function has the following expansion
\begin{equation}\nonumber
  \begin{aligned}
    \,&-v^{11}-\chi^{F}_{(60)\oplus(06)\oplus(44)}v^{13}+(\mathbf{133}\cdot\chi^{F}_{(42)\oplus(24)}-\chi^{F}_{(48)\oplus(84)})v^{15}\\
    \,&\phantom{--}-(\mathbf{912}\cdot\chi^{F}_{(33)}-\mathbf{56}\cdot\chi^{F}_{(73)\oplus(37)})v^{16}+\mathcal{O}(v^{17}),
  \end{aligned}
\end{equation}
which agrees with the (A.20) of \cite{Kim:2019uqw}. In fact, we find
an exact formula for the reduced 5d Nekrasov partition function which
will be given in Appendix (\ref{eq:exactn4E7Z1}). For the subleading
$\Qtau$ order we obtain the following expansion
\begin{equation}\nonumber
  \begin{aligned}
    \,&\chi_{(22)}^Fv^9+(\chi^F_{(26)\oplus(62)} - \chi^F_{(02)\oplus(20)}- 1-\mathbf{133})v^{11}-(\chi_{(64)\oplus(46)\oplus(80)\oplus(08)\oplus(62)\oplus(26)\oplus(40)\oplus(04)}^F\\
    &+(\mathbf{133}+3)\chi^F_{(44)}
    +(\mathbf{133}+2)\chi^F_{(60)\oplus(06)}+(\mathbf{133}+1)\chi^F_{(42)\oplus(24)})v^{13}
    +\mathcal{O}(v^{15}).
  \end{aligned}
\end{equation}

\subsubsection*{$\mathbf{n=2,\, G=E_7,\,F=\mf{so}(6)}$}
There are 16 unity blowup equations with $\lambda_F\in\mathbf{4}$ or
$\bar{\bf 4}$. Noticing the flavor symmetry
$\mf{so}(6)\cong \mf{su}(4)$, we can turn on the fugacity of a
sub-algebra $\mf{su}(2)$ to perform the computation on elliptic genus
easily. Using the Weyl orbit expansion method, we computed the
one-string elliptic genus with $\mf{su}(2)$ flavor fugacities to
$\mathcal{O}(\Qtau^2)$. For example, denote the reduced one-string
elliptic genus as
\begin{equation}\label{n2E7E1}
  \mathbb{E}_{h_{2,E_7}^{(1)}}(\Qtau,v,m_{E_7}=0,m_{\mf{so}(6)}=0)=\Qtau^{1/6}v^{-1}\sum_{n=0}^{\infty}\Qtau^n\frac{(1 - v)^2P_n(v)}{ (1 + v)^{34}},
\end{equation}
we obtain
\begin{align}
  P_0(v)=
  &(1 + 36 v + 632 v^2 + 7212 v^3 + 60168 v^4 + 391380 v^5 +
    2067496 v^6 + 9123228 v^7+ 34335094 v^8\nn
  &  + 111995836 v^9 + 320744719 v^{10} + 815144896
    v^{11} + 1854166712 v^{12} +
    3796415104 v^{13}\nn
  & + 6997399845 v^{14} + 11475775012 v^{15} + 16204920073 v^{16} +
    18551114752 v^{17} + \dots + v^{34}),\nn
  P_1(v)=\,
  &v^{-2} (133 + 4452 v + 72109 v^2 + 752208 v^3 + 5673385
    v^4 +  32915460 v^5\nn
  & + 152504980 v^6 + 577794348 v^7 + 1815737068 v^8 + 4761819476
    v^9 + 10385374307 v^{10}\nn
  & + 18472471608 v^{11} + 25278998607
    v^{12} + 21455489108 v^{13} - 5924034231 v^{14}\nn
  & - 61899269488
    v^{15} - 122152636908 v^{16} - 122341883440 v^{17} -
    16307972890 v^{18}\nn
  & + 84187540856 v^{19} +\dots+ 133 v^{38}),\nn
  P_2(v)=\,
  &v^{-4} (7371 + 226476 v + 3331334 v^2 + 31148940 v^3 +
    207151332 v^4 + 1037448756 v^5\nn
  & + 4032702373 v^6 + 12310917456
    v^7 + 29294737640 v^8 + 52132350336 v^9 + 59789988702 v^{10}\nn
  & + 9188063128 v^{11} - 131878217677 v^{12} - 303150457484 v^{13} -
    293119312706 v^{14}\nn
  & + 124103122340 v^{15} + 762220055405 v^{16}
    + 700512400544 v^{17} - 879753089280 v^{18}\nn
  & - 2366194285936 v^{19} + 148716225866 v^{20} + 4344623389448 v^{21} +\dots+ 7371 v^{42}).\nonumber
\end{align}

If we turn on all gauge $E_7$ and flavor $\mf{su}(4)$ fugacities, we
find the leading $\Qtau$ order of reduced one-string elliptic genus
has the following expansion
\begin{equation}\nonumber
  \begin{aligned}
    \,&v^{-1}-(\chi^{\mf{su}(4)}_{(400)}+\chi^{\mf{su}(4)}_{(004)})v^7-\chi^{\mf{su}(4)}_{(222)}v^9-\mathbf{56}\cdot\chi^{\mf{su}(4)}_{(030)}v^{10}\\
    -(\chi^{\mf{su}(4)}_{(602)}+\,&\chi^{\mf{su}(4)}_{(206)}+\chi^{\mf{su}(4)}_{(323)}
    +\chi^{\mf{su}(4)}_{(060)}-\mathbf{133}\cdot\chi^{\mf{su}(4)}_{(121)}+\mathbf{1463})v^{11}+\mathbf{6480}\cdot\chi^{\mf{su}(4)}_{(010)}v^{12}+\mathcal{O}(v^{13}).
  \end{aligned}
\end{equation}
The subleading $\Qtau$ order has expansion as
\begin{equation}\nonumber
  \begin{aligned}
    \,&\mathbf{133}\,v^{-3}-\mathbf{56}\cdot\chi^{\mf{su}(4)}_{(010)}v^{-2}+(\mathbf{133}+\chi^{\mf{su}(4)}_{(101)}+1)v^{-1}+\chi^{\mf{su}(4)}_{(040)}v\\
    &\phantom{--}+\chi^{\mf{su}(4)}_{(303)}v^3
    +(\chi^{\mf{su}(4)}_{(420)}+\chi^{\mf{su}(4)}_{(024)})v^5+\mathcal{O}(v^{6}).
  \end{aligned}
\end{equation}

Let us further denote
\begin{equation}
  \mathbb{E}_{h_{2,E_7}^{(1)}}(\Qtau,v,m_{E_7}=0,m_{\mf{so}(6)}=0)=\Qtau^{1/6}\sum_{i,j}c_{i,j} v^j(\Qtau/v^2)^i.
\end{equation}
Then we have the following Table \ref{tb:n2E7one} for the coefficients
$c_{ij}$. Note the red numbers in the first column are just the
dimensions of representations ${k\theta}$ of $E_7$ where $\theta$ is
the adjoint representation. The blue numbers in the second column are
six times the dimensions of representations $\square+k\theta$ of
$E_7$, where the six is the dimension of the matter representation
$\bf 6$ of flavor symmetry $\mf{so}(6)$. The orange number 149 in the
third column is given by
$\mathrm{dim}(E_7)+\dim(\mf{so}(6))+1=133+15+1=149$. These are the
constraints given in \cite{DelZotto:2018tcj} by analyzing the
spectral flow to NSR elliptic genus, which our result satisfies perfectly. By
combining our result and the constraints from NSR elliptic genus at
even higher $q$ order, we are able to determine the modular ansatz of
$\mathbb{E}_{h_{2,E_7}^{(1)}}(\Qtau,v)$, which will be given in the \texttt{Mathematica} file on the website \cite{kl}.
\begin{table}[h]
 \begin{center}
\resizebox{\linewidth}{!}{
  \begin{tabular}{c| *{11}{>{$}c<{$}} }
    $i,j$& -1 & 0 & 1&2&3&4&5&6 & 7 & 8 & 9  \\
    \hline
0& \color{red}1 & 0 & 0 & 0 & 0 & 0 & 0 & 0  & -70 & 0 & -729\\
1& \color{red}133 & \color{blue}-336 & \color{orange}149 & 0 & 105 & 0 & 300 & 0& 720& 7840& -20777\\
2& \color{red}7371 & \color{blue}-38880 & 72542& -50064 & 11324 & -21504 & 15645 & -30240 & 47340 & -146106 & 1938800\\
\hline
\end{tabular}}
\caption{Series coefficients $ c_{i,j} $ for the one-string elliptic
  genus of the $\fn =2$ $E_7$ model.}\label{tb:n2E7one}
\end{center}
\end{table}

\section{Three higher rank non-Higgsable clusters}
\label{sec:3nhc}
The three non-Higgsable clusters in Table \ref{tb:threeNHC} are the
simplest higher-rank 6d $(1,0)$ SCFTs and building blocks for more
complicated higher-rank theories \cite{Morrison:2012np}.
\begin{table}[h]
\centering
\begin{tabular}{|c|c|c|c|}
  \hline
  base
  & $3,2$
  & $3,2,2$
  & $2,3,2$\\
  \hline
  gauge symmetry &  $G_2\times \mf{su}(2)$
  & $G_2\times \mf{su}(2)\times\{\ \}$
  & $\mf{su}(2)\times \mf{so}(7)\times \mf{su}(2)$ \\
  \hline matter  &$\frac{1}{2}({\bf 7}+{\bf 1},{\bf 2})$
  & $\frac{1}{2}({\bf 7}+{\bf 1},{\bf 2})$
  & $\frac{1}{2}({\bf 2},{\bf 8},{\bf 1})+
    \frac{1}{2}({\bf 1},{\bf 8},{\bf 2})$ \\
  \hline
\end{tabular}
\caption{Three higher-rank NHCs.}\label{tb:threeNHC}
\end{table}
The 2d quiver gauge theories corresponding to these three NHCs have
been constructed in \cite{Kim:2018gjo}. Using Jeffrey-Kirwan residue,
the elliptic genera can be explicitly computed as formulae involving
Jacobi theta functions. It is interesting to see how blowup equations
work for these higher rank theories. The most prominent feature here
is that there only exist \emph{vanishing} blowup equations for these
three NHCs. We study them with two approaches: from the viewpoint of
gauge theory, to which end we derive the \emph{higher-rank elliptic
  blowup equations}, and from the viewpoint of \localelliptic
Calabi-Yau. In particular, we give the toric constructions for the
\localelliptic Calabi-Yau threefolds associated with the three NHCs,
which to our knowledge are new.

We first introduce a special kind of higher dimension Riemann theta
function associated to a $N\times N$ matrix $\Omega$. It turns out that
the polynomial part of the higher rank 6d theories contributes to the
blowup equation as this type of Riemann theta function after
de-affinization. We define
\begin{equation}\label{Rthetawitha}
  \Theta^{[a]}_{\Omega}(\tau,z)=\sum_{k\in\IZ^N+a}
  (-1)^{k\cdot\mathrm{diag}(\Omega)}\exp\(\frac{1}{2}k\cdot\Omega\cdot k\,
\tau+k\cdot\Omega\cdot z
\).
\end{equation}
Here the characteristic $a=(a_1,a_2,\dots,a_N)$ takes the following values
\begin{equation}\label{generaldefa}
  a_i = \sum_{j}(\Omega^{-1})_{ij} \Big(\frac{1}{2}\Omega_{jj}+
  m_j\Big),
  \quad m_j\in\IZ.
\end{equation}
The number of different such characteristics is $\Det(\Omega)$. This
kind of Riemann theta function is the proper generalization of
$\theta_i^{[a]}(n\tau,nz)$ appearing in rank one elliptic blowup
equations. As in the rank one cases, when the
characteristic $a$ is trivial, we suppress the superscript
$\Theta_\Omega = \Theta_\Omega^{[0]}$.

\subsection{NHC $3,2$}
\label{sec:32}

\subsubsection*{Elliptic genera from quiver gauge theory}
The elliptic genera of the NHC $3,2$ can be obtained from the elliptic
genera $\IE_{k_1,k_2,k_3}$ of the NHC $2,3,2$, which are given in
\cite{Kim:2018gjo}, via decompactification and Higgsing
mechanism. Recall that the NHC $2,3,2$ carries the product of gauge
groups $\mf{su}(2)_1\times \mf{so}(7)\times \mf{su}(2)_2$ over the
three compact curves, and the matter content is
$\tfrac{1}{2}(\md 1,\md 8, \md 2) + \tfrac{1}{2}(\md 2,\md 8, \md
1)$. In the formula for $\IE_{k_1,k_2,k_3}$ given in
\cite{Kim:2018gjo}, the gauge fugacities of
$\mf{su}(2)_1,\mf{so}(7), \mf{su}(2)_2$ are respectively
$\nu, v_\ell, \tilde{\nu}$ for $\ell=1,\ldots,4$ with the constraint
$\sum_\ell v_\ell = 0$.

We first decompactify the first $(-2)$ curve, and arrive at the
$(-3,-2)$ geometry with gauge group $\mf{so}(7)\times \mf{su}(2)_2$
and matter content $\tfrac{1}{2}(\md 8, \md 2) + (\md 8, \md 1)$. The
gauge group $\mf{su}(2)_1$ becomes flavor symmetry after
decompactification. The hyper $(\md 8, \md 1)$ is nontrivially charged
under this flavor symmetry, and its mass is $\nu$. This step
corresponds to setting $k_1 = 0$ in the elliptic genus
$\IE_{k_1,k_2,k_3}$. Then following the discussion on page 18 of
\cite{Kim:2018gjo}, we can use the hyper $(\md 8, \md 1)$ to Higgs
$\mf{so}(7)$ down to $G_2$. Given the branching rule
$\md 8 \to \md 7+\md 1$ for $G_2\subset \mf{so}(7)$, we can give vev
to the hyper in $\md 1$, which becomes massive and decouples in the
IR. The hypers in $\md 7$ get eaten by $7$ copies of vector multiplets
from the adjoint $\md{21}$ of $\mf{so}(7)$, which also become massive
and decouple. The remaining $14$ copies of vector multiplets form the
adjoint of $G_2$. The hypers in $\tfrac{1}{2}(\md 8, \md 2)$ decompose
to $\tfrac{1}{2}(\md 7+\md 1,\md 2)$, which is precisely the matter
content of the $(-3,-2)$ NHC. This step of Higgsing is realised by
\cite{Kim:2018gjo} $\nu = \epsilon_+, v_4 = 0$ in the elliptic
genus. In the end, we find the elliptic genus $\IE_{k_1,k_2}$ of the
NHC $3,2$ should be ($k_1$ degree on $(-3)$ curve, $k_2$ degree on
$(-2)$ curve)
\begin{align}
  (-1)^{k_1+k_2}
  &\IE_{k_1,k_2} (\tau,v_\ell,\tilde{\nu})=\nn
    \sum_{\substack{Y_{(1,2)}\\|Y_{(a)}|=k_a}}
  &\prod_{i=1}^3
    \prod_{s_1\in Y_{(1)_i}}
    \frac{\theta(2\phi(s_1))\theta(2\phi(s_1)-2\epsilon_+)
    \theta(\tilde{\nu}\pm\phi(s_1))}
    {\left(\prod_{j=1}^3\theta(E_{ij}(s_1))\theta(E_{ij}(s_1)-2\epsilon_+)
    \theta(\epsilon_+-\phi(s_1)-v_j) \right) \theta(\epsilon_+-\phi(s_1))}
    \nonumber\\[-3mm]
  &\times\prod_{i\leq j}^3
    \prod_{\substack{s_1,\tilde{s}_1\in Y_{(1)_{i,j}}\\ s_1<\tilde{s}_1}}
  \frac{\theta(\phi(s_1)+\phi(\tilde{s}_1))
  \theta(\phi(s_1)+\phi(\tilde{s}_1)-2\epsilon_+)}
  {\theta(\epsilon_{1,2}-\phi(s_1)-\phi(\tilde{s}_1))}\nonumber \\[-3mm]
 \times \prod_{i=1}^3\prod_{j=1}^2&\prod_{s_1\in Y_{(1)_i}}\prod_{s_2\in Y_{(2)_j} }
    \frac{\theta(\epsilon_-\pm(\phi(s_1)-\phi(s_2)))}
    {\theta(\epsilon_+\pm(\phi(s_1)-\phi(s_2)))}
    \cdot\prod_{i=1}^2\prod_{s_2\in Y_{(2)_i}}
    \frac{\theta(\phi(s_2))\prod_{\ell=1}^3\theta(v_\ell-\phi(s_2))}
    {\prod_{j=1}^2\theta(E_{ij}(s_2))\theta(E_{ij}(s_2)-2\epsilon_+)}.
    \label{eq:EG-32}
\end{align}
Here $Y_{(1)}, Y_{(2)}$ are tuples of three and two Young diagrams
respectively, and $|Y_{(a)}| = k_a$ means the total number of boxes in
the tuple of Young diagrams is $k_a$. $v_\ell$ and $\tilde{\nu}$ are
fugacities of $G_2$ and $\mf{su}(2)$ with the constraint
$\sum_{\ell=1}^3 v_\ell = 0$. We define that for
$s_a = (m_a,n_a) \in Y_{(a)_i}$
\begin{align}
  &E_{ij}(s_a) = v_{(a)_i} - v_{(a)_j} - \eq h_i(s_a) + \et(v_j(s_a)+1)
    \ , \\
  &\phi(s_a) = v_{(a)_i} - \epsilon_+ - (n_a-1)\eq - (m_a-1)\et \ .
\end{align}
Here $h_i(s_a)$ denotes the distance from $s_a$ to the right end of
the diagram $Y_{(a)_i}$, and $v_j(s_a)$ denotes the distance from
$s_a$ to the bottom of the \emph{diagram $Y_{(a)_j}$}. Concretely
\begin{equation}
  h_i(s_a) = Y_{(a)_i}(m) - n \ ,\quad v_j(s_a) = Y^t_{(a)j}(n) - m \
  ,\quad s = (m,n) \ .
\end{equation}
Besides, $s_i<s_j$ means $(i<j)$ or $(i=j, m_i<m_j)$ or
$(i=j, m_i = m_j, n_i<n_j)$. We also use the notation
$v_{(1)_{1,2,3}} = (v_1,v_2,v_3)$ and
$v_{(2)_{1,2}} = \pm \tilde{\nu}$. In the derivation of the Higgsing
formula, we need to use the identity that
\begin{equation}
  \prod_{s\in Y_i} \theta(E_{i4}(s))\theta(E_{i4}(s)-2\epsilon_+)
  = -\prod_{s\in Y_i} \theta(\epsilon_+\pm \phi(s))) \ ,\quad
  \text{with}\; v_4 = 0,\; Y_4 = \emptyset, \;i=1,2,3
  \ ,
\end{equation}
which can be easily proved by a change of summation order in
$s\in Y_i$.

We checked that indeed when $k_2 = 0$, \eqref{eq:EG-32} reduces to the
elliptic genus $\IE_{k_1}$ of $G_2$ with one hyper in $\md 7$ over
$(-3)$ curve \cite{Kim:2018gjo}, where $\tilde{\nu}$ is identified
with the flavor mass. Furthermore the genus zero GV invariants with base
degrees $(1,0)$, $(0,1)$, $(1,1)$ agree with mirror symmetry calculations.

\subsubsection*{Elliptic blowup equations}
The geometric construction of this NHC has been given in
(\ref{NHC32polytope}). Let us denote the intersection matrix between
the two base curves as $-\Omega$, i.e.
\begin{equation}
  \Omega=\left(
    \begin{array}{cc}
      3 & -1 \\
      -1 &  2 \\
    \end{array}
  \right).
\end{equation}
Note $\det\Omega=5$. It turns out there are in total five vanishing
$\lF$ fields and no unity $\lF$ fields. To see this, one can simply
look at the matter representation $({\bf 7+1},{\tfrac{1}{2}\bf
  2})$. Note $G_2$ can only bear unity equations due to the Lie
algebra fact $P^\vee\cong Q^{\vee}$. On the other hand, the unpaired
half-hyper on the $\mf{su}(2)$ node indicates only vanishing
equations. Thus combining unity and vanishing equations naturally
results in vanishing equations. The idea can be roughly expressed as
\begin{equation}
  \rm{U}\star\rm{V}=\rm{V}.
\end{equation}

To derive the elliptic blowup equations, we apply the similar
\emph{de-affinization} procedure \cite{Gu:2018gmy} as in the rank one
cases. As there are two base curves and each has a nontrivial gauge
symmetry, now we have \emph{two} degeneration directions. From the
polynomial part given in (\ref{32triple}), we can compute the
coefficients of $t_{{\rm ell}_1}$ and $t_{{\rm ell}_2}$ in the blowup
equations as
\begin{equation}
  P_1(n_0,n_1,n_2)=P_{\widehat{G}_2}(n_0,n_1,n_2)=
  3n_1^2-3n_1n_2+n^2_2-n_0n_2+n^2_0,
\end{equation}
and
\begin{equation}
  P_2(n_3,n_4)=P_{\widehat{\mf{su}(2)}}(n_3,n_4)=n^2_3-2n_3n_4+n_4^2.
\end{equation}
It is important that $P_1$ and $P_2$ satisfy the following shift
invariance
\begin{equation}
\begin{aligned}
  P_1(n_0+k,n_1+k,n_2+2k)
  &=P_1(n_0,n_1,n_2),\quad\quad k\in\IZ,\\
  P_2(n_3+l,n_4+l) &=P_2(n_3,n_4),\quad\quad l\in\IZ.
\end{aligned}
\end{equation}
Besides, the $R$ shifts satisfy
\begin{equation}
  R(n_0+k,n_1+k,n_2+2k,n_3+l,n_4+l)-R(n_0,n_1,n_2,n_3,n_4)=(-3k+l,k-2l,0,0,0,0),
\end{equation}
which is crucial for modularity. By careful calculations, we find
the polynomial part contributes to blowup equations as a special
Riemann theta function defined in (\ref{Rthetawitha}). The
contributions from vector and hypermultiplets remain invariant under
the shift. Finally, after careful calculations, the five vanishing
elliptic blowup equations can be written as
\begin{equation}\label{32bldeg}
  \begin{aligned}
    \sum_{\displaystyle
      \substack{\alpha^\vee\in Q_{G_2}^\vee, d_{1,2} \\
        d_0=\frac{1}{2}||\alpha^{\vee}||^2}}^{d_0+d_1+d_2=k_1}
    &\sum_{\displaystyle \substack{\lambda\in(P^\vee\backslash
        Q^\vee)_{\mf{su}(2)},
        d'_{1,2}\\
        d'_0+1/4=\frac{1}{2}||\lambda||^2}}^{d'_0+d'_1+d'_2=k_2}
    (-1)^{|\alpha^{\vee}|+|\lambda|}
    \Theta^{[a]}_{\Omega}\(\tau,\left(
      \begin{array}{c}
        \alpha^{\vee}\cdot m_{G_2}+(\bar{y}_1-d_0)(\eq+\et) -d_1\eq-d_2\et\\
        \lambda\cdot m_{\mf{su}(2)}+(\bar{y}_2-d_0')(\eq+\et) -d_1'\eq-d_2'\et\\
      \end{array}
    \right)\)\\[-2mm]
    &\times A_{V}^{G_2}(\alpha^{\vee},\tau,m_{G_2})
    A_{V}^{\mf{su}(2)}(\lambda,\tau,m_{\mf{su}(2)}) A^{({\bf 7+1},{\tfrac{1}{2}\bf
        2})}_{H}
    (\alpha^{\vee},\lambda,\tau,m_{G_2},m_{\mf{su}(2)})\\
    &\times \IE_{d_1,d'_1}
    \big(\tau,m_{G_2}-\eq\alpha^{\vee},m_{\mf{su}(2)}-\eq\lambda,\eq,\et-\eq\big)\\
    &\times \IE_{d_2,d'_2}
    \big(\tau,m_{G_2}-\et\alpha^{\vee},m_{\mf{su}(2)}-\et\lambda,\eq-\et,\et\big)=\,0,\qquad\qquad
    \text{for fixed}\;\; k_{1,2} \in \IZ_{\geq 0},
  \end{aligned}
\end{equation}
where the summation indices $d_{0,1,2},d'_{0,1,2}\in \IZ_{\geq 0}$.
The parameters $y_{1,2}$ are
$\bar{y}_1=3/5,\bar{y}_2=3/10$, and
\begin{equation}\label{eq:a32}
  a=\left(
    \begin{array}{c}
      a_k \\
      a_l \\
    \end{array}
  \right)=\left(
    \begin{array}{c}
      2j/5 \\
      -1/2+j/5 \\
    \end{array}
  \right),\quad\quad j=-2,-1,0,1,2.
\end{equation}
This is our starting point to prove the modularity. Note
$\bar{y}_1,\bar{y}_2$ satisfy the following relation
\begin{equation}
  \Omega \left(
    \begin{array}{c}
      \bar{y}_1 \\
      \bar{y}_2 \\
    \end{array}
  \right)=\left(
    \begin{array}{c}
      3/2 \\
      0 \\
    \end{array}
  \right)=\left(
    \begin{array}{c}
      \bar{y}_{\rm u} \text{ of rank one $n=3$ $G_2$ theory} \\
      \bar{y}_{\rm v} \text{ of rank one $n=2$ $\mf{su}(2)$ theory}\\
    \end{array}
  \right).
\end{equation}
It can be shown this is necessary to be consistent with the
established elliptic blowup equations for rank one theories when
decompactifying one of the base curves.

The leading base degree of the vanishing blowup equations,
i.e. $d_0=d_1=d_2=d_0'=d_1'=d_2'=0$ can be simply written
as\footnote{The $\mf{su}(2)$ vector multiplets do contribute to the
  blowup equation here. However, the contribution to each of the two
  terms can be factored out.}
\begin{equation}\label{nhc32vid}
\Theta^{[a]}_{\Omega}\(\tau,\left(\begin{array}{c}
6\epsilon_+/5\\
m_{\mf{su}(2)}+3\epsilon_+/5\\
\end{array}
\right)\)-\Theta^{[a]}_{\Omega}\(\tau,\left(\begin{array}{c}
6\epsilon_+/5\\
-m_{\mf{su}(2)}+3\epsilon_+/5\\
\end{array}
\right)\)=0.
\end{equation}
It is easy to check that the above identity is correct. For higher base
degrees, the vanishing blowup equations (\ref{32bldeg}) involve
nontrivial elliptic genera. We have checked them from the Calabi-Yau
setting to high degrees of K\"ahler classes. Besides, we find the five
vanishing blowup equations are not sufficient to solve all refined BPS
invariants. This is not surprising since vanishing blowup equations
give less constraints just like in the rank one theories.

\subsubsection*{Modularity}
The index of the elliptic genus $\IE_{k_1,k_2}$ is known to be
\begin{align}\label{32Eindex}
  \text{Ind}_{\IE_{k_1,k_2}} =
  &-\frac{(\eq+\et)^2}{4}(3k_1+2k_2)
    +\frac{\eq\et}{2}(3k_1^2+2k_2^2-2k_1k_2-k_1) \nn
  &+(-3k_1+k_2)\frac{({m},{m})_{G_2}}{2}
    +(-2k_2+k_1)\frac{({m},{m})_{\mf{su}(2)}}{2}\ .
\end{align}
Let us use this to prove the modularity of (\ref{32bldeg}). First, it
is easy to derive from the general theory of Riemann theta functions
that the index quadratic form of $\Theta^{[a]}_{\Omega}(\tau,z)$ under
special modular transformation $\tau\rightarrow-1/\tau$ is just
\begin{equation}
\frac{1}{2}z\cdot\Omega\cdot z.
\end{equation}
This fact is useful when computing the index of the polynomial
contribution. Indeed, the index of polynomial part in (\ref{32bldeg})
is
\begin{equation}\nonumber
\begin{aligned}
  {\rm{Ind}}_{\rm poly}=&\,\frac{3}{2}(\alpha^{\vee}\cdot
  m_{G_2}+(y_1-d_0)(\eq+\et) -d_1\eq-d_2\et)^2\\
&+(\lambda\cdot m_{\mf{su}(2)}+(y_2-d_0')(\eq+\et) -d_1'\eq-d_2'\et)^2\\
-(\alpha^{\vee}\cdot m_{G_2}+(y_1-d_0)&(\eq+\et) -d_1\eq-d_2\et)(\lambda\cdot m_{\mf{su}(2)}+(y_2-d_0')(\eq+\et) -d_1'\eq-d_2'\et).
\end{aligned}
\end{equation}
The $G_2$ vector multiplet contributes to the index as
\begin{equation}\nonumber
\begin{aligned}
{\rm{Ind}}_{V}^{G_2}=&-\frac{5}{3}\Big((\alpha^{\vee}\cdot m_{G_2})^2+d_0m_{G_2}\cdot m_{G_2}\Big)+\frac{2}{3}(5d_0-2)(\eq+\et)(\alpha^{\vee}\cdot m_{G_2})\\
&-\frac{1}{3}(5d_0^2-2d_0)(\epsilon_1^2+\eq\et+\epsilon_2^2).
\end{aligned}
\end{equation}
and the $\mf{su}(2)$ vector multiplet contributes to the index as
\begin{equation}\nonumber
\begin{aligned}
{\rm{Ind}}_{V}^{\mf{su}(2)}=&-\frac{4}{3}\Big((\lambda\cdot m_{\mf{su}(2)})^2+(d_0'+\frac{1}{4})m_{\mf{su}(2)}\cdot m_{\mf{su}(2)}\Big)+\frac{8}{3}d_0'(\eq+\et)(\lambda\cdot m_{\mf{su}(2)})\\
&-\frac{1}{3}(4{d_0'}^2+d_0')(\epsilon_1^2+\eq\et+\epsilon_2^2).
\end{aligned}
\end{equation}
The hypermultiplet in the representation
$({\bf 7+1},{\tfrac{1}{2}\bf 2})$ contributes to the index as
\begin{equation}\nonumber
\begin{aligned}
{\rm{Ind}}_{H}^{({\bf 7+1},{\tfrac{1}{2}\bf 2})}&
=\frac{1}{4}\bigg(\frac{2}{3}\Big((\alpha^{\vee}\cdot m_{G_2})^2+d_0m_{G_2}\cdot m_{G_2}\Big)+2d_0m_{\mf{su}(2)}\cdot m_{\mf{su}(2)}+2(d_0'+\frac{1}{4})m_{G_2}\cdot m_{G_2}\\
&+\frac{4}{3}\Big((\lambda\cdot m_{\mf{su}(2)})^2+(d_0'+\frac{1}{4})m_{\mf{su}(2)}\cdot m_{\mf{su}(2)}\Big)+4(\alpha^{\vee}\cdot m_{G_2})(\lambda\cdot m_{\mf{su}(2)})\bigg)\\
&-\frac{1}{8}(m_{G_2}\cdot m_{G_2}+2m_{\mf{su}(2)}\cdot m_{\mf{su}(2)})-\frac{1}{6}\bigg(2d_0\alpha^{\vee}\cdot m_{G_2}+6(d_0'+\frac{1}{4})\alpha^{\vee}\cdot m_{G_2}+6d_0\lambda\cdot m_{\mf{su}(2)}\\
&+4(d_0'+\frac{1}{4})\lambda\cdot m_{\mf{su}(2)}\bigg)+\frac{1}{12}(\alpha^{\vee}\cdot m_{G_2} +2\lambda\cdot m_{\mf{su}(2)})+\cdots.
\end{aligned}
\end{equation}

Using (\ref{32Eindex}), we can also easily compute the index of
$\IE_{d_1,d'_1}\big(\tau,m_{G_2}-\eq\alpha^{\vee},m_{\mf{su}(2)}-\eq\lambda,\eq,\et-\eq\big)$
as
\begin{equation}\nonumber
  \begin{aligned}
    {\rm{Ind}}_{\IE_{d_1,d'_1}}=&-\frac{\epsilon_2^2}{4}(3d_1+2d_1')
    +\frac{\eq(\et-\eq)}{2}(3d_1^2+2{d_1'}^2-2d_1d_1'-d_1) \\
    &+(-3d_1+d_1')\Big(\frac{({m},{m})_{G_2}}{2}-\eq\alpha^{\vee}\cdot m_{G_2}+d_0\epsilon_1^2\Big)\\
    &  +(d_1-2d_1')\Big(\frac{({m},{m})_{A_1}}{2}-\eq\lambda\cdot m_{\mf{su}(2)}+d_0\epsilon_1^2\Big)\ .
\end{aligned}
\end{equation}
and the index of
$\IE_{d_2,d'_2}\big(\tau,m_{G_2}-\et\alpha^{\vee},m_{\mf{su}(2)}-\et\lambda,\eq-\et,\et\big)$
as
\begin{equation}\nonumber
\begin{aligned}
{\rm{Ind}}_{\IE_{d_2,d'_2}}=&-\frac{\epsilon_1^2}{4}(3d_2+2d_2')
    +\frac{(\eq-\et)\et}{2}(3d_2^2+2{d_2'}^2-2d_2d_2'-d_2) \\
  &+(-3d_2+d_2')\Big(\frac{({m},{m})_{G_2}}{2}-\et\alpha^{\vee}\cdot m_{G_2}+d_0\epsilon_2^2\Big)\\
  &  +(d_1-2d_1')\Big(\frac{({m},{m})_{\mf{su}(2)}}{2}-\et\lambda\cdot m_{\mf{su}(2)}+d_0\epsilon_2^2\Big)\ .
\end{aligned}
\end{equation}
Finally, by directly adding all contributions together and using the
constraints $d_0+d_1+d_2=k_1$ and $d'_0+d'_1+d'_2=k_2$, we obtain
\begin{equation}
  \begin{aligned}
    {\rm{Ind}}_{\rm poly}+{\rm{Ind}}_{V}^{G_2}+
    &{\rm{Ind}}_{V}^{\mf{su}(2)}+{\rm{Ind}}_{H}^{({\bf 7+1},{\tfrac{1}{2}\bf 2})}+{\rm{Ind}}_{\IE_{d_1,d'_1}}+{\rm{Ind}}_{\IE_{d_2,d'_2}}\\
    =-\frac{1}{2}\left(
      \begin{array}{cc} k_1 & k_2
      \end{array}
    \right)\Omega\left(
      \begin{array}{c}
        m_{G_2}\cdot m_{G_2} \\
        m_{\mf{su}(2)}\cdot m_{\mf{su}(2)} \\
      \end{array} \right)
    &-\frac{\epsilon_1^2+\epsilon_2^2}{4}(3k_1+2k_2)
    +\frac{\eq\et}{2}(3k_1^2-2k_1k_2+2k_2^2-4k_1)+\frac{9}{5}\epsilon_+^2.
\end{aligned}
\end{equation}
The final sum is \emph{independent} from
$\alpha^{\vee},\lambda,d_1,d_1',d_2,d_2'$ themselves, but only depends
on their combination $(k_1,k_2)$! This concludes the modularity of
elliptic blowup equations, which serves as the most nontrivial check
to arbitrary base degrees.

\subsubsection*{Limit to rank one theories}
By taking the node $2$ to zero limit, one obtains the $n=3$ $G_2$
theory with $n_{\bf 7}=1$. The ungauged $\mf{su}(2)$ becomes the
$\mf{sp}(1)$ flavor symmetry, thus $t_{\mf{su}(2)}$ becomes the mass
$m$ of matter $\bf 7$. As shown in Section \ref{sec:g2},
there are six unity elliptic blowup equations for the $n=3$ $G_2$
theory. In the following, we analyze how they can be obtained from the
five vanishing blowup equations of $3,2$ NHC. In fact, it is not hard
to find that under the limit $Q_{\rm{ell}_2}\to 0$, the vanishing
blowup equation (\ref{32bldeg}) with characteristic \eqref{eq:a32}
labeled with $j$ reduces to
\begin{equation}
  \theta_4^{[\frac{1}{6}+\frac{2j}{5}]}(15\tau,3\ep_+)
  \mathcal{U}_{G_2}^{[-\frac{1}{6}]}
  +\theta_4^{[-\frac{1}{6}+\frac{2j}{5}]}(15\tau,3\ep_+)
  \mathcal{U}_{G_2}^{[\frac{1}{6}]}
  +\theta_4^{[-\frac{1}{2}+\frac{2j}{5}]}(15\tau,3\ep_+)
  \mathcal{U}_{G_2}^{[\frac{1}{2}]}
  =0.
\end{equation}
where we define
\begin{equation}
  \mathcal{U}_{G_2}^{[a]}=\mathrm{U}_{G_2}^{[a]}
  ({r_{\mf{su}(2)}=1})-\mathrm{U}_{G_2}^{[a]}({r_{\mf{su}(2)}=-1}),
\end{equation}
and $\mathrm{U}_{G_2}^{[a]}$ denotes the l.h.s of unity blowup
equations of the $n=3$ $G_2$ theory with characteristic $a$. Since
$j=-2,-1,0,1,2$, clearly, one can conclude
\begin{equation}
  \mathcal{U}_{G_2}^{[a]}=0,\quad\mathrm{for }\quad a=-1/6,1/6,1/2,
\end{equation}
which are
\begin{equation}
  \mathrm{U}_{G_2}^{[a]}({r_{\mf{su}(2)}=1})=\mathrm{U}_{G_2}^{[a]}
  ({r_{\mf{su}(2)}=-1}),\quad\mathrm{for }\quad a=-1/6,1/6,1/2.
\end{equation}
By adding the r.h.s of the unity blowup equations, these give exactly
the six unity blowup equations as we already knew.

On the other hand, by taking the node $3$ to zero limit, one obtains
the $n=2$ $\mf{su}(2)$ theory with 8 half-hypers
transforming in $\md{2}$ of $\mf{su}(2)$.
There are two vanishing elliptic blowup equations for the $n=2$
$\mf{su}(2)$ theory. In fact, it is not hard to find that under the
limit $Q_{\rm{ell}_1}\to 0$, the vanishing blowup equation
(\ref{32bldeg}) with characteristic \eqref{eq:a32} labeled with $j$
reduces to
\begin{equation}
  \theta_3^{[\frac{j}{5}]}(10\tau,6\ep_+)\mathrm{V}_{\mf{su}(2)}^{[-\frac{1}{2}]}-\theta_3^{[-\frac{1}{2}+\frac{2j}{5}]}(10\tau,6\ep_+)\mathrm{V}_{\mf{su}(2)}^{[0]}=0,
\end{equation}
where $\mathrm{V}_{\mf{su}(2)}^{[a]}$ denotes the l.h.s of vanishing
blowup equations of the $n=2$ $\mf{su}(2)$ theory. Since
$j=-2,-1,0,1,2$, clearly, one can conclude
\begin{equation}
  \mathrm{V}_{\mf{su}(2)}^{[-\frac{1}{2}]}=\mathrm{V}_{\mf{su}(2)}^{[0]}=0.
\end{equation}
These are just the two vanishing blowup equations of the $n=2$
$\mf{su}(2)$ theory as we already knew.

\subsection{NHC $3,2,2$}
\label{sec:322}

NHC $3,2,2$ can be understood as coupling a M-string node 2 to NHC
$3,2$ from the right. The 2d quiver construction was conjectured in
\cite{Kim:2018gjo}, therefore the elliptic genera are exactly
computable. We give a geometric construction for the local Calabi-Yau
associated to this theory in Appendix \ref{app:cy}.
%

\subsubsection*{Elliptic blowup equations}
There are in total seven vanishing blowup equations and no unity
blowup equations, which is as expected since the M-string only have
unity blowup equations, while the NHC $3,2$ has only
vanishing equations. The idea can be roughly expressed as
\begin{equation}
  \rm{V}\star\rm{U}=\rm{V}.
\end{equation}

To derive the elliptic blowup equations, we apply the similar
de-affinization procedure as in rank one cases. As there are three base
curves $(-3,-2,-2)$ and only the first two have nontrivial gauge
symmetry, we have \emph{two} degeneration directions.
Let us denote the intersection matrix between the three base curves as
$-\Omega$, i.e.
\begin{equation}
  \Omega=\left(
    \begin{array}{ccc}
      3 & -1 & 0\\
      -1 &  2 & -1\\
      0 & -1 & 2
    \end{array}
  \right).
\end{equation}
Note $\det \Omega=7$ gives the number of non-equivalent vanishing
blowup equations.
We find the seven vanishing elliptic blowup equations can be written as
\begin{equation}\label{322bldeg}
  \begin{aligned}
    0\,&=\sum_{\displaystyle
      \substack{\alpha^\vee\in Q^\vee_{G_2},d_{1,2}\\
        d_0=\frac{1}{2}||\alpha^{\vee}||^2}}^{d_0+d_1+d_2=k_1}
    \sum_{\displaystyle \substack{\lambda\in(P^\vee\backslash
        Q^\vee)_{\mf{su}(2)},d'_{1,2}\\d'_0+1/4=\frac{1}{2}||\lambda||^2}}^{d'_0+d'_1+d'_2=k_2}
    \sum^{d''_1+d''_2=k_3}_{d''_{1,2}}(-1)^{|\alpha^{\vee}|+|\lambda|}\\
    &\times\Theta^{[a]}_{\Omega}\(\tau,\left(
      \begin{array}{c}
        \alpha^{\vee}\cdot m_{G_2}+(\bar{y}_1-d_0)(\eq+\et) -d_1\eq-d_2\et\\
        \lambda\cdot m_{\mf{su}(2)}+(\bar{y}_2-d_0')(\eq+\et) -d_1'\eq-d_2'\et\\
        \bar{y}_3(\eq+\et) -d_1'\eq-d_2'\et
      \end{array}
    \right)\)\\[-1mm]
    &\times A_{V}^{G_2}(\alpha^{\vee},\tau,m_{G_2})
    A_{V}^{\mf{su}(2)}(\lambda,\tau,m_{\mf{su}(2)}) A^{({\bf 7+1},{\tfrac{1}{2}\bf
        2},{\emptyset})}_{H}(\alpha^{\vee},\lambda,\tau,m_{G_2},m_{\mf{su}(2)})\\
    &\times \IE_{d_1,d'_1,d''_1}
    \big(\tau,m_{G_2}-\eq\alpha^{\vee},m_{\mf{su}(2)}-\eq\lambda,\eq,\et-\eq\big)\\
    &\times
    \IE_{d_2,d'_2,d''_2}\big(\tau,m_{G_2}-\et\alpha^{\vee},m_{\mf{su}(2)}-\et\lambda,\eq-\et,\et\big),\quad\qquad
    \text{for fixed}\;\; k_{1,2,3}\in \IZ_{\geq 0},
  \end{aligned}
\end{equation}
where the summation indices
$d_{0,1,2},d'_{0,1,2},d''_{1,2}\in\IZ_{\geq 0}$.
The parameters $(\bar{y}_1,\bar{y}_2,\bar{y}_3)=(5/7,9/14,4/7)$, and
\begin{equation}\label{eq:a322}
  a=\left(
    \begin{array}{c}
      a_k \\
      a_l \\
      a_s
    \end{array}
  \right)=\left(
    \begin{array}{c}
      3j/7 \\
      -1/2+2j/7 \\
      j/7 \\
    \end{array}
  \right),\quad\quad j=-3,-2,-1,0,1,2,3.
\end{equation}
Note $\bar{y}_1,\bar{y}_2,\bar{y}_3$ satisfy the following relation
\begin{equation}
  \Omega \left(
    \begin{array}{c}
      \bar{y}_1 \\
      \bar{y}_2 \\
      \bar{y}_3 \\
    \end{array}
  \right)=\left(
    \begin{array}{c}
      3/2 \\
      0 \\
      1/2\\
    \end{array}
  \right)=\left(
    \begin{array}{c}
      \bar{y}_{\rm u} \text{ of rank one $n=3$ $G_2$ theory} \\
      \bar{y}_{\rm v} \text{ of rank one $n=2$ $\mf{su}(2)$ theory}\\
      \bar{y}_{\rm u} \text{ of $n=2$ M-string theory} \\
    \end{array}
  \right).
\end{equation}
This is necessary to be consistent with the rank one elliptic blowup
equations when decompactifying one of the base curves.

The leading base degree of the vanishing blowup equations,
i.e. $d_0=d_1=d_2=d_0'=d_1'=d_2'=0$ can be simply written as
\begin{equation}\label{nhc322vid}
\Theta^{[a]}_{\Omega}\(\tau,\left(\begin{array}{c}
10\epsilon_+/7\\
m_{\mf{su}(2)}+9\epsilon_+/7\\
8\epsilon_+/7\\
\end{array}
\right)\)-\Theta^{[a]}_{\Omega}\(\tau,\left(\begin{array}{c}
10\epsilon_+/7\\
-m_{\mf{su}(2)}+9\epsilon_+/7\\
8\epsilon_+/7\\
\end{array}
\right)\)=0.
\end{equation}
It is easy to check the above identity is correct. For higher base
degrees, we have checked the seven vanishing blowup equations from the
Calabi-Yau setting to substantial degrees of K\"ahler classes.

\subsubsection*{Modularity}
The index of the elliptic genus $\IE_{k_1,k_2,k_3}$ is known to be
\begin{align}\label{322Eindex}
  \text{Ind}_{\IE_{k_1,k_2,k_3}} =
  &-\frac{(\eq+\et)^2}{4}(3k_1+2k_2+k_3)
    +\frac{\eq\et}{2}(3k_1^2+2k_2^2+2k_3^2-2k_1k_2-2k_2k_3-k_1) \nn
  &+(-3k_1+k_2)\frac{({m},{m})_{G_2}}{2}
    +(-2k_2+k_1+k_3)\frac{({m},{m})_{\mf{su}(2)}}{2}\ .
\end{align}
To prove modularity, we need to calculate the index of each term in
the elliptic blowup equations (\ref{322bldeg}). After lengthy
computations similar with the NHC $3,2$ case, by directly adding all
contributions together and using the constraints $d_0+d_1+d_2=k_1$ and
$d'_0+d'_1+d'_2=k_2$ and $d''_1+d''_2=k_3$, we obtain
\begin{equation}
\begin{aligned}
\phantom{=}&{\rm{Ind}}_{\rm poly}+{\rm{Ind}}_{V}^{G_2}+{\rm{Ind}}_{V}^{\mf{su}(2)}+{\rm{Ind}}_{H}^{({\bf 7+1},{\tfrac{1}{2}\bf 2},{\emptyset})}+{\rm{Ind}}_{\IE_{d_1,d'_1,d''_1}}+{\rm{Ind}}_{\IE_{d_2,d'_2,d''_2}}\\[-1mm]
=&-\frac{1}{2}\left(\begin{array}{ccc}
k_1 & k_2 & k_3
\end{array}
\right)\Omega\left(\begin{array}{c}
m_{G_2}\cdot m_{G_2} \\
m_{\mf{su}(2)}\cdot m_{\mf{su}(2)} \\
0\\
\end{array}
\right)-\frac{\epsilon_1^2+\epsilon_2^2}{4}(3k_1+2k_2+k_3)\\[-1mm]
&+\frac{\eq\et}{2}(3k_1^2-2k_1k_2+2k_2^2-2k_2k_3+2k_3^2-4k_1)+\frac{19}{28}(\eq+\et)^2.
\end{aligned}
\end{equation}
This final sum is \emph{independent} from
$\alpha^{\vee},\lambda,d_1,d_1',d_2,d_2',d_1'',d_2''$ themselves, but
only depends on their combination $(k_1,k_2,k_3)$! This concludes the
modularity of elliptic blowup equations, which serves as the most
nontrivial check to arbitrary base degrees.

\subsubsection*{Limits}
It is well-known by dropping the last $-2$ base curve, i.e. taking
$k_3=0$, one goes back to the $-3,-2$ NHC. By dropping the left
$-3,-2$ base curves, i.e. taking $k_1=k_2=0$, one obtains the M-string
theory. By dropping the left $-3$ base curve, i.e. taking $k_1=0$, one
obtains a rank-two Higgsable theory with three vanishing blowup
equations. Such theory has
\begin{equation}
  \Omega=\left(
    \begin{array}{cc}
      2 & -1 \\
      -1 & 2 \\
    \end{array}
  \right).
\end{equation}
This theory can be obtained in the following way: one can take the
$n=2,G=\mf{su}(2)$ theory, restrict the flavor $\mf{so}(7)\to G_2$,
and make the gauge $\mf{su}(2)$ coincide with the flavor $\mf{su}(2)$
of an M-string theory. It is easy to write down the three vanishing
blowup equations at degree $(k_2,k_3)$ as
\begin{equation}\label{22bldeg}
  \begin{aligned}
    \sum_{\displaystyle
      \substack{d'_0+1/4=\frac{1}{2}||\lambda||^2}}^{d'_0+d'_1+d'_2=k_2}
    \sum_{d''_1+d''_2=k_3}
    &(-1)^{|\lambda|}\Theta^{[a]}_{\Omega}\(\tau,\left(
      \begin{array}{c}
        \lambda\cdot m_{\mf{su}(2)}+(\bar{y}_2-d_0')(\eq+\et) -d_1'\eq-d_2'\et\\
        \bar{y}_3(\eq+\et) -d_1'\eq-d_2'\et\\
      \end{array}
    \right)\)\\[-2mm]
    &\times A_{V}^{\mf{su}(2)}(\lambda,\tau,m_{\mf{su}(2)}) A^{({\bf
        7+1},{\tfrac{1}{2}\bf 2},\emptyset)}_{H}
    (\lambda,\tau,m_{G_2},m_{\mf{su}(2)})\\
    &\times \IE_{d'_1,d''_1}\big(\tau,m_{G_2},m_{\mf{su}(2)}-\eq\lambda,\eq,\et-\eq\big)\\
    &\times \IE_{d'_2,d''_2}\big
    (\tau,m_{G_2},m_{\mf{su}(2)}-\et\lambda,\eq-\et,\et\big)=\,0.
  \end{aligned}
\end{equation}
where $\bar{y}_2=1/6,\bar{y}_3=1/3$.

\subsection{NHC $2,3,2$}
\label{sec:232}

NHC $2,3,2$ can be understood as coupling two $2_{\mf{su}(2)}$
theories to the rank one theory $3_{\mf{so}(7)}$. The 2d quiver
construction of this theory was given in \cite{Kim:2018gjo}. Besides,
this model has an orbifold construction \cite{DelZotto:2015rca}, where
the underlying geometry $T^2\times\IC^2/\Gamma$ has discrete action
$\Gamma$ generated by $(\omega^{-6},\omega,\omega^5)$, where $\omega$
is a root of unity with $\omega^8=1$. The $S^1$ compactification to 5d
has been studied with topological vertex in \cite{Hayashi:2017jze}. We
give a toric geometric construction for the local Calabi-Yau
associated to this theory in Appendix \ref{app:cy}.

\subsubsection*{Elliptic blowup equations}
Let us denote the intersection matrix between the three base curves as
$-\Omega$, i.e.
\begin{equation}
  \Omega=\left(
    \begin{array}{ccc}
      2 & -1 & 0\\
      -1 &  3 & -1\\
      0 & -1 & 2
    \end{array}
  \right).
\end{equation}
Note $\det \Omega=8$. It turns out there exist in total 16 vanishing
blowup equations and no unity blowup equation. These vanishing
equations are divided to two types, each consists of eight
equations. One type comes from the configuration
\begin{equation}
  \rm{V}\star\rm{U}\star\rm{V}=\rm{V},
\end{equation}
which means the unity equations of $3_{\mf{so}(7)}$ theory coupled
with the vanishing equations of two $2_{\mf{su}(2)}$ theories. The
other comes from the configuration
\begin{equation}
  \rm{U}\star\rm{V}\star\rm{U}=\rm{V},
\end{equation}
which means the vanishing equations of $3_{\mf{so}(7)}$ theory coupled
with the unity equations of two $2_{\mf{su}(2)}$ theories.

To precisely derive the elliptic blowup equations, we apply the
similar de-affinization procedure as in rank one cases. As there are
three base curves $(-2,-3,-2)$ and each has a nontrivial gauge
symmetry, we have \emph{three} degeneration directions. After
careful computations, we find the eight VUV type vanishing elliptic
blowup equations can be written as
\begin{equation}\label{232bldeg}
  \begin{aligned}  \phantom{\,}
    &\phantom{----}0=\ \sum_{\displaystyle
      \substack{\lambda\in(P^\vee\backslash Q^\vee)_{\mf{su}(2)},d_{1,2}\\
        d_0+1/4=\frac{1}{2}||\lambda||^2}}^{d_0+d_1+d_2=k_1}\
    \sum_{\displaystyle
      \substack{\alpha^\vee\in Q^\vee_{\mf{so}(7)}, d'_{1,2}\\d'_0=\frac{1}{2}||\alpha^{\vee}||^2}}^{d'_0+d'_1+d'_2=k_2}\
    \sum_{\displaystyle
      \substack{\lambda'\in(P^\vee\backslash Q^\vee)_{\mf{su}(2)}, d''_{1,2}\\
        d''_0+1/4=\frac{1}{2}||\lambda'||^2}}^{d''_0+d''_1+d''_2=k_3}\
    (-1)^{|\alpha^{\vee}|+|\lambda|+|\lambda'|}\\
    &\phantom{-----}\times\Theta^{[a]}_{\Omega}\(\tau,\left(
      \begin{array}{c}
        \lambda\cdot m_{\mf{su}(2)}+(\bar{y}_1-d_0)(\eq+\et) -d_1\eq-d_2\et\\
        \alpha^{\vee}\cdot m_{\mf{so}(7)}+(\bar{y}_2-d_0')(\eq+\et) -d_1'\eq-d_2'\et\\
        \lambda'\cdot m_{\mf{su}(2)}'+(\bar{y}_3-d_0'')(\eq+\et) -d_1''\eq-d_2''\et\\
      \end{array}
    \right)\)\\[-0mm]
    &\times A_{V}^{\mf{so}(7)}(\alpha^{\vee},\tau,m_{\mf{so}(7)}) A_{V}^{\mf{su}(2)}(\lambda,\tau,m_{\mf{su}(2)})A_{V}^{\mf{su}(2)}(\lambda',\tau,m_{\mf{su}(2)}')
    A^{\mathfrak{R}}_{H}(\lambda,\alpha^{\vee},\lambda',\tau,m_{\mf{so}(7)},m_{\mf{su}(2)},m_{\mf{su}(2)}')\\
    &\phantom{-----}\times \IE_{d_1,d'_1,d''_1}\big(\tau,m_{\mf{su}(2)}-\eq\lambda,m_{\mf{so}(7)}-\eq\alpha^{\vee},m_{\mf{su}(2)}'-\eq\lambda',\eq,\et-\eq\big)\\
    &\phantom{-----}\times
    \IE_{d_2,d'_2,d''_2}\big(\tau,m_{\mf{su}(2)}-\et\lambda,m_{\mf{so}(7)}-\et\alpha^{\vee},m_{\mf{su}(2)}'-\et\lambda',\eq-\et,\et\big),
    \quad \text{fixed}\;\; k_{1,2,3} \in \IZ_{\geq 0},
  \end{aligned}
\end{equation}
where the summation indices
$d_{0,1,2},d'_{0,1,2},d''_{0,1,2}\in \IZ_{\geq 0}$.  The parameters
$\bar{y}_{1,2,3}$ are $\bar{y}_1=1/2,\bar{y}_2=1,\bar{y}_3=1/2$,
$\mathfrak{R}=({\bf 1},{\bf 8},{\tfrac{1}{2}\bf 2})+({\tfrac{1}{2}\bf
  2},{\bf 8},{\bf 1})$, and
\begin{equation}\label{a232}
  a=\left(
    \begin{array}{c}
      (2j-1)/8 \\
      (2j-1)/4 \\
      (2j-1)/8
    \end{array}
  \right),\quad\quad j=-3,-2,-1,0,1,2,3,4.
\end{equation}
Note $\bar{y}_1,\bar{y}_2,\bar{y}_3$ satisfy the following relation
\begin{equation}
  \Omega \left(
    \begin{array}{c}
      \bar{y}_1 \\
      \bar{y}_2 \\
      \bar{y}_3 \\
    \end{array}
  \right)=\left(
    \begin{array}{c}
      0 \\
      2 \\
      0\\
    \end{array}
  \right)=\left(
    \begin{array}{c}
      \bar{y}_{\rm v} \text{ of rank one $n=2$ $\mf{su}(2)$ theory} \\
      \bar{y}_{\rm u} \text{ of rank one $n=3$ $\mf{so}(7)$ theory}\\
      \bar{y}_{\rm v} \text{ of rank one $n=2$ $\mf{su}(2)$ theory} \\
    \end{array}
  \right).
\end{equation}
This is necessary to be consistent with the established elliptic
blowup equations for rank one theories when decompactifying some of
the base curves.

The leading base degree of the vanishing blowup equations
(\ref{232bldeg}), i.e. $d_0=d_1=d_2=d_0'=d_1'=d_2'=0$ can be simply
written as
\begin{equation}\label{nhc232vid}
  \begin{aligned}
    \phantom{-}&\Theta^{[a]}_{\Omega}\(\tau,\left(
      \begin{array}{c}
        m'_{\mf{su}(2)}+\epsilon_+\\
        2\epsilon_+\\
        m'_{\mf{su}(2)}+\epsilon_+\\
      \end{array}
    \right)\)+\Theta^{[a]}_{\Omega}\(\tau,\left(
      \begin{array}{c}
        -m'_{\mf{su}(2)}+\epsilon_+\\
        2\epsilon_+\\
        -m'_{\mf{su}(2)}+\epsilon_+\\
      \end{array}
    \right)\)\\
    -&\Theta^{[a]}_{\Omega}\(\tau,\left(
      \begin{array}{c}
        m_{\mf{su}(2)}+\epsilon_+\\
        2\epsilon_+\\
        -m'_{\mf{su}(2)}+\epsilon_+\\
      \end{array}
    \right)\)-\Theta^{[a]}_{\Omega}\(\tau,\left(
      \begin{array}{c}
        -m_{\mf{su}(2)}+\epsilon_+\\
        2\epsilon_+\\
        m'_{\mf{su}(2)}+\epsilon_+\\
      \end{array}
    \right)\)=0.
  \end{aligned}
\end{equation}
It is easy to check the above identity is correct.

By similar de-affinazation procedure, the eight UVU type vanishing
elliptic blowup equations can be written as
\begin{equation}\label{232bldegUVU}\nonumber
\begin{aligned}
  \phantom{\,}
  &\phantom{----}0=\ \sum_{\displaystyle \substack{d_0=\frac{1}{2}||\alpha||^2}}^{d_0+d_1+d_2=k_1}\ \sum_{\displaystyle \substack{d'_0+1/2=\frac{1}{2}||\lambda||^2}}^{d'_0+d'_1+d'_2=k_2}\ \sum_{\displaystyle \substack{d''_0=\frac{1}{2}||\alpha'||^2}}^{d''_0+d''_1+d''_2=k_3}\ (-1)^{|\alpha|+|\lambda|+|\alpha'|}\\
  &\phantom{-----}\times\Theta^{[a]}_{\Omega}\(\tau,\left(
    \begin{array}{c}
      \alpha\cdot m_{\mf{su}(2)}+(\bar{y}_1-d_0)(\eq+\et) -d_1\eq-d_2\et\\
      \lambda\cdot m_{\mf{so}(7)}+(\bar{y}_2-d_0')(\eq+\et) -d_1'\eq-d_2'\et\\
      \alpha'\cdot m_{\mf{su}(2)}'+(\bar{y}_3-d_0'')(\eq+\et) -d_1''\eq-d_2''\et\\
    \end{array}
  \right)\)\\[-0mm]
  &\times A_{V}^{\mf{so}(7)}(\lambda,\tau,m_{\mf{so}(7)}) A_{V}^{\mf{su}(2)}(\alpha,\tau,m_{\mf{su}(2)})A_{V}^{\mf{su}(2)}(\alpha',\tau,m_{\mf{su}(2)}')A^{\mathfrak{R}}_{H}(\alpha,\lambda,\alpha',\tau,m_{\mf{so}(7)},m_{\mf{su}(2)},m_{\mf{su}(2)}')\\
  &\phantom{-----}\times \IE_{d_1,d'_1,d''_1}\big(\tau,m_{\mf{su}(2)}-\eq\alpha,m_{\mf{so}(7)}-\eq\lambda,m_{\mf{su}(2)}'-\eq\alpha',\eq,\et-\eq\big)\\
  &\phantom{-----}\times
  \IE_{d_2,d'_2,d''_2}\big(\tau,m_{\mf{su}(2)}-\et\alpha,m_{\mf{so}(7)}-\et\lambda,m_{\mf{su}(2)}'-\et\alpha',\eq-\et,\et\big),\quad
  \text{fixed}\;\; k_{1,2,3}\in\IZ_{\geq 0},
\end{aligned}
\end{equation}
where $\lambda\in (P^\vee\backslash Q^{\vee})_{\mf{so}(7)}$ and
$\alpha,\alpha'\in Q^{\vee}_{\mf{su}(2)}$, and
$\bar{y}_1=\bar{y}_3=3/4,\bar{y}_2=1/2$. The characteristics $a$ are
still those defined in (\ref{a232}). Note
$\bar{y}_1,\bar{y}_2,\bar{y}_3$ satisfy the following relation
\begin{equation}
  \Omega \left(
    \begin{array}{c}
      \bar{y}_1 \\
      \bar{y}_2 \\
      \bar{y}_3 \\
    \end{array}
  \right)=\left(
    \begin{array}{c}
      1 \\
      0 \\
      1\\
    \end{array}
  \right)=\left(
    \begin{array}{c}
      \bar{y}_{\rm u} \text{ of rank one $n=2$ $\mf{su}(2)$ theory} \\
      \bar{y}_{\rm v} \text{ of rank one $n=3$ $\mf{so}(7)$ theory}\\
      \bar{y}_{\rm u} \text{ of rank one $n=2$ $\mf{su}(2)$ theory} \\
    \end{array}
  \right).
\end{equation}
Since the smallest Weyl orbit in
$(P^\vee\backslash Q^{\vee})_{\mf{so}(7)}$ is
$\mathcal{O}_{6}$,\footnote{Note the vector representation
  ${\bf 7}_{\bf v }^{\mf{so}(7)}=1+\mathcal{O}_{6}$.} the leading base
degree of the vanishing blowup equations (\ref{232bldegUVU}) can be
simply written as
\begin{equation}\label{nhc232vid2}
  \begin{aligned}
    \sum_{w\in\mathcal{O}_{6}}(-1)^{|w|}\Theta^{[a]}_{\Omega}\(\tau,\left(
      \begin{array}{c}
        3\epsilon_+/2\\
        m_{\omega}+\epsilon_+\\
        3\epsilon_+/2\\
      \end{array}
    \right)\)\times\prod_{\beta\in\Delta(\mf{so}(7))}^{w\cdot\beta=
      1}\frac{1}{\theta_1(\tau,m_{\beta})}=0.
  \end{aligned}
\end{equation}
We have checked this identity up to $\mathcal{O}(q^{10})$. For higher
base degrees, we have checked all the 16 vanishing blowup equations
from the Calabi-Yau setting to substantial degrees of K\"ahler
classes.


\subsubsection*{Modularity}
The index of the elliptic genus $\IE_{k_1,k_2,k_3}$ is known to be
\begin{equation}
\begin{aligned}\label{232Eindex}\nonumber
  \text{Ind}_{\IE_{k_1,k_2,k_3}} &=
  -\frac{(\eq+\et)^2}{2}(k_1+2k_2+k_3)
    +\frac{\eq\et}{2}(2k_1^2+3k_2^2+2k_3^2-2k_1k_2-2k_2k_3-k_2) \nn
  +\,&(-2k_1+k_2)\frac{({m}_1,{m}_1)_{\mf{su}(2)}}{2}
    +(-3k_2+k_1+k_3)\frac{({m}_2,{m}_2)_{\mf{so}(7)}}{2}
    +(-2k_3+k_2)\frac{({m}_3,{m}_3)_{\mf{su}(2)}}{2} \ .
\end{aligned}
\end{equation}
Let us just show the modularity of the VUV type equations here. We need
to calculate the index of each term in the vanishing elliptic blowup
equations (\ref{232bldeg}). After lengthy computations similar with
the NHC $3,2$ case, by directly adding all contributions together and
using the constraints $d_0+d_1+d_2=k_1$ and $d'_0+d'_1+d'_2=k_2$ and
$d''_0+d''_1+d''_2=k_3$, we obtain
\begin{equation}
\begin{aligned}
\phantom{=}&{\rm{Ind}}_{\rm poly}+{\rm{Ind}}_{V}^{\mf{so}(7)}+{\rm{Ind}}_{V}^{\mf{su}(2)}+{\rm{Ind}}_{V}^{\mf{su}(2)'}+{\rm{Ind}}_{H}^{\mathfrak{R}}+{\rm{Ind}}_{\IE_{d_1,d'_1,d''_1}}+{\rm{Ind}}_{\IE_{d_2,d'_2,d''_2}}\\
=&-\frac{1}{2}\left(\begin{array}{ccc}
k_1 & k_2 & k_3
\end{array}
\right)\Omega\left(\begin{array}{c}
m_{\mf{su}(2)}\cdot m_{\mf{su}(2)} \\
m_{\mf{so}(7)}\cdot m_{\mf{so}(7)} \\
m_{\mf{su}(2)'}\cdot m_{\mf{su}(2)'} \\
\end{array}
\right)-\frac{\epsilon_1^2+\epsilon_2^2}{4}(k_1+2k_2+k_3)\\
&+\frac{\eq\et}{2}(2k_1^2+3k_2^2+2k_3^2-2k_1k_2-2k_2k_3-5k_2)+\frac{7}{8}(\eq+\et)^2.
\end{aligned}
\end{equation}
This final sum is \emph{independent} from
$\alpha^{\vee},\lambda_1,\lambda_2,d_1,d_1',d_2,d_2',d_1'',d_2''$
themselves, but only depends on their combination $(k_1,k_2,k_3)$!
This concludes the modularity of elliptic blowup equations, which
serves as the most nontrivial check to arbitrary base degrees.

\section{Arbitrary rank}
\label{sec:arbitraryrank}

In this section we study the most general 6d $(1,0)$ SCFTs in the
atomic classification. We first propose a simple set of rules to glue
together the blowup equations of rank one theories to the blowup
equations of higher-rank theories. With these gluing rules at hand, we
write down the precise form of the elliptic blowup equations for arbitrary
6d $(1,0)$ SCFTs and prove their modularity. We then present the
admissible blowup equations for a lot of examples including the ADE
chain of $-2$ curves with gauge symmetry, all conformal matter
theories and the blown-ups of some $-n$ curves in particular
$-9,-10,-11$ curves. The prominent feature here is that for
higher-rank theories, most of their blowup equations are of
vanishing type.

\subsection{Gluing rules}
\label{sec:glue}

One of the key steps to write down the blowup equations
for a higher rank theory is to fix the parameters $\lG$ and $\lF$ in
the gauge and the flavor symmetry sectors.  They are in fact both
components of the $r$-field in the blowup equations of topological
string theory~\cite{Gu:2017ccq,Huang:2017mis}.  Besides constructing higher
rank theories from rank one theories involves gauging the flavor symmetry.
Therefore we can view $\lG,\lF$ on an equal footing, and here we
consider them collectively as the $r$-field $(\lG,\lF)$.

Based on the gluing rules of higher rank 6d $(1,0)$ SCFTs
\cite{Heckman:2013pva,Heckman:2015bfa}, we propose the following
gluing rules for higher rank elliptic blowup equations, which are
simple criteria to determine which $r$ fields of one node can be
coupled to which $r$ fields of the adjacent nodes.
\begin{itemize}
\item For a node $(G,F)$ with blowup equations
  labeled by the $r$-field $(\lambda,\omega)$ and all adjacent nodes
  $(G_i,F_i)$ with blowup equations labeled by the $r$-field
  $(\lambda_i,\omega_i)$, $i=1,2,\dots,s$, $s\le 3$ and possibly an
  adherent free hyper with flavor $F_{f}$ with $r$-field $\lambda_f$,
  the admissible coupling for the node $(G,F)$ is such that
  $\cO_{\lambda_1}\times\cO_{\lambda_2}\times\ldots\times\cO_{\lambda_s}
  \times\cO_{\lambda_{f}} \subset \cO_w$ according to decomposition
  $\prod_i^s G_i\times F_f\subset F$, where $\cO_w$ is the Weyl orbit
  containing $w$.
\item The admissible blowup equations for a higher-rank theory is such
  that all its nodes satisfy the above criteria.
\end{itemize}
A few comments are in order. Note that a node may bear no gauge group such
as the E-string theory, in which case $G=\emptyset$ and
$\lambda \in \bf 1$. The concept of nodes in the criteria can be
generalized to molecules in the atomic classification, which makes it
easier to find all admissible blowup equations when lots of molecules
are involved. These criteria actually guarantee the consistency with
the blowup equations of lower-rank theories when decoupling nodes.

Also note that in this section, we will use the notation $\md{n}_p$ to
denote a Weyl orbit consisting of $n$ weights which all have norm
square $p$.  Very often we will suppress the subscript $p$ if $p$ is
minimal and there is no cause for confusion. We sometimes also use the
conjugate bar and subscripts $s,c$ to distinguish orbits of the same
length just like in the notation of irreducible representations.

Now let us demonstrate the above criteria for NHC $2,3,2$.
We recall the $r$-fields of the individual nodes from
Tables~\ref{tb:ubeq-1},\ref{tb:ubeq-2},\ref{tb:vbeq-1},\ref{tb:vbeq-2}
\begin{align}
  &\fn=2, (G,F)=(\mf{su}(2), \mf{so}(7)):\;
  \begin{cases}
    \text{unity}\ r\text{-fields} \in (\md{1}_0,\md{6}_1)\\
    \text{vanishing}\ r\text{-fields} \in (\md{2}_{1/2},\md{1}_0)
  \end{cases}\\
  &\fn=3, (G,F)=(\mf{so}(7), \mf{sp}(2)):\;
  \begin{cases}
    \text{unity}\ r\text{-fields} \in (\md{1}_0,\md{4}_1)\\
    \text{vanishing}\ r\text{-fields} \in (\md{6}_1,\md{1}_0),
    (\md{6}_1,\md{4}_{1/2})
  \end{cases}
\end{align}
First, to couple the central node $G=\mf{so}(7),F=\mf{sp}(2)$ of the
NHC 2,3,2 with the two side nodes
$G_{1,2}=\mf{su}(2),F_{1,2}=\mf{so}(7)$, the flavor group $F$ must
decompose as $\mf{sp}(2)\to \mf{su}(2)\times \mf{su}(2)$.  As we have
seen, the unity $r$ fields
$(\lambda_{\mf{so}(7)},\omega_{\mf{sp}(2)})$ of the central node
$3_{\mf{so}(7)}$ are elements of $({\bf 1}_0,{\bf 4}_1)$. Under the
flavor $F$ decomposition, we have
${\bf 4}_1\to
(\md{2}_{1/2},\md{2}_{1/2})$. 
Since
$(\lambda_{\mf{su}(2)},\omega_{\mf{so}(7)})\in(\md{2}_{1/2},\md{1}_0)$
is indeed a correct set of vanishing $r$ fields of the node
$2_{\mf{su}(2)}$, we find one set of admissible $r$ fields for the NHC
$2,3,2$ with the parameter $\lF$ of the entire $2,3,2$ chain belonging
to $(\md{2}_{1/2},\md{1}_0,\md{2}_{1/2})$, which give rise to
vanishing blowup equations. On the other hand, the vanishing $r$
fields of the central node $3_{\mf{so}(7)}$ are elements of
$({\bf 6}_1,{\bf 1}_0)$ or $({\bf 6}_1,{\bf 4}_{1/2})$. Under flavor
$F$ decomposition, ${\bf 1}_0\to ({\bf 1}_0,{\bf 1}_0)$ and
${\bf 4}_{1/2}\to ({\bf 2}_{1/2},{\bf 1}_0)+({\bf 1}_0,{\bf
  2}_{1/2})$.  The combination $(\md{2}_{1/2},\md{6}_1)$ does not
contain $r$ fields of the node $2_{\mf{su}(2)}$, but the combination
$(\md{1}_0,\md{6}_1)$ does contain valid $r$-fields of the unity type.
Clearly, the overall $\lF$ parameters belonging to
$(\md{1}_0,\md{6}_1,\md{1}_0)$ give rise to the other set of vanishing
blowup equations
for NHC $2,3,2$ and there is no other possible admissible $\lF$.
These simple analysis confirms our blowup equations in Section
\ref{sec:232}.

For more examples with adherent free hypers, we refer to Section
\ref{sec:exhigh}.

\subsection{Arbitrary rank elliptic blowup equations}
\label{sc:arebe}

Let us consider F-theory compactifications on an elliptic Calabi-Yau
threefold, whose non-compact base contains $r$ compact curves with a
negative definite intersection matrix $-\Omega_{ij}=A_{ij}$.  Recall
the symmetry algebras and the massless fields which can arise in this
theory.  Over the $i$-th compact curve $C_i$ there could be singular
elliptic fibers corresponding to a symmetry algebra $G_i$.  In addition
$C_i$ could intersect with a non-compact curve $N_i$ with intersection
number $k_{F_i}$, and the latter supports singular elliptic fibers
corresponding to symmetry algebra $F_i$.

The resulting field theory is a 6d SCFT in its $r$ dimensional tensor
branch 
with total gauge symmetry $\prod_i G_i$ and flavor symmetry
$\prod_i F_i$.  If we compactify the 6d SCFT on a torus, we can also
turn on the gauge and flavor fugacities $m_{G_i},m_{F_i}$.  There are
also charged matter fields localised at intersections of curves.  At
the intersection locus of two compact base curves $C_i,C_j$ there are
hypermultiplets charged under both gauge groups $G_i,G_j$.  We also
consider hypermultiplets localised at the intersection locus of
compact and non-compact curves.  Finally BPS strings arise from
D3-branes wrapping compact base curves.  The number of times a string
wraps each base curve is interpreted as the charge of this string.
The string charges form a rank $r$ lattice $\Lambda$ with the negative
definite bilinear form defined by $-\Omega_{ij}=A_{ij}$.

Now let us write down the blowup equations for the elliptic genera
$\IE_{d_i}(\tau,m_{G_i},m_{F_i},\eq,\et)$ of the 6d SCFT.
\begin{align}
  &\sum_{\alpha_i \in \phi_i (Q^\vee(G_i)),d'_i,d''_i\in\IN}^{||\alpha_i||^2/2+d'_i+d''_i=d_i+\delta_i/2}
    (-1)^{\sum_i|\phi_i^{-1}(\alpha_i)|} \nonumber\\[-2mm]
  &\phantom{===}\times\Theta_{\Omega}^{[a_i]}(\tau,-(\alpha_i\cdot m_{G_i}) +
    \sum_j(\Omega^{-1})_{ij}k_{F_j}(\lambda_j\cdot m_{F_j})+
    (y_i -
    \frac{1}{2}(\alpha_i\cdot\alpha_i))(\eq+\et)-d'_i\eq-d''_i\et)
    \nonumber\\[-1mm]
  &\phantom{===}\times\prod_i A_{V_i}(\tau,m_{G_i},\alpha_i)
    \prod_{ij} A_{H_{ij}}(\tau,m_{G_i},\mu_j,\alpha_i,\alpha_j)
    \prod_{i} A_{H_i}(\tau,m_{G_i},m_{F_i},\alpha_i,\lambda_i)\nonumber\\[-2mm]
  &\phantom{===}\times
    \IE_{d'_i}(\tau,m_{G_i}+\alpha_i\eq,m_{F_i}+\lambda_i\eq,\eq,\et-\eq)
    \IE_{d''_i}(\tau,m_{G_i}+\alpha_i\et,m_{F_i}+\lambda_i\et,\eq-\et,\et)\nonumber\\[+1mm]
  &\phantom{==}=
    \Lambda(\delta_i)
    \Theta_{\Omega}^{[a_i]}(\tau,\sum_j(\Omega^{-1})_{ij}k_{F_j}(\lambda_j\cdot\nu_j)+
    y_i(\eq+\et)) \IE_{d_i}(\tau,m_{G_i},m_{F_i},\eq,\et),
         \label{eq:rblp}
\end{align}
with
\begin{equation}
  y_i =
  \sum_j(\Omega^{-1})_{ij}(\frac{1}{4}(-2+\Omega_{jj}+h^\vee_{G_j})+
  \frac{1}{2}k_{F_j}(\lambda_j\cdot\lambda_j)),
\end{equation}
and
\begin{equation}
  \Lambda(\delta_i)  =
  \begin{cases}
    1, & \forall i,\delta_i = 0,\\
    0, & \exists i,\delta_i >0\,.
  \end{cases}
\end{equation}
Here as in the rank one case, $\phi_i$ is an embedding of the coroot
lattice $Q^\vee(G_i)$ in the coweight lattice of $G_i$ by an overall
shift of a coweight vector.  $\delta_i$ is the smallest norm in the
image $\phi_i(Q^\vee(G_i))$; $\delta_i$ is zero if the embedding is
unshifted so that $\phi_i(Q^\vee(G_i)) = Q^\vee(G_i)$ and positive
otherwise.  $\phi_i^{-1}(\alpha_i)$ gives back a coroot vector, and
$|\bullet|$ is the sum of the coefficients in its decomposition in
terms of simple coroots.  $A_{V_i}$ is the contribution of vector
multiplets transforming in the adjoint representation of $G_i$, and
$A_{H_{ij}}, A_{H_i}$ are respectively the contributions of
hypermultiplets charged in the mixed representation of two gauge
groups, and in the representation of one gauge group.  Their
expressions have been given in \eqref{eq:AV},\eqref{eq:AH}.  Finally
the parameters $\lambda_i$ are the components of $r$-fields associated
to the flavor symmetries.  They take value in the coweight lattice and
they are determined by the gluing rules discussed in the previous
subsection.  One important consistency condition is that if we turn
off all string charges except for the one indexed by $i$, i.e.~we set
$d_j = 0$ for $j\neq i$, and consequently $d'_j = d''_j = 0$,
$\alpha_j =0$ for $j\neq i$ as well, \eqref{eq:rblp} must reduce to
the blowup equations for a rank one 6d SCFT with $\fn = \Omega_{ii}$,
gauge group $G_i$, and the surviving $\lambda_i$ should be the $\lF$
parameter worked out in Section~\ref{sec:ellipt-blow-equat-1}.  We
have seen some examples of this decompactification in action in
Sections~\ref{sec:32}, \ref{sec:322}, and \ref{sec:232}.

\subsection{Modularity}\label{Ssec:ModularityII}

In this section we illustrate that the elliptic blowup equations
\eqref{eq:rblp} satisfy the modularity consistency conditions.  For
ease of presentation, we divide each term on the l.h.s.~by the r.h.s.
except for the $\Lambda(\delta_i)$ factor.  The modularity consistency
conditions then require that the modular index of each term on the
l.h.s.  after division is independent of the summation indices
$\alpha_i,d'_i,d''_i$, and in the case of unity equations where
$\Lambda(\delta_i) = 1$ be equal to zero.

We prove the modularity in two steps.  First, we isolate an individual
string charge indexed $i$, and turn off all the other string charges
by setting $d_j = 0$ for $j\neq i$.  As we argued before,
\eqref{eq:rblp} reduces to those of a rank one 6d SCFT, and by the way
we choose $\lambda_i$, the modularity condition is automatically
satisfied.  Second, we consider the modular index of mixed string
charges.  Let us pick a pair of string charges, say $d_1,d_2$, whose
associated base curves intersect.  Then we turn off all the other
string charges $d_j = 0$ ($j\neq 1,2$), and check the modularity
conditions for the following components of the modular index for each
term
\begin{equation}
  \Ind^{\text{mix}}(d_1,d_2) = \Ind(d_1,d_2) - \Ind(0,d_2) - \Ind(d_1,0) + \Ind(0,0).
\end{equation}

The modular index polynomial of the generalised theta function
$\Theta_{\Omega}^{[a_i]}(\tau,z_i)$ is
\begin{equation}
  \Ind\Theta_{\Omega}(z_i) = \frac{1}{2}\sum_{i,j}z_i\Omega_{ij}z_j.
\end{equation}
Its contribution to the modular index of the blowup equations is
\begin{align}
  \Ind {\Theta}(d_i) =
  -\sum_{i}
  &((\alpha_i\cdot m_{G_i})+\frac{1}{2}||\alpha_i||^2(\eq+\et)+d_i'\eq+d_i''\et)\nn
  &\times(k_{F_i}(\lambda_i\cdot m_{F_i}) +
    (y_i+\frac{1}{2}k_{F_i}||\lambda_i||^2)(\eq+\et))\nn
    +\frac{1}{2}\sum_{i,j}
  &(\alpha_i\cdot m_{G_i}+\frac{1}{2}||\alpha_i||^2(\eq+\et)+d_i'\eq+d_i''\et)\nn
  &\times(\alpha_j\cdot\mu_j+\frac{1}{2}||\alpha_j||^2(\eq+\et)+d_j'\eq+d_j''\et)\Omega_{ij}.
\end{align}
The components of mixed string charges $d_1,d_2$ are
\begin{align}
  \Ind^{\text{mix}}\Theta(d_1,d_2) =
  &((\alpha_1-\beta_1)\cdot m_{G_1}+\frac{1}{2}(||\alpha_1||^2-||\beta_1||^2)(\eq+\et)+d_1'\eq+d_1''\et)\nn
  &\times((\alpha_2-\beta_2)\cdot m_{G_2}+\frac{1}{2}(||\alpha_2||^2-||\beta_2||^2)(\eq+\et)+d_2'\eq+d_2''\et)\Omega_{12}
\end{align}
where $\beta_i$ is a vector in $\phi_i(Q^\vee(G_i))$ whose norm equals
$\delta_i$.  The modular index polynomial of the elliptic genus with
degrees $d_i$ is given in \eqref{eq:IndEd}, which we reproduce here
\begin{align}
  \Ind &\IE_{d_i}(\eq,\et,m_{G_i},m_{F_i},d_i) =\nn
  &-\frac{1}{4}(\eq+\et)^2\sum_{i}(2-\Omega_{ii}+h^\vee_{G_i})d_i
    +\frac{1}{2}\eq\et
    \Big(\sum_i(2-\Omega_{ii})d_i + \sum_{ij}d_i d_j \Omega_{ij}\Big)\nn
  &-\frac{1}{2}\sum_{ij} d_j(m_{G_i},m_{G_i})\Omega_{ij} +\frac{1}{2}\sum_i
    k_{F_i} d_i (m_{F_i},m_{F_i}).
\end{align}
The components of its contribution associated to mixed string charges
are
\begin{align}
  \Ind^{\text{mix}}\IE(d_1,d_2) =
  &-\frac{1}{2}d'_1((||\alpha_2||^2-||\beta_2||^2)\eq^2+2((\alpha_2-\beta_2)\cdot
    m_{G_2})\eq)\Omega_{12} \nn
  &-\frac{1}{2}d''_1((||\alpha_2||^2-||\beta_2||^2)\et^2+2((\alpha_2-\beta_2)\cdot
    m_{G_2})\et)\Omega_{12} \nn
  &-\frac{1}{2}d'_2((||\alpha_1||^2-||\beta_1||^2)\eq^2+2((\alpha_1-\beta_1)\cdot
    m_{G_1})\eq)\Omega_{21} \nn
  &-\frac{1}{2}d''_2((||\alpha_1||^2-||\beta_1||^2)\et^2+2((\alpha_1-\beta_1)\cdot
    m_{G_1})\et)\Omega_{21}.
\end{align}
Neither $A_{V_i}$ nor $A_{H_i}$ would contribute to the modular index
of mixed string charges, and we only have to consider $A_{H_{12}}$,
whose modular index polynomial can be read off from \eqref{eq:AH},
\eqref{eq:thetaH*}.  Combining all these ingredients together, the
modular index polynomial of an arbitrary term on the l.h.s.~after
division for mixed string charge $d_1,d_2$ is
\begin{align}
  \Ind^{\text{mix}}(d_1,d_2) =
  \frac{1}{4}
  &\Omega_{12}(4\ind_{R_{G_1}}\ind_{R_{G_2}}n_{12}-1)
    \Big(4(\alpha_1-\beta_1)\cdot m_{G_1}\times(1 \to 2) \nn
  &+2(\eq+\et)((||\alpha_1||^2-||\beta_1||^2)(\alpha_2-\beta_2)\cdot m_{G_2}+(1,2
    \to 2,1))\nn
  &+(\eq^2+\eq\et+\et^2)(||\alpha_1||^2-||\beta_1||^2)\times(1 \to 2)
    \Big).
\end{align}
This contribution vanishes identically thanks to the anomaly
cancellation condition \eqref{eq:anm-1}.

\subsection{Examples}
\label{sec:exhigh}
With the gluing rules given in Section~\ref{sec:glue}, we can
efficiently write down all admissible blowup equations for any
higher-rank theory once the gauge groups, flavor groups and matter
representation are known. We use a simple quiver diagram to denote
blowup equations with the following rules:
\begin{itemize}
\item We use a circle for a compact base curve and a rectangle for a
  noncompact one.
\item For each base curve with associated gauge/flavor symmetry $G$,
  we mark it with a Weyl orbit $\md{n}_p$ ($p$ is often suppressed if
  it is minimal) of $G$ to denote the $r$ field of $G$ fugacities. If
  a compact curve has no associated gauge symmetry, we leave the
  circle blank.
\end{itemize}
For example, for the $3,2$ NHC with associated gauge symmetry
$G_2,\mf{su}(2)$, we denote the five vanishing blowup equations simply
as
\begin{center}
    \begin{tikzpicture}
        \node at (1,0) [circle,draw] (b) {$\bf 1$};
        \node at (2,0) [circle,draw] (c) {$\bf 2$};
        \draw (c) -- (b);
    \end{tikzpicture}
\end{center}
Meanwhile, the vanishing blowup equations of $3,2,2$ NHC can be denoted respectively as
\begin{center}
    \begin{tikzpicture}
    \node at (0,0) [circle,draw] (a) {$\bf{1}$};
        \node at (1,0) [circle,draw] (b) {$\bf 2$};
        \node at (2,0) [circle,draw] (c) {$\phantom{1}$};
        \draw (c) -- (b);
        \draw (a) -- (b);
    \end{tikzpicture}
\end{center}
and those of $2,3,2$
NHC as
\begin{center}
    \begin{tikzpicture}
    \node at (0,0) [circle,draw] (a) {$\bf 2$};
        \node at (1,0) [circle,draw] (b) {$\bf 1$};
        \node at (2,0) [circle,draw] (c) {$\bf 2$};
        \draw (c) -- (b);
        \draw (a) -- (b);
        \node at (4,0) [circle,draw] (e) {$\bf 1$};
        \node at (5,0) [circle,draw] (f) {$\bf 6$};
        \node at (6,0) [circle,draw] (g) {$\bf 1$};
        \draw (e) -- (f);
        \draw (g) -- (f);
    \end{tikzpicture}
\end{center}
Note each quiver diagram above represents $\det(\Omega)$
non-equivalent blowup equations where $-\Omega$ is the intersection
matrix of compact base curves. In the following, we will show the
blowup equations for some most interesting examples of higher-rank
theories including ADE chains of $-2$ curves, conformal matters and
the blowups of $-9,-10,-11$ curves.

\subsubsection{ADE chains of $-2$ curves}
The 2d quiver construction and elliptic genera for this type of 6d $(1,0)$ quiver SCFTs are given in
\cite{Gadde:2015tra}, see also another form in
\cite{Haghighat:2017vch}.
A crucial property of simply-laced Dynkin diagrams is needed in order
to achieve admissible gluing of the blowup equations of individual
nodes: the mark of each node has to be the average of the marks of all
its adjacent nodes. Besides, when a node is at the end, its mark is
half of the mark of its adjacent node. The problem of finding all
admissible blowup equations then reduces to the decomposition of Weyl
orbits of the special unitary algebra to its subalgebras.

In the following we demonstrate the application of gluing rules for
some typical examples including $A_{2,3}$, $D_{4,5}$ and $E_{6,7,8}$
quivers. The gluing here often involves decomposition like $SU(N+M)\to SU(N)\times SU(M)\times U(1)$. In the following, for simplicity, we do not explicitly writ down the $U(1)$ part. But it should be emphasized that the shifts of $U(1)$ indeed play a role during the gluing. 
For $A$ type quiver, there exist one more global $U(1)$ symmetry, while for $D,E$ type quivers, such $U(1)$ is anomalous \cite{Gadde:2015tra}. This also makes a difference when counting the total number of admissible blowup equations. 
\begin{itemize}
\item We first demonstrate the gluing for a simple example which is an
  $A$ type quiver with gauge group $\mf{su}(2)$. Note when two $n=2$
  $\mf{su}(2)$ gauge theories are coupled together, the flavor
  symmetry $\mf{su}(4)$ (or equivalently $\mf{so}(7)$) breaks down to
  $\mf{su}(2)\times \mf{su}(2)$. Then one of the flavor symmetry
  $\mf{su}(2)$ becomes the gauge symmetry $\mf{su}(2)$ for the other
  theory. For rank one $n=2$ $\mf{su}(2)$ theory, the unity $\lF$ is
  in $\bf 1$, while the vanishing $\lF$ is in $\bf 6$. Under the
  flavor group splitting,
  $\mathbf{6}=2(\md{1},\md{1})+(\mathbf{2},\mathbf{2})$. Note also
  $\mathbf{2}\subset(P^\vee\backslash Q^{\vee})_{\mf{su}(2)}$. This means for
  a unity $-2$ node, the adjacent two $-2$ nodes must be both unity or
  both vanishing. On the other hand, for a vanishing $-2$ node, the
  adjacent two $-2$ nodes can only be both unity.

  For example, for the $A_2$ quiver, we find the following structure
  or the blowup equations
  \begin{equation}
    \begin{aligned}
      \mathrm{U}\star\mathrm{U}&=\mathrm{U},\\
      \mathrm{V}\star\mathrm{U}&=\mathrm{V},\\
      \mathrm{U}\star\mathrm{V}&=\mathrm{V}.
    \end{aligned}
  \end{equation}
  Keep in mind there are two $\mf{su}(2)$ fundamental matters at the
  two ends of the $A_2$ quiver. Therefore, the above blowup equations
  can be expressed in quiver diagrams as
  \begin{center}
    \begin{tikzpicture}
        \node at (0,0) [rectangle,draw] (a) {$\bf 1$};
        \node at (1,0) [circle,draw] (b) {$\bf 1$};
        \node at (2,0) [circle,draw] (c) {$\bf 1$};
        \node at (3,0) [rectangle,draw] (d) {$\bf 1$};
        \draw (a) -- (b);
        \draw (b) -- (c);
        \draw (c) -- (d);
      \end{tikzpicture}
    \end{center}

    \begin{center}
    \begin{tikzpicture}
        \node at (0,0) [rectangle,draw] (a) {$\bf {1}$};
        \node at (1,0) [circle,draw] (b) {$\bf 2$};
        \node at (2,0) [circle,draw] (c) {$\bf 1$};
        \node at (3,0) [rectangle,draw] (d) {$\bf 2$};
        \draw (a) -- (b);
        \draw (b) -- (c);
        \draw (c) -- (d);
    \end{tikzpicture}
\end{center}

\begin{center}
    \begin{tikzpicture}
        \node at (0,0) [rectangle,draw] (a) {$\bf {2}$};
        \node at (1,0) [circle,draw] (b) {$\bf 1$};
        \node at (2,0) [circle,draw] (c) {$\bf 2$};
        \node at (3,0) [rectangle,draw] (d) {$\bf 1$};
        \draw (a) -- (b);
        \draw (b) -- (c);
        \draw (c) -- (d);
    \end{tikzpicture}
  \end{center}

  The first quiver diagram represents unity equations, while the other
  two represent vanishing equations.  
  The number of equations with  fixed characteristic represented by each quiver diagram is the  product of numbers in square nodes, while the number of characteristics is the determinant of the Cartan matrix $C$ of the quiver diagram. We find $\det(C_{A_2})=3$. The first diagram actually represents two possible shifts of the global $U(1)$ flavor $\lambda_{U(1)}=\pm 2$, while the second and third diagram have $\lambda_{U(1)}=0$. Thus there are in total  $3\times 2 = 6$ unity equations and $3\times(2+2)=12$ vanishing  equations.

  For $A_3$ quiver, there are following blowup equations
  \begin{equation}
    \begin{aligned}
      \mathrm{U}\star\mathrm{U}\star\mathrm{U}&=\mathrm{U},\\
      \mathrm{U}\star\mathrm{V}\star\mathrm{U}&=\mathrm{V},\\
      \mathrm{V}\star\mathrm{U}\star\mathrm{V}&=\mathrm{V},\\
    \end{aligned}
  \end{equation}
  or in quiver diagrams as
  \begin{center}
    \begin{tikzpicture}
        \node at (0,0) [rectangle,draw] (a) {$\bf 1$};
        \node at (1,0) [circle,draw] (b) {$\bf 1$};
        \node at (2,0) [circle,draw] (c) {$\bf 1$};
        \node at (3,0) [circle,draw] (d) {$\bf 1$};
        \node at (4,0) [rectangle,draw] (e) {$\bf 1$};
        \draw (a) -- (b);
        \draw (b) -- (c);
        \draw (c) -- (d);
        \draw (d) -- (e);
      \end{tikzpicture}
    \end{center}

    \begin{center}
      \begin{tikzpicture}
        \node at (0,0) [rectangle,draw] (a) {$\bf 2$};
        \node at (1,0) [circle,draw] (b) {$\bf 1$};
        \node at (2,0) [circle,draw] (c) {$\bf 2$};
        \node at (3,0) [circle,draw] (d) {$\bf 1$};
        \node at (4,0) [rectangle,draw] (e) {$\bf 2$};
        \draw (a) -- (b);
        \draw (b) -- (c);
        \draw (c) -- (d);
        \draw (d) -- (e);
      \end{tikzpicture}
    \end{center}

    \begin{center}
      \begin{tikzpicture}
        \node at (0,0) [rectangle,draw] (a) {$\bf 1$};
        \node at (1,0) [circle,draw] (b) {$\bf 2$};
        \node at (2,0) [circle,draw] (c) {$\bf 1$};
        \node at (3,0) [circle,draw] (d) {$\bf 2$};
        \node at (4,0) [rectangle,draw] (e) {$\bf 1$};
        \draw (a) -- (b);
        \draw (b) -- (c);
        \draw (c) -- (d);
        \draw (d) -- (e);
      \end{tikzpicture}
    \end{center}

  \item Consider $A$ type quiver theories with $\mf{su}(3)$
    symmetry. When two $n=2$ $\mf{su}(3)$ gauge theories are coupled
    together, the flavor symmetry $\mf{su}(6)$ breaks down to
    $\mf{su}(3)\times \mf{su}(3)$. Then one of the flavor symmetry
    $\mf{su}(3)$ becomes the gauge symmetry $\mf{su}(3)$ for the other
    theory. We summarize the $r$ fields behavior of rank one $n=2$
    $\mf{su}(3)$ theory under the flavor group splitting in the
    following table,
    \begin{table}[h]
      \begin{center}
        \begin{tabular}{|c|c|c|c|}
          \hline
          & $\lG$
          & $\lF$
          & branching rules of $\lF$ \\
          \hline
          & $\mf{su}(3)$ & $\mf{su}(6)$ & $\mf{su}(3)\times \mf{su}(3)$ \\
          \hline
          unity &  $\bf 1$
          & $\cO_{\omega_3} = {\bf 20}$ & $2{\bf (1,1)}+\mathbf{(3,\bar 3)}+\mathbf{(\bar{ 3},3)}$ \\
          \hline
          vanishing & $\bf 3$
          & $\cO_{\omega_5} = \bf \overline{6}$
          & ${\bf (\bar{3},1)}+{\bf (1,\bar{3})}$ \\
          \hline
          vanishing & $\bf \bar{3}$
          & $\cO_{\omega_1}=\bf 6$ & ${\bf (3,1)}+{\bf (1,3)}$ \\
          \hline
        \end{tabular}
      \end{center}
    \end{table}
    where $\cO_{\omega_i}$ is the Weyl orbit generated by the $i$-th
    fundamental coweight.

    For example, for $A_2$ quiver, read from the table above and
    gluing rules, we find there are following blowup equations
    \begin{center}
      \begin{tikzpicture}
        \node at (0,0) [rectangle,draw] (a) {$\bf 1$};
        \node at (1,0) [circle,draw] (b) {$\bf 1$};
        \node at (2,0) [circle,draw] (c) {$\bf 1$};
        \node at (3,0) [rectangle,draw] (d) {$\bf 1$};
        \draw (a) -- (b);
        \draw (b) -- (c);
        \draw (c) -- (d);
      \end{tikzpicture}
    \end{center}

    \begin{center}
      \begin{tikzpicture}
        \node at (0,0) [rectangle,draw] (a) {$\bf \bar{3}$};
        \node at (1,0) [circle,draw] (b) {$\bf 1$};
        \node at (2,0) [circle,draw] (c) {$\bf 3$};
        \node at (3,0) [rectangle,draw] (d) {$\bf \bar{3}$};
        \draw (a) -- (b);
        \draw (b) -- (c);
        \draw (c) -- (d);
        
                \node at (5,0) [rectangle,draw] (a1) {$\bf \bar{3}$};
        \node at (6,0) [circle,draw] (b1) {$\bf 3$};
        \node at (7,0) [circle,draw] (c1) {$\bf 1$};
        \node at (8,0) [rectangle,draw] (d1) {$\bf \bar3$};
        \draw (a1) -- (b1);
        \draw (b1) -- (c1);
        \draw (c1) -- (d1);
      \end{tikzpicture}
    \end{center}

    \begin{center}
      \begin{tikzpicture}
        \node at (0,0) [rectangle,draw] (a) {$\bf {3}$};
        \node at (1,0) [circle,draw] (b) {$\bf 1$};
        \node at (2,0) [circle,draw] (c) {$\bf \bar3$};
        \node at (3,0) [rectangle,draw] (d) {$\bf 3$};
        \draw (a) -- (b);
        \draw (b) -- (c);
        \draw (c) -- (d);
        
                \node at (5,0) [rectangle,draw] (a1) {$\bf {3}$};
        \node at (6,0) [circle,draw] (b1) {$\bf \bar3$};
        \node at (7,0) [circle,draw] (c1) {$\bf 1$};
        \node at (8,0) [rectangle,draw] (d1) {$\bf 3$};
         \draw (a1) -- (b1);
        \draw (b1) -- (c1);
        \draw (c1) -- (d1);
      \end{tikzpicture}
    \end{center}

    \begin{center}
      \begin{tikzpicture}
        \node at (0,0) [rectangle,draw] (a) {$\bf {1}$};
        \node at (1,0) [circle,draw] (b) {$\bf 3$};
        \node at (2,0) [circle,draw] (c) {$\bf \bar3$};
        \node at (3,0) [rectangle,draw] (d) {$\bf 1$};
        \draw (a) -- (b);
        \draw (b) -- (c);
        \draw (c) -- (d);
        
                \node at (5,0) [rectangle,draw] (a1) {$\bf {1}$};
        \node at (6,0) [circle,draw] (b1) {$\bf \bar3$};
        \node at (7,0) [circle,draw] (c1) {$\bf 3$};
        \node at (8,0) [rectangle,draw] (d1) {$\bf 1$};
         \draw (a1) -- (b1);
        \draw (b1) -- (c1);
        \draw (c1) -- (d1);
      \end{tikzpicture}
    \end{center}

 The first quiver represents unity equations, while all
    the other quivers represent vanishing equations.  

    For $A_3$ quiver, there are following blowup equations
    \begin{center}
      \begin{tikzpicture}
        \node at (0,0) [rectangle,draw] (a) {$\bf 1$};
        \node at (1,0) [circle,draw] (b) {$\bf 1$};
        \node at (2,0) [circle,draw] (c) {$\bf 1$};
        \node at (3,0) [circle,draw] (d) {$\bf 1$};
        \node at (4,0) [rectangle,draw] (e) {$\bf 1$};
        \draw (a) -- (b);
        \draw (b) -- (c);
        \draw (c) -- (d);
        \draw (d) -- (e);
    \end{tikzpicture}
    \end{center}

\begin{center}
\begin{tikzpicture}
        \node at (0,0) [rectangle,draw] (a) {$\bf \bar3$};
        \node at (1,0) [circle,draw] (b) {$\bf 1$};
        \node at (2,0) [circle,draw] (c) {$\bf 3$};
        \node at (3,0) [circle,draw] (d) {$\bf \bar3$};
        \node at (4,0) [rectangle,draw] (e) {$\bf 1$};
        \draw (a) -- (b);
        \draw (b) -- (c);
        \draw (c) -- (d);
        \draw (d) -- (e);
        
         \node at (6,0) [rectangle,draw] (a1) {$\bf 1$};
        \node at (7,0) [circle,draw] (b1) {$\bf \bar3$};
        \node at (8,0) [circle,draw] (c1) {$\bf 3$};
        \node at (9,0) [circle,draw] (d1) {$\bf 1$};
        \node at (10,0) [rectangle,draw] (e1) {$\bf \bar3$};
        \draw (a1) -- (b1);
        \draw (b1) -- (c1);
        \draw (c1) -- (d1);
        \draw (d1) -- (e1);
    \end{tikzpicture}
    \end{center}

\begin{center}
\begin{tikzpicture}
        \node at (0,0) [rectangle,draw] (a) {$\bf 3$};
        \node at (1,0) [circle,draw] (b) {$\bf 1$};
        \node at (2,0) [circle,draw] (c) {$\bf \bar3$};
        \node at (3,0) [circle,draw] (d) {$\bf 3$};
        \node at (4,0) [rectangle,draw] (e) {$\bf 1$};
        \draw (a) -- (b);
        \draw (b) -- (c);
        \draw (c) -- (d);
        \draw (d) -- (e);
        
         \node at (6,0) [rectangle,draw] (a1) {$\bf 1$};
        \node at (7,0) [circle,draw] (b1) {$\bf 3$};
        \node at (8,0) [circle,draw] (c1) {$\bf \bar3$};
        \node at (9,0) [circle,draw] (d1) {$\bf 1$};
        \node at (10,0) [rectangle,draw] (e1) {$\bf 3$};
        \draw (a1) -- (b1);
        \draw (b1) -- (c1);
        \draw (c1) -- (d1);
        \draw (d1) -- (e1);
    \end{tikzpicture}
    \end{center}

\begin{center}
\begin{tikzpicture}
        \node at (0,0) [rectangle,draw] (a) {$\bf \bar3$};
        \node at (1,0) [circle,draw] (b) {$\bf 3$};
        \node at (2,0) [circle,draw] (c) {$\bf 1$};
        \node at (3,0) [circle,draw] (d) {$\bf \bar3$};
        \node at (4,0) [rectangle,draw] (e) {$\bf 3$};
        \draw (a) -- (b);
        \draw (b) -- (c);
        \draw (c) -- (d);
        \draw (d) -- (e);
        
         \node at (6,0) [rectangle,draw] (a1) {$\bf 3$};
        \node at (7,0) [circle,draw] (b1) {$\bf \bar3$};
        \node at (8,0) [circle,draw] (c1) {$\bf 1$};
        \node at (9,0) [circle,draw] (d1) {$\bf 3$};
        \node at (10,0) [rectangle,draw] (e1) {$\bf \bar3$};
        \draw (a1) -- (b1);
        \draw (b1) -- (c1);
        \draw (c1) -- (d1);
        \draw (d1) -- (e1);
    \end{tikzpicture}
    \end{center}

 The first quiver represents unity equations, while the
    remaining quivers represent vanishing equations.  
  \item Consider $A$ type quiver theories with $\mf{su}(4)$
    symmetry. When two $n=2$ $\mf{su}(4)$ gauge theories are coupled
    together, the flavor symmetry $\mf{su}(8)$ breaks down to
    $\mf{su}(4)\times \mf{su}(4)$. Then one of the flavor $\mf{su}(4)$
    becomes the gauge $\mf{su}(4)$ for the other theory. Note
    $(P^\vee\slash Q^{\vee})_{A_3}=\IZ_4$. We summarize the $r$ fields
    behavior under the flavor group splitting in the following table.
    \begin{table}[h]
      \begin{center}
        \begin{tabular}{|c|c|c|c|}
          \hline
          &	$\lG$ & $\lF$ & branching rules of $\lF$ \\
          \hline
          &   $\mf{su}(4)$ & $\mf{su}(8)$ & $\mf{su}(4)\times \mf{su}(4)$ \\
          \hline
          u &  $\bf 1$ & $\cO_{\omega_4}=\bf 70$ & $2{\bf (1,1)}+\mathbf{(4,\bar 4)}+\mathbf{(\bar{ 4},4)}+ \mathbf{(6,6)}$ \\
          \hline
          v & $\bf 4$ & $\cO_{\omega_6}=\bf \overline{28}$ & ${\bf (6,1)}+{\bf (1,6)}+{\bf (\bar{4},\bar{4})}$  \\ \hline
          v  & $\bf 6$ & $\cO_{0}=\bf 1$ & ${\bf (1,1)}$ \\
          \hline
          v & $\bf \bar{4}$ & $\cO_{\omega_2}=\bf 28$ & ${\bf (6,1)}+{\bf (1,6)}+{\bf (4,4)}$ \\
          \hline
        \end{tabular}
      \end{center}
    \end{table}

    Now based on the general gluing procedure, we can directly write
    down all admissible blowup equations. For example, for $A_2$
    quiver, there are following blowup equations
    \begin{center}
      \begin{tikzpicture}
        \node at (0,0) [rectangle,draw] (a) {$\bf 1$};
        \node at (1,0) [circle,draw] (b) {$\bf 1$};
        \node at (2,0) [circle,draw] (c) {$\bf 1$};
        \node at (3,0) [rectangle,draw] (d) {$\bf 1$};
        \draw (a) -- (b);
        \draw (b) -- (c);
        \draw (c) -- (d);
    \end{tikzpicture}
    \end{center}

    \begin{center}
    \begin{tikzpicture}
        \node at (0,0) [rectangle,draw] (a) {$\bf \bar{4}$};
        \node at (1,0) [circle,draw] (b) {$\bf 1$};
        \node at (2,0) [circle,draw] (c) {$\bf 4$};
        \node at (3,0) [rectangle,draw] (d) {$\bf 6$};
        \draw (a) -- (b);
        \draw (b) -- (c);
        \draw (c) -- (d);
        
               \node at (5,0) [rectangle,draw] (a1) {$\bf {4}$};
        \node at (6,0) [circle,draw] (b1) {$\bf 1$};
        \node at (7,0) [circle,draw] (c1) {$\bf \bar{4}$};
        \node at (8,0) [rectangle,draw] (d1) {$\bf 6$};
        \draw (a1) -- (b1);
        \draw (b1) -- (c1);
        \draw (c1) -- (d1);
    \end{tikzpicture}
\end{center}

    \begin{center}
    \begin{tikzpicture}
        \node at (0,0) [rectangle,draw] (a) {$\bf {6}$};
        \node at (1,0) [circle,draw] (b) {$\bf 4$};
        \node at (2,0) [circle,draw] (c) {$\bf 1$};
        \node at (3,0) [rectangle,draw] (d) {$\bf \bar{4}$};
        \draw (a) -- (b);
        \draw (b) -- (c);
        \draw (c) -- (d);
        
        \node at (5,0) [rectangle,draw] (a1) {$\bf {6}$};
        \node at (6,0) [circle,draw] (b1) {$\bf \bar{4}$};
        \node at (7,0) [circle,draw] (c1) {$\bf 1$};
        \node at (8,0) [rectangle,draw] (d1) {$\bf {4}$};
        \draw (a1) -- (b1);
        \draw (b1) -- (c1);
        \draw (c1) -- (d1);
    \end{tikzpicture}
\end{center}

   \begin{center}
    \begin{tikzpicture}
        \node at (0,0) [rectangle,draw] (a) {$\bf \bar{4}$};
        \node at (1,0) [circle,draw] (b) {$\bf 4$};
        \node at (2,0) [circle,draw] (c) {$\bf \bar{4}$};
        \node at (3,0) [rectangle,draw] (d) {$\bf {4}$};
        \draw (a) -- (b);
        \draw (b) -- (c);
        \draw (c) -- (d);
        
        \node at (5,0) [rectangle,draw] (a1) {$\bf {4}$};
        \node at (6,0) [circle,draw] (b1) {$\bf \bar{4}$};
        \node at (7,0) [circle,draw] (c1) {$\bf 4$};
        \node at (8,0) [rectangle,draw] (d1) {$\bf \bar{4}$};
        \draw (a1) -- (b1);
        \draw (b1) -- (c1);
        \draw (c1) -- (d1);
    \end{tikzpicture}
\end{center}

    \begin{center}
    \begin{tikzpicture}
        \node at (0,0) [rectangle,draw] (a) {$\bf {6}$};
        \node at (1,0) [circle,draw] (b) {$\bf 1$};
        \node at (2,0) [circle,draw] (c) {$\bf 6$};
        \node at (3,0) [rectangle,draw] (d) {$\bf {1}$};
        \draw (a) -- (b);
        \draw (b) -- (c);
        \draw (c) -- (d);
        
        \node at (5,0) [rectangle,draw] (a1) {$\bf 1$};
        \node at (6,0) [circle,draw] (b1) {$\bf 6$};
        \node at (7,0) [circle,draw] (c1) {$\bf 1$};
        \node at (8,0) [rectangle,draw] (d1) {$\bf {6}$};
         \draw (a1) -- (b1);
        \draw (b1) -- (c1);
        \draw (c1) -- (d1);
    \end{tikzpicture}
\end{center}

  The first quiver diagram represents unity equations,
  while the other quiver diagrams represent vanishing equations.  

For the $A_3$ quiver, there are the following blowup equations:
\begin{center}
\begin{tikzpicture}
        \node at (0,0) [rectangle,draw] (a) {$\bf 1$};
        \node at (1,0) [circle,draw] (b) {$\bf 1$};
        \node at (2,0) [circle,draw] (c) {$\bf 1$};
        \node at (3,0) [circle,draw] (d) {$\bf 1$};
        \node at (4,0) [rectangle,draw] (e) {$\bf 1$};
        \draw (a) -- (b);
        \draw (b) -- (c);
        \draw (c) -- (d);
        \draw (d) -- (e);
    \end{tikzpicture}
    \end{center}

\begin{center}
\begin{tikzpicture}
        \node at (0,0) [rectangle,draw] (a) {$\bf 6$};
        \node at (1,0) [circle,draw] (b) {$\bf 4$};
        \node at (2,0) [circle,draw] (c) {$\bf 1$};
        \node at (3,0) [circle,draw] (d) {$\bf \bar{4}$};
        \node at (4,0) [rectangle,draw] (e) {$\bf 6$};
        \draw (a) -- (b);
        \draw (b) -- (c);
        \draw (c) -- (d);
        \draw (d) -- (e);
        
             \node at (6,0) [rectangle,draw] (a1) {$\bf 6$};
        \node at (7,0) [circle,draw] (b1) {$\bf \bar{4}$};
        \node at (8,0) [circle,draw] (c1) {$\bf 1$};
        \node at (9,0) [circle,draw] (d1) {$\bf {4}$};
        \node at (10,0) [rectangle,draw] (e1) {$\bf 6$};
        \draw (a1) -- (b1);
        \draw (b1) -- (c1);
        \draw (c1) -- (d1);
        \draw (d1) -- (e1);
    \end{tikzpicture}
    \end{center}

\begin{center}
\begin{tikzpicture}
        \node at (0,0) [rectangle,draw] (a) {$\bf \bar{4}$};
        \node at (1,0) [circle,draw] (b) {$\bf 4$};
        \node at (2,0) [circle,draw] (c) {$\bf \bar{4}$};
        \node at (3,0) [circle,draw] (d) {$\bf {4}$};
        \node at (4,0) [rectangle,draw] (e) {$\bf \bar{4}$};
        \draw (a) -- (b);
        \draw (b) -- (c);
        \draw (c) -- (d);
        \draw (d) -- (e);
        
             \node at (6,0) [rectangle,draw] (a1) {$\bf 4$};
        \node at (7,0) [circle,draw] (b1) {$\bf \bar{4}$};
        \node at (8,0) [circle,draw] (c1) {$\bf 4$};
        \node at (9,0) [circle,draw] (d1) {$\bf \bar{4}$};
        \node at (10,0) [rectangle,draw] (e1) {$\bf 4$};
        \draw (a1) -- (b1);
        \draw (b1) -- (c1);
        \draw (c1) -- (d1);
        \draw (d1) -- (e1);
    \end{tikzpicture}
    \end{center}

\begin{center}
\begin{tikzpicture}
        \node at (0,0) [rectangle,draw] (a) {$\bf 1$};
        \node at (1,0) [circle,draw] (b) {$\bf {6}$};
        \node at (2,0) [circle,draw] (c) {$\bf 1$};
        \node at (3,0) [circle,draw] (d) {$\bf {6}$};
        \node at (4,0) [rectangle,draw] (e) {$\bf 1$};
        \draw (a) -- (b);
        \draw (b) -- (c);
        \draw (c) -- (d);
        \draw (d) -- (e);
        
               \node at (6,0) [rectangle,draw] (a1) {$\bf 6$};
        \node at (7,0) [circle,draw] (b1) {$\bf {1}$};
        \node at (8,0) [circle,draw] (c1) {$\bf 6$};
        \node at (9,0) [circle,draw] (d1) {$\bf {1}$};
        \node at (10,0) [rectangle,draw] (e1) {$\bf 6$};
        \draw (a1) -- (b1);
        \draw (b1) -- (c1);
        \draw (c1) -- (d1);
        \draw (d1) -- (e1);
    \end{tikzpicture}
    \end{center}
The first quiver diagram represents unity equations,
  while the other quiver diagrams represent vanishing equations.  
\item Now consider a $D_4$ quiver with gauge group $\mf{su}(2d_i)$. We
  want to couple a $n=2$ $\mf{su}(4)$ gauge theory with three $n=2$
  $\mf{su}(2)$ gauge theories and a extra $\mf{su}(2)$
  fundamental. Note the flavor symmetry $\mf{su}(8)$ of the center
  node breaks down to $\mf{su}(2)^4$. Note
  $(P^\vee\slash Q^{\vee})_{A_3}=\IZ_4$. We summarize the $r$ fields
  behavior under the flavor group splitting in the following table,
 where $\cO_{\omega_i}$ is the Weyl orbit of the
  $i$-th fundamental coweight.
  \begin{table}[h]
    \begin{center}
      \begin{tabular}{|c|c|c|c|}
        \hline
        &	$\lG$ & $\lF$ & branching rules of $\lF$ \\
        \hline
        &   $\mf{su}(4)$ & $\mf{su}(8)$ & $\mf{su}(2)^4$ \\
        \hline
        u &  $\bf 1$ & $\cO_{\omega_4}=\bf 70$ & $6{\bf (1,1,1,1)}+2(\mathbf{(2,2,1,1)}\textrm{ and permutations})+\mathbf{(2,2,2,2)}$ \\
        \hline
        v & $\bf 4$ & $\cO_{\omega_6}=\bf \overline{28}$ & $4{\bf (1,1,1,1)}+(\mathbf{(2,2,1,1)}\textrm{ and permutations})$  \\
        \hline
        v  & $\bf 6$ & $\cO_{0}=\bf 1$ & ${\bf (1,1,1,1)}$ \\
        \hline
        v & $\bf \bar{4}$ & $\cO_{\omega_2}=\bf 28$ & $4{\bf (1,1,1,1)}+(\mathbf{(2,2,1,1)}\textrm{ and permutations})$ \\
        \hline
      \end{tabular}
    \end{center}
  \end{table}

  Now based on the general gluing procedure, we can directly write
  down all admissible blowup equations as:
\begin{center}
\begin{tikzpicture}
        \node at (1,1) [rectangle,draw] (a) {$\bf 1$};
        \node at (0,0) [circle,draw] (b) {$\bf 1$};
        \node at (1,-1) [circle,draw] (c) {$\bf 1$};
        \node at (-1,-1) [circle,draw] (d) {$\bf 1$};
        \node at (-1,1) [circle,draw] (e) {$\bf 1$};
        \draw (a) -- (b);
        \draw (b) -- (c);
        \draw (b) -- (d);
        \draw (b) -- (e);
        
                \node at (5,1) [rectangle,draw] (a1) {$\bf 2$};
        \node at (4,0) [circle,draw] (b1) {$\bf 1$};
        \node at (5,-1) [circle,draw] (c1) {$\bf 2$};
        \node at (3,-1) [circle,draw] (d1) {$\bf 2$};
        \node at (3,1) [circle,draw] (e1) {$\bf 2$};
        \draw (a1) -- (b1);
        \draw (b1) -- (c1);
        \draw (b1) -- (d1);
        \draw (b1) -- (e1);
        
         \node at (9,1) [rectangle,draw] (a2) {$\bf 1$};
        \node at (8,0) [circle,draw] (b2) {$\bf 6$};
        \node at (9,-1) [circle,draw] (c2) {$\bf 1$};
        \node at (7,-1) [circle,draw] (d2) {$\bf 1$};
        \node at (7,1) [circle,draw] (e2) {$\bf 1$};
        \draw (a2) -- (b2);
        \draw (b2) -- (c2);
        \draw (b2) -- (d2);
        \draw (b2) -- (e2);
    \end{tikzpicture}
    \end{center}

    The first quiver diagram represents unity equations, while the
    remaining two diagrams represent vanishing equations.  Note
    $\det(C_{D_4})=4$.
    Thus there are in total 4 unity and
    $4\times(2+1)=12$ vanishing blowup equations. Let us show the
    leading base degree identities for the two types of vanishing
    blowup equations. The intersection matrix among base curves
    $-\Omega$ is just the negative of the Cartan matrix of $D_4$, i.e.
    \begin{equation} \Omega=\left(
        \begin{array}{cccc}
          2 & -1 & 0 & 0 \\
          -1 & 2 & -1 & -1 \\
          0 & -1 & 2  & 0\\
          0 & -1 & 0 & 2
        \end{array}
      \right).
    \end{equation}
    Then we find the first type of vanishing blowup equations has the
    following leading degree vanishing identities
    \begin{equation}
      \sum_{\lambda_{a,b,c}=\pm
        1/2}(-1)^{\lambda_a+\lambda_b+\lambda_c}\Theta^{[a]}_{\Omega}\(\tau,\left(
        \begin{array}{c}
          -\lambda_a m_{a}+2(\epsilon_1+\epsilon_2)\\
          4(\epsilon_1+\epsilon_2)\\
          -\lambda_b m_{b}+2(\epsilon_1+\epsilon_2)\\
          -\lambda_c m_{c}+2(\epsilon_1+\epsilon_2)\\
        \end{array}
      \right)\)=0.
      \label{D4quiverthetaindent1}
    \end{equation}
    where
    $m_{a,b,c}$ are the fugacities associated to the three
    $\mf{su}(2)$ gauge node. Here the contributions from vector and
    hyper multiplets do not depend on the summation indices
    $\lambda_{a,b,c}$ and thus we have factored them out. The four
    possible characteristics
    $a$ are defined according to (\ref{generaldefa}). The second type
    of vanish blowup equation has leading base degree as
    \begin{equation}
      \sum_{1\le i< j\le 4}\Theta^{[a]}_{\Omega}\(\tau,\left(
        \begin{array}{c}
          2(\epsilon_1+\epsilon_2)\\
          -m_i-m_j+3(\epsilon_1+\epsilon_2)\\
          2(\epsilon_1+\epsilon_2)\\
          2(\epsilon_1+\epsilon_2)\\
        \end{array}
      \right)\)\frac{1}{\prod_{k\neq i,j}\theta_1(m_i-m_k)\theta_1(m_j-m_k)}=0.
      \label{D4quiverthetaindent2}
    \end{equation}
    Here $m_i,i=1,2,3,4$ are the $\mf{su}(4)$ fugacities of the
    central node with $\sum_{i=1}^4m_i=0$. We have checked these
    identities up to order $\mathcal{O}(q^{10})$.

  \item Consider a $D_5$ quiver with gauge group $\mf{su}(2d_i)$. We
    want to couple two $n=2$ $\mf{su}(4)$ gauge theories together with
    three $n=2$ $\mf{su}(2)$ gauge theories and an extra $\mf{su}(2)$
    fundamental. Note the flavor symmetry $\mf{su}(8)$ of the
    $\mf{su}(4)$ node breaks down to $\mf{su}(4)\times
    \mf{su}(2)^2$. Note also $(P^\vee\slash Q^{\vee})_{A_3}=\IZ_4$. We
    summarize the $r$ fields behavior under the flavor group splitting
    in the following table, where $\cO_{\omega_i}$ is
    the Weyl orbit generated by the $i$-th fundamental coweight.
    \begin{table}[h]
      \begin{center}
        \resizebox{\linewidth}{!}{
        \begin{tabular}{|c|c|c|c|}
          \hline
          &	$\lG$ & $\lF$ & branching rules of $\lF$ \\
          \hline
          &   $\mf{su}(4)$ & $\mf{su}(8)$ & $\mf{su}(4)\times \mf{su}(2)\times \mf{su}(2)$ \\
          \hline
          u &  $\bf 1$ & $\cO_{\omega_4}=\bf 70$ & $2{\bf (1,1,1)}+\mathbf{(6,2,2)}+2\mathbf{(6,1,1)}+\mathbf{(4,2,1)}+\mathbf{(4,1,2)}+\mathbf{(\bar4,2,1)}+\mathbf{(\bar4,1,2)}$ \\
          \hline
          v & $\bf 4$ & $\cO_{\omega_6}=\bf \overline{28}$ & $2{\bf (1,1,1)}+\mathbf{(6,1,1)}+\mathbf{(\bar4,2,1)}+\mathbf{(\bar4,1,2)}+\mathbf{(1,2,1)}+\mathbf{(1,1,2)}$ \\
          \hline
          v  & $\bf 6$ & $\cO_{0}=\bf 1$ & ${\bf (1,1,1)}$ \\
          \hline
          v & $\bf \bar{4}$ & $\cO_{\omega_2}=\bf 28$ & $2{\bf (1,1,1)}+\mathbf{(6,1,1)}+\mathbf{(4,2,1)}+\mathbf{(4,1,2)}+\mathbf{(1,2,1)}+\mathbf{(1,1,2)}$ \\
          \hline
        \end{tabular}}
      \end{center}
    \end{table}

    Now based on the general gluing procedure, we can directly write
    down all admissible blowup equations as:
\begin{center}
\begin{tikzpicture}
        \node at (1,1) [rectangle,draw] (a) {$\bf 1$};
        \node at (0,0) [circle,draw] (b) {$\bf 1$};
        \node at (1,-1) [circle,draw] (c) {$\bf 1$};
        \node at (-1.5,0) [circle,draw] (f) {$\bf 1$};
        \node at (-2.5,-1) [circle,draw] (d) {$\bf 1$};
        \node at (-2.5,1) [circle,draw] (e) {$\bf 1$};
        \draw (a) -- (b);
        \draw (b) -- (c);
        \draw (b) -- (f);
        \draw (f) -- (d);
        \draw (f) -- (e);
    \end{tikzpicture}
    \end{center}
    \begin{center}
\begin{tikzpicture}
        \node at (1,1) [rectangle,draw] (a) {$\bf 1$};
        \node at (0,0) [circle,draw] (b) {$\bf 6$};
        \node at (1,-1) [circle,draw] (c) {$\bf 1$};
        \node at (-1.5,0) [circle,draw] (f) {$\bf 1$};
        \node at (-2.5,-1) [circle,draw] (d) {$\bf 2$};
        \node at (-2.5,1) [circle,draw] (e) {$\bf 2$};
        \draw (a) -- (b);
        \draw (b) -- (c);
        \draw (b) -- (f);
        \draw (f) -- (d);
        \draw (f) -- (e);
        
            \node at (6,1) [rectangle,draw] (a1) {$\bf 2$};
        \node at (5,0) [circle,draw] (b1) {$\bf 1$};
        \node at (6,-1) [circle,draw] (c1) {$\bf 2$};
        \node at (3.5,0) [circle,draw] (f1) {$\bf 6$};
        \node at (2.5,-1) [circle,draw] (d1) {$\bf 1$};
        \node at (2.5,1) [circle,draw] (e1) {$\bf 1$};
        \draw (a1) -- (b1);
        \draw (b1) -- (c1);
        \draw (b1) -- (f1);
        \draw (f1) -- (d1);
        \draw (f1) -- (e1);
    \end{tikzpicture}
    \end{center}

    The first quiver diagram represents unity equations, while the
    remaining two diagrams represent vanishing equations.  Note
     $\det(C_{D_5})=4$.
    Thus there are in total 4
    unity and $4\times(1+2)=12$ vanishing blowup equations.
  \item Consider the $E_6$ quiver with gauge group $\mf{su}(2d_i)$. We
    want to couple an $n=2$ $\mf{su}(6)$ gauge theory with three $n=2$
    $\mf{su}(4)$ gauge theories and two of the $\mf{su}(4)$ theories
    each with an $\mf{su}(2)$ theory and the other $\mf{su}(4)$ theory to
    an extra $\mf{su}(2)$ fundamental hypermultiplet. All the nodes together then
    form the Dynkin diagram of affine $E_6$. Note the flavor symmetry
    $\mf{su}(12)$ of the center node breaks down to $\mf{su}(4)^3$,
    and the flavor symmetry $\mf{su}(8)$ of the $\mf{su}(4)$ node
    breaks down to $\mf{su}(2)\times \mf{su}(6)$. Besides,
    $(P^\vee\slash Q^{\vee})_{A_5}=\IZ_6$. We summarize the $r$ fields
    behavior under the flavor group splitting in the following tables.
    \begin{table}[h]
      \begin{center}
        \begin{tabular}{|c|c|c|c|}
          \hline
          &	$\lG$ & $\lF$ & branching rules of $\lF$ \\
          \hline
          &   $\mf{su}(6)$ & $\mf{su}(12)$ & $\mf{su}(4)^3$ \\
          \hline
          u &  $\bf 1$ & $\cO_{\omega_6}=\bf 924$ & $2{\mathbf {(6,1,1)}}+\mathbf{(4,4,1)}+\mathbf{(\bar{4},\bar4,1)}+\mathbf{(4,6,\bar4)}+\mathbf{(6,6,6)}+\textrm{permutations}$ \\
          \hline
          v & $\bf 6$ & $\cO_{\omega_8}=\bf \overline{495}$ & $3{\bf (1,1,1)}+\mathbf{(4,\bar4,1)}+\mathbf{(6,6,1)}+\mathbf{(\bar4,\bar4,6)}+\textrm{permutations}$ \\
          \hline
          v  & $\bf 15$ & $\cO_{\omega_{10}}=\bf \overline{66}$ & ${\bf (6,1,1)}+{\bf (\bar4,\bar4,1)}+\textrm{permutations}$ \\
          \hline
          v  & $\bf 20$ & $\cO_{0}=\bf 1$ & ${\bf (1,1,1)}$ \\
          \hline
          v & $\bf \overline{15}$ & $\cO_{\omega_2}=\bf 66$ & ${\bf (6,1,1)}+{\bf (4,4,1)}+\textrm{permutations}$  \\
          \hline
          v & $\bf \bar{6}$ & $\cO_{\omega_4}=\bf 495$ & $3{\bf (1,1,1)}+\mathbf{(4,\bar4,1)}+\mathbf{(6,6,1)}+\mathbf{(4,4,6)}+\textrm{permutations}$  \\
          \hline
        \end{tabular}
      \end{center}
    \end{table}

    \begin{table}[h]
      \begin{center}
        \begin{tabular}{|c|c|c|c|}
          \hline
          &	$\lG$ & $\lF$ & branching rules of $\lF$ \\
          \hline
          &   $\mf{su}(4)$ & $\mf{su}(8)$ & $\mf{su}(2)\times \mf{su}(6)$ \\
          \hline
          u &  $\bf 1$ & $\cO_{\omega_4}=\bf 70$ & ${\bf (1,\overline{15})}+\mathbf{(2,20)}+\mathbf{(1,15)}$ \\
          \hline
          v & $\bf 4$ & $\cO_{\omega_6}=\bf \overline{28}$ & ${\bf (1,1)}+\mathbf{(2,6)}+\mathbf{(1,\overline{15})}$  \\
          \hline
          v & $\bf 6$ & $\cO_0=\bf 1$ & ${\bf (1,1)}$ \\
          \hline
          v & $\bf \bar{4}$ & $\cO_{\omega_2}=\bf 28$ & ${\bf (1,1)}+\mathbf{(2,6)}+\mathbf{(1,15)}$ \\
          \hline
        \end{tabular}
      \end{center}
    \end{table}

    Now based on the general gluing procedure, we can directly write
    down all admissible blowup equations as:
    \begin{center}
      \begin{tikzpicture}
\node at (-1,0) [circle,draw] (e) {$\bf 1$};
        \node at (0,0) [circle,draw] (a) {$\bf 6$};
        \node at (1,0) [circle,draw] (b) {$\bf 1$};
        \node at (2,0) [circle,draw] (c) {$\bf 6$};
        \node at (3,0) [circle,draw] (d) {$\bf 1$};
        \node at (1,1) [circle,draw] (f) {$\bf 6$};
        \node at (1,2) [rectangle,draw] (g) {$\bf 1$};
        \draw (e) -- (a);
        \draw (a) -- (b);
        \draw (b) -- (c);
        \draw (c) -- (d);
        \draw (b) -- (f);
        \draw (f) -- (g);
        
        \node at (5,0) [circle,draw] (e1) {$\bf 2$};
        \node at (6,0) [circle,draw] (a1) {$\bf 1$};
        \node at (7,0) [circle,draw] (b1) {$\bf 20$};
        \node at (8,0) [circle,draw] (c1) {$\bf 1$};
        \node at (9,0) [circle,draw] (d1) {$\bf 2$};
        \node at (7,1) [circle,draw] (f1) {$\bf 1$};
        \node at (7,2) [rectangle,draw] (g1) {$\bf 2$};
        \draw (e1) -- (a1);
        \draw (a1) -- (b1);
        \draw (b1) -- (c1);
        \draw (c1) -- (d1);
        \draw (b1) -- (f1);
        \draw (f1) -- (g1);
    \end{tikzpicture}
    \end{center}

    Both quiver diagrams represent vanishing equations.  Note
 $\det(C_{E_6})=3$.
    Thus there are in
    total $3\times(1+2)=9$ vanishing blowup equations.

  \item Consider the $E_7$ quiver with gauge group $\mf{su}(2d_i)$. In
    this case, the flavor symmetry $\mf{su}(16)$ of the center node
    breaks down to $\mf{su}(6)^2\times \mf{su}(4)$, and the flavor
    symmetry $\mf{su}(12)$ of the $\mf{su}(6)$ node breaks down to
    $\mf{su}(8)\times \mf{su}(4)$, and the flavor symmetry
    $\mf{su}(8)$ of the $\mf{su}(4)$ node breaks down to
    $\mf{su}(6)\times \mf{su}(2)$. Besides,
    $(P^\vee\slash Q^{\vee})_{A_7}=\IZ_8$. We summarize the $r$ fields
    behavior under the flavor group splitting in the following
    tables.
    \begin{table}[h]
      \begin{center}
      \resizebox{\linewidth}{!}{
        \begin{tabular}{|c|c|c|c|}
          \hline
          &	$\lG$ & $\lF$ & branching rules of $\lF$ \\
          \hline
          &   $\mf{su}(8)$ & $\mf{su}(16)$ & $\mf{su}(6)\times \mf{su}(6)\times \mf{su}(4)$ \\
          \hline
          u &  $\bf 1$ & $\cO_{\omega_8}=\bf 12870$ & \tabincell{c}{${\mathbf {(1,15,1)}}+\mathbf{(\bar{6},20,1)}+\mathbf{(\overline{15},\overline{15},1)}+\mathbf{(1,6,4)}+\mathbf{(\bar6,15,4)}+\mathbf{(\overline{15},20,4)}$ \\
          $+{\mathbf {(1,1,6)}}+{\mathbf {(\bar6,6,6)}}+{\mathbf {(\overline{15},15,6)}}+{\mathbf {(20,20,6)}}+\cdots$} \\
          \hline
          v & $\bf 8$ & $\cO_{\omega_{10}}=\bf \overline{8008}$  & \tabincell{c}{${\bf (1,1,1)}+\mathbf{(\bar6,6,1)}+\mathbf{(\overline{15},{15},1)}+\mathbf{(20,20,1)}+{\bf (6,1,\bar4)}+{\bf ({15},\bar{6},\bar4)}$ \\
          $+{\bf (20,\overline{15},\bar4)}+{\bf ({15},1,6)}+{\bf (20,\bar6,6)}+{\bf (\overline{15},\overline{15},6)}+\cdots$} \\
          \hline
          v  & $\bf 28$ & $\cO_{\omega_{12}}=\bf \overline{1820}$ & \tabincell{c}{${\bf ({15},1,1)}+{\bf (20,\bar6,1)}+{\bf (\overline{15},\overline{15},1)}+{\bf (20,1,\bar4)}+{\bf (\overline{15},\bar6,\bar4)}$\\
          $+{\bf (\overline{15},1,6)}+{\bf (\bar6,\bar6,6)}+\cdots$} \\
          \hline
          v  & $\bf 56$ & $\cO_{\omega_{14}}=\bf \overline{120}$  & ${\bf (\overline{15},1,1)}+{\bf (\bar6,\bar6,1)}+{\bf (\bar6,1,\bar4)}+{\bf (1,1,6)}+\cdots$ \\
          \hline
          v  & $\bf 70$ & $\cO_0=\bf 1$ & ${\bf (1,1,1)}$ \\
          \hline
          v & $\bf \overline{56}$ & $\cO_{\omega_2}=\bf 120$ & conjugate \\
          \hline
          v & $\bf \overline{28}$ & $\cO_{\omega_4}=\bf 1820$  & conjugate \\
          \hline
          v & $\bf \bar{8}$ &  $\cO_{\omega_6}=\bf 8008$ & conjugate \\
          \hline
        \end{tabular}}
      \end{center}
    \end{table}
    Note here the $\dots$ means conjugate representations and
    permutations over the first two $\mf{su}(6)$.
    \begin{table}[h]
      \begin{center}
        \begin{tabular}{|c|c|c|c|}
          \hline
          &	$\lG$ & $\lF$ & branching rules of $\lF$ \\
          \hline
          &   $\mf{su}(6)$ & $\mf{su}(12)$ & $\mf{su}(8)\times \mf{su}(4)$ \\
          \hline
          u &  $\bf 1$ & $\cO_{\omega_6}=\bf 924$ & ${\mathbf {(\overline{28},1)}}+\mathbf{(\overline{56},4)}+\mathbf{(70,6)}+\mathbf{(56,\bar4)}+\mathbf{(28,1)}$ \\
          \hline
          v & $\bf 6$ & $\cO_{\omega_8}=\bf \overline{495}$ & ${\bf (70,1)}+\mathbf{(\overline{56},\overline{4})}+\mathbf{(\overline{28},\overline{6})}+\mathbf{(\overline{8},4)}+{\bf (1,1)}$ \\
          \hline
          v  & $\bf 15$ & $\cO_{\omega_{10}}=\bf \overline{66}$ & ${\bf (\overline{28},1)}+{\bf (\overline{8},\overline{4})}+{\bf (1,\overline{6})}$  \\
          \hline
          v  & $\bf 20$ & $\cO_0=\bf 1$ & ${\bf (1,1,1)}$ \\
          \hline
          v & $\bf \overline{15}$ &  $\cO_{\omega_2}=\bf 66$ & ${\bf (28,1)}+{\bf (8,4)}+{\bf (1,6)}$\\
          \hline
          v & $\bf \bar{6}$ & $\cO_{\omega_4}=\bf 495$ & ${\bf (70,1)}+\mathbf{(56,4)}+\mathbf{(28,6)}+\mathbf{(8,\bar4)}+{\bf (1,1)}$  \\
          \hline
        \end{tabular}
      \end{center}
    \end{table}
    \begin{table}[h]
      \begin{center}
        \begin{tabular}{|c|c|c|c|}
          \hline
          &	$\lG$ & $\lF$ & branching rules of $\lF$ \\
          \hline
          &   $\mf{su}(4)$ & $\mf{su}(8)$ & $\mf{su}(2)\times \mf{su}(6)$ \\
          \hline
          u &  $\bf 1$ & $\cO_{\omega_4}=\bf 70$ & ${\bf (1,\overline{15})}+\mathbf{(2,20)}+\mathbf{(1,15)}$ \\
          \hline
          v & $\bf 4$ & $\cO_{\omega_6}=\bf \overline{28}$ & ${\bf (1,1)}+\mathbf{(2,6)}+\mathbf{(1,\overline{15})}$ \\
          \hline
          v  & $\bf 6$ & $\cO_0=\bf 1$ & ${\bf (1,1)}$ \\
          \hline
          v & $\bf \bar{4}$ & $\cO_{\omega_2}=\bf 28$ & ${\bf (1,1)}+\mathbf{(2,6)}+\mathbf{(1,15)}$ \\
          \hline
        \end{tabular}
      \end{center}
    \end{table}

    Now based on the general gluing procedure, we can directly write
    down all admissible blowup equations as:
    \begin{center}
      \begin{tikzpicture}
\node at (-2,0) [circle,draw] (h) {$\bf 2$};
\node at (-1,0) [circle,draw] (e) {$\bf 1$};
        \node at (0,0) [circle,draw] (a) {$\bf 20$};
        \node at (1,0) [circle,draw] (b) {$\bf 1$};
        \node at (2,0) [circle,draw] (c) {$\bf 20$};
        \node at (3,0) [circle,draw] (d) {$\bf 1$};
        \node at (1,1) [circle,draw] (f) {$\bf 6$};
        \node at (4,0) [rectangle,draw] (g) {$\bf 2$};
        \draw (e) -- (h);
        \draw (e) -- (a);
        \draw (a) -- (b);
        \draw (b) -- (c);
        \draw (c) -- (d);
        \draw (b) -- (f);
        \draw (d) -- (g);
    \end{tikzpicture}
    \end{center}

\begin{center}
\begin{tikzpicture}
\node at (-2,0) [circle,draw] (h) {$\bf 1$};
\node at (-1,0) [circle,draw] (e) {$\bf 6$};
        \node at (0,0) [circle,draw] (a) {$\bf 1$};
        \node at (1,0) [circle,draw] (b) {$\bf 70$};
        \node at (2,0) [circle,draw] (c) {$\bf 1$};
        \node at (3,0) [circle,draw] (d) {$\bf 6$};
        \node at (1,1) [circle,draw] (f) {$\bf 1$};
        \node at (4,0) [rectangle,draw] (g) {$\bf 1$};
        \draw (e) -- (h);
        \draw (e) -- (a);
        \draw (a) -- (b);
        \draw (b) -- (c);
        \draw (c) -- (d);
        \draw (b) -- (f);
        \draw (d) -- (g);
    \end{tikzpicture}
    \end{center}

    Both quiver diagrams represent vanishing equations.
    Note $\det(C_{E_7})=2$.
    Thus there are in
    total $2\times(2+1)=6$ vanishing blowup equations.

  \item Consider the $E_8$ quiver with gauge group $\mf{su}(2d_i)$. After a
    long but elementary computation on the representation
    decomposition like in the cases above, and based on the general
    gluing procedure, we can directly write down all admissible
    blowup equations as:
    \begin{center}
      \begin{tikzpicture}
\node at (-1,0) [circle,draw] (e) {$\bf 1$};
        \node at (0,0) [circle,draw] (a) {$\bf 70$};
        \node at (1,0) [circle,draw] (b) {$\bf 1$};
        \node at (2,0) [circle,draw] (c) {$\bf 252$};
        \node at (3,0) [circle,draw] (d) {$\bf 1$};
        \node at (1,1) [circle,draw] (f) {$\bf 20$};
        \node at (4,0) [circle,draw] (g) {$\bf 20$};
        \node at (5,0) [circle,draw] (h) {$\bf 1$};
        \node at (6,0) [rectangle,draw] (i) {$\bf 2$};
        \draw (i) -- (h);
        \draw (g) -- (h);
        \draw (e) -- (a);
        \draw (a) -- (b);
        \draw (b) -- (c);
        \draw (c) -- (d);
        \draw (b) -- (f);
        \draw (d) -- (g);
    \end{tikzpicture}
    \end{center}

\begin{center}
\begin{tikzpicture}
\node at (-1,0) [circle,draw] (e) {$\bf 6$};
        \node at (0,0) [circle,draw] (a) {$\bf 1$};
        \node at (1,0) [circle,draw] (b) {$\bf 924$};
        \node at (2,0) [circle,draw] (c) {$\bf 1$};
        \node at (3,0) [circle,draw] (d) {$\bf 70$};
        \node at (1,1) [circle,draw] (f) {$\bf 1$};
        \node at (4,0) [circle,draw] (g) {$\bf 1$};
        \node at (5,0) [circle,draw] (h) {$\bf 6$};
        \node at (6,0) [rectangle,draw] (i) {$\bf 1$};
        \draw (i) -- (h);
        \draw (g) -- (h);
        \draw (e) -- (a);
        \draw (a) -- (b);
        \draw (b) -- (c);
        \draw (c) -- (d);
        \draw (b) -- (f);
        \draw (d) -- (g);
    \end{tikzpicture}
    \end{center}

    Both quiver diagrams represent vanishing equations.
    Note $\det(C_{E_8})=1$.
    Thus there are in total
    $2+1=3$ vanishing blowup equations.
  \end{itemize}

  \subsubsection{Conformal matter theories}
  6d conformal matter theories are interesting SCFTs coming from
  M5-branes probing an ADE singularity in M-theory or intersecting an
  ADE singularity with a Horava-Witten M9-wall
  \cite{DelZotto:2014hpa}. The elliptic genera of these theories are
  rarely known except for a few rank one cases such as the $(D_N,D_N)$
  models. In the following, we present the blowup equations for all
  notable conformal matter theories. Note for all conformal matter
  theories except for $(\mf{sp}(n)$,$\mf{sp}(n))$ theory, the
  determinant of the intersection matrix of base curves is
  $\det(\Omega)=1$.\footnote{This property can be easily deduced from
    the fact that all these conformal matter theories can be blown
    down successively to one single $-1$ curve.} Therefore the number
  of non-equivalent blowup equations for each of these theories is
  just the number of non-equivalent admissible $r$ fields for the
  nodes.
\begin{itemize}
\item $(D_4,D_4)$ conformal matter theory is often denoted as
  $[D_4],1,[D_4]$. The elliptic genera of this theory can be computed
  from 2d quiver gauge theory
  \cite{Hayashi:2019fsa}. 
  This model is actually a special case of the E-string theory. The
  $E_8$ flavor group of node $1$ splits to
  $\mf{so}(8)\times \mf{so}(8)$. Since the vanishing $r$
  field of E-string theory decomposes as ${\bf 1}\to {\bf (1,1)}$, we
  obtain the following vanishing equation for $(D_4,D_4)$:
\begin{center}
  \begin{tikzpicture}
        \node at (0,0) [rectangle,draw] (a) {$\bf 1$};
        \node at (1,0) [circle,draw] (b) {$\phantom{o}$};
        \node at (2,0) [rectangle,draw] (c) {$\bf 1$};
        \draw (a) -- (b);
        \draw (b) -- (c);
    \end{tikzpicture}
\end{center}
On the other hand, the unity $r$ fields of E-string theory decompose
as
\begin{equation}
  {\bf 240}_2\to (\md{24}_2,\md{1})+ (\md{1},\md{24}_2)+{\bf (8_{\rm
      v},8_{\rm v})}+{\bf (8_{\rm c},8_{\rm s})}+{\bf (8_{\rm
      s},8_{\rm c})}.
\end{equation}
Apply the gluing rules, we find the following five types of unity
blowup equations:
\begin{center}
    \begin{tikzpicture}
        \node at (0,0) [rectangle,draw] (a) {$\bf 1$};
        \node at (1,0) [circle,draw] (b) {$\phantom{o}$};
        \node at (2,0) [rectangle,draw] (c) {$\md{24}$};
        \draw (a) -- (b);
        \draw (b) -- (c);
        
            \node at (4,0) [rectangle,draw] (a1) {$\md{24}$};
        \node at (5,0) [circle,draw] (b1) {$\phantom{o}$};
        \node at (6,0) [rectangle,draw] (c1) {$\bf 1$};
        \draw (a1) -- (b1);
        \draw (b1) -- (c1);
    \end{tikzpicture}
\end{center}

    \begin{center}
    \begin{tikzpicture}
        \node at (0,0) [rectangle,draw] (a) {$\bf 8_{\rm v}$};
        \node at (1,0) [circle,draw] (b) {$\phantom{o}$};
        \node at (2,0) [rectangle,draw] (c) {$\bf 8_{\rm v}$};
        \draw (a) -- (b);
        \draw (b) -- (c);
        
         \node at (4,0) [rectangle,draw] (a1) {$\bf 8_{\rm c}$};
        \node at (5,0) [circle,draw] (b1) {$\phantom{o}$};
        \node at (6,0) [rectangle,draw] (c1) {$\bf 8_{\rm s}$};
         \draw (a1) -- (b1);
        \draw (b1) -- (c1);
        
        \node at (8,0) [rectangle,draw] (a2) {$\bf 8_{\rm s}$};
        \node at (9,0) [circle,draw] (b2) {$\phantom{o}$};
        \node at (10,0) [rectangle,draw] (c2) {$\bf 8_{\rm c}$};
         \draw (a2) -- (b2);
        \draw (b2) -- (c2);
    \end{tikzpicture}
\end{center}

\item $(D_{N+4},D_{N+4})$ theories are often denoted as
  $[D_{N+4}],1_{\mf{sp}(N)},[D_{N+4}]$. For $N\ge 1$, the $D_{2(N+4)}$
  flavor group of the $n=1$ node splits to $D_{N+4}\times D_{N+4}$.
Under splitting $D_{2(N+4)}\to D_{N+4}\times D_{N+4}$,
\begin{equation}
S_{D_{2(N+4)}}\to (S_{D_{N+4}},C_{D_{N+4}})+(C_{D_{N+4}},S_{D_{N+4}}).
\end{equation}
Denote $\cO_N$ as the Weyl orbit of $\mf{sp}(N)$ generated by
weight $[0,0,\dots,0,1]$, and $V,S,C$ as the Weyl orbits
of $\mf{so}(N+4)$ generated by weights $[1,0,\ldots,0,0]$,
$[0,0,\ldots,0,1]$ and $[0,0,\ldots,1,0]$. Apply the gluing rules, we
find two types of unity blowup equations
    \begin{center}
    \begin{tikzpicture}
        \node at (0,0) [rectangle,draw] (a) {${S}$};
        \node at (1,0) [circle,draw] (b) {$\bf 1$};
        \node at (2,0) [rectangle,draw] (c) {$C$};
        \draw (a) -- (b);
        \draw (b) -- (c);
        
          \node at (4,0) [rectangle,draw] (a1) {${C}$};
        \node at (5,0) [circle,draw] (b1) {$\bf 1$};
        \node at (6,0) [rectangle,draw] (c1) {$S$};
        \draw (a1) -- (b1);
        \draw (b1) -- (c1);
    \end{tikzpicture}
\end{center}
There also exist numerous vanishing blowup equations. For example, for $N=1$ case, i.e. the $(\mf{so}(10),\mf{so}(10))$ model, the vanishing blowup equations are
\begin{center}
    \begin{tikzpicture}
        \node at (-0.1,0) [rectangle,draw] (a) {$\bf 10$};
        \node at (1,0) [circle,draw] (b) {$\cO_1$};
        \node at (2,0) [rectangle,draw] (c) {$\bf 1$};
        \draw (a) -- (b);
        \draw (b) -- (c);
        
         \node at (4,0) [rectangle,draw] (a1) {$\bf 1$};
        \node at (5,0) [circle,draw] (b1) {$\cO_1$};
        \node at (6.1,0) [rectangle,draw] (c1) {$\bf 10$};
        \draw (a1) -- (b1);
        \draw (b1) -- (c1);
    \end{tikzpicture}
\end{center}
For $N\ge2$, there exist many vanishing blowup equations including
\begin{center}
    \begin{tikzpicture}
        \node at (0,0) [rectangle,draw] (a) {$\bf 1$};
        \node at (1,0) [circle,draw] (b) {$\cO_N$};
        \node at (2,0) [rectangle,draw] (c) {$\bf 1$};
        \draw (a) -- (b);
        \draw (b) -- (c);
    \end{tikzpicture}
\end{center}

\item $(E_6,E_6)$ conformal matter theory is often denoted as
  $[E_6],1,3_{\mf{su}(3)},1,[E_6]$. The base curve intersection matrix
  $-\Omega$ has
\begin{equation}
\Omega=\left(
\begin{array}{ccc}
  1 & -1 & 0  \\
  -1 & 3 & -1  \\
  0 & -1 & 1  \\
\end{array}
\right).
\end{equation}
Note $\Det(\Omega)=1$. The $E_8$ flavor group of node $1$ splits to
$E_6\times \mf{su}(3)$ when coupled with $n=6$ $E_6$ gauge theory and
$n=3$ $\mf{su}(3)$ gauge theory. Since ${\bf 1}\to {\bf (1,1)}$ and
\begin{equation}
  {\bf 240}_2\to (\md{72}_2,\md{1})+(\md{27}_{4/3},\md{3}_{2/3})+
  (\overline{\md{27}}_{4/3},\overline{\md{3}}_{2/3})+ (\md{1},\md{6}_2),
\end{equation}
apply the gluing rule, we find one type of unity blowup equations
\begin{center}
    \begin{tikzpicture}
        \node at (0,0) [rectangle,draw] (a) {$\md{72}$};
        \node at (1,0) [circle,draw] (b) {$\phantom{o}$};
        \node at (2,0) [circle,draw] (c) {$\bf 1$};
        \node at (3,0) [circle,draw] (d) {$\phantom{o}$};
        \node at (4,0) [rectangle,draw] (e) {$\md{72}$};
        \draw (a) -- (b);
        \draw (b) -- (c);
        \draw (d) -- (c);
        \draw (d) -- (e);
    \end{tikzpicture}
\end{center}
and five types of vanishing blowup equations
    \begin{center}
    \begin{tikzpicture}
        \node at (0,0) [rectangle,draw] (a) {$\bf 1$};
        \node at (1,0) [circle,draw] (b) {$\phantom{o}$};
        \node at (2,0) [circle,draw] (c) {$\bf 1$};
        \node at (3,0) [circle,draw] (d) {$\phantom{o}$};
        \node at (4,0) [rectangle,draw] (e) {$\bf 1$};
        \draw (a) -- (b);
        \draw (b) -- (c);
        \draw (d) -- (c);
        \draw (d) -- (e);
    \end{tikzpicture}
\end{center}

\begin{center}
    \begin{tikzpicture}
        \node at (0,0) [rectangle,draw] (a) {$\bf 1$};
        \node at (1,0) [circle,draw] (b) {$\phantom{o}$};
        \node at (2,0) [circle,draw] (c) {$\bf 1$};
        \node at (3,0) [circle,draw] (d) {$\phantom{o}$};
        \node at (4,0) [rectangle,draw] (e) {$\md{72}$};
        \draw (a) -- (b);
        \draw (b) -- (c);
        \draw (d) -- (c);
        \draw (d) -- (e);
        
                \node at (6,0) [rectangle,draw] (a1) {$\md{72}$};
        \node at (7,0) [circle,draw] (b1) {$\phantom{o}$};
        \node at (8,0) [circle,draw] (c1) {$\bf 1$};
        \node at (9,0) [circle,draw] (d1) {$\phantom{o}$};
        \node at (10,0) [rectangle,draw] (e1) {$\bf 1$};
        \draw (a1) -- (b1);
        \draw (b1) -- (c1);
        \draw (d1) -- (c1);
        \draw (d1) -- (e1);
    \end{tikzpicture}
\end{center}

 \begin{center}
    \begin{tikzpicture}
        \node at (0,0) [rectangle,draw] (a) {$\bf 27$};
        \node at (1,0) [circle,draw] (b) {$\phantom{o}$};
        \node at (2,0) [circle,draw] (c) {$\bf 3$};
        \node at (3,0) [circle,draw] (d) {$\phantom{o}$};
        \node at (4,0) [rectangle,draw] (e) {$\bf 27$};
        \draw (a) -- (b);
        \draw (b) -- (c);
        \draw (d) -- (c);
        \draw (d) -- (e);
        
        \node at (6,0) [rectangle,draw] (a1) {$\bf \overline{27}$};
        \node at (7,0) [circle,draw] (b1) {$\phantom{o}$};
        \node at (8,0) [circle,draw] (c1) {$\bf \overline{3}$};
        \node at (9,0) [circle,draw] (d1) {$\phantom{o}$};
        \node at (10,0) [rectangle,draw] (e1) {$\bf \overline{27}$};
        \draw (a1) -- (b1);
        \draw (b1) -- (c1);
        \draw (d1) -- (c1);
        \draw (d1) -- (e1);
    \end{tikzpicture}
\end{center}

One can easily check the leading degree vanishing identities. For
example, the first vanish blowup equation has leading base degree as
\begin{equation}
  \Theta^{[a]}_{\Omega}\(\tau,\left(
    \begin{array}{c}
      \epsilon_1+\epsilon_2\\
      \epsilon_1+\epsilon_2\\
      \epsilon_1+\epsilon_2\\
    \end{array}
  \right)\)=0,
\label{thetaindentVIIVII1}
\end{equation}
while the second vanish blowup equation has leading base degree as
\begin{equation}\label{thetaindentVIIVII2}
  \Theta^{[a]}_{\Omega}\(\tau,\Omega^{-1}\left(
    \begin{array}{c}
      0\\
      \epsilon_1+\epsilon_2\\
      m_{\alpha}^{E_6}+\epsilon_1+\epsilon_2\\
    \end{array}
  \right)\)=0,
\end{equation}
and the forth
vanish blowup equation has leading base degree as
\begin{equation}
  \sum_{i=1,2,3}\Theta^{[a]}_{\Omega}\(\tau,\left(
    \begin{array}{c}
      0\\
      -m_i\\
     0\\
    \end{array}
  \right)+\Omega^{-1}\left(
    \begin{array}{c}
      m_{w}^{E_6}+\epsilon_1+\epsilon_2\\
      0\\
      m_{w'}^{E_6}+\epsilon_1+\epsilon_2\\
    \end{array}
  \right)\)\frac{1}{\prod_{j\neq i}\theta_1(m_i-m_j)}=0.
\label{thetaindentVIIVII3}
\end{equation}
Here the characteristic $a=(0,1/2,0)$ and $\alpha,\alpha'$ are
arbitrary roots of $E_6$, and $w,w'$ are
arbitrary weights of the fundamental representation $\bf 27$. Besides, $m_i,i=1,2,3$ are the $\mf{su}(3)$
fugacities satisfying $ m_1+m_2+m_3=0$. It is easy to check these identities are correct.
\item $(E_7,E_7)$ conformal matter theory is often denoted as
  $[E_7],1,2_{\mf{su}(2)},3_{\mf{so}(7)},2_{\mf{su}(2)},1,[E_7]$. The
  base curve intersection matrix $-\Omega$ has
  \begin{equation}
    \Omega=\left(
      \begin{array}{ccccc}
        1 & -1 & 0 & 0 & 0  \\
        -1 & 2 & -1 & 0 & 0  \\
        0 & -1 & 3 & -1 & 0  \\
        0 & 0 & -1 & 2 & -1  \\
        0 & 0 & 0 & -1 & 1  \\
      \end{array}
    \right).
  \end{equation}
  Note $\Det(\Omega)=1$. The $E_8$ flavor group of node $1$ splits to
  $E_7\times \mf{su}(2)$ when coupled with $n=8$ $E_7$ gauge theories
  and $n=2$ $\mf{su}(2)$ gauge theory. Since
  \begin{equation}
    {\bf 240}_2\to (\md{126}_2,\md{1})+(\md{56}_{3/2},\md{2}_{1/2})
    +(\md{1},\md{3}_2),
  \end{equation}
  apply the gluing rules, we find the following possible blowup
  equations which are all vanishing:

    \begin{center}
    \begin{tikzpicture}
        \node at (-0.5,0) [rectangle,draw] (a) {$\md{126/1}$};
        \node at (1,0) [circle,draw] (b) {$\phantom{o}$};
        \node at (2,0) [circle,draw] (c) {$\bf 1$};
        \node at (3,0) [circle,draw] (d) {$\bf 7$};
        \node at (4,0) [circle,draw] (e) {$\bf 1$};
        \node at (5,0) [circle,draw] (f) {$\phantom{o}$};
        \node at (6.5,0) [rectangle,draw] (g) {$\md{126/1}$};
        \draw (a) -- (b);
        \draw (b) -- (c);
        \draw (d) -- (c);
        \draw (d) -- (e);
        \draw (f) -- (e);
        \draw (f) -- (g);
    \end{tikzpicture}
\end{center}

 \begin{center}
    \begin{tikzpicture}
        \node at (0,0) [rectangle,draw] (a) {$\md{56}$};
        \node at (1,0) [circle,draw] (b) {$\phantom{o}$};
        \node at (2,0) [circle,draw] (c) {$\bf 2$};
        \node at (3,0) [circle,draw] (d) {$\bf 1$};
        \node at (4,0) [circle,draw] (e) {$\bf 2$};
        \node at (5,0) [circle,draw] (f) {$\phantom{o}$};
        \node at (6,0) [rectangle,draw] (g) {$\md{56}$};
        \draw (a) -- (b);
        \draw (b) -- (c);
        \draw (d) -- (c);
        \draw (d) -- (e);
        \draw (f) -- (e);
        \draw (f) -- (g);
    \end{tikzpicture}
\end{center}

For example, the second
type vanish blowup equation has leading base degree as
\begin{equation}
\sum_{\lambda_{a,b}=\pm 1/2}(-1)^{\lambda_a+\lambda_b}\Theta^{[a]}_{\Omega}\(\tau,\left(\begin{array}{c}
0\\
-\lambda_a m_a^{\mf{su}(2)}\\
0\\
-\lambda_b m_b^{\mf{su}(2)} \\
0\\
\end{array}
\right)+\Omega^{-1}\left(\begin{array}{c}
m_{w}+\epsilon_1+\epsilon_2\\
0\\
2(\epsilon_1+\epsilon_2)\\
0\\
m_{w'}+\epsilon_1+\epsilon_2\\
\end{array}
\right)\)=0.
\label{E7E7vanish2}
\end{equation}
Here the characteristic $a=(1/2,0,1/2,0,1/2)$, $w,w'\in\mathbf{56}$ of $E_7$, and $\mathcal{O}_{1/2,6}$ is the Weyl orbit $\mathcal{O}^{\mf{so}(7)}_{(100)}$. We have checked this identity
is correct.

\item $(E_8,E_8)$ theory is often denoted as
  $[E_8],1,2,2_{\mf{su}(2)},3_{G_2},1,5_{F_4},1,3_{G_2},2_{\mf{su}(2)},2,1,[E_8]$. The
  base curve intersection matrix $-\Omega$ has
  $\Det(\Omega)=1$. Apply the gluing rule, we find the following
  possible vanishing blowup equations:

  \begin{center}
    \begin{tikzpicture}
        \node at (-0.5,0) [rectangle,draw] (a) {${\bf 1}/{\bf 240}$};
        \node at (1,0) [circle,draw] (b) {$\phantom{o}$};
        \node at (2,0) [circle,draw] (c) {$\phantom{o}$};
        \node at (3,0) [circle,draw] (d) {$\bf 2$};
        \node at (4,0) [circle,draw] (e) {${\bf 1}$};
        \node at (5,0) [circle,draw] (f) {$\phantom{o}$};
        \node at (6,0) [circle,draw] (g) {$\bf 1$};
        \node at (7,0) [circle,draw] (h) {$\phantom{o}$};
        \node at (8,0) [circle,draw] (i) {$\bf 1$};
        \node at (9,0) [circle,draw] (j) {$\bf 2$};
        \node at (10,0) [circle,draw] (k) {$\phantom{o}$};
        \node at (11,0) [circle,draw] (l) {$\phantom{o}$};
        \node at (12.5,0) [rectangle,draw] (m) {${\bf 1}/{\bf 240}$};
        \draw (a) -- (b);
        \draw (b) -- (c);
        \draw (d) -- (c);
        \draw (d) -- (e);
        \draw (f) -- (e);
        \draw (f) -- (g);
        \draw (h) -- (g);
        \draw (h) -- (i);
        \draw (i) -- (j);
        \draw (j) -- (k);
        \draw (k) -- (l);
        \draw (l) -- (m);
    \end{tikzpicture}
\end{center}

Thus there is no unity and just one type of vanishing blowup
equations.

  \item $(E_7,\mf{so}(7))$ conformal matter theory is often denoted as
    $[E_7],1,2_{\mf{su}(2)},[\mf{so}(7)]$. Apply the gluing rule, we
    find the following possible unity blowup equation
    \begin{center}
    \begin{tikzpicture}
        \node at (-0.3,0) [rectangle,draw] (a) {$\bf 126$};
        \node at (1,0) [circle,draw] (b) {$\phantom{o}$};
        \node at (2,0) [circle,draw] (c) {$\bf 1$};
        \node at (3,0) [rectangle,draw] (d) {$\bf 6$};
        \draw (a) -- (b);
        \draw (b) -- (c);
        \draw (d) -- (c);
        \end{tikzpicture}
  \end{center}
    and vanishing blowup equations:
    \begin{center}
    \begin{tikzpicture}
        \node at (0,0) [rectangle,draw] (a) {$\bf 1$};
        \node at (1,0) [circle,draw] (b) {$\phantom{o}$};
        \node at (2,0) [circle,draw] (c) {$\bf 1$};
        \node at (3,0) [rectangle,draw] (d) {$\bf 6$};
        \draw (a) -- (b);
        \draw (b) -- (c);
        \draw (d) -- (c);
        
            \node at (5,0) [rectangle,draw] (a1) {$\bf 56$};
        \node at (6,0) [circle,draw] (b1) {$\phantom{o}$};
        \node at (7,0) [circle,draw] (c1) {$\bf 2$};
        \node at (8,0) [rectangle,draw] (d1) {$\bf 1$};
        \draw (a1) -- (b1);
        \draw (b1) -- (c1);
        \draw (d1) -- (c1);
    \end{tikzpicture}
  \end{center}

\item $(E_8,G_2)$ conformal matter theory is often denoted as
  $[E_8],1,2,2_{\mf{su}(2)},[{G_2}]$. Apply the gluing rule, we find
  the following possible vanishing blowup equations:

  \begin{center}
    \begin{tikzpicture}
        \node at (-0.5,0) [rectangle,draw] (a) {${\bf 1}/{\bf 240}$};
        \node at (1,0) [circle,draw] (b) {$\phantom{o}$};
        \node at (2,0) [circle,draw] (c) {$\phantom{o}$};
        \node at (3,0) [circle,draw] (d) {$\bf 2$};
        \node at (4,0) [rectangle,draw] (e) {$\bf 1$};
        \draw (a) -- (b);
        \draw (b) -- (c);
        \draw (d) -- (c);
        \draw (d) -- (e);
    \end{tikzpicture}
  \end{center}
\item $(E_8,F_4)$ conformal matter theory is often denoted as
  $[E_8],1,2,2_{\mf{su}(2)},3_{G_2},1,[{F_4}]$. Apply the gluing rule,
  we find the following possible blowup equations:

  \begin{center}
    \begin{tikzpicture}
        \node at (-0.5,0) [rectangle,draw] (a) {${\bf 1}/{\bf 240}$};
        \node at (1,0) [circle,draw] (b) {$\phantom{o}$};
        \node at (2,0) [circle,draw] (c) {$\phantom{o}$};
        \node at (3,0) [circle,draw] (d) {$\bf 2$};
        \node at (4,0) [circle,draw] (e) {$\bf 1$};
        \node at (5,0) [circle,draw] (f) {$\phantom{o}$};
        \node at (6,0) [rectangle,draw] (g) {$\bf 1$};
        \draw (a) -- (b);
        \draw (b) -- (c);
        \draw (d) -- (c);
        \draw (d) -- (e);
        \draw (f) -- (e);
        \draw (f) -- (g);
    \end{tikzpicture}
  \end{center}
\item $(\mf{sp}(N),\mf{sp}(N))$ conformal matter theory is often
  denoted as $[\mf{sp}(N)],4_{\mf{so}(2N+8)},[\mf{sp}(N)]$. The flavor
  $\mf{sp}(2N)$ of node 4 splits to $\mf{sp}(N)\times \mf{sp}(N)$.
  Apply the gluing rule, we find the following possible blowup
  equations: one type of unity equation
  \begin{center}
    \begin{tikzpicture}
        \node at (0,0) [rectangle,draw] (a) {$\cO_N$};
        \node at (1,0) [circle,draw] (b) {$\bf 1$};
        \node at (2,0) [rectangle,draw] (c) {$\cO_N$};
        \draw (a) -- (b);
        \draw (b) -- (c);
    \end{tikzpicture}
  \end{center}
  and lots of vanishing ones including
  \begin{center}
    \begin{tikzpicture}
        \node at (0,0) [rectangle,draw] (a) {$\cO_N$};
        \node at (1,0) [circle,draw] (b) {$V$};
        \node at (2,0) [rectangle,draw] (c) {$\cO_N$};
        \draw (a) -- (b);
        \draw (b) -- (c);
       
         \node at (4,0) [rectangle,draw] (a1) {$\bf 1$};
        \node at (5,0) [circle,draw] (b1) {$C$};
        \node at (6,0) [rectangle,draw] (c1) {$\bf 1$};
        \draw (a1) -- (b1);
        \draw (b1) -- (c1);
        
          \node at (8,0) [rectangle,draw] (a2) {$\bf 1$};
        \node at (9,0) [circle,draw] (b2) {$S$};
        \node at (10,0) [rectangle,draw] (c2) {$\bf 1$};
        \draw (a2) -- (b2);
        \draw (b2) -- (c2);
    \end{tikzpicture}
  \end{center}

\item $(E_8,\mf{su}(N))$ conformal matter theory is often denoted as
\begin{align} [E_{8}] \,\, 1 \,\, \overset{\mathfrak{su}(1)}{2}\,\,
  \overset{\mathfrak{su}(2)}{2 }\,\, ...\,\,
  \overset{\mathfrak{su}(N-1)}{2} \,\, [\mf{su}(N)]\, .
\end{align}
Apply the gluing rule, we find the following possible blowup
equations:
\begin{center}
    \begin{tikzpicture}
    \node at (-1.5,0) [rectangle,draw] (d) {${\bf 1}/{\bf 240}$};
        \node at (0,0) [circle,draw] (a) {$\phantom{o}$};
        \node at (1,0) [circle,draw] (b) {$\phantom{o}$};
        \node at (2,0) [circle,draw] (c) {$\bf 2$};
        \node at (3,0) [circle,draw] (e) {$\bf 1$};
        \node at (4,0) [circle,draw] (f) {$\bf 6$};
        \node at (5,0) [circle,draw] (g) {$\bf 1$};
        \node at (6,0) [circle,draw] (h) {$\bf 20$};
        \node at (7,0) [circle,draw] (i) {$\bf 1$};
        \node at (9,0) [circle,draw] (j) {$\cdots$};
        \node at (10,0) [rectangle,draw] (k) {$\cdots$};
        \draw (a) -- (d);
        \draw (a) -- (b);
        \draw (b) -- (c);
        \draw (c) -- (e);
        \draw (f) -- (e);
        \draw (f) -- (g);
        \draw (h) -- (g);
        \draw (h) -- (i);
        \draw[dotted] (j) -- (i);
        \draw (j) -- (k);
    \end{tikzpicture}
  \end{center}

  The $\lambda_G/\lambda_F$ field associated to the
  circular/rectangular node carrying gauge/flavor symmetry
  $\mf{su}(k)$ $(k=1,\ldots,N)$ is trivial if $k$ is odd and is a
  non-trivial weight vector belonging to the Weyl orbit $\cO_{k/2}$ if
  $k$ is even.
\item $(E_8,B_k/D_k)$ conformal matter theory is often denoted as
  \begin{align}
    [E_{8}] \,\, 1 \,\, 2 \,\, \overset{\mathfrak{su}(2)}{2}
    \,\, \overset{\mathfrak{g}_2}{3} \,\, 1 \,\,
    \overset{\mathfrak{so}(9)}{4}\,\, \overset{\mathfrak{sp}(1)}{1 }\,\,
    \overset{\mathfrak{so}(11)}{4}\,\, ...\,\,
    \overset{\mathfrak{sp}(k-4)}{1} \,\, [\mf{so}(2k)/\mf{so}(2k+1)]\, .
\end{align}
Apply the gluing rule, we find the following possible blowup
equations:
\begin{center}
    \begin{tikzpicture}
    \node at (-1.5,0) [rectangle,draw] (d) {${\bf 1}/{\bf 240}$};
        \node at (0,0) [circle,draw] (a) {$\phantom{o}$};
        \node at (1,0) [circle,draw] (b) {$\phantom{o}$};
        \node at (2,0) [circle,draw] (c) {$\bf 2$};
        \node at (3,0) [circle,draw] (e) {$\bf 1$};
        \node at (4,0) [circle,draw] (f) {$\phantom{o}$};
        \node at (5,0) [circle,draw] (g) {$\bf 1$};
        \node at (6,0) [circle,draw] (h) {$\omega_1$};
        \node at (7,0) [circle,draw] (i) {$\bf 1$};
        \node at (9,0) [circle,draw] (j) {$\omega_{k-4}$};
        \node at (10,0) [rectangle,draw] (k) {$\bf 1$};
        \draw (a) -- (d);
        \draw (a) -- (b);
        \draw (b) -- (c);
        \draw (c) -- (e);
        \draw (f) -- (e);
        \draw (f) -- (g);
        \draw (h) -- (g);
        \draw (h) -- (i);
        \draw[dotted] (j) -- (i);
        \draw (j) -- (k);
    \end{tikzpicture}
\end{center}
\item $(E_8,E_7)$ theory is often denoted as
  $[E_8],1,2,2_{\mf{su}(2)},3_{G_2},1,5_{F_4},1,3_{G_2},2_{\mf{su}(2)},1,[E_7]$.
  Apply the gluing rule, we find the following possible blowup
  equations:

    \begin{center}
    \begin{tikzpicture}
        \node at (-0.5,0) [rectangle,draw] (a) {${\bf 1}/{\bf 240}$};
        \node at (1,0) [circle,draw] (b) {$\phantom{o}$};
        \node at (2,0) [circle,draw] (c) {$\phantom{o}$};
        \node at (3,0) [circle,draw] (d) {$\bf 2$};
        \node at (4,0) [circle,draw] (e) {$\bf 1$};
        \node at (5,0) [circle,draw] (f) {$\phantom{o}$};
        \node at (6,0) [circle,draw] (g) {$\bf 1$};
        \node at (7,0) [circle,draw] (h) {$\phantom{o}$};
        \node at (8,0) [circle,draw] (i) {$\bf 1$};
        \node at (9,0) [circle,draw] (j) {$\bf 2$};
        \node at (10,0) [circle,draw] (k) {$\phantom{o}$};
        \node at (11,0) [rectangle,draw] (l) {$\bf 1$};
        \draw (a) -- (b);
        \draw (b) -- (c);
        \draw (d) -- (c);
        \draw (d) -- (e);
        \draw (f) -- (e);
        \draw (f) -- (g);
        \draw (h) -- (g);
        \draw (h) -- (i);
        \draw (i) -- (j);
        \draw (j) -- (k);
        \draw (k) -- (l);
    \end{tikzpicture}
  \end{center}

\item $(E_8,E_6)$ conformal matter theory is often denoted as
  \begin{equation}
    [E_8] \,\,1 \,\, 2 \,\,
    \overset{\mathfrak{su}(2)}{2 }\,\, \overset{\mathfrak{g}_{2}}{3
    }\,\, 1 \,\, \overset{\mathfrak{f}_{4}}{5 }\,\, 1 \,\,
    \overset{\mathfrak{su}(3)}{3 }\,\, 1 \,\, [E_6]\, .
  \end{equation}
  Apply the gluing rule, we find the following possible blowup
  equations:
  \begin{center}
    \begin{tikzpicture}
        \node at (-0.5,0) [rectangle,draw] (a) {${\bf 1}/{\bf 240}$};
        \node at (1,0) [circle,draw] (b) {$\phantom{o}$};
        \node at (2,0) [circle,draw] (c) {$\phantom{o}$};
        \node at (3,0) [circle,draw] (d) {$\bf 2$};
        \node at (4,0) [circle,draw] (e) {$\bf 1$};
        \node at (5,0) [circle,draw] (f) {$\phantom{o}$};
        \node at (6,0) [circle,draw] (g) {$\bf 1$};
        \node at (7,0) [circle,draw] (h) {$\phantom{o}$};
        \node at (8,0) [circle,draw] (i) {$\bf 1$};
        \node at (9,0) [circle,draw] (j) {$\phantom{o}$};
        \node at (10,0) [rectangle,draw] (k) {$\bf 1$};
        \draw (a) -- (b);
        \draw (b) -- (c);
        \draw (d) -- (c);
        \draw (d) -- (e);
        \draw (f) -- (e);
        \draw (f) -- (g);
        \draw (h) -- (g);
        \draw (h) -- (i);
        \draw (i) -- (j);
        \draw (j) -- (k);
    \end{tikzpicture}
  \end{center}

\item $(E_7,D_4)$ conformal matter theory is often denoted as
  \begin{equation}
    [E_7] \,\,1 \,\, \overset{\mathfrak{su}_{2}}{2 }
    \,\, \overset{\mathfrak{g}_2}{3 }\,\, 1 \,\, [\mf{so}(8)]\, .
  \end{equation}
  Apply the gluing rule, we find the following possible blowup
  equations:
  \begin{center}
    \begin{tikzpicture}
        \node at (0,0) [rectangle,draw] (a) {$\bf 56$};
        \node at (1,0) [circle,draw] (b) {$\phantom{o}$};
        \node at (2,0) [circle,draw] (c) {$\bf 2$};
        \node at (3,0) [circle,draw] (d) {$\bf 1$};
        \node at (4,0) [circle,draw] (e) {$\phantom{o}$};
        \node at (5,0) [rectangle,draw] (f) {$\bf 1$};
        \draw (a) -- (b);
        \draw (b) -- (c);
        \draw (d) -- (c);
        \draw (d) -- (e);
        \draw (f) -- (e);
    \end{tikzpicture}
  \end{center}
\end{itemize}

\subsubsection{Blowups of $n=9, 10, 11$ curves}
The rank one theories with $n=9, 10, 11$ do not admit Kodaira-Tate
elliptic fibers. One needs to do further blowups which result in
higher dimensional tensor branches. There are normally several ways to
do this, see for example \cite{Heckman:2018jxk}. The toric
construction of some blown-up Calabi-Yau geometries were given in
\cite{Haghighat:2014vxa}. For $n=11$ curve, one blows up once and gets
theory $12_{E_8},1$. It is easy to find the following vanishing blowup
equation for it:
\begin{center}
  \begin{tikzpicture}
        \node at (1,0) [circle,draw] (b) {$\bf 1$};
        \node at (2,0) [circle,draw] (c) {$\phantom{1}$};
        \draw (c) -- (b);
      \end{tikzpicture}
\end{center}

For $n=10$, one blows up twice and gets $1,12_{E_8},1$ with vanishing
blowup equation
\begin{center}
    \begin{tikzpicture}
    \node at (0,0) [circle,draw] (a) {$\phantom{1}$};
        \node at (1,0) [circle,draw] (b) {$\bf 1$};
        \node at (2,0) [circle,draw] (c) {$\phantom{1}$};
        \draw (c) -- (b);
        \draw (a) -- (b);
    \end{tikzpicture}
\end{center}
or $12_{E_8},1,2$ with vanishing blowup equations
\begin{center}
    \begin{tikzpicture}
        \node at (1,0) [circle,draw] (b) {$\bf 1$};
        \node at (2,0) [circle,draw] (c) {$\phantom{1}$};
         \node at (3,0) [circle,draw] (a) {$\phantom{1}$};
         \node at (4,0) [rectangle,draw] (d) {$\bf 2$};
        \draw (c) -- (b);
        \draw (a) -- (c);
        \draw (d) -- (a);
    \end{tikzpicture}
\end{center}

For $n=9$, one blows up twice and gets $1,\overset{1}{12}_{E_8},1$
with vanishing blowup equation
\begin{center}
    \begin{tikzpicture}
    \node at (0,0) [circle,draw] (a) {$\phantom{1}$};
        \node at (1,0) [circle,draw] (b) {$\bf 1$};
        \node at (2,0) [circle,draw] (c) {$\phantom{1}$};
        \node at (1,1) [circle,draw] (d) {$\phantom{1}$};
        \draw (c) -- (b);
        \draw (a) -- (b);
        \draw (d) -- (b);
    \end{tikzpicture}
\end{center}
or $1,12_{E_8},1,2$ with vanishing blowup equations
\begin{center}
    \begin{tikzpicture}
    \node at (0,0) [circle,draw] (e) {$\phantom{1}$};
        \node at (1,0) [circle,draw] (b) {$\bf 1$};
        \node at (2,0) [circle,draw] (c) {$\phantom{1}$};
         \node at (3,0) [circle,draw] (a) {$\phantom{1}$};
         \node at (4,0) [rectangle,draw] (d) {$\bf 2$};
         \draw (e) -- (b);
        \draw (c) -- (b);
        \draw (a) -- (c);
        \draw (d) -- (a);
    \end{tikzpicture}
\end{center}
or $12_{E_8},1,2,2$ with vanishing blowup equation
\begin{center}
    \begin{tikzpicture}
    \node at (5,0) [rectangle,draw] (e) {$\bf{1}$};
        \node at (1,0) [circle,draw] (b) {$\bf 1$};
        \node at (2,0) [circle,draw] (c) {$\phantom{1}$};
         \node at (3,0) [circle,draw] (a) {$\phantom{1}$};
         \node at (4,0) [circle,draw] (d) {$\bf 2$};
         \draw (e) -- (d);
        \draw (c) -- (b);
        \draw (a) -- (c);
        \draw (d) -- (a);
    \end{tikzpicture}
\end{center}

Let us now take a closer look at the first example the $12_{E_8},1$
theory.
The intersection matrix between the two base curves is just
\begin{equation}
  \Omega=\left(
    \begin{array}{ccccc}
      12 & -1  \\
      -1 & 1  \\
    \end{array}
  \right),
\end{equation}
thus we have $\det(\Omega)=11$ vanishing blowup equations. Since there
is only one $E_8$ vector multiplet and no hypermultiplet, the leading
base degree of the vanishing blowup equations can be simply written as
\begin{equation}\label{vanishingblowup11}
\Theta^{[a]}_{\Omega}\(\tau,\left(
\begin{array}{ccccc}
  5/33  \\
 5/33  \\
\end{array}
\right)\epsilon_+\)=0.
\end{equation}
We have checked this identity up to $\Qtau^{30}$. Remember here
characteristics $a$ are associated to $\Omega$ as defined in
(\ref{generaldefa}).


As a similar example, we consider the blown-up of a $-7$ curve, which
can be represented as $8_{E_7},1,[\mf{su}(2)]$. There are two types of
vanishing blowup equations:
\begin{center}
    \begin{tikzpicture}
        \node at (2,0) [circle,draw] (c) {$\bf{1}$};
         \node at (3,0) [circle,draw] (a) {$\phantom{1}$};
         \node at (4,0) [rectangle,draw] (d) {$\bf 1$};
        \draw (a) -- (c);
        \draw (d) -- (a);

        \node at (6,0) [circle,draw] (c1) {$\bf{56}$};
         \node at (7,0) [circle,draw] (a1) {$\phantom{1}$};
         \node at (8,0) [rectangle,draw] (d1) {$\bf 2$};
        \draw (a1) -- (c1);
        \draw (d1) -- (a1);
    \end{tikzpicture}
\end{center}
In fact, for any of $-2,-3,-4,-5,-7,-11$ curves, one can blowup once
and obtain a rank two theory which is the coupling between a pure
gauge minimal 6d SCFT and the E-string theory. For these rank two
theories, there always exists one type of vanishing blowup equations
represented as
\begin{center}
    \begin{tikzpicture}
        \node at (2,0) [circle,draw] (c) {$\bf{1}$};
         \node at (3,0) [circle,draw] (a) {$\phantom{1}$};
         \node at (4,0) [rectangle,draw] (d) {$\bf 1$};
        \draw (a) -- (c);
        \draw (d) -- (a);
    \end{tikzpicture}
\end{center}
The leading base degree of the vanishing equations are due to the
following identity:
\begin{equation}\label{vanishingblowupn}
\Theta^{[a]}_{\Omega}\(\tau,\left(
\begin{array}{ccccc}
  z  \\
 z  \\
\end{array}
\right)\)=0,
\end{equation}
where
\begin{equation}
  \Omega=\left(
    \begin{array}{ccccc}
      n & -1  \\
      -1 & 1  \\
    \end{array}
  \right).
\end{equation}
In fact, this identity holds for arbitrary $n\ge 2$.

\subsection{Remarks on solving elliptic genera}
For higher rank theories, there in general seems to be no efficient
way to solve elliptic genera from elliptic blowup equations. The main
reason as mentioned before is that there usually only exist vanishing
blowup equations for higher rank theories which do not give enough
constraints. Besides, even in the rare cases where exist unity blowup
equations, we can hardly make use of the equations to solve elliptic
genera.\footnote{The higher rank theories with unity blowup equations
  include for example all $A,D$ type chain of $-2$ curves with gauge
  symmetry and $(E_6,E_6)$ conformal matter theory.} Naively, one may
think there could exist some explicit higher dimensional recursion
formulas analogous to the rank one cases as long as there exist three
or more unity blowup equations. Unfortunately, because any such
higher-rank theory involves $-2$ or $-1$ curves, the recursion fails
when one of these curves is left but all other base curves are
decompactified. Therefore, in some sense, \emph{all higher-rank
  theories with unity blowup equations are in class $ \mathbf{ B}$ as
  in Section~\ref{sc:sol}, and all those with only vanishing blowup
  equations are in class $ \mathbf{ C}$.} Let us consider a good
example, the $A_2$ chain with gauge symmetry $\mf{su}(N)$ on each
node. For arbitrary $N$, there always exist unity blowup equations:
\begin{center}
\begin{tikzpicture}
        \node at (0,0) [rectangle,draw] (a) {$\bf 1$};
        \node at (1,0) [circle,draw] (b) {$\bf 1$};
        \node at (2,0) [circle,draw] (c) {$\bf 1$};
        \node at (3,0) [rectangle,draw] (d) {$\bf 1$};
        \draw (a) -- (b);
        \draw (b) -- (c);
        \draw (c) -- (d);
    \end{tikzpicture}
    \end{center}
Since the intersection matrix between the base classes is
\begin{equation}
  \Omega=\left(
    \begin{array}{ccccc}
      2 & -1  \\
      -1 & 2  \\
    \end{array}
  \right),
\end{equation}
we have in total $\det(\Omega)=3$ non-equivalent unity blowup
equations. To solve elliptic genus say $\IE_{2,1}$ by recursion, one
need to know $\IE_{2,0},\IE_{1,1},\IE_{1,0},\IE_{0,1}$ as initial
data. However, all the essentially rank one elliptic genera
$\IE_{n,0}$ and $\IE_{0,n}$ are not possible to solve by recursion as
they are in class $ \mathbf{ B}$ of rank one theories. In fact, when
one decompactifies the right $-2$ curve, the three unity blowup
equations will reduce to just two non-equivalent unity equations of
the left $-2$ curve which are just the two unity equations of the
$n=2$, $G=\mf{su}(N)$ theory. Thus there are not enough unity equations
to proceed with the recursion. See the detailed analysis for the
degeneration of M-M string chain in Section 3.3 of
\cite{Gu:2019pqj}. Nevertheless, from the perspective of the
$\eq,\et $ expansion, the refined BPS expansion or the Weyl orbit
expansion, one can still get some constraints. We do not pursue this
direction further since the perfect 2d quiver description were already
found for these higher-rank theories.

\section{Conclusions and Outlook}

In this paper, we obtained the elliptic blowup equations for all 6d
$(1,0)$ SCFTs in the atomic classification. In particular, we studied
extensively all the rank one theories which are labeled by an integer
$n$, a gauge symmetry $G$ and a flavor symmetry $F$. We divide these
theories into three classes: class $\bf A$ ($n\ge 3$) and class
$\bf B$ ($n=1,2$) contain theories without unpaired half
hypermultiplet, which make up the most of the rank one list, while
class $\bf C$ contains the remaining 12 theories with unpaired half
hypermultiplets. We find that for classes $\bf A$ and $\bf B$, there
always exist unity blowup equations and possibly also vanishing blowup
equations, while for class $\bf C$, there only exist vanishing blowup
equations. This has the following implications for the solvability of
the elliptic genera from the blowup equations: For class $\bf A$, we
obtain a recursion formula that determines the elliptic genera
completely, i.e.\ for arbitrary numbers of strings from the unity
blowup equations, which is the ideal situation. For class $\bf B$, we
can solve the elliptic genera and the refined BPS invariants order by
order from the Weyl orbit expansion, the refined BPS expansion or the
$\eq,\et$ expansion. For class $\bf C$, we do not have a universal
description how to solve elliptic genera from vanishing blowup
equations. For the classes $\bf A$ and $\bf B$, we have checked that
our results for elliptic genera agree with all previous partial
results from 2d quiver gauge theories, 5d partition functions and the
modular ansatz. For class $\bf C$, although we could not solve the
full elliptic genera. However we checked theta identities in leading
base degree and showed that part of the refined BPS invariants can be
determined. We expect that these theories can still be completely
solved by combining the constraints from the vanishing blowup
equations, the modular ansatz and the Higgsing conditions.

We also propose the elliptic blowup equations for three higher rank
non-Higgsable clusters, which only have vanishing equations due to the
presence of unpaired half hypermultiplets. We checked our blowup
equations by using the elliptic genera computed by localization
formulas in the 2d quiver construction in \cite{Kim:2018gjo} and by
analysing the base curve decompactification limits where they reduce
correctly to the blowup equations of rank one theories. We further
give {\emph {gluing rules}} which make it easy to write down all
admissible blowup equations for any 6d $(1,0)$ SCFTs in the atomic
classification. In particular, we explicitly present the blowup
equations for ADE chains of $-2$ curves, conformal matter theories and
the blown-ups of $-9,-10,-11$ curves. Most of the higher rank theories
only have vanishing blowup equations. One can even consider the blowup
equations for little string theories. We leave this for future study.

The solution of the theories with the blowup equations proceeds in two
steps: first one establishes the validity of blowup equations by
demonstrating that the partition functions or equivalently the
elliptic genera satisfy these equations; in the second step one has to
develop efficient procedures to solve the blowup equations.  Albeit
the first step has been very successful for all kinds of theories
including all 6d $(1,0)$ SCFTs in the atomic classification, there is
still uncertainty in the second step. In particular we do not know
what precise conditions or inputs are needed for a complete solution
in general.  In practice, we have developed several efficient
techniques to extract information from blowup equations, such as
recursion formulas, Weyl orbit expansions, refined BPS expansions, and
the $\eq,\et$ expansion.  Each method typically requires different
inputs. For the theories of class $\bf C$ with only vanishing blowup
equations, we do not know in general what the minimal inputs should
be. This may cast some doubts on the solvability of blowup equations
in the general situation, since even the seemingly simple $n=7,G=E_7$
theory can not be solved completely. One remedy may be to combine the
blowup equations and modular ansatz together. Another possible remedy
draws inspiration from the massless E-string theory, which corresponds
to a naturally realised \localelliptic Calabi-Yau 3-fold with two
K\"ahler parameters. Although the theory itself has only one vanishing
blowup equation and is therefore not solvable, once one of the eight
possible mass parameters is turned on, there are enough unity blowup
equations to allow for the complete solution of the theory.  The
example suggests that in some cases one can recover the necessary
unity blowup equations after deforming the theory with additional
natural parameters with a mass scale.

The 6d $(1,0)$ SCFTs we are considering have to be compactified on a
torus in order for elliptic genera to be defined.  In the low energy
limit, these theories can be equivalently seen as 5d KK/marginal
theories \cite{Jefferson:2018irk} compactified on a circle, where the
radius of the other circle in the 6d theory is identified with the KK
scale.  They can then be reduced to 5d SCFTs either by decoupling mass
deformed hypermultiplets which corresponds to flopping $(-1)$ curves
out of compact
surfaces~\cite{Jefferson:2018irk,Bhardwaj:2018yhy,Bhardwaj:2018vuu,Apruzzi:2019vpe,Apruzzi:2019opn,Apruzzi:2019enx},
or by decoupling a gauge sector, which corresponds to decompactifying
the surfaces
themselves\cite{DelZotto:2017pti,Bhardwaj:2019xeg,Apruzzi:2019kgb}.
The simplest 5d SCFTs are the infinite coupling limit of 5d gauge
theories, possibly with matter.  On the one hand, the blowup equations
of 5d gauge theories have been studied for all simple Lie groups in
\cite{Keller:2012da} and for all possible matter contents in
\cite{Kim:2019uqw}.  On the other hand, we have developed techniques
in our previous papers to reduce blowup equations of 6d SCFTs to those
of 5d SCFTs through either of the two methods
\cite{Gu:2018gmy,Gu:2019pqj}.  In this paper we do not go into details
about reducing the large collection of 6d blowup equations to 5d
equations. Nevertheless, we do point out and present many new blowup
equations for 5d gauge theories beyond those found
in~\cite{Keller:2012da} and~\cite{Kim:2019uqw}.  The blowup equations
for 5d SCFTs obtained in this way could be helpful for solving the BPS
states of these theories \cite{Huang:2013yta,Apruzzi:2019opn}.

There are many open problems.  First of all, we expect blowup
equations to exist for many other field theories, for instance 6d
SCFTs with ``frozen singularity''
\cite{Tachikawa:2015wka,Bhardwaj:2018jgp} not covered in the atomic
classification, 6d SCFTs with twisted compactification on circle
\cite{Bhardwaj:2019fzv}, and little string theories
\cite{Bhardwaj:2015oru}.  Furthermore, in the second paper of this
series \cite{Gu:2019dan}, we studied a surprising conjectural relation
\cite{DelZotto:2017mee} between the elliptic genera of pure gauge 6d
$(1,0)$ SCFTs and the Schur indices of 4d $\mathcal{N}=2$ $H_G$ SCFTs,
and generalized it from one string elliptic genera to higher
strings. For theories with matter, it was identified in
\cite{DelZotto:2018tcj} that the worldsheet $(0,4)$ theories also
correspond to some 4d $\mathcal{N}=2$ SCFTs but with some $(0,4)$
surface defects. The Schur indices of such configurations have rarely
been studied, see some relevant results in \cite{Pan:2017zie}. It is
interesting to see if the Schur indices of such 4d SCFTs with $(0,4)$
defects are also related to the elliptic genera of 6d $(1,0)$ SCFTs
with matter.

\section*{Acknowledgement}

We would like to thank Michele Del Zotto, Hirotaka Hayashi, Amir-Kian Kashani-Poor,
Joonho Kim, Sung-Soo Kim, Sheng Meng, Yiwen Pan, Hiraku Nakajima, Satoshi Nawata, Haowu Wang, Rui-Dong Zhu and especially Guglielmo Lockhart for valuable
discussions. Part of this work has been presented in Oxford, ICTP and
Naples in 2019.  The work of J.G. is supported in part by the Fonds
National Suisse, subsidy 200021-175539 and by the NCCR 51NF40-182902
``The Mathematics of Physics'' (SwissMAP).
\appendix

\section{Lie algebraic convention}
\label{sc:Lie}

We collect some definitions in (affine) Lie algebras and fix our
convention used throughout the paper.

\subsection{Definitions and convention}
\label{sc:con}


Given a simple Lie algebra $\fg$ of rank $r$, there are four important
$r$-dimensional lattices: the root and coroot lattices $Q, Q^{\vee}$,
as well as the weight and coweight lattices\footnote{The coweight
  lattice is sometimes called the magnetic weight lattice in the
  literature, e.g.~\cite{Kapustin:2005py}.} $P, P^{\vee}$. They are
related to each other by
\begin{gather}
  Q^\vee \subset P^\vee \subset \fh_{\IC} \ ,\\
  Q \subset P \subset \fh^*_{\IC} \ ,
\end{gather}
where $\fh_{\IC}, \fh_{\IC}^* \cong \IC^r$ denote the complexified
Cartan subalgebra and its dual equipped with the natural pairing
\begin{equation}
  \vev{\bullet,\bullet}: \fh_{\IC}^* \times \fh_{\IC} \to \IC \ .
\end{equation}
The root and coroot lattices $Q,\Qv$ are spanned by the simple roots
$\alpha_i$ and the simple coroots $\av_j$, whose pairings are entries
of the Cartan matrix $A$
\begin{equation}
  \vev{\alpha_i,\av_j} = A_{ij}.
\end{equation}
The weight and coweight lattices $P,\Pv$ are spanned by the
fundamental weights $\w_i$ and the fundamental coweights $\wv_i$,
defined through
\begin{equation}
  \vev{\alpha_i,\wv_j} = \vev{\w_i, \av_j} = \delta_{ij}\ ,
\end{equation}
in other words, they are the duals of the coroot and the
root lattices respectively.  Every weight vector $\w$ can be
represented by the coefficients $\lambda_i$ in its decomposition in
terms of the fundamental weights, which are called the Dynkin labels
\begin{equation}
  \w = \sum_i \lambda_i \w_i.
\end{equation}
A weight vector is said to be \emph{dominant} if all of its Dynkin
labels are non-negative integers.
Likewise, we can represent a coweight vector $\wv$ by the coefficients
$\lambda_i^\vee$ in its decomposition in terms of the fundamental
coweights
\begin{equation}
  \wv = \sum_i \lambda^\vee_i \wv_i.
\end{equation}
We will also call $\lambda^\vee_i$ the Dynkin labels of the coweight
$\w^\vee$ and say the coweight vector is dominant if all
$\lambda^\vee_i$ are non-negative.  Dominant (co)weight vectors can be
used to label Weyl orbits as each Weyl orbit of (co)weight vectors has
one and only one dominant element.

We define the Weyl invariant bilinear form $(\bullet,\bullet)$ on
$\fh_\IC$ by
\begin{equation}\label{eq:bilin-A}
  (k,\ell) := \frac{1}{2\hg}\sum_{\alpha\in\Delta}
  \vev{\alpha,k}\vev{\alpha,\ell},
  \quad k,\ell\in\fh_\IC,
\end{equation}
where $\hg$ is the dual Coxeter number of $\fg$.  It has the nice
property that the norm $||k||^2 = (k,k)$ of any coroot is an even
integer, and in particular the norm of the shortest non-zero coroot
$\theta^\vee$ is two.  Note that the dual Coxeter number $\hg$ can be
interpreted as the Dynkin index of the adjoint representation
$\mf{adj}$, while for an arbitrary representation $R$ its Dynkin index
$\ind_R$ is defined by \cite{DiFrancesco:1997nk}
\begin{equation}
  \tr_R(\mc{R}(J^a)\mc{R}(J^b)) = 2\ind_R \delta_{ab},
\end{equation}
where $\mc{R}(J^a)$ is the matrix representation of the generator
$J^a$ of $\fg$.  Consequently the bilinear form \eqref{eq:bilin-A} can
be expressed in terms of any representation $R$ of $\fg$ through
\begin{equation}\label{eq:bilin-R}
    (k,\ell) = \frac{1}{2\ind_R}\sum_{\w\in R}
  \vev{\w,k}\vev{\w,\ell},  \quad k,\ell\in\fh_\IC,
\end{equation}
where we have used the same symbol $R$ for the weight space of the
representation.

\begin{figure}
\center
\includegraphics[width=0.8\textwidth]{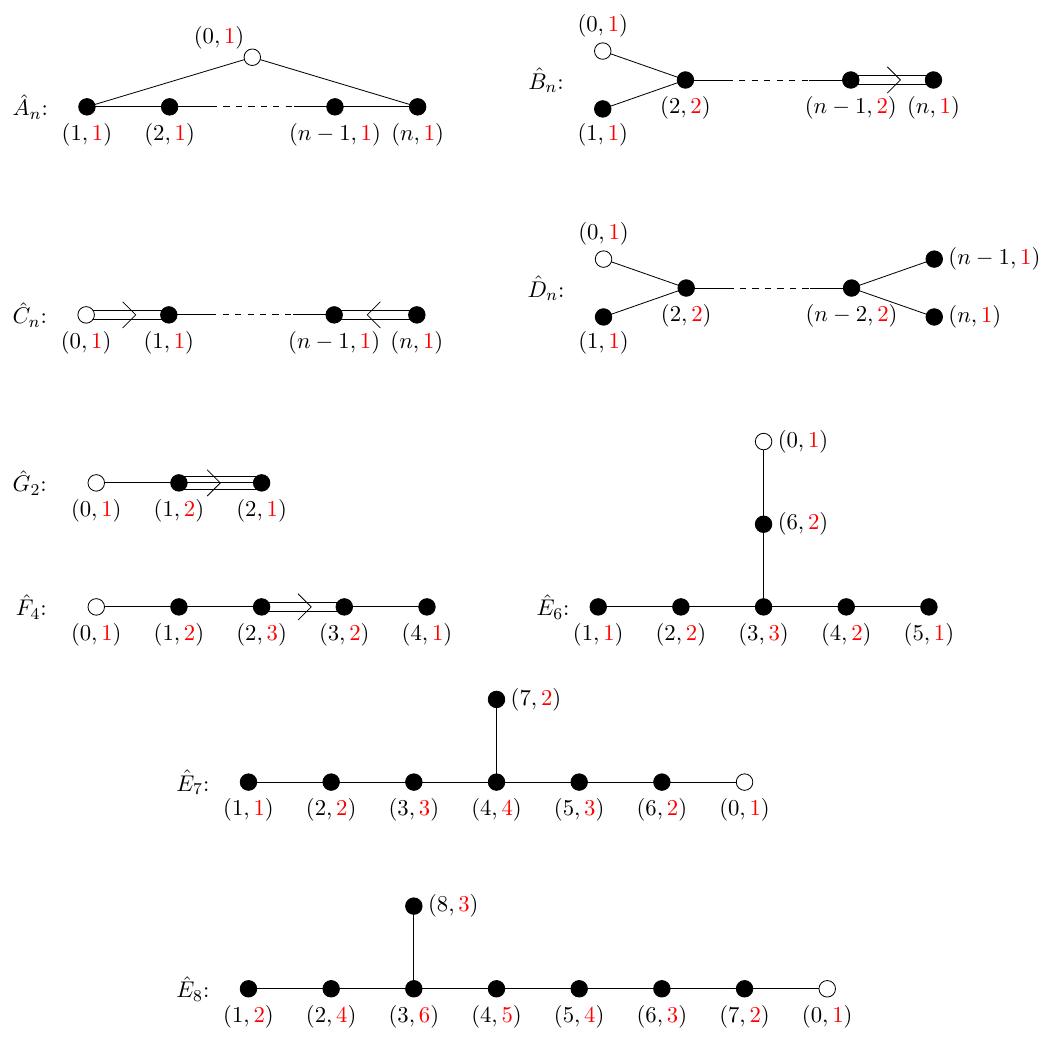}
\caption{Affine Dynkin diagrams associated to simple Lie algebras. The
  $i$-th node with comark $m_i$ is labeled by the pair $(i,m_i)$ where
  the number $m_i$ is colored in red.  In each diagram, the white node
  is the affine node, and the black nodes are nodes of simple Lie
  algebra. The arrows point from short coroots to long coroots.  We
  follow the same node order and same representation names as in the
  \texttt{LieART} package \cite{Feger:2012bs,Feger:2019tvk} of
  \texttt{Mathematica}.}\label{fg:dynkin}
\end{figure}

The bilinear form $(\bullet,\bullet)$ is symmetric and non-degenerate.
It then defines an isomorphism from $\fh_\IC$ to $\fh_\IC^*$ by
\begin{align}
  \varphi: \fh_{\IC}
  &\xrightarrow{\sim} \fh_\IC^*\nn
    k &\mapsto \varphi(k) =  (k,\bullet);
                           \label{eq:iso}
\end{align}
in other words, we have
\begin{equation}
  \vev{\varphi(k),\ell} = (k,\ell),\quad \forall \ell \in \fh_\IC.
\end{equation}
The isomorphism then induces a Weyl invariant bilinear form on
$\fh_\IC^*$
\begin{equation}
  (\omega,\eta) = \vev{\omega,\varphi^{-1}(\eta)} =
  (\varphi^{-1}(\omega),\varphi^{-1}(\eta)),\quad \omega,\eta\in\fh_{\IC}^{*}.
\end{equation}
Concretely we have
\begin{equation}\label{eq:isoaw}
  \varphi(\av_i) = \frac{||\av_i||^2}{2}\alpha_i,\quad
  \varphi(\wv_i) = \frac{||\av_i||^2}{2} \w_i.
\end{equation}
It is easy to see that the Dynkin labels $\lambda^\vee_i$ of a
coweight $\w^\vee$ and the Dynkin labels $\lambda_i$ of its isomorphic
weight vector $\w = \varphi(\w^\vee)$ are related by
\begin{equation}\label{eq:isolamb}
  \lambda_i = \lambda^\vee_i \frac{||\alpha^\vee_i||^2}{2}.
\end{equation}
We list below the norms of simple coroots of simple Lie algebras used
in this paper.
\begin{itemize}
\item $A_n,D_n,E_{6,7,8}$: These are simply laced Lie algebras and all
  the simple coroots have norm 2.
\item $B_n (n\geq 2)$:
  \begin{equation}
    ||\av_i||^2 = 2,\;\; i=1,\ldots,n-1, \quad ||\av_n||^2  =4.
  \end{equation}
\item $C_n (n\geq 2)$:
  \begin{equation}
    ||\av_i||^2 = 4,\;\; i=1,\ldots,n-1, \quad ||\av_n||^2  =2.
  \end{equation}
\item $G_2$: 
  \begin{equation}
    ||\av_1||^2 = 2, \quad ||\av_2||^2  =6.
  \end{equation}
\item $F_4$:
  \begin{equation}
    ||\av_1||^2=||\av_2||^2 = 2, \quad ||\av_3||^2=||\av_4||^2  =4.
  \end{equation}
\end{itemize}
 We give in
Fig.~\ref{fg:dynkin} the affine Dynkin diagrams of simple Lie algebras
and the ordering of nodes used in our paper.

In the main text, to lighten notation we use $\cdot$ to represent both
the pairing $\vev{\bullet,\bullet}$ and the bilinear form
$(\bullet,\bullet)$.  Hopefully the actual meaning of $\cdot$ will be
clear from the context.

\subsection{Lie sub-algebra decomposition}
\label{sc:br}

Let $\fg'$ be a Lie sub-algebra of $\fg$.  The weight lattice $P$ of
$\fg$ can be mapped to the weight lattice $P'$ of $\fg'$ by a
surjective map
\begin{equation}\label{eq:f}
  f: P\twoheadrightarrow P'.
\end{equation}
This map induces an injective map from $Q^{\vee,'}$ to $Q^\vee$
\begin{equation}\label{eq:f*}
  f^*: Q^{\vee,'} \hookrightarrow Q^\vee
\end{equation}
defined by
\begin{equation}\label{eq:fZ}
  \vev{\w,f^*(\alpha^{\vee,'})} = \vev{f(\w),\alpha^{\vee,'}} \in \IZ,\quad \forall
  \w\in P,
\end{equation}
for $\alpha^{\vee,'}\in Q^{\vee,'}$.
It is easy to see that $f^*$ is indeed an injection.  If
$f^*(\alpha^{\vee,'}) = f^*(\beta^{\vee,'})$, we have
\begin{equation}
  \vev{f(\w),\alpha^{\vee,'}-\beta^{\vee,'}} =
  \vev{w,f^*(\alpha^{\vee,'})-f^*(\beta^{\vee,'})} = 0,\quad \forall \w\in P
\end{equation}
Since $f(\w)$ runs over $P'$, we must have $\alpha^{\vee,'}-\beta^{\vee,'}=0$.
We can also show that
\begin{equation}\label{eq:imagf*}
  \imag(f^*) = \{\alpha^\vee\in Q^\vee \,|\, \vev{\w,\alpha^\vee} = 0
  ,\;\forall \w\in \Ker P\}.
\end{equation}
First of all $\imag(f^*)$ is a subset of the r.h.s., since if
$\alpha^\vee\in \imag(f^*)$, there exists $\alpha^{\vee,'}$ such that
$\forall \w\in\Ker P$
\begin{equation}
  \vev{\w,\alpha^\vee} = \vev{f(\w),\alpha^{\vee,'}} =
  \vev{0,\alpha^{\vee,'}} = 0.
\end{equation}
On the other hand, both the l.h.s.~and r.h.s.~of \eqref{eq:imagf*} are
linear spaces, and they have the same dimension.  Therefore they must
be identical.  Finally, by definition, the pair of maps $f,f^*$
preserve the natural pairing $\vev{\bullet,\bullet}$.  We can also
deduce that they preserve the bilinear form $(\bullet,\bullet)$ on
coroot lattices as well.  This is because given the images
$f^*(\alpha^{\vee,'}),f^*(\beta^{\vee,'}) \in Q^{\vee}$ of two vectors
$\alpha^{\vee,'},\beta^{\vee,'}$ in $Q^{\vee,'}$, we can define their
bilinear form by an arbitrary representation $R$ in $\fg$
\begin{align}
  (f^*(\alpha^{\vee,'}),f^*(\beta^{\vee,'})) =
  &\frac{1}{2\ind_R}
    \sum_{\w\in R}
    \vev{\w,f^*(\alpha^{\vee,'})}\vev{\w,f^*(\beta^{\vee,'})}\nn =
  &\frac{1}{2\ind_{f(R)}}\sum_{f(\w)\in f(R)}
    \vev{f(\w),\alpha^{\vee,'}}\vev{f(\w),\beta^{\vee,'}}\nn =
  &(\alpha^{\vee,'},\beta^{\vee,'}),
    \label{eq:proj-()}
\end{align}
where in the second step, we used the definition of the pull-back map
$f^*$ and that the representation index does not change under the
projection of weight spaces\footnote{If an irreducible representation
  decomposes to multiple irreducible representations after the
  projection of weight space, the index of the composite
  representation is the sum of the indices of the individual
  irreducible representations.}.

We comment that the injective map $f^*$ \emph{cannot} be extended to a
map from $P^{\vee,'}$ to $P$.  One important reason the map between
coroot lattices can be defined is that the pairing between $P$ and
$Q^\vee$ always takes value in $\IZ$, and thus the definition
\eqref{eq:fZ} makes sense.  On the other hand, the pairing between $P$
and $P^\vee$ takes value in different domains for different Lie
algebras and \eqref{eq:fZ} will no longer make sense.  For instance,
it takes value in $\IZ/2$ for $E_7$ and in $\IZ/3$ for $E_6$.

\section{List of results}\label{app:list}
For the convenience of readers, we make a list for the computational
results in this paper.
\begin{itemize}
\item Elliptic genera\\[+2mm]
  Although our computation on the elliptic genera of rank one 6d
  $(1,0)$ SCFTs mostly contain all gauge and flavor fugacities, for
  some $\mf{so}(N)$ theories we only present the results with all
  fugacities turned off. For most theories especially the exceptional
  theories, we not only show the elliptic genera with fugacities
  turned off, but also the $v$ expansion with gauge and flavor
  fugacities turned
  on.\\[+2mm]
  Class $\bf A$
  \begin{itemize}
\item $n=3$, $G=\mf{so}(7)$, $\IE_1$ (\ref{n3SO7E1}), $\IE_2$ (\ref{n3SO7E2})\\
\phantom{$n=3$,} $G=\mf{so}(8)$, $\IE_1$ (\ref{n3SO8E1}), $\IE_2$ (\ref{n3SO8E2})\\
\phantom{$n=3$,} $G=\mf{so}(9)$, $\IE_1$ (\ref{n3SO9E1}), $\IE_2$ (\ref{n3SO9E2})\\
\phantom{$n=3$,} $G=\mf{so}(10)$, $\IE_1$ (\ref{n3SO10E1}), $\IE_2$ (\ref{n3SO10E2})\\
\phantom{$n=3$,} $G=G_2$, $\IE_1$ (\ref{n3G2E1}), $\IE_2$ (\ref{n3G2E2})\\
\phantom{$n=3$,} $G=F_4$, $\IE_1$ (\ref{n3F4E1}), $\IE_2$ (\ref{n3F4E2})\\
\phantom{$n=3$,} $G=E_6$, $\IE_1$ (\ref{n3E6E1}), $\IE_2$ (\ref{n3E6E2})
\item $n=4$, $G=\mf{so}(9)$, $\IE_1$ (\ref{n4SO9E1}), $\IE_2$ (\ref{n4SO9E2})\\
\phantom{$n=3$,} $G=\mf{so}(10)$, $\IE_1$ (\ref{n4SO10E1}), $\IE_2$ (\ref{n4SO10E2})\\
\phantom{$n=3$,} $G=F_4$, $\IE_1$ (\ref{n4F4E1}), $\IE_2$ (\ref{n4F4E2})\\
\phantom{$n=3$,} $G=E_6$, $\IE_1$ (\ref{n4E6E1}), $\IE_2$ (\ref{n4E6E2})\\
\phantom{$n=3$,} $G=E_7$, $\IE_1$ (\ref{n4E7E1})
\item $n=5$, $G=E_6$, $\IE_1$ (\ref{n5E6E1}), $\IE_2$ (\ref{n5E6E2})
\item $n=6$, $G=E_7$, $\IE_1$ (\ref{n6E7E1})
\end{itemize}
Class $\bf B$
\begin{itemize}
\item $n=1$, $G=\mf{su}(3)$, $\IE_1$ (\ref{n1su3E1})\\
\phantom{$n=1$,} $G=\mf{su}(4)$, $\IE_1$ (\ref{n1su4E1})\\
\phantom{$n=1$,} $G=\mf{so}(7)$, $\IE_1$ (\ref{n1so7E1})\\
\phantom{$n=1$,} $G=\mf{so}(8)$, $\IE_1$ (\ref{n1so8E1}) \\
\phantom{$n=1$,} $G=\mf{so}(9)$, $\IE_1$ (\ref{n1so9E1})\\
\phantom{$n=1$,} $G=G_2$, $\IE_1$ (\ref{n1g2E1})\\
\phantom{$n=1$,} $G=F_4$, $\IE_1$ (\ref{n1F4E1})\\
\phantom{$n=1$,} $G=E_6$, $\IE_1$ (\ref{n1E6E1})
\item $n=2$, $G=\mf{so}(9)$, $\IE_1$ (\ref{n2so9E1})\\
\phantom{$n=2$,} $G=\mf{so}(10)$, $\IE_1$ (\ref{n2so10E1})\\
\phantom{$n=2$,} $G=\mf{so}(11)$, $\IE_1$ (\ref{n2so11E1})\\
\phantom{$n=2$,} $G=\mf{so}(12)_a$, $\IE_1$ (\ref{n2so12E1})\\
\phantom{$n=2$,} $G=G_2$, $\IE_1$ (\ref{n2g2E1}) \\
\phantom{$n=2$,} $G=F_4$, $\IE_1$ (\ref{n2F4E1})\\
\phantom{$n=2$,} $G=E_6$, $\IE_1$ (\ref{n2E6E1})\\
\phantom{$n=2$,} $G=E_7$, $\IE_1$ (\ref{n2E7E1})
\end{itemize}
\item Exact $v$ expansion formulas for 5d one-instanton Nekrasov partition functions\\[+2mm]
  For a lot of rank-one theories with matters, the exact formulas for
  the $v$ expansion of 5d one-instanton partition function have been
  proposed in \cite{DelZotto:2018tcj} and \cite{Kim:2019uqw}. In this
  paper, we further obtain the exact formulas for the following new
  theories
\begin{itemize}
\item $n=1$, $G=\mf{su}(3)$ (\ref{n1SU3HS}), $\mf{su}(4)$ (\ref{n1SU4HS}), $\mf{so}(7)$ (\ref{n1SO7HS}), $\mf{so}(8)$ (\ref{n1SO8HS}), $\mf{so}(9)$ (\ref{n1SO9HS})\\
  \phantom{$n=1$, $G=$} $G_2$ (\ref{n1g2hs}), $F_4$
  (\ref{n1F4exactE1}), $E_6$ (\ref{n1E6exactE1})
\item $n=2$, $G=\mf{so}(9)$ (\ref{n2so9HS}), $\mf{so}(10)$ (\ref{n2so10HS}), $\mf{so}(11)$ (\ref{n2SO11HS}), $\mf{so}(12)_a$ (\ref{n2so12HS}),\\
  \phantom{$n=2$, $G=$} $E_6$ (\ref{n2E6HS}), $E_7$ (\ref{n2E7HS})
\item $n=3$, $G=\mf{so}(12)$ (\ref{n3SO12HS}), $E_6$ (\ref{n3E6exactE1})
\item $n=4$, $G=E_6$ (\ref{n4E6HS}), $E_7$ (\ref{eq:exactn4E7Z1})
\end{itemize}
\item Modular ansatz\\[+2mm]
  Among the ten theories whose modular ansatz for reduced one-string
  elliptic genus were not fixed in \cite{DelZotto:2018tcj}, five of
  them listed below belong to class $\bf A$ or $\bf B$. Benefitting
  from blowup equations, we are able to determine their modular
  ansatz. See results in the \texttt{Mathematica} file
  \texttt{ModularAnsatzAppendix.nb} on the website \cite{kl}.
\begin{itemize}
\item $n=1$, $G=E_6$
\item $n=2$, $G=\mf{so}(11)$, $E_6$, $E_7$
\item $n=4$, $G=E_7$
\end{itemize}
\item Calabi-Yau construction and triple intersection numbers\\[+2mm]
  We give the polytope, the Mori cone generators and triple
  intersection ring for the non-compact elliptic Calabi-Yau threefolds associated to
  the following theories
\begin{itemize}
\item $n=1$, $G=G_2$ (\ref{n1G2polytope}, \ref{n1G2triple})
\item $n=2$, $G=G_2$ (\ref{n2G2polytope}, \ref{n2G2triple})
\item $n=3$, $G=G_2$ (\ref{n3G2polytope2}, \ref{n3G2triple})
\item $n=3$, $G=\mf{so}(7)$ (\ref{n3SO7polytope}, \ref{n3SO7triple})
\item $n=7$, $G=E_7$ (\ref{n7E7polytope}, \ref{n7E7triple})
\item NHC $3,2$ (\ref{NHC32polytope}, \ref{32triple})
\item NHC $3,2,2$ (\ref{322polytope}, \ref{322triple})
\item NHC $2,3,2$ (\ref{232polytope}, \ref{232triple})
\end{itemize}
\item Refined BPS invariants
\begin{itemize}
\item $n=1$, $G=G_2$ (Table \ref{tb:n1,G_2-BPS})
\item $n=2$, $G=G_2$ (Table \ref{tb:n2,G_2-BPS})
\item $n=3$, $G=G_2$ (Table \ref{tb:n3,G_2-BPS})
\item $n=3$, $G=\mf{so}(7)$ (Table \ref{tb:n3,SO(7)-BPS})
\item $n=7$, $G=E_7$  (Table \ref{tb:n7, E_7-BPS})
\end{itemize}
\item Vanishing theta identities\\[+2mm]
We checked the leading degree identities for all the vanishing blowup equations in Table \ref{tb:vbeq-1} and \ref{tb:vbeq-2} up to $\mathcal{O}(q^{20})$. We write down the explicit form of the vanishing identities for the following theories:
\begin{itemize}
\item $n=1$, $G=\mf{su}(3)$ (\ref{thetaindentI}, \ref{thetaindentIeasy})
\item $n=1$, $G=\mf{su}(N)$ (\ref{thetaindentn1suN}, \ref{thetaindentn1suNw2}, \ref{thetaindentn1suNwk}, \ref{thetaindentn1suNwmiddle})
\item $n=1$, $G=\mf{sp}(N)$ (\ref{spvanishj0}, \ref{sp1vanish})
\item $n=2$, $G=\mf{su}(N)$ (\ref{suvanishodd1}, \ref{suvanishodd2}, \ref{suvanish2}, \ref{suvanishomega1})
\item $n=3$, $G=\mf{so}(7)$ (\ref{n3so7v1}, \ref{n3so7v2})
\item $n=4$, $G=\mf{so}(8+N)$ (\ref{n4sov}, \ref{n4sov2})
\item $n=1,2,\dots,6$, $G=E_6$ (\ref{E6vanish0}, \ref{E6vanish00}, \ref{n5E6vanish0more}, \ref{allnE6vanish0more})
\item $n=1,2,\dots,8$, $G=E_7$ (\ref{E7vanish0}, \ref{E7vanish00}, \ref{n7vanish0})
\item NHC $3,2$ (\ref{nhc32vid})
\item NHC $3,2,2$ (\ref{nhc322vid})
\item NHC $2,3,2$ (\ref{nhc232vid}, \ref{nhc232vid2})
\item $D_4$ quiver of $-2$ curves (\ref{D4quiverthetaindent1}, \ref{D4quiverthetaindent2})
\item $(E_6,E_6)$ conformal matter (\ref{thetaindentVIIVII1}, \ref{thetaindentVIIVII2}, \ref{thetaindentVIIVII3})
\item $(E_7,E_7)$ conformal matter (\ref{E7E7vanish1}, \ref{E7E7vanish2})
\item blown-up of $-n$ curve with $n=2,3,4,5,7,9,10,11$ (\ref{vanishingblowup11}, \ref{vanishingblowupn})
\end{itemize}
\end{itemize}

\section{Semi-classical free energy for higher rank theories}
\label{sc:hrp}

In this section, we give the genus zero and genus one free energies at
large volume limit for a generic higher rank SCFT from gluing of rank
one Calabi-Yau three-folds. When we say gluing
Calabi-Yau three-folds, we put their cycles together, which means the
B-periods should have the same expression as they were in the rank one
theories. Simply integrating over periods in rank one
theories, we get the tree level prepotential of higher rank
theories. By adding the one loop contributions, for a
generic 6d (1,0) SCFT, the classical prepotential at the large volume
limit is
\begin{equation}\label{conjF0highrank}
  \begin{split}
    {F}^{(0,0)}=&-\frac{1}{6}\sum_i\sum_{\alpha\in\Delta_{i,+}}(\alpha\cdot\und{m}_{G_i})^3
    +
    \frac{1}{12}\sum_{i,j}\sum_{\w_{G_i,G_j}\in \mfr{R}_{G_i,G_j}^{+}}
    (\w_{G_i}\cdot\und{m}_{G_i}+\w_{G_j}\cdot\und{m}_{G_j})^3\\[-2mm]
    &-\frac{1}{2}(t_{\text{ell},i}-(\fn_i-2)\tau/2)\Omega_{ij}^{-1}(-k_{F,j}m_{F_j} \cdot m_{F_j} +\sum_{k}\Omega_{jk}m_{G_k} \cdot m_{G_k})\\[-2mm]
&-\frac{1}{2}t_{\text{ell},i}\Omega_{ij}^{-1}t_{\text{ell},j}\tau+\mathcal{O}(\tau^3),
\end{split}
\end{equation}
where $-\Omega_{ij}$ is the intersection matrix of the base,
$\Omega_{ij}^{-1}=(\Omega^{-1})_{ij}$. Note here $\mfr{R}_{G_i,G_j}$ are all the possible
bi-representations, also including cases when one of the group is
flavor group $F_j$.

For genus one part, we have 
\begin{align}
  b_i^{(1,0)}t_i=
  &\frac{1}{12}\sum_i\sum_{\alpha \in
    \Delta_{i,+}}{\alpha} \cdot t+\frac{1}{48}\sum_{i,j}\sum_{{\omega
    \in \mfr{R}_{G_i,F_j}^{+}}}{\omega} \cdot
    t+\frac{1}{4}\sum_{i,j} \Omega_{ij}^{-1}
    (\fn_j-2-{\hG}_j)t_{\text{ell},i},
    \label{bi1hr}\\
  b_i^{(0,1)}t_i=
  &-\frac{1}{12}\sum_i\sum_{\alpha \in
    \Delta_{i,+}}{\alpha} \cdot t+\frac{1}{24}\sum_{i,j}\sum_{{\omega
    \in \mfr{R}_{G_i,F_j}^{+}}}{\omega} \cdot
    t+\frac{1}{2}\sum_{i,j} \Omega_{ij}^{-1}
    ({\fn}_j-2)t_{\text{ell},i}.
    \label{bi2hr}
\end{align}
Here, the
$b_{\text{ell},i}^{(0,1)}=\frac{1}{24}\int c_2\wedge J_{\text{ell},i}$
has a geometric meaning.  For complete intersetion Calabi-Yau
threefolds, it is proved that \cite{Hosono:1993qy,Hosono:1994ax}
\begin{equation}\label{biterm}
  \int c_2\wedge J_a=\frac{1}{2}
  \sum_{bc}{\kappa}_{abc}(l_0^bl_0^c-\sum_{i>0}l^b_il^c_i),
\end{equation}
where ${\kappa}_{abc}$ are the triple intersection
numbers, and $l^a_i$ the components of the Mori cone vector $l^a$. It
seems this formula is correct for both compact and
non-compact Calabi-Yau hypersurfaces \cite{Gu:2019dan}. By turning off all the gauge
fugacities, the Mori cone vector for $\tau, t_{\text{ell},i}$ can be
write down effectively as
\begin{equation}\label{eq:ap:ltau}
l_{\tau}=\{-6;2,3,1,0,\cdots\},
\end{equation}
\begin{equation}\label{eq:ap:lB}
l_{B,i}=\{0;0,0,{\fn}_i-2,1,-\fn_i,1,0,0,\cdots\},
\end{equation}
with $l_{B,i}=l_{\text{ell},i}+l_{\tau}({\fn}_i-2)/2$.  Strictly
speaking, (\ref{eq:ap:ltau}) (\ref{eq:ap:lB}) are correct only for
A-type bases, which are realized by a smooth toric surface with self
intersection number $-\fn_i$. Together with (\ref{conjF0highrank}) and
(\ref{biterm}), if we want to compute $b_{\text{ell},i}^{(0,1)}$, we
fix $a$ in (\ref{biterm}) to the base direction then we only encounter
products of $l$ vector between $l_{\tau}$ and $l_{\text{ell},i}$,
which means we can always compute the product locally, as a $A_1$ type
base and therefore we conclude that they are effectively true even for
$D,\, E$ type bases when we compute $b_{\text{ell},i}^{(0,1)}$.  Then
we have
\begin{equation}
  b_{\text{ell},i}^{(0,1)}=\frac{1}{24}\int c_2\wedge
  J_{\text{ell},i}=
  \frac{1}{24}(\sum_{j} \Omega_{ij}^{-1}(\fn_j-2)+\frac{1}{2}
  \sum_{j} \Omega_{ij}^{-1} ({\fn}_j-2)(36-9-4-1))=
  \frac{1}{2}\sum_{j} \Omega_{ij}^{-1}(\fn_j-2),
\end{equation}
as we predicted in (\ref{bi2hr}).

\section{Derivation of the elliptic blowup equations}\label{sc:dbe}

In this appendix, we derive the elliptic blowup equations from the
blowup equations for refined topological strings. This procedure is
called de-affinazation which we have elaborated in length in
\cite{Gu:2019dan}. Therefore we will be very brief here. Remember that
in the blowup equations for refined topological strings, there is
always a ${B}$ field shift to the instanton part
\cite{Huang:2017mis}. This shift is trivial for 6d pure gauge
theories, but play a crucial rule in theory with matters.

It is well-known the hypermultiplets in representation $\mathfrak{R}$
contributes to $Z_{\text{1-loop}}$ as
\begin{equation}\label{Ghyper}
Z_{H}^{\mathfrak{R}}=\mathrm{PE}\bigg(f_{(0,0)}(q_1,q_2)\Big(\sum_{w\in \mathfrak{R}_+(G)}\Big(Q^{w}+\frac{\Qtau}{Q^{w}}\Big)\Big)
\Big(\frac{1}{1-\Qtau}\Big)\bigg).
\end{equation}
Recall the definition (\ref{fjljr}), here
$f_{(0,0)}(q_1,q_2)=((q_1^{1/2}-q_1^{-1/2})(q_2^{1/2}-q_2^{-1/2}))^{-1}$,
which represents the spin $(0,0)$ nature of hypermultiplets. For
general local Calabi-Yau, the spin $(j_L,j_R)$ of the non-vanishing
refined BPS invariants at degree $d$ have a checkerboard pattern,
i.e. satisfy the $B$ field condition. The ${B}$ field is defined by
\begin{equation}
2j_L+2j_R+1={B}\cdot d \mod 2.
\end{equation}
For spin $(0,0)$, the $B$ fields always belong to $\IZ+1/2$. Then
recall the definition of $Bl$ function in (\ref{Bldef}), the $Z_{H}$
contributes to the blowup equations as
\begin{equation}\label{Z0hyp}
\begin{aligned}
  \mathrm{PE}&\(\Big(Bl_{(0,0,R_z)}(q_1,q_2)(-Q_z)+Bl_{(0,0,-R_z)}(q_1,q_2){\Qtau\over -Q_z}\Big)\Big({1\over {1-\Qtau}}\Big)\)\\
  =\,
  \Big(\frac{Q_z^{1/2}}{\Qtau^{1/12}}&\Big)^{\frac{(R_z^2-1/4)}{2}}(q_1q_2)^{\frac{(R_z^2-1/4)R_z}{12}}\prod_{\displaystyle
    \substack{0\leq m,n\\m+n\leq
      R_z-3/2}}\frac{\theta_2(z+(m+1/2)\eq+(n+1/2)\et)}{\eta}.
\end{aligned}
\end{equation}
Here $R_z>0$ and $\mathrm{PE}$ is the plethystic exponent operator defined as
$ \mathrm{PE}[f(x)]=\exp[\,\sum_{n=1}^\infty\frac{1}{n}f(x^n)]$. Note
in (\ref{Z0hyp}), the Jacobi theta function is $\theta_2$ rather than
$\theta_1$ because of the ${B}$ field shift. Shifting back the $B$ field gives a factor $(-1)^{R_z}$ as is already shown in  \eqref{eq:thetaH*}. On the other hand, the
vector multiplet with spin $(0,1/2)$ has $B$ field as integers and
contributes to blowup equations as \cite{Gu:2018gmy,Gu:2019dan}
\begin{equation}\label{Z0vec}
\begin{aligned}
  \,&\mathrm{PE}\(-\Big(Bl_{(0,1/2,R_z)}(q_1,q_2)Q_z+Bl_{(0,1/2,-R_z)}(q_1,q_2){\Qtau\over
    Q_z}\Big)\Big({1\over {1-\Qtau}}\Big)\)
  =\big(\ri\Qtau^{1/12} Q_z^{-1/2}\big)^{R_z^2}\\
  \times&(q_1q_2)^{-\frac{(R_z-1)R_z(R_z+1)}{6}}\prod_{\displaystyle
    \substack{0\leq m,n\\m+n\leq
      R_z-1}}\frac{\eta}{\theta_1(z+m\eq+n\et)}\prod_{\displaystyle
    \substack{0\leq m,n\\m+n\leq
      R_z-2}}\frac{\eta}{\theta_1(z+(m+1)\eq+(n+1)\et)}.
\end{aligned}
\end{equation}
Here $R_z\ge 0$. The tensor multiplet does contribute to
$Z_{\text{1-loop}}$, but decouples from elliptic
blowup equations. See more in \cite{Gu:2019dan}.

Consider only the $R_z^3(\ep_1+\ep_2)$ terms, it is easy to see they
cancel with the summation over the positive roots and half weights in
(\ref{conjF0highrank}) if the half weights belong to the same set. The
remaining $\theta_2$ part is not sensitive to the choices of half
weights, which indicates that different Calabi-Yau phases give the
same blowup equation.

In (\ref{Z0vec}) and (\ref{Z0hyp}), there are $\tau$ linear part, this
part will break the modularity in our blowup equation. The
cancellation of the $\tau$ linear term gives a
constraint on the representations of the theory  
\begin{small}
\begin{equation}\label{D4}
  \sum_i \frac{3(\fn_i-2)-{\hG}_i}{{\hG}_i} \sum_{\alpha \in
    \Delta^+_i}({\alpha} \cdot t)^2=-\frac{1}{2}
  \sum_{i,j}\sum_{(\omega_i,\omega_j) \in
    \mfr{R}_{G_i,G_j}^{+}}(\omega_i \cdot m_{G_i}+\omega_j \cdot m_{G_j})^2-\frac{1}{2}  \sum_{i,j}\sum_{(\omega_i,\omega_j) \in \mfr{R}_{G_i,F_j}^{+}}({\omega_i } \cdot m_{G_i})^2.
\end{equation}
\end{small}
In rank one theories, the above constraint can be simplified as
\begin{equation}\label{constraintrep}
  \frac{\hG-3(\fn-2)}{12\hG} \sum_{\alpha \in \Delta}({\alpha} \cdot t)^2=\frac{\dim{\frak{R}_F}}{24} \sum_{\omega \in \frak{R}_{G}}({\omega } \cdot t)^2.
\end{equation}
By basic properties of root lattice and weight lattice, we obtain the
following useful constraint for arbitrary rank-one 6d $(1,0)$ SCFTs:
\begin{equation}\label{id1}
   2\hG- 6({\fn-2})={\dim{\frak{R}_F}} \text{ind} \frak{R}_G.
\end{equation}
With the constraint (\ref{D4}), one can check the non-modular part of
(\ref{Z0vec}) and (\ref{Z0hyp}) combined with the
polynomial contributions indeed give the theta function in
(\ref{eq:rblp}), with
\begin{equation}
  y_{u/v,i}=\frac{1}{4}\Omega^{-1}_{ij}\left({{\fn}_j-2}+{\hG}_j+2k_{F_j}
    {\lambda}_{F_j}\cdot {\lambda}_{F_j}\right).
\end{equation}
Finally, we shift back the ${B}$ field in the instanton part, the
$\theta_2$ in (\ref{Z0hyp}) becomes $\theta_1$, and we arrive at the
elliptic blowup equations which are functional equations of the
conventional RR elliptic genera.

\section{More on elliptic genera}\label{app:D}
Here we record more results on the one-string and two-string elliptic
genera for certain rank one theories which we obtain from blowup
equations. Note all ``$\dots$'' in the polynomial of $v$ means
palindromic. More detailed results can be found on the website \cite{kl}.

\subsection*{$\mathbf{n=1,\, G=\mf{so}(7),\,F=\mf{sp}(2)_a\times \mf{sp}(6)_b}$}
Using the Weyl orbit expansion, we turn on a diagonal subgroup
$\mf{sp}(1)\times \mf{sp}(1)$ of the flavor group to compute the elliptic
genus. We obtain the reduced one-string elliptic genus as
\begin{equation}\label{n1so7E1}
\mathbb{E}_{h_{1,\mf{so}(7)}^{(1)}}(\Qtau,v,m_{\mf{so}(7)}=0,m_F=0)=\Qtau^{-1/3}+\Qtau^{2/3}v^{-2}\sum_{n=0}^{\infty}\Qtau^n\frac{P_n(v)}{(1 + v)^{8}},
\end{equation}
where
\begin{equation}\nonumber
\begin{aligned}
P_0(v)=21 + 44 v - 294 v^2 - 1156 v^3 + 475 v^4 + 13400 v^5 + 38508 v^6 +
 13400 v^7 + \dots+ 21 v^{12}.
\end{aligned}
\end{equation}
This agrees with the modular ansatz in \cite{DelZotto:2018tcj}.
Using the result with flavor fugacities turned on, we obtain the following exact $v$ expansion  formula for the subleading $q$ order coefficient, which contains the 5d one-instanton Nekrasov partition function:
\begin{align} \nonumber
\,&\chi_{(010)}^{\mf{so}(7)}v^{-2}-(\chi_{(100)}^{\mf{so}(7)}\chi^F_{(10)_a}+\chi_{(001)}^{\mf{so}(7)}\chi^F_{(100000)_b})v^{-1}
+(\chi_{(002)}^{\mf{so}(7)}+\chi^F_{(20)_a}+\chi^F_{(200000)_b}+1)\\ \nonumber
&+ \chi^F_{(10)_a\otimes(010000)_b}v  +(\chi^F_{(000100)_b}+\chi^F_{(01)_a\otimes(010000)_b}+\chi_{(100)}^{\mf{so}(7)}\chi^F_{(01)_a})v^2  \\ \nonumber
&+ (\chi^F_{(10)_a\otimes(000100)_b}-\chi_{(001)}^{\mf{so}(7)}\chi^F_{(01)_a\otimes(100000)_b}-\chi_{(100)}^{\mf{so}(7)}\chi^F_{(100000)_b})v^3\\ \nonumber
&-(\chi_{(100)}^{\mf{so}(7)}\chi^F_{(000100)_b}+\chi_{(001)}^{\mf{so}(7)}\chi^F_{(10)_a\otimes(001000)_b}+\chi_{(010)}^{\mf{so}(7)}\chi^F_{(010000)_b}
-\chi_{(002)}^{\mf{so}(7)}\chi^F_{(01)_a})v^4\\ \nonumber
&+(\chi_{(101)}^{\mf{so}(7)}\chi^F_{(001000)_b}+\chi_{(002)}^{\mf{so}(7)}\chi^F_{(10)_a\otimes(010000)_b}+\chi_{(011)}^{\mf{so}(7)}\chi^F_{(100000)_b}
)v^5 \\ \nonumber
&-(\chi_{(102)}^{\mf{so}(7)}\chi^F_{(010000)_b}+\chi_{(003)}^{\mf{so}(7)}\chi^F_{(10)_a\otimes(100000)_b}+\chi_{(012)}^{\mf{so}(7)}
)v^6 \\ \nonumber
&+(\chi_{(103)}^{\mf{so}(7)}\chi^F_{(100000)_b}+\chi_{(004)}^{\mf{so}(7)}\chi^F_{(10)_a}
)v^7 -\chi_{(104)}^{\mf{so}(7)}v^8+ \\ \nonumber
\sum_{n=0}^{\infty}&\,\Big[\chi_{(0n0)}^{\mf{so}(7)}\chi^F_{(01)_a\otimes(000001)_b}v^{4+2n}
-(\chi_{(1n0)}^{\mf{so}(7)}\chi^F_{(10)_a\otimes(000001)_b}+\chi_{(0n1)}^{\mf{so}(7)}\chi^F_{(01)_a\otimes(000010)_b})v^{5+2n}\\[-1mm] \nonumber
&+(\chi_{(2n0)}^{\mf{so}(7)}\chi^F_{(000001)_b}+\chi_{(1n1)}^{\mf{so}(7)}\chi^F_{(10)_a\otimes(000010)_b}
+\chi_{(0n2)}^{\mf{so}(7)}\chi^F_{(01)_a\otimes(000100)_b})v^{6+2n}\\ \nonumber
&-(\chi_{(2n1)}^{\mf{so}(7)}\chi^F_{(000010)_b}+\chi_{(1n2)}^{\mf{so}(7)}\chi^F_{(10)_a\otimes(000100)_b}
+\chi_{(0n3)}^{\mf{so}(7)}\chi^F_{(01)_a\otimes(001000)_b})v^{7+2n}\\ \nonumber
&+(\chi_{(2n2)}^{\mf{so}(7)}\chi^F_{(000100)_b}+\chi_{(1n3)}^{\mf{so}(7)}\chi^F_{(10)_a\otimes(001000)_b}
+\chi_{(0n4)}^{\mf{so}(7)}\chi^F_{(01)_a\otimes(010000)_b})v^{8+2n}\\ \nonumber
&-(\chi_{(2n3)}^{\mf{so}(7)}\chi^F_{(001000)_b}+\chi_{(1n4)}^{\mf{so}(7)}\chi^F_{(10)_a\otimes(010000)_b}
+\chi_{(0n5)}^{\mf{so}(7)}\chi^F_{(01)_a\otimes(100000)_b})v^{9+2n}\\ \nonumber
&+(\chi_{(2n4)}^{\mf{so}(7)}\chi^F_{(010000)_b}+\chi_{(1n5)}^{\mf{so}(7)}\chi^F_{(10)_a\otimes(100000)_b}
+\chi_{(0n6)}^{\mf{so}(7)}\chi^F_{(01)_a})v^{10+2n}\\
&-(\chi_{(2n5)}^{\mf{so}(7)}\chi^F_{(100000)_b}+\chi_{(1n6)}^{\mf{so}(7)}\chi^F_{(10)_a})v^{11+2n}+\chi_{(2n6)}^{\mf{so}(7)}v^{12+2n}
\Big].\label{n1SO7HS}
\end{align}
After turning off all gauge and flavor fugacities, this goes back to
the rational function of $v$ by Weyl dimension formulas.

\subsection*{$\mathbf{n=1,\, G=\mf{so}(8),\,F=\mf{sp}(3)_a\times \mf{sp}(3)_b\times \mf{sp}(3)_c}$}

Using the Weyl orbit expansion, we turn on a subgroup $\mf{sp}(1)\times \mf{sp}(1)\times \mf{sp}(1)$ of the flavor group to compute the elliptic genus. We obtain the reduced one-string elliptic genus as
\begin{equation}\label{n1so8E1}
  \mathbb{E}_{h_{1,\mf{so}(8)}^{(1)}}(\Qtau,v,m_{\mf{so}(8)}=0,m_F=0)=\Qtau^{-1/3}+\Qtau^{2/3}v^{-2}\sum_{n=0}^{\infty}\Qtau^n\frac{P_n(v)}{(1 + v)^{10}},
\end{equation}
where
\begin{equation}\nonumber
\begin{aligned}
P_0(v)=4 (7 + 34 v - 22 v^2 - 496 v^3 - 1128 v^4 + 1326 v^5 + 14327 v^6 +
   35392 v^7 + 14327 v^8 + \dots+ 7 v^{14}).
\end{aligned}
\end{equation}
This agrees with the modular ansatz in \cite{DelZotto:2018tcj}.
Using the result with flavor fugacities turned on, we find the following exact formula for the subleading $q$ order coefficient, which contains the 5d one-instanton Nekrasov partition function:
\begin{align}
\,&\chi^F_{(010)_{a}\otimes(010)_b\otimes(010)_c}v^4-(\chi^G_{(1000)}\chi^F_{(100)_{a}\otimes(010)_b\otimes(010)_c}+\text{tri.})v^5
+(\chi^F_{(100)_{a}\otimes(100)_b\otimes(001)_c}+\text{tri.})v^3\nn
&-(\chi^G_{(1000)}\chi^F_{(100)_b\otimes(001)_c}+\text{tri.})v^4+(\chi^F_{(010)_b\otimes(010)_c}+\text{tri.})v^2
+(\chi^G_{(2000)}\chi^F_{(010)_b\otimes(010)_c}+\text{tri.})v^6  \nn
&+(\chi^G_{(0011)}\chi^F_{(010)_{a}\otimes(100)_b\otimes(100)_c}+\text{tri.})v^6+(v-\chi^G_{(0100)}v^5-\chi^G_{(1011)}v^7)\chi^F_{(100)_{a}\otimes(100)_b\otimes(100)_c} \nn
&-(\chi^G_{(2010)}\chi^F_{(100)_b\otimes(010)_c}+\text{tri.})v^7+(\chi^G_{(1000)}\chi^F_{(001)_{a}}+\text{tri.})v^3
+(\chi^G_{(0011)}\chi^F_{(001)_{a}}+\text{tri.})v^5 \nn
&- (\chi^G_{(0100)}\chi^F_{(010)_{a}}+\text{tri.})v^4+(\chi^G_{(0022)}\chi^F_{(010)_{a}}+\text{tri.})v^8
+(\chi^G_{(1100)}\chi^F_{(100)_b\otimes(100)_c}+\text{tri.})v^6\nn
&+(\chi^G_{(2011)}\chi^F_{(100)_b\otimes(100)_c}+\text{tri.})v^8- (\chi^G_{(1000)}\chi^F_{(100)_{a}}+\text{tri.})v^{-1}
- (\chi^G_{(0111)}\chi^F_{(100)_{a}}+\text{tri.})v^{7}\nn
&- (\chi^G_{(1022)}\chi^F_{(100)_{a}}+\text{tri.})v^{9}+\chi^G_{(0100)}v^{-2}+\chi^G_{(0100)}+\chi^F_{(200)_a\oplus(200)_b\oplus(200)_c}+1 \nn
&+\chi^G_{(0200)}v^6+\chi^G_{(1111)}v^8+\chi^G_{(2022)}v^{10}+\nn
\sum_{n=0}^\infty& \Big[\chi^G_{(0n00)}\chi^F_{(001)_a\otimes(001)_b\otimes(001)_c}v^{5+2n}-  (\chi^G_{(1n00)}\chi^F_{(010)_a\otimes(001)_{b}\otimes (001)_{c}}+\text{tri.})v^{6+2n}\nn
&+ (\chi^G_{(1n10)}\chi^F_{(010)_{a}\otimes(010)_b\otimes(001)_c}+\text{tri.})v^{7+2n}- \chi^G_{(1n11)}\chi^F_{(010)_{a}\otimes(010)_b\otimes(010)_c}v^{8+2n}\nn
&-(\chi^G_{(2n10)}\chi^F_{(100)_{a}\otimes(010)_b\otimes(001)_c}+\text{tri.})v^{8+2n}
-(\chi^G_{(3n00)}\chi^F_{(001)_b\otimes(001)_c}+\text{tri.})v^{8+2n}\nn
&+(\chi^G_{(2n11)}\chi^F_{(100)_{a}\otimes(010)_b\otimes(010)_c}+\text{tri.})v^{9+2n}
+(\chi^G_{(3n10)}\chi^F_{(010)_b\otimes(001)_c}+\text{tri.})v^{9+2n}\nn
&+(\chi^G_{(2n20)}\chi^F_{(100)_{a}\otimes(100)_b\otimes(001)_c}+\text{tri.})v^{9+2n}
-(\chi^G_{(3n20)}\chi^F_{(100)_b\otimes(001)_c}+\text{tri.})v^{10+2n}\nn
&-(\chi^G_{(3n11)}\chi^F_{(010)_b\otimes(010)_c}+\text{tri.})v^{10+2n}
-(\chi^G_{(1n22)}\chi^F_{(010)_{a}\otimes(100)_b\otimes(100)_c}+\text{tri.})v^{10+2n}
\nn
&+\chi^G_{(2n22)}\chi^F_{(100)_a\otimes(100)_b\otimes(100)_c}v^{11+2n}+(\chi^G_{(3n21)}\chi^F_{(100)_b\otimes(010)_c}+\text{tri.})v^{11+2n}  \nn
&+(\chi^G_{(0n33)}\chi^F_{(001)_a}+\text{tri.})v^{11+2n}-(\chi^G_{(1n33)}\chi^F_{(010)_a}+\text{tri.})v^{12+2n}\nn
&-(\chi^G_{(3n22)}\chi^F_{(100)_b\otimes(100)_c}+\text{tri.})v^{12+2n}+(\chi^G_{(2n33)}\chi^F_{(100)_a}+\text{tri.})v^{13+2n}-\chi^G_{(3n33)}v^{14+2n}
\Big].\label{n1SO8HS}
 \end{align}
 Here ``tri.'' means the two or five more terms implied by triality of
 both $\mf{so}(8)$ and the three $\mf{sp}(3)$ flavor groups
 together. We represent the $v $ expansion terms both inside and
 outside the infinite summation in a descending order of the flavor
 representations. By Weyl dimension formulas of $\mf{so}(8)$ and
 $\mf{sp}(3)$, the above exact formula goes back to the rational
 function of $v$ after turning off the gauge and flavor fugacities.

\subsection*{$\mathbf{n=1,\, G=\mf{so}(9),\,F=\mf{sp}(4)_a\times \mf{sp}(3)_b}$}
Using the Weyl orbit expansion, we turn on the subgroup
$\mf{sp}(1)\times \mf{sp}(1)$ of the flavor group to compute the
elliptic genus. We obtain the reduced one-string elliptic genus as
\begin{equation}\label{n1so9E1}
\mathbb{E}_{h_{1,\mf{so}(9)}^{(1)}}(\Qtau,v,m_{\mf{so}(9)}=0,m_F=0)=\Qtau^{-1/3}+\Qtau^{2/3}v^{-2}\sum_{n=0}^{\infty}\Qtau^n\frac{P_n(v)}{(1 + v)^{12}},
\end{equation}
where
\begin{equation}\nonumber
\begin{aligned}
P_0(v)=2 (&18 + 132 v + 227 v^2 - 936 v^3 - 5226 v^4 - 7904 v^5 + 17037 v^6\\ &\,+
   118788 v^7 + 263632 v^8 + 118788 v^9 + \dots+ 18 v^{16}).
 \end{aligned}
\end{equation}
This agrees with the modular ansatz in \cite{DelZotto:2018tcj}.
Using the result with flavor fugacities turned on, we obtain the following exact formula for the subleading $q$ order coefficient, which contains the 5d one-instanton Nekrasov partition function:
\begin{align} \nonumber
\,& \chi_{(0100)}^{\mf{so}(7)}\chi^F_{(1000)_a}v^{-2}-(\chi_{(1000)}^{\mf{so}(7)}\chi^F_{(1000)_a}+\chi_{(0001)}^{\mf{so}(7)}\chi^F_{(002)_b})v^{-1}
+\chi_{(0100)}^{\mf{so}(7)}+\chi^F_{(2000)_a}+\chi^F_{(200)_b} +1 \nn
&+\chi^F_{(1000)_a\otimes(200)_b}v +(\chi^F_{(020)_b}+\chi^F_{(0100)_a\otimes(010)_b})v^2\nn
&+(\chi^F_{(1000)_a\otimes(101)_b}+\chi^F_{(0010)_a\otimes(010)_b}+
\chi_{(0001)}^{\mf{so}(7)}\chi^F_{(001)_b})v^3+\dots \nn
+&\sum_{n=0}^{\infty}\Big[\chi_{(0n00)}^{\mf{so}(7)}\chi^F_{(0001)_a\otimes(002)_b}v^{6+2n}-
(\chi_{(0n01)}^{\mf{so}(7)}\chi^F_{(0001)_a\otimes(011)_b}+\chi_{(1n00)}^{\mf{so}(7)}\chi^F_{(0010)_a\otimes(002)_b})v^{7+2n}\nn
+(&\chi_{(0n02)}^{\mf{so}(7)}\chi^F_{(0001)_a\otimes(101)_b}+\chi_{(0n10)}^{\mf{so}(7)}\chi^F_{(0001)_a\otimes(020)_b}+
\chi_{(1n01)}^{\mf{so}(7)}\chi^F_{(0010)_a\otimes(011)_b}+\chi_{(2n00)}^{\mf{so}(7)}\chi^F_{(0100)_a\otimes(002)_b})v^{8+2n}\nn
-(&\chi_{(0n03)}^{\mf{so}(7)}\chi^F_{(0001)_a\otimes(001)_b}+\chi_{(0n11)}^{\mf{so}(7)}\chi^F_{(0001)_a\otimes(110)_b}
+
\chi_{(1n02)}^{\mf{so}(7)}\chi^F_{(0010)_a\otimes(101)_b}+
\chi_{(1n10)}^{\mf{so}(7)}\chi^F_{(0010)_a\otimes(020)_b}\nn
&+\chi_{(2n01)}^{\mf{so}(7)}\chi^F_{(0100)_a\otimes(011)_b}+\chi_{(3n00)}^{\mf{so}(7)}\chi^F_{(1000)_a\otimes(002)_b})v^{9+2n}\nn
+(&\chi_{(0n20)}^{\mf{so}(7)}\chi^F_{(0001)_a\otimes(200)_b}+\chi_{(0n12)}^{\mf{so}(7)}\chi^F_{(0001)_a\otimes(010)_b}
+
\chi_{(1n03)}^{\mf{so}(7)}\chi^F_{(0010)_a\otimes(001)_b}+
\chi_{(1n11)}^{\mf{so}(7)}\chi^F_{(0010)_a\otimes(110)_b}\nn
&+\chi_{(2n02)}^{\mf{so}(7)}\chi^F_{(0100)_a\otimes(101)_b}+\chi_{(2n10)}^{\mf{so}(7)}\chi^F_{(0100)_a\otimes(020)_b}
+\chi_{(3n01)}^{\mf{so}(7)}\chi^F_{(1000)_a\otimes(011)_b}+\chi_{(4n00)}^{\mf{so}(7)}\chi^F_{(002)_b})v^{10+2n}\nn
-(&\chi_{(0n21)}^{\mf{so}(7)}\chi^F_{(0001)_a\otimes(100)_b}
+
\chi_{(1n20)}^{\mf{so}(7)}\chi^F_{(0010)_a\otimes(200)_b}+
\chi_{(1n12)}^{\mf{so}(7)}\chi^F_{(0010)_a\otimes(010)_b}+\chi_{(2n03)}^{\mf{so}(7)}\chi^F_{(0100)_a\otimes(001)_b}\nn
&+\chi_{(2n11)}^{\mf{so}(7)}\chi^F_{(0100)_a\otimes(110)_b}
+\chi_{(3n02)}^{\mf{so}(7)}\chi^F_{(1000)_a\otimes(101)_b}+\chi_{(3n10)}^{\mf{so}(7)}\chi^F_{(1000)_a\otimes(020)_b}
+\chi_{(4n01)}^{\mf{so}(7)}\chi^F_{(011)_b})v^{11+2n}\nn
+(&\chi_{(0n30)}^{\mf{so}(7)}\chi^F_{(0001)_a}
+
\chi_{(1n21)}^{\mf{so}(7)}\chi^F_{(0010)_a\otimes(100)_b}+
\chi_{(2n20)}^{\mf{so}(7)}\chi^F_{(0100)_a\otimes(200)_b}+\chi_{(2n12)}^{\mf{so}(7)}\chi^F_{(0100)_a\otimes(010)_b}\nn
&
+\chi_{(3n03)}^{\mf{so}(7)}\chi^F_{(1000)_a\otimes(001)_b}+\chi_{(3n11)}^{\mf{so}(7)}\chi^F_{(1000)_a\otimes(110)_b}
+\chi_{(4n02)}^{\mf{so}(7)}\chi^F_{(101)_b}+\chi_{(4n10)}^{\mf{so}(7)}\chi^F_{(020)_b})v^{12+2n}\nn
-(&
\chi_{(1n30)}^{\mf{so}(7)}\chi^F_{(0010)_a}+
\chi_{(2n21)}^{\mf{so}(7)}\chi^F_{(0100)_a\otimes(100)_b}
+\chi_{(3n20)}^{\mf{so}(7)}\chi^F_{(1000)_a\otimes(200)_b}+\chi_{(3n12)}^{\mf{so}(7)}\chi^F_{(1000)_a\otimes(010)_b}\nn
&+\chi_{(4n03)}^{\mf{so}(7)}\chi^F_{(001)_b}+\chi_{(4n11)}^{\mf{so}(7)}\chi^F_{(110)_b})v^{13+2n}\nn
+(&
\chi_{(2n30)}^{\mf{so}(7)}\chi^F_{(0100)_a}
+\chi_{(3n21)}^{\mf{so}(7)}\chi^F_{(1000)_a\otimes(100)_b}+\chi_{(4n20)}^{\mf{so}(7)}\chi^F_{(200)_b}+\chi_{(4n12)}^{\mf{so}(7)}\chi^F_{(010)_b})v^{14+2n}\nn
-(&
\chi_{(3n30)}^{\mf{so}(7)}\chi^F_{(1000)_a}+\chi_{(4n21)}^{\mf{so}(7)}\chi^F_{(100)_b})v^{15+2n}+\chi_{(4n30)}^{\mf{so}(7)}v^{16+2n}
\Big].\label{n1SO9HS}
\end{align}
The sporadic terms outside the infinite summations are too long to
present, thus here we only present those in a few leading
orders.

\subsection*{$\mathbf{n=1,\, G=F_4,\,F=\mf{sp}(4)}$}
Using $v$ expansion method, we turn on all flavor $\mf{sp}(4)$
fugacities to compute the reduced one-string elliptic genus. The 5d
one-instanton Nekrasov partition function is contained in the
subleading $q$ order, for which we find the following exact formula
\begin{align} \nonumber
&\chi_{(0030)}^{\mf{sp}(4)}v^7+\chi_{(0201)}^{\mf{sp}(4)}v^6-\chi_{(0001)}^{F_4}\chi_{(0120)}^{\mf{sp}(4)}v^8+\chi_{(0002)}^{F_4}\chi_{(1020)}^{\mf{sp}(4)}v^9
+\chi_{(1101)}^{\mf{sp}(4)}(v^5-\chi_{(0001)}^{F_4}v^7)\\ \nonumber
+&\chi_{(0010)}^{F_4}\chi_{(0210)}^{\mf{sp}(4)}v^9 +\chi_{(0101)}^{\mf{sp}(4)}(-\chi_{(0001)}^{F_4}v^6+\chi_{(0002)}^{F_4}v^8)
+\chi_{(0020)}^{\mf{sp}(4)}(v^4-\chi_{(0003)}^{F_4}v^{10}) \\ \nonumber
+&\chi_{(2001)}^{\mf{sp}(4)}(v^4+\chi_{(0010)}^{F_4}v^{8})- \chi_{(0011)}^{F_4}\chi_{(1110)}^{\mf{sp}(4)}v^{10}-\chi_{(0300)}^{\mf{sp}(4)}(\chi_{(1000)}^{F_4}v^8+\chi_{(0100)}^{F_4}v^{10}) \\ \nonumber
-&\chi_{(0011)}^{F_4}\chi_{(1001)}^{\mf{sp}(4)}v^{9}+\chi_{(0110)}^{\mf{sp}(4)}(v^3-\chi_{(0012)}^{F_4}v^{11})
+\chi_{(2010)}^{\mf{sp}(4)}(-\chi_{(1000)}^{F_4}v^7+\chi_{(0020)}^{F_4}v^{11}) \\ \nonumber
+&\chi_{(1200)}^{\mf{sp}(4)}(\chi_{(1001)}^{F_4}v^9+\chi_{(0101)}^{F_4}v^{11})+ \chi_{(0001)}^{\mf{sp}(4)}(\chi_{(0001)}^{F_4}v^4+\chi_{(0100)}^{F_4}v^{8}+\chi_{(0020)}^{F_4}v^{10})     \\ \nonumber
-&\chi_{(1010)}^{\mf{sp}(4)}(\chi_{(1000)}^{F_4}v^6-\chi_{(1001)}^{F_4}v^{8}+\chi_{(0021)}^{F_4}v^{12})
+\chi_{(0200)}^{\mf{sp}(4)}(v^2-\chi_{(1002)}^{F_4}v^{10}-\chi_{(0102)}^{F_4}v^{12})     \\ \nonumber
-&\chi_{(2100)}^{\mf{sp}(4)}(\chi_{(1010)}^{F_4}v^{10}+\chi_{(0110)}^{F_4}v^{12})
-    \chi_{(0010)}^{\mf{sp}(4)}(\chi_{(1000)}^{F_4}v^5+\chi_{(1010)}^{F_4}v^{9}-\chi_{(0030)}^{F_4}v^{13})        \\ \nonumber
+&\chi_{(1100)}^{\mf{sp}(4)}(\chi_{(1011)}^{F_4}v^{11}+\chi_{(0111)}^{F_4}v^{13})
+\chi_{(3000)}^{\mf{sp}(4)}(v+\chi_{(2000)}^{F_4}v^{9}+\chi_{(1100)}^{F_4}v^{11}+\chi_{(0200)}^{F_4}v^{13}) \\ \nonumber
+&\chi_{(0100)}^{\mf{sp}(4)}(\chi_{(2000)}^{F_4}v^8-\chi_{(1020)}^{F_4}v^{12}-\chi_{(0120)}^{F_4}v^{14})
+\chi_{(2000)}^{\mf{sp}(4)}(1-\chi_{(2001)}^{F_4}v^{10}-\chi_{(1101)}^{F_4}v^{12}-\chi_{(0201)}^{F_4}v^{14})   \\ \nonumber
+&\chi_{(1000)}^{\mf{sp}(4)}(-\chi_{(0001)}^{F_4}v^{-1}+\chi_{(2010)}^{F_4}v^{11}+\chi_{(1110)}^{F_4}v^{13}+\chi_{(0210)}^{F_4}v^{15})   \\ \nonumber
+&(\chi_{(1000)}^{F_4}v^{-2}+\chi_{(1000)}^{F_4}+1-\chi_{(3000)}^{F_4}v^{10}-\chi_{(2100)}^{F_4}v^{12}-\chi_{(1200)}^{F_4}v^{14}-\chi_{(0300)}^{F_4}v^{16}) \\ \nonumber
\sum_{n=0}^{\infty}\Big[&\chi_{(n000)}^{F_4}\chi_{(0003)}^{\mf{sp}(4)}v^{8+2n}-\chi_{(n001)}^{F_4}\chi_{(0012)}^{\mf{sp}(4)}v^{9+2n}
+(\chi_{(n010)}^{F_4}\chi_{(0021)}^{\mf{sp}(4)}+\chi_{(n002)}^{F_4}\chi_{(0102)}^{\mf{sp}(4)})v^{10+2n}\\[-1mm] \nonumber
&-(\chi_{(n100)}^{F_4}\chi_{(0030)}^{\mf{sp}(4)}+\chi_{(n011)}^{F_4}\chi_{(0111)}^{\mf{sp}(4)}+\chi_{(n003)}^{F_4}\chi_{(1002)}^{\mf{sp}(4)})v^{11+2n}\\ \nonumber
&+(\chi_{(n012)}^{F_4}\chi_{(1011)}^{\mf{sp}(4)}+\chi_{(n020)}^{F_4}\chi_{(0201)}^{\mf{sp}(4)}+\chi_{(n101)}^{F_4}\chi_{(0120)}^{\mf{sp}(4)}
+\chi_{(n004)}^{F_4}\chi_{(0002)}^{\mf{sp}(4)})v^{12+2n}\\ \nonumber
&-(\chi_{(n102)}^{F_4}\chi_{(1020)}^{\mf{sp}(4)}+\chi_{(n013)}^{F_4}\chi_{(0011)}^{\mf{sp}(4)}+\chi_{(n021)}^{F_4}\chi_{(1101)}^{\mf{sp}(4)}
+\chi_{(n110)}^{F_4}\chi_{(0210)}^{\mf{sp}(4)})v^{13+2n}\\ \nonumber
&+(\chi_{(n022)}^{F_4}\chi_{(0101)}^{\mf{sp}(4)}+\chi_{(n103)}^{F_4}\chi_{(0020)}^{\mf{sp}(4)}+\chi_{(n030)}^{F_4}\chi_{(2001)}^{\mf{sp}(4)}
+\chi_{(n111)}^{F_4}\chi_{(1110)}^{\mf{sp}(4)}+\chi_{(n200)}^{F_4}\chi_{(0300)}^{\mf{sp}(4)})v^{14+2n}\\ \nonumber
&-(\chi_{(n031)}^{F_4}\chi_{(1001)}^{\mf{sp}(4)}+\chi_{(n112)}^{F_4}\chi_{(0110)}^{\mf{sp}(4)}+\chi_{(n120)}^{F_4}\chi_{(2010)}^{\mf{sp}(4)}
+\chi_{(n201)}^{F_4}\chi_{(1200)}^{\mf{sp}(4)})v^{15+2n}\\ \nonumber
&+(\chi_{(n040)}^{F_4}\chi_{(0001)}^{\mf{sp}(4)}+\chi_{(n121)}^{F_4}\chi_{(1010)}^{\mf{sp}(4)}+\chi_{(n202)}^{F_4}\chi_{(0200)}^{\mf{sp}(4)}
+\chi_{(n210)}^{F_4}\chi_{(2100)}^{\mf{sp}(4)})v^{16+2n}\\ \nonumber
&-(\chi_{(n130)}^{F_4}\chi_{(0010)}^{\mf{sp}(4)}+\chi_{(n211)}^{F_4}\chi_{(1100)}^{\mf{sp}(4)}+\chi_{(n300)}^{F_4}\chi_{(3000)}^{\mf{sp}(4)}
)v^{17+2n}\\
&+(\chi_{(n220)}^{F_4}\chi_{(0100)}^{\mf{sp}(4)}+\chi_{(n301)}^{F_4}\chi_{(2000)}^{\mf{sp}(4)}
)v^{18+2n}-\chi_{(n310)}^{F_4}\chi_{(1000)}^{\mf{sp}(4)}v^{19+2n}+\chi_{(n400)}^{F_4}v^{20+2n}\Big].\label{n1F4exactE1}
\end{align}
Turning off all $F_4$ and $\mf{sp}(4)$ fugacities, the above exact formula reduces to the rational function of $v$ in (\ref{n1F4E1}) by Weyl dimension formulas.

\subsection*{$\mathbf{n=2,\, G=\mf{so}(9),\,F=\mf{sp}(3)_a\times \mf{sp}(2)_b}$}
Using the Weyl orbit expansion, we turn on the subgroup $\mf{sp}(1)\times \mf{sp}(1)$ of the flavor group to compute the elliptic genus. We obtain the reduced one-string elliptic genus as
\be\label{n2so9E1}
\mathbb{E}_{h_{2,\mf{so}(9)}^{(1)}}(\Qtau,v,m_{\mf{so}(9)}=0,m_F=0)=\Qtau^{1/6}v^{-1}\sum_{n=0}^{\infty}\Qtau^n\frac{P_n(v)}{(1 + v)^{12}},
\ee
where
\be\nonumber
\ba
P_0(v)=(1 - v)^2 (1 + 14 v + 93 v^2 + 392 v^3 + 1181 v^4 + 2658 v^5 +
   4106 v^6 + 2658 v^7 + \dots +
   v^{12}).
\ea
\ee
This agrees with the modular ansatz in \cite{DelZotto:2018tcj}.
Using the result with flavor fugacities turned on, we obtain the following exact $v$ expansion formula for the leading $q$ order coefficient, which contains the reduced 5d one-instanton Nekrasov partition function:
\begin{align} \nonumber
\,&-  \chi^F_{(001)_a}v^4-\chi^F_{(010)_a}(\chi^F_{(20)_b}v^5-\chi^{\mf{so}(9)}_{(0001)}\chi^F_{(10)_b}v^6+\chi^{\mf{so}(9)}_{(0010)}v^7)
+\chi^F_{(100)_a}(-\chi^F_{(01)_b}v^4\nn
&+ \chi^{\mf{so}(9)}_{(1000)}\chi^F_{(20)_b}v^6-\chi^{\mf{so}(9)}_{(1001)}\chi^F_{(10)_b}v^7
+\chi^{\mf{so}(9)}_{(1010)}v^8+ \chi^{\mf{so}(9)}_{(0100)}v^6)+\chi^F_{(01)_b}(-v^3+\chi^{\mf{so}(9)}_{(1000)} v^5)   \nn
&-\chi^{\mf{so}(9)}_{(2000)}\chi^F_{(20)_b}v^7+\chi^{\mf{so}(9)}_{(2001)}\chi^F_{(10)_b}v^8+v^{-1}-\chi^{\mf{so}(9)}_{(1100)}v^7-\chi^{\mf{so}(9)}_{(2010)}v^9 + \nn
\sum_{n=0}^\infty\Big[&\chi^F_{(001)_a}\Big(-\chi^{\mf{so}(9)}_{(0n00)}\chi^F_{(02)_b}v^{6+2n}+\chi^{\mf{so}(9)}_{(0n01)}\chi^F_{(11)_b}v^{7+2n}
-(\chi^{\mf{so}(9)}_{(0n02)}\chi^F_{(01)_b}+\chi^{\mf{so}(9)}_{(0n10)}\chi^F_{(20)_b})v^{8+2n}\nn
&+\chi^{\mf{so}(9)}_{(0n11)}\chi^F_{(10)_b}v^{9+2n}
-\chi^{\mf{so}(9)}_{(0n20)}v^{10+2n}\Big)+\chi^F_{(010)_a}\Big(\chi^{\mf{so}(9)}_{(1n00)}\chi^F_{(02)_b}v^{7+2n}-\chi^{\mf{so}(9)}_{(1n01)}\chi^F_{(11)_b}v^{8+2n}\nn
&
+(\chi^{\mf{so}(9)}_{(1n02)}\chi^F_{(01)_b}+\chi^{\mf{so}(9)}_{(1n10)}\chi^F_{(20)_b})v^{9+2n}-\chi^{\mf{so}(9)}_{(1n11)}\chi^F_{(10)_b}v^{10+2n}
+\chi^{\mf{so}(9)}_{(1n20)}v^{11+2n}\Big)\nn
&+\chi^F_{(100)_a}\Big(-\chi^{\mf{so}(9)}_{(2n00)}\chi^F_{(02)_b}v^{8+2n}+\chi^{\mf{so}(9)}_{(2n01)}\chi^F_{(11)_b}v^{9+2n}
-(\chi^{\mf{so}(9)}_{(2n02)}\chi^F_{(01)_b}+\chi^{\mf{so}(9)}_{(2n10)}\chi^F_{(20)_b})v^{10+2n}\nn
&+\chi^{\mf{so}(9)}_{(2n11)}\chi^F_{(10)_b}v^{11+2n}
-\chi^{\mf{so}(9)}_{(2n20)}v^{12+2n}\Big)+\Big(\chi^{\mf{so}(9)}_{(3n00)}\chi^F_{(02)_b}v^{9+2n}-\chi^{\mf{so}(9)}_{(3n01)}\chi^F_{(11)_b}v^{10+2n}\nn
&
+(\chi^{\mf{so}(9)}_{(3n02)}\chi^F_{(01)_b}+\chi^{\mf{so}(9)}_{(3n10)}\chi^F_{(20)_b})v^{11+2n}-\chi^{\mf{so}(9)}_{(3n11)}\chi^F_{(10)_b}v^{12+2n}
+\chi^{\mf{so}(9)}_{(3n20)}v^{13+2n}\Big)
\Big].\label{n2so9HS}
\end{align}
A few leading terms in the $v$ expansion has been determined in (H.20) of \cite{DelZotto:2018tcj}.

\subsection*{$\mathbf{n=2,\, G=\mf{so}(10),\,F=\mf{sp}(4)_a\times \mf{su}(2)_b\times\mf{u}(1)_c}$}
Using the Weyl orbit expansion, we turn on the subgroup $\mf{sp}(1)\times \mf{su}(2)\times\mf{u}(1)$ of the flavor group to compute the elliptic genus. We obtain the reduced one-string elliptic genus as
\be\label{n2so10E1}
\mathbb{E}_{h_{2,\mf{so}(10)}^{(1)}}(\Qtau,v,m_{\mf{so}(10)}=0,m_F=0)=\Qtau^{1/6}v^{-1}\sum_{n=0}^{\infty}\Qtau^n\frac{P_n(v)}{(1 + v)^{14}},
\ee
where
\be\nonumber
\ba
P_0(v)=&\,(1 - v)^2 (1 + 16 v + 122 v^2 + 592 v^3 + 2060 v^4 + 5472 v^5 + 11287 v^6 +
 16496 v^7\\ &\ \,+ 11287 v^8 + 5472 v^9 + 2060 v^{10} + 592 v^{11} + 122 v^{12} +
 16 v^{13} + v^{14}).
\ea
\ee
This agrees with the modular ansatz in \cite{DelZotto:2018tcj}.
Using the result with flavor fugacities turned on, we obtain the following exact $v$ expansion formula for the leading $q$ order coefficient, which contains the reduced 5d one-instanton Nekrasov partition function:
\begin{align} \nonumber
\,&v^{-1}-\chi^F_{(2)_b}v^3-\chi^F_{(1000)_a\otimes((2)_c\oplus(-2)_c)}v^{4}+(\chi^{\mf{so}(10)}_{(10000)}\chi^F_{(2)_c\oplus(-2)_c}
-\chi^F_{(0100)_a\otimes(2)_b}-\chi^F_{(0001)_a})v^5\nn
&+(\chi^F_{(0010)_a\otimes((2)_c\oplus(-2)_c)}-\chi^{\mf{so}(10)}_{(10000)}\chi^F_{(1000)_a})\chi^F_{(2)_b}v^6  +\dots+        \nn
\sum_{n=0}^\infty\Big[&(\chi^{\mf{u}(1)}_{(-4)\oplus(4)}+\chi^{\mf{su}(2)}_{(4)})\Big(-\chi^{\mf{so}(10)}_{(0n000)}\chi^{\mf{sp}(4)}_{(0001)}v^{7+2n}
+\chi^{\mf{so}(10)}_{(1n000)}\chi^{\mf{sp}(4)}_{(0010)}v^{8+2n}-\chi^{\mf{so}(10)}_{(2n000)}\chi^{\mf{sp}(4)}_{(0100)}v^{9+2n}\nn
&+\chi^{\mf{so}(10)}_{(3n000)}\chi^{\mf{sp}(4)}_{(1000)}v^{10+2n}-\chi^{\mf{so}(10)}_{(4n000)}v^{11+2n}\Big)+\Big(
  (\chi^{\mf{u}(1)}_{(-3)}\chi^{\mf{su}(2)}_{(1)}+\chi^{\mf{u}(1)}_{(1)}\chi^{\mf{su}(2)}_{(3)})(\chi^{\mf{so}(10)}_{(0n001)}\chi^{\mf{sp}(4)}_{(0001)}v^{8+2n}\nn
 & -\chi^{\mf{so}(10)}_{(1n001)}\chi^{\mf{sp}(4)}_{(0010)}v^{9+2n}+\chi^{\mf{so}(10)}_{(2n001)}\chi^{\mf{sp}(4)}_{(0100)}v^{10+2n}
  -\chi^{\mf{so}(10)}_{(3n001)}\chi^{\mf{sp}(4)}_{(1000)}v^{11+2n}+\chi^{\mf{so}(10)}_{(4n001)}v^{12+2n})\nn
  & +c.c.\Big)+\chi^{\mf{u}(1)}_{(-2)\oplus(2)}\chi^{\mf{su}(2)}_{(2)}\Big(-\chi^{\mf{so}(10)}_{(0n100)}\chi^{\mf{sp}(4)}_{(0001)}v^{9+2n}
+\chi^{\mf{so}(10)}_{(1n100)}\chi^{\mf{sp}(4)}_{(0010)}v^{10+2n}\nn
&-\chi^{\mf{so}(10)}_{(2n100)}\chi^{\mf{sp}(4)}_{(0100)}v^{11+2n}+\chi^{\mf{so}(10)}_{(3n100)}\chi^{\mf{sp}(4)}_{(1000)}v^{12+2n}
-\chi^{\mf{so}(10)}_{(4n100)}v^{13+2n}\Big)\nn
&-\Big(
  \chi^{\mf{u}(1)}_{(2)}(\chi^{\mf{so}(10)}_{(0n020)}\chi^{\mf{sp}(4)}_{(0001)}v^{9+2n} -\chi^{\mf{so}(10)}_{(1n020)}\chi^{\mf{sp}(4)}_{(0010)}v^{10+2n}+\chi^{\mf{so}(10)}_{(2n020)}\chi^{\mf{sp}(4)}_{(0100)}v^{11+2n}\nn
&  -\chi^{\mf{so}(10)}_{(3n020)}\chi^{\mf{sp}(4)}_{(1000)}v^{12+2n}+\chi^{\mf{so}(10)}_{(4n020)}v^{13+2n}) +c.c.\Big)+\Big(\chi^{\mf{u}(1)}_{(-1)}\chi^{\mf{su}(2)}_{(1)}(\chi^{\mf{so}(10)}_{(0n101)}\chi^{\mf{sp}(4)}_{(0001)}v^{10+2n}\nn
 & -\chi^{\mf{so}(10)}_{(1n101)}\chi^{\mf{sp}(4)}_{(0010)}v^{11+2n}+\chi^{\mf{so}(10)}_{(2n101)}\chi^{\mf{sp}(4)}_{(0100)}v^{12+2n}
  -\chi^{\mf{so}(10)}_{(3n101)}\chi^{\mf{sp}(4)}_{(1000)}v^{13+2n}+\chi^{\mf{so}(10)}_{(4n101)}v^{14+2n})\nn
  &+c.c.
 \Big)-\chi^{\mf{su}(2)}_{(2)}(\chi^{\mf{so}(10)}_{(0n011)}\chi^{\mf{sp}(4)}_{(0001)}v^{9+2n} -\chi^{\mf{so}(10)}_{(1n011)}\chi^{\mf{sp}(4)}_{(0010)}v^{10+2n}+\chi^{\mf{so}(10)}_{(2n011)}\chi^{\mf{sp}(4)}_{(0100)}v^{11+2n}\nn
 &  -\chi^{\mf{so}(10)}_{(3n011)}\chi^{\mf{sp}(4)}_{(1000)}v^{12+2n}+\chi^{\mf{so}(10)}_{(4n011)}v^{13+2n})
 +\Big(-\chi^{\mf{so}(10)}_{(0n200)}\chi^{\mf{sp}(4)}_{(0001)}v^{11+2n}
+\chi^{\mf{so}(10)}_{(1n200)}\chi^{\mf{sp}(4)}_{(0010)}v^{12+2n}\nn
&-\chi^{\mf{so}(10)}_{(2n200)}\chi^{\mf{sp}(4)}_{(0100)}v^{13+2n}
+\chi^{\mf{so}(10)}_{(3n200)}\chi^{\mf{sp}(4)}_{(1000)}v^{14+2n}
-\chi^{\mf{so}(10)}_{(4n200)}v^{15+2n}\Big)
\Big].\label{n2so10HS}
\end{align}
The sporadic terms outside the infinite summation are too long to
present, thus here we only show some in leading orders. In general they can
be recovered from the terms inside the infinite summation.  Note the
complex conjugate $c.c.$ interchanges the Dynkin labels of spinor and
conjugate spinor representations of gauge $\mf{so}(10)$ and reverses
the charge of $\mf{u}(1)$ flavor simultaneously. We also checked this expression from 5d blowup equations. A few leading terms
in the $v$ expansion has been determined in (H.21) of
\cite{DelZotto:2018tcj}.

\subsection*{$\mathbf{n=2,\, G=\mf{so}(11),\,F=\mf{sp}(5)_a\times \mf{so}(2)_b}$}

There are 128 unity blowup equations in total. Let us regard the
flavor subgroup $\mf{so}(2)$ as $\mf{u}(1)$. The $r$ fields
$\lambda_{\mf{sp}(5)}$ takes value in
$\mathcal{O}_{[00001]}^{\mf{sp}(5)}$, while
$\lambda_{\mf{u}(1)}=\pm 1/2$. Using the Weyl orbit expansion method,
we turn on a subgroup $\mf{sp}(1)\times \mf{u}(1)$ of the flavor and
compute the one-string elliptic genus to $\mathcal{O}(\Qtau^2)$. For
example, with gauge and flavor fugacities turned off we obtain the
reduced one-string elliptic genus as\footnote{In
  \cite{DelZotto:2018tcj}, the modular ansatz for the reduced
  one-string elliptic genus of this theory is determined up to two
  unfixed parameters. Using our result from blowup equations, we are
  able to determine their two unfixed parameters as
\begin{equation}
a_1=\frac{16291}{1283918464548864},a_2=\frac{9983}{7703510787293184}.
\end{equation}
}
\begin{equation}\label{n2so11E1}
\mathbb{E}_{h_{2,\mf{so}(11)}^{(1)}}(\Qtau,v,m_{\mf{so}(11)}=0,m_{F}=0)=\Qtau^{1/6}\sum_{n=0}^{\infty}\Qtau^n\frac{(1-v)^2 P_n(v)}{v (1 + v)^{16}},
\end{equation}
where
\begin{equation}\nonumber
\begin{aligned}
P_0(v)=1 + 18 v + 155 v^2 + 852 v^3 + 3367 v^4 + 10208 v^5 + 24624 v^6 +
 47390 v^7 + 66362 v^8  + \dots + v^{16},
 \end{aligned}
\end{equation}
and
\begin{equation}\nonumber
\begin{aligned}
P_1(v)=&\,v^{-2} (55 + 816 v + 5505 v^2 + 21936 v^3 + 55038 v^4 + 79650 v^5 +
 18864 v^6 - 193544 v^7\\
 & - 427293 v^8 - 245690 v^9 + 410958 v^{10} -\dots + 55 v^{20}).\\
 \end{aligned}
\end{equation}
If we turn on all gauge and flavor fugacities, we find the leading $\Qtau$ order of reduced one-string elliptic genus is
\begin{equation}
\begin{aligned}
\,&v^{-1}-\chi^F_{(-2)_b\oplus(2)_b}v^3-\chi^F_{(10000)_a}v^4+(\mathbf{11}-\chi^F_{(01000)_a})v^5\\
+\,&(\mathbf{11}\cdot\chi^F_{(10000)_a}-\chi^F_{(00100)_a\otimes((-2)_b\oplus(2)_b)}-\chi^F_{(00001)_a})v^6\\
+\,&\Big((\mathbf{11}\cdot\chi^F_{(01000)_a}-\chi^F_{(00010)_a})\chi^F_{(-2)_b\oplus(2)_b}-\chi^F_{(00010)_a}-\mathbf{65}\Big)v^7\\
+\,&\Big(-\chi^F_{(00001)_a\otimes((-4)_b\oplus(4)_b)}+(\mathbf{11}\cdot\chi^F_{(00100)_a}-\mathbf{65}\cdot\chi^F_{(10000)_a})\chi^F_{(-2)_b\oplus(2)_b}\\
&+\mathbf{32}\cdot\chi^F_{(00010)_a\otimes((-1)_b\oplus(1)_b)}-\chi^F_{(00001)_a}+(\mathbf{11}+\mathbf{55})\chi^F_{(00100)_a}\Big)v^8+\mathcal{O}(v^{9})
\end{aligned}
\end{equation}
In fact, we find the following exact formula:
\begin{align}\nonumber
\,&\chi^F_{(-2)_b\oplus(2)_b}\Big(-v^3-\chi^F_{(00100)_{a}}v^6+(\chi^G_{(10000)}\chi^F_{(01000)_{a}}-\chi^F_{(00010)_{a}})v^7\\ \nonumber
&\phantom{-----}+(\chi^G_{(10000)}\chi^F_{(00100)_{a}}-\chi^G_{(20000)}\chi^F_{(10000)_{a}})v^8+(\chi^G_{(30000)}-\chi^G_{(20000)}\chi^F_{(01000)_{a}})v^9\\ \nonumber
&\phantom{-----}+\chi^G_{(30000)}\chi^F_{(10000)_{a}}v^{10}-\chi^G_{(40000)}v^{11}
\Big)\\ \nonumber
\,&\chi^F_{(-1)_b\oplus(1)_b}\Big(\chi^G_{(00001)}\chi^F_{(00010)_{a}}v^8-\chi^G_{(10001)}\chi^F_{(00100)_{a}}v^9+\chi^G_{(20001)}\chi^F_{(01000)_{a}}v^{10}\\ \nonumber
&\phantom{-----}-\chi^G_{(30001)}\chi^F_{(10000)_{a}}v^{11}+\chi^G_{(40001)}v^{12}
\Big)\\ \nonumber
\,&+\Big(v^{-1}-\chi^F_{(10000)_a}v^4+(\chi^G_{(10000)}-\chi^F_{(01000)_a})v^5+
(\chi^G_{(10000)}\cdot\chi^F_{(10000)_a}-\chi^F_{(00001)_a})v^6\\ \nonumber
&\phantom{-}-(\chi^G_{(20000)}+\chi^F_{(00010)_a})v^7+\chi^G_{(10000)\oplus(01000)}\chi^F_{(00100)_a}v^8\\ \nonumber
&\phantom{-}-(\chi^G_{(00100)}\chi^F_{(00010)_a}+\chi^G_{(20000)\oplus(11000)}\chi^F_{(01000)_a})v^9\\ \nonumber
&\phantom{-}+(\chi^G_{(10100)}\chi^F_{(00100)_a}+\chi^G_{(30000)\oplus(21000)}\chi^F_{(10000)_a})v^{10}\\ \nonumber
&\phantom{-}-(\chi^G_{(20100)}\chi^F_{(01000)_a}+\chi^G_{(40000)\oplus(31000)})v^{11}
+\chi^G_{(30100)}\chi^F_{(10000)_a}v^{12}-\chi^G_{(40100)}v^{13}\Big) \\ \nonumber
+\sum_{n=0}^\infty\bigg[&\chi^F_{(-4)_b\oplus(4)_b}\Big(-v^{8+2n}\chi^G_{(0n000)}\chi^F_{(00001)_{a}}+v^{9+2n}\chi^G_{(1n000)}\chi^F_{(00010)_{a}}
-v^{10+2n}\chi^G_{(2n000)}\chi^F_{(00100)_{a}}\\ \nonumber
&+v^{11+2n}\chi^G_{(3n000)}\chi^F_{(01000)_{a}}-v^{12+2n}\chi^G_{(4n000)}\chi^F_{(10000)_{a}}+v^{13+2n}\chi^G_{(5n000)} \Big)\\ \nonumber
+&\,\chi^F_{(-3)_b\oplus(3)_b}\Big(v^{9+2n}\chi^G_{(0n001)}\chi^F_{(00001)_{a}}-v^{10+2n}\chi^G_{(1n001)}\chi^F_{(00010)_{a}}
+v^{11+2n}\chi^G_{(2n001)}\chi^F_{(00100)_{a}}\\ \nonumber
&-v^{12+2n}\chi^G_{(3n001)}\chi^F_{(01000)_{a}}+v^{13+2n}\chi^G_{(4n001)}\chi^F_{(10000)_{a}}-v^{14+2n}\chi^G_{(5n001)} \Big)\\ \nonumber
+&\,\chi^F_{(-2)_b\oplus(2)_b}\Big(-v^{10+2n}(\chi^G_{(0n100)}+\chi^G_{(0n010)})\chi^F_{(00001)_{a}}
+v^{11+2n}(\chi^G_{(1n100)}+\chi^G_{(1n010)})\chi^F_{(00010)_{a}}\\ \nonumber
&-v^{12+2n}(\chi^G_{(2n100)}+\chi^G_{(2n010)})\chi^F_{(00100)_{a}}+v^{13+2n}(\chi^G_{(3n100)}+\chi^G_{(3n010)})\chi^F_{(01000)_{a}}\\ \nonumber
&-v^{14+2n}(\chi^G_{(4n100)}+\chi^G_{(4n010)})\chi^F_{(10000)_{a}}+v^{15+2n}(\chi^G_{(5n100)}+\chi^G_{(5n010)})\Big)\\ \nonumber
+&\,\chi^F_{(-1)_b\oplus(1)_b}\Big(v^{9+2n}(\chi^G_{(0n001)}+v^2\chi^G_{(0n101)})\chi^F_{(00001)_{a}}-
v^{10+2n}(\chi^G_{(1n001)}+v^2\chi^G_{(1n101)})\chi^F_{(00010)_{a}}\\ \nonumber
&+v^{11+2n}(\chi^G_{(2n001)}+v^2\chi^G_{(2n101)})\chi^F_{(00100)_{a}}-
v^{12+2n}(\chi^G_{(3n001)}+v^2\chi^G_{(3n101)})\chi^F_{(01000)_{a}}\\ \nonumber
&+v^{13+2n}(\chi^G_{(4n001)}+v^2\chi^G_{(4n101)})\chi^F_{(10000)_{a}}-
v^{14+2n}(\chi^G_{(5n001)}+v^2\chi^G_{(5n101)})\Big)\\ \nonumber
+&\,\Big(-v^{8+2n}(\chi^G_{(0n000)}+v^2(\chi^G_{(0n100)}+\chi^G_{(0n002)})+v^4\chi^G_{(0n200)})\chi^F_{(00001)_{a}}\\ \nonumber
&\phantom{-}+v^{9+2n}(\chi^G_{(1n000)}+v^2(\chi^G_{(1n100)}+\chi^G_{(1n002)})+v^4\chi^G_{(1n200)})\chi^F_{(00010)_{a}}\\ \nonumber
&\phantom{-}-v^{10+2n}(\chi^G_{(2n000)}+v^2(\chi^G_{(2n100)}+\chi^G_{(2n002)})+v^4\chi^G_{(2n200)})\chi^F_{(00100)_{a}}\\ \nonumber
&\phantom{-}+v^{11+2n}(\chi^G_{(3n000)}+v^2(\chi^G_{(3n100)}+\chi^G_{(3n002)})+v^4\chi^G_{(3n200)})\chi^F_{(01000)_{a}}\\ \nonumber
&\phantom{-}-v^{12+2n}(\chi^G_{(4n000)}+v^2(\chi^G_{(4n100)}+\chi^G_{(4n002)})+v^4\chi^G_{(4n200)})\chi^F_{(10000)_{a}}\\
&\phantom{-}+v^{13+2n}(\chi^G_{(5n000)}+v^2(\chi^G_{(5n100)}+\chi^G_{(5n002)})+v^4\chi^G_{(5n200)})
\Big)\bigg].\label{n2SO11HS}
\end{align}
The subleading $\Qtau$ order of reduced one-string elliptic genus is
\begin{equation}
\begin{aligned}
\,&\mathbf{55}v^{-3}-(\mathbf{11}\cdot\chi^F_{(10000)_{a}}+\mathbf{32}\cdot\chi^F_{(-1)_b\oplus (1)_b})v^{-2}+(\mathbf{55}+\chi^F_{(20000)_{a}}+2)v^{-1}\\[-1mm]
+&\,\chi^F_{(10000)_{a}\otimes((-1)_b\oplus(1)_b\oplus(0)_b)}+(\chi^F_{(01000)_{a}\otimes((-2)_b\oplus(2)_b)}+\chi^F_{(-4)_b\oplus(4)_b}+1)v\\
+&\,(\chi^F_{(00100)_{a}}+\mathbf{32}\cdot\chi^F_{(-1)_b\oplus (1)_b} )v^2+\mathcal{O}(v^3)
\end{aligned}
\end{equation}

Let us further denote
\begin{equation}
\mathbb{E}_{h_{2,\mf{so}(11)}^{(1)}}(\Qtau,v,m_{\mf{so}(11)}=0,m_{F}=0)=\Qtau^{1/6}v^{-1}\sum_{i,j}c_{i,j} v^j(\Qtau/v^2)^i.
\end{equation}
Then we have the following table \ref{tb:n2so11one} for the
coefficients $c_{ij}$. Note the red numbers in the first column are
just the dimensions of representations ${k\theta}$ of $\mf{so}(11)$
where $\theta$ is the adjoint representation. The blue numbers in the
second column are given by
$-10\dim(\chi_{[1n000]}^{\mf{so}(11)})-2\dim(\chi_{[0n001]}^{\mf{so}(11)})$
with $n=i-1$, consistent with the fact that the matter is in
representation $(\md{11},\md{10}^a)\oplus(\md{32},\md 2^b)$. The
orange number 112 in the third column is given by
$\mathrm{dim}(\mf{so}(11))+\dim(\mf{sp}(5)\times
\mf{u}(1))+1=55+55+1+1=112$. These are the constraints given in
\cite{DelZotto:2018tcj} by analyzing the spectral flow to NSR elliptic
genus, which our result satisfies perfectly.
\begin{table}[h]
 \begin{center}
\begin{small}
\begin{tabular}{c| cccccccccccccccc }$i,j$&  0 & 1&2&3&4&5&6 & 7 & 8 & 9 & 10 \\
\hline
0& \color{red}1 & 0 & 0 & 0 &-2& -10 & -33 & -242 & 408 & 18544 & -102190 \\
1& \color{red}55 & \color{blue}-174 & \color{orange}112 &30 & 91 & 174 & -150 & -686 & -651 & -33420 & 21765 \\
2& \color{red}1144 & \color{blue}-7106 & 17037 & -17196 & 2998 & 330 & 6602 & 15822 & -16128 & -16234 & 116549 \\
\hline
\end{tabular}
\end{small}
\caption{Series coefficients $ c_{i,j} $ for the one-string elliptic genus of $\fn =2$ $\mf{so}(11)$ model.}\label{tb:n2so11one}
\end{center}
\end{table}

\subsection*{$\mathbf{n=2,\, G=\mf{so}(12)_a,\,F=\mf{sp}(6)_a\times\mf{so}(2)_b}$}
This is a chiral theory in the sense that the spinor and conjugate
spinor representations of $\mf{so}(12)$ are not on an equal
footing. The chirality comes from the matter representation
$(\mathbf{32_s},\mathbf{2}_b)$. This is reflected in the vanishing $r$
fields in Table \ref{tb:ubeq-2} and also the exact $v$ expansion
formula below (\ref{n2so12HS}). Using the Weyl orbit expansion, we
turn on the subgroup $\mf{sp}(1)\times\mf{u}(1)$ of the flavor group
to compute the elliptic genus. We obtain the reduced one-string
elliptic genus as
\begin{equation}\label{n2so12E1}
\mathbb{E}_{h_{2,\mf{so}(12)}^{(1)}}(\Qtau,v,m_{\mf{so}(12)}=0,m_F=0)=\Qtau^{1/6}v^{-1}\sum_{n=0}^{\infty}\Qtau^n\frac{P_n(v)}{(1 + v)^{18}},
\end{equation}
where
\be\nonumber
\ba
P_0(v)=\,&(1 - v)^2 (1 + 20 v + 192 v^2 + 1180 v^3 + 5226 v^4 + 17804 v^5 + 48575 v^6 +
 108512 v^7\\&\quad + 197370 v^8 + 267144 v^9 + 197370 v^{10} + \dots + v^{18}).
\ea
\ee
This agrees with the modular ansatz in \cite{DelZotto:2018tcj}.
Using the result with flavor fugacities turned on, we obtain the following exact $v$ expansion formula for the leading $q$ order coefficient, which contains the reduced 5d one-instanton Nekrasov partition function:
\begin{align} \nonumber
\,&v^{-1}-\chi^{F}_{(2)_b\oplus(-2)_b}v^3-\chi^{F}_{(010000)_a}v^5+\chi^{\mf{so}(12)}_{(100000)}\chi^{F}_{(100000)_a}v^6 -(\chi^G_{(200000)}\chi^F_{(01)_b}\nn
&+\chi^F_{(000100)_a\otimes((2)_b\oplus(-2)_b)}+\chi^F_{(000001)_a})v^7+\chi^{\mf{so}(12)}_{(100000)}\chi^F_{(001000)_a\otimes((2)_b\oplus(-2)_b)}v^8 \nn
-(\,&\chi^{\mf{so}(12)}_{(200000)}\chi^F_{(010000)_a\otimes((2)_b\oplus(-2)_b)}
-\chi^{\mf{so}(12)}_{(000010)}\chi^F_{(000010)_a\otimes((1)_b\oplus(-1)_b)}
-\chi^{\mf{so}(12)}_{(010000)}\chi^F_{(000100)_a})v^9+\dots+\nn
\sum_{n=0}^\infty\Big[&\chi^{\mf{u}(1)}_{(-4)\oplus(4)}\Big(-\chi^{\mf{so}(12)}_{(0n0000)}\chi^{\mf{sp}(6)}_{(000001)}v^{9+2n}
+\chi^{\mf{so}(12)}_{(1n0000)}\chi^{\mf{sp}(6)}_{(000010)}v^{10+2n}-\chi^{\mf{so}(12)}_{(2n0000)}\chi^{\mf{sp}(6)}_{(000100)}v^{11+2n}\nonumber\\[-1mm]
&+\chi^{\mf{so}(12)}_{(3n0000)}\chi^{\mf{sp}(6)}_{(001000)}v^{12+2n}
-\chi^{\mf{so}(12)}_{(4n0000)}\chi^{\mf{sp}(6)}_{(010000)}v^{13+2n}+\chi^{\mf{so}(12)}_{(5n0000)}\chi^{\mf{sp}(6)}_{(100000)}v^{14+2n}\nn
&-\chi^{\mf{so}(12)}_{(6n0000)}v^{15+2n}\Big)+\chi^{\mf{u}(1)}_{(-3)\oplus(3)}\Big(\chi^{\mf{so}(12)}_{(0n0001)}\chi^{\mf{sp}(6)}_{(000001)}v^{10+2n}
- \chi^{\mf{so}(12)}_{(1n0001)}\chi^{\mf{sp}(6)}_{(000010)}v^{11+2n} \nn
&+\chi^{\mf{so}(12)}_{(2n0001)}\chi^{\mf{sp}(6)}_{(000100)}v^{12+2n}-\chi^{\mf{so}(12)}_{(3n0001)}\chi^{\mf{sp}(6)}_{(001000)}v^{13+2n}
+\chi^{\mf{so}(12)}_{(4n0001)}\chi^{\mf{sp}(6)}_{(010000)}v^{14+2n}\nn
&-\chi^{\mf{so}(12)}_{(5n0001)}\chi^{\mf{sp}(6)}_{(100000)}v^{15+2n}+\chi^{\mf{so}(12)}_{(6n0001)}v^{16+2n}\Big)-
\chi^{\mf{u}(1)}_{(-2)\oplus(2)}\Big(\chi^{\mf{so}(12)}_{(0n0100)}\chi^{\mf{sp}(6)}_{(000001)}v^{11+2n}\nn
&- \chi^{\mf{so}(12)}_{(1n0100)}\chi^{\mf{sp}(6)}_{(000010)}v^{12+2n} +\chi^{\mf{so}(12)}_{(2n0100)}\chi^{\mf{sp}(6)}_{(000100)}v^{13+2n}-\chi^{\mf{so}(12)}_{(3n0100)}\chi^{\mf{sp}(6)}_{(001000)}v^{14+2n}\nn
&+\chi^{\mf{so}(12)}_{(4n0100)}\chi^{\mf{sp}(6)}_{(010000)}v^{15+2n}-\chi^{\mf{so}(12)}_{(5n0100)}\chi^{\mf{sp}(6)}_{(100000)}v^{16+2n}
+\chi^{\mf{so}(12)}_{(6n0100)}v^{17+2n}\Big)\nn
&+\chi^{\mf{u}(1)}_{(-1)\oplus(1)}\Big(\chi^{\mf{so}(12)}_{(0n1010)}\chi^{\mf{sp}(6)}_{(000001)}v^{12+2n}
- \chi^{\mf{so}(12)}_{(1n1010)}\chi^{\mf{sp}(6)}_{(000010)}v^{13+2n} +\chi^{\mf{so}(12)}_{(2n1010)}\chi^{\mf{sp}(6)}_{(000100)}v^{14+2n}\nn &-\chi^{\mf{so}(12)}_{(3n1010)}\chi^{\mf{sp}(6)}_{(001000)}v^{15+2n}
+\chi^{\mf{so}(12)}_{(4n1010)}\chi^{\mf{sp}(6)}_{(010000)}v^{16+2n}-\chi^{\mf{so}(12)}_{(5n1010)}\chi^{\mf{sp}(6)}_{(100000)}v^{17+2n}\nn
&+\chi^{\mf{so}(12)}_{(6n1010)}v^{18+2n}\Big)
-\Big((\chi^{\mf{so}(12)}_{(0n0020)}+\chi^{\mf{so}(12)}_{(0n2000)}v^2)\chi^{\mf{sp}(6)}_{(000001)}v^{11+2n}\nn
&- (\chi^{\mf{so}(12)}_{(1n0020)}+\chi^{\mf{so}(12)}_{(1n2000)}v^2)\chi^{\mf{sp}(6)}_{(000010)}v^{12+2n} +(\chi^{\mf{so}(12)}_{(2n0020)}+\chi^{\mf{so}(12)}_{(2n2000)}v^2)\chi^{\mf{sp}(6)}_{(000100)}v^{13+2n}\nn
&-(\chi^{\mf{so}(12)}_{(3n0020)}+\chi^{\mf{so}(12)}_{(3n2000)}v^2)\chi^{\mf{sp}(6)}_{(001000)}v^{14+2n}
+(\chi^{\mf{so}(12)}_{(4n0020)}+\chi^{\mf{so}(12)}_{(4n2000)}v^2)\chi^{\mf{sp}(6)}_{(010000)}v^{15+2n}\nn
&-(\chi^{\mf{so}(12)}_{(5n0020)}+\chi^{\mf{so}(12)}_{(5n2000)}v^2)\chi^{\mf{sp}(6)}_{(100000)}v^{16+2n}
+(\chi^{\mf{so}(12)}_{(6n0020)}+\chi^{\mf{so}(12)}_{(6n2000)}v^2)v^{17+2n}\Big)
\Big].\label{n2so12HS}
\end{align}
The sporadic terms outside the infinite summation are too long to
present, thus here we only show some in leading orders. In general they can
be recovered from the terms inside the infinite summation. We also checked this expression from 5d blowup equations. A few
leading terms in the $v$ expansion has been determined in (H.22) of
\cite{DelZotto:2018tcj}.

\subsection*{$\mathbf{n=2,\, G=E_7,\,F=\mf{so}(6)}$}\label{sec:n2E7E1}
Let us regard the flavor group as $\mf{su}(4)$ to present the elliptic
genus. We use both the $v$ expansion method and the recursion formula
from 5d blowup equations to compute the leading $q$ order of the
reduced one-string elliptic genus, and find the following exact
formula:
\begin{align}\nonumber
-\,&\chi^{su(4)}_{(8,0,4)\oplus(4,0,8)}v^{15}-\chi^{su(4)}_{(7,0,5)\oplus(5,0,7)}\chi^{E_7}_{(n000001)}v^{18}
+\chi^{su(4)}_{(6,0,6)}(\chi^{E_7}_{(1000000)}v^{17}+\chi^{E_7}_{(0100000)}v^{19}) \\ \nonumber
+\,&\chi^{su(4)}_{(7,1,3)\oplus(3,1,7)}\chi^{E_7}_{(0000010)}v^{16}+\chi^{su(4)}_{(6,1,4)\oplus(4,1,6)}\chi^{E_7}_{(0000011)}v^{19} \\ \nonumber
-\,&\chi^{su(4)}_{(5,1,5)}(\chi^{E_7}_{(1000010)}v^{18}+\chi^{E_7}_{(0100010)}v^{20})+\dots \\ \nonumber
+\sum_{n=0}^\infty\Big[&-\chi^{su(4)}_{(12,0,0)\oplus(0,0,12)}\chi^{E_7}_{(n000000)}v^{17+2n}+\chi^{su(4)}_{(11,0,1)\oplus(1,0,11)}\chi^{E_7}_{(n000010)}v^{18+2n}
\\[-2mm] \nonumber
&-\chi^{su(4)}_{(10,0,2)\oplus(2,0,10)}\chi^{E_7}_{(n000100)}v^{19+2n}+\chi^{su(4)}_{(9,0,3)\oplus(3,0,9)}\chi^{E_7}_{(n001000)}v^{20+2n}\\ \nonumber
&-\chi^{su(4)}_{(8,0,4)\oplus(4,0,8)}\chi^{E_7}_{(n010000)}v^{21+2n}
+\chi^{su(4)}_{(7,0,5)\oplus(5,0,7)}\chi^{E_7}_{(n100001)}v^{22+2n}\\ \nonumber
&-\chi^{su(4)}_{(6,0,6)}(\chi^{E_7}_{(n000002)}v^{21+2n}+\chi^{E_7}_{(n200000)}v^{23+n})-\chi^{su(4)}_{(10,1,0)\oplus(0,1,10)}\chi^{E_7}_{(n000020)}v^{19+2n}\\ \nonumber
&+\chi^{su(4)}_{(9,1,1)\oplus(1,1,9)}\chi^{E_7}_{(n000110)}v^{20+2n}-\chi^{su(4)}_{(8,1,2)\oplus(2,1,8)}\chi^{E_7}_{(n001010)}v^{21+2n}\\ \nonumber
&+\chi^{su(4)}_{(7,1,3)\oplus(3,1,7)}\chi^{E_7}_{(n010010)}v^{22+2n}-\chi^{su(4)}_{(6,1,4)\oplus(4,1,6)}\chi^{E_7}_{(n100011)}v^{23+2n}\\
&+\chi^{su(4)}_{(5,1,5)}(\chi^{E_7}_{(n000012)}v^{22+2n}+\chi^{E_7}_{(n200010)}v^{24+n})+\dots
\Big].\label{n2E7HS}
\end{align}
The full dependence on flavor representations are too long to
present. Here we only show the terms involving the largest
representations of $\mf{su}(4)$ with Dynkin label $(b_1,b_2,b_3)$
satisfying $b_1+2b_2+b_3=12$ and $b_2=0,1$.

\subsection*{$\mathbf{n=3,\, G=\mf{so}(8),\,F=\mf{sp}(1)_a\times \mf{sp}(1)_b\times \mf{sp}(1)_c}$}\label{sec:n3SO8}
Denote the reduced one-string elliptic genus with all gauge and flavor
fugacities turned off as
\begin{equation}\label{n3SO8E1}
\mathbb{E}_{h_{3,\mf{so}(8)}^{(1)}}(\Qtau,v,m_{\mf{so}(8)}=0,m_F=0)=\Qtau^{-1/3}v^{4}\sum_{n=0}^{\infty}\Qtau^n\frac{P_n(v)}{(1 - v)^{4} (1 + v)^{10}}.
\end{equation}
From the recursion formula from blowup equations, we obtain
\begin{equation}\label{n3SO8E1P0}
\begin{aligned}
P_0&(v)=1 + 14 v - 37 v^2 + 68 v^3 - 37 v^4 + 14 v^5 + v^6,
 \end{aligned}
\end{equation}
\begin{equation}\label{n3SO8E1P1}
\begin{aligned}
P_1&(v)=v^{-6}(1 + 6 v + 11 v^2 - 4 v^3 - 41 v^4 - 50 v^5 + 43 v^6 + 564 v^7 -
  1310 v^8 + 1752 v^9 - \dots + v^{18}).\nonumber
 \end{aligned}
\end{equation}
These agree with the modular ansatz in \cite{DelZotto:2018tcj}. With
all gauge and flavor fugacities turned on, we reobtain the exact
formula for the leading $q$ order of the reduced one-string elliptic
genus in \cite{DelZotto:2018tcj} and \cite{Kim:2019uqw} as
\begin{align}
v^4+&\sum_{n=0}^\infty\Big[ \chi^{\mf{so}(8)}_{(0n00)}\chi^F_{(1)_a\otimes(1)_b\otimes(1)_c}v^{5+2n}-  (\chi^{\mf{so}(8)}_{(1n00)}\chi^F_{(1)_{b}\otimes (1)_{c}}+\chi^{\mf{so}(8)}_{(0n10)}\chi^F_{(1)_{a}\otimes (1)_{c}}+\chi^{\mf{so}(8)}_{(0n01)}\chi^F_{(1)_{a}\otimes (1)_{b}})v^{6+2n}\nonumber\\
&+ (\chi^{\mf{so}(8)}_{(1n10)}\chi^F_{(1)_c}+\chi^G_{(1n01)}\chi^F_{(1)_b}+\chi^{\mf{so}(8)}_{(0n11)}\chi^F_{(1)_a})v^{7+2n}- \chi^{\mf{so}(8)}_{(1n11)}v^{8+2n}\Big].
 \end{align}
We also obtain the subleading $q$ order as
\begin{align}\nonumber
v^{-2}-2v^2+(\chi^F_{(2)_{a}\oplus(2)_{b}\oplus(2)_{c}}+1+\chi^{\mf{so}(8)}_{(0100)})v^4+\chi^F_{(1)_a\otimes(1)_b\otimes(1)_c}(
\chi^{\mf{so}(8)}_{(0100)}+4)v^5+\mathcal{O}(v^6).
 \end{align}

Denote the reduced two-string elliptic genus with all gauge and flavor fugacities turned off as
\begin{equation}\label{n3SO8E2}
\mathbb{E}_{h_{3,\mf{so}(8)}^{(2)}}(\Qtau,v,x=1,m_{\mf{so}(8)}=0,m_F=0)=-\Qtau^{-5/6}v^{9}\sum_{n=0}^{\infty}\Qtau^n\frac{P^{(2)}_n(v)}{(1 - v)^{10} (1 + v)^{10} (1 + v + v^2)^{11}}.
\end{equation}
We obtain
\begin{align}
P_0^{(2)}(v)=\,&1 + 19 v + 94 v^2 + 77 v^3 + 31 v^4 + 592 v^5 + 1681 v^6 + 1395 v^7 +
 942 v^8 + 3775 v^9 \nn
 &+ 7249 v^{10} + 5434 v^{11} + 3008 v^{12} + \dots + v^{24},\nn
P_1^{(2)}(v)=&\,v^{-6} (1 + 24 v + 152 v^2 + 541 v^3 + 1377 v^4 + 2582 v^5 + 3949 v^6 +
 5335 v^7 + 9170 v^8\nn
 & + 13009 v^9 + 6362 v^{10} - 5437 v^{11} +
 23841 v^{12} + 92713 v^{13} + 134067 v^{14} + 169449 v^{15}\nn
 & + 309565 v^{16} +
 451272 v^{17} + 425964 v^{18} + 359168 v^{19} + \dots + v^{38}).\nonumber
 \end{align}

\subsection*{$\mathbf{n=3,\, G=\mf{so}(9),\,F=\mf{sp}(2)\times \mf{sp}(1)}$}\label{sec:n3SO9}
Denote the reduced one-string elliptic genus with all gauge and flavor fugacities turned off as
\begin{equation}\label{n3SO9E1}
\mathbb{E}_{h_{3,\mf{so}(9)}^{(1)}}(\Qtau,v,m_{\mf{so}(9)}=0,m_F=0)=\Qtau^{-1/3}v^{5}\sum_{n=0}^{\infty}\Qtau^n\frac{P_n(v)}{(1 - v)^{4} (1 + v)^{12}}.
\end{equation}
We obtain
\begin{equation}\label{n3SO9E1P0}
\begin{aligned}
P_0&(v)=-2  (2 + 19 v - 62 v^2 + 106 v^3 - 62 v^4 + 19 v^5 + 2 v^6),
 \end{aligned}
\end{equation}
\begin{equation}
\begin{aligned}
P_1(v)=-v^{-7}(&\,1 + 8 v + 24 v^2 + 24 v^3 - 37 v^4 - 132 v^5 - 144 v^6 + 180 v^7 +
  2004 v^8\\
  & - 5264 v^9 + 7056 v^{10} - \dots + v^{20}).\nonumber
 \end{aligned}
\end{equation}
Denote the reduced two-string elliptic genus with all gauge and flavor fugacities turned off as
\begin{equation}\label{n3SO9E2}
\mathbb{E}_{h_{3,\mf{so}(9)}^{(2)}}(\Qtau,v,x=1,m_{\mf{so}(9)}=0,m_F=0)=-\Qtau^{-5/6}v^{11}\sum_{n=0}^{\infty}\Qtau^n\frac{P^{(2)}_n(v)}{(1 - v)^{10} (1 + v)^{12} (1 + v + v^2)^{13}}.
\end{equation}
We obtain
\begin{align}
P_0^{(2)}(v)=&\,10 + 174 v + 707 v^{2 }+ 851 v^{3 }- 109 v^{4 }+ 1860 v^{5 }+ 11190 v^{6 }+
 16610 v^{7 }+ 6728 v^{8 }\nn
 &+ 7008 v^{9 }+ 43183 v^{10 }+ 70861 v^{11 }+
 45001 v^{12 }+ 18164 v^{13 }+ \dots + 10 v^{26},\nn
P_1^{(2)}(v)=&\,v^{-7}(4 + 74 v + 398 v^2 + 1414 v^3 + 3488 v^4 + 6697 v^5 + 9871 v^6 +
 12142 v^7 + 18585 v^8\nn
 & + 43069 v^9 + 55702 v^{10} - 10441 v^{11} -
 73597 v^{12} + 105935 v^{13} + 359120 v^{14} + 239627 v^{15}\nn
 & + 114575 v^{16} +
 750264 v^{17} + 1400325 v^{18} + 990699 v^{19} + 470338 v^{20}+ \dots+ 4 v^{40}).
 \end{align}
\subsection*{$\mathbf{n=3,\, G=\mf{so}(10),\,F=\mf{sp}(3)\times \mf{u}(1)}$}\label{sec:n3SO10}
Denote the reduced one-string elliptic genus with all gauge and flavor fugacities turned off as
\begin{equation}\label{n3SO10E1}
\mathbb{E}_{h_{3,\mf{so}(10)}^{(1)}}(\Qtau,v,m_{\mf{so}(10)}=0,m_F=0)=\Qtau^{-1/3}v^{6}\sum_{n=0}^{\infty}\Qtau^n\frac{P_n(v)}{(1 - v)^{4} (1 + v)^{14}}.
\end{equation}
We obtain
\begin{equation}\label{n3SO10E1P0}
\begin{aligned}
P_0(v)=&\,2 (7 + 54 v - 210 v^2 + 344 v^3 - 210 v^4 + 54 v^5 + 7 v^6),\\
P_1(v)=&\,v^{-8}(1 + 10 v + 41 v^2 + 80 v^3 + 35 v^4 - 178 v^5 - 419 v^6 - 428 v^7\\& +
  676 v^8 + 7284 v^9 - 20742 v^{10} + 28016 v^{11} - \dots + v^{22}).
 \end{aligned}
\end{equation}
Denote the reduced two-string elliptic genus with all gauge and flavor fugacities turned off as
\begin{equation}\label{n3SO10E2}\nonumber
\mathbb{E}_{h_{3,\mf{so}(10)}^{(2)}}(\Qtau,v,x=1,m_{\mf{so}(10)}=0,m_F=0)=-\Qtau^{-5/6}v^{13}\sum_{n=0}^{\infty}\Qtau^n\frac{P^{(2)}_n(v)}{(1 - v)^{10} (1 + v)^{18} (1 + v + v^2)^{15}}.
\end{equation}
We obtain
\begin{align}
P_0^{(2)}(v)=&\,2(45 + 932 v + 6264 v^{2}+ 21096 v^{3}+ 37801 v^{4}+ 32448 v^{5}+
 31299 v^{6}+ 178325 v^{7}\nn
 &+ 549579 v^{8}+ 838987 v^{9}+ 682443 v^{10}+
 561148 v^{11}+ 1511348 v^{12}+ 3259788 v^{13}\nn
 &+ 3952706 v^{14}+
 2932464 v^{15}+ 2106794 v^{16}+ \dots+ 45 v^{32}),\nn
P_1^{(2)}(v)=&\,2v^{-8}(7 + 159 v + 1412 v^{2}+ 7693 v^{3}+ 29780 v^{4}+ 87899 v^{5}+
 205494 v^{6}+ 388084 v^{7}\nn
 &+ 597939 v^{8}+ 790211 v^{9}+ 1104282 v^{10}+
 1974138 v^{11}+ 3342747 v^{12}+ 3399917 v^{13}\nn
 &+ 355771 v^{14} -
 2250673 v^{15}+ 2821724 v^{16}+ 13232633 v^{17}+ 15679593 v^{18}+
 11581039 v^{19}\nn
 +\,& 25206981 v^{20}+ 61068134 v^{21}+ 81796560 v^{22}+
 66229422 v^{23}+ 50700908 v^{24}+ \dots+ 7 v^{48}).
 \end{align}

\subsection*{$\mathbf{n=3,\, G=\mf{so}(12),\,F=\mf{sp}(5)}$}
This theory belongs to class $\mathbf C$ which only has vanishing blowup equations. The leading $q$ order of reduced one-string elliptic genus, i.e. the reduced 5d one-instanton partition function was partially determined in \cite{DelZotto:2018tcj}. Using the vanishing blowup equations, we are able to fix it as
\begin{align}
\,& v^8\chi^{\mf{sp}(5)}_{(00010)}-v^9\chi^{\mf{so}(12)}_{(100000)}\chi^{\mf{sp}(5)}_{(00100)}+v^{10}\chi^{\mf{so}(12)}_{(200000)}\chi^{\mf{sp}(5)}_{(01000)}-v^{11}\chi^{\mf{so}(12)}_{(300000)}\chi^{\mf{sp}(5)}_{(10000)}
+v^{12}\chi^{\mf{so}(12)}_{(400000)}\nonumber\\ \nonumber
&+\sum_{n=0}^\infty\Big[ -v^{10+2n}\chi^{\mf{so}(12)}_{(0n0010)}\chi^{\mf{sp}(5)}_{(00001)}+ v^{11+2n}(\chi^{\mf{so}(12)}_{(1n0010)}\chi^{\mf{sp}(5)}_{(00010)}+\chi^{\mf{so}(12)}_{(0n1000)}\chi^{\mf{sp}(5)}_{(00001)})\\[-1mm] \nonumber
&- v^{12+2n}(\chi^{\mf{so}(12)}_{(2n0010)}\chi^{\mf{sp}(5)}_{(00100)}+\chi^{\mf{so}(12)}_{(1n1000)}\chi^{\mf{sp}(5)}_{(00010)})+ v^{13+2n}(\chi^{\mf{so}(12)}_{(3n0010)}\chi^{\mf{sp}(5)}_{(01000)}+\chi^{\mf{so}(12)}_{(2n1000)}\chi^{\mf{sp}(5)}_{(00100)})\\ \nonumber &- v^{14+2n}(\chi^{\mf{so}(12)}_{(4n0010)}\chi^{\mf{sp}(5)}_{(10000)}+\chi^{\mf{so}(12)}_{(3n1000)}\chi^{\mf{sp}(5)}_{(01000)})+ v^{15+2n}(\chi^{\mf{so}(12)}_{(5n0010)}+\chi^{\mf{so}(12)}_{(4n1000)}\chi^{\mf{sp}(5)}_{(10000)})\\ &-v^{16+2n}\chi^{\mf{so}(12)}_{(5n1000)}\Big].\label{n3SO12HS}
 \end{align}

\subsection*{$\mathbf{n=3,\, G=E_6,\,F=\mf{su}(3)_a\times \mf{u}(1)_b}$}\label{sec:n3E6E2}
From the recursion formula, we obtain the following exact formula for the leading $q$ order of reduced one-string elliptic genus, i.e. the reduced 5d one-instanton partition function:
\begin{align}\nonumber
\ &\Big(v^{10}\chi^F_{(03)_a\oplus(6)_b}-v^{11}\chi^{E_6}_{(100000)}\chi^F_{(12)_a\oplus(5)_b}+v^{12}\chi^{E_6}_{(010000)}\chi^F_{(21)_a\oplus(4)_b}
+v^{12}\chi^{E_6}_{(200000)}\chi^F_{(02)_a\oplus(4)_b}\\ \nonumber
&v^9\chi^F_{(06)_a\oplus(3)_b}-v^{11}\chi^{E_6}_{(000001)}\chi^F_{(30)_a\oplus(3)_b}-v^{13}\chi^{E_6}_{(001000)}\chi^F_{(30)_a\oplus(3)_b}v^{11}
-\chi^{E_6}_{(110000)}\chi^F_{(11)_a\oplus(3)_b}\\ \nonumber
&+\chi^F_{(3)_b}v^7+\chi^{E_6}_{(000100)}\chi^F_{(31)_a\oplus(2)_b}v^{12}-\chi^{E_6}_{(100000)}\chi^F_{(12)_a\oplus(2)_b}v^{10}
+\chi^{E_6}_{(100001)}\chi^F_{(20)_a\oplus(2)_b}v^{12}\\ \nonumber
&+\chi^{E_6}_{(101000)}\chi^F_{(20)_a\oplus(2)_b}v^{14}+\chi^{E_6}_{(020000)}\chi^F_{(01)_a\oplus(2)_b}v^{14}-\chi^{E_6}_{(000010)}\chi^F_{(32)_a\oplus(1)_b}v^{11}
\\ \nonumber
&-\chi^{E_6}_{(100100)}\chi^F_{(21)_a\oplus(1)_b}v^{13}+\chi^{E_6}_{(010000)}\chi^F_{(02)_a\oplus(1)_b}v^{11}-\chi^{E_6}_{(010001)}\chi^F_{(10)_a\oplus(1)_b}v^{13}
\\ \nonumber
&-\chi^{E_6}_{(011000)}\chi^F_{(10)_a\oplus(1)_b}v^{15}+v^8\chi^F_{(03)_a}
+c.c.\Big)+\chi^F_{(33)_a}v^{10}+\chi^{E_6}_{(101010)}\chi^F_{(22)_a}v^{12}\\ \nonumber
&-\chi^{E_6}_{(000001)}\chi^F_{(11)_a}v^{10}+\chi^{E_6}_{(010100)}\chi^F_{(11)_a}v^{14}+\chi^{E_6}_{(000002)}v^{12}
+\chi^{E_6}_{(001001)}v^{14}+\chi^{E_6}_{(002000)}v^{16}
\\ \nonumber
\sum_{n=0}^{\infty}&\Big[v^{11+2n}\chi^{E_6}_{(00000n)}\chi^F_{(9)_b\oplus(-9)_b}
-v^{12+2n}(\chi^{E_6}_{(10000n)}\chi^F_{(01)_a\oplus(8)_b}+c.c.)\\[-2mm] \nonumber
&\,+v^{13+2n}(\chi^{E_6}_{(01000n)}\chi^F_{(02)_a\oplus(7)_b}+\chi^{E_6}_{(20000n)}\chi^F_{(10)_a\oplus(7)_b}+c.c.)\\ \nonumber
&\,-v^{14+2n}(\chi^{E_6}_{(00100n)}\chi^F_{(03)_a\oplus(6)_b}+\chi^{E_6}_{(11000n)}\chi^F_{(11)_a\oplus(6)_b}+\chi^{E_6}_{(30000n)}\chi^F_{(00)_a\oplus(6)_b}+c.c.)\\ \nonumber
&\,+v^{13+2n}(\chi^{E_6}_{(00010n)}\chi^F_{(04)_a\oplus(5)_b}+c.c.)\\ \nonumber
&\,+v^{15+2n}(\chi^{E_6}_{(10100n)}\chi^F_{(12)_a\oplus(5)_b}+\chi^{E_6}_{(02000n)}\chi^F_{(20)_a\oplus(5)_b}+\chi^{E_6}_{(21000n)}\chi^F_{(01)_a\oplus(5)_b}+c.c.)\\ \nonumber
&\,-v^{12+2n}(\chi^{E_6}_{(00001n)}\chi^F_{(05)_a\oplus(4)_b}+c.c.)-v^{14+2n}(\chi^{E_6}_{(10010n)}\chi^F_{(13)_a\oplus(4)_b}+c.c.)\\ \nonumber
&\,-v^{16+2n}(\chi^{E_6}_{(01100n)}\chi^F_{(21)_a\oplus(4)_b}+\chi^{E_6}_{(20100n)}\chi^F_{(02)_a\oplus(4)_b}+\chi^{E_6}_{(12000n)}
\chi^F_{(10)_a\oplus(4)_b}+c.c.)\\ \nonumber
&\,+v^{11+2n}(\chi^{E_6}_{(00000n)}\chi^F_{(06)_a\oplus(3)_b}+c.c.)+v^{13+2n}(\chi^{E_6}_{(10001n)}\chi^F_{(14)_a\oplus(3)_b}+c.c.)\\ \nonumber
&\,+v^{15+2n}(\chi^{E_6}_{(01010n)}\chi^F_{(22)_a\oplus(3)_b}+\chi^{E_6}_{(20010n)}\chi^F_{(03)_a\oplus(3)_b}+c.c.)\\ \nonumber
&\,+v^{17+2n}(\chi^{E_6}_{(00200n)}\chi^F_{(30)_a\oplus(3)_b}+\chi^{E_6}_{(11100n)}\chi^F_{(11)_a\oplus(3)_b}+\chi^{E_6}_{(03000n)}\chi^F_{(3)_b}+c.c.)\\ \nonumber
&\,-v^{12+2n}(\chi^{E_6}_{(10000n)}\chi^F_{(15)_a\oplus(2)_b}+c.c.)-v^{14+2n}(\chi^{E_6}_{(01001n)}\chi^F_{(23)_a\oplus(2)_b}
+\chi^{E_6}_{(20001n)}\chi^F_{(04)_a\oplus(2)_b}+c.c.)\\ \nonumber
&\,-v^{16+2n}(\chi^{E_6}_{(00110n)}\chi^F_{(31)_a\oplus(2)_b}
+\chi^{E_6}_{(11010n)}\chi^F_{(12)_a\oplus(2)_b}+c.c.)\\ \nonumber
&\,-v^{18+2n}(\chi^{E_6}_{(10200n)}\chi^F_{(20)_a\oplus(2)_b}
+\chi^{E_6}_{(02100n)}\chi^F_{(01)_a\oplus(2)_b}+c.c.)\\ \nonumber
&\,+v^{13+2n}(\chi^{E_6}_{(01000n)}\chi^F_{(24)_a\oplus(1)_b}
+\chi^{E_6}_{(20000n)}\chi^F_{(05)_a\oplus(1)_b}+c.c.)\\ \nonumber
&\,+v^{15+2n}(\chi^{E_6}_{(00101n)}\chi^F_{(32)_a\oplus(1)_b}
+\chi^{E_6}_{(11001n)}\chi^F_{(13)_a\oplus(1)_b}+\chi^{E_6}_{(00020n)}\chi^F_{(40)_a\oplus(1)_b}+c.c.)\\ \nonumber
&\,+v^{17+2n}(\chi^{E_6}_{(10110n)}\chi^F_{(21)_a\oplus(1)_b}
+\chi^{E_6}_{(02010n)}\chi^F_{(02)_a\oplus(1)_b}+c.c.)  +v^{19+2n}\chi^{E_6}_{(01200n)}(\chi^F_{(10)_a\oplus(1)_b}+ c.c.)\\ \nonumber
&\,-v^{14+2n}\chi^{E_6}_{(00100n)}\chi^F_{(33)_a}-v^{14+2n}(\chi^{E_6}_{(11000n)}\chi^F_{(14)_a}+c.c.)\\ \nonumber
&\,-v^{16+2n}\chi^{E_6}_{(10101n)}\chi^F_{(22)_a}-v^{16+2n}(\chi^{E_6}_{(02001n)}\chi^F_{(03)_a}+c.c.)\\
&\,-v^{18+2n}\chi^{E_6}_{(01110n)}\chi^F_{(11)_a}-v^{20+2n}\chi^{E_6}_{(00300n)}
\Big]\label{n3E6exactE1}
\end{align}
After turning off all $E_6$ gauge fugacities, the above exact formula reduces to the result (A.17) of \cite{Kim:2019uqw} by Weyl dimension formula of representations of $E_6$. Further turning off all flavor fugacities, one obtains the rational function of $v$ in (\ref{n3E6E1}).

Denote the reduced two-string elliptic genus with all gauge and flavor fugacities turned off as
\begin{equation}\label{n3E6E2}\nonumber
\mathbb{E}_{h_{3,E_6}^{(2)}}(\Qtau,v,x=1,m_{E_6}=0,m_F=0)=-\Qtau^{-5/6}v^{15}\sum_{n=0}^{\infty}\Qtau^n\frac{P^{(2)}_n(v)}{(1 - v)^{10} (1 + v)^{26} (1 + v + v^2)^{23}},
\end{equation}
we obtain
\begin{align}
P_0^{(2)}&(v)=3 + 159 v + 4245 v^{2 }+ 72622 v^{3 }+ 863819 v^{4 }+ 7446591 v^{5 }+
 47902516 v^{6 }+ 235241313 v^{7 }\nn
 &+ 896085222 v^{8 }+ 2671738023 v^{9 }+
 6257280290 v^{10 }+ 11565342413 v^{11 }+ 17441014579 v^{12 }\nn
 &+
 24757146408 v^{13 }+ 43167107703 v^{14 }+ 92340625269 v^{15 }+
 184446978968 v^{16 }\nn
 &+ 297014465909 v^{17 }+ 380602273913 v^{18 }+
 427769333206 v^{19 }+ 533426305310 v^{20 }\nn
 &+ 825794587232 v^{21 }+
 1287690035763 v^{22 }+ 1693325870657 v^{23 }+ 1815742557209 v^{24 }\nn
 &+
 1695462175970 v^{25 }+ 1602451245554 v^{26 }+\dots+
 3 v^{52}.\label{n3E6E2P0}
 \end{align}

\subsection*{$\mathbf{n=4,\, G=E_6,\,F=\mf{su}(2)\times \mf{u}(1)}$}\label{sec:n4E6E2}
Denote the reduced two-string elliptic genus with all gauge and flavor fugacities turned off as
\begin{equation}\label{n4E6E2}
\mathbb{E}_{h_{4,E_6}^{(2)}}(\Qtau,v,x=1,m_G=0,m_F=0)=-\Qtau^{-11/6}v^{19}\sum_{n=0}^{\infty}\Qtau^n\frac{P^{(2)}_n(v)}{(1 - v)^{22} (1 + v)^{28} (1 + v + v^2)^{23}},
\end{equation}
we obtain
\begin{align}\nonumber
P_0^{(2)}&(v)=6 + 200 v + 2632 v^{2 }+ 17758 v^{3 }+ 75489 v^{4 }+ 243367 v^{5 }+
 760467 v^{6 }+ 2577888 v^{7 }\nn&+ 8317316 v^{8 }+ 23236506 v^{9 }+
 58513940 v^{10 }+ 143767140 v^{11 }+ 347390848 v^{12 }+ 786032254 v^{13 }\nn&+
 1633105895 v^{14 }+ 3195818881 v^{15 }+ 6041990014 v^{16 }+
 10959026237 v^{17 }+ 18715741117 v^{18 }\nn&+ 30093383834 v^{19 }+
 46262367433 v^{20 }+ 68471264635 v^{21 }+ 96730928747 v^{22 }\nn&+
 129436722092 v^{23 }+ 164888050451 v^{24 }+ 201811431341 v^{25 }+
 237209409984 v^{26 }\nn&+ 265667738531 v^{27 }+ 282914996487 v^{28 }+
 288440699594 v^{29 }+ \dots+ 6 v^{58},
 \end{align}
and
\begin{align}\nonumber
P_1^{(2)}&(v)=-v^{-2}(8 + 262 v + 2954 v^{2 }+ 6882 v^{3 }- 125701 v^{4 }- 1314279 v^{5 }-
 6621327 v^{6 }- 23006770 v^{7 }\nn
 &- 69417453 v^{8 }- 213977845 v^{9 }-
 651520698 v^{10 }- 1773023963 v^{11 }- 4276376371 v^{12 }\nn
 &-
 9730496854 v^{13 }- 21781260461 v^{14 }- 46890358519 v^{15 }-
 94029008670 v^{16 }- 176640724111 v^{17 }\nn
 &- 318761640562 v^{18 }-
 556066823089 v^{19 }- 924340036971 v^{20 }- 1452988495522 v^{21 }\nn
 &-
 2179171428592 v^{22 }- 3147790892042 v^{23 }- 4365630688208 v^{24 }-
 5770440288994 v^{25 }\nn
 &- 7276423650370 v^{26 }- 8812083976234 v^{27 }-
 10262845252021 v^{28 }- 11435602558269 v^{29 }\nn
 &- 12163726096281 v^{30 }-
 12402928893114 v^{31} +\dots+ 20 v^{62}).
 \end{align}
\subsection*{$\mathbf{n=4,\, G=E_7,\,F=\mf{so}(4)}$}\label{sec:n4E7E1}

Regarding the flavor symmetry $F$ as $\mf{su}(2)\times\mf{su}(2) $, we obtain the following exact formula for the leading $q$ order of reduced one-string elliptic genus, i.e. the reduced 5d one-instanton partition function:
\begin{align}\nonumber
-&\chi^F_{(8,4)\oplus(4,8)}v^{15}-\chi^F_{(7,5)\oplus(5,7)}\chi^{E_7}_{(0000001)}v^{18}+\chi^F_{(6,6)}(\chi^{E_7}_{(1000000)}v^{17}+\chi^{E_7}_{(0100000)}v^{19})\\ \nonumber
+&\chi^F_{(7,3)\oplus(3,7)}\chi^{E_7}_{(0000010)}v^{16}+\chi^F_{(6,4)\oplus(4,6)}\chi^{E_7}_{(0000011)}v^{19} -\chi^F_{(5,5)}(\chi^{E_7}_{(1000010)}v^{18}+\chi^{E_7}_{(0100010)}v^{20})  \\ \nonumber
-&\chi^F_{(6,2)\oplus(2,6)}\chi^{E_7}_{(0000100)}v^{17}-\chi^F_{(5,3)\oplus(3,5)}\chi^{E_7}_{(0000101)}v^{20} +\chi^F_{(4,4)}(-v^{13}+\chi^{E_7}_{(10000100)}v^{19}+\chi^{E_7}_{(0100100)}v^{21})   \\ \nonumber
-&\chi^F_{(6,0)\oplus(0,6)}v^{13}+\chi^F_{(5,1)\oplus(1,5)}\chi^{E_7}_{(0001000)}v^{18}+\chi^F_{(4,2)\oplus(2,4)}(\chi^{E_7}_{(1000000)}v^{15}+\chi^{E_7}_{(0001001)}v^{21})\\ \nonumber
-&\chi^F_{(3,3)}(\chi^{E_7}_{(0000001)}v^{16}+\chi^{E_7}_{(1001000)}v^{20}+\chi^{E_7}_{(0101000)}v^{22}) -\chi^F_{(4,0)\oplus(0,4)}(\chi^{E_7}_{(0100000)}v^{17}+\chi^{E_7}_{(0010000)}v^{19})  \\ \nonumber
+&\chi^F_{(3,1)\oplus(1,3)}(\chi^{E_7}_{(0000101)}v^{18}-\chi^{E_7}_{(0010001)}v^{22})+\chi^F_{(2,2)}(-\chi^{E_7}_{(2000000)}v^{17}-\chi^{E_7}_{(0000002)}v^{19}\\ \nonumber
\phantom{-}&+\chi^{E_7}_{(1010000)}v^{21}+\chi^{E_7}_{(0110000)}v^{23})+ \chi^F_{(2,0)\oplus(0,2)}(\chi^{E_7}_{(1000002)}v^{21}+\chi^{E_7}_{(0100002)}v^{23})   \\ \nonumber
-&\,\chi^F_{(1,1)}(\chi^{E_7}_{(2000001)}v^{20}+(\chi^{E_7}_{(1100001)}+\chi^{E_7}_{(0000003)})v^{22}+\chi^{E_7}_{(0200001)}v^{24})\\ \nonumber
-&\,v^{11}+\chi^{E_7}_{(3000000)}v^{19}+\chi^{E_7}_{(2100000)}v^{21}+\chi^{E_7}_{(1200000)}v^{23}+\chi^{E_7}_{(0300000)}v^{25}+\\[-1mm] \nonumber
\sum_{n=0}^{\infty}&\,\Big[-\chi^F_{(12,0)\oplus(0,12)}\chi^{E_7}_{(n000000)}v^{17+2n}+\chi^F_{(11,1)\oplus(1,11)}\chi^{E_7}_{(n000010)}v^{18+2n}
-\chi^F_{(10,2)\oplus(2,10)}\chi^{E_7}_{(n000100)}v^{19+2n}\\[-1mm] \nonumber
&+\chi^F_{(9,3)\oplus(3,9)}\chi^{E_7}_{(n001000)}v^{20+2n}-\chi^F_{(8,4)\oplus(4,8)}\chi^{E_7}_{(n010000)}v^{21+2n}+\chi^F_{(7,5)\oplus(5,7)}\chi^{E_7}_{(n100001)}v^{22+2n}\\ \nonumber
&-\chi^F_{(6,6)}v^{21+2n}(\chi^{E_7}_{(n000002)}+v^2\chi^{E_7}_{(n200000)})-\chi^F_{(10,0)\oplus(0,10)}\chi^{E_7}_{(n000020)}v^{19+2n}\\ \nonumber
&+\chi^F_{(9,1)\oplus(1,9)}\chi^{E_7}_{(n000110)}v^{20+2n}-\chi^F_{(8,2)\oplus(2,8)}\chi^{E_7}_{(n001010)}v^{21+2n}+\chi^F_{(7,3)\oplus(3,7)}\chi^{E_7}_{(n010010)}v^{22+2n}\\ \nonumber
&-\chi^F_{(6,4)\oplus(4,6)}\chi^{E_7}_{(n10011)}v^{23+2n}+\chi^F_{(5,5)}v^{22+2n}(\chi^{E_7}_{(n000012)}+v^2\chi^{E_7}_{(n200010)})\\ \nonumber
&-\chi^F_{(8,0)\oplus(0,8)}\chi^{E_7}_{(n000200)}v^{21+2n}+\chi^F_{(7,1)\oplus(1,7)}\chi^{E_7}_{(n001100)}v^{22+2n}-\chi^F_{(6,2)\oplus(2,6)}\chi^{E_7}_{(n010100)}v^{23+2n}\\ \nonumber
&+\chi^F_{(5,3)\oplus(3,5)}\chi^{E_7}_{(n100101)}v^{24+2n}-\chi^F_{(4,4)}v^{23+2n}(\chi^{E_7}_{(n000102)}+v^2\chi^{E_7}_{(n200100)})\\ \nonumber
&-\chi^F_{(6,0)\oplus(0,6)}\chi^{E_7}_{(n002000)}v^{23+2n}+\chi^F_{(5,1)\oplus(1,5)}\chi^{E_7}_{(n011000)}v^{24+2n}-\chi^F_{(4,2)\oplus(2,4)}\chi^{E_7}_{(n101001)}v^{25+2n}\\ \nonumber
&+\chi^F_{(3,3)}v^{24+2n}(\chi^{E_7}_{(n001002)}+v^2\chi^{E_7}_{(n201000)}) -\chi^F_{(4,0)\oplus(0,4)}\chi^{E_7}_{(n020000)}v^{25+2n}\\ \nonumber
&+\chi^F_{(3,1)\oplus(1,3)}\chi^{E_7}_{(n110001)}v^{26+2n}-\chi^F_{(2,2)}v^{25+2n}(\chi^{E_7}_{(n010002)}+v^2\chi^{E_7}_{(n210000)}) \\ \nonumber
& -\chi^F_{(2,0)\oplus(0,2)}\chi^{E_7}_{(n200002)}v^{27+2n}+\chi^F_{(1,1)}v^{26+2n}(\chi^{E_7}_{(n100003)}+v^2\chi^{E_7}_{(n300001)}) \\
&-(\chi^{E_7}_{(n000004)}+v^4\chi^{E_7}_{(n400000)})v^{25+2n}
\Big].\label{eq:exactn4E7Z1}
\end{align}
After turning off all $E_7$ gauge fugacities, the above exact formula
reduces to the result (A.20) of \cite{Kim:2019uqw} by Weyl dimension
formula of representations of $E_7$. Further turning off all flavor
fugacities, one obtains the rational function of $v$ in
(\ref{n4E7E1}).

For higher $q$ order, recall the notation (\ref{n4E7E1}) for the
reduced one-string elliptic genus with all gauge and flavor fugacities
turned off, we record some higher order results for the numerators
here:
\begin{align}
P_2(v)=\,&v^{-14}(1 + 24 v + 266 v^2 + 1784 v^3 + 7911 v^4 + 23344 v^5 + 40636 v^6 +
 9264 v^7 - 165238 v^8\nn
 & - 478912 v^9 - 556440 v^{10 }+ 298240 v^{11 }+
 2128520 v^{12 }+ 3082560 v^{13 }+ 550409 v^{14 }\nn
 &- 3596776 v^{15 }+
 700889 v^{16 }+ 4057056 v^{17 }- 152281616 v^{18 }- 456584704 v^{19 }\nn
 &+
 2336222907 v^{20 }+ 3518750120 v^{21 }- 39318682643 v^{22 }+
 116568917840 v^{23 }\nn
 &- 205266204842 v^{24 }+ 245300442560 v^{25} -\dots+v^{50}).\label{n4E7E1P2}
 \end{align}
\begin{align}
P_3(v)=\,&v^{-16}(133 + 2968 v + 30142 v^{2 }+ 181048 v^{3 }+ 689812 v^{4 }+ 1583048 v^{5 }+
 1289736 v^{6 }\nn
 &- 4557256 v^{7 }- 18160096 v^{8 }- 24750520 v^{9 }+
 11527121 v^{10 }+ 99398192 v^{11 }\nn
 &+ 140533368 v^{12 }- 28415888 v^{13 }-
 371622426 v^{14 }- 468305664 v^{15 }+ 101124822 v^{16 }\nn
 &+ 1079768624 v^{17 }+
 1758536636 v^{18 }+ 1201001616 v^{19 }- 8948098843 v^{20 }\nn
 &-
 35689905400 v^{21 }+ 94419997502 v^{22 }+ 281480553480 v^{23 }-
 1852610135319 v^{24 }\nn
 &+ 4805915623088 v^{25 }- 7955500120108 v^{26 }+
 9332969375888 v^{27 }-\dots+v^{54}).\label{n4E7E1P3}
 \end{align}
\begin{align}
P_4(v)=\,&v^{-18}(7371 + 150984 v + 1379855 v^{2 }+ 7213760 v^{3 }+ 22300305 v^{4 }+
 32499512 v^{5 }- 32572277 v^{6 }\nn
 &- 252945648 v^{7 }- 438572547 v^{8 }+
 80757800 v^{9 }+ 1590185774 v^{10 }+ 2182830856 v^{11 }\nn
 &- 1301117169 v^{12 }-
 6862074336 v^{13 }- 4256604675 v^{14 }+ 10850256216 v^{15 }\nn
 &+
 18344377949 v^{16 }- 7908490560 v^{17 }- 46555176815 v^{18 }-
 23990305416 v^{19 }\nn
 &+ 98055494050 v^{20 }+ 216546552760 v^{21 }-
 184743607501 v^{22 }- 1788205999184 v^{23 }\nn
 &+ 2239774014885 v^{24 }+
 13565904866280 v^{25 }- 64302524207535 v^{26 }\nn
 &+ 149664342880240 v^{27 }-
 235464619399970 v^{28 }+ 272014399573792 v^{29 }-\dots+v^{58}).\label{n4E7E1P4}
 \end{align}
\subsection*{$\mathbf{n=5,\, G=E_6,\,F=\mf{u}(1)}$}\label{sec:n5E6E2}
Denote the reduced two-string elliptic genus with all gauge and flavor fugacities turned off as
\begin{equation}\label{n5E6E2}
\mathbb{E}_{h_{5,E_6}^{(2)}}(\Qtau,v,x=1,m_{E_6}=0,m_{\mf{u}(1)}=0)=-\Qtau^{-17/6}v^{21}\sum_{n=0}^{\infty}\Qtau^n\frac{P^{(2)}_n(v)}{(1 - v)^{34} (1 + v)^{30} (1 + v + v^2)^{23}},
\end{equation}
we obtain
\begin{align}\nonumber
P_0^{(2)}&(v)=1 + 21 v + 153 v^{2 }+ 904 v^{3 }+ 5116 v^{4 }+ 25914 v^{5 }+ 116029 v^{6 }+
 477409 v^{7 }+ 1823569 v^{8 }\nn
 &+ 6443864 v^{9 }+ 21148972 v^{10 }+
 64945868 v^{11 }+ 187225307 v^{12 }+ 507470579 v^{13 }\nn
 &+ 1296690701 v^{14 }+
 3132384316 v^{15 }+ 7167102255 v^{16 }+ 15555191149 v^{17 }\nn
 &+
 32075501088 v^{18 }+ 62937552731 v^{19 }+ 117653600727 v^{20 }+
 209750655294 v^{21 }\nn
 &+ 356983566607 v^{22 }+ 580561108791 v^{23 }+
 902887841711 v^{24 }+ 1343669144748 v^{25 }\nn
 &+ 1914685757018 v^{26 }+
 2613923784990 v^{27 }+ 3420367203355 v^{28 }+ 4291402109101 v^{29 }\nn
 &+
 5164404456225 v^{30 }+ 5962900573462 v^{31 }+ 6606847822339 v^{32 }+
 7025662161955 v^{33 }\nn
 &+ 7170987830896 v^{34 }+ \dots+ v^{68},
 \end{align}
and
\begin{align}
P_1^{(2)}&(v)=84 + 1870 v + 15150 v^{2 }+ 92382 v^{3 }+ 509942 v^{4 }+ 2529414 v^{5 }+
 11170010 v^{6 }+ 45018822 v^{7 }\nn
 &+ 167914134 v^{8 }+ 580737756 v^{9 }+
 1867913107 v^{10 }+ 5619089721 v^{11 }+ 15872495069 v^{12 }\nn
 &+
 42199602702 v^{13 }+ 105848677375 v^{14 }+ 251124006621 v^{15 }+
 564703393888 v^{16 }\nn
 &+ 1205575234175 v^{17 }+ 2447284329306 v^{18 }+
 4730834408879 v^{19 }+ 8719854968064 v^{20 }\nn
 &+ 15341684421093 v^{21 }+
 25790951006163 v^{22 }+ 41466404452278 v^{23 }+ 63813198389587 v^{24 }\nn
 &+
 94061792487301 v^{25 }+ 132885858904299 v^{26 }+ 180032677369322 v^{27 }+
 234011514454012 v^{28 }\nn
 &+ 291950610885280 v^{29 }+ 349716381424128 v^{30 }+
 402326438406440 v^{31 }\nn
 &+ 444618538975344 v^{32 }+ 472069443334672 v^{33 }+
 481585928612732 v^{34}+ \dots+ 2 v^{68}).
 \end{align}

\section{More on Calabi-Yau construction}\label{app:cy}
In this section, we list the dual polytope $\nu_i^*$, Mori cone generators $l^{(i)}$ and the triple intersection ring $\mathcal{R}$ in terms of K\"ahler classes $J_i$ for the geometries we have constructed. We also give the relation between the geometric bases and Lie bases.

\subsection*{$\mathbf{\fn=7, G=E_7}$}
\begin{equation} \label{n7E7polytope}
 \begin{array}{rrrr|rrrrrrrrrrr|}
    \multicolumn{4}{c}{\nu_i^*}     &l^{(1)} & l^{(2)} & l^{(3)} & l^{(4)} & l^{(5)} & l^{(6)} & l^{(7)} & l^{(8)} & l^{(9)} &  \\
    0 & 0 & 0 & 0 \phantom{,} & -1 & 0 & 0 & 0 & 0 & 0 & 0 & 0 & -1 & \\
-1 & 0 & 0 & 0 \phantom{,} & 0 & 0 & 0 & 0 & 0 & 0 & 0 & 0 & 1 & \\
0 & -1 & 0 & 0 \phantom{,} & 0 & 0 & 0 & 0 & 0 & 0 & 0 & 1 & 0 & \\
0 & 0 & 0 & -1 \phantom{,} & 0 & 0 & 0 & 0 & 0 & 0 & 1 & -2 & 0 & \\
0 & 1 & 0 & -2 \phantom{,} & 0 & 0 & 0 & 0 & 0 & 1 & -1 & 1 & -1 & \\
1 & 1 & 0 & -2 \phantom{,} & 1 & 0 & 0 & 0 & 0 & 0 & -1 & 0 & 1 & \\
1 & 2 & 0 & -3 \phantom{,} & 1 & 0 & 0 & 0 & 0 & -2 & 1 & 0 & 0 & \\
2 & 3 & 0 & -4 \phantom{,} & -2 & 1 & -1 & 0 & 0 & 1 & 0 & 0 & 0 & \\
2 & 3 & 0 & -3 \phantom{,} & 1 & -2 & -1 & 0 & 1 & 0 & 0 & 0 & 0 & \\
2 & 3 & 0 & -2 \phantom{,} & 0 & 1 & 0 & 1 & -2 & 0 & 0 & 0 & 0 & \\
2 & 3 & 0 & -1 \phantom{,} & 0 & 0 & 0 & -2 & 1 & 0 & 0 & 0 & 0 & \\
2 & 3 & -1 & -7 \phantom{,} & 0 & 0 & 1 & 0 & 0 & 0 & 0 & 0 & 0 & \\
2 & 3 & 0 & 0 \phantom{,} & 0 & 0 & 0 & 1 & 0 & 0 & 0 & 0 & 0 & \\
2 & 3 & 1 & 0 \phantom{,} & 0 & 0 & 1 & 0 & 0 & 0 & 0 & 0 & 0 & \\
\end{array} \
\end{equation}
The triple intersection ring corresponding to the above Mori cone generators is
\begin{align}\nonumber
\mathcal{R}&=-\frac{1}{7}(4 J_1^3+4 J_3 J_1^2+5 J_6 J_1^2+6 J_7 J_1^2+3 J_8 J_1^2+2 J_9 J_1^2+4 J_3^2 J_1+15 J_6^2 J_1+30 J_7^2 J_1+11 J_8^2 J_1\\[-1mm]
+\,&8 J_9^2 J_1+5 J_3 J_6 J_1+6 J_3 J_7 J_1+18 J_6 J_7 J_1+3 J_3 J_8 J_1+9 J_6 J_8 J_1+15 J_7 J_8 J_1+2 J_3 J_9 J_1+6 J_6 J_9 J_1\nn
+\,&10 J_7 J_9 J_1+5 J_8 J_9 J_1+3 J_2^3+25 J_4^3+18 J_5^3+45 J_6^3+150 J_7^3+52 J_8^3+32 J_9^3+3 J_2 J_3^2+5 J_2 J_4^2\nn
+\,&5 J_3 J_4^2
+6 J_2 J_5^2+6 J_3 J_5^2+9 J_4 J_5^2+15 J_3 J_6^2+30 J_3 J_7^2+90 J_6 J_7^2+11 J_3 J_8^2+33 J_6 J_8^2+55 J_7 J_8^2\nn
+\,&8 J_3 J_9^2
+24 J_6 J_9^2+40 J_7 J_9^2+20 J_8 J_9^2+3 J_2^2 J_3+J_2^2 J_4+J_3^2 J_4+J_2 J_3 J_4+2 J_2^2 J_5+2 J_3^2 J_5\nn
+\,&15 J_4^2 J_5+2 J_2 J_3 J_5+3 J_2 J_4 J_5+3 J_3 J_4 J_5+5 J_3^2 J_6+6 J_3^2 J_7+54 J_6^2 J_7+18 J_3 J_6 J_7+3 J_3^2 J_8\nn
+\,&27 J_6^2 J_8+75 J_7^2 J_8+9 J_3 J_6 J_8+15 J_3 J_7 J_8+45 J_6 J_7 J_8+2 J_3^2 J_9+18 J_6^2 J_9+50 J_7^2 J_9+30 J_8^2 J_9\nn
+\,&6 J_3 J_6 J_9+10 J_3 J_7 J_9+30 J_6 J_7 J_9+5 J_3 J_8 J_9+15 J_6 J_8 J_9+25 J_7 J_8 J_9).\label{n7E7triple}
\end{align}
The relation between the above geometric bases and the Lie algebra bases is
\begin{align}
l_{E_7}^{(1)} &= l^{(5)},\quad l_{E_7}^{(1)} = l^{(2)},\quad
l_{E_7}^{(1)} = l^{(1)},\quad
l_{E_7}^{(1)} = l^{(6)},\quad
l_{E_7}^{(1)} = l^{(7)}+l^{(9)},\quad
l_{E_7}^{(1)}= l^{(8)},\quad\nn
l_{E_7}^{(1)} &= l^{(6)}+2l^{(7)}+l^{(8)},\quad
l_{B}= l^{(2)}+l^{(3)}+5l^{(4)}+3l^{(5)},\nn
l_{\tau}\,&= 4l^{(1)}+3l^{(2)}+l^{(4)}+2l^{(5)}+5l^{(6)}+6l^{(7)}+3l^{(8)}+2l^{(9)}
\end{align}

\subsection*{NHC $2,3,2$}
\begin{equation} \label{232polytope}
 \begin{array}{rrrr|rrrrrrrrrrr|}
    \multicolumn{4}{c}{\nu_i^*}     &l^{(1)} & l^{(2)} & l^{(3)} & l^{(4)} & l^{(5)} & l^{(6)} & l^{(7)} & l^{(8)} & l^{(9)} &  \\
    0 & 0 & 0 & 0\phantom{,} & 0 & 0 & -1 & 0 & 0 & 0 & 0 & 0 & 0 & \\
-1 & 0 & 0 & 0\phantom{,} & 0 & 0 & 0 & 0 & 0 & 0 & 0 & 0 & 1 & \\
0 & -1 & 0 & 0\phantom{,} & 0 & 0 & 0 & 0 & 0 & 0 & 0 & 1 & 0 & \\
2 & 3 & 0 & 0\phantom{,} & 0 & 1 & 0 & 0 & 0 & 0 & 0 & 0 & 0 & \\
2 & 3 & 1 & 0\phantom{,} & 0 & 0 & -1 & 0 & 0 & 0 & 1 & 0 & 0 & \\
2 & 3 & 0 & -1\phantom{,} & -1 & 0 & 1 & 0 & 0 & 1 & 0 & 0 & -1 & \\
1 & 2 & 0 & -1\phantom{,} & 1 & 0 & 1 & 0 & 0 & 0 & -2 & 0 & 1 & \\
2 & 3 & -1 & -2\phantom{,} & 0 & -2 & 0 & 0 & 1 & -1 & 0 & 0 & 0 & \\
1 & 1 & -1 & -2\phantom{,} & 0 & 0 & 1 & 0 & 0 & 0 & 0 & -2 & 0 & \\
2 & 3 & -2 & -4\phantom{,} & 0 & 1 & -1 & 0 & 0 & -1 & 0 & 1 & 1 & \\
0 & 1 & -1 & -2\phantom{,} & -1 & 0 & 0 & 1 & 0 & 0 & 1 & 0 & -2 & \\
2 & 3 & -3 & -5\phantom{,} & 1 & 0 & 0 & 0 & -2 & 1 & 0 & 0 & 0 & \\
1 & 2 & -3 & -5\phantom{,} & 1 & 0 & 0 & -2 & 0 & 0 & 0 & 0 & 0 & \\
2 & 3 & -5 & -8\phantom{,} & -1 & 0 & 0 & 1 & 1 & 0 & 0 & 0 & 0 & \\
\end{array} \
\end{equation}
The triple intersection ring corresponding to the above Mori cone generators is
\begin{align}\nonumber
\mathcal{R}&=-\frac{1}{8}(112 J_1^3+20 J_2 J_1^2+24 J_3 J_1^2+56 J_4 J_1^2+72 J_5 J_1^2+32 J_6 J_1^2+72 J_7 J_1^2+12 J_8 J_1^2+8 J_9 J_1^2\\[-1mm]
+\,&4 J_2^2 J_1+16 J_3^2 J_1+28 J_4^2 J_1+48 J_5^2 J_1+16 J_6^2 J_1+60 J_7^2 J_1+12 J_8^2 J_1+16 J_9^2 J_1+10 J_2 J_4 J_1\nn
+\,&12 J_3 J_4 J_1+12 J_2 J_5 J_1+24 J_3 J_5 J_1+36 J_4 J_5 J_1+4 J_2 J_6 J_1+24 J_3 J_6 J_1+16 J_4 J_6 J_1+24 J_5 J_6 J_1\nn
+\,&10 J_2 J_7 J_1+28 J_3 J_7 J_1+36 J_4 J_7 J_1+52 J_5 J_7 J_1+32 J_6 J_7 J_1+8 J_3 J_8 J_1+6 J_4 J_8 J_1+12 J_5 J_8 J_1\nn
+\,&12 J_6 J_8 J_1+14 J_7 J_8 J_1+16 J_3 J_9 J_1+4 J_4 J_9 J_1+8 J_5 J_9 J_1+8 J_6 J_9 J_1+20 J_7 J_9 J_1+8 J_8 J_9 J_1\nn
+\,&4 J_2^3+32 J_3^3+22 J_4^3+40 J_5^3+114 J_7^3+12 J_8^3+32 J_9^3+5 J_2 J_4^2+14 J_3 J_4^2+12 J_2 J_5^2+24 J_3 J_5^2\nn
+\,&24 J_4 J_5^2+4 J_2 J_6^2+24 J_3 J_6^2+8 J_4 J_6^2+8 J_5 J_6^2+5 J_2 J_7^2+62 J_3 J_7^2+30 J_4 J_7^2+50 J_5 J_7^2+40 J_6 J_7^2\nn
+\,&8 J_3 J_8^2+6 J_4 J_8^2+12 J_5 J_8^2+12 J_6 J_8^2+14 J_7 J_8^2+32 J_3 J_9^2+8 J_4 J_9^2+16 J_5 J_9^2+16 J_6 J_9^2+40 J_7 J_9^2\nn
+\,&16 J_8 J_9^2+2 J_2^2 J_4+8 J_3^2 J_4+4 J_2^2 J_5+16 J_3^2 J_5+18 J_4^2 J_5+6 J_2 J_4 J_5+12 J_3 J_4 J_5+4 J_2^2 J_6\nn
+\,&16 J_3^2 J_6+8 J_4^2 J_6+16 J_5^2 J_6+2 J_2 J_4 J_6+12 J_3 J_4 J_6+4 J_2 J_5 J_6+24 J_3 J_5 J_6+12 J_4 J_5 J_6+2 J_2^2 J_7\nn
+\,&40 J_3^2 J_7+26 J_4^2 J_7+40 J_5^2 J_7+24 J_6^2 J_7+5 J_2 J_4 J_7+14 J_3 J_4 J_7+6 J_2 J_5 J_7+28 J_3 J_5 J_7\nn
+\,&26 J_4 J_5 J_7+2 J_2 J_6 J_7+28 J_3 J_6 J_7+16 J_4 J_6 J_7+28 J_5 J_6 J_7+16 J_3^2 J_8+7 J_4^2 J_8+12 J_5^2 J_8\nn
+\,&12 J_6^2 J_8+31 J_7^2 J_8+4 J_3 J_4 J_8+8 J_3 J_5 J_8+6 J_4 J_5 J_8+8 J_3 J_6 J_8+6 J_4 J_6 J_8+12 J_5 J_6 J_8\nn
+\,&20 J_3 J_7 J_8+7 J_4 J_7 J_8+14 J_5 J_7 J_8+14 J_6 J_7 J_8+32 J_3^2 J_9+10 J_4^2 J_9+8 J_5^2 J_9+8 J_6^2 J_9+58 J_7^2 J_9\nn
+\,&8 J_8^2 J_9+8 J_3 J_4 J_9+16 J_3 J_5 J_9+4 J_4 J_5 J_9+16 J_3 J_6 J_9+4 J_4 J_6 J_9+8 J_5 J_6 J_9+40 J_3 J_7 J_9\nn
+\,&10 J_4 J_7 J_9+20 J_5 J_7 J_9+20 J_6 J_7 J_9+16 J_3 J_8 J_9+4 J_4 J_8 J_9+8 J_5 J_8 J_9+8 J_6 J_8 J_9+20 J_7 J_8 J_9).\label{232triple}
\end{align}
The relation between the above geometric bases and the Lie algebra bases is
\begin{align}
l_{\mf{su}(2)} &= 4l^{(3)}+4l^{(7)}+2l^{(8)}+2l^{(9)},\quad l_{\mf{su}(2)'} = 2l^{(1)}+4l^{(3)}+2l^{(4)}+4l^{(7)}+2l^{(8)}+2l^{(9)}\nn
l_{\mf{so}(7)}^{(1)} &= 2l^{(3)}+2l^{(7)}+l^{(8)}+2l^{(9)},\quad
l_{\mf{so}(7)}^{(2)} = 2l^{(1)}+2l^{(3)}+l^{(4)}+l^{(5)}+2l^{(7)},\quad
l_{\mf{so}(7)}^{(3)} = l^{(8)},\quad \nn
l_{B_1} &= 2l^{(1)}+l^{(4)}+l^{(5)}+l^{(7)},\quad
l_{B_2} = l^{(2)}+l^{(6)},\quad
l_{B_3} = l^{(5)},\quad \nn
l_{\tau} &= 4l^{(1)}+l^{(2)}+6l^{(3)}+2l^{(4)}+2l^{(5)}+6l^{(7)}+3l^{(8)}+2l^{(9)}.
\end{align}
The instanton partition function of refined topological string on the above local Calabi-Yau threefold differs from the elliptic genera of NHC 2,3,2 by the following four degree flipping of refined BPS invariants with spin $(0,0)$:
\be
\ba
-&\frac{{t_{\mf{su}(2)}}}{2}+\frac{t_{\mf{so}(7)}^{(1)}}{2}+t_{\mf{so}(7)}^{(2)}+\frac{t_{\mf{so}(7)}^{(3)}}{2}-t_{B_1},\quad
-\frac{{t_{\mf{su}(2)}}}{2}+\frac{t_{\mf{so}(7)}^{(1)}}{2}+t_{\mf{so}(7)}^{(2)}+\frac{3{t_{\mf{so}(7)}^{(3)}}}{2}-t_{B_1},\\
-&\frac{{t_{\mf{su}(2)'}}}{2}+\frac{t_{\mf{so}(7)}^{(1)}}{2}+t_{\mf{so}(7)}^{(2)}+\frac{t_{\mf{so}(7)}^{(3)}}{2}-t_{B_3},\quad
-\frac{{t_{\mf{su}(2)'}}}{2}+\frac{t_{\mf{so}(7)}^{(1)}}{2}+t_{\mf{so}(7)}^{(2)}+\frac{3 {t_{\mf{so}(7)}^{(3)}}}{2}-t_{B_3}.
\ea
\ee
\subsection*{NHC $3,2,2$}
\begin{equation} \label{322polytope}
 \begin{array}{rrrr|rrrrrrrrr|}
    \multicolumn{4}{c}{\nu_i^*}     &l^{(1)} & l^{(2)} & l^{(3)} & l^{(4)} & l^{(5)} & l^{(6)} & l^{(7)} &  \\
    0 & 0 & 0 & 0\phantom{,} & 0 & 0 & -1 & 0 & 0 & 0 & 0 & \\
-1 & 0 & 0 & 0\phantom{,} & 0 & 0 & 0 & 0 & 0 & 0 & 1 & \\
0 & -1 & 0 & 0\phantom{,} & 0 & 0 & 0 & 0 & 0 & 1 & 0 & \\
2 & 3 & 0 & 0\phantom{,} & 0 & 1 & 0 & 0 & 0 & 0 & 0 & \\
2 & 3 & 1 & 0\phantom{,} & 0 & 0 & 0 & 0 & 1 & 0 & 0 & \\
2 & 3 & 0 & -1\phantom{,} & 0 & -2 & 0 & 1 & -1 & 0 & 0 & \\
2 & 3 & 0 & -2\phantom{,} & 0 & 1 & -1 & 0 & -1 & 1 & 1 & \\
1 & 1 & 0 & -1\phantom{,} & 0 & 0 & 1 & 0 & 0 & -2 & 0 & \\
2 & 3 & -1 & -3\phantom{,} & 1 & 0 & 0 & -2 & 1 & 0 & 0 & \\
1 & 2 & -1 & -3\phantom{,} & 0 & 0 & 1 & 0 & 0 & 0 & -3 & \\
2 & 3 & -2 & -5\phantom{,} & -2 & 0 & 1 & 1 & 0 & 0 & 0 & \\
2 & 3 & -3 & -7\phantom{,} & 1 & 0 & -1 & 0 & 0 & 0 & 1 & \\
\end{array} \
\end{equation}
The triple intersection ring corresponding to the above Mori cone generators is
\begin{align}\nonumber
\mathcal{R}&=-\frac{1}{7}(36 J_1^3+5 J_2 J_1^2+18 J_3 J_1^2+26 J_4 J_1^2+16 J_5 J_1^2+9 J_6 J_1^2+6 J_7 J_1^2+J_2^2 J_1+18 J_3^2 J_1+20 J_4^2 J_1\\[-1mm]
+\,&12 J_5^2 J_1+15 J_6^2 J_1+2 J_7^2 J_1+3 J_2 J_4 J_1+18 J_3 J_4 J_1+J_2 J_5 J_1+18 J_3 J_5 J_1+14 J_4 J_5 J_1+9 J_3 J_6 J_1\nn
+\,&9 J_4 J_6 J_1+9 J_5 J_6 J_1+6 J_3 J_7 J_1+6 J_4 J_7 J_1+6 J_5 J_7 J_1+3 J_6 J_7 J_1+3 J_2^3+18 J_3^3+22 J_4^3+25 J_6^3\nn
+\,&3 J_7^3+6 J_2 J_4^2+18 J_3 J_4^2+3 J_2 J_5^2+18 J_3 J_5^2+6 J_4 J_5^2+15 J_3 J_6^2+15 J_4 J_6^2+15 J_5 J_6^2+2 J_3 J_7^2\nn
+\,&2 J_4 J_7^2+2 J_5 J_7^2+J_6 J_7^2+2 J_2^2 J_4+18 J_3^2 J_4+3 J_2^2 J_5+18 J_3^2 J_5+10 J_4^2 J_5+2 J_2 J_4 J_5+18 J_3 J_4 J_5\nn
+\,&9 J_3^2 J_6+9 J_4^2 J_6+9 J_5^2 J_6+9 J_3 J_4 J_6+9 J_3 J_5 J_6+9 J_4 J_5 J_6+6 J_3^2 J_7+6 J_4^2 J_7+6 J_5^2 J_7+5 J_6^2 J_7\nn
+\,&6 J_3 J_4 J_7+6 J_3 J_5 J_7+6 J_4 J_5 J_7+3 J_3 J_6 J_7+3 J_4 J_6 J_7+3 J_5 J_6 J_7).\label{322triple}
\end{align}
The relation between the above geometric bases and the Lie algebra bases is
\begin{align}
l_{G_2}^{(1)} &= 2l^{(1)}+3l^{(3)}+l^{(4)}+l^{(7)},\quad
l_{G_2}^{(2)} = l^{(6)},\quad
l_{\mf{su}(2)} = 2l^{(1)}+4l^{(3)}+2l^{(6)}+2l^{(7)},\quad
l_{B_1} = l^{(2)}+l^{(5)},\quad \nn
l_{B_2} &= l^{(4)},\quad
l_{B_3} = l^{(1)},\quad
l_{\tau} = 4l^{(1)}+l^{(2)}+6l^{(3)}+2l^{(4)}+3l^{(6)}+2l^{(7)}.
\end{align}
The instanton partition function of refined topological string on the above local Calabi-Yau threefold differs from the elliptic genera of NHC 3,2,2 by the following five degree flipping of refined BPS invariants with spin $(0,0)$:
\be
\ba
-&\frac{{t_{\mf{su}(2)}}}{2}-t_{B_2}-t_{B_3}+t_{G_2}^{(1)}+t_{G_2}^{(2)},\quad
-\frac{{t_{\mf{su}(2)}}}{2}-t_{B_2}-t_{B_3}+t_{G_2}^{(1)}+2 t_{G_2}^{(2)},\quad
\frac{{t_{\mf{su}(2)}}}{2}-t_{B_3},\\
-&\frac{{t_{\mf{su}(2)}}}{2}-t_{B_2}+t_{G_2}^{(1)}+t_{G_2}^{(2)},\quad
-\frac{{t_{\mf{su}(2)}}}{2}-t_{B_2}+t_{G_2}^{(1)}+2 t_{G_2}^{(2)}.
\ea
\ee
\subsection*{$\mathbf{\fn=3, G=\mf{so}(7)}$}
\begin{equation} \label{n3SO7polytope}
 \begin{array}{rrrr|rrrrrrrrr|}
    \multicolumn{4}{c}{\nu_i^*}     &l^{(1)} & l^{(2)} & l^{(3)} & l^{(4)} & l^{(5)} & l^{(6)} & l^{(7)} &  \\
    0 & 0 & 0 & 0\phantom{,} & 0 & 0 & 0 & -1 & 0 & 0 & 0 & \\
-1 & 0 & 0 & 0\phantom{,} & 0 & 0 & 0 & 0 & 0 & 0 & 1 & \\
0 & -1 & 0 & 0\phantom{,} & 0 & 0 & 0 & 0 & 0 & 1 & 0 & \\
2 & 3 & 0 & 0\phantom{,} & 0 & 0 & 1 & 0 & 0 & 0 & 0 & \\
2 & 3 & 1 & 0\phantom{,} & -1 & -1 & 0 & 1 & 1 & 0 & -1 & \\
1 & 2 & 1 & 0\phantom{,} & 1 & 1 & 0 & -1 & 0 & 0 & 1 & \\
2 & 3 & 0 & -1\phantom{,} & 0 & 1 & -2 & 0 & -1 & 0 & 0 & \\
1 & 1 & 0 & -1\phantom{,} & 0 & 0 & 0 & 1 & 0 & -2 & 0 & \\
2 & 3 & 0 & -2\phantom{,} & 0 & 0 & 1 & -1 & -1 & 1 & 1 & \\
0 & 1 & 0 & -1\phantom{,} & 0 & -1 & 0 & 1 & 0 & 0 & -2 & \\
2 & 3 & -1 & -3\phantom{,} & 1 & -1 & 0 & 0 & 1 & 0 & 0 & \\
1 & 2 & -1 & -3\phantom{,} & -1 & 1 & 0 & 0 & 0 & 0 & 0 & \\
\end{array} \
\end{equation}
The triple intersection ring corresponding to the above Mori cone generators is
\begin{align}\nonumber
\mathcal{R}&=-\frac{1}{3}(2 J_2^3+2 J_1 J_2^2+6 J_4 J_2^2+2 J_5 J_2^2+3 J_6 J_2^2+2 J_7 J_2^2+8 J_4^2 J_2+2 J_5^2
   J_2+5 J_6^2 J_2+4 J_7^2 J_2\\[-1mm]
   &+2 J_1 J_4 J_2+2 J_1 J_5 J_2+6 J_4 J_5 J_2+J_1 J_6 J_2+4
   J_4 J_6 J_2+3 J_5 J_6 J_2+4 J_4 J_7 J_2+2 J_5 J_7 J_2\nn
   &+2 J_6 J_7 J_2+J_3^3+12 J_4^3+9
   J_6^3+8 J_7^3+2 J_1 J_4^2+2 J_1 J_5^2+J_3 J_5^2+6 J_4 J_5^2+2 J_1 J_6^2+6 J_4 J_6^2\nn
   &+5
   J_5 J_6^2+8 J_4 J_7^2+4 J_5 J_7^2+4 J_6 J_7^2+J_3^2 J_5+8 J_4^2 J_5+2 J_1 J_4 J_5+6
   J_4^2 J_6+3 J_5^2 J_6+J_1 J_4 J_6\nn
   &+J_1 J_5 J_6+4 J_4 J_5 J_6+8 J_4^2 J_7+2 J_5^2 J_7+2
   J_6^2 J_7+4 J_4 J_5 J_7+4 J_4 J_6 J_7+2 J_5 J_6 J_7).\label{n3SO7triple}
   \end{align}
The relation between the above geometric bases and the Lie algebra bases is
\begin{align}
l_{\mf{su}(2)} &= 4l^{(4)}+2l^{(6)}+2l^{(7)},\quad
l_{\mf{so}(7)}^{(1)} = 2l^{(4)}+l^{(6)}+2l^{(7)},\quad
l_{\mf{so}(7)}^{(2)} = l^{(1)}+l^{(2)}+2l^{(4)},\quad
l_{\mf{so}(7)}^{(3)} = l^{(6)},\nn
l_{\mf{su}(2)'} &= 2l^{(1)}+4l^{(4)}+2l^{(6)}+2l^{(7)},\quad
l_{B} = l^{(3)}+l^{(5)},\quad
l_{\tau} = 2l^{(1)}+2l^{(2)}+l^{(3)}+6l^{(4)}+3l^{(6)}+2l^{(7)}.
\end{align}

\subsection*{$\mathbf{\fn=2, G=G_2}$}
Here we only turn on a subgroup $\mf{su}(2)$ of the full flavor group $\mf{sp}(4)$.
\begin{equation} \label{n2G2polytope}
 \begin{array}{rrrr|rrrrrrr|}
    \multicolumn{4}{c}{\nu_i^*}     &l^{(1)} & l^{(2)} & l^{(3)} & l^{(4)} & l^{(5)} &  \\
    0 & 0 & 0 & 0\phantom{,} & 0 & -1 & 0 & 0 & -2 & \\
-1 & 0 & 0 & 0\phantom{,} & 0 & 0 & 0 & 0 & 1 & \\
0 & -1 & 0 & 0\phantom{,} & 0 & 0 & 0 & 1 & 0 & \\
2 & 3 & 0 & 0\phantom{,} & 1 & 0 & 0 & 0 & 0 & \\
2 & 3 & 1 & 0\phantom{,} & 0 & 0 & 1 & 0 & 0 & \\
2 & 3 & 0 & -1\phantom{,} & -2 & 1 & -2 & 0 & 0 & \\
2 & 3 & 0 & -2\phantom{,} & 1 & -1 & 0 & 1 & -1 & \\
1 & 1 & 0 & -1\phantom{,} & 0 & 1 & 0 & -2 & 2 & \\
2 & 3 & -1 & -2\phantom{,} & 0 & -1 & 1 & 0 & 1 & \\
1 & 2 & -1 & -2\phantom{,} & 0 & 1 & 0 & 0 & -1 & \\
\end{array} \
\end{equation}
The triple intersection ring corresponding to the above Mori cone generators is
\begin{equation}\label{n2G2triple}
\begin{aligned}
\mathcal{R}&=-\frac{1}{2}(8 J_2^3+4 J_3 J_2^2+12 J_4 J_2^2+8 J_5 J_2^2+2 J_3^2 J_2+20 J_4^2 J_2+4 J_5^2 J_2+6 J_3 J_4 J_2+4 J_3 J_5 J_2\\
+&\,8 J_4 J_5 J_2+32 J_4^3+2 J_5^3+J_1 J_3^2+10 J_3 J_4^2+2 J_3 J_5^2+4 J_4 J_5^2+3 J_3^2 J_4+2 J_3^2 J_5+12 J_4^2 J_5+4 J_3 J_4 J_5)
\end{aligned}
\end{equation}
The relation between the above geometric bases and the Lie algebra bases is
\be\begin{split}
l_{G_2}^{(1)} &= l^{(2)}+l^{(5)},\quad
l_{G_2}^{(2)} = l^{(4)},\quad
l_{\mf{su}(2)} = 2l^{(4)}+2l^{(5)},\quad \\
l_{B} &= l^{(3)},\quad
l_{\tau} = l^{(1)}+2l^{(2)}+3l^{(4)}+2l^{(5)}
\end{split}\ee

\subsection*{$\mathbf{\fn=1, G=G_2}$}
Here we only turn on a subgroup $\mf{su}(2)$ of the full flavor group $\mf{sp}(7)$.

\begin{equation} \label{n1G2polytope}
 \begin{array}{rrrr|rrrrrrr|}
    \multicolumn{4}{c}{\nu_i^*}     &l^{(1)} & l^{(2)} & l^{(3)} & l^{(4)} & l^{(5)} &  \\
    0 & 0 & 0 & 0\phantom{,} & 0 & -1 & 0 & 0 & -2 & \\
-1 & 0 & 0 & 0\phantom{,} & 0 & 0 & 0 & 0 & 1 & \\
0 & -1 & 0 & 0\phantom{,} & 0 & 0 & 0 & 1 & 0 & \\
2 & 3 & 0 & 0\phantom{,} & 1 & 0 & -1 & 0 & 0 & \\
2 & 3 & 1 & 0\phantom{,} & 0 & 0 & 1 & 0 & 0 & \\
2 & 3 & 0 & -1\phantom{,} & -2 & 1 & -1 & 0 & 0 & \\
2 & 3 & 0 & -2\phantom{,} & 1 & -1 & 0 & 1 & -1 & \\
1 & 1 & 0 & -1\phantom{,} & 0 & 1 & 0 & -2 & 2 & \\
2 & 3 & -1 & -1\phantom{,} & 0 & -1 & 1 & 0 & 1 & \\
1 & 2 & -1 & -1\phantom{,} & 0 & 1 & 0 & 0 & -1 & \\
\end{array} \
\end{equation}
The triple intersection ring corresponding to the above Mori cone generators is
\begin{align}
\mathcal{R}&=-(J_1^3+2 J_2 J_1^2+J_3 J_1^2+3 J_4 J_1^2+2 J_5 J_1^2+6 J_2^2 J_1+J_3^2 J_1+13 J_4^2 J_1+4 J_5^2 J_1+2 J_2 J_3 J_1\nn
+&\,9 J_2 J_4 J_1+3 J_3 J_4 J_1+6 J_2 J_5 J_1+2 J_3 J_5 J_1+7 J_4 J_5 J_1+18 J_2^3+54 J_4^3+8 J_5^3+2 J_2 J_3^2+39 J_2 J_4^2\nn
+&\,13 J_3 J_4^2+12 J_2 J_5^2+4 J_3 J_5^2+14 J_4 J_5^2+6 J_2^2 J_3+27 J_2^2 J_4+3 J_3^2 J_4+9 J_2 J_3 J_4+18 J_2^2 J_5+2 J_3^2 J_5\nn
+&\,28 J_4^2 J_5+6 J_2 J_3 J_5+21 J_2 J_4 J_5+7 J_3 J_4 J_5).\label{n1G2triple}
\end{align}
The relation between the above geometric bases and the Lie algebra bases is
\be\begin{split}
l_{G_2}^{(1)} &= l^{(2)}+l^{(5)},\quad
l_{G_2}^{(2)} = l^{(4)},\quad
l_{\mf{su}(2)} = 2l^{(4)}+2l^{(5)},\quad \\
l_{B} &= l^{(3)},\quad
l_{\tau} = l^{(1)}+2l^{(2)}+3l^{(4)}+2l^{(5)}
\end{split}\ee
\section{Refined BPS invariants}\label{app:bps}
In this appendix, we list the refined BPS invariants solved from the blowup equations up to total degree 7. We drop the base degree zero invariants for all the models as they are exactly known from tensor, vector and hyper multiplets. More results for higher total degrees can be found on the website \cite{kl}.
\footnotesize{

\tablehead{Header of first column & Header of second column \\}
 \tablefirsthead{%
   \hline
   \multicolumn{1}{|c}{ ${\beta}$} &
   \multicolumn{1}{|c||}{$\oplus N^{\mathbf{d}}_{j_l,j_r}(j_l,j_r)$} &
   $\und{\beta}$ &
   \multicolumn{1}{c|}{$\oplus N^{\mathbf{d}}_{j_l,j_r}(j_l,j_r)$} \\
   \hline}
 \tablehead{%
   \hline
   \multicolumn{1}{|c}{ ${\beta}$} &
   \multicolumn{1}{|c||}{$\oplus N^{\mathbf{d}}_{j_l,j_r}(j_l,j_r)$} &
   $\und{\beta}$ &
   \multicolumn{1}{c|}{$\oplus N^{\mathbf{d}}_{j_l,j_r}(j_l,j_r)$} \\
   \hline}
 \tabletail{%
   \hline
   \multicolumn{4}{r}{\small\emph{continued on next page}}\\
   }
 \tablelasttail{\hline}
 \bottomcaption{Refined BPS invariants of 6d $\fn=1,G_2$ model, with a $\mf{su}(2)$ mass turned on.}
\center{
 \begin{supertabular}{|c|p{5.7cm}||c|p{5.7cm}|}\label{tb:n1,G_2-BPS}$(0, 0, 1, 0, 0)$&$(0,0)$&$(1, 0, 1, 0, 0)$&$(0,1)$\\ \hline
 $(1, 1, 1, 0, 0)$&$(0,1/2)$&$(1, 1, 1, 1, 0)$&$(0,1/2)$\\ \hline
 $(1, 2, 1, 1, 0)$&$(0,0)$&$(2, 0, 1, 0, 0)$&$(0,2)$\\ \hline
 $(2, 0, 2, 0, 0)$&$(0,5/2)$&$(2, 1, 1, 0, 0)$&$(0,3/2)$\\ \hline
 $(2, 1, 1, 1, 0)$&$(0,3/2)$&$(2, 1, 2, 0, 0)$&$(0,2)$\\ \hline
 $(2, 1, 2, 1, 0)$&$(0,2)$&$(2, 2, 1, 1, 0)$&$(0,1)$\\ \hline
 $(2, 2, 2, 1, 0)$&$(0,3/2)$&$(3, 0, 1, 0, 0)$&$(0,3)$\\ \hline
 $(3, 0, 2, 0, 0)$&$(0,5/2)\oplus(0,7/2)\oplus(1/2,4)$&$(3, 0, 3, 0, 0)$&$(0,3)\oplus(1/2,9/2)$\\ \hline
 $(3, 1, 1, 0, 0)$&$(0,5/2)$&$(3, 1, 1, 1, 0)$&$(0,5/2)$\\ \hline
 $(3, 1, 2, 0, 0)$&$(0,2)\oplus2(0,3)\oplus(1/2,7/2)$&$(3, 1, 2, 1, 0)$&$(0,2)\oplus2(0,3)\oplus(1/2,7/2)$\\ \hline
 $(3, 1, 3, 0, 0)$&$(0,5/2)\oplus(0,7/2)\oplus(1/2,4)$&$(3, 2, 1, 1, 0)$&$(0,2)$\\ \hline
 $(3, 2, 2, 0, 0)$&$(0,5/2)$&$(4, 0, 1, 0, 0)$&$(0,4)$\\ \hline
 $(4, 0, 2, 0, 0)$&$(0,5/2)\oplus(0,7/2)\oplus2(0,9/2)\oplus(1/2,4)\oplus(1/2,5)\oplus(1,11/2)$&$(4, 0, 3, 0, 0)$&$(0,2)\oplus(0,3)\oplus2(0,4)\oplus(0,5)\oplus(0,6)\oplus(1/2,7/2)\oplus2(1/2,9/2)\oplus2(1/2,11/2)\oplus(1,5)\oplus(1,6)\oplus(3/2,13/2)$\\ \hline
 $(4, 1, 1, 0, 0)$&$(0,7/2)$&$(4, 1, 1, 1, 0)$&$(0,7/2)$\\ \hline
 $(4, 1, 2, 0, 0)$&$(0,2)\oplus2(0,3)\oplus3(0,4)\oplus(1/2,7/2)\oplus2(1/2,9/2)\oplus(1,5)$&$(5, 0, 1, 0, 0)$&$(0,5)$\\ \hline
 $(5, 0, 2, 0, 0)$&$(0,5/2)\oplus(0,7/2)\oplus2(0,9/2)\oplus2(0,11/2)\oplus(1/2,4)\oplus(1/2,5)\oplus2(1/2,6)\oplus(1,11/2)\oplus(1,13/2)\oplus(3/2,7)$&$(5, 1, 1, 0, 0)$&$(0,9/2)$\\ \hline
 $(6, 0, 1, 0, 0)$&$(0,6)$&$(1, 1, 1, 0, 1)$&$(0,0)\oplus(0,1)$\\ \hline
 $(1, 1, 1, 1, 1)$&$5(0,0)\oplus5(0,1)$&$(1, 1, 1, 2, 1)$&$5(0,0)\oplus5(0,1)$\\ \hline
 $(1, 1, 1, 3, 1)$&$(0,0)\oplus(0,1)$&$(1, 2, 1, 0, 1)$&$(0,1/2)$\\ \hline
 $(1, 2, 1, 1, 1)$&$4(0,1/2)$&$(1, 2, 1, 2, 1)$&$10(0,1/2)$\\ \hline
 $(2, 1, 1, 0, 1)$&$(0,1)\oplus(0,2)$&$(2, 1, 1, 1, 1)$&$5(0,1)\oplus5(0,2)$\\ \hline
 $(2, 1, 1, 2, 1)$&$5(0,1)\oplus5(0,2)$&$(2, 1, 2, 0, 1)$&$(0,3/2)\oplus(0,5/2)$\\ \hline
 $(2, 1, 2, 1, 1)$&$5(0,3/2)\oplus5(0,5/2)$&$(2, 2, 1, 0, 1)$&$(0,1/2)\oplus(0,3/2)$\\ \hline
 $(2, 2, 1, 1, 1)$&$4(0,1/2)\oplus4(0,3/2)$&$(2, 2, 2, 0, 1)$&$(0,1)\oplus(0,2)$\\ \hline
 $(3, 1, 1, 0, 1)$&$(0,2)\oplus(0,3)$&$(3, 1, 1, 1, 1)$&$5(0,2)\oplus5(0,3)$\\ \hline
 $(3, 1, 2, 0, 1)$&$(0,3/2)\oplus3(0,5/2)\oplus2(0,7/2)\oplus(1/2,3)\oplus(1/2,4)$&$(3, 2, 1, 0, 1)$&$(0,3/2)\oplus(0,5/2)$\\ \hline
 $(4, 1, 1, 0, 1)$&$(0,3)\oplus(0,4)$&$(1, 1, 1, 2, 2)$&$(0,1/2)$\\ \hline
 $(1, 2, 1, 0, 2)$&$(0,1)$&$(1, 2, 1, 1, 2)$&$5(0,0)\oplus5(0,1)$\\ \hline
 $(1, 3, 1, 0, 2)$&$(0,3/2)$&$(2, 2, 1, 0, 2)$&$(0,0)\oplus(0,1)\oplus(0,2)$\\ \hline
  \end{supertabular}}
}
\footnotesize{

\tablehead{Header of first column & Header of second column \\}
 \tablefirsthead{%
   \hline
   \multicolumn{1}{|c}{ ${\beta}$} &
   \multicolumn{1}{|c||}{$\oplus N^{\mathbf{d}}_{j_l,j_r}(j_l,j_r)$} &
   $\und{\beta}$ &
   \multicolumn{1}{c|}{$\oplus N^{\mathbf{d}}_{j_l,j_r}(j_l,j_r)$} \\
   \hline}
 \tablehead{%
   \hline
   \multicolumn{1}{|c}{ ${\beta}$} &
   \multicolumn{1}{|c||}{$\oplus N^{\mathbf{d}}_{j_l,j_r}(j_l,j_r)$} &
   $\und{\beta}$ &
   \multicolumn{1}{c|}{$\oplus N^{\mathbf{d}}_{j_l,j_r}(j_l,j_r)$} \\
   \hline}
 \tabletail{%
   \hline
   \multicolumn{4}{r}{\small\emph{continued on next page}}\\
   }
 \tablelasttail{\hline}
 \bottomcaption{Refined BPS invariants of 6d $\fn=2,G_2$ model, with a
   $\mf{su}(2)$ mass turned on.}
\center{
 \begin{supertabular}{|c|p{5.7cm}||c|p{5.7cm}|}\label{tb:n2,G_2-BPS}$(0, 0, 1, 0, 0)$&$(0,1/2)$&$(0, 1, 1, 0, 0)$&$(0,0)$\\ \hline
 $(0, 1, 1, 1, 0)$&$(0,0)$&$(1, 0, 1, 0, 0)$&$(0,3/2)$\\ \hline
 $(1, 0, 2, 0, 0)$&$(0,5/2)$&$(1, 0, 3, 0, 0)$&$(0,7/2)$\\ \hline
 $(1, 0, 4, 0, 0)$&$(0,9/2)$&$(1, 0, 5, 0, 0)$&$(0,11/2)$\\ \hline
 $(1, 0, 6, 0, 0)$&$(0,13/2)$&$(1, 1, 1, 0, 0)$&$(0,1)$\\ \hline
 $(1, 1, 1, 1, 0)$&$(0,1)$&$(1, 1, 2, 0, 0)$&$(0,2)$\\ \hline
 $(1, 1, 2, 1, 0)$&$(0,2)$&$(1, 1, 3, 0, 0)$&$(0,3)$\\ \hline
 $(1, 1, 3, 1, 0)$&$(0,3)$&$(1, 1, 4, 0, 0)$&$(0,4)$\\ \hline
 $(1, 1, 4, 1, 0)$&$(0,4)$&$(1, 1, 5, 0, 0)$&$(0,5)$\\ \hline
 $(1, 2, 1, 1, 0)$&$(0,1/2)$&$(1, 2, 2, 1, 0)$&$(0,3/2)$\\ \hline
 $(1, 2, 3, 1, 0)$&$(0,5/2)$&$(2, 0, 1, 0, 0)$&$(0,5/2)$\\ \hline
 $(2, 0, 2, 0, 0)$&$(0,5/2)\oplus(0,7/2)\oplus(1/2,4)$&$(2, 0, 3, 0, 0)$&$(0,5/2)\oplus(0,7/2)\oplus2(0,9/2)\oplus(1/2,4)\oplus(1/2,5)\oplus(1,11/2)$\\ \hline
 $(2, 0, 4, 0, 0)$&$(0,5/2)\oplus(0,7/2)\oplus2(0,9/2)\oplus2(0,11/2)\oplus(1/2,4)\oplus(1/2,5)\oplus2(1/2,6)\oplus(1,11/2)\oplus(1,13/2)\oplus(3/2,7)$&$(2, 0, 5, 0, 0)$&$(0,5/2)\oplus(0,7/2)\oplus2(0,9/2)\oplus2(0,11/2)\oplus3(0,13/2)\oplus(1/2,4)\oplus(1/2,5)\oplus2(1/2,6)\oplus2(1/2,7)\oplus(1,11/2)\oplus(1,13/2)\oplus2(1,15/2)\oplus(3/2,7)\oplus(3/2,8)\oplus(2,17/2)$\\ \hline
 $(2, 1, 1, 0, 0)$&$(0,2)$&$(2, 1, 1, 1, 0)$&$(0,2)$\\ \hline
 $(2, 1, 2, 0, 0)$&$(0,2)\oplus2(0,3)\oplus(1/2,7/2)$&$(2, 1, 2, 1, 0)$&$(0,2)\oplus2(0,3)\oplus(1/2,7/2)$\\ \hline
 $(2, 1, 3, 0, 0)$&$(0,2)\oplus2(0,3)\oplus3(0,4)\oplus(1/2,7/2)\oplus2(1/2,9/2)\oplus(1,5)$&$(2, 1, 3, 1, 0)$&$(0,2)\oplus2(0,3)\oplus3(0,4)\oplus(1/2,7/2)\oplus2(1/2,9/2)\oplus(1,5)$\\ \hline
 $(2, 1, 4, 0, 0)$&$(0,2)\oplus2(0,3)\oplus3(0,4)\oplus4(0,5)\oplus(1/2,7/2)\oplus2(1/2,9/2)\oplus3(1/2,11/2)\oplus(1,5)\oplus2(1,6)\oplus(3/2,13/2)$&$(2, 2, 1, 1, 0)$&$(0,3/2)$\\ \hline
 $(2, 2, 2, 0, 0)$&$(0,5/2)$&$(2, 2, 2, 1, 0)$&$(0,3/2)\oplus3(0,5/2)\oplus(1/2,3)$\\ \hline
 $(2, 2, 3, 0, 0)$&$(0,5/2)\oplus(0,7/2)\oplus(1/2,4)$&$(3, 0, 1, 0, 0)$&$(0,7/2)$\\ \hline
 $(3, 0, 2, 0, 0)$&$(0,5/2)\oplus(0,7/2)\oplus2(0,9/2)\oplus(1/2,4)\oplus(1/2,5)\oplus(1,11/2)$&$(3, 0, 3, 0, 0)$&$(0,3/2)\oplus(0,5/2)\oplus3(0,7/2)\oplus3(0,9/2)\oplus4(0,11/2)\oplus(1/2,3)\oplus2(1/2,4)\oplus3(1/2,5)\oplus3(1/2,6)\oplus(1/2,7)\oplus(1,9/2)\oplus2(1,11/2)\oplus3(1,13/2)\oplus(3/2,6)\oplus(3/2,7)\oplus(2,15/2)$\\ \hline
 $(3, 0, 4, 0, 0)$&$(0,1/2)\oplus(0,3/2)\oplus3(0,5/2)\oplus4(0,7/2)\oplus7(0,9/2)\oplus6(0,11/2)\oplus7(0,13/2)\oplus(0,15/2)\oplus(0,17/2)\oplus(1/2,2)\oplus2(1/2,3)\oplus4(1/2,4)\oplus6(1/2,5)\oplus8(1/2,6)\oplus7(1/2,7)\oplus2(1/2,8)\oplus(1,7/2)\oplus2(1,9/2)\oplus5(1,11/2)\oplus6(1,13/2)\oplus7(1,15/2)\oplus(1,17/2)\oplus(3/2,5)\oplus2(3/2,6)\oplus4(3/2,7)\oplus4(3/2,8)\oplus(3/2,9)\oplus(2,13/2)\oplus2(2,15/2)\oplus3(2,17/2)\oplus(5/2,8)\oplus(5/2,9)\oplus(3,19/2)$&$(3, 1, 1, 0, 0)$&$(0,3)$\\ \hline
 $(3, 1, 1, 1, 0)$&$(0,3)$&$(3, 1, 2, 0, 0)$&$(0,2)\oplus2(0,3)\oplus3(0,4)\oplus(1/2,7/2)\oplus2(1/2,9/2)\oplus(1,5)$\\ \hline
 $(3, 1, 2, 1, 0)$&$(0,2)\oplus2(0,3)\oplus3(0,4)\oplus(1/2,7/2)\oplus2(1/2,9/2)\oplus(1,5)$&$(3, 1, 3, 0, 0)$&$(0,1)\oplus2(0,2)\oplus5(0,3)\oplus6(0,4)\oplus7(0,5)\oplus(0,6)\oplus(1/2,5/2)\oplus3(1/2,7/2)\oplus6(1/2,9/2)\oplus6(1/2,11/2)\oplus(1/2,13/2)\oplus(1,4)\oplus3(1,5)\oplus5(1,6)\oplus(3/2,11/2)\oplus2(3/2,13/2)\oplus(2,7)$\\ \hline
 $(3, 2, 1, 1, 0)$&$(0,5/2)$&$(3, 2, 2, 0, 0)$&$(0,5/2)\oplus(0,7/2)\oplus(1/2,4)$\\ \hline
 $(4, 0, 1, 0, 0)$&$(0,9/2)$&$(4, 0, 2, 0, 0)$&$(0,5/2)\oplus(0,7/2)\oplus2(0,9/2)\oplus2(0,11/2)\oplus(1/2,4)\oplus(1/2,5)\oplus2(1/2,6)\oplus(1,11/2)\oplus(1,13/2)\oplus(3/2,7)$\\ \hline
 $(4, 0, 3, 0, 0)$&$(0,1/2)\oplus(0,3/2)\oplus3(0,5/2)\oplus4(0,7/2)\oplus7(0,9/2)\oplus6(0,11/2)\oplus7(0,13/2)\oplus(0,15/2)\oplus(0,17/2)\oplus(1/2,2)\oplus2(1/2,3)\oplus4(1/2,4)\oplus6(1/2,5)\oplus8(1/2,6)\oplus7(1/2,7)\oplus2(1/2,8)\oplus(1,7/2)\oplus2(1,9/2)\oplus5(1,11/2)\oplus6(1,13/2)\oplus7(1,15/2)\oplus(1,17/2)\oplus(3/2,5)\oplus2(3/2,6)\oplus4(3/2,7)\oplus4(3/2,8)\oplus(3/2,9)\oplus(2,13/2)\oplus2(2,15/2)\oplus3(2,17/2)\oplus(5/2,8)\oplus(5/2,9)\oplus(3,19/2)$&$(4, 1, 1, 0, 0)$&$(0,4)$\\ \hline
 $(4, 1, 1, 1, 0)$&$(0,4)$&$(4, 1, 2, 0, 0)$&$(0,2)\oplus2(0,3)\oplus3(0,4)\oplus4(0,5)\oplus(1/2,7/2)\oplus2(1/2,9/2)\oplus3(1/2,11/2)\oplus(1,5)\oplus2(1,6)\oplus(3/2,13/2)$\\ \hline
 $(5, 0, 1, 0, 0)$&$(0,11/2)$&$(5, 0, 2, 0, 0)$&$(0,5/2)\oplus(0,7/2)\oplus2(0,9/2)\oplus2(0,11/2)\oplus3(0,13/2)\oplus(1/2,4)\oplus(1/2,5)\oplus2(1/2,6)\oplus2(1/2,7)\oplus(1,11/2)\oplus(1,13/2)\oplus2(1,15/2)\oplus(3/2,7)\oplus(3/2,8)\oplus(2,17/2)$\\ \hline
 $(5, 1, 1, 0, 0)$&$(0,5)$&$(6, 0, 1, 0, 0)$&$(0,13/2)$\\ \hline
 $(0, 1, 1, 0, 1)$&$(0,1/2)$&$(0, 1, 1, 1, 1)$&$2(0,1/2)$\\ \hline
 $(0, 1, 1, 2, 1)$&$2(0,1/2)$&$(0, 1, 1, 3, 1)$&$(0,1/2)$\\ \hline
 $(0, 2, 1, 0, 1)$&$(0,1)$&$(0, 2, 1, 1, 1)$&$2(0,0)\oplus2(0,1)$\\ \hline
 $(0, 2, 1, 2, 1)$&$3(0,0)\oplus(0,1)$&$(0, 2, 1, 3, 1)$&$14(0,0)\oplus2(0,1)$\\ \hline
 $(1, 1, 1, 0, 1)$&$(0,1/2)\oplus(0,3/2)$&$(1, 1, 1, 1, 1)$&$2(0,1/2)\oplus2(0,3/2)$\\ \hline
 $(1, 1, 1, 2, 1)$&$2(0,1/2)\oplus2(0,3/2)$&$(1, 1, 1, 3, 1)$&$(0,1/2)\oplus(0,3/2)$\\ \hline
 $(1, 1, 2, 0, 1)$&$(0,3/2)\oplus(0,5/2)$&$(1, 1, 2, 1, 1)$&$2(0,3/2)\oplus2(0,5/2)$\\ \hline
 $(1, 1, 2, 2, 1)$&$2(0,3/2)\oplus2(0,5/2)$&$(1, 1, 3, 0, 1)$&$(0,5/2)\oplus(0,7/2)$\\ \hline
 $(1, 1, 3, 1, 1)$&$2(0,5/2)\oplus2(0,7/2)$&$(1, 1, 4, 0, 1)$&$(0,7/2)\oplus(0,9/2)$\\ \hline
 $(1, 2, 1, 0, 1)$&$(0,0)\oplus(0,1)$&$(1, 2, 1, 1, 1)$&$(0,0)\oplus(0,1)$\\ \hline
 $(1, 2, 1, 2, 1)$&$4(0,0)\oplus4(0,1)$&$(1, 2, 2, 0, 1)$&$(0,1)\oplus(0,2)$\\ \hline
 $(1, 2, 2, 1, 1)$&$(0,1)\oplus(0,2)$&$(1, 2, 3, 0, 1)$&$(0,2)\oplus(0,3)$\\ \hline
 $(1, 3, 1, 1, 1)$&$(0,1/2)$&$(2, 1, 1, 0, 1)$&$(0,3/2)\oplus(0,5/2)$\\ \hline
 $(2, 1, 1, 1, 1)$&$2(0,3/2)\oplus2(0,5/2)$&$(2, 1, 1, 2, 1)$&$2(0,3/2)\oplus2(0,5/2)$\\ \hline
 $(2, 1, 2, 0, 1)$&$(0,3/2)\oplus3(0,5/2)\oplus2(0,7/2)\oplus(1/2,3)\oplus(1/2,4)$&$(2, 1, 2, 1, 1)$&$2(0,3/2)\oplus6(0,5/2)\oplus4(0,7/2)\oplus2(1/2,3)\oplus2(1/2,4)$\\ \hline
 $(2, 1, 3, 0, 1)$&$(0,3/2)\oplus3(0,5/2)\oplus5(0,7/2)\oplus3(0,9/2)\oplus(1/2,3)\oplus3(1/2,4)\oplus2(1/2,5)\oplus(1,9/2)\oplus(1,11/2)$&$(2, 2, 1, 0, 1)$&$(0,1)\oplus(0,2)$\\ \hline
 $(2, 2, 1, 1, 1)$&$(0,1)\oplus(0,2)$&$(2, 2, 2, 0, 1)$&$(0,1)\oplus4(0,2)\oplus3(0,3)\oplus(1/2,5/2)\oplus(1/2,7/2)$\\ \hline
 $(3, 1, 1, 0, 1)$&$(0,5/2)\oplus(0,7/2)$&$(3, 1, 1, 1, 1)$&$2(0,5/2)\oplus2(0,7/2)$\\ \hline
 $(3, 1, 2, 0, 1)$&$(0,3/2)\oplus3(0,5/2)\oplus5(0,7/2)\oplus3(0,9/2)\oplus(1/2,3)\oplus3(1/2,4)\oplus2(1/2,5)\oplus(1,9/2)\oplus(1,11/2)$&$(3, 2, 1, 0, 1)$&$(0,2)\oplus(0,3)$\\ \hline
 $(4, 1, 1, 0, 1)$&$(0,7/2)\oplus(0,9/2)$&$(0, 1, 1, 2, 2)$&$(0,0)$\\ \hline
 $(0, 1, 1, 3, 2)$&$(0,0)$&$(0, 2, 1, 0, 2)$&$(0,3/2)$\\ \hline
 $(0, 2, 1, 1, 2)$&$2(0,1/2)\oplus2(0,3/2)$&$(0, 2, 1, 2, 2)$&$5(0,1/2)\oplus(0,3/2)$\\ \hline
 $(0, 3, 1, 0, 2)$&$(0,2)$&$(0, 3, 1, 1, 2)$&$2(0,1)\oplus2(0,2)$\\ \hline
 $(0, 3, 2, 0, 2)$&$(0,2)$&$(1, 1, 1, 2, 2)$&$(0,1)$\\ \hline
 $(1, 2, 1, 0, 2)$&$(0,1/2)\oplus(0,3/2)$&$(1, 2, 1, 1, 2)$&$4(0,1/2)\oplus2(0,3/2)$\\ \hline
 $(1, 2, 2, 0, 2)$&$(0,1/2)\oplus(0,3/2)\oplus(0,5/2)$&$(1, 3, 1, 0, 2)$&$(0,1)\oplus(0,2)$\\ \hline
 $(2, 2, 1, 0, 2)$&$(0,1/2)\oplus(0,3/2)\oplus(0,5/2)$&$(0, 3, 1, 0, 3)$&$(0,5/2)$\\ \hline
  \end{supertabular}}
}
\footnotesize{

\tablehead{Header of first column & Header of second column \\}
 \tablefirsthead{%
   \hline
   \multicolumn{1}{|c}{ ${\beta}$} &
   \multicolumn{1}{|c||}{$\oplus N^{\mathbf{d}}_{j_l,j_r}(j_l,j_r)$} &
   $\und{\beta}$ &
   \multicolumn{1}{c|}{$\oplus N^{\mathbf{d}}_{j_l,j_r}(j_l,j_r)$} \\
   \hline}
 \tablehead{%
   \hline
   \multicolumn{1}{|c}{ ${\beta}$} &
   \multicolumn{1}{|c||}{$\oplus N^{\mathbf{d}}_{j_l,j_r}(j_l,j_r)$} &
   $\und{\beta}$ &
   \multicolumn{1}{c|}{$\oplus N^{\mathbf{d}}_{j_l,j_r}(j_l,j_r)$} \\
   \hline}
 \tabletail{%
   \hline
   \multicolumn{4}{r}{\small\emph{continued on next page}}\\
   }
 \tablelasttail{\hline}
 \bottomcaption{Refined BPS invariants of 6d $\fn=3,G_2$ model, with a $\mf{su}(2)$ mass turned on.}
\center{
 \begin{supertabular}{|c|p{5.7cm}||c|p{5.7cm}|}\label{tb:n3,G_2-BPS}$(0, 0, 1, 0, 0)$&$(0,0)$&$(0, 1, 1, 0, 0)$&$(0,1/2)$\\ \hline
 $(0, 1, 1, 1, 0)$&$(0,1/2)$&$(1, 0, 1, 0, 0)$&$(0,1)$\\ \hline
 $(1, 1, 1, 0, 0)$&$(0,1/2)$&$(1, 1, 1, 1, 0)$&$(0,1/2)$\\ \hline
 $(1, 2, 1, 1, 0)$&$(0,0)$&$(2, 0, 1, 0, 0)$&$(0,2)$\\ \hline
 $(2, 0, 2, 0, 0)$&$(0,5/2)$&$(2, 1, 1, 0, 0)$&$(0,3/2)$\\ \hline
 $(2, 1, 1, 1, 0)$&$(0,3/2)$&$(2, 1, 2, 0, 0)$&$(0,2)$\\ \hline
 $(2, 1, 2, 1, 0)$&$(0,2)$&$(2, 2, 1, 1, 0)$&$(0,1)$\\ \hline
 $(2, 2, 2, 1, 0)$&$(0,3/2)$&$(3, 0, 1, 0, 0)$&$(0,3)$\\ \hline
 $(3, 0, 2, 0, 0)$&$(0,5/2)\oplus(0,7/2)\oplus(1/2,4)$&$(3, 0, 3, 0, 0)$&$(0,3)\oplus(1/2,9/2)$\\ \hline
 $(3, 1, 1, 0, 0)$&$(0,5/2)$&$(3, 1, 1, 1, 0)$&$(0,5/2)$\\ \hline
 $(3, 1, 2, 0, 0)$&$(0,2)\oplus2(0,3)\oplus(1/2,7/2)$&$(3, 1, 2, 1, 0)$&$(0,2)\oplus2(0,3)\oplus(1/2,7/2)$\\ \hline
 $(3, 1, 3, 0, 0)$&$(0,5/2)\oplus(0,7/2)\oplus(1/2,4)$&$(3, 2, 1, 1, 0)$&$(0,2)$\\ \hline
 $(3, 2, 2, 0, 0)$&$(0,5/2)$&$(4, 0, 1, 0, 0)$&$(0,4)$\\ \hline
 $(4, 0, 2, 0, 0)$&$(0,5/2)\oplus(0,7/2)\oplus2(0,9/2)\oplus(1/2,4)\oplus(1/2,5)\oplus(1,11/2)$&$(4, 0, 3, 0, 0)$&$(0,2)\oplus(0,3)\oplus2(0,4)\oplus(0,5)\oplus(0,6)\oplus(1/2,7/2)\oplus2(1/2,9/2)\oplus2(1/2,11/2)\oplus(1,5)\oplus(1,6)\oplus(3/2,13/2)$\\ \hline
 $(4, 1, 1, 0, 0)$&$(0,7/2)$&$(4, 1, 1, 1, 0)$&$(0,7/2)$\\ \hline
 $(4, 1, 2, 0, 0)$&$(0,2)\oplus2(0,3)\oplus3(0,4)\oplus(1/2,7/2)\oplus2(1/2,9/2)\oplus(1,5)$&$(5, 0, 1, 0, 0)$&$(0,5)$\\ \hline
 $(5, 0, 2, 0, 0)$&$(0,5/2)\oplus(0,7/2)\oplus2(0,9/2)\oplus2(0,11/2)\oplus(1/2,4)\oplus(1/2,5)\oplus2(1/2,6)\oplus(1,11/2)\oplus(1,13/2)\oplus(3/2,7)$&$(5, 1, 1, 0, 0)$&$(0,9/2)$\\ \hline
 $(6, 0, 1, 0, 0)$&$(0,6)$&$(0, 1, 1, 0, 1)$&$(0,1)$\\ \hline
 $(0, 1, 1, 1, 1)$&$(0,0)\oplus(0,1)$&$(0, 1, 1, 2, 1)$&$(0,0)\oplus(0,1)$\\ \hline
 $(0, 1, 1, 3, 1)$&$(0,1)$&$(0, 2, 1, 0, 1)$&$(0,3/2)$\\ \hline
 $(0, 2, 1, 1, 1)$&$(0,1/2)\oplus(0,3/2)$&$(0, 2, 1, 2, 1)$&$(0,1/2)\oplus(0,3/2)$\\ \hline
 $(0, 2, 1, 3, 1)$&$(0,1/2)\oplus(0,3/2)$&$(0, 2, 2, 0, 1)$&$(0,2)$\\ \hline
 $(0, 2, 2, 1, 1)$&$(0,1)\oplus(0,2)$&$(0, 2, 2, 2, 1)$&$(0,0)\oplus(0,1)\oplus(0,2)$\\ \hline
 $(0, 3, 2, 0, 1)$&$(0,5/2)$&$(0, 3, 2, 1, 1)$&$(0,3/2)\oplus(0,5/2)$\\ \hline
 $(0, 3, 3, 0, 1)$&$(0,3)$&$(1, 1, 1, 0, 1)$&$(0,0)\oplus(0,1)$\\ \hline
 $(1, 1, 1, 1, 1)$&$(0,0)\oplus(0,1)$&$(1, 1, 1, 2, 1)$&$(0,0)\oplus(0,1)$\\ \hline
 $(1, 1, 1, 3, 1)$&$(0,0)\oplus(0,1)$&$(1, 2, 1, 0, 1)$&$(0,1/2)\oplus(0,3/2)$\\ \hline
 $(1, 2, 1, 1, 1)$&$3(0,1/2)\oplus(0,3/2)$&$(1, 2, 1, 2, 1)$&$3(0,1/2)\oplus(0,3/2)$\\ \hline
 $(1, 2, 2, 0, 1)$&$(0,1)\oplus(0,2)$&$(1, 2, 2, 1, 1)$&$(0,0)\oplus2(0,1)\oplus(0,2)$\\ \hline
 $(1, 3, 1, 1, 1)$&$(0,1)$&$(1, 3, 2, 0, 1)$&$(0,3/2)\oplus(0,5/2)$\\ \hline
 $(2, 1, 1, 0, 1)$&$(0,1)\oplus(0,2)$&$(2, 1, 1, 1, 1)$&$(0,1)\oplus(0,2)$\\ \hline
 $(2, 1, 1, 2, 1)$&$(0,1)\oplus(0,2)$&$(2, 1, 2, 0, 1)$&$(0,3/2)\oplus(0,5/2)$\\ \hline
 $(2, 1, 2, 1, 1)$&$(0,3/2)\oplus(0,5/2)$&$(2, 2, 1, 0, 1)$&$(0,1/2)\oplus(0,3/2)$\\ \hline
 $(2, 2, 1, 1, 1)$&$2(0,1/2)\oplus2(0,3/2)$&$(2, 2, 2, 0, 1)$&$(0,0)\oplus2(0,1)\oplus2(0,2)\oplus(0,3)$\\ \hline
 $(3, 1, 1, 0, 1)$&$(0,2)\oplus(0,3)$&$(3, 1, 1, 1, 1)$&$(0,2)\oplus(0,3)$\\ \hline
 $(3, 1, 2, 0, 1)$&$(0,3/2)\oplus3(0,5/2)\oplus2(0,7/2)\oplus(1/2,3)\oplus(1/2,4)$&$(3, 2, 1, 0, 1)$&$(0,3/2)\oplus(0,5/2)$\\ \hline
 $(4, 1, 1, 0, 1)$&$(0,3)\oplus(0,4)$&$(0, 1, 1, 2, 2)$&$(0,1/2)$\\ \hline
 $(0, 1, 1, 3, 2)$&$(0,1/2)$&$(0, 2, 1, 0, 2)$&$(0,2)$\\ \hline
 $(0, 2, 1, 1, 2)$&$(0,1)\oplus(0,2)$&$(0, 2, 1, 2, 2)$&$(0,0)\oplus2(0,1)\oplus(0,2)$\\ \hline
 $(0, 2, 2, 0, 2)$&$(0,5/2)$&$(0, 2, 2, 1, 2)$&$(0,3/2)\oplus(0,5/2)$\\ \hline
 $(0, 3, 1, 0, 2)$&$(0,5/2)$&$(0, 3, 1, 1, 2)$&$(0,3/2)\oplus(0,5/2)$\\ \hline
 $(0, 3, 2, 0, 2)$&$(0,2)\oplus2(0,3)\oplus(1/2,7/2)$&$(1, 1, 1, 2, 2)$&$(0,1/2)$\\ \hline
 $(1, 2, 1, 0, 2)$&$(0,1)\oplus(0,2)$&$(1, 2, 1, 1, 2)$&$(0,0)\oplus2(0,1)\oplus(0,2)$\\ \hline
 $(1, 2, 2, 0, 2)$&$(0,3/2)\oplus(0,5/2)$&$(1, 3, 1, 0, 2)$&$(0,3/2)\oplus(0,5/2)$\\ \hline
 $(2, 2, 1, 0, 2)$&$(0,0)\oplus(0,1)\oplus(0,2)$&$(0, 3, 1, 0, 3)$&$(0,3)$\\ \hline
  \end{supertabular}}
}
\footnotesize{

\tablehead{Header of first column & Header of second column \\}
 \tablefirsthead{%
   \hline
   \multicolumn{1}{|c}{ ${\beta}$} &
   \multicolumn{1}{|c||}{$\oplus N^{\mathbf{d}}_{j_l,j_r}(j_l,j_r)$} &
   $\und{\beta}$ &
   \multicolumn{1}{c|}{$\oplus N^{\mathbf{d}}_{j_l,j_r}(j_l,j_r)$} \\
   \hline}
 \tablehead{%
   \hline
   \multicolumn{1}{|c}{ ${\beta}$} &
   \multicolumn{1}{|c||}{$\oplus N^{\mathbf{d}}_{j_l,j_r}(j_l,j_r)$} &
   $\und{\beta}$ &
   \multicolumn{1}{c|}{$\oplus N^{\mathbf{d}}_{j_l,j_r}(j_l,j_r)$} \\
   \hline}
 \tabletail{%
   \hline
   \multicolumn{4}{r}{\small\emph{continued on next page}}\\
   }
 \tablelasttail{\hline}
 \bottomcaption{Refined BPS invariants of 6d $\fn=3,\mf{so}(7)$ model, with $\mf{su}(2)\times \mf{su}(2)$ masses turned on.}
\center{
 \begin{supertabular}{|c|p{4.7cm}||c|p{4.7cm}|}\label{tb:n3,SO(7)-BPS}$(0, 0, 0, 0, 1, 0, 0)$&$(0,0)$&$(0, 0, 1, 0, 1, 0, 0)$&$(0,1)$\\ \hline
 $(0, 0, 2, 0, 1, 0, 0)$&$(0,2)$&$(0, 0, 2, 0, 2, 0, 0)$&$(0,5/2)$\\ \hline
 $(0, 0, 3, 0, 1, 0, 0)$&$(0,3)$&$(0, 0, 3, 0, 2, 0, 0)$&$(0,5/2)\oplus(0,7/2)\oplus(1/2,4)$\\ \hline
 $(0, 0, 3, 0, 3, 0, 0)$&$(0,3)\oplus(1/2,9/2)$&$(0, 0, 4, 0, 1, 0, 0)$&$(0,4)$\\ \hline
 $(0, 0, 4, 0, 2, 0, 0)$&$(0,5/2)\oplus(0,7/2)\oplus2(0,9/2)\oplus(1/2,4)\oplus(1/2,5)\oplus(1,11/2)$&$(0, 0, 4, 0, 3, 0, 0)$&$(0,2)\oplus(0,3)\oplus2(0,4)\oplus(0,5)\oplus(0,6)\oplus(1/2,7/2)\oplus2(1/2,9/2)\oplus2(1/2,11/2)\oplus(1,5)\oplus(1,6)\oplus(3/2,13/2)$\\ \hline
 $(0, 0, 5, 0, 1, 0, 0)$&$(0,5)$&$(0, 0, 5, 0, 2, 0, 0)$&$(0,5/2)\oplus(0,7/2)\oplus2(0,9/2)\oplus2(0,11/2)\oplus(1/2,4)\oplus(1/2,5)\oplus2(1/2,6)\oplus(1,11/2)\oplus(1,13/2)\oplus(3/2,7)$\\ \hline
 $(0, 0, 6, 0, 1, 0, 0)$&$(0,6)$&$(0, 1, 0, 0, 1, 0, 0)$&$(0,0)$\\ \hline
 $(0, 1, 0, 1, 1, 0, 0)$&$(0,1/2)$&$(0, 1, 0, 1, 1, 1, 0)$&$(0,1/2)$\\ \hline
 $(0, 1, 1, 1, 1, 0, 0)$&$(0,1/2)$&$(0, 1, 1, 1, 1, 1, 0)$&$(0,1/2)$\\ \hline
 $(0, 1, 2, 1, 1, 0, 0)$&$(0,3/2)$&$(0, 1, 2, 1, 1, 1, 0)$&$(0,3/2)$\\ \hline
 $(0, 1, 2, 1, 2, 0, 0)$&$(0,2)$&$(0, 1, 2, 1, 2, 1, 0)$&$(0,2)$\\ \hline
 $(0, 1, 3, 1, 1, 0, 0)$&$(0,5/2)$&$(0, 1, 3, 1, 1, 1, 0)$&$(0,5/2)$\\ \hline
 $(0, 1, 3, 1, 2, 0, 0)$&$(0,2)\oplus2(0,3)\oplus(1/2,7/2)$&$(0, 1, 4, 1, 1, 0, 0)$&$(0,7/2)$\\ \hline
 $(0, 2, 1, 2, 1, 1, 0)$&$(0,0)$&$(1, 1, 0, 1, 1, 0, 0)$&$(0,1/2)$\\ \hline
 $(1, 1, 0, 1, 1, 1, 0)$&$(0,1/2)$&$(1, 1, 0, 2, 1, 0, 0)$&$(0,1)$\\ \hline
 $(1, 1, 0, 2, 1, 1, 0)$&$(0,0)\oplus(0,1)$&$(1, 1, 0, 2, 1, 2, 0)$&$(0,1)$\\ \hline
 $(1, 1, 1, 1, 1, 0, 0)$&$(0,1/2)$&$(1, 1, 1, 1, 1, 1, 0)$&$(0,1/2)$\\ \hline
 $(1, 1, 1, 2, 1, 0, 0)$&$(0,0)\oplus(0,1)$&$(1, 1, 1, 2, 1, 1, 0)$&$(0,0)\oplus(0,1)$\\ \hline
 $(1, 1, 2, 1, 1, 0, 0)$&$(0,3/2)$&$(1, 1, 2, 1, 1, 1, 0)$&$(0,3/2)$\\ \hline
 $(1, 1, 2, 1, 2, 0, 0)$&$(0,2)$&$(1, 1, 2, 2, 1, 0, 0)$&$(0,1)\oplus(0,2)$\\ \hline
 $(1, 1, 3, 1, 1, 0, 0)$&$(0,5/2)$&$(1, 2, 0, 2, 1, 0, 0)$&$(0,1)$\\ \hline
 $(1, 2, 0, 2, 1, 1, 0)$&$(0,0)\oplus(0,1)$&$(1, 2, 0, 2, 2, 0, 0)$&$(0,3/2)$\\ \hline
 $(1, 2, 0, 3, 1, 0, 0)$&$(0,3/2)$&$(1, 2, 1, 2, 1, 0, 0)$&$(0,0)\oplus(0,1)$\\ \hline
 $(0, 1, 0, 2, 1, 1, 2)$&$(0,1)$& &\\ \end{supertabular}}
}
\footnotesize{

\tablehead{Header of first column & Header of second column \\}
 \tablefirsthead{%
   \hline
   \multicolumn{1}{|c}{ ${\beta}$} &
   \multicolumn{1}{|c||}{$\oplus N^{\mathbf{d}}_{j_l,j_r}(j_l,j_r)$} &
   $\und{\beta}$ &
   \multicolumn{1}{c|}{$\oplus N^{\mathbf{d}}_{j_l,j_r}(j_l,j_r)$} \\
   \hline}
 \tablehead{%
   \hline
   \multicolumn{1}{|c}{ ${\beta}$} &
   \multicolumn{1}{|c||}{$\oplus N^{\mathbf{d}}_{j_l,j_r}(j_l,j_r)$} &
   $\und{\beta}$ &
   \multicolumn{1}{c|}{$\oplus N^{\mathbf{d}}_{j_l,j_r}(j_l,j_r)$} \\
   \hline}
 \tabletail{%
   \hline
   \multicolumn{4}{r}{\small\emph{continued on next page}}\\
   }
 \tablelasttail{\hline}
 \bottomcaption{Refined BPS invariants of 6d $\fn=7, E_7$ model up to total degree 8. For those that are not determined by the vanishing blowup equations, we mark them with ``?".}
\center{
 \begin{supertabular}{|c|p{4.3cm}||c|p{4.3cm}|}\label{tb:n7, E_7-BPS}$(0, 0, 1, 0, 0, 0, 0, 0, 0)$&$(0,0)$&$(0, 1, 1, 0, 0, 0, 0, 0, 0)$&$(0,1)$\\ \hline
 $(0, 1, 1, 0, 1, 0, 0, 0, 0)$&$(0,0)\oplus(0,1)$&$(0, 1, 1, 0, 2, 0, 0, 0, 0)$&$(0,1)$\\ \hline
 $(0, 1, 1, 0, 3, 0, 0, 0, 0)$&$?$&$(0, 1, 1, 0, 4, 0, 0, 0, 0)$&$?$\\ \hline
 $(0, 1, 1, 0, 5, 0, 0, 0, 0)$&$?$&$(0, 1, 1, 1, 1, 0, 0, 0, 0)$&$(0,0)\oplus(0,1)$\\ \hline
 $(0, 1, 1, 1, 2, 0, 0, 0, 0)$&$(0,0)\oplus(0,1)$&$(0, 1, 1, 1, 3, 0, 0, 0, 0)$&$?$\\ \hline
 $(0, 1, 1, 1, 4, 0, 0, 0, 0)$&$?$&$(0, 1, 1, 2, 2, 0, 0, 0, 0)$&$(0,1)$\\ \hline
 $(0, 1, 1, 2, 3, 0, 0, 0, 0)$&$?$&$(0, 2, 1, 0, 0, 0, 0, 0, 0)$&$(0,2)$\\ \hline
 $(0, 2, 1, 0, 1, 0, 0, 0, 0)$&$(0,1)\oplus(0,2)$&$(0, 2, 1, 0, 2, 0, 0, 0, 0)$&$(0,0)\oplus(0,1)\oplus(0,2)$\\ \hline
 $(0, 2, 1, 0, 3, 0, 0, 0, 0)$&$(0,1)\oplus(0,2)$&$(0, 2, 1, 0, 4, 0, 0, 0, 0)$&$?$\\ \hline
 $(0, 2, 1, 1, 1, 0, 0, 0, 0)$&$(0,1)\oplus(0,2)$&$(0, 2, 1, 1, 2, 0, 0, 0, 0)$&$(0,0)\oplus2(0,1)\oplus(0,2)$\\ \hline
 $(0, 2, 1, 1, 3, 0, 0, 0, 0)$&$(0,0)\oplus2(0,1)\oplus(0,2)$&$(0, 2, 1, 2, 2, 0, 0, 0, 0)$&$(0,0)\oplus(0,1)\oplus(0,2)$\\ \hline
 $(0, 2, 2, 0, 0, 0, 0, 0, 0)$&$(0,5/2)$&$(0, 2, 2, 0, 1, 0, 0, 0, 0)$&$(0,3/2)\oplus(0,5/2)$\\ \hline
 $(0, 2, 2, 0, 2, 0, 0, 0, 0)$&$(0,1/2)\oplus(0,3/2)\oplus(0,5/2)$&$(0, 2, 2, 0, 3, 0, 0, 0, 0)$&$?$\\ \hline
 $(0, 2, 2, 1, 1, 0, 0, 0, 0)$&$(0,3/2)\oplus(0,5/2)$&$(0, 2, 2, 1, 2, 0, 0, 0, 0)$&$(0,1/2)\oplus2(0,3/2)\oplus(0,5/2)$\\ \hline
 $(0, 3, 1, 0, 0, 0, 0, 0, 0)$&$(0,3)$&$(0, 3, 1, 0, 1, 0, 0, 0, 0)$&$(0,2)\oplus(0,3)$\\ \hline
 $(0, 3, 1, 0, 2, 0, 0, 0, 0)$&$(0,1)\oplus(0,2)\oplus(0,3)$&$(0, 3, 1, 0, 3, 0, 0, 0, 0)$&$(0,0)\oplus(0,1)\oplus(0,2)\oplus(0,3)$\\ \hline
 $(0, 3, 1, 1, 1, 0, 0, 0, 0)$&$(0,2)\oplus(0,3)$&$(0, 3, 1, 1, 2, 0, 0, 0, 0)$&$(0,1)\oplus2(0,2)\oplus(0,3)$\\ \hline
 $(0, 3, 2, 0, 0, 0, 0, 0, 0)$&$(0,5/2)\oplus(0,7/2)\oplus(1/2,4)$&$(0, 3, 2, 0, 1, 0, 0, 0, 0)$&$(0,3/2)\oplus3(0,5/2)\oplus2(0,7/2)\oplus(1/2,3)\oplus(1/2,4)$\\ \hline
 $(0, 3, 2, 0, 2, 0, 0, 0, 0)$&$(0,1/2)\oplus3(0,3/2)\oplus4(0,5/2)\oplus2(0,7/2)\oplus(1/2,2)\oplus(1/2,3)\oplus(1/2,4)$&$(0, 3, 2, 1, 1, 0, 0, 0, 0)$&$(0,3/2)\oplus3(0,5/2)\oplus2(0,7/2)\oplus(1/2,3)\oplus(1/2,4)$\\ \hline
 $(0, 3, 3, 0, 0, 0, 0, 0, 0)$&$(0,3)\oplus(1/2,9/2)$&$(0, 3, 3, 0, 1, 0, 0, 0, 0)$&$?$\\ \hline
 $(0, 4, 1, 0, 0, 0, 0, 0, 0)$&$(0,4)$&$(0, 4, 1, 0, 1, 0, 0, 0, 0)$&$(0,3)\oplus(0,4)$\\ \hline
 $(0, 4, 1, 0, 2, 0, 0, 0, 0)$&$(0,2)\oplus(0,3)\oplus(0,4)$&$(0, 4, 1, 1, 1, 0, 0, 0, 0)$&$(0,3)\oplus(0,4)$\\ \hline
 $(0, 4, 2, 0, 0, 0, 0, 0, 0)$&$(0,5/2)\oplus(0,7/2)\oplus2(0,9/2)\oplus(1/2,4)\oplus(1/2,5)\oplus(1,11/2)$&$(0, 4, 2, 0, 1, 0, 0, 0, 0)$&$(0,3/2)\oplus3(0,5/2)\oplus5(0,7/2)\oplus3(0,9/2)\oplus(1/2,3)\oplus3(1/2,4)\oplus2(1/2,5)\oplus(1,9/2)\oplus(1,11/2)$\\ \hline
 $(0, 4, 3, 0, 0, 0, 0, 0, 0)$&$(0,2)\oplus(0,3)\oplus2(0,4)\oplus(0,5)\oplus(0,6)\oplus(1/2,7/2)\oplus2(1/2,9/2)\oplus2(1/2,11/2)\oplus(1,5)\oplus(1,6)\oplus(3/2,13/2)$&$(0, 5, 1, 0, 0, 0, 0, 0, 0)$&$(0,5)$\\ \hline
 $(0, 5, 1, 0, 1, 0, 0, 0, 0)$&$(0,4)\oplus(0,5)$&$(0, 5, 2, 0, 0, 0, 0, 0, 0)$&$(0,5/2)\oplus(0,7/2)\oplus2(0,9/2)\oplus2(0,11/2)\oplus(1/2,4)\oplus(1/2,5)\oplus2(1/2,6)\oplus(1,11/2)\oplus(1,13/2)\oplus(3/2,7)$\\ \hline
 $(0, 6, 1, 0, 0, 0, 0, 0, 0)$&$(0,6)$&$(1, 0, 1, 0, 0, 0, 0, 0, 0)$&$(0,1)$\\ \hline
 $(1, 0, 1, 0, 0, 1, 0, 0, 0)$&$(0,0)\oplus(0,1)$&$(1, 0, 1, 0, 0, 1, 1, 0, 0)$&$(0,1/2)$\\ \hline
 $(1, 0, 1, 0, 0, 1, 1, 1, 0)$&$(0,1/2)$&$(1, 0, 1, 0, 0, 1, 2, 1, 0)$&$?$\\ \hline
 $(1, 0, 1, 0, 0, 2, 0, 0, 0)$&$(0,1)$&$(1, 0, 1, 0, 0, 2, 1, 0, 0)$&$(0,1/2)$\\ \hline
 $(1, 0, 1, 0, 0, 2, 1, 1, 0)$&$(0,1/2)$&$(1, 0, 1, 0, 0, 2, 2, 1, 0)$&$?$\\ \hline
 $(1, 0, 1, 0, 0, 3, 0, 0, 0)$&$(0,2)$&$(1, 0, 1, 0, 0, 3, 1, 0, 0)$&$(0,3/2)$\\ \hline
 $(1, 0, 1, 0, 0, 3, 1, 1, 0)$&$?$&$(1, 0, 1, 0, 0, 4, 0, 0, 0)$&$(0,3)$\\ \hline
 $(1, 0, 1, 0, 0, 4, 1, 0, 0)$&$(0,5/2)$&$(1, 0, 1, 0, 0, 5, 0, 0, 0)$&$(0,4)$\\ \hline
 $(1, 1, 1, 0, 0, 0, 0, 0, 0)$&$(0,0)\oplus(0,1)$&$(1, 1, 1, 0, 0, 1, 0, 0, 0)$&$(0,0)\oplus(0,1)$\\ \hline
 $(1, 1, 1, 0, 0, 1, 1, 0, 0)$&$(0,1/2)$&$(1, 1, 1, 0, 0, 1, 1, 1, 0)$&$(0,1/2)$\\ \hline
 $(1, 1, 1, 0, 0, 1, 2, 1, 0)$&$(0,0)\oplus(0,1)$&$(1, 1, 1, 0, 1, 0, 0, 0, 0)$&$(0,0)\oplus(0,1)$\\ \hline
 $(1, 1, 1, 0, 1, 1, 0, 0, 0)$&$(0,0)\oplus(0,1)$&$(1, 1, 1, 0, 1, 1, 1, 0, 0)$&$(0,1/2)$\\ \hline
 $(1, 1, 1, 0, 1, 1, 1, 1, 0)$&$(0,1/2)$&$(1, 1, 1, 1, 1, 0, 0, 0, 0)$&$(0,0)\oplus(0,1)$\\ \hline
 $(1, 1, 1, 1, 1, 1, 0, 0, 0)$&$(0,0)\oplus(0,1)$&$(1, 1, 1, 1, 1, 1, 1, 0, 0)$&$(0,1/2)$\\ \hline
 $(1, 2, 1, 0, 0, 0, 0, 0, 0)$&$(0,1)\oplus(0,2)$&$(1, 2, 1, 0, 0, 1, 0, 0, 0)$&$(0,1)\oplus(0,2)$\\ \hline
 $(1, 2, 1, 0, 0, 1, 1, 0, 0)$&$(0,3/2)$&$(1, 2, 1, 0, 0, 1, 1, 1, 0)$&$(0,3/2)$\\ \hline
 $(1, 2, 1, 0, 1, 0, 0, 0, 0)$&$(0,0)\oplus2(0,1)\oplus(0,2)$&$(1, 2, 1, 0, 1, 1, 0, 0, 0)$&$(0,0)\oplus2(0,1)\oplus(0,2)$\\ \hline
 $(1, 2, 1, 0, 1, 1, 1, 0, 0)$&$(0,1/2)\oplus(0,3/2)$&$(1, 2, 1, 0, 2, 0, 0, 0, 0)$&$(0,0)\oplus2(0,1)\oplus(0,2)$\\ \hline
 $(1, 2, 1, 0, 2, 1, 0, 0, 0)$&$(0,0)\oplus2(0,1)\oplus(0,2)$&$(1, 2, 1, 0, 3, 0, 0, 0, 0)$&$(0,1)\oplus(0,2)$\\ \hline
 $(1, 2, 1, 1, 1, 0, 0, 0, 0)$&$(0,0)\oplus2(0,1)\oplus(0,2)$&$(1, 2, 1, 1, 1, 1, 0, 0, 0)$&$(0,0)\oplus2(0,1)\oplus(0,2)$\\ \hline
 $(1, 2, 1, 1, 2, 0, 0, 0, 0)$&$2(0,0)\oplus3(0,1)\oplus(0,2)$&$(1, 2, 2, 0, 0, 0, 0, 0, 0)$&$(0,3/2)\oplus(0,5/2)$\\ \hline
 $(1, 2, 2, 0, 0, 1, 0, 0, 0)$&$(0,3/2)\oplus(0,5/2)$&$(1, 2, 2, 0, 0, 1, 1, 0, 0)$&$(0,2)$\\ \hline
 $(1, 2, 2, 0, 1, 0, 0, 0, 0)$&$(0,1/2)\oplus2(0,3/2)\oplus(0,5/2)$&$(1, 2, 2, 0, 1, 1, 0, 0, 0)$&$(0,1/2)\oplus2(0,3/2)\oplus(0,5/2)$\\ \hline
 $(1, 2, 2, 0, 2, 0, 0, 0, 0)$&$2(0,1/2)\oplus2(0,3/2)\oplus(0,5/2)$&$(1, 2, 2, 1, 1, 0, 0, 0, 0)$&$(0,1/2)\oplus2(0,3/2)\oplus(0,5/2)$\\ \hline
 $(1, 3, 1, 0, 0, 0, 0, 0, 0)$&$(0,2)\oplus(0,3)$&$(1, 3, 1, 0, 0, 1, 0, 0, 0)$&$(0,2)\oplus(0,3)$\\ \hline
 $(1, 3, 1, 0, 0, 1, 1, 0, 0)$&$(0,5/2)$&$(1, 3, 1, 0, 1, 0, 0, 0, 0)$&$(0,1)\oplus2(0,2)\oplus(0,3)$\\ \hline
 $(1, 3, 1, 0, 1, 1, 0, 0, 0)$&$(0,1)\oplus2(0,2)\oplus(0,3)$&$(1, 3, 1, 0, 2, 0, 0, 0, 0)$&$(0,0)\oplus2(0,1)\oplus2(0,2)\oplus(0,3)$\\ \hline
 $(1, 3, 1, 1, 1, 0, 0, 0, 0)$&$(0,1)\oplus2(0,2)\oplus(0,3)$&$(1, 3, 2, 0, 0, 0, 0, 0, 0)$&$(0,3/2)\oplus3(0,5/2)\oplus2(0,7/2)\oplus(1/2,3)\oplus(1/2,4)$\\ \hline
 $(1, 3, 2, 0, 0, 1, 0, 0, 0)$&$(0,3/2)\oplus3(0,5/2)\oplus2(0,7/2)\oplus(1/2,3)\oplus(1/2,4)$&$(1, 3, 2, 0, 1, 0, 0, 0, 0)$&$(0,1/2)\oplus5(0,3/2)\oplus7(0,5/2)\oplus3(0,7/2)\oplus(1/2,2)\oplus2(1/2,3)\oplus(1/2,4)$\\ \hline
 $(1, 3, 3, 0, 0, 0, 0, 0, 0)$&$(0,2)\oplus2(0,3)\oplus(0,4)\oplus(1/2,7/2)\oplus(1/2,9/2)$&$(1, 4, 1, 0, 0, 0, 0, 0, 0)$&$(0,3)\oplus(0,4)$\\ \hline
 $(1, 4, 1, 0, 0, 1, 0, 0, 0)$&$(0,3)\oplus(0,4)$&$(1, 4, 1, 0, 1, 0, 0, 0, 0)$&$(0,2)\oplus2(0,3)\oplus(0,4)$\\ \hline
 $(1, 4, 2, 0, 0, 0, 0, 0, 0)$&$(0,3/2)\oplus3(0,5/2)\oplus5(0,7/2)\oplus3(0,9/2)\oplus(1/2,3)\oplus3(1/2,4)\oplus2(1/2,5)\oplus(1,9/2)\oplus(1,11/2)$&$(1, 5, 1, 0, 0, 0, 0, 0, 0)$&$(0,4)\oplus(0,5)$\\ \hline
 $(2, 0, 1, 0, 0, 0, 0, 0, 0)$&$(0,2)$&$(2, 0, 1, 0, 0, 1, 0, 0, 0)$&$(0,1)\oplus(0,2)$\\ \hline
 $(2, 0, 1, 0, 0, 1, 1, 0, 0)$&$(0,3/2)$&$(2, 0, 1, 0, 0, 1, 1, 1, 0)$&$(0,3/2)$\\ \hline
 $(2, 0, 1, 0, 0, 1, 2, 1, 0)$&$?$&$(2, 0, 1, 0, 0, 2, 0, 0, 0)$&$(0,0)\oplus(0,1)\oplus(0,2)$\\ \hline
 $(2, 0, 1, 0, 0, 2, 1, 0, 0)$&$(0,1/2)\oplus(0,3/2)$&$(2, 0, 1, 0, 0, 2, 1, 1, 0)$&$(0,1/2)\oplus(0,3/2)$\\ \hline
 $(2, 0, 1, 0, 0, 3, 0, 0, 0)$&$(0,1)\oplus(0,2)$&$(2, 0, 1, 0, 0, 3, 1, 0, 0)$&$(0,1/2)\oplus(0,3/2)$\\ \hline
 $(2, 0, 1, 0, 0, 4, 0, 0, 0)$&$(0,2)\oplus(0,3)$&$(2, 0, 2, 0, 0, 0, 0, 0, 0)$&$(0,5/2)$\\ \hline
 $(2, 0, 2, 0, 0, 1, 0, 0, 0)$&$(0,3/2)\oplus(0,5/2)$&$(2, 0, 2, 0, 0, 1, 1, 0, 0)$&$(0,2)$\\ \hline
 $(2, 0, 2, 0, 0, 1, 1, 1, 0)$&$(0,2)$&$(2, 0, 2, 0, 0, 2, 0, 0, 0)$&$(0,1/2)\oplus(0,3/2)\oplus(0,5/2)$\\ \hline
 $(2, 0, 2, 0, 0, 2, 1, 0, 0)$&$(0,1)\oplus(0,2)$&$(2, 0, 2, 0, 0, 3, 0, 0, 0)$&$(0,1/2)\oplus(0,3/2)\oplus(0,5/2)$\\ \hline
 $(2, 1, 1, 0, 0, 0, 0, 0, 0)$&$(0,1)\oplus(0,2)$&$(2, 1, 1, 0, 0, 1, 0, 0, 0)$&$(0,0)\oplus2(0,1)\oplus(0,2)$\\ \hline
 $(2, 1, 1, 0, 0, 1, 1, 0, 0)$&$(0,1/2)\oplus(0,3/2)$&$(2, 1, 1, 0, 0, 1, 1, 1, 0)$&$(0,1/2)\oplus(0,3/2)$\\ \hline
 $(2, 1, 1, 0, 0, 2, 0, 0, 0)$&$(0,0)\oplus2(0,1)\oplus(0,2)$&$(2, 1, 1, 0, 0, 2, 1, 0, 0)$&$2(0,1/2)\oplus(0,3/2)$\\ \hline
 $(2, 1, 1, 0, 0, 3, 0, 0, 0)$&$(0,1)\oplus(0,2)$&$(2, 1, 1, 0, 1, 0, 0, 0, 0)$&$(0,1)\oplus(0,2)$\\ \hline
 $(2, 1, 1, 0, 1, 1, 0, 0, 0)$&$(0,0)\oplus2(0,1)\oplus(0,2)$&$(2, 1, 1, 0, 1, 1, 1, 0, 0)$&$(0,1/2)\oplus(0,3/2)$\\ \hline
 $(2, 1, 1, 0, 1, 2, 0, 0, 0)$&$(0,0)\oplus2(0,1)\oplus(0,2)$&$(2, 1, 1, 1, 1, 0, 0, 0, 0)$&$(0,1)\oplus(0,2)$\\ \hline
 $(2, 1, 1, 1, 1, 1, 0, 0, 0)$&$(0,0)\oplus2(0,1)\oplus(0,2)$&$(2, 1, 2, 0, 0, 0, 0, 0, 0)$&$(0,3/2)\oplus(0,5/2)$\\ \hline
 $(2, 1, 2, 0, 0, 1, 0, 0, 0)$&$(0,1/2)\oplus2(0,3/2)\oplus(0,5/2)$&$(2, 1, 2, 0, 0, 1, 1, 0, 0)$&$(0,1)\oplus(0,2)$\\ \hline
 $(2, 1, 2, 0, 0, 2, 0, 0, 0)$&$2(0,1/2)\oplus2(0,3/2)\oplus(0,5/2)$&$(2, 1, 2, 0, 1, 0, 0, 0, 0)$&$(0,3/2)\oplus(0,5/2)$\\ \hline
 $(2, 1, 2, 0, 1, 1, 0, 0, 0)$&$(0,1/2)\oplus2(0,3/2)\oplus(0,5/2)$&$(2, 1, 2, 1, 1, 0, 0, 0, 0)$&$(0,3/2)\oplus(0,5/2)$\\ \hline
 $(2, 2, 1, 0, 0, 0, 0, 0, 0)$&$(0,0)\oplus(0,1)\oplus(0,2)$&$(2, 2, 1, 0, 0, 1, 0, 0, 0)$&$(0,0)\oplus2(0,1)\oplus(0,2)$\\ \hline
 $(2, 2, 1, 0, 0, 1, 1, 0, 0)$&$(0,1/2)\oplus(0,3/2)$&$(2, 2, 1, 0, 0, 2, 0, 0, 0)$&$(0,0)\oplus(0,1)\oplus(0,2)$\\ \hline
 $(2, 2, 1, 0, 1, 0, 0, 0, 0)$&$(0,0)\oplus2(0,1)\oplus(0,2)$&$(2, 2, 1, 0, 1, 1, 0, 0, 0)$&$2(0,0)\oplus3(0,1)\oplus(0,2)$\\ \hline
 $(2, 2, 1, 0, 2, 0, 0, 0, 0)$&$(0,0)\oplus(0,1)\oplus(0,2)$&$(2, 2, 1, 1, 1, 0, 0, 0, 0)$&$(0,0)\oplus2(0,1)\oplus(0,2)$\\ \hline
 $(2, 2, 2, 0, 0, 0, 0, 0, 0)$&$2(0,1/2)\oplus2(0,3/2)\oplus2(0,5/2)\oplus(0,7/2)$&$(2, 2, 2, 0, 0, 1, 0, 0, 0)$&$3(0,1/2)\oplus4(0,3/2)\oplus3(0,5/2)\oplus(0,7/2)$\\ \hline
 $(2, 2, 2, 0, 1, 0, 0, 0, 0)$&$3(0,1/2)\oplus4(0,3/2)\oplus3(0,5/2)\oplus(0,7/2)$&$(2, 2, 3, 0, 0, 0, 0, 0, 0)$&$(0,0)\oplus(0,1)\oplus(0,2)\oplus(0,3)\oplus(0,4)$\\ \hline
 $(2, 3, 1, 0, 0, 0, 0, 0, 0)$&$(0,1)\oplus(0,2)\oplus(0,3)$&$(2, 3, 1, 0, 0, 1, 0, 0, 0)$&$(0,1)\oplus2(0,2)\oplus(0,3)$\\ \hline
 $(2, 3, 1, 0, 1, 0, 0, 0, 0)$&$(0,0)\oplus2(0,1)\oplus2(0,2)\oplus(0,3)$&$(2, 3, 2, 0, 0, 0, 0, 0, 0)$&$2(0,1/2)\oplus4(0,3/2)\oplus5(0,5/2)\oplus3(0,7/2)\oplus(0,9/2)\oplus(1/2,2)\oplus(1/2,3)\oplus(1/2,4)$\\ \hline
 $(2, 4, 1, 0, 0, 0, 0, 0, 0)$&$(0,2)\oplus(0,3)\oplus(0,4)$&$(3, 0, 1, 0, 0, 0, 0, 0, 0)$&$(0,3)$\\ \hline
 $(3, 0, 1, 0, 0, 1, 0, 0, 0)$&$(0,2)\oplus(0,3)$&$(3, 0, 1, 0, 0, 1, 1, 0, 0)$&$(0,5/2)$\\ \hline
 $(3, 0, 1, 0, 0, 1, 1, 1, 0)$&$(0,5/2)$&$(3, 0, 1, 0, 0, 2, 0, 0, 0)$&$(0,1)\oplus(0,2)\oplus(0,3)$\\ \hline
 $(3, 0, 1, 0, 0, 2, 1, 0, 0)$&$(0,3/2)\oplus(0,5/2)$&$(3, 0, 1, 0, 0, 3, 0, 0, 0)$&$(0,0)\oplus(0,1)\oplus(0,2)\oplus(0,3)$\\ \hline
 $(3, 0, 2, 0, 0, 0, 0, 0, 0)$&$(0,5/2)\oplus(0,7/2)\oplus(1/2,4)$&$(3, 0, 2, 0, 0, 1, 0, 0, 0)$&$(0,3/2)\oplus3(0,5/2)\oplus2(0,7/2)\oplus(1/2,3)\oplus(1/2,4)$\\ \hline
 $(3, 0, 2, 0, 0, 1, 1, 0, 0)$&$(0,2)\oplus2(0,3)\oplus(1/2,7/2)$&$(3, 0, 2, 0, 0, 2, 0, 0, 0)$&$(0,1/2)\oplus3(0,3/2)\oplus4(0,5/2)\oplus2(0,7/2)\oplus(1/2,2)\oplus(1/2,3)\oplus(1/2,4)$\\ \hline
 $(3, 0, 3, 0, 0, 0, 0, 0, 0)$&$(0,3)\oplus(1/2,9/2)$&$(3, 0, 3, 0, 0, 1, 0, 0, 0)$&$(0,2)\oplus2(0,3)\oplus(0,4)\oplus(1/2,7/2)\oplus(1/2,9/2)$\\ \hline
 $(3, 1, 1, 0, 0, 0, 0, 0, 0)$&$(0,2)\oplus(0,3)$&$(3, 1, 1, 0, 0, 1, 0, 0, 0)$&$(0,1)\oplus2(0,2)\oplus(0,3)$\\ \hline
 $(3, 1, 1, 0, 0, 1, 1, 0, 0)$&$(0,3/2)\oplus(0,5/2)$&$(3, 1, 1, 0, 0, 2, 0, 0, 0)$&$(0,0)\oplus2(0,1)\oplus2(0,2)\oplus(0,3)$\\ \hline
 $(3, 1, 1, 0, 1, 0, 0, 0, 0)$&$(0,2)\oplus(0,3)$&$(3, 1, 1, 0, 1, 1, 0, 0, 0)$&$(0,1)\oplus2(0,2)\oplus(0,3)$\\ \hline
 $(3, 1, 1, 1, 1, 0, 0, 0, 0)$&$(0,2)\oplus(0,3)$&$(3, 1, 2, 0, 0, 0, 0, 0, 0)$&$(0,3/2)\oplus3(0,5/2)\oplus2(0,7/2)\oplus(1/2,3)\oplus(1/2,4)$\\ \hline
 $(3, 1, 2, 0, 0, 1, 0, 0, 0)$&$(0,1/2)\oplus5(0,3/2)\oplus7(0,5/2)\oplus3(0,7/2)\oplus(1/2,2)\oplus2(1/2,3)\oplus(1/2,4)$&$(3, 1, 2, 0, 1, 0, 0, 0, 0)$&$(0,3/2)\oplus3(0,5/2)\oplus2(0,7/2)\oplus(1/2,3)\oplus(1/2,4)$\\ \hline
 $(3, 1, 3, 0, 0, 0, 0, 0, 0)$&$(0,2)\oplus2(0,3)\oplus(0,4)\oplus(1/2,7/2)\oplus(1/2,9/2)$&$(3, 2, 1, 0, 0, 0, 0, 0, 0)$&$(0,1)\oplus(0,2)\oplus(0,3)$\\ \hline
 $(3, 2, 1, 0, 0, 1, 0, 0, 0)$&$(0,0)\oplus2(0,1)\oplus2(0,2)\oplus(0,3)$&$(3, 2, 1, 0, 1, 0, 0, 0, 0)$&$(0,1)\oplus2(0,2)\oplus(0,3)$\\ \hline
 $(3, 2, 2, 0, 0, 0, 0, 0, 0)$&$2(0,1/2)\oplus4(0,3/2)\oplus5(0,5/2)\oplus3(0,7/2)\oplus(0,9/2)\oplus(1/2,2)\oplus(1/2,3)\oplus(1/2,4)$&$(3, 3, 1, 0, 0, 0, 0, 0, 0)$&$(0,0)\oplus(0,1)\oplus(0,2)\oplus(0,3)$\\ \hline
 $(4, 0, 1, 0, 0, 0, 0, 0, 0)$&$(0,4)$&$(4, 0, 1, 0, 0, 1, 0, 0, 0)$&$(0,3)\oplus(0,4)$\\ \hline
 $(4, 0, 1, 0, 0, 1, 1, 0, 0)$&$(0,7/2)$&$(4, 0, 1, 0, 0, 2, 0, 0, 0)$&$(0,2)\oplus(0,3)\oplus(0,4)$\\ \hline
 $(4, 0, 2, 0, 0, 0, 0, 0, 0)$&$(0,5/2)\oplus(0,7/2)\oplus2(0,9/2)\oplus(1/2,4)\oplus(1/2,5)\oplus(1,11/2)$&$(4, 0, 2, 0, 0, 1, 0, 0, 0)$&$(0,3/2)\oplus3(0,5/2)\oplus5(0,7/2)\oplus3(0,9/2)\oplus(1/2,3)\oplus3(1/2,4)\oplus2(1/2,5)\oplus(1,9/2)\oplus(1,11/2)$\\ \hline
 $(4, 0, 3, 0, 0, 0, 0, 0, 0)$&$(0,2)\oplus(0,3)\oplus2(0,4)\oplus(0,5)\oplus(0,6)\oplus(1/2,7/2)\oplus2(1/2,9/2)\oplus2(1/2,11/2)\oplus(1,5)\oplus(1,6)\oplus(3/2,13/2)$&$(4, 1, 1, 0, 0, 0, 0, 0, 0)$&$(0,3)\oplus(0,4)$\\ \hline
 $(4, 1, 1, 0, 0, 1, 0, 0, 0)$&$(0,2)\oplus2(0,3)\oplus(0,4)$&$(4, 1, 1, 0, 1, 0, 0, 0, 0)$&$(0,3)\oplus(0,4)$\\ \hline
 $(4, 1, 2, 0, 0, 0, 0, 0, 0)$&$(0,3/2)\oplus3(0,5/2)\oplus5(0,7/2)\oplus3(0,9/2)\oplus(1/2,3)\oplus3(1/2,4)\oplus2(1/2,5)\oplus(1,9/2)\oplus(1,11/2)$&$(4, 2, 1, 0, 0, 0, 0, 0, 0)$&$(0,2)\oplus(0,3)\oplus(0,4)$\\ \hline
 $(5, 0, 1, 0, 0, 0, 0, 0, 0)$&$(0,5)$&$(5, 0, 1, 0, 0, 1, 0, 0, 0)$&$(0,4)\oplus(0,5)$\\ \hline
 $(5, 0, 2, 0, 0, 0, 0, 0, 0)$&$(0,5/2)\oplus(0,7/2)\oplus2(0,9/2)\oplus2(0,11/2)\oplus(1/2,4)\oplus(1/2,5)\oplus2(1/2,6)\oplus(1,11/2)\oplus(1,13/2)\oplus(3/2,7)$&$(5, 1, 1, 0, 0, 0, 0, 0, 0)$&$(0,4)\oplus(0,5)$\\ \hline
 $(6, 0, 1, 0, 0, 0, 0, 0, 0)$&$(0,6)$&$(1, 0, 1, 0, 0, 1, 1, 0, 1)$&$(0,0)\oplus(0,1)$\\ \hline
 $(1, 0, 1, 0, 0, 1, 1, 1, 1)$&$(0,0)\oplus(0,1)$&$(1, 0, 1, 0, 0, 1, 2, 1, 1)$&$(0,1/2)$\\ \hline
 $(1, 0, 1, 0, 0, 2, 1, 0, 1)$&$(0,0)\oplus(0,1)$&$(1, 0, 1, 0, 0, 2, 1, 1, 1)$&$(0,0)\oplus(0,1)$\\ \hline
 $(1, 0, 1, 0, 0, 2, 2, 0, 1)$&$(0,1/2)$&$(1, 0, 1, 0, 0, 3, 1, 0, 1)$&$(0,1)\oplus(0,2)$\\ \hline
 $(1, 1, 1, 0, 0, 1, 1, 0, 1)$&$(0,0)\oplus(0,1)$&$(1, 1, 1, 0, 0, 1, 1, 1, 1)$&$(0,0)\oplus(0,1)$\\ \hline
 $(1, 1, 1, 0, 1, 1, 1, 0, 1)$&$(0,0)\oplus(0,1)$&$(1, 2, 1, 0, 0, 1, 1, 0, 1)$&$(0,1)\oplus(0,2)$\\ \hline
 $(2, 0, 1, 0, 0, 1, 1, 0, 1)$&$(0,1)\oplus(0,2)$&$(2, 0, 1, 0, 0, 1, 1, 1, 1)$&$(0,1)\oplus(0,2)$\\ \hline
 $(2, 0, 1, 0, 0, 2, 1, 0, 1)$&$(0,0)\oplus2(0,1)\oplus(0,2)$&$(2, 0, 2, 0, 0, 1, 1, 0, 1)$&$(0,3/2)\oplus(0,5/2)$\\ \hline
 $(2, 1, 1, 0, 0, 1, 1, 0, 1)$&$(0,0)\oplus2(0,1)\oplus(0,2)$&$(3, 0, 1, 0, 0, 1, 1, 0, 1)$&$(0,2)\oplus(0,3)$\\ \hline
  \end{supertabular}}
}


\bibliographystyle{JHEP}
\bibliography{bibupdated}

\end{document}